%% file: main.tex
\newcommand{\ispreprint}{}
\ifdefined\ispreprint
  \documentclass[manuscript,screen]{acmart}
\else
  \documentclass[manuscript,screen,review]{acmart}
\fi

\usepackage{sosy-paper}
\usepackage{fancyvrb}
\usepackage{colortbl}
\usepackage{xfp}
\usepackage[xcolor]{changebar}
\usepackage{tcolorbox}
\usepackage{bm}

\usetikzlibrary{external}
\tikzexternaldisable 

\newcommand{\disablenotes}{}
\ifdefined\disablenotes
  \usepackage[disable]{todonotes}
  \newcommand{\inlinetodo}[1]{}  
  \nochangebars
\else
  \usepackage[textsize=scriptsize]{todonotes}
  \newcommand{\inlinetodo}[1]{\textcolor{red}{\bfseries #1}}
\fi

\cbcolor{green}

\newcommand{\btortwo}{\textsc{Btor2}\xspace}
\newcommand{\aiger}{\textsc{Aiger}\xspace}

\definetool{\abc}{ABC}
\definetool{\avr}{AVR}
\definetool{\btortoaiger}{Btor2Aiger}
\definetool{\btorsim}{BtorSim}
\definetool{\coveriteam}{CoVeriTeam}
\definetool{\cpv}{CPV}
\definetool{\kratostwo}{Kratos2}
\definetool{\pono}{Pono}
\definetool{\ricthree}{rIC3}
\definetool{\symwitch}{Symbiotic-Witch}
\definetool{\transver}{TransVer}
\definetool{\transvercpv}{TransVer-CPV}
\definetool{\witch}{Witch}

\newcommand{\boldnum}[1]{\textbf{\tablenum[table-format=5]{#1}}}
\newcommand{\best}[1]{\cellcolor{yellow!40}{\boldnum{#1}}}
\newcommand{\second}[1]{\cellcolor{gray!20}{\boldnum{#1}}}
\newcommand{\wrong}[1]{\cellcolor{red!20}{~\tablenum[table-format=4]{#1}}}
\newcommand{\greencmark}{{\color{green}{\cmark}}\xspace}

\newcommand{\shortcommit}[1]{%
  \StrLeft{#1}{7}%
}
\newcommand{\commiturl}[2]{\href{#1/#2}{\texttt{\shortcommit{#2}}}}

\newlength{\acmparindent}
\newlength{\acmparskip}
\lstdefinestyle{cstyle}{
  language=C,
  basicstyle=\ttfamily\small,
  keywordstyle=\color{blue}\bfseries,
  commentstyle=\color{green!50!black}\itshape,
  stringstyle=\color{red!70!black},
  numberstyle=\tiny\color{darkgray},
  numbers=left,
  stepnumber=1,
  numbersep=8pt,
  showstringspaces=false,
  tabsize=2,
}

\newcommand{\mypaperkeywords}{
    C programs,
    Sequential circuits,
    \btortwo{},
    Intermediate representation,
    Verification witnesses
}

\setcopyright{cc}
\setcctype[4.0]{by}
\acmYear{2026}
\acmMonth{8}

\title{Circuit-Based Program Verification}
\subtitle{Sequential Circuits as an Intermediate Representation for Verifying C Programs}
\ifdefined\ispreprint
  \thanks{A preliminary version of this work was published in the proceedings of TACAS~2024~\cite{CPV-SVCOMP24}.}
\fi

\begin{document}

\author{Po-Chun Chien}
\email{po-chun.chien@sosy.ifi.lmu.de}
\orcid{0000-0001-5139-5178}
\affiliation{
  \institution{LMU Munich}
  \country{Germany}
}

\author{Nian-Ze Lee}
\email{nzlee@ntu.edu.tw}
\orcid{0000-0002-8096-5595}
\affiliation{
  \institution{National Taiwan University}
  \country{Taiwan}
}
\additionalaffiliation{
  \institution{LMU Munich}
  \country{Germany}
}

\author{Armin Biere}
\email{biere@cs.uni-freiburg.de}
\orcid{0000-0001-7170-9242}
\affiliation{
  \institution{University of Freiburg}
  \country{Germany}
}

\author{Dirk Beyer}
\email{dirk.beyer@lmu.de}
\orcid{0000-0003-4832-7662}
\affiliation{
  \institution{LMU Munich}
  \country{Germany}
}

\input{eval-results/tex/data-commands.tex}

\begin{abstract}
    \input{abstract}
\end{abstract}



\maketitle

\input{introduction}
\input{related-work}

\input{background}
\input{pipeline}
\input{encoding}
\input{reachsafety}
\input{termination}
\input{implementation}

\input{evaluation}
\input{conclusion}

\subsection*{Data-Availability Statement}
\cpv is an open-source project maintained on \href{https://gitlab.com/sosy-lab/software/cpv}{GitLab},
with stable releases archived on Zenodo~\cite{CPV-latest}.
The benchmark set~\cite{SVCOMP26-SVBENCHMARKS-artifact},
verifiers~\cite{CPV-1.1,CPAchecker-4.2.2,ESBMC-SVCOMP26-archive,Symbiotic-SVCOMP26-archive,UAutomizer-SVCOMP26-archive,Kratos2-CAV23-artifact},
and validators~\cite{CPAchecker-4.2.2,Witch-SVCOMP26-archive,SWitch-SVCOMP25-archive}
used in the evaluation are available on Zenodo as well.

\subsection*{Funding Statement}
This project was funded in part by the Deutsche Forschungsgemeinschaft~(DFG)
under grants \href{http://gepris.dfg.de/gepris/projekt/378803395}{378803395}~(ConVeY)
and \href{http://gepris.dfg.de/gepris/projekt/536040111}{536040111}~(Bridge),
the LMU Postdoc Support Fund,
and a fellowship from the German Academic Exchange Service~(DAAD).

\interlinepenalty=10000
\bibliography{bibs/dbeyer,bibs/artifacts,bibs/sw,bibs/svcomp,bibs/svcomp-artifacts,bibs/websites,bibs/cpv,bibs/biere}





\end{document}

%% file: eval-results/tex/data-commands.tex
\input{eval-results/tex/data-commands.translation}

\input{eval-results/tex/data-commands.encoding}
\input{eval-results/tex/data-commands.gates}
\input{eval-results/tex/data-commands.scorr}

\input{eval-results/tex/data-commands.l2s-simp}
\input{eval-results/tex/data-commands.transver}

\input{eval-results/tex/data-commands.wit-val}

\input{eval-results/tex/data-commands.svcomp-reach}
\input{eval-results/tex/data-commands.svcomp-term}

%% file: eval-results/tex/data-commands.translation.tex
\providecommand\StoreBenchExecResult[7]{\expandafter\newcommand\csname#1#2#3#4#5#6\endcsname{#7}}%
\StoreBenchExecResult{CpvTranslate}{FuncReachSafetyArrays}{Status}{All}{}{Score}{0}%
\StoreBenchExecResult{CpvTranslate}{FuncReachSafetyArrays}{Status}{All}{}{Count}{440}%
\StoreBenchExecResult{CpvTranslate}{FuncReachSafetyArrays}{Status}{Correct}{}{Count}{0}%
\StoreBenchExecResult{CpvTranslate}{FuncReachSafetyArrays}{Status}{Correct}{True}{Count}{0}%
\StoreBenchExecResult{CpvTranslate}{FuncReachSafetyArrays}{Status}{Correct}{False}{Count}{0}%
\StoreBenchExecResult{CpvTranslate}{FuncReachSafetyArrays}{Status}{Wrong}{}{Count}{0}%
\StoreBenchExecResult{CpvTranslate}{FuncReachSafetyArrays}{Status}{Wrong}{True}{Count}{0}%
\StoreBenchExecResult{CpvTranslate}{FuncReachSafetyArrays}{Status}{Wrong}{False}{Count}{0}%
\StoreBenchExecResult{CpvTranslate}{FuncReachSafetyArrays}{KII}{All}{}{Sum}{438}%
\StoreBenchExecResult{CpvTranslate}{FuncReachSafetyArrays}{KII}{All}{}{Min}{0}%
\StoreBenchExecResult{CpvTranslate}{FuncReachSafetyArrays}{KII}{All}{}{Max}{1}%
\StoreBenchExecResult{CpvTranslate}{FuncReachSafetyArrays}{KII}{All}{}{Avg}{0.9954545454545454545454545455}%
\StoreBenchExecResult{CpvTranslate}{FuncReachSafetyArrays}{KII}{All}{}{Median}{1}%
\StoreBenchExecResult{CpvTranslate}{FuncReachSafetyArrays}{KII}{All}{}{Stdev}{0.06726658448613064571401974056}%
\StoreBenchExecResult{CpvTranslate}{FuncReachSafetyArrays}{BtorII}{All}{}{Sum}{421}%
\StoreBenchExecResult{CpvTranslate}{FuncReachSafetyArrays}{BtorII}{All}{}{Min}{0}%
\StoreBenchExecResult{CpvTranslate}{FuncReachSafetyArrays}{BtorII}{All}{}{Max}{1}%
\StoreBenchExecResult{CpvTranslate}{FuncReachSafetyArrays}{BtorII}{All}{}{Avg}{0.9568181818181818181818181818}%
\StoreBenchExecResult{CpvTranslate}{FuncReachSafetyArrays}{BtorII}{All}{}{Median}{1}%
\StoreBenchExecResult{CpvTranslate}{FuncReachSafetyArrays}{BtorII}{All}{}{Stdev}{0.2032662017166911556504922279}%
\providecommand\StoreBenchExecResult[7]{\expandafter\newcommand\csname#1#2#3#4#5#6\endcsname{#7}}%
\StoreBenchExecResult{CpvTranslate}{FuncReachSafetyBitVectors}{Status}{All}{}{Score}{0}%
\StoreBenchExecResult{CpvTranslate}{FuncReachSafetyBitVectors}{Status}{All}{}{Count}{48}%
\StoreBenchExecResult{CpvTranslate}{FuncReachSafetyBitVectors}{Status}{Correct}{}{Count}{0}%
\StoreBenchExecResult{CpvTranslate}{FuncReachSafetyBitVectors}{Status}{Correct}{True}{Count}{0}%
\StoreBenchExecResult{CpvTranslate}{FuncReachSafetyBitVectors}{Status}{Correct}{False}{Count}{0}%
\StoreBenchExecResult{CpvTranslate}{FuncReachSafetyBitVectors}{Status}{Wrong}{}{Count}{0}%
\StoreBenchExecResult{CpvTranslate}{FuncReachSafetyBitVectors}{Status}{Wrong}{True}{Count}{0}%
\StoreBenchExecResult{CpvTranslate}{FuncReachSafetyBitVectors}{Status}{Wrong}{False}{Count}{0}%
\StoreBenchExecResult{CpvTranslate}{FuncReachSafetyBitVectors}{KII}{All}{}{Sum}{48}%
\StoreBenchExecResult{CpvTranslate}{FuncReachSafetyBitVectors}{KII}{All}{}{Min}{1}%
\StoreBenchExecResult{CpvTranslate}{FuncReachSafetyBitVectors}{KII}{All}{}{Max}{1}%
\StoreBenchExecResult{CpvTranslate}{FuncReachSafetyBitVectors}{KII}{All}{}{Avg}{1}%
\StoreBenchExecResult{CpvTranslate}{FuncReachSafetyBitVectors}{KII}{All}{}{Median}{1}%
\StoreBenchExecResult{CpvTranslate}{FuncReachSafetyBitVectors}{KII}{All}{}{Stdev}{0.00000000000000}%
\StoreBenchExecResult{CpvTranslate}{FuncReachSafetyBitVectors}{BtorII}{All}{}{Sum}{48}%
\StoreBenchExecResult{CpvTranslate}{FuncReachSafetyBitVectors}{BtorII}{All}{}{Min}{1}%
\StoreBenchExecResult{CpvTranslate}{FuncReachSafetyBitVectors}{BtorII}{All}{}{Max}{1}%
\StoreBenchExecResult{CpvTranslate}{FuncReachSafetyBitVectors}{BtorII}{All}{}{Avg}{1}%
\StoreBenchExecResult{CpvTranslate}{FuncReachSafetyBitVectors}{BtorII}{All}{}{Median}{1}%
\StoreBenchExecResult{CpvTranslate}{FuncReachSafetyBitVectors}{BtorII}{All}{}{Stdev}{0.00000000000000}%
\providecommand\StoreBenchExecResult[7]{\expandafter\newcommand\csname#1#2#3#4#5#6\endcsname{#7}}%
\StoreBenchExecResult{CpvTranslate}{FuncReachSafetyCombinations}{Status}{All}{}{Score}{0}%
\StoreBenchExecResult{CpvTranslate}{FuncReachSafetyCombinations}{Status}{All}{}{Count}{671}%
\StoreBenchExecResult{CpvTranslate}{FuncReachSafetyCombinations}{Status}{Correct}{}{Count}{0}%
\StoreBenchExecResult{CpvTranslate}{FuncReachSafetyCombinations}{Status}{Correct}{True}{Count}{0}%
\StoreBenchExecResult{CpvTranslate}{FuncReachSafetyCombinations}{Status}{Correct}{False}{Count}{0}%
\StoreBenchExecResult{CpvTranslate}{FuncReachSafetyCombinations}{Status}{Wrong}{}{Count}{0}%
\StoreBenchExecResult{CpvTranslate}{FuncReachSafetyCombinations}{Status}{Wrong}{True}{Count}{0}%
\StoreBenchExecResult{CpvTranslate}{FuncReachSafetyCombinations}{Status}{Wrong}{False}{Count}{0}%
\StoreBenchExecResult{CpvTranslate}{FuncReachSafetyCombinations}{KII}{All}{}{Sum}{636}%
\StoreBenchExecResult{CpvTranslate}{FuncReachSafetyCombinations}{KII}{All}{}{Min}{0}%
\StoreBenchExecResult{CpvTranslate}{FuncReachSafetyCombinations}{KII}{All}{}{Max}{1}%
\StoreBenchExecResult{CpvTranslate}{FuncReachSafetyCombinations}{KII}{All}{}{Avg}{0.9478390461997019374068554396}%
\StoreBenchExecResult{CpvTranslate}{FuncReachSafetyCombinations}{KII}{All}{}{Median}{1}%
\StoreBenchExecResult{CpvTranslate}{FuncReachSafetyCombinations}{KII}{All}{}{Stdev}{0.2223514980811715459822416092}%
\StoreBenchExecResult{CpvTranslate}{FuncReachSafetyCombinations}{BtorII}{All}{}{Sum}{110}%
\StoreBenchExecResult{CpvTranslate}{FuncReachSafetyCombinations}{BtorII}{All}{}{Min}{0}%
\StoreBenchExecResult{CpvTranslate}{FuncReachSafetyCombinations}{BtorII}{All}{}{Max}{1}%
\StoreBenchExecResult{CpvTranslate}{FuncReachSafetyCombinations}{BtorII}{All}{}{Avg}{0.1639344262295081967213114754}%
\StoreBenchExecResult{CpvTranslate}{FuncReachSafetyCombinations}{BtorII}{All}{}{Median}{0}%
\StoreBenchExecResult{CpvTranslate}{FuncReachSafetyCombinations}{BtorII}{All}{}{Stdev}{0.3702160587093840958970573753}%
\providecommand\StoreBenchExecResult[7]{\expandafter\newcommand\csname#1#2#3#4#5#6\endcsname{#7}}%
\StoreBenchExecResult{CpvTranslate}{FuncReachSafetyControlFlow}{Status}{All}{}{Score}{0}%
\StoreBenchExecResult{CpvTranslate}{FuncReachSafetyControlFlow}{Status}{All}{}{Count}{73}%
\StoreBenchExecResult{CpvTranslate}{FuncReachSafetyControlFlow}{Status}{Correct}{}{Count}{0}%
\StoreBenchExecResult{CpvTranslate}{FuncReachSafetyControlFlow}{Status}{Correct}{True}{Count}{0}%
\StoreBenchExecResult{CpvTranslate}{FuncReachSafetyControlFlow}{Status}{Correct}{False}{Count}{0}%
\StoreBenchExecResult{CpvTranslate}{FuncReachSafetyControlFlow}{Status}{Wrong}{}{Count}{0}%
\StoreBenchExecResult{CpvTranslate}{FuncReachSafetyControlFlow}{Status}{Wrong}{True}{Count}{0}%
\StoreBenchExecResult{CpvTranslate}{FuncReachSafetyControlFlow}{Status}{Wrong}{False}{Count}{0}%
\StoreBenchExecResult{CpvTranslate}{FuncReachSafetyControlFlow}{KII}{All}{}{Sum}{35}%
\StoreBenchExecResult{CpvTranslate}{FuncReachSafetyControlFlow}{KII}{All}{}{Min}{0}%
\StoreBenchExecResult{CpvTranslate}{FuncReachSafetyControlFlow}{KII}{All}{}{Max}{1}%
\StoreBenchExecResult{CpvTranslate}{FuncReachSafetyControlFlow}{KII}{All}{}{Avg}{0.4794520547945205479452054795}%
\StoreBenchExecResult{CpvTranslate}{FuncReachSafetyControlFlow}{KII}{All}{}{Median}{0}%
\StoreBenchExecResult{CpvTranslate}{FuncReachSafetyControlFlow}{KII}{All}{}{Stdev}{0.4995776035290539463185022016}%
\StoreBenchExecResult{CpvTranslate}{FuncReachSafetyControlFlow}{BtorII}{All}{}{Sum}{33}%
\StoreBenchExecResult{CpvTranslate}{FuncReachSafetyControlFlow}{BtorII}{All}{}{Min}{0}%
\StoreBenchExecResult{CpvTranslate}{FuncReachSafetyControlFlow}{BtorII}{All}{}{Max}{1}%
\StoreBenchExecResult{CpvTranslate}{FuncReachSafetyControlFlow}{BtorII}{All}{}{Avg}{0.4520547945205479452054794521}%
\StoreBenchExecResult{CpvTranslate}{FuncReachSafetyControlFlow}{BtorII}{All}{}{Median}{0}%
\StoreBenchExecResult{CpvTranslate}{FuncReachSafetyControlFlow}{BtorII}{All}{}{Stdev}{0.4976959486187657532201423789}%
\providecommand\StoreBenchExecResult[7]{\expandafter\newcommand\csname#1#2#3#4#5#6\endcsname{#7}}%
\StoreBenchExecResult{CpvTranslate}{FuncReachSafetyECA}{Status}{All}{}{Score}{0}%
\StoreBenchExecResult{CpvTranslate}{FuncReachSafetyECA}{Status}{All}{}{Count}{1263}%
\StoreBenchExecResult{CpvTranslate}{FuncReachSafetyECA}{Status}{Correct}{}{Count}{0}%
\StoreBenchExecResult{CpvTranslate}{FuncReachSafetyECA}{Status}{Correct}{True}{Count}{0}%
\StoreBenchExecResult{CpvTranslate}{FuncReachSafetyECA}{Status}{Correct}{False}{Count}{0}%
\StoreBenchExecResult{CpvTranslate}{FuncReachSafetyECA}{Status}{Wrong}{}{Count}{0}%
\StoreBenchExecResult{CpvTranslate}{FuncReachSafetyECA}{Status}{Wrong}{True}{Count}{0}%
\StoreBenchExecResult{CpvTranslate}{FuncReachSafetyECA}{Status}{Wrong}{False}{Count}{0}%
\StoreBenchExecResult{CpvTranslate}{FuncReachSafetyECA}{KII}{All}{}{Sum}{1252}%
\StoreBenchExecResult{CpvTranslate}{FuncReachSafetyECA}{KII}{All}{}{Min}{0}%
\StoreBenchExecResult{CpvTranslate}{FuncReachSafetyECA}{KII}{All}{}{Max}{1}%
\StoreBenchExecResult{CpvTranslate}{FuncReachSafetyECA}{KII}{All}{}{Avg}{0.9912905779889152810768012668}%
\StoreBenchExecResult{CpvTranslate}{FuncReachSafetyECA}{KII}{All}{}{Median}{1}%
\StoreBenchExecResult{CpvTranslate}{FuncReachSafetyECA}{KII}{All}{}{Stdev}{0.09291699510486524517055301934}%
\StoreBenchExecResult{CpvTranslate}{FuncReachSafetyECA}{BtorII}{All}{}{Sum}{784}%
\StoreBenchExecResult{CpvTranslate}{FuncReachSafetyECA}{BtorII}{All}{}{Min}{0}%
\StoreBenchExecResult{CpvTranslate}{FuncReachSafetyECA}{BtorII}{All}{}{Max}{1}%
\StoreBenchExecResult{CpvTranslate}{FuncReachSafetyECA}{BtorII}{All}{}{Avg}{0.6207442596991290577988915281}%
\StoreBenchExecResult{CpvTranslate}{FuncReachSafetyECA}{BtorII}{All}{}{Median}{1}%
\StoreBenchExecResult{CpvTranslate}{FuncReachSafetyECA}{BtorII}{All}{}{Stdev}{0.4852018381557403770885682966}%
\providecommand\StoreBenchExecResult[7]{\expandafter\newcommand\csname#1#2#3#4#5#6\endcsname{#7}}%
\StoreBenchExecResult{CpvTranslate}{FuncReachSafetyFloats}{Status}{All}{}{Score}{0}%
\StoreBenchExecResult{CpvTranslate}{FuncReachSafetyFloats}{Status}{All}{}{Count}{1400}%
\StoreBenchExecResult{CpvTranslate}{FuncReachSafetyFloats}{Status}{Correct}{}{Count}{0}%
\StoreBenchExecResult{CpvTranslate}{FuncReachSafetyFloats}{Status}{Correct}{True}{Count}{0}%
\StoreBenchExecResult{CpvTranslate}{FuncReachSafetyFloats}{Status}{Correct}{False}{Count}{0}%
\StoreBenchExecResult{CpvTranslate}{FuncReachSafetyFloats}{Status}{Wrong}{}{Count}{0}%
\StoreBenchExecResult{CpvTranslate}{FuncReachSafetyFloats}{Status}{Wrong}{True}{Count}{0}%
\StoreBenchExecResult{CpvTranslate}{FuncReachSafetyFloats}{Status}{Wrong}{False}{Count}{0}%
\StoreBenchExecResult{CpvTranslate}{FuncReachSafetyFloats}{KII}{All}{}{Sum}{321}%
\StoreBenchExecResult{CpvTranslate}{FuncReachSafetyFloats}{KII}{All}{}{Min}{0}%
\StoreBenchExecResult{CpvTranslate}{FuncReachSafetyFloats}{KII}{All}{}{Max}{1}%
\StoreBenchExecResult{CpvTranslate}{FuncReachSafetyFloats}{KII}{All}{}{Avg}{0.2292857142857142857142857143}%
\StoreBenchExecResult{CpvTranslate}{FuncReachSafetyFloats}{KII}{All}{}{Median}{0}%
\StoreBenchExecResult{CpvTranslate}{FuncReachSafetyFloats}{KII}{All}{}{Stdev}{0.4203733763099229404297488451}%
\StoreBenchExecResult{CpvTranslate}{FuncReachSafetyFloats}{BtorII}{All}{}{Sum}{10}%
\StoreBenchExecResult{CpvTranslate}{FuncReachSafetyFloats}{BtorII}{All}{}{Min}{0}%
\StoreBenchExecResult{CpvTranslate}{FuncReachSafetyFloats}{BtorII}{All}{}{Max}{1}%
\StoreBenchExecResult{CpvTranslate}{FuncReachSafetyFloats}{BtorII}{All}{}{Avg}{0.007142857142857142857142857143}%
\StoreBenchExecResult{CpvTranslate}{FuncReachSafetyFloats}{BtorII}{All}{}{Median}{0}%
\StoreBenchExecResult{CpvTranslate}{FuncReachSafetyFloats}{BtorII}{All}{}{Stdev}{0.08421304373251139977477988395}%
\providecommand\StoreBenchExecResult[7]{\expandafter\newcommand\csname#1#2#3#4#5#6\endcsname{#7}}%
\StoreBenchExecResult{CpvTranslate}{FuncReachSafetyHardness}{Status}{All}{}{Score}{0}%
\StoreBenchExecResult{CpvTranslate}{FuncReachSafetyHardness}{Status}{All}{}{Count}{6789}%
\StoreBenchExecResult{CpvTranslate}{FuncReachSafetyHardness}{Status}{Correct}{}{Count}{0}%
\StoreBenchExecResult{CpvTranslate}{FuncReachSafetyHardness}{Status}{Correct}{True}{Count}{0}%
\StoreBenchExecResult{CpvTranslate}{FuncReachSafetyHardness}{Status}{Correct}{False}{Count}{0}%
\StoreBenchExecResult{CpvTranslate}{FuncReachSafetyHardness}{Status}{Wrong}{}{Count}{0}%
\StoreBenchExecResult{CpvTranslate}{FuncReachSafetyHardness}{Status}{Wrong}{True}{Count}{0}%
\StoreBenchExecResult{CpvTranslate}{FuncReachSafetyHardness}{Status}{Wrong}{False}{Count}{0}%
\StoreBenchExecResult{CpvTranslate}{FuncReachSafetyHardness}{KII}{All}{}{Sum}{6525}%
\StoreBenchExecResult{CpvTranslate}{FuncReachSafetyHardness}{KII}{All}{}{Min}{0}%
\StoreBenchExecResult{CpvTranslate}{FuncReachSafetyHardness}{KII}{All}{}{Max}{1}%
\StoreBenchExecResult{CpvTranslate}{FuncReachSafetyHardness}{KII}{All}{}{Avg}{0.9611135660627485638532920901}%
\StoreBenchExecResult{CpvTranslate}{FuncReachSafetyHardness}{KII}{All}{}{Median}{1}%
\StoreBenchExecResult{CpvTranslate}{FuncReachSafetyHardness}{KII}{All}{}{Stdev}{0.1933242850572457591482753537}%
\StoreBenchExecResult{CpvTranslate}{FuncReachSafetyHardness}{BtorII}{All}{}{Sum}{742}%
\StoreBenchExecResult{CpvTranslate}{FuncReachSafetyHardness}{BtorII}{All}{}{Min}{0}%
\StoreBenchExecResult{CpvTranslate}{FuncReachSafetyHardness}{BtorII}{All}{}{Max}{1}%
\StoreBenchExecResult{CpvTranslate}{FuncReachSafetyHardness}{BtorII}{All}{}{Avg}{0.1092944468993960818971866254}%
\StoreBenchExecResult{CpvTranslate}{FuncReachSafetyHardness}{BtorII}{All}{}{Median}{0}%
\StoreBenchExecResult{CpvTranslate}{FuncReachSafetyHardness}{BtorII}{All}{}{Stdev}{0.3120082863905238848764374134}%
\providecommand\StoreBenchExecResult[7]{\expandafter\newcommand\csname#1#2#3#4#5#6\endcsname{#7}}%
\StoreBenchExecResult{CpvTranslate}{FuncReachSafetyHardware}{Status}{All}{}{Score}{0}%
\StoreBenchExecResult{CpvTranslate}{FuncReachSafetyHardware}{Status}{All}{}{Count}{1224}%
\StoreBenchExecResult{CpvTranslate}{FuncReachSafetyHardware}{Status}{Correct}{}{Count}{0}%
\StoreBenchExecResult{CpvTranslate}{FuncReachSafetyHardware}{Status}{Correct}{True}{Count}{0}%
\StoreBenchExecResult{CpvTranslate}{FuncReachSafetyHardware}{Status}{Correct}{False}{Count}{0}%
\StoreBenchExecResult{CpvTranslate}{FuncReachSafetyHardware}{Status}{Wrong}{}{Count}{0}%
\StoreBenchExecResult{CpvTranslate}{FuncReachSafetyHardware}{Status}{Wrong}{True}{Count}{0}%
\StoreBenchExecResult{CpvTranslate}{FuncReachSafetyHardware}{Status}{Wrong}{False}{Count}{0}%
\StoreBenchExecResult{CpvTranslate}{FuncReachSafetyHardware}{KII}{All}{}{Sum}{1182}%
\StoreBenchExecResult{CpvTranslate}{FuncReachSafetyHardware}{KII}{All}{}{Min}{0}%
\StoreBenchExecResult{CpvTranslate}{FuncReachSafetyHardware}{KII}{All}{}{Max}{1}%
\StoreBenchExecResult{CpvTranslate}{FuncReachSafetyHardware}{KII}{All}{}{Avg}{0.9656862745098039215686274510}%
\StoreBenchExecResult{CpvTranslate}{FuncReachSafetyHardware}{KII}{All}{}{Median}{1}%
\StoreBenchExecResult{CpvTranslate}{FuncReachSafetyHardware}{KII}{All}{}{Stdev}{0.1820337708590896227395067208}%
\StoreBenchExecResult{CpvTranslate}{FuncReachSafetyHardware}{BtorII}{All}{}{Sum}{1167}%
\StoreBenchExecResult{CpvTranslate}{FuncReachSafetyHardware}{BtorII}{All}{}{Min}{0}%
\StoreBenchExecResult{CpvTranslate}{FuncReachSafetyHardware}{BtorII}{All}{}{Max}{1}%
\StoreBenchExecResult{CpvTranslate}{FuncReachSafetyHardware}{BtorII}{All}{}{Avg}{0.9534313725490196078431372549}%
\StoreBenchExecResult{CpvTranslate}{FuncReachSafetyHardware}{BtorII}{All}{}{Median}{1}%
\StoreBenchExecResult{CpvTranslate}{FuncReachSafetyHardware}{BtorII}{All}{}{Stdev}{0.2107130522495276931347423855}%
\providecommand\StoreBenchExecResult[7]{\expandafter\newcommand\csname#1#2#3#4#5#6\endcsname{#7}}%
\StoreBenchExecResult{CpvTranslate}{FuncReachSafetyHeap}{Status}{All}{}{Score}{0}%
\StoreBenchExecResult{CpvTranslate}{FuncReachSafetyHeap}{Status}{All}{}{Count}{237}%
\StoreBenchExecResult{CpvTranslate}{FuncReachSafetyHeap}{Status}{Correct}{}{Count}{0}%
\StoreBenchExecResult{CpvTranslate}{FuncReachSafetyHeap}{Status}{Correct}{True}{Count}{0}%
\StoreBenchExecResult{CpvTranslate}{FuncReachSafetyHeap}{Status}{Correct}{False}{Count}{0}%
\StoreBenchExecResult{CpvTranslate}{FuncReachSafetyHeap}{Status}{Wrong}{}{Count}{0}%
\StoreBenchExecResult{CpvTranslate}{FuncReachSafetyHeap}{Status}{Wrong}{True}{Count}{0}%
\StoreBenchExecResult{CpvTranslate}{FuncReachSafetyHeap}{Status}{Wrong}{False}{Count}{0}%
\StoreBenchExecResult{CpvTranslate}{FuncReachSafetyHeap}{KII}{All}{}{Sum}{78}%
\StoreBenchExecResult{CpvTranslate}{FuncReachSafetyHeap}{KII}{All}{}{Min}{0}%
\StoreBenchExecResult{CpvTranslate}{FuncReachSafetyHeap}{KII}{All}{}{Max}{1}%
\StoreBenchExecResult{CpvTranslate}{FuncReachSafetyHeap}{KII}{All}{}{Avg}{0.3291139240506329113924050633}%
\StoreBenchExecResult{CpvTranslate}{FuncReachSafetyHeap}{KII}{All}{}{Median}{0}%
\StoreBenchExecResult{CpvTranslate}{FuncReachSafetyHeap}{KII}{All}{}{Stdev}{0.4698914226144452012453432239}%
\StoreBenchExecResult{CpvTranslate}{FuncReachSafetyHeap}{BtorII}{All}{}{Sum}{53}%
\StoreBenchExecResult{CpvTranslate}{FuncReachSafetyHeap}{BtorII}{All}{}{Min}{0}%
\StoreBenchExecResult{CpvTranslate}{FuncReachSafetyHeap}{BtorII}{All}{}{Max}{1}%
\StoreBenchExecResult{CpvTranslate}{FuncReachSafetyHeap}{BtorII}{All}{}{Avg}{0.2236286919831223628691983122}%
\StoreBenchExecResult{CpvTranslate}{FuncReachSafetyHeap}{BtorII}{All}{}{Median}{0}%
\StoreBenchExecResult{CpvTranslate}{FuncReachSafetyHeap}{BtorII}{All}{}{Stdev}{0.4166760133545488201338244783}%
\providecommand\StoreBenchExecResult[7]{\expandafter\newcommand\csname#1#2#3#4#5#6\endcsname{#7}}%
\StoreBenchExecResult{CpvTranslate}{FuncReachSafetyLoops}{Status}{All}{}{Score}{0}%
\StoreBenchExecResult{CpvTranslate}{FuncReachSafetyLoops}{Status}{All}{}{Count}{762}%
\StoreBenchExecResult{CpvTranslate}{FuncReachSafetyLoops}{Status}{Correct}{}{Count}{0}%
\StoreBenchExecResult{CpvTranslate}{FuncReachSafetyLoops}{Status}{Correct}{True}{Count}{0}%
\StoreBenchExecResult{CpvTranslate}{FuncReachSafetyLoops}{Status}{Correct}{False}{Count}{0}%
\StoreBenchExecResult{CpvTranslate}{FuncReachSafetyLoops}{Status}{Wrong}{}{Count}{0}%
\StoreBenchExecResult{CpvTranslate}{FuncReachSafetyLoops}{Status}{Wrong}{True}{Count}{0}%
\StoreBenchExecResult{CpvTranslate}{FuncReachSafetyLoops}{Status}{Wrong}{False}{Count}{0}%
\StoreBenchExecResult{CpvTranslate}{FuncReachSafetyLoops}{KII}{All}{}{Sum}{753}%
\StoreBenchExecResult{CpvTranslate}{FuncReachSafetyLoops}{KII}{All}{}{Min}{0}%
\StoreBenchExecResult{CpvTranslate}{FuncReachSafetyLoops}{KII}{All}{}{Max}{1}%
\StoreBenchExecResult{CpvTranslate}{FuncReachSafetyLoops}{KII}{All}{}{Avg}{0.9881889763779527559055118110}%
\StoreBenchExecResult{CpvTranslate}{FuncReachSafetyLoops}{KII}{All}{}{Median}{1}%
\StoreBenchExecResult{CpvTranslate}{FuncReachSafetyLoops}{KII}{All}{}{Stdev}{0.1080348246772617579063612756}%
\StoreBenchExecResult{CpvTranslate}{FuncReachSafetyLoops}{BtorII}{All}{}{Sum}{713}%
\StoreBenchExecResult{CpvTranslate}{FuncReachSafetyLoops}{BtorII}{All}{}{Min}{0}%
\StoreBenchExecResult{CpvTranslate}{FuncReachSafetyLoops}{BtorII}{All}{}{Max}{1}%
\StoreBenchExecResult{CpvTranslate}{FuncReachSafetyLoops}{BtorII}{All}{}{Avg}{0.9356955380577427821522309711}%
\StoreBenchExecResult{CpvTranslate}{FuncReachSafetyLoops}{BtorII}{All}{}{Median}{1}%
\StoreBenchExecResult{CpvTranslate}{FuncReachSafetyLoops}{BtorII}{All}{}{Stdev}{0.2452945130176661960748537724}%
\providecommand\StoreBenchExecResult[7]{\expandafter\newcommand\csname#1#2#3#4#5#6\endcsname{#7}}%
\StoreBenchExecResult{CpvTranslate}{FuncReachSafetyProductLines}{Status}{All}{}{Score}{0}%
\StoreBenchExecResult{CpvTranslate}{FuncReachSafetyProductLines}{Status}{All}{}{Count}{597}%
\StoreBenchExecResult{CpvTranslate}{FuncReachSafetyProductLines}{Status}{Correct}{}{Count}{0}%
\StoreBenchExecResult{CpvTranslate}{FuncReachSafetyProductLines}{Status}{Correct}{True}{Count}{0}%
\StoreBenchExecResult{CpvTranslate}{FuncReachSafetyProductLines}{Status}{Correct}{False}{Count}{0}%
\StoreBenchExecResult{CpvTranslate}{FuncReachSafetyProductLines}{Status}{Wrong}{}{Count}{0}%
\StoreBenchExecResult{CpvTranslate}{FuncReachSafetyProductLines}{Status}{Wrong}{True}{Count}{0}%
\StoreBenchExecResult{CpvTranslate}{FuncReachSafetyProductLines}{Status}{Wrong}{False}{Count}{0}%
\StoreBenchExecResult{CpvTranslate}{FuncReachSafetyProductLines}{KII}{All}{}{Sum}{597}%
\StoreBenchExecResult{CpvTranslate}{FuncReachSafetyProductLines}{KII}{All}{}{Min}{1}%
\StoreBenchExecResult{CpvTranslate}{FuncReachSafetyProductLines}{KII}{All}{}{Max}{1}%
\StoreBenchExecResult{CpvTranslate}{FuncReachSafetyProductLines}{KII}{All}{}{Avg}{1}%
\StoreBenchExecResult{CpvTranslate}{FuncReachSafetyProductLines}{KII}{All}{}{Median}{1}%
\StoreBenchExecResult{CpvTranslate}{FuncReachSafetyProductLines}{KII}{All}{}{Stdev}{0.00000000000000}%
\StoreBenchExecResult{CpvTranslate}{FuncReachSafetyProductLines}{BtorII}{All}{}{Sum}{419}%
\StoreBenchExecResult{CpvTranslate}{FuncReachSafetyProductLines}{BtorII}{All}{}{Min}{0}%
\StoreBenchExecResult{CpvTranslate}{FuncReachSafetyProductLines}{BtorII}{All}{}{Max}{1}%
\StoreBenchExecResult{CpvTranslate}{FuncReachSafetyProductLines}{BtorII}{All}{}{Avg}{0.7018425460636515912897822446}%
\StoreBenchExecResult{CpvTranslate}{FuncReachSafetyProductLines}{BtorII}{All}{}{Median}{1}%
\StoreBenchExecResult{CpvTranslate}{FuncReachSafetyProductLines}{BtorII}{All}{}{Stdev}{0.4574489989042960901757564879}%
\providecommand\StoreBenchExecResult[7]{\expandafter\newcommand\csname#1#2#3#4#5#6\endcsname{#7}}%
\StoreBenchExecResult{CpvTranslate}{FuncReachSafetySequentialized}{Status}{All}{}{Score}{0}%
\StoreBenchExecResult{CpvTranslate}{FuncReachSafetySequentialized}{Status}{All}{}{Count}{585}%
\StoreBenchExecResult{CpvTranslate}{FuncReachSafetySequentialized}{Status}{Correct}{}{Count}{0}%
\StoreBenchExecResult{CpvTranslate}{FuncReachSafetySequentialized}{Status}{Correct}{True}{Count}{0}%
\StoreBenchExecResult{CpvTranslate}{FuncReachSafetySequentialized}{Status}{Correct}{False}{Count}{0}%
\StoreBenchExecResult{CpvTranslate}{FuncReachSafetySequentialized}{Status}{Wrong}{}{Count}{0}%
\StoreBenchExecResult{CpvTranslate}{FuncReachSafetySequentialized}{Status}{Wrong}{True}{Count}{0}%
\StoreBenchExecResult{CpvTranslate}{FuncReachSafetySequentialized}{Status}{Wrong}{False}{Count}{0}%
\StoreBenchExecResult{CpvTranslate}{FuncReachSafetySequentialized}{KII}{All}{}{Sum}{245}%
\StoreBenchExecResult{CpvTranslate}{FuncReachSafetySequentialized}{KII}{All}{}{Min}{0}%
\StoreBenchExecResult{CpvTranslate}{FuncReachSafetySequentialized}{KII}{All}{}{Max}{1}%
\StoreBenchExecResult{CpvTranslate}{FuncReachSafetySequentialized}{KII}{All}{}{Avg}{0.4188034188034188034188034188}%
\StoreBenchExecResult{CpvTranslate}{FuncReachSafetySequentialized}{KII}{All}{}{Median}{0}%
\StoreBenchExecResult{CpvTranslate}{FuncReachSafetySequentialized}{KII}{All}{}{Stdev}{0.4933630663132243392948695669}%
\StoreBenchExecResult{CpvTranslate}{FuncReachSafetySequentialized}{BtorII}{All}{}{Sum}{183}%
\StoreBenchExecResult{CpvTranslate}{FuncReachSafetySequentialized}{BtorII}{All}{}{Min}{0}%
\StoreBenchExecResult{CpvTranslate}{FuncReachSafetySequentialized}{BtorII}{All}{}{Max}{1}%
\StoreBenchExecResult{CpvTranslate}{FuncReachSafetySequentialized}{BtorII}{All}{}{Avg}{0.3128205128205128205128205128}%
\StoreBenchExecResult{CpvTranslate}{FuncReachSafetySequentialized}{BtorII}{All}{}{Median}{0}%
\StoreBenchExecResult{CpvTranslate}{FuncReachSafetySequentialized}{BtorII}{All}{}{Stdev}{0.4636419303505930691879986926}%
\providecommand\StoreBenchExecResult[7]{\expandafter\newcommand\csname#1#2#3#4#5#6\endcsname{#7}}%
\StoreBenchExecResult{CpvTranslate}{FuncReachSafetyXCSP}{Status}{All}{}{Score}{0}%
\StoreBenchExecResult{CpvTranslate}{FuncReachSafetyXCSP}{Status}{All}{}{Count}{119}%
\StoreBenchExecResult{CpvTranslate}{FuncReachSafetyXCSP}{Status}{Correct}{}{Count}{0}%
\StoreBenchExecResult{CpvTranslate}{FuncReachSafetyXCSP}{Status}{Correct}{True}{Count}{0}%
\StoreBenchExecResult{CpvTranslate}{FuncReachSafetyXCSP}{Status}{Correct}{False}{Count}{0}%
\StoreBenchExecResult{CpvTranslate}{FuncReachSafetyXCSP}{Status}{Wrong}{}{Count}{0}%
\StoreBenchExecResult{CpvTranslate}{FuncReachSafetyXCSP}{Status}{Wrong}{True}{Count}{0}%
\StoreBenchExecResult{CpvTranslate}{FuncReachSafetyXCSP}{Status}{Wrong}{False}{Count}{0}%
\StoreBenchExecResult{CpvTranslate}{FuncReachSafetyXCSP}{KII}{All}{}{Sum}{119}%
\StoreBenchExecResult{CpvTranslate}{FuncReachSafetyXCSP}{KII}{All}{}{Min}{1}%
\StoreBenchExecResult{CpvTranslate}{FuncReachSafetyXCSP}{KII}{All}{}{Max}{1}%
\StoreBenchExecResult{CpvTranslate}{FuncReachSafetyXCSP}{KII}{All}{}{Avg}{1}%
\StoreBenchExecResult{CpvTranslate}{FuncReachSafetyXCSP}{KII}{All}{}{Median}{1}%
\StoreBenchExecResult{CpvTranslate}{FuncReachSafetyXCSP}{KII}{All}{}{Stdev}{0.00000000000000}%
\StoreBenchExecResult{CpvTranslate}{FuncReachSafetyXCSP}{BtorII}{All}{}{Sum}{98}%
\StoreBenchExecResult{CpvTranslate}{FuncReachSafetyXCSP}{BtorII}{All}{}{Min}{0}%
\StoreBenchExecResult{CpvTranslate}{FuncReachSafetyXCSP}{BtorII}{All}{}{Max}{1}%
\StoreBenchExecResult{CpvTranslate}{FuncReachSafetyXCSP}{BtorII}{All}{}{Avg}{0.8235294117647058823529411765}%
\StoreBenchExecResult{CpvTranslate}{FuncReachSafetyXCSP}{BtorII}{All}{}{Median}{1}%
\StoreBenchExecResult{CpvTranslate}{FuncReachSafetyXCSP}{BtorII}{All}{}{Stdev}{0.3812200410828153077038804419}%
\edef\CpvTranslateFuncReachSafetyStatusAllCount{\the\numexpr\CpvTranslateFuncReachSafetyArraysStatusAllCount+\CpvTranslateFuncReachSafetyBitVectorsStatusAllCount+\CpvTranslateFuncReachSafetyCombinationsStatusAllCount+\CpvTranslateFuncReachSafetyControlFlowStatusAllCount+\CpvTranslateFuncReachSafetyECAStatusAllCount+\CpvTranslateFuncReachSafetyFloatsStatusAllCount+\CpvTranslateFuncReachSafetyHardnessStatusAllCount+\CpvTranslateFuncReachSafetyHardwareStatusAllCount+\CpvTranslateFuncReachSafetyHeapStatusAllCount+\CpvTranslateFuncReachSafetyLoopsStatusAllCount+\CpvTranslateFuncReachSafetyProductLinesStatusAllCount+\CpvTranslateFuncReachSafetySequentializedStatusAllCount+\CpvTranslateFuncReachSafetyXCSPStatusAllCount}
\edef\CpvTranslateFuncReachSafetyKIIAllSum{\the\numexpr\CpvTranslateFuncReachSafetyArraysKIIAllSum+\CpvTranslateFuncReachSafetyBitVectorsKIIAllSum+\CpvTranslateFuncReachSafetyCombinationsKIIAllSum+\CpvTranslateFuncReachSafetyControlFlowKIIAllSum+\CpvTranslateFuncReachSafetyECAKIIAllSum+\CpvTranslateFuncReachSafetyFloatsKIIAllSum+\CpvTranslateFuncReachSafetyHardnessKIIAllSum+\CpvTranslateFuncReachSafetyHardwareKIIAllSum+\CpvTranslateFuncReachSafetyHeapKIIAllSum+\CpvTranslateFuncReachSafetyLoopsKIIAllSum+\CpvTranslateFuncReachSafetyProductLinesKIIAllSum+\CpvTranslateFuncReachSafetySequentializedKIIAllSum+\CpvTranslateFuncReachSafetyXCSPKIIAllSum}
\edef\CpvTranslateFuncReachSafetyBtorIIAllSum{\the\numexpr\CpvTranslateFuncReachSafetyArraysBtorIIAllSum+\CpvTranslateFuncReachSafetyBitVectorsBtorIIAllSum+\CpvTranslateFuncReachSafetyCombinationsBtorIIAllSum+\CpvTranslateFuncReachSafetyControlFlowBtorIIAllSum+\CpvTranslateFuncReachSafetyECABtorIIAllSum+\CpvTranslateFuncReachSafetyFloatsBtorIIAllSum+\CpvTranslateFuncReachSafetyHardnessBtorIIAllSum+\CpvTranslateFuncReachSafetyHardwareBtorIIAllSum+\CpvTranslateFuncReachSafetyHeapBtorIIAllSum+\CpvTranslateFuncReachSafetyLoopsBtorIIAllSum+\CpvTranslateFuncReachSafetyProductLinesBtorIIAllSum+\CpvTranslateFuncReachSafetySequentializedBtorIIAllSum+\CpvTranslateFuncReachSafetyXCSPBtorIIAllSum}
\providecommand\StoreBenchExecResult[7]{\expandafter\newcommand\csname#1#2#3#4#5#6\endcsname{#7}}%
\StoreBenchExecResult{CpvTranslate}{FuncTranslationDoneReachSafetyFunc}{Cputime}{All}{}{Sum}{25284.622495}%
\StoreBenchExecResult{CpvTranslate}{FuncTranslationDoneReachSafetyFunc}{Cputime}{All}{}{Min}{2.239883}%
\StoreBenchExecResult{CpvTranslate}{FuncTranslationDoneReachSafetyFunc}{Cputime}{All}{}{Max}{161.980664}%
\StoreBenchExecResult{CpvTranslate}{FuncTranslationDoneReachSafetyFunc}{Cputime}{All}{}{Avg}{5.288563583978247228613260824}%
\StoreBenchExecResult{CpvTranslate}{FuncTranslationDoneReachSafetyFunc}{Cputime}{All}{}{Median}{2.884259}%
\StoreBenchExecResult{CpvTranslate}{FuncTranslationDoneReachSafetyFunc}{Cputime}{All}{}{Stdev}{6.962426286915656550950467411}%
\StoreBenchExecResult{CpvTranslate}{FuncTranslationDoneReachSafetyFunc}{Cputime}{All}{}{Unit}{s}%
\providecommand\StoreBenchExecResult[7]{\expandafter\newcommand\csname#1#2#3#4#5#6\endcsname{#7}}%
\StoreBenchExecResult{CpvTranslate}{FuncLIIsTerminationBitVectors}{Status}{All}{}{Score}{0}%
\StoreBenchExecResult{CpvTranslate}{FuncLIIsTerminationBitVectors}{Status}{All}{}{Count}{34}%
\StoreBenchExecResult{CpvTranslate}{FuncLIIsTerminationBitVectors}{Status}{Correct}{}{Count}{0}%
\StoreBenchExecResult{CpvTranslate}{FuncLIIsTerminationBitVectors}{Status}{Correct}{True}{Count}{0}%
\StoreBenchExecResult{CpvTranslate}{FuncLIIsTerminationBitVectors}{Status}{Correct}{False}{Count}{0}%
\StoreBenchExecResult{CpvTranslate}{FuncLIIsTerminationBitVectors}{Status}{Wrong}{}{Count}{0}%
\StoreBenchExecResult{CpvTranslate}{FuncLIIsTerminationBitVectors}{Status}{Wrong}{True}{Count}{0}%
\StoreBenchExecResult{CpvTranslate}{FuncLIIsTerminationBitVectors}{Status}{Wrong}{False}{Count}{0}%
\StoreBenchExecResult{CpvTranslate}{FuncLIIsTerminationBitVectors}{KII}{All}{}{Sum}{32}%
\StoreBenchExecResult{CpvTranslate}{FuncLIIsTerminationBitVectors}{KII}{All}{}{Min}{0}%
\StoreBenchExecResult{CpvTranslate}{FuncLIIsTerminationBitVectors}{KII}{All}{}{Max}{1}%
\StoreBenchExecResult{CpvTranslate}{FuncLIIsTerminationBitVectors}{KII}{All}{}{Avg}{0.9411764705882352941176470588}%
\StoreBenchExecResult{CpvTranslate}{FuncLIIsTerminationBitVectors}{KII}{All}{}{Median}{1}%
\StoreBenchExecResult{CpvTranslate}{FuncLIIsTerminationBitVectors}{KII}{All}{}{Stdev}{0.2352941176470588235294117654}%
\StoreBenchExecResult{CpvTranslate}{FuncLIIsTerminationBitVectors}{BtorII}{All}{}{Sum}{32}%
\StoreBenchExecResult{CpvTranslate}{FuncLIIsTerminationBitVectors}{BtorII}{All}{}{Min}{0}%
\StoreBenchExecResult{CpvTranslate}{FuncLIIsTerminationBitVectors}{BtorII}{All}{}{Max}{1}%
\StoreBenchExecResult{CpvTranslate}{FuncLIIsTerminationBitVectors}{BtorII}{All}{}{Avg}{0.9411764705882352941176470588}%
\StoreBenchExecResult{CpvTranslate}{FuncLIIsTerminationBitVectors}{BtorII}{All}{}{Median}{1}%
\StoreBenchExecResult{CpvTranslate}{FuncLIIsTerminationBitVectors}{BtorII}{All}{}{Stdev}{0.2352941176470588235294117654}%
\providecommand\StoreBenchExecResult[7]{\expandafter\newcommand\csname#1#2#3#4#5#6\endcsname{#7}}%
\StoreBenchExecResult{CpvTranslate}{FuncLIIsTerminationMainControlFlow}{Status}{All}{}{Score}{0}%
\StoreBenchExecResult{CpvTranslate}{FuncLIIsTerminationMainControlFlow}{Status}{All}{}{Count}{253}%
\StoreBenchExecResult{CpvTranslate}{FuncLIIsTerminationMainControlFlow}{Status}{Correct}{}{Count}{0}%
\StoreBenchExecResult{CpvTranslate}{FuncLIIsTerminationMainControlFlow}{Status}{Correct}{True}{Count}{0}%
\StoreBenchExecResult{CpvTranslate}{FuncLIIsTerminationMainControlFlow}{Status}{Correct}{False}{Count}{0}%
\StoreBenchExecResult{CpvTranslate}{FuncLIIsTerminationMainControlFlow}{Status}{Wrong}{}{Count}{0}%
\StoreBenchExecResult{CpvTranslate}{FuncLIIsTerminationMainControlFlow}{Status}{Wrong}{True}{Count}{0}%
\StoreBenchExecResult{CpvTranslate}{FuncLIIsTerminationMainControlFlow}{Status}{Wrong}{False}{Count}{0}%
\StoreBenchExecResult{CpvTranslate}{FuncLIIsTerminationMainControlFlow}{KII}{All}{}{Sum}{243}%
\StoreBenchExecResult{CpvTranslate}{FuncLIIsTerminationMainControlFlow}{KII}{All}{}{Min}{0}%
\StoreBenchExecResult{CpvTranslate}{FuncLIIsTerminationMainControlFlow}{KII}{All}{}{Max}{1}%
\StoreBenchExecResult{CpvTranslate}{FuncLIIsTerminationMainControlFlow}{KII}{All}{}{Avg}{0.9604743083003952569169960474}%
\StoreBenchExecResult{CpvTranslate}{FuncLIIsTerminationMainControlFlow}{KII}{All}{}{Median}{1}%
\StoreBenchExecResult{CpvTranslate}{FuncLIIsTerminationMainControlFlow}{KII}{All}{}{Stdev}{0.1948420165038140324550485401}%
\StoreBenchExecResult{CpvTranslate}{FuncLIIsTerminationMainControlFlow}{BtorII}{All}{}{Sum}{242}%
\StoreBenchExecResult{CpvTranslate}{FuncLIIsTerminationMainControlFlow}{BtorII}{All}{}{Min}{0}%
\StoreBenchExecResult{CpvTranslate}{FuncLIIsTerminationMainControlFlow}{BtorII}{All}{}{Max}{1}%
\StoreBenchExecResult{CpvTranslate}{FuncLIIsTerminationMainControlFlow}{BtorII}{All}{}{Avg}{0.9565217391304347826086956522}%
\StoreBenchExecResult{CpvTranslate}{FuncLIIsTerminationMainControlFlow}{BtorII}{All}{}{Median}{1}%
\StoreBenchExecResult{CpvTranslate}{FuncLIIsTerminationMainControlFlow}{BtorII}{All}{}{Stdev}{0.2039311199923230241115491315}%
\providecommand\StoreBenchExecResult[7]{\expandafter\newcommand\csname#1#2#3#4#5#6\endcsname{#7}}%
\StoreBenchExecResult{CpvTranslate}{FuncLIIsTerminationMainHeap}{Status}{All}{}{Score}{0}%
\StoreBenchExecResult{CpvTranslate}{FuncLIIsTerminationMainHeap}{Status}{All}{}{Count}{180}%
\StoreBenchExecResult{CpvTranslate}{FuncLIIsTerminationMainHeap}{Status}{Correct}{}{Count}{0}%
\StoreBenchExecResult{CpvTranslate}{FuncLIIsTerminationMainHeap}{Status}{Correct}{True}{Count}{0}%
\StoreBenchExecResult{CpvTranslate}{FuncLIIsTerminationMainHeap}{Status}{Correct}{False}{Count}{0}%
\StoreBenchExecResult{CpvTranslate}{FuncLIIsTerminationMainHeap}{Status}{Wrong}{}{Count}{0}%
\StoreBenchExecResult{CpvTranslate}{FuncLIIsTerminationMainHeap}{Status}{Wrong}{True}{Count}{0}%
\StoreBenchExecResult{CpvTranslate}{FuncLIIsTerminationMainHeap}{Status}{Wrong}{False}{Count}{0}%
\StoreBenchExecResult{CpvTranslate}{FuncLIIsTerminationMainHeap}{KII}{All}{}{Sum}{0}%
\StoreBenchExecResult{CpvTranslate}{FuncLIIsTerminationMainHeap}{KII}{All}{}{Min}{0}%
\StoreBenchExecResult{CpvTranslate}{FuncLIIsTerminationMainHeap}{KII}{All}{}{Max}{0}%
\StoreBenchExecResult{CpvTranslate}{FuncLIIsTerminationMainHeap}{KII}{All}{}{Avg}{0}%
\StoreBenchExecResult{CpvTranslate}{FuncLIIsTerminationMainHeap}{KII}{All}{}{Median}{0}%
\StoreBenchExecResult{CpvTranslate}{FuncLIIsTerminationMainHeap}{KII}{All}{}{Stdev}{0.00000000000000}%
\StoreBenchExecResult{CpvTranslate}{FuncLIIsTerminationMainHeap}{BtorII}{All}{}{Sum}{0}%
\StoreBenchExecResult{CpvTranslate}{FuncLIIsTerminationMainHeap}{BtorII}{All}{}{Min}{0}%
\StoreBenchExecResult{CpvTranslate}{FuncLIIsTerminationMainHeap}{BtorII}{All}{}{Max}{0}%
\StoreBenchExecResult{CpvTranslate}{FuncLIIsTerminationMainHeap}{BtorII}{All}{}{Avg}{0}%
\StoreBenchExecResult{CpvTranslate}{FuncLIIsTerminationMainHeap}{BtorII}{All}{}{Median}{0}%
\StoreBenchExecResult{CpvTranslate}{FuncLIIsTerminationMainHeap}{BtorII}{All}{}{Stdev}{0.00000000000000}%
\providecommand\StoreBenchExecResult[7]{\expandafter\newcommand\csname#1#2#3#4#5#6\endcsname{#7}}%
\StoreBenchExecResult{CpvTranslate}{FuncLIIsTerminationOther}{Status}{All}{}{Score}{0}%
\StoreBenchExecResult{CpvTranslate}{FuncLIIsTerminationOther}{Status}{All}{}{Count}{1526}%
\StoreBenchExecResult{CpvTranslate}{FuncLIIsTerminationOther}{Status}{Correct}{}{Count}{0}%
\StoreBenchExecResult{CpvTranslate}{FuncLIIsTerminationOther}{Status}{Correct}{True}{Count}{0}%
\StoreBenchExecResult{CpvTranslate}{FuncLIIsTerminationOther}{Status}{Correct}{False}{Count}{0}%
\StoreBenchExecResult{CpvTranslate}{FuncLIIsTerminationOther}{Status}{Wrong}{}{Count}{0}%
\StoreBenchExecResult{CpvTranslate}{FuncLIIsTerminationOther}{Status}{Wrong}{True}{Count}{0}%
\StoreBenchExecResult{CpvTranslate}{FuncLIIsTerminationOther}{Status}{Wrong}{False}{Count}{0}%
\StoreBenchExecResult{CpvTranslate}{FuncLIIsTerminationOther}{KII}{All}{}{Sum}{1391}%
\StoreBenchExecResult{CpvTranslate}{FuncLIIsTerminationOther}{KII}{All}{}{Min}{0}%
\StoreBenchExecResult{CpvTranslate}{FuncLIIsTerminationOther}{KII}{All}{}{Max}{1}%
\StoreBenchExecResult{CpvTranslate}{FuncLIIsTerminationOther}{KII}{All}{}{Avg}{0.9115334207077326343381389253}%
\StoreBenchExecResult{CpvTranslate}{FuncLIIsTerminationOther}{KII}{All}{}{Median}{1}%
\StoreBenchExecResult{CpvTranslate}{FuncLIIsTerminationOther}{KII}{All}{}{Stdev}{0.2839722585757142873056637163}%
\StoreBenchExecResult{CpvTranslate}{FuncLIIsTerminationOther}{BtorII}{All}{}{Sum}{1080}%
\StoreBenchExecResult{CpvTranslate}{FuncLIIsTerminationOther}{BtorII}{All}{}{Min}{0}%
\StoreBenchExecResult{CpvTranslate}{FuncLIIsTerminationOther}{BtorII}{All}{}{Max}{1}%
\StoreBenchExecResult{CpvTranslate}{FuncLIIsTerminationOther}{BtorII}{All}{}{Avg}{0.7077326343381389252948885976}%
\StoreBenchExecResult{CpvTranslate}{FuncLIIsTerminationOther}{BtorII}{All}{}{Median}{1}%
\StoreBenchExecResult{CpvTranslate}{FuncLIIsTerminationOther}{BtorII}{All}{}{Stdev}{0.4548045213395938356904165643}%
\edef\CpvTranslateFuncLIIsTerminationStatusAllCount{\the\numexpr\CpvTranslateFuncLIIsTerminationBitVectorsStatusAllCount+\CpvTranslateFuncLIIsTerminationMainControlFlowStatusAllCount+\CpvTranslateFuncLIIsTerminationMainHeapStatusAllCount+\CpvTranslateFuncLIIsTerminationOtherStatusAllCount}
\edef\CpvTranslateFuncLIIsTerminationKIIAllSum{\the\numexpr\CpvTranslateFuncLIIsTerminationBitVectorsKIIAllSum+\CpvTranslateFuncLIIsTerminationMainControlFlowKIIAllSum+\CpvTranslateFuncLIIsTerminationMainHeapKIIAllSum+\CpvTranslateFuncLIIsTerminationOtherKIIAllSum}
\edef\CpvTranslateFuncLIIsTerminationBtorIIAllSum{\the\numexpr\CpvTranslateFuncLIIsTerminationBitVectorsBtorIIAllSum+\CpvTranslateFuncLIIsTerminationMainControlFlowBtorIIAllSum+\CpvTranslateFuncLIIsTerminationMainHeapBtorIIAllSum+\CpvTranslateFuncLIIsTerminationOtherBtorIIAllSum}
\providecommand\StoreBenchExecResult[7]{\expandafter\newcommand\csname#1#2#3#4#5#6\endcsname{#7}}%
\StoreBenchExecResult{CpvTranslate}{FuncLIIsTranslationDoneTerminationFunc}{Cputime}{All}{}{Sum}{5508.561813}%
\StoreBenchExecResult{CpvTranslate}{FuncLIIsTranslationDoneTerminationFunc}{Cputime}{All}{}{Min}{2.257194}%
\StoreBenchExecResult{CpvTranslate}{FuncLIIsTranslationDoneTerminationFunc}{Cputime}{All}{}{Max}{164.102593}%
\StoreBenchExecResult{CpvTranslate}{FuncLIIsTranslationDoneTerminationFunc}{Cputime}{All}{}{Avg}{4.068361752584933530280649926}%
\StoreBenchExecResult{CpvTranslate}{FuncLIIsTranslationDoneTerminationFunc}{Cputime}{All}{}{Median}{2.533891}%
\StoreBenchExecResult{CpvTranslate}{FuncLIIsTranslationDoneTerminationFunc}{Cputime}{All}{}{Stdev}{7.333374603142336358906734650}%
\StoreBenchExecResult{CpvTranslate}{FuncLIIsTranslationDoneTerminationFunc}{Cputime}{All}{}{Unit}{s}%
\providecommand\StoreBenchExecResult[7]{\expandafter\newcommand\csname#1#2#3#4#5#6\endcsname{#7}}%
\StoreBenchExecResult{CpvTranslate}{RelReachSafetyArrays}{Status}{All}{}{Score}{0}%
\StoreBenchExecResult{CpvTranslate}{RelReachSafetyArrays}{Status}{All}{}{Count}{440}%
\StoreBenchExecResult{CpvTranslate}{RelReachSafetyArrays}{Status}{Correct}{}{Count}{0}%
\StoreBenchExecResult{CpvTranslate}{RelReachSafetyArrays}{Status}{Correct}{True}{Count}{0}%
\StoreBenchExecResult{CpvTranslate}{RelReachSafetyArrays}{Status}{Correct}{False}{Count}{0}%
\StoreBenchExecResult{CpvTranslate}{RelReachSafetyArrays}{Status}{Wrong}{}{Count}{0}%
\StoreBenchExecResult{CpvTranslate}{RelReachSafetyArrays}{Status}{Wrong}{True}{Count}{0}%
\StoreBenchExecResult{CpvTranslate}{RelReachSafetyArrays}{Status}{Wrong}{False}{Count}{0}%
\StoreBenchExecResult{CpvTranslate}{RelReachSafetyArrays}{KII}{All}{}{Sum}{438}%
\StoreBenchExecResult{CpvTranslate}{RelReachSafetyArrays}{KII}{All}{}{Min}{0}%
\StoreBenchExecResult{CpvTranslate}{RelReachSafetyArrays}{KII}{All}{}{Max}{1}%
\StoreBenchExecResult{CpvTranslate}{RelReachSafetyArrays}{KII}{All}{}{Avg}{0.9954545454545454545454545455}%
\StoreBenchExecResult{CpvTranslate}{RelReachSafetyArrays}{KII}{All}{}{Median}{1}%
\StoreBenchExecResult{CpvTranslate}{RelReachSafetyArrays}{KII}{All}{}{Stdev}{0.06726658448613064571401974056}%
\StoreBenchExecResult{CpvTranslate}{RelReachSafetyArrays}{BtorII}{All}{}{Sum}{432}%
\StoreBenchExecResult{CpvTranslate}{RelReachSafetyArrays}{BtorII}{All}{}{Min}{0}%
\StoreBenchExecResult{CpvTranslate}{RelReachSafetyArrays}{BtorII}{All}{}{Max}{1}%
\StoreBenchExecResult{CpvTranslate}{RelReachSafetyArrays}{BtorII}{All}{}{Avg}{0.9818181818181818181818181818}%
\StoreBenchExecResult{CpvTranslate}{RelReachSafetyArrays}{BtorII}{All}{}{Median}{1}%
\StoreBenchExecResult{CpvTranslate}{RelReachSafetyArrays}{BtorII}{All}{}{Stdev}{0.1336085314245369871743973130}%
\providecommand\StoreBenchExecResult[7]{\expandafter\newcommand\csname#1#2#3#4#5#6\endcsname{#7}}%
\StoreBenchExecResult{CpvTranslate}{RelReachSafetyBitVectors}{Status}{All}{}{Score}{0}%
\StoreBenchExecResult{CpvTranslate}{RelReachSafetyBitVectors}{Status}{All}{}{Count}{48}%
\StoreBenchExecResult{CpvTranslate}{RelReachSafetyBitVectors}{Status}{Correct}{}{Count}{0}%
\StoreBenchExecResult{CpvTranslate}{RelReachSafetyBitVectors}{Status}{Correct}{True}{Count}{0}%
\StoreBenchExecResult{CpvTranslate}{RelReachSafetyBitVectors}{Status}{Correct}{False}{Count}{0}%
\StoreBenchExecResult{CpvTranslate}{RelReachSafetyBitVectors}{Status}{Wrong}{}{Count}{0}%
\StoreBenchExecResult{CpvTranslate}{RelReachSafetyBitVectors}{Status}{Wrong}{True}{Count}{0}%
\StoreBenchExecResult{CpvTranslate}{RelReachSafetyBitVectors}{Status}{Wrong}{False}{Count}{0}%
\StoreBenchExecResult{CpvTranslate}{RelReachSafetyBitVectors}{KII}{All}{}{Sum}{48}%
\StoreBenchExecResult{CpvTranslate}{RelReachSafetyBitVectors}{KII}{All}{}{Min}{1}%
\StoreBenchExecResult{CpvTranslate}{RelReachSafetyBitVectors}{KII}{All}{}{Max}{1}%
\StoreBenchExecResult{CpvTranslate}{RelReachSafetyBitVectors}{KII}{All}{}{Avg}{1}%
\StoreBenchExecResult{CpvTranslate}{RelReachSafetyBitVectors}{KII}{All}{}{Median}{1}%
\StoreBenchExecResult{CpvTranslate}{RelReachSafetyBitVectors}{KII}{All}{}{Stdev}{0.00000000000000}%
\StoreBenchExecResult{CpvTranslate}{RelReachSafetyBitVectors}{BtorII}{All}{}{Sum}{48}%
\StoreBenchExecResult{CpvTranslate}{RelReachSafetyBitVectors}{BtorII}{All}{}{Min}{1}%
\StoreBenchExecResult{CpvTranslate}{RelReachSafetyBitVectors}{BtorII}{All}{}{Max}{1}%
\StoreBenchExecResult{CpvTranslate}{RelReachSafetyBitVectors}{BtorII}{All}{}{Avg}{1}%
\StoreBenchExecResult{CpvTranslate}{RelReachSafetyBitVectors}{BtorII}{All}{}{Median}{1}%
\StoreBenchExecResult{CpvTranslate}{RelReachSafetyBitVectors}{BtorII}{All}{}{Stdev}{0.00000000000000}%
\providecommand\StoreBenchExecResult[7]{\expandafter\newcommand\csname#1#2#3#4#5#6\endcsname{#7}}%
\StoreBenchExecResult{CpvTranslate}{RelReachSafetyCombinations}{Status}{All}{}{Score}{0}%
\StoreBenchExecResult{CpvTranslate}{RelReachSafetyCombinations}{Status}{All}{}{Count}{671}%
\StoreBenchExecResult{CpvTranslate}{RelReachSafetyCombinations}{Status}{Correct}{}{Count}{0}%
\StoreBenchExecResult{CpvTranslate}{RelReachSafetyCombinations}{Status}{Correct}{True}{Count}{0}%
\StoreBenchExecResult{CpvTranslate}{RelReachSafetyCombinations}{Status}{Correct}{False}{Count}{0}%
\StoreBenchExecResult{CpvTranslate}{RelReachSafetyCombinations}{Status}{Wrong}{}{Count}{0}%
\StoreBenchExecResult{CpvTranslate}{RelReachSafetyCombinations}{Status}{Wrong}{True}{Count}{0}%
\StoreBenchExecResult{CpvTranslate}{RelReachSafetyCombinations}{Status}{Wrong}{False}{Count}{0}%
\StoreBenchExecResult{CpvTranslate}{RelReachSafetyCombinations}{KII}{All}{}{Sum}{636}%
\StoreBenchExecResult{CpvTranslate}{RelReachSafetyCombinations}{KII}{All}{}{Min}{0}%
\StoreBenchExecResult{CpvTranslate}{RelReachSafetyCombinations}{KII}{All}{}{Max}{1}%
\StoreBenchExecResult{CpvTranslate}{RelReachSafetyCombinations}{KII}{All}{}{Avg}{0.9478390461997019374068554396}%
\StoreBenchExecResult{CpvTranslate}{RelReachSafetyCombinations}{KII}{All}{}{Median}{1}%
\StoreBenchExecResult{CpvTranslate}{RelReachSafetyCombinations}{KII}{All}{}{Stdev}{0.2223514980811715459822416092}%
\StoreBenchExecResult{CpvTranslate}{RelReachSafetyCombinations}{BtorII}{All}{}{Sum}{636}%
\StoreBenchExecResult{CpvTranslate}{RelReachSafetyCombinations}{BtorII}{All}{}{Min}{0}%
\StoreBenchExecResult{CpvTranslate}{RelReachSafetyCombinations}{BtorII}{All}{}{Max}{1}%
\StoreBenchExecResult{CpvTranslate}{RelReachSafetyCombinations}{BtorII}{All}{}{Avg}{0.9478390461997019374068554396}%
\StoreBenchExecResult{CpvTranslate}{RelReachSafetyCombinations}{BtorII}{All}{}{Median}{1}%
\StoreBenchExecResult{CpvTranslate}{RelReachSafetyCombinations}{BtorII}{All}{}{Stdev}{0.2223514980811715459822416092}%
\providecommand\StoreBenchExecResult[7]{\expandafter\newcommand\csname#1#2#3#4#5#6\endcsname{#7}}%
\StoreBenchExecResult{CpvTranslate}{RelReachSafetyControlFlow}{Status}{All}{}{Score}{0}%
\StoreBenchExecResult{CpvTranslate}{RelReachSafetyControlFlow}{Status}{All}{}{Count}{73}%
\StoreBenchExecResult{CpvTranslate}{RelReachSafetyControlFlow}{Status}{Correct}{}{Count}{0}%
\StoreBenchExecResult{CpvTranslate}{RelReachSafetyControlFlow}{Status}{Correct}{True}{Count}{0}%
\StoreBenchExecResult{CpvTranslate}{RelReachSafetyControlFlow}{Status}{Correct}{False}{Count}{0}%
\StoreBenchExecResult{CpvTranslate}{RelReachSafetyControlFlow}{Status}{Wrong}{}{Count}{0}%
\StoreBenchExecResult{CpvTranslate}{RelReachSafetyControlFlow}{Status}{Wrong}{True}{Count}{0}%
\StoreBenchExecResult{CpvTranslate}{RelReachSafetyControlFlow}{Status}{Wrong}{False}{Count}{0}%
\StoreBenchExecResult{CpvTranslate}{RelReachSafetyControlFlow}{KII}{All}{}{Sum}{35}%
\StoreBenchExecResult{CpvTranslate}{RelReachSafetyControlFlow}{KII}{All}{}{Min}{0}%
\StoreBenchExecResult{CpvTranslate}{RelReachSafetyControlFlow}{KII}{All}{}{Max}{1}%
\StoreBenchExecResult{CpvTranslate}{RelReachSafetyControlFlow}{KII}{All}{}{Avg}{0.4794520547945205479452054795}%
\StoreBenchExecResult{CpvTranslate}{RelReachSafetyControlFlow}{KII}{All}{}{Median}{0}%
\StoreBenchExecResult{CpvTranslate}{RelReachSafetyControlFlow}{KII}{All}{}{Stdev}{0.4995776035290539463185022016}%
\StoreBenchExecResult{CpvTranslate}{RelReachSafetyControlFlow}{BtorII}{All}{}{Sum}{33}%
\StoreBenchExecResult{CpvTranslate}{RelReachSafetyControlFlow}{BtorII}{All}{}{Min}{0}%
\StoreBenchExecResult{CpvTranslate}{RelReachSafetyControlFlow}{BtorII}{All}{}{Max}{1}%
\StoreBenchExecResult{CpvTranslate}{RelReachSafetyControlFlow}{BtorII}{All}{}{Avg}{0.4520547945205479452054794521}%
\StoreBenchExecResult{CpvTranslate}{RelReachSafetyControlFlow}{BtorII}{All}{}{Median}{0}%
\StoreBenchExecResult{CpvTranslate}{RelReachSafetyControlFlow}{BtorII}{All}{}{Stdev}{0.4976959486187657532201423789}%
\providecommand\StoreBenchExecResult[7]{\expandafter\newcommand\csname#1#2#3#4#5#6\endcsname{#7}}%
\StoreBenchExecResult{CpvTranslate}{RelReachSafetyECA}{Status}{All}{}{Score}{0}%
\StoreBenchExecResult{CpvTranslate}{RelReachSafetyECA}{Status}{All}{}{Count}{1263}%
\StoreBenchExecResult{CpvTranslate}{RelReachSafetyECA}{Status}{Correct}{}{Count}{0}%
\StoreBenchExecResult{CpvTranslate}{RelReachSafetyECA}{Status}{Correct}{True}{Count}{0}%
\StoreBenchExecResult{CpvTranslate}{RelReachSafetyECA}{Status}{Correct}{False}{Count}{0}%
\StoreBenchExecResult{CpvTranslate}{RelReachSafetyECA}{Status}{Wrong}{}{Count}{0}%
\StoreBenchExecResult{CpvTranslate}{RelReachSafetyECA}{Status}{Wrong}{True}{Count}{0}%
\StoreBenchExecResult{CpvTranslate}{RelReachSafetyECA}{Status}{Wrong}{False}{Count}{0}%
\StoreBenchExecResult{CpvTranslate}{RelReachSafetyECA}{KII}{All}{}{Sum}{1252}%
\StoreBenchExecResult{CpvTranslate}{RelReachSafetyECA}{KII}{All}{}{Min}{0}%
\StoreBenchExecResult{CpvTranslate}{RelReachSafetyECA}{KII}{All}{}{Max}{1}%
\StoreBenchExecResult{CpvTranslate}{RelReachSafetyECA}{KII}{All}{}{Avg}{0.9912905779889152810768012668}%
\StoreBenchExecResult{CpvTranslate}{RelReachSafetyECA}{KII}{All}{}{Median}{1}%
\StoreBenchExecResult{CpvTranslate}{RelReachSafetyECA}{KII}{All}{}{Stdev}{0.09291699510486524517055301934}%
\StoreBenchExecResult{CpvTranslate}{RelReachSafetyECA}{BtorII}{All}{}{Sum}{1050}%
\StoreBenchExecResult{CpvTranslate}{RelReachSafetyECA}{BtorII}{All}{}{Min}{0}%
\StoreBenchExecResult{CpvTranslate}{RelReachSafetyECA}{BtorII}{All}{}{Max}{1}%
\StoreBenchExecResult{CpvTranslate}{RelReachSafetyECA}{BtorII}{All}{}{Avg}{0.8313539192399049881235154394}%
\StoreBenchExecResult{CpvTranslate}{RelReachSafetyECA}{BtorII}{All}{}{Median}{1}%
\StoreBenchExecResult{CpvTranslate}{RelReachSafetyECA}{BtorII}{All}{}{Stdev}{0.3744390206754025294713228064}%
\providecommand\StoreBenchExecResult[7]{\expandafter\newcommand\csname#1#2#3#4#5#6\endcsname{#7}}%
\StoreBenchExecResult{CpvTranslate}{RelReachSafetyFloats}{Status}{All}{}{Score}{0}%
\StoreBenchExecResult{CpvTranslate}{RelReachSafetyFloats}{Status}{All}{}{Count}{1400}%
\StoreBenchExecResult{CpvTranslate}{RelReachSafetyFloats}{Status}{Correct}{}{Count}{0}%
\StoreBenchExecResult{CpvTranslate}{RelReachSafetyFloats}{Status}{Correct}{True}{Count}{0}%
\StoreBenchExecResult{CpvTranslate}{RelReachSafetyFloats}{Status}{Correct}{False}{Count}{0}%
\StoreBenchExecResult{CpvTranslate}{RelReachSafetyFloats}{Status}{Wrong}{}{Count}{0}%
\StoreBenchExecResult{CpvTranslate}{RelReachSafetyFloats}{Status}{Wrong}{True}{Count}{0}%
\StoreBenchExecResult{CpvTranslate}{RelReachSafetyFloats}{Status}{Wrong}{False}{Count}{0}%
\StoreBenchExecResult{CpvTranslate}{RelReachSafetyFloats}{KII}{All}{}{Sum}{321}%
\StoreBenchExecResult{CpvTranslate}{RelReachSafetyFloats}{KII}{All}{}{Min}{0}%
\StoreBenchExecResult{CpvTranslate}{RelReachSafetyFloats}{KII}{All}{}{Max}{1}%
\StoreBenchExecResult{CpvTranslate}{RelReachSafetyFloats}{KII}{All}{}{Avg}{0.2292857142857142857142857143}%
\StoreBenchExecResult{CpvTranslate}{RelReachSafetyFloats}{KII}{All}{}{Median}{0}%
\StoreBenchExecResult{CpvTranslate}{RelReachSafetyFloats}{KII}{All}{}{Stdev}{0.4203733763099229404297488451}%
\StoreBenchExecResult{CpvTranslate}{RelReachSafetyFloats}{BtorII}{All}{}{Sum}{233}%
\StoreBenchExecResult{CpvTranslate}{RelReachSafetyFloats}{BtorII}{All}{}{Min}{0}%
\StoreBenchExecResult{CpvTranslate}{RelReachSafetyFloats}{BtorII}{All}{}{Max}{1}%
\StoreBenchExecResult{CpvTranslate}{RelReachSafetyFloats}{BtorII}{All}{}{Avg}{0.1664285714285714285714285714}%
\StoreBenchExecResult{CpvTranslate}{RelReachSafetyFloats}{BtorII}{All}{}{Median}{0}%
\StoreBenchExecResult{CpvTranslate}{RelReachSafetyFloats}{BtorII}{All}{}{Stdev}{0.3724649004145441978006053140}%
\providecommand\StoreBenchExecResult[7]{\expandafter\newcommand\csname#1#2#3#4#5#6\endcsname{#7}}%
\StoreBenchExecResult{CpvTranslate}{RelReachSafetyHardness}{Status}{All}{}{Score}{0}%
\StoreBenchExecResult{CpvTranslate}{RelReachSafetyHardness}{Status}{All}{}{Count}{6789}%
\StoreBenchExecResult{CpvTranslate}{RelReachSafetyHardness}{Status}{Correct}{}{Count}{0}%
\StoreBenchExecResult{CpvTranslate}{RelReachSafetyHardness}{Status}{Correct}{True}{Count}{0}%
\StoreBenchExecResult{CpvTranslate}{RelReachSafetyHardness}{Status}{Correct}{False}{Count}{0}%
\StoreBenchExecResult{CpvTranslate}{RelReachSafetyHardness}{Status}{Wrong}{}{Count}{0}%
\StoreBenchExecResult{CpvTranslate}{RelReachSafetyHardness}{Status}{Wrong}{True}{Count}{0}%
\StoreBenchExecResult{CpvTranslate}{RelReachSafetyHardness}{Status}{Wrong}{False}{Count}{0}%
\StoreBenchExecResult{CpvTranslate}{RelReachSafetyHardness}{KII}{All}{}{Sum}{6525}%
\StoreBenchExecResult{CpvTranslate}{RelReachSafetyHardness}{KII}{All}{}{Min}{0}%
\StoreBenchExecResult{CpvTranslate}{RelReachSafetyHardness}{KII}{All}{}{Max}{1}%
\StoreBenchExecResult{CpvTranslate}{RelReachSafetyHardness}{KII}{All}{}{Avg}{0.9611135660627485638532920901}%
\StoreBenchExecResult{CpvTranslate}{RelReachSafetyHardness}{KII}{All}{}{Median}{1}%
\StoreBenchExecResult{CpvTranslate}{RelReachSafetyHardness}{KII}{All}{}{Stdev}{0.1933242850572457591482753537}%
\StoreBenchExecResult{CpvTranslate}{RelReachSafetyHardness}{BtorII}{All}{}{Sum}{6079}%
\StoreBenchExecResult{CpvTranslate}{RelReachSafetyHardness}{BtorII}{All}{}{Min}{0}%
\StoreBenchExecResult{CpvTranslate}{RelReachSafetyHardness}{BtorII}{All}{}{Max}{1}%
\StoreBenchExecResult{CpvTranslate}{RelReachSafetyHardness}{BtorII}{All}{}{Avg}{0.8954190602445131830902931212}%
\StoreBenchExecResult{CpvTranslate}{RelReachSafetyHardness}{BtorII}{All}{}{Median}{1}%
\StoreBenchExecResult{CpvTranslate}{RelReachSafetyHardness}{BtorII}{All}{}{Stdev}{0.3060126905789138247415963416}%
\providecommand\StoreBenchExecResult[7]{\expandafter\newcommand\csname#1#2#3#4#5#6\endcsname{#7}}%
\StoreBenchExecResult{CpvTranslate}{RelReachSafetyHardware}{Status}{All}{}{Score}{0}%
\StoreBenchExecResult{CpvTranslate}{RelReachSafetyHardware}{Status}{All}{}{Count}{1224}%
\StoreBenchExecResult{CpvTranslate}{RelReachSafetyHardware}{Status}{Correct}{}{Count}{0}%
\StoreBenchExecResult{CpvTranslate}{RelReachSafetyHardware}{Status}{Correct}{True}{Count}{0}%
\StoreBenchExecResult{CpvTranslate}{RelReachSafetyHardware}{Status}{Correct}{False}{Count}{0}%
\StoreBenchExecResult{CpvTranslate}{RelReachSafetyHardware}{Status}{Wrong}{}{Count}{0}%
\StoreBenchExecResult{CpvTranslate}{RelReachSafetyHardware}{Status}{Wrong}{True}{Count}{0}%
\StoreBenchExecResult{CpvTranslate}{RelReachSafetyHardware}{Status}{Wrong}{False}{Count}{0}%
\StoreBenchExecResult{CpvTranslate}{RelReachSafetyHardware}{KII}{All}{}{Sum}{1182}%
\StoreBenchExecResult{CpvTranslate}{RelReachSafetyHardware}{KII}{All}{}{Min}{0}%
\StoreBenchExecResult{CpvTranslate}{RelReachSafetyHardware}{KII}{All}{}{Max}{1}%
\StoreBenchExecResult{CpvTranslate}{RelReachSafetyHardware}{KII}{All}{}{Avg}{0.9656862745098039215686274510}%
\StoreBenchExecResult{CpvTranslate}{RelReachSafetyHardware}{KII}{All}{}{Median}{1}%
\StoreBenchExecResult{CpvTranslate}{RelReachSafetyHardware}{KII}{All}{}{Stdev}{0.1820337708590896227395067208}%
\StoreBenchExecResult{CpvTranslate}{RelReachSafetyHardware}{BtorII}{All}{}{Sum}{1181}%
\StoreBenchExecResult{CpvTranslate}{RelReachSafetyHardware}{BtorII}{All}{}{Min}{0}%
\StoreBenchExecResult{CpvTranslate}{RelReachSafetyHardware}{BtorII}{All}{}{Max}{1}%
\StoreBenchExecResult{CpvTranslate}{RelReachSafetyHardware}{BtorII}{All}{}{Avg}{0.9648692810457516339869281046}%
\StoreBenchExecResult{CpvTranslate}{RelReachSafetyHardware}{BtorII}{All}{}{Median}{1}%
\StoreBenchExecResult{CpvTranslate}{RelReachSafetyHardware}{BtorII}{All}{}{Stdev}{0.1841101614251803881290711669}%
\providecommand\StoreBenchExecResult[7]{\expandafter\newcommand\csname#1#2#3#4#5#6\endcsname{#7}}%
\StoreBenchExecResult{CpvTranslate}{RelReachSafetyHeap}{Status}{All}{}{Score}{0}%
\StoreBenchExecResult{CpvTranslate}{RelReachSafetyHeap}{Status}{All}{}{Count}{237}%
\StoreBenchExecResult{CpvTranslate}{RelReachSafetyHeap}{Status}{Correct}{}{Count}{0}%
\StoreBenchExecResult{CpvTranslate}{RelReachSafetyHeap}{Status}{Correct}{True}{Count}{0}%
\StoreBenchExecResult{CpvTranslate}{RelReachSafetyHeap}{Status}{Correct}{False}{Count}{0}%
\StoreBenchExecResult{CpvTranslate}{RelReachSafetyHeap}{Status}{Wrong}{}{Count}{0}%
\StoreBenchExecResult{CpvTranslate}{RelReachSafetyHeap}{Status}{Wrong}{True}{Count}{0}%
\StoreBenchExecResult{CpvTranslate}{RelReachSafetyHeap}{Status}{Wrong}{False}{Count}{0}%
\StoreBenchExecResult{CpvTranslate}{RelReachSafetyHeap}{KII}{All}{}{Sum}{78}%
\StoreBenchExecResult{CpvTranslate}{RelReachSafetyHeap}{KII}{All}{}{Min}{0}%
\StoreBenchExecResult{CpvTranslate}{RelReachSafetyHeap}{KII}{All}{}{Max}{1}%
\StoreBenchExecResult{CpvTranslate}{RelReachSafetyHeap}{KII}{All}{}{Avg}{0.3291139240506329113924050633}%
\StoreBenchExecResult{CpvTranslate}{RelReachSafetyHeap}{KII}{All}{}{Median}{0}%
\StoreBenchExecResult{CpvTranslate}{RelReachSafetyHeap}{KII}{All}{}{Stdev}{0.4698914226144452012453432239}%
\StoreBenchExecResult{CpvTranslate}{RelReachSafetyHeap}{BtorII}{All}{}{Sum}{53}%
\StoreBenchExecResult{CpvTranslate}{RelReachSafetyHeap}{BtorII}{All}{}{Min}{0}%
\StoreBenchExecResult{CpvTranslate}{RelReachSafetyHeap}{BtorII}{All}{}{Max}{1}%
\StoreBenchExecResult{CpvTranslate}{RelReachSafetyHeap}{BtorII}{All}{}{Avg}{0.2236286919831223628691983122}%
\StoreBenchExecResult{CpvTranslate}{RelReachSafetyHeap}{BtorII}{All}{}{Median}{0}%
\StoreBenchExecResult{CpvTranslate}{RelReachSafetyHeap}{BtorII}{All}{}{Stdev}{0.4166760133545488201338244783}%
\providecommand\StoreBenchExecResult[7]{\expandafter\newcommand\csname#1#2#3#4#5#6\endcsname{#7}}%
\StoreBenchExecResult{CpvTranslate}{RelReachSafetyLoops}{Status}{All}{}{Score}{0}%
\StoreBenchExecResult{CpvTranslate}{RelReachSafetyLoops}{Status}{All}{}{Count}{762}%
\StoreBenchExecResult{CpvTranslate}{RelReachSafetyLoops}{Status}{Correct}{}{Count}{0}%
\StoreBenchExecResult{CpvTranslate}{RelReachSafetyLoops}{Status}{Correct}{True}{Count}{0}%
\StoreBenchExecResult{CpvTranslate}{RelReachSafetyLoops}{Status}{Correct}{False}{Count}{0}%
\StoreBenchExecResult{CpvTranslate}{RelReachSafetyLoops}{Status}{Wrong}{}{Count}{0}%
\StoreBenchExecResult{CpvTranslate}{RelReachSafetyLoops}{Status}{Wrong}{True}{Count}{0}%
\StoreBenchExecResult{CpvTranslate}{RelReachSafetyLoops}{Status}{Wrong}{False}{Count}{0}%
\StoreBenchExecResult{CpvTranslate}{RelReachSafetyLoops}{KII}{All}{}{Sum}{753}%
\StoreBenchExecResult{CpvTranslate}{RelReachSafetyLoops}{KII}{All}{}{Min}{0}%
\StoreBenchExecResult{CpvTranslate}{RelReachSafetyLoops}{KII}{All}{}{Max}{1}%
\StoreBenchExecResult{CpvTranslate}{RelReachSafetyLoops}{KII}{All}{}{Avg}{0.9881889763779527559055118110}%
\StoreBenchExecResult{CpvTranslate}{RelReachSafetyLoops}{KII}{All}{}{Median}{1}%
\StoreBenchExecResult{CpvTranslate}{RelReachSafetyLoops}{KII}{All}{}{Stdev}{0.1080348246772617579063612756}%
\StoreBenchExecResult{CpvTranslate}{RelReachSafetyLoops}{BtorII}{All}{}{Sum}{742}%
\StoreBenchExecResult{CpvTranslate}{RelReachSafetyLoops}{BtorII}{All}{}{Min}{0}%
\StoreBenchExecResult{CpvTranslate}{RelReachSafetyLoops}{BtorII}{All}{}{Max}{1}%
\StoreBenchExecResult{CpvTranslate}{RelReachSafetyLoops}{BtorII}{All}{}{Avg}{0.9737532808398950131233595801}%
\StoreBenchExecResult{CpvTranslate}{RelReachSafetyLoops}{BtorII}{All}{}{Median}{1}%
\StoreBenchExecResult{CpvTranslate}{RelReachSafetyLoops}{BtorII}{All}{}{Stdev}{0.1598681609747092970196468690}%
\providecommand\StoreBenchExecResult[7]{\expandafter\newcommand\csname#1#2#3#4#5#6\endcsname{#7}}%
\StoreBenchExecResult{CpvTranslate}{RelReachSafetyProductLines}{Status}{All}{}{Score}{0}%
\StoreBenchExecResult{CpvTranslate}{RelReachSafetyProductLines}{Status}{All}{}{Count}{597}%
\StoreBenchExecResult{CpvTranslate}{RelReachSafetyProductLines}{Status}{Correct}{}{Count}{0}%
\StoreBenchExecResult{CpvTranslate}{RelReachSafetyProductLines}{Status}{Correct}{True}{Count}{0}%
\StoreBenchExecResult{CpvTranslate}{RelReachSafetyProductLines}{Status}{Correct}{False}{Count}{0}%
\StoreBenchExecResult{CpvTranslate}{RelReachSafetyProductLines}{Status}{Wrong}{}{Count}{0}%
\StoreBenchExecResult{CpvTranslate}{RelReachSafetyProductLines}{Status}{Wrong}{True}{Count}{0}%
\StoreBenchExecResult{CpvTranslate}{RelReachSafetyProductLines}{Status}{Wrong}{False}{Count}{0}%
\StoreBenchExecResult{CpvTranslate}{RelReachSafetyProductLines}{KII}{All}{}{Sum}{597}%
\StoreBenchExecResult{CpvTranslate}{RelReachSafetyProductLines}{KII}{All}{}{Min}{1}%
\StoreBenchExecResult{CpvTranslate}{RelReachSafetyProductLines}{KII}{All}{}{Max}{1}%
\StoreBenchExecResult{CpvTranslate}{RelReachSafetyProductLines}{KII}{All}{}{Avg}{1}%
\StoreBenchExecResult{CpvTranslate}{RelReachSafetyProductLines}{KII}{All}{}{Median}{1}%
\StoreBenchExecResult{CpvTranslate}{RelReachSafetyProductLines}{KII}{All}{}{Stdev}{0.00000000000000}%
\StoreBenchExecResult{CpvTranslate}{RelReachSafetyProductLines}{BtorII}{All}{}{Sum}{419}%
\StoreBenchExecResult{CpvTranslate}{RelReachSafetyProductLines}{BtorII}{All}{}{Min}{0}%
\StoreBenchExecResult{CpvTranslate}{RelReachSafetyProductLines}{BtorII}{All}{}{Max}{1}%
\StoreBenchExecResult{CpvTranslate}{RelReachSafetyProductLines}{BtorII}{All}{}{Avg}{0.7018425460636515912897822446}%
\StoreBenchExecResult{CpvTranslate}{RelReachSafetyProductLines}{BtorII}{All}{}{Median}{1}%
\StoreBenchExecResult{CpvTranslate}{RelReachSafetyProductLines}{BtorII}{All}{}{Stdev}{0.4574489989042960901757564879}%
\providecommand\StoreBenchExecResult[7]{\expandafter\newcommand\csname#1#2#3#4#5#6\endcsname{#7}}%
\StoreBenchExecResult{CpvTranslate}{RelReachSafetySequentialized}{Status}{All}{}{Score}{0}%
\StoreBenchExecResult{CpvTranslate}{RelReachSafetySequentialized}{Status}{All}{}{Count}{585}%
\StoreBenchExecResult{CpvTranslate}{RelReachSafetySequentialized}{Status}{Correct}{}{Count}{0}%
\StoreBenchExecResult{CpvTranslate}{RelReachSafetySequentialized}{Status}{Correct}{True}{Count}{0}%
\StoreBenchExecResult{CpvTranslate}{RelReachSafetySequentialized}{Status}{Correct}{False}{Count}{0}%
\StoreBenchExecResult{CpvTranslate}{RelReachSafetySequentialized}{Status}{Wrong}{}{Count}{0}%
\StoreBenchExecResult{CpvTranslate}{RelReachSafetySequentialized}{Status}{Wrong}{True}{Count}{0}%
\StoreBenchExecResult{CpvTranslate}{RelReachSafetySequentialized}{Status}{Wrong}{False}{Count}{0}%
\StoreBenchExecResult{CpvTranslate}{RelReachSafetySequentialized}{KII}{All}{}{Sum}{245}%
\StoreBenchExecResult{CpvTranslate}{RelReachSafetySequentialized}{KII}{All}{}{Min}{0}%
\StoreBenchExecResult{CpvTranslate}{RelReachSafetySequentialized}{KII}{All}{}{Max}{1}%
\StoreBenchExecResult{CpvTranslate}{RelReachSafetySequentialized}{KII}{All}{}{Avg}{0.4188034188034188034188034188}%
\StoreBenchExecResult{CpvTranslate}{RelReachSafetySequentialized}{KII}{All}{}{Median}{0}%
\StoreBenchExecResult{CpvTranslate}{RelReachSafetySequentialized}{KII}{All}{}{Stdev}{0.4933630663132243392948695669}%
\StoreBenchExecResult{CpvTranslate}{RelReachSafetySequentialized}{BtorII}{All}{}{Sum}{217}%
\StoreBenchExecResult{CpvTranslate}{RelReachSafetySequentialized}{BtorII}{All}{}{Min}{0}%
\StoreBenchExecResult{CpvTranslate}{RelReachSafetySequentialized}{BtorII}{All}{}{Max}{1}%
\StoreBenchExecResult{CpvTranslate}{RelReachSafetySequentialized}{BtorII}{All}{}{Avg}{0.3709401709401709401709401709}%
\StoreBenchExecResult{CpvTranslate}{RelReachSafetySequentialized}{BtorII}{All}{}{Median}{0}%
\StoreBenchExecResult{CpvTranslate}{RelReachSafetySequentialized}{BtorII}{All}{}{Stdev}{0.4830564775707367559105739696}%
\providecommand\StoreBenchExecResult[7]{\expandafter\newcommand\csname#1#2#3#4#5#6\endcsname{#7}}%
\StoreBenchExecResult{CpvTranslate}{RelReachSafetyXCSP}{Status}{All}{}{Score}{0}%
\StoreBenchExecResult{CpvTranslate}{RelReachSafetyXCSP}{Status}{All}{}{Count}{119}%
\StoreBenchExecResult{CpvTranslate}{RelReachSafetyXCSP}{Status}{Correct}{}{Count}{0}%
\StoreBenchExecResult{CpvTranslate}{RelReachSafetyXCSP}{Status}{Correct}{True}{Count}{0}%
\StoreBenchExecResult{CpvTranslate}{RelReachSafetyXCSP}{Status}{Correct}{False}{Count}{0}%
\StoreBenchExecResult{CpvTranslate}{RelReachSafetyXCSP}{Status}{Wrong}{}{Count}{0}%
\StoreBenchExecResult{CpvTranslate}{RelReachSafetyXCSP}{Status}{Wrong}{True}{Count}{0}%
\StoreBenchExecResult{CpvTranslate}{RelReachSafetyXCSP}{Status}{Wrong}{False}{Count}{0}%
\StoreBenchExecResult{CpvTranslate}{RelReachSafetyXCSP}{KII}{All}{}{Sum}{119}%
\StoreBenchExecResult{CpvTranslate}{RelReachSafetyXCSP}{KII}{All}{}{Min}{1}%
\StoreBenchExecResult{CpvTranslate}{RelReachSafetyXCSP}{KII}{All}{}{Max}{1}%
\StoreBenchExecResult{CpvTranslate}{RelReachSafetyXCSP}{KII}{All}{}{Avg}{1}%
\StoreBenchExecResult{CpvTranslate}{RelReachSafetyXCSP}{KII}{All}{}{Median}{1}%
\StoreBenchExecResult{CpvTranslate}{RelReachSafetyXCSP}{KII}{All}{}{Stdev}{0.00000000000000}%
\StoreBenchExecResult{CpvTranslate}{RelReachSafetyXCSP}{BtorII}{All}{}{Sum}{119}%
\StoreBenchExecResult{CpvTranslate}{RelReachSafetyXCSP}{BtorII}{All}{}{Min}{1}%
\StoreBenchExecResult{CpvTranslate}{RelReachSafetyXCSP}{BtorII}{All}{}{Max}{1}%
\StoreBenchExecResult{CpvTranslate}{RelReachSafetyXCSP}{BtorII}{All}{}{Avg}{1}%
\StoreBenchExecResult{CpvTranslate}{RelReachSafetyXCSP}{BtorII}{All}{}{Median}{1}%
\StoreBenchExecResult{CpvTranslate}{RelReachSafetyXCSP}{BtorII}{All}{}{Stdev}{0.00000000000000}%
\edef\CpvTranslateRelReachSafetyStatusAllCount{\the\numexpr\CpvTranslateRelReachSafetyArraysStatusAllCount+\CpvTranslateRelReachSafetyBitVectorsStatusAllCount+\CpvTranslateRelReachSafetyCombinationsStatusAllCount+\CpvTranslateRelReachSafetyControlFlowStatusAllCount+\CpvTranslateRelReachSafetyECAStatusAllCount+\CpvTranslateRelReachSafetyFloatsStatusAllCount+\CpvTranslateRelReachSafetyHardnessStatusAllCount+\CpvTranslateRelReachSafetyHardwareStatusAllCount+\CpvTranslateRelReachSafetyHeapStatusAllCount+\CpvTranslateRelReachSafetyLoopsStatusAllCount+\CpvTranslateRelReachSafetyProductLinesStatusAllCount+\CpvTranslateRelReachSafetySequentializedStatusAllCount+\CpvTranslateRelReachSafetyXCSPStatusAllCount}
\edef\CpvTranslateRelReachSafetyKIIAllSum{\the\numexpr\CpvTranslateRelReachSafetyArraysKIIAllSum+\CpvTranslateRelReachSafetyBitVectorsKIIAllSum+\CpvTranslateRelReachSafetyCombinationsKIIAllSum+\CpvTranslateRelReachSafetyControlFlowKIIAllSum+\CpvTranslateRelReachSafetyECAKIIAllSum+\CpvTranslateRelReachSafetyFloatsKIIAllSum+\CpvTranslateRelReachSafetyHardnessKIIAllSum+\CpvTranslateRelReachSafetyHardwareKIIAllSum+\CpvTranslateRelReachSafetyHeapKIIAllSum+\CpvTranslateRelReachSafetyLoopsKIIAllSum+\CpvTranslateRelReachSafetyProductLinesKIIAllSum+\CpvTranslateRelReachSafetySequentializedKIIAllSum+\CpvTranslateRelReachSafetyXCSPKIIAllSum}
\edef\CpvTranslateRelReachSafetyBtorIIAllSum{\the\numexpr\CpvTranslateRelReachSafetyArraysBtorIIAllSum+\CpvTranslateRelReachSafetyBitVectorsBtorIIAllSum+\CpvTranslateRelReachSafetyCombinationsBtorIIAllSum+\CpvTranslateRelReachSafetyControlFlowBtorIIAllSum+\CpvTranslateRelReachSafetyECABtorIIAllSum+\CpvTranslateRelReachSafetyFloatsBtorIIAllSum+\CpvTranslateRelReachSafetyHardnessBtorIIAllSum+\CpvTranslateRelReachSafetyHardwareBtorIIAllSum+\CpvTranslateRelReachSafetyHeapBtorIIAllSum+\CpvTranslateRelReachSafetyLoopsBtorIIAllSum+\CpvTranslateRelReachSafetyProductLinesBtorIIAllSum+\CpvTranslateRelReachSafetySequentializedBtorIIAllSum+\CpvTranslateRelReachSafetyXCSPBtorIIAllSum}
\providecommand\StoreBenchExecResult[7]{\expandafter\newcommand\csname#1#2#3#4#5#6\endcsname{#7}}%
\StoreBenchExecResult{CpvTranslate}{RelTranslationDoneReachSafetyRel}{Cputime}{All}{}{Sum}{81818.492030}%
\StoreBenchExecResult{CpvTranslate}{RelTranslationDoneReachSafetyRel}{Cputime}{All}{}{Min}{2.237242}%
\StoreBenchExecResult{CpvTranslate}{RelTranslationDoneReachSafetyRel}{Cputime}{All}{}{Max}{780.886326}%
\StoreBenchExecResult{CpvTranslate}{RelTranslationDoneReachSafetyRel}{Cputime}{All}{}{Avg}{7.277930264187866927592954990}%
\StoreBenchExecResult{CpvTranslate}{RelTranslationDoneReachSafetyRel}{Cputime}{All}{}{Median}{2.6230535}%
\StoreBenchExecResult{CpvTranslate}{RelTranslationDoneReachSafetyRel}{Cputime}{All}{}{Stdev}{31.86968573975640939755865496}%
\StoreBenchExecResult{CpvTranslate}{RelTranslationDoneReachSafetyRel}{Cputime}{All}{}{Unit}{s}%
\providecommand\StoreBenchExecResult[7]{\expandafter\newcommand\csname#1#2#3#4#5#6\endcsname{#7}}%
\StoreBenchExecResult{CpvTranslate}{RelLIIsTerminationBitVectors}{Status}{All}{}{Score}{0}%
\StoreBenchExecResult{CpvTranslate}{RelLIIsTerminationBitVectors}{Status}{All}{}{Count}{34}%
\StoreBenchExecResult{CpvTranslate}{RelLIIsTerminationBitVectors}{Status}{Correct}{}{Count}{0}%
\StoreBenchExecResult{CpvTranslate}{RelLIIsTerminationBitVectors}{Status}{Correct}{True}{Count}{0}%
\StoreBenchExecResult{CpvTranslate}{RelLIIsTerminationBitVectors}{Status}{Correct}{False}{Count}{0}%
\StoreBenchExecResult{CpvTranslate}{RelLIIsTerminationBitVectors}{Status}{Wrong}{}{Count}{0}%
\StoreBenchExecResult{CpvTranslate}{RelLIIsTerminationBitVectors}{Status}{Wrong}{True}{Count}{0}%
\StoreBenchExecResult{CpvTranslate}{RelLIIsTerminationBitVectors}{Status}{Wrong}{False}{Count}{0}%
\StoreBenchExecResult{CpvTranslate}{RelLIIsTerminationBitVectors}{KII}{All}{}{Sum}{32}%
\StoreBenchExecResult{CpvTranslate}{RelLIIsTerminationBitVectors}{KII}{All}{}{Min}{0}%
\StoreBenchExecResult{CpvTranslate}{RelLIIsTerminationBitVectors}{KII}{All}{}{Max}{1}%
\StoreBenchExecResult{CpvTranslate}{RelLIIsTerminationBitVectors}{KII}{All}{}{Avg}{0.9411764705882352941176470588}%
\StoreBenchExecResult{CpvTranslate}{RelLIIsTerminationBitVectors}{KII}{All}{}{Median}{1}%
\StoreBenchExecResult{CpvTranslate}{RelLIIsTerminationBitVectors}{KII}{All}{}{Stdev}{0.2352941176470588235294117654}%
\StoreBenchExecResult{CpvTranslate}{RelLIIsTerminationBitVectors}{BtorII}{All}{}{Sum}{32}%
\StoreBenchExecResult{CpvTranslate}{RelLIIsTerminationBitVectors}{BtorII}{All}{}{Min}{0}%
\StoreBenchExecResult{CpvTranslate}{RelLIIsTerminationBitVectors}{BtorII}{All}{}{Max}{1}%
\StoreBenchExecResult{CpvTranslate}{RelLIIsTerminationBitVectors}{BtorII}{All}{}{Avg}{0.9411764705882352941176470588}%
\StoreBenchExecResult{CpvTranslate}{RelLIIsTerminationBitVectors}{BtorII}{All}{}{Median}{1}%
\StoreBenchExecResult{CpvTranslate}{RelLIIsTerminationBitVectors}{BtorII}{All}{}{Stdev}{0.2352941176470588235294117654}%
\providecommand\StoreBenchExecResult[7]{\expandafter\newcommand\csname#1#2#3#4#5#6\endcsname{#7}}%
\StoreBenchExecResult{CpvTranslate}{RelLIIsTerminationMainControlFlow}{Status}{All}{}{Score}{0}%
\StoreBenchExecResult{CpvTranslate}{RelLIIsTerminationMainControlFlow}{Status}{All}{}{Count}{253}%
\StoreBenchExecResult{CpvTranslate}{RelLIIsTerminationMainControlFlow}{Status}{Correct}{}{Count}{0}%
\StoreBenchExecResult{CpvTranslate}{RelLIIsTerminationMainControlFlow}{Status}{Correct}{True}{Count}{0}%
\StoreBenchExecResult{CpvTranslate}{RelLIIsTerminationMainControlFlow}{Status}{Correct}{False}{Count}{0}%
\StoreBenchExecResult{CpvTranslate}{RelLIIsTerminationMainControlFlow}{Status}{Wrong}{}{Count}{0}%
\StoreBenchExecResult{CpvTranslate}{RelLIIsTerminationMainControlFlow}{Status}{Wrong}{True}{Count}{0}%
\StoreBenchExecResult{CpvTranslate}{RelLIIsTerminationMainControlFlow}{Status}{Wrong}{False}{Count}{0}%
\StoreBenchExecResult{CpvTranslate}{RelLIIsTerminationMainControlFlow}{KII}{All}{}{Sum}{243}%
\StoreBenchExecResult{CpvTranslate}{RelLIIsTerminationMainControlFlow}{KII}{All}{}{Min}{0}%
\StoreBenchExecResult{CpvTranslate}{RelLIIsTerminationMainControlFlow}{KII}{All}{}{Max}{1}%
\StoreBenchExecResult{CpvTranslate}{RelLIIsTerminationMainControlFlow}{KII}{All}{}{Avg}{0.9604743083003952569169960474}%
\StoreBenchExecResult{CpvTranslate}{RelLIIsTerminationMainControlFlow}{KII}{All}{}{Median}{1}%
\StoreBenchExecResult{CpvTranslate}{RelLIIsTerminationMainControlFlow}{KII}{All}{}{Stdev}{0.1948420165038140324550485401}%
\StoreBenchExecResult{CpvTranslate}{RelLIIsTerminationMainControlFlow}{BtorII}{All}{}{Sum}{243}%
\StoreBenchExecResult{CpvTranslate}{RelLIIsTerminationMainControlFlow}{BtorII}{All}{}{Min}{0}%
\StoreBenchExecResult{CpvTranslate}{RelLIIsTerminationMainControlFlow}{BtorII}{All}{}{Max}{1}%
\StoreBenchExecResult{CpvTranslate}{RelLIIsTerminationMainControlFlow}{BtorII}{All}{}{Avg}{0.9604743083003952569169960474}%
\StoreBenchExecResult{CpvTranslate}{RelLIIsTerminationMainControlFlow}{BtorII}{All}{}{Median}{1}%
\StoreBenchExecResult{CpvTranslate}{RelLIIsTerminationMainControlFlow}{BtorII}{All}{}{Stdev}{0.1948420165038140324550485401}%
\providecommand\StoreBenchExecResult[7]{\expandafter\newcommand\csname#1#2#3#4#5#6\endcsname{#7}}%
\StoreBenchExecResult{CpvTranslate}{RelLIIsTerminationMainHeap}{Status}{All}{}{Score}{0}%
\StoreBenchExecResult{CpvTranslate}{RelLIIsTerminationMainHeap}{Status}{All}{}{Count}{180}%
\StoreBenchExecResult{CpvTranslate}{RelLIIsTerminationMainHeap}{Status}{Correct}{}{Count}{0}%
\StoreBenchExecResult{CpvTranslate}{RelLIIsTerminationMainHeap}{Status}{Correct}{True}{Count}{0}%
\StoreBenchExecResult{CpvTranslate}{RelLIIsTerminationMainHeap}{Status}{Correct}{False}{Count}{0}%
\StoreBenchExecResult{CpvTranslate}{RelLIIsTerminationMainHeap}{Status}{Wrong}{}{Count}{0}%
\StoreBenchExecResult{CpvTranslate}{RelLIIsTerminationMainHeap}{Status}{Wrong}{True}{Count}{0}%
\StoreBenchExecResult{CpvTranslate}{RelLIIsTerminationMainHeap}{Status}{Wrong}{False}{Count}{0}%
\StoreBenchExecResult{CpvTranslate}{RelLIIsTerminationMainHeap}{KII}{All}{}{Sum}{0}%
\StoreBenchExecResult{CpvTranslate}{RelLIIsTerminationMainHeap}{KII}{All}{}{Min}{0}%
\StoreBenchExecResult{CpvTranslate}{RelLIIsTerminationMainHeap}{KII}{All}{}{Max}{0}%
\StoreBenchExecResult{CpvTranslate}{RelLIIsTerminationMainHeap}{KII}{All}{}{Avg}{0}%
\StoreBenchExecResult{CpvTranslate}{RelLIIsTerminationMainHeap}{KII}{All}{}{Median}{0}%
\StoreBenchExecResult{CpvTranslate}{RelLIIsTerminationMainHeap}{KII}{All}{}{Stdev}{0.00000000000000}%
\StoreBenchExecResult{CpvTranslate}{RelLIIsTerminationMainHeap}{BtorII}{All}{}{Sum}{0}%
\StoreBenchExecResult{CpvTranslate}{RelLIIsTerminationMainHeap}{BtorII}{All}{}{Min}{0}%
\StoreBenchExecResult{CpvTranslate}{RelLIIsTerminationMainHeap}{BtorII}{All}{}{Max}{0}%
\StoreBenchExecResult{CpvTranslate}{RelLIIsTerminationMainHeap}{BtorII}{All}{}{Avg}{0}%
\StoreBenchExecResult{CpvTranslate}{RelLIIsTerminationMainHeap}{BtorII}{All}{}{Median}{0}%
\StoreBenchExecResult{CpvTranslate}{RelLIIsTerminationMainHeap}{BtorII}{All}{}{Stdev}{0.00000000000000}%
\providecommand\StoreBenchExecResult[7]{\expandafter\newcommand\csname#1#2#3#4#5#6\endcsname{#7}}%
\StoreBenchExecResult{CpvTranslate}{RelLIIsTerminationOther}{Status}{All}{}{Score}{0}%
\StoreBenchExecResult{CpvTranslate}{RelLIIsTerminationOther}{Status}{All}{}{Count}{1526}%
\StoreBenchExecResult{CpvTranslate}{RelLIIsTerminationOther}{Status}{Correct}{}{Count}{0}%
\StoreBenchExecResult{CpvTranslate}{RelLIIsTerminationOther}{Status}{Correct}{True}{Count}{0}%
\StoreBenchExecResult{CpvTranslate}{RelLIIsTerminationOther}{Status}{Correct}{False}{Count}{0}%
\StoreBenchExecResult{CpvTranslate}{RelLIIsTerminationOther}{Status}{Wrong}{}{Count}{0}%
\StoreBenchExecResult{CpvTranslate}{RelLIIsTerminationOther}{Status}{Wrong}{True}{Count}{0}%
\StoreBenchExecResult{CpvTranslate}{RelLIIsTerminationOther}{Status}{Wrong}{False}{Count}{0}%
\StoreBenchExecResult{CpvTranslate}{RelLIIsTerminationOther}{KII}{All}{}{Sum}{1391}%
\StoreBenchExecResult{CpvTranslate}{RelLIIsTerminationOther}{KII}{All}{}{Min}{0}%
\StoreBenchExecResult{CpvTranslate}{RelLIIsTerminationOther}{KII}{All}{}{Max}{1}%
\StoreBenchExecResult{CpvTranslate}{RelLIIsTerminationOther}{KII}{All}{}{Avg}{0.9115334207077326343381389253}%
\StoreBenchExecResult{CpvTranslate}{RelLIIsTerminationOther}{KII}{All}{}{Median}{1}%
\StoreBenchExecResult{CpvTranslate}{RelLIIsTerminationOther}{KII}{All}{}{Stdev}{0.2839722585757142873056637163}%
\StoreBenchExecResult{CpvTranslate}{RelLIIsTerminationOther}{BtorII}{All}{}{Sum}{1166}%
\StoreBenchExecResult{CpvTranslate}{RelLIIsTerminationOther}{BtorII}{All}{}{Min}{0}%
\StoreBenchExecResult{CpvTranslate}{RelLIIsTerminationOther}{BtorII}{All}{}{Max}{1}%
\StoreBenchExecResult{CpvTranslate}{RelLIIsTerminationOther}{BtorII}{All}{}{Avg}{0.7640891218872870249017038008}%
\StoreBenchExecResult{CpvTranslate}{RelLIIsTerminationOther}{BtorII}{All}{}{Median}{1}%
\StoreBenchExecResult{CpvTranslate}{RelLIIsTerminationOther}{BtorII}{All}{}{Stdev}{0.4245667623599398733916585847}%
\edef\CpvTranslateRelLIIsTerminationStatusAllCount{\the\numexpr\CpvTranslateRelLIIsTerminationBitVectorsStatusAllCount+\CpvTranslateRelLIIsTerminationMainControlFlowStatusAllCount+\CpvTranslateRelLIIsTerminationMainHeapStatusAllCount+\CpvTranslateRelLIIsTerminationOtherStatusAllCount}
\edef\CpvTranslateRelLIIsTerminationKIIAllSum{\the\numexpr\CpvTranslateRelLIIsTerminationBitVectorsKIIAllSum+\CpvTranslateRelLIIsTerminationMainControlFlowKIIAllSum+\CpvTranslateRelLIIsTerminationMainHeapKIIAllSum+\CpvTranslateRelLIIsTerminationOtherKIIAllSum}
\edef\CpvTranslateRelLIIsTerminationBtorIIAllSum{\the\numexpr\CpvTranslateRelLIIsTerminationBitVectorsBtorIIAllSum+\CpvTranslateRelLIIsTerminationMainControlFlowBtorIIAllSum+\CpvTranslateRelLIIsTerminationMainHeapBtorIIAllSum+\CpvTranslateRelLIIsTerminationOtherBtorIIAllSum}
\providecommand\StoreBenchExecResult[7]{\expandafter\newcommand\csname#1#2#3#4#5#6\endcsname{#7}}%
\StoreBenchExecResult{CpvTranslate}{RelLIIsTranslationDoneTerminationRel}{Cputime}{All}{}{Sum}{5534.631471}%
\StoreBenchExecResult{CpvTranslate}{RelLIIsTranslationDoneTerminationRel}{Cputime}{All}{}{Min}{2.258121}%
\StoreBenchExecResult{CpvTranslate}{RelLIIsTranslationDoneTerminationRel}{Cputime}{All}{}{Max}{415.856574}%
\StoreBenchExecResult{CpvTranslate}{RelLIIsTranslationDoneTerminationRel}{Cputime}{All}{}{Avg}{3.840826836224843858431644691}%
\StoreBenchExecResult{CpvTranslate}{RelLIIsTranslationDoneTerminationRel}{Cputime}{All}{}{Median}{2.500825}%
\StoreBenchExecResult{CpvTranslate}{RelLIIsTranslationDoneTerminationRel}{Cputime}{All}{}{Stdev}{12.02745451624343295909353971}%
\StoreBenchExecResult{CpvTranslate}{RelLIIsTranslationDoneTerminationRel}{Cputime}{All}{}{Unit}{s}%

%% file: eval-results/tex/data-commands.encoding.tex
\providecommand\StoreBenchExecResult[7]{\expandafter\newcommand\csname#1#2#3#4#5#6\endcsname{#7}}%
\StoreBenchExecResult{Abc}{ImcReachSafetyBitVectorsSampledBvFunc}{Status}{All}{}{Score}{0}%
\StoreBenchExecResult{Abc}{ImcReachSafetyBitVectorsSampledBvFunc}{Status}{All}{}{Count}{47}%
\StoreBenchExecResult{Abc}{ImcReachSafetyBitVectorsSampledBvFunc}{Status}{Correct}{}{Count}{40}%
\StoreBenchExecResult{Abc}{ImcReachSafetyBitVectorsSampledBvFunc}{Status}{Correct}{True}{Count}{29}%
\StoreBenchExecResult{Abc}{ImcReachSafetyBitVectorsSampledBvFunc}{Status}{Correct}{False}{Count}{11}%
\StoreBenchExecResult{Abc}{ImcReachSafetyBitVectorsSampledBvFunc}{Status}{Wrong}{}{Count}{0}%
\StoreBenchExecResult{Abc}{ImcReachSafetyBitVectorsSampledBvFunc}{Status}{Wrong}{True}{Count}{0}%
\StoreBenchExecResult{Abc}{ImcReachSafetyBitVectorsSampledBvFunc}{Status}{Wrong}{False}{Count}{0}%
\providecommand\StoreBenchExecResult[7]{\expandafter\newcommand\csname#1#2#3#4#5#6\endcsname{#7}}%
\StoreBenchExecResult{Abc}{ImcReachSafetyCombinationsSampledBvFunc}{Status}{All}{}{Score}{0}%
\StoreBenchExecResult{Abc}{ImcReachSafetyCombinationsSampledBvFunc}{Status}{All}{}{Count}{110}%
\StoreBenchExecResult{Abc}{ImcReachSafetyCombinationsSampledBvFunc}{Status}{Correct}{}{Count}{72}%
\StoreBenchExecResult{Abc}{ImcReachSafetyCombinationsSampledBvFunc}{Status}{Correct}{True}{Count}{1}%
\StoreBenchExecResult{Abc}{ImcReachSafetyCombinationsSampledBvFunc}{Status}{Correct}{False}{Count}{71}%
\StoreBenchExecResult{Abc}{ImcReachSafetyCombinationsSampledBvFunc}{Status}{Wrong}{}{Count}{0}%
\StoreBenchExecResult{Abc}{ImcReachSafetyCombinationsSampledBvFunc}{Status}{Wrong}{True}{Count}{0}%
\StoreBenchExecResult{Abc}{ImcReachSafetyCombinationsSampledBvFunc}{Status}{Wrong}{False}{Count}{0}%
\providecommand\StoreBenchExecResult[7]{\expandafter\newcommand\csname#1#2#3#4#5#6\endcsname{#7}}%
\StoreBenchExecResult{Abc}{ImcReachSafetyControlFlowSampledBvFunc}{Status}{All}{}{Score}{0}%
\StoreBenchExecResult{Abc}{ImcReachSafetyControlFlowSampledBvFunc}{Status}{All}{}{Count}{29}%
\StoreBenchExecResult{Abc}{ImcReachSafetyControlFlowSampledBvFunc}{Status}{Correct}{}{Count}{26}%
\StoreBenchExecResult{Abc}{ImcReachSafetyControlFlowSampledBvFunc}{Status}{Correct}{True}{Count}{24}%
\StoreBenchExecResult{Abc}{ImcReachSafetyControlFlowSampledBvFunc}{Status}{Correct}{False}{Count}{2}%
\StoreBenchExecResult{Abc}{ImcReachSafetyControlFlowSampledBvFunc}{Status}{Wrong}{}{Count}{0}%
\StoreBenchExecResult{Abc}{ImcReachSafetyControlFlowSampledBvFunc}{Status}{Wrong}{True}{Count}{0}%
\StoreBenchExecResult{Abc}{ImcReachSafetyControlFlowSampledBvFunc}{Status}{Wrong}{False}{Count}{0}%
\providecommand\StoreBenchExecResult[7]{\expandafter\newcommand\csname#1#2#3#4#5#6\endcsname{#7}}%
\StoreBenchExecResult{Abc}{ImcReachSafetyECASampledBvFunc}{Status}{All}{}{Score}{0}%
\StoreBenchExecResult{Abc}{ImcReachSafetyECASampledBvFunc}{Status}{All}{}{Count}{150}%
\StoreBenchExecResult{Abc}{ImcReachSafetyECASampledBvFunc}{Status}{Correct}{}{Count}{111}%
\StoreBenchExecResult{Abc}{ImcReachSafetyECASampledBvFunc}{Status}{Correct}{True}{Count}{49}%
\StoreBenchExecResult{Abc}{ImcReachSafetyECASampledBvFunc}{Status}{Correct}{False}{Count}{62}%
\StoreBenchExecResult{Abc}{ImcReachSafetyECASampledBvFunc}{Status}{Wrong}{}{Count}{0}%
\StoreBenchExecResult{Abc}{ImcReachSafetyECASampledBvFunc}{Status}{Wrong}{True}{Count}{0}%
\StoreBenchExecResult{Abc}{ImcReachSafetyECASampledBvFunc}{Status}{Wrong}{False}{Count}{0}%
\providecommand\StoreBenchExecResult[7]{\expandafter\newcommand\csname#1#2#3#4#5#6\endcsname{#7}}%
\StoreBenchExecResult{Abc}{ImcReachSafetyFloatsSampledBvFunc}{Status}{All}{}{Score}{0}%
\StoreBenchExecResult{Abc}{ImcReachSafetyFloatsSampledBvFunc}{Status}{All}{}{Count}{10}%
\StoreBenchExecResult{Abc}{ImcReachSafetyFloatsSampledBvFunc}{Status}{Correct}{}{Count}{10}%
\StoreBenchExecResult{Abc}{ImcReachSafetyFloatsSampledBvFunc}{Status}{Correct}{True}{Count}{10}%
\StoreBenchExecResult{Abc}{ImcReachSafetyFloatsSampledBvFunc}{Status}{Correct}{False}{Count}{0}%
\StoreBenchExecResult{Abc}{ImcReachSafetyFloatsSampledBvFunc}{Status}{Wrong}{}{Count}{0}%
\StoreBenchExecResult{Abc}{ImcReachSafetyFloatsSampledBvFunc}{Status}{Wrong}{True}{Count}{0}%
\StoreBenchExecResult{Abc}{ImcReachSafetyFloatsSampledBvFunc}{Status}{Wrong}{False}{Count}{0}%
\providecommand\StoreBenchExecResult[7]{\expandafter\newcommand\csname#1#2#3#4#5#6\endcsname{#7}}%
\StoreBenchExecResult{Abc}{ImcReachSafetyHardnessSampledBvFunc}{Status}{All}{}{Score}{0}%
\StoreBenchExecResult{Abc}{ImcReachSafetyHardnessSampledBvFunc}{Status}{All}{}{Count}{132}%
\StoreBenchExecResult{Abc}{ImcReachSafetyHardnessSampledBvFunc}{Status}{Correct}{}{Count}{132}%
\StoreBenchExecResult{Abc}{ImcReachSafetyHardnessSampledBvFunc}{Status}{Correct}{True}{Count}{132}%
\StoreBenchExecResult{Abc}{ImcReachSafetyHardnessSampledBvFunc}{Status}{Correct}{False}{Count}{0}%
\StoreBenchExecResult{Abc}{ImcReachSafetyHardnessSampledBvFunc}{Status}{Wrong}{}{Count}{0}%
\StoreBenchExecResult{Abc}{ImcReachSafetyHardnessSampledBvFunc}{Status}{Wrong}{True}{Count}{0}%
\StoreBenchExecResult{Abc}{ImcReachSafetyHardnessSampledBvFunc}{Status}{Wrong}{False}{Count}{0}%
\providecommand\StoreBenchExecResult[7]{\expandafter\newcommand\csname#1#2#3#4#5#6\endcsname{#7}}%
\StoreBenchExecResult{Abc}{ImcReachSafetyHardwareSampledBvFunc}{Status}{All}{}{Score}{0}%
\StoreBenchExecResult{Abc}{ImcReachSafetyHardwareSampledBvFunc}{Status}{All}{}{Count}{148}%
\StoreBenchExecResult{Abc}{ImcReachSafetyHardwareSampledBvFunc}{Status}{Correct}{}{Count}{64}%
\StoreBenchExecResult{Abc}{ImcReachSafetyHardwareSampledBvFunc}{Status}{Correct}{True}{Count}{21}%
\StoreBenchExecResult{Abc}{ImcReachSafetyHardwareSampledBvFunc}{Status}{Correct}{False}{Count}{43}%
\StoreBenchExecResult{Abc}{ImcReachSafetyHardwareSampledBvFunc}{Status}{Wrong}{}{Count}{0}%
\StoreBenchExecResult{Abc}{ImcReachSafetyHardwareSampledBvFunc}{Status}{Wrong}{True}{Count}{0}%
\StoreBenchExecResult{Abc}{ImcReachSafetyHardwareSampledBvFunc}{Status}{Wrong}{False}{Count}{0}%
\providecommand\StoreBenchExecResult[7]{\expandafter\newcommand\csname#1#2#3#4#5#6\endcsname{#7}}%
\StoreBenchExecResult{Abc}{ImcReachSafetyHeapSampledBvFunc}{Status}{All}{}{Score}{0}%
\StoreBenchExecResult{Abc}{ImcReachSafetyHeapSampledBvFunc}{Status}{All}{}{Count}{6}%
\StoreBenchExecResult{Abc}{ImcReachSafetyHeapSampledBvFunc}{Status}{Correct}{}{Count}{6}%
\StoreBenchExecResult{Abc}{ImcReachSafetyHeapSampledBvFunc}{Status}{Correct}{True}{Count}{4}%
\StoreBenchExecResult{Abc}{ImcReachSafetyHeapSampledBvFunc}{Status}{Correct}{False}{Count}{2}%
\StoreBenchExecResult{Abc}{ImcReachSafetyHeapSampledBvFunc}{Status}{Wrong}{}{Count}{0}%
\StoreBenchExecResult{Abc}{ImcReachSafetyHeapSampledBvFunc}{Status}{Wrong}{True}{Count}{0}%
\StoreBenchExecResult{Abc}{ImcReachSafetyHeapSampledBvFunc}{Status}{Wrong}{False}{Count}{0}%
\providecommand\StoreBenchExecResult[7]{\expandafter\newcommand\csname#1#2#3#4#5#6\endcsname{#7}}%
\StoreBenchExecResult{Abc}{ImcReachSafetyLoopsSampledBvFunc}{Status}{All}{}{Score}{0}%
\StoreBenchExecResult{Abc}{ImcReachSafetyLoopsSampledBvFunc}{Status}{All}{}{Count}{137}%
\StoreBenchExecResult{Abc}{ImcReachSafetyLoopsSampledBvFunc}{Status}{Correct}{}{Count}{57}%
\StoreBenchExecResult{Abc}{ImcReachSafetyLoopsSampledBvFunc}{Status}{Correct}{True}{Count}{30}%
\StoreBenchExecResult{Abc}{ImcReachSafetyLoopsSampledBvFunc}{Status}{Correct}{False}{Count}{27}%
\StoreBenchExecResult{Abc}{ImcReachSafetyLoopsSampledBvFunc}{Status}{Wrong}{}{Count}{0}%
\StoreBenchExecResult{Abc}{ImcReachSafetyLoopsSampledBvFunc}{Status}{Wrong}{True}{Count}{0}%
\StoreBenchExecResult{Abc}{ImcReachSafetyLoopsSampledBvFunc}{Status}{Wrong}{False}{Count}{0}%
\providecommand\StoreBenchExecResult[7]{\expandafter\newcommand\csname#1#2#3#4#5#6\endcsname{#7}}%
\StoreBenchExecResult{Abc}{ImcReachSafetyProductLinesSampledBvFunc}{Status}{All}{}{Score}{0}%
\StoreBenchExecResult{Abc}{ImcReachSafetyProductLinesSampledBvFunc}{Status}{All}{}{Count}{150}%
\StoreBenchExecResult{Abc}{ImcReachSafetyProductLinesSampledBvFunc}{Status}{Correct}{}{Count}{150}%
\StoreBenchExecResult{Abc}{ImcReachSafetyProductLinesSampledBvFunc}{Status}{Correct}{True}{Count}{75}%
\StoreBenchExecResult{Abc}{ImcReachSafetyProductLinesSampledBvFunc}{Status}{Correct}{False}{Count}{75}%
\StoreBenchExecResult{Abc}{ImcReachSafetyProductLinesSampledBvFunc}{Status}{Wrong}{}{Count}{0}%
\StoreBenchExecResult{Abc}{ImcReachSafetyProductLinesSampledBvFunc}{Status}{Wrong}{True}{Count}{0}%
\StoreBenchExecResult{Abc}{ImcReachSafetyProductLinesSampledBvFunc}{Status}{Wrong}{False}{Count}{0}%
\providecommand\StoreBenchExecResult[7]{\expandafter\newcommand\csname#1#2#3#4#5#6\endcsname{#7}}%
\StoreBenchExecResult{Abc}{ImcReachSafetySequentializedSampledBvFunc}{Status}{All}{}{Score}{0}%
\StoreBenchExecResult{Abc}{ImcReachSafetySequentializedSampledBvFunc}{Status}{All}{}{Count}{150}%
\StoreBenchExecResult{Abc}{ImcReachSafetySequentializedSampledBvFunc}{Status}{Correct}{}{Count}{105}%
\StoreBenchExecResult{Abc}{ImcReachSafetySequentializedSampledBvFunc}{Status}{Correct}{True}{Count}{18}%
\StoreBenchExecResult{Abc}{ImcReachSafetySequentializedSampledBvFunc}{Status}{Correct}{False}{Count}{87}%
\StoreBenchExecResult{Abc}{ImcReachSafetySequentializedSampledBvFunc}{Status}{Wrong}{}{Count}{0}%
\StoreBenchExecResult{Abc}{ImcReachSafetySequentializedSampledBvFunc}{Status}{Wrong}{True}{Count}{0}%
\StoreBenchExecResult{Abc}{ImcReachSafetySequentializedSampledBvFunc}{Status}{Wrong}{False}{Count}{0}%
\providecommand\StoreBenchExecResult[7]{\expandafter\newcommand\csname#1#2#3#4#5#6\endcsname{#7}}%
\StoreBenchExecResult{Abc}{ImcReachSafetyXCSPSampledBvFunc}{Status}{All}{}{Score}{0}%
\StoreBenchExecResult{Abc}{ImcReachSafetyXCSPSampledBvFunc}{Status}{All}{}{Count}{98}%
\StoreBenchExecResult{Abc}{ImcReachSafetyXCSPSampledBvFunc}{Status}{Correct}{}{Count}{93}%
\StoreBenchExecResult{Abc}{ImcReachSafetyXCSPSampledBvFunc}{Status}{Correct}{True}{Count}{51}%
\StoreBenchExecResult{Abc}{ImcReachSafetyXCSPSampledBvFunc}{Status}{Correct}{False}{Count}{42}%
\StoreBenchExecResult{Abc}{ImcReachSafetyXCSPSampledBvFunc}{Status}{Wrong}{}{Count}{0}%
\StoreBenchExecResult{Abc}{ImcReachSafetyXCSPSampledBvFunc}{Status}{Wrong}{True}{Count}{0}%
\StoreBenchExecResult{Abc}{ImcReachSafetyXCSPSampledBvFunc}{Status}{Wrong}{False}{Count}{0}%
\providecommand\StoreBenchExecResult[7]{\expandafter\newcommand\csname#1#2#3#4#5#6\endcsname{#7}}%
\StoreBenchExecResult{Abc}{ImcTerminationBitVectorsSampledBvFunc}{Status}{All}{}{Score}{0}%
\StoreBenchExecResult{Abc}{ImcTerminationBitVectorsSampledBvFunc}{Status}{All}{}{Count}{32}%
\StoreBenchExecResult{Abc}{ImcTerminationBitVectorsSampledBvFunc}{Status}{Correct}{}{Count}{24}%
\StoreBenchExecResult{Abc}{ImcTerminationBitVectorsSampledBvFunc}{Status}{Correct}{True}{Count}{13}%
\StoreBenchExecResult{Abc}{ImcTerminationBitVectorsSampledBvFunc}{Status}{Correct}{False}{Count}{11}%
\StoreBenchExecResult{Abc}{ImcTerminationBitVectorsSampledBvFunc}{Status}{Wrong}{}{Count}{0}%
\StoreBenchExecResult{Abc}{ImcTerminationBitVectorsSampledBvFunc}{Status}{Wrong}{True}{Count}{0}%
\StoreBenchExecResult{Abc}{ImcTerminationBitVectorsSampledBvFunc}{Status}{Wrong}{False}{Count}{0}%
\providecommand\StoreBenchExecResult[7]{\expandafter\newcommand\csname#1#2#3#4#5#6\endcsname{#7}}%
\StoreBenchExecResult{Abc}{ImcTerminationMainControlFlowSampledBvFunc}{Status}{All}{}{Score}{0}%
\StoreBenchExecResult{Abc}{ImcTerminationMainControlFlowSampledBvFunc}{Status}{All}{}{Count}{236}%
\StoreBenchExecResult{Abc}{ImcTerminationMainControlFlowSampledBvFunc}{Status}{Correct}{}{Count}{75}%
\StoreBenchExecResult{Abc}{ImcTerminationMainControlFlowSampledBvFunc}{Status}{Correct}{True}{Count}{27}%
\StoreBenchExecResult{Abc}{ImcTerminationMainControlFlowSampledBvFunc}{Status}{Correct}{False}{Count}{48}%
\StoreBenchExecResult{Abc}{ImcTerminationMainControlFlowSampledBvFunc}{Status}{Wrong}{}{Count}{0}%
\StoreBenchExecResult{Abc}{ImcTerminationMainControlFlowSampledBvFunc}{Status}{Wrong}{True}{Count}{0}%
\StoreBenchExecResult{Abc}{ImcTerminationMainControlFlowSampledBvFunc}{Status}{Wrong}{False}{Count}{0}%
\providecommand\StoreBenchExecResult[7]{\expandafter\newcommand\csname#1#2#3#4#5#6\endcsname{#7}}%
\StoreBenchExecResult{Abc}{ImcTerminationOtherSampledBvFunc}{Status}{All}{}{Score}{0}%
\StoreBenchExecResult{Abc}{ImcTerminationOtherSampledBvFunc}{Status}{All}{}{Count}{973}%
\StoreBenchExecResult{Abc}{ImcTerminationOtherSampledBvFunc}{Status}{Correct}{}{Count}{851}%
\StoreBenchExecResult{Abc}{ImcTerminationOtherSampledBvFunc}{Status}{Correct}{True}{Count}{210}%
\StoreBenchExecResult{Abc}{ImcTerminationOtherSampledBvFunc}{Status}{Correct}{False}{Count}{641}%
\StoreBenchExecResult{Abc}{ImcTerminationOtherSampledBvFunc}{Status}{Wrong}{}{Count}{0}%
\StoreBenchExecResult{Abc}{ImcTerminationOtherSampledBvFunc}{Status}{Wrong}{True}{Count}{0}%
\StoreBenchExecResult{Abc}{ImcTerminationOtherSampledBvFunc}{Status}{Wrong}{False}{Count}{0}%
\edef\AbcImcReachSafetySampledBvFuncStatusAllCount{\the\numexpr\AbcImcReachSafetyBitVectorsSampledBvFuncStatusAllCount+\AbcImcReachSafetyCombinationsSampledBvFuncStatusAllCount+\AbcImcReachSafetyControlFlowSampledBvFuncStatusAllCount+\AbcImcReachSafetyECASampledBvFuncStatusAllCount+\AbcImcReachSafetyFloatsSampledBvFuncStatusAllCount+\AbcImcReachSafetyHardnessSampledBvFuncStatusAllCount+\AbcImcReachSafetyHardwareSampledBvFuncStatusAllCount+\AbcImcReachSafetyHeapSampledBvFuncStatusAllCount+\AbcImcReachSafetyLoopsSampledBvFuncStatusAllCount+\AbcImcReachSafetyProductLinesSampledBvFuncStatusAllCount+\AbcImcReachSafetySequentializedSampledBvFuncStatusAllCount+\AbcImcReachSafetyXCSPSampledBvFuncStatusAllCount}
\edef\AbcImcReachSafetySampledBvFuncStatusCorrectCount{\the\numexpr\AbcImcReachSafetyBitVectorsSampledBvFuncStatusCorrectCount+\AbcImcReachSafetyCombinationsSampledBvFuncStatusCorrectCount+\AbcImcReachSafetyControlFlowSampledBvFuncStatusCorrectCount+\AbcImcReachSafetyECASampledBvFuncStatusCorrectCount+\AbcImcReachSafetyFloatsSampledBvFuncStatusCorrectCount+\AbcImcReachSafetyHardnessSampledBvFuncStatusCorrectCount+\AbcImcReachSafetyHardwareSampledBvFuncStatusCorrectCount+\AbcImcReachSafetyHeapSampledBvFuncStatusCorrectCount+\AbcImcReachSafetyLoopsSampledBvFuncStatusCorrectCount+\AbcImcReachSafetyProductLinesSampledBvFuncStatusCorrectCount+\AbcImcReachSafetySequentializedSampledBvFuncStatusCorrectCount+\AbcImcReachSafetyXCSPSampledBvFuncStatusCorrectCount}
\edef\AbcImcReachSafetySampledBvFuncStatusCorrectTrueCount{\the\numexpr\AbcImcReachSafetyBitVectorsSampledBvFuncStatusCorrectTrueCount+\AbcImcReachSafetyCombinationsSampledBvFuncStatusCorrectTrueCount+\AbcImcReachSafetyControlFlowSampledBvFuncStatusCorrectTrueCount+\AbcImcReachSafetyECASampledBvFuncStatusCorrectTrueCount+\AbcImcReachSafetyFloatsSampledBvFuncStatusCorrectTrueCount+\AbcImcReachSafetyHardnessSampledBvFuncStatusCorrectTrueCount+\AbcImcReachSafetyHardwareSampledBvFuncStatusCorrectTrueCount+\AbcImcReachSafetyHeapSampledBvFuncStatusCorrectTrueCount+\AbcImcReachSafetyLoopsSampledBvFuncStatusCorrectTrueCount+\AbcImcReachSafetyProductLinesSampledBvFuncStatusCorrectTrueCount+\AbcImcReachSafetySequentializedSampledBvFuncStatusCorrectTrueCount+\AbcImcReachSafetyXCSPSampledBvFuncStatusCorrectTrueCount}
\edef\AbcImcReachSafetySampledBvFuncStatusCorrectFalseCount{\the\numexpr\AbcImcReachSafetyBitVectorsSampledBvFuncStatusCorrectFalseCount+\AbcImcReachSafetyCombinationsSampledBvFuncStatusCorrectFalseCount+\AbcImcReachSafetyControlFlowSampledBvFuncStatusCorrectFalseCount+\AbcImcReachSafetyECASampledBvFuncStatusCorrectFalseCount+\AbcImcReachSafetyFloatsSampledBvFuncStatusCorrectFalseCount+\AbcImcReachSafetyHardnessSampledBvFuncStatusCorrectFalseCount+\AbcImcReachSafetyHardwareSampledBvFuncStatusCorrectFalseCount+\AbcImcReachSafetyHeapSampledBvFuncStatusCorrectFalseCount+\AbcImcReachSafetyLoopsSampledBvFuncStatusCorrectFalseCount+\AbcImcReachSafetyProductLinesSampledBvFuncStatusCorrectFalseCount+\AbcImcReachSafetySequentializedSampledBvFuncStatusCorrectFalseCount+\AbcImcReachSafetyXCSPSampledBvFuncStatusCorrectFalseCount}
\edef\AbcImcReachSafetySampledBvFuncStatusWrongCount{\the\numexpr\AbcImcReachSafetyBitVectorsSampledBvFuncStatusWrongCount+\AbcImcReachSafetyCombinationsSampledBvFuncStatusWrongCount+\AbcImcReachSafetyControlFlowSampledBvFuncStatusWrongCount+\AbcImcReachSafetyECASampledBvFuncStatusWrongCount+\AbcImcReachSafetyFloatsSampledBvFuncStatusWrongCount+\AbcImcReachSafetyHardnessSampledBvFuncStatusWrongCount+\AbcImcReachSafetyHardwareSampledBvFuncStatusWrongCount+\AbcImcReachSafetyHeapSampledBvFuncStatusWrongCount+\AbcImcReachSafetyLoopsSampledBvFuncStatusWrongCount+\AbcImcReachSafetyProductLinesSampledBvFuncStatusWrongCount+\AbcImcReachSafetySequentializedSampledBvFuncStatusWrongCount+\AbcImcReachSafetyXCSPSampledBvFuncStatusWrongCount}
\edef\AbcImcTerminationSampledBvFuncStatusAllCount{\the\numexpr\AbcImcTerminationBitVectorsSampledBvFuncStatusAllCount+\AbcImcTerminationMainControlFlowSampledBvFuncStatusAllCount+\AbcImcTerminationOtherSampledBvFuncStatusAllCount}
\edef\AbcImcTerminationSampledBvFuncStatusCorrectCount{\the\numexpr\AbcImcTerminationBitVectorsSampledBvFuncStatusCorrectCount+\AbcImcTerminationMainControlFlowSampledBvFuncStatusCorrectCount+\AbcImcTerminationOtherSampledBvFuncStatusCorrectCount}
\edef\AbcImcTerminationSampledBvFuncStatusCorrectTrueCount{\the\numexpr\AbcImcTerminationBitVectorsSampledBvFuncStatusCorrectTrueCount+\AbcImcTerminationMainControlFlowSampledBvFuncStatusCorrectTrueCount+\AbcImcTerminationOtherSampledBvFuncStatusCorrectTrueCount}
\edef\AbcImcTerminationSampledBvFuncStatusCorrectFalseCount{\the\numexpr\AbcImcTerminationBitVectorsSampledBvFuncStatusCorrectFalseCount+\AbcImcTerminationMainControlFlowSampledBvFuncStatusCorrectFalseCount+\AbcImcTerminationOtherSampledBvFuncStatusCorrectFalseCount}
\edef\AbcImcTerminationSampledBvFuncStatusWrongCount{\the\numexpr\AbcImcTerminationBitVectorsSampledBvFuncStatusWrongCount+\AbcImcTerminationMainControlFlowSampledBvFuncStatusWrongCount+\AbcImcTerminationOtherSampledBvFuncStatusWrongCount}
\providecommand\StoreBenchExecResult[7]{\expandafter\newcommand\csname#1#2#3#4#5#6\endcsname{#7}}%
\StoreBenchExecResult{Abc}{PdrReachSafetyBitVectorsSampledBvFunc}{Status}{All}{}{Score}{0}%
\StoreBenchExecResult{Abc}{PdrReachSafetyBitVectorsSampledBvFunc}{Status}{All}{}{Count}{47}%
\StoreBenchExecResult{Abc}{PdrReachSafetyBitVectorsSampledBvFunc}{Status}{Correct}{}{Count}{40}%
\StoreBenchExecResult{Abc}{PdrReachSafetyBitVectorsSampledBvFunc}{Status}{Correct}{True}{Count}{29}%
\StoreBenchExecResult{Abc}{PdrReachSafetyBitVectorsSampledBvFunc}{Status}{Correct}{False}{Count}{11}%
\StoreBenchExecResult{Abc}{PdrReachSafetyBitVectorsSampledBvFunc}{Status}{Wrong}{}{Count}{0}%
\StoreBenchExecResult{Abc}{PdrReachSafetyBitVectorsSampledBvFunc}{Status}{Wrong}{True}{Count}{0}%
\StoreBenchExecResult{Abc}{PdrReachSafetyBitVectorsSampledBvFunc}{Status}{Wrong}{False}{Count}{0}%
\providecommand\StoreBenchExecResult[7]{\expandafter\newcommand\csname#1#2#3#4#5#6\endcsname{#7}}%
\StoreBenchExecResult{Abc}{PdrReachSafetyCombinationsSampledBvFunc}{Status}{All}{}{Score}{0}%
\StoreBenchExecResult{Abc}{PdrReachSafetyCombinationsSampledBvFunc}{Status}{All}{}{Count}{110}%
\StoreBenchExecResult{Abc}{PdrReachSafetyCombinationsSampledBvFunc}{Status}{Correct}{}{Count}{75}%
\StoreBenchExecResult{Abc}{PdrReachSafetyCombinationsSampledBvFunc}{Status}{Correct}{True}{Count}{10}%
\StoreBenchExecResult{Abc}{PdrReachSafetyCombinationsSampledBvFunc}{Status}{Correct}{False}{Count}{65}%
\StoreBenchExecResult{Abc}{PdrReachSafetyCombinationsSampledBvFunc}{Status}{Wrong}{}{Count}{0}%
\StoreBenchExecResult{Abc}{PdrReachSafetyCombinationsSampledBvFunc}{Status}{Wrong}{True}{Count}{0}%
\StoreBenchExecResult{Abc}{PdrReachSafetyCombinationsSampledBvFunc}{Status}{Wrong}{False}{Count}{0}%
\providecommand\StoreBenchExecResult[7]{\expandafter\newcommand\csname#1#2#3#4#5#6\endcsname{#7}}%
\StoreBenchExecResult{Abc}{PdrReachSafetyControlFlowSampledBvFunc}{Status}{All}{}{Score}{0}%
\StoreBenchExecResult{Abc}{PdrReachSafetyControlFlowSampledBvFunc}{Status}{All}{}{Count}{29}%
\StoreBenchExecResult{Abc}{PdrReachSafetyControlFlowSampledBvFunc}{Status}{Correct}{}{Count}{26}%
\StoreBenchExecResult{Abc}{PdrReachSafetyControlFlowSampledBvFunc}{Status}{Correct}{True}{Count}{24}%
\StoreBenchExecResult{Abc}{PdrReachSafetyControlFlowSampledBvFunc}{Status}{Correct}{False}{Count}{2}%
\StoreBenchExecResult{Abc}{PdrReachSafetyControlFlowSampledBvFunc}{Status}{Wrong}{}{Count}{0}%
\StoreBenchExecResult{Abc}{PdrReachSafetyControlFlowSampledBvFunc}{Status}{Wrong}{True}{Count}{0}%
\StoreBenchExecResult{Abc}{PdrReachSafetyControlFlowSampledBvFunc}{Status}{Wrong}{False}{Count}{0}%
\providecommand\StoreBenchExecResult[7]{\expandafter\newcommand\csname#1#2#3#4#5#6\endcsname{#7}}%
\StoreBenchExecResult{Abc}{PdrReachSafetyECASampledBvFunc}{Status}{All}{}{Score}{0}%
\StoreBenchExecResult{Abc}{PdrReachSafetyECASampledBvFunc}{Status}{All}{}{Count}{150}%
\StoreBenchExecResult{Abc}{PdrReachSafetyECASampledBvFunc}{Status}{Correct}{}{Count}{60}%
\StoreBenchExecResult{Abc}{PdrReachSafetyECASampledBvFunc}{Status}{Correct}{True}{Count}{39}%
\StoreBenchExecResult{Abc}{PdrReachSafetyECASampledBvFunc}{Status}{Correct}{False}{Count}{21}%
\StoreBenchExecResult{Abc}{PdrReachSafetyECASampledBvFunc}{Status}{Wrong}{}{Count}{0}%
\StoreBenchExecResult{Abc}{PdrReachSafetyECASampledBvFunc}{Status}{Wrong}{True}{Count}{0}%
\StoreBenchExecResult{Abc}{PdrReachSafetyECASampledBvFunc}{Status}{Wrong}{False}{Count}{0}%
\providecommand\StoreBenchExecResult[7]{\expandafter\newcommand\csname#1#2#3#4#5#6\endcsname{#7}}%
\StoreBenchExecResult{Abc}{PdrReachSafetyFloatsSampledBvFunc}{Status}{All}{}{Score}{0}%
\StoreBenchExecResult{Abc}{PdrReachSafetyFloatsSampledBvFunc}{Status}{All}{}{Count}{10}%
\StoreBenchExecResult{Abc}{PdrReachSafetyFloatsSampledBvFunc}{Status}{Correct}{}{Count}{10}%
\StoreBenchExecResult{Abc}{PdrReachSafetyFloatsSampledBvFunc}{Status}{Correct}{True}{Count}{10}%
\StoreBenchExecResult{Abc}{PdrReachSafetyFloatsSampledBvFunc}{Status}{Correct}{False}{Count}{0}%
\StoreBenchExecResult{Abc}{PdrReachSafetyFloatsSampledBvFunc}{Status}{Wrong}{}{Count}{0}%
\StoreBenchExecResult{Abc}{PdrReachSafetyFloatsSampledBvFunc}{Status}{Wrong}{True}{Count}{0}%
\StoreBenchExecResult{Abc}{PdrReachSafetyFloatsSampledBvFunc}{Status}{Wrong}{False}{Count}{0}%
\providecommand\StoreBenchExecResult[7]{\expandafter\newcommand\csname#1#2#3#4#5#6\endcsname{#7}}%
\StoreBenchExecResult{Abc}{PdrReachSafetyHardnessSampledBvFunc}{Status}{All}{}{Score}{0}%
\StoreBenchExecResult{Abc}{PdrReachSafetyHardnessSampledBvFunc}{Status}{All}{}{Count}{132}%
\StoreBenchExecResult{Abc}{PdrReachSafetyHardnessSampledBvFunc}{Status}{Correct}{}{Count}{132}%
\StoreBenchExecResult{Abc}{PdrReachSafetyHardnessSampledBvFunc}{Status}{Correct}{True}{Count}{132}%
\StoreBenchExecResult{Abc}{PdrReachSafetyHardnessSampledBvFunc}{Status}{Correct}{False}{Count}{0}%
\StoreBenchExecResult{Abc}{PdrReachSafetyHardnessSampledBvFunc}{Status}{Wrong}{}{Count}{0}%
\StoreBenchExecResult{Abc}{PdrReachSafetyHardnessSampledBvFunc}{Status}{Wrong}{True}{Count}{0}%
\StoreBenchExecResult{Abc}{PdrReachSafetyHardnessSampledBvFunc}{Status}{Wrong}{False}{Count}{0}%
\providecommand\StoreBenchExecResult[7]{\expandafter\newcommand\csname#1#2#3#4#5#6\endcsname{#7}}%
\StoreBenchExecResult{Abc}{PdrReachSafetyHardwareSampledBvFunc}{Status}{All}{}{Score}{0}%
\StoreBenchExecResult{Abc}{PdrReachSafetyHardwareSampledBvFunc}{Status}{All}{}{Count}{148}%
\StoreBenchExecResult{Abc}{PdrReachSafetyHardwareSampledBvFunc}{Status}{Correct}{}{Count}{68}%
\StoreBenchExecResult{Abc}{PdrReachSafetyHardwareSampledBvFunc}{Status}{Correct}{True}{Count}{27}%
\StoreBenchExecResult{Abc}{PdrReachSafetyHardwareSampledBvFunc}{Status}{Correct}{False}{Count}{41}%
\StoreBenchExecResult{Abc}{PdrReachSafetyHardwareSampledBvFunc}{Status}{Wrong}{}{Count}{0}%
\StoreBenchExecResult{Abc}{PdrReachSafetyHardwareSampledBvFunc}{Status}{Wrong}{True}{Count}{0}%
\StoreBenchExecResult{Abc}{PdrReachSafetyHardwareSampledBvFunc}{Status}{Wrong}{False}{Count}{0}%
\providecommand\StoreBenchExecResult[7]{\expandafter\newcommand\csname#1#2#3#4#5#6\endcsname{#7}}%
\StoreBenchExecResult{Abc}{PdrReachSafetyHeapSampledBvFunc}{Status}{All}{}{Score}{0}%
\StoreBenchExecResult{Abc}{PdrReachSafetyHeapSampledBvFunc}{Status}{All}{}{Count}{6}%
\StoreBenchExecResult{Abc}{PdrReachSafetyHeapSampledBvFunc}{Status}{Correct}{}{Count}{6}%
\StoreBenchExecResult{Abc}{PdrReachSafetyHeapSampledBvFunc}{Status}{Correct}{True}{Count}{4}%
\StoreBenchExecResult{Abc}{PdrReachSafetyHeapSampledBvFunc}{Status}{Correct}{False}{Count}{2}%
\StoreBenchExecResult{Abc}{PdrReachSafetyHeapSampledBvFunc}{Status}{Wrong}{}{Count}{0}%
\StoreBenchExecResult{Abc}{PdrReachSafetyHeapSampledBvFunc}{Status}{Wrong}{True}{Count}{0}%
\StoreBenchExecResult{Abc}{PdrReachSafetyHeapSampledBvFunc}{Status}{Wrong}{False}{Count}{0}%
\providecommand\StoreBenchExecResult[7]{\expandafter\newcommand\csname#1#2#3#4#5#6\endcsname{#7}}%
\StoreBenchExecResult{Abc}{PdrReachSafetyLoopsSampledBvFunc}{Status}{All}{}{Score}{0}%
\StoreBenchExecResult{Abc}{PdrReachSafetyLoopsSampledBvFunc}{Status}{All}{}{Count}{137}%
\StoreBenchExecResult{Abc}{PdrReachSafetyLoopsSampledBvFunc}{Status}{Correct}{}{Count}{52}%
\StoreBenchExecResult{Abc}{PdrReachSafetyLoopsSampledBvFunc}{Status}{Correct}{True}{Count}{23}%
\StoreBenchExecResult{Abc}{PdrReachSafetyLoopsSampledBvFunc}{Status}{Correct}{False}{Count}{29}%
\StoreBenchExecResult{Abc}{PdrReachSafetyLoopsSampledBvFunc}{Status}{Wrong}{}{Count}{0}%
\StoreBenchExecResult{Abc}{PdrReachSafetyLoopsSampledBvFunc}{Status}{Wrong}{True}{Count}{0}%
\StoreBenchExecResult{Abc}{PdrReachSafetyLoopsSampledBvFunc}{Status}{Wrong}{False}{Count}{0}%
\providecommand\StoreBenchExecResult[7]{\expandafter\newcommand\csname#1#2#3#4#5#6\endcsname{#7}}%
\StoreBenchExecResult{Abc}{PdrReachSafetyProductLinesSampledBvFunc}{Status}{All}{}{Score}{0}%
\StoreBenchExecResult{Abc}{PdrReachSafetyProductLinesSampledBvFunc}{Status}{All}{}{Count}{150}%
\StoreBenchExecResult{Abc}{PdrReachSafetyProductLinesSampledBvFunc}{Status}{Correct}{}{Count}{144}%
\StoreBenchExecResult{Abc}{PdrReachSafetyProductLinesSampledBvFunc}{Status}{Correct}{True}{Count}{69}%
\StoreBenchExecResult{Abc}{PdrReachSafetyProductLinesSampledBvFunc}{Status}{Correct}{False}{Count}{75}%
\StoreBenchExecResult{Abc}{PdrReachSafetyProductLinesSampledBvFunc}{Status}{Wrong}{}{Count}{0}%
\StoreBenchExecResult{Abc}{PdrReachSafetyProductLinesSampledBvFunc}{Status}{Wrong}{True}{Count}{0}%
\StoreBenchExecResult{Abc}{PdrReachSafetyProductLinesSampledBvFunc}{Status}{Wrong}{False}{Count}{0}%
\providecommand\StoreBenchExecResult[7]{\expandafter\newcommand\csname#1#2#3#4#5#6\endcsname{#7}}%
\StoreBenchExecResult{Abc}{PdrReachSafetySequentializedSampledBvFunc}{Status}{All}{}{Score}{0}%
\StoreBenchExecResult{Abc}{PdrReachSafetySequentializedSampledBvFunc}{Status}{All}{}{Count}{150}%
\StoreBenchExecResult{Abc}{PdrReachSafetySequentializedSampledBvFunc}{Status}{Correct}{}{Count}{85}%
\StoreBenchExecResult{Abc}{PdrReachSafetySequentializedSampledBvFunc}{Status}{Correct}{True}{Count}{15}%
\StoreBenchExecResult{Abc}{PdrReachSafetySequentializedSampledBvFunc}{Status}{Correct}{False}{Count}{70}%
\StoreBenchExecResult{Abc}{PdrReachSafetySequentializedSampledBvFunc}{Status}{Wrong}{}{Count}{0}%
\StoreBenchExecResult{Abc}{PdrReachSafetySequentializedSampledBvFunc}{Status}{Wrong}{True}{Count}{0}%
\StoreBenchExecResult{Abc}{PdrReachSafetySequentializedSampledBvFunc}{Status}{Wrong}{False}{Count}{0}%
\providecommand\StoreBenchExecResult[7]{\expandafter\newcommand\csname#1#2#3#4#5#6\endcsname{#7}}%
\StoreBenchExecResult{Abc}{PdrReachSafetyXCSPSampledBvFunc}{Status}{All}{}{Score}{0}%
\StoreBenchExecResult{Abc}{PdrReachSafetyXCSPSampledBvFunc}{Status}{All}{}{Count}{98}%
\StoreBenchExecResult{Abc}{PdrReachSafetyXCSPSampledBvFunc}{Status}{Correct}{}{Count}{96}%
\StoreBenchExecResult{Abc}{PdrReachSafetyXCSPSampledBvFunc}{Status}{Correct}{True}{Count}{51}%
\StoreBenchExecResult{Abc}{PdrReachSafetyXCSPSampledBvFunc}{Status}{Correct}{False}{Count}{45}%
\StoreBenchExecResult{Abc}{PdrReachSafetyXCSPSampledBvFunc}{Status}{Wrong}{}{Count}{0}%
\StoreBenchExecResult{Abc}{PdrReachSafetyXCSPSampledBvFunc}{Status}{Wrong}{True}{Count}{0}%
\StoreBenchExecResult{Abc}{PdrReachSafetyXCSPSampledBvFunc}{Status}{Wrong}{False}{Count}{0}%
\providecommand\StoreBenchExecResult[7]{\expandafter\newcommand\csname#1#2#3#4#5#6\endcsname{#7}}%
\StoreBenchExecResult{Abc}{PdrTerminationBitVectorsSampledBvFunc}{Status}{All}{}{Score}{0}%
\StoreBenchExecResult{Abc}{PdrTerminationBitVectorsSampledBvFunc}{Status}{All}{}{Count}{32}%
\StoreBenchExecResult{Abc}{PdrTerminationBitVectorsSampledBvFunc}{Status}{Correct}{}{Count}{18}%
\StoreBenchExecResult{Abc}{PdrTerminationBitVectorsSampledBvFunc}{Status}{Correct}{True}{Count}{8}%
\StoreBenchExecResult{Abc}{PdrTerminationBitVectorsSampledBvFunc}{Status}{Correct}{False}{Count}{10}%
\StoreBenchExecResult{Abc}{PdrTerminationBitVectorsSampledBvFunc}{Status}{Wrong}{}{Count}{0}%
\StoreBenchExecResult{Abc}{PdrTerminationBitVectorsSampledBvFunc}{Status}{Wrong}{True}{Count}{0}%
\StoreBenchExecResult{Abc}{PdrTerminationBitVectorsSampledBvFunc}{Status}{Wrong}{False}{Count}{0}%
\providecommand\StoreBenchExecResult[7]{\expandafter\newcommand\csname#1#2#3#4#5#6\endcsname{#7}}%
\StoreBenchExecResult{Abc}{PdrTerminationMainControlFlowSampledBvFunc}{Status}{All}{}{Score}{0}%
\StoreBenchExecResult{Abc}{PdrTerminationMainControlFlowSampledBvFunc}{Status}{All}{}{Count}{236}%
\StoreBenchExecResult{Abc}{PdrTerminationMainControlFlowSampledBvFunc}{Status}{Correct}{}{Count}{52}%
\StoreBenchExecResult{Abc}{PdrTerminationMainControlFlowSampledBvFunc}{Status}{Correct}{True}{Count}{10}%
\StoreBenchExecResult{Abc}{PdrTerminationMainControlFlowSampledBvFunc}{Status}{Correct}{False}{Count}{42}%
\StoreBenchExecResult{Abc}{PdrTerminationMainControlFlowSampledBvFunc}{Status}{Wrong}{}{Count}{0}%
\StoreBenchExecResult{Abc}{PdrTerminationMainControlFlowSampledBvFunc}{Status}{Wrong}{True}{Count}{0}%
\StoreBenchExecResult{Abc}{PdrTerminationMainControlFlowSampledBvFunc}{Status}{Wrong}{False}{Count}{0}%
\providecommand\StoreBenchExecResult[7]{\expandafter\newcommand\csname#1#2#3#4#5#6\endcsname{#7}}%
\StoreBenchExecResult{Abc}{PdrTerminationOtherSampledBvFunc}{Status}{All}{}{Score}{0}%
\StoreBenchExecResult{Abc}{PdrTerminationOtherSampledBvFunc}{Status}{All}{}{Count}{973}%
\StoreBenchExecResult{Abc}{PdrTerminationOtherSampledBvFunc}{Status}{Correct}{}{Count}{604}%
\StoreBenchExecResult{Abc}{PdrTerminationOtherSampledBvFunc}{Status}{Correct}{True}{Count}{91}%
\StoreBenchExecResult{Abc}{PdrTerminationOtherSampledBvFunc}{Status}{Correct}{False}{Count}{513}%
\StoreBenchExecResult{Abc}{PdrTerminationOtherSampledBvFunc}{Status}{Wrong}{}{Count}{0}%
\StoreBenchExecResult{Abc}{PdrTerminationOtherSampledBvFunc}{Status}{Wrong}{True}{Count}{0}%
\StoreBenchExecResult{Abc}{PdrTerminationOtherSampledBvFunc}{Status}{Wrong}{False}{Count}{0}%
\edef\AbcPdrReachSafetySampledBvFuncStatusAllCount{\the\numexpr\AbcPdrReachSafetyBitVectorsSampledBvFuncStatusAllCount+\AbcPdrReachSafetyCombinationsSampledBvFuncStatusAllCount+\AbcPdrReachSafetyControlFlowSampledBvFuncStatusAllCount+\AbcPdrReachSafetyECASampledBvFuncStatusAllCount+\AbcPdrReachSafetyFloatsSampledBvFuncStatusAllCount+\AbcPdrReachSafetyHardnessSampledBvFuncStatusAllCount+\AbcPdrReachSafetyHardwareSampledBvFuncStatusAllCount+\AbcPdrReachSafetyHeapSampledBvFuncStatusAllCount+\AbcPdrReachSafetyLoopsSampledBvFuncStatusAllCount+\AbcPdrReachSafetyProductLinesSampledBvFuncStatusAllCount+\AbcPdrReachSafetySequentializedSampledBvFuncStatusAllCount+\AbcPdrReachSafetyXCSPSampledBvFuncStatusAllCount}
\edef\AbcPdrReachSafetySampledBvFuncStatusCorrectCount{\the\numexpr\AbcPdrReachSafetyBitVectorsSampledBvFuncStatusCorrectCount+\AbcPdrReachSafetyCombinationsSampledBvFuncStatusCorrectCount+\AbcPdrReachSafetyControlFlowSampledBvFuncStatusCorrectCount+\AbcPdrReachSafetyECASampledBvFuncStatusCorrectCount+\AbcPdrReachSafetyFloatsSampledBvFuncStatusCorrectCount+\AbcPdrReachSafetyHardnessSampledBvFuncStatusCorrectCount+\AbcPdrReachSafetyHardwareSampledBvFuncStatusCorrectCount+\AbcPdrReachSafetyHeapSampledBvFuncStatusCorrectCount+\AbcPdrReachSafetyLoopsSampledBvFuncStatusCorrectCount+\AbcPdrReachSafetyProductLinesSampledBvFuncStatusCorrectCount+\AbcPdrReachSafetySequentializedSampledBvFuncStatusCorrectCount+\AbcPdrReachSafetyXCSPSampledBvFuncStatusCorrectCount}
\edef\AbcPdrReachSafetySampledBvFuncStatusCorrectTrueCount{\the\numexpr\AbcPdrReachSafetyBitVectorsSampledBvFuncStatusCorrectTrueCount+\AbcPdrReachSafetyCombinationsSampledBvFuncStatusCorrectTrueCount+\AbcPdrReachSafetyControlFlowSampledBvFuncStatusCorrectTrueCount+\AbcPdrReachSafetyECASampledBvFuncStatusCorrectTrueCount+\AbcPdrReachSafetyFloatsSampledBvFuncStatusCorrectTrueCount+\AbcPdrReachSafetyHardnessSampledBvFuncStatusCorrectTrueCount+\AbcPdrReachSafetyHardwareSampledBvFuncStatusCorrectTrueCount+\AbcPdrReachSafetyHeapSampledBvFuncStatusCorrectTrueCount+\AbcPdrReachSafetyLoopsSampledBvFuncStatusCorrectTrueCount+\AbcPdrReachSafetyProductLinesSampledBvFuncStatusCorrectTrueCount+\AbcPdrReachSafetySequentializedSampledBvFuncStatusCorrectTrueCount+\AbcPdrReachSafetyXCSPSampledBvFuncStatusCorrectTrueCount}
\edef\AbcPdrReachSafetySampledBvFuncStatusCorrectFalseCount{\the\numexpr\AbcPdrReachSafetyBitVectorsSampledBvFuncStatusCorrectFalseCount+\AbcPdrReachSafetyCombinationsSampledBvFuncStatusCorrectFalseCount+\AbcPdrReachSafetyControlFlowSampledBvFuncStatusCorrectFalseCount+\AbcPdrReachSafetyECASampledBvFuncStatusCorrectFalseCount+\AbcPdrReachSafetyFloatsSampledBvFuncStatusCorrectFalseCount+\AbcPdrReachSafetyHardnessSampledBvFuncStatusCorrectFalseCount+\AbcPdrReachSafetyHardwareSampledBvFuncStatusCorrectFalseCount+\AbcPdrReachSafetyHeapSampledBvFuncStatusCorrectFalseCount+\AbcPdrReachSafetyLoopsSampledBvFuncStatusCorrectFalseCount+\AbcPdrReachSafetyProductLinesSampledBvFuncStatusCorrectFalseCount+\AbcPdrReachSafetySequentializedSampledBvFuncStatusCorrectFalseCount+\AbcPdrReachSafetyXCSPSampledBvFuncStatusCorrectFalseCount}
\edef\AbcPdrReachSafetySampledBvFuncStatusWrongCount{\the\numexpr\AbcPdrReachSafetyBitVectorsSampledBvFuncStatusWrongCount+\AbcPdrReachSafetyCombinationsSampledBvFuncStatusWrongCount+\AbcPdrReachSafetyControlFlowSampledBvFuncStatusWrongCount+\AbcPdrReachSafetyECASampledBvFuncStatusWrongCount+\AbcPdrReachSafetyFloatsSampledBvFuncStatusWrongCount+\AbcPdrReachSafetyHardnessSampledBvFuncStatusWrongCount+\AbcPdrReachSafetyHardwareSampledBvFuncStatusWrongCount+\AbcPdrReachSafetyHeapSampledBvFuncStatusWrongCount+\AbcPdrReachSafetyLoopsSampledBvFuncStatusWrongCount+\AbcPdrReachSafetyProductLinesSampledBvFuncStatusWrongCount+\AbcPdrReachSafetySequentializedSampledBvFuncStatusWrongCount+\AbcPdrReachSafetyXCSPSampledBvFuncStatusWrongCount}
\edef\AbcPdrTerminationSampledBvFuncStatusAllCount{\the\numexpr\AbcPdrTerminationBitVectorsSampledBvFuncStatusAllCount+\AbcPdrTerminationMainControlFlowSampledBvFuncStatusAllCount+\AbcPdrTerminationOtherSampledBvFuncStatusAllCount}
\edef\AbcPdrTerminationSampledBvFuncStatusCorrectCount{\the\numexpr\AbcPdrTerminationBitVectorsSampledBvFuncStatusCorrectCount+\AbcPdrTerminationMainControlFlowSampledBvFuncStatusCorrectCount+\AbcPdrTerminationOtherSampledBvFuncStatusCorrectCount}
\edef\AbcPdrTerminationSampledBvFuncStatusCorrectTrueCount{\the\numexpr\AbcPdrTerminationBitVectorsSampledBvFuncStatusCorrectTrueCount+\AbcPdrTerminationMainControlFlowSampledBvFuncStatusCorrectTrueCount+\AbcPdrTerminationOtherSampledBvFuncStatusCorrectTrueCount}
\edef\AbcPdrTerminationSampledBvFuncStatusCorrectFalseCount{\the\numexpr\AbcPdrTerminationBitVectorsSampledBvFuncStatusCorrectFalseCount+\AbcPdrTerminationMainControlFlowSampledBvFuncStatusCorrectFalseCount+\AbcPdrTerminationOtherSampledBvFuncStatusCorrectFalseCount}
\edef\AbcPdrTerminationSampledBvFuncStatusWrongCount{\the\numexpr\AbcPdrTerminationBitVectorsSampledBvFuncStatusWrongCount+\AbcPdrTerminationMainControlFlowSampledBvFuncStatusWrongCount+\AbcPdrTerminationOtherSampledBvFuncStatusWrongCount}
\providecommand\StoreBenchExecResult[7]{\expandafter\newcommand\csname#1#2#3#4#5#6\endcsname{#7}}%
\StoreBenchExecResult{Avr}{IcIIIsaReachSafetyArraysSampledFunc}{Status}{All}{}{Score}{0}%
\StoreBenchExecResult{Avr}{IcIIIsaReachSafetyArraysSampledFunc}{Status}{All}{}{Count}{150}%
\StoreBenchExecResult{Avr}{IcIIIsaReachSafetyArraysSampledFunc}{Status}{Correct}{}{Count}{12}%
\StoreBenchExecResult{Avr}{IcIIIsaReachSafetyArraysSampledFunc}{Status}{Correct}{True}{Count}{0}%
\StoreBenchExecResult{Avr}{IcIIIsaReachSafetyArraysSampledFunc}{Status}{Correct}{False}{Count}{12}%
\StoreBenchExecResult{Avr}{IcIIIsaReachSafetyArraysSampledFunc}{Status}{Wrong}{}{Count}{0}%
\StoreBenchExecResult{Avr}{IcIIIsaReachSafetyArraysSampledFunc}{Status}{Wrong}{True}{Count}{0}%
\StoreBenchExecResult{Avr}{IcIIIsaReachSafetyArraysSampledFunc}{Status}{Wrong}{False}{Count}{0}%
\providecommand\StoreBenchExecResult[7]{\expandafter\newcommand\csname#1#2#3#4#5#6\endcsname{#7}}%
\StoreBenchExecResult{Avr}{IcIIIsaReachSafetyBitVectorsSampledFunc}{Status}{All}{}{Score}{0}%
\StoreBenchExecResult{Avr}{IcIIIsaReachSafetyBitVectorsSampledFunc}{Status}{All}{}{Count}{48}%
\StoreBenchExecResult{Avr}{IcIIIsaReachSafetyBitVectorsSampledFunc}{Status}{Correct}{}{Count}{34}%
\StoreBenchExecResult{Avr}{IcIIIsaReachSafetyBitVectorsSampledFunc}{Status}{Correct}{True}{Count}{23}%
\StoreBenchExecResult{Avr}{IcIIIsaReachSafetyBitVectorsSampledFunc}{Status}{Correct}{False}{Count}{11}%
\StoreBenchExecResult{Avr}{IcIIIsaReachSafetyBitVectorsSampledFunc}{Status}{Wrong}{}{Count}{0}%
\StoreBenchExecResult{Avr}{IcIIIsaReachSafetyBitVectorsSampledFunc}{Status}{Wrong}{True}{Count}{0}%
\StoreBenchExecResult{Avr}{IcIIIsaReachSafetyBitVectorsSampledFunc}{Status}{Wrong}{False}{Count}{0}%
\providecommand\StoreBenchExecResult[7]{\expandafter\newcommand\csname#1#2#3#4#5#6\endcsname{#7}}%
\StoreBenchExecResult{Avr}{IcIIIsaReachSafetyCombinationsSampledFunc}{Status}{All}{}{Score}{0}%
\StoreBenchExecResult{Avr}{IcIIIsaReachSafetyCombinationsSampledFunc}{Status}{All}{}{Count}{110}%
\StoreBenchExecResult{Avr}{IcIIIsaReachSafetyCombinationsSampledFunc}{Status}{Correct}{}{Count}{78}%
\StoreBenchExecResult{Avr}{IcIIIsaReachSafetyCombinationsSampledFunc}{Status}{Correct}{True}{Count}{7}%
\StoreBenchExecResult{Avr}{IcIIIsaReachSafetyCombinationsSampledFunc}{Status}{Correct}{False}{Count}{71}%
\StoreBenchExecResult{Avr}{IcIIIsaReachSafetyCombinationsSampledFunc}{Status}{Wrong}{}{Count}{0}%
\StoreBenchExecResult{Avr}{IcIIIsaReachSafetyCombinationsSampledFunc}{Status}{Wrong}{True}{Count}{0}%
\StoreBenchExecResult{Avr}{IcIIIsaReachSafetyCombinationsSampledFunc}{Status}{Wrong}{False}{Count}{0}%
\providecommand\StoreBenchExecResult[7]{\expandafter\newcommand\csname#1#2#3#4#5#6\endcsname{#7}}%
\StoreBenchExecResult{Avr}{IcIIIsaReachSafetyControlFlowSampledFunc}{Status}{All}{}{Score}{0}%
\StoreBenchExecResult{Avr}{IcIIIsaReachSafetyControlFlowSampledFunc}{Status}{All}{}{Count}{33}%
\StoreBenchExecResult{Avr}{IcIIIsaReachSafetyControlFlowSampledFunc}{Status}{Correct}{}{Count}{27}%
\StoreBenchExecResult{Avr}{IcIIIsaReachSafetyControlFlowSampledFunc}{Status}{Correct}{True}{Count}{25}%
\StoreBenchExecResult{Avr}{IcIIIsaReachSafetyControlFlowSampledFunc}{Status}{Correct}{False}{Count}{2}%
\StoreBenchExecResult{Avr}{IcIIIsaReachSafetyControlFlowSampledFunc}{Status}{Wrong}{}{Count}{0}%
\StoreBenchExecResult{Avr}{IcIIIsaReachSafetyControlFlowSampledFunc}{Status}{Wrong}{True}{Count}{0}%
\StoreBenchExecResult{Avr}{IcIIIsaReachSafetyControlFlowSampledFunc}{Status}{Wrong}{False}{Count}{0}%
\providecommand\StoreBenchExecResult[7]{\expandafter\newcommand\csname#1#2#3#4#5#6\endcsname{#7}}%
\StoreBenchExecResult{Avr}{IcIIIsaReachSafetyECASampledFunc}{Status}{All}{}{Score}{0}%
\StoreBenchExecResult{Avr}{IcIIIsaReachSafetyECASampledFunc}{Status}{All}{}{Count}{150}%
\StoreBenchExecResult{Avr}{IcIIIsaReachSafetyECASampledFunc}{Status}{Correct}{}{Count}{81}%
\StoreBenchExecResult{Avr}{IcIIIsaReachSafetyECASampledFunc}{Status}{Correct}{True}{Count}{47}%
\StoreBenchExecResult{Avr}{IcIIIsaReachSafetyECASampledFunc}{Status}{Correct}{False}{Count}{34}%
\StoreBenchExecResult{Avr}{IcIIIsaReachSafetyECASampledFunc}{Status}{Wrong}{}{Count}{0}%
\StoreBenchExecResult{Avr}{IcIIIsaReachSafetyECASampledFunc}{Status}{Wrong}{True}{Count}{0}%
\StoreBenchExecResult{Avr}{IcIIIsaReachSafetyECASampledFunc}{Status}{Wrong}{False}{Count}{0}%
\providecommand\StoreBenchExecResult[7]{\expandafter\newcommand\csname#1#2#3#4#5#6\endcsname{#7}}%
\StoreBenchExecResult{Avr}{IcIIIsaReachSafetyFloatsSampledFunc}{Status}{All}{}{Score}{0}%
\StoreBenchExecResult{Avr}{IcIIIsaReachSafetyFloatsSampledFunc}{Status}{All}{}{Count}{10}%
\StoreBenchExecResult{Avr}{IcIIIsaReachSafetyFloatsSampledFunc}{Status}{Correct}{}{Count}{10}%
\StoreBenchExecResult{Avr}{IcIIIsaReachSafetyFloatsSampledFunc}{Status}{Correct}{True}{Count}{10}%
\StoreBenchExecResult{Avr}{IcIIIsaReachSafetyFloatsSampledFunc}{Status}{Correct}{False}{Count}{0}%
\StoreBenchExecResult{Avr}{IcIIIsaReachSafetyFloatsSampledFunc}{Status}{Wrong}{}{Count}{0}%
\StoreBenchExecResult{Avr}{IcIIIsaReachSafetyFloatsSampledFunc}{Status}{Wrong}{True}{Count}{0}%
\StoreBenchExecResult{Avr}{IcIIIsaReachSafetyFloatsSampledFunc}{Status}{Wrong}{False}{Count}{0}%
\providecommand\StoreBenchExecResult[7]{\expandafter\newcommand\csname#1#2#3#4#5#6\endcsname{#7}}%
\StoreBenchExecResult{Avr}{IcIIIsaReachSafetyHardnessSampledFunc}{Status}{All}{}{Score}{0}%
\StoreBenchExecResult{Avr}{IcIIIsaReachSafetyHardnessSampledFunc}{Status}{All}{}{Count}{150}%
\StoreBenchExecResult{Avr}{IcIIIsaReachSafetyHardnessSampledFunc}{Status}{Correct}{}{Count}{147}%
\StoreBenchExecResult{Avr}{IcIIIsaReachSafetyHardnessSampledFunc}{Status}{Correct}{True}{Count}{147}%
\StoreBenchExecResult{Avr}{IcIIIsaReachSafetyHardnessSampledFunc}{Status}{Correct}{False}{Count}{0}%
\StoreBenchExecResult{Avr}{IcIIIsaReachSafetyHardnessSampledFunc}{Status}{Wrong}{}{Count}{0}%
\StoreBenchExecResult{Avr}{IcIIIsaReachSafetyHardnessSampledFunc}{Status}{Wrong}{True}{Count}{0}%
\StoreBenchExecResult{Avr}{IcIIIsaReachSafetyHardnessSampledFunc}{Status}{Wrong}{False}{Count}{0}%
\providecommand\StoreBenchExecResult[7]{\expandafter\newcommand\csname#1#2#3#4#5#6\endcsname{#7}}%
\StoreBenchExecResult{Avr}{IcIIIsaReachSafetyHardwareSampledFunc}{Status}{All}{}{Score}{0}%
\StoreBenchExecResult{Avr}{IcIIIsaReachSafetyHardwareSampledFunc}{Status}{All}{}{Count}{150}%
\StoreBenchExecResult{Avr}{IcIIIsaReachSafetyHardwareSampledFunc}{Status}{Correct}{}{Count}{36}%
\StoreBenchExecResult{Avr}{IcIIIsaReachSafetyHardwareSampledFunc}{Status}{Correct}{True}{Count}{17}%
\StoreBenchExecResult{Avr}{IcIIIsaReachSafetyHardwareSampledFunc}{Status}{Correct}{False}{Count}{19}%
\StoreBenchExecResult{Avr}{IcIIIsaReachSafetyHardwareSampledFunc}{Status}{Wrong}{}{Count}{0}%
\StoreBenchExecResult{Avr}{IcIIIsaReachSafetyHardwareSampledFunc}{Status}{Wrong}{True}{Count}{0}%
\StoreBenchExecResult{Avr}{IcIIIsaReachSafetyHardwareSampledFunc}{Status}{Wrong}{False}{Count}{0}%
\providecommand\StoreBenchExecResult[7]{\expandafter\newcommand\csname#1#2#3#4#5#6\endcsname{#7}}%
\StoreBenchExecResult{Avr}{IcIIIsaReachSafetyHeapSampledFunc}{Status}{All}{}{Score}{0}%
\StoreBenchExecResult{Avr}{IcIIIsaReachSafetyHeapSampledFunc}{Status}{All}{}{Count}{53}%
\StoreBenchExecResult{Avr}{IcIIIsaReachSafetyHeapSampledFunc}{Status}{Correct}{}{Count}{45}%
\StoreBenchExecResult{Avr}{IcIIIsaReachSafetyHeapSampledFunc}{Status}{Correct}{True}{Count}{32}%
\StoreBenchExecResult{Avr}{IcIIIsaReachSafetyHeapSampledFunc}{Status}{Correct}{False}{Count}{13}%
\StoreBenchExecResult{Avr}{IcIIIsaReachSafetyHeapSampledFunc}{Status}{Wrong}{}{Count}{0}%
\StoreBenchExecResult{Avr}{IcIIIsaReachSafetyHeapSampledFunc}{Status}{Wrong}{True}{Count}{0}%
\StoreBenchExecResult{Avr}{IcIIIsaReachSafetyHeapSampledFunc}{Status}{Wrong}{False}{Count}{0}%
\providecommand\StoreBenchExecResult[7]{\expandafter\newcommand\csname#1#2#3#4#5#6\endcsname{#7}}%
\StoreBenchExecResult{Avr}{IcIIIsaReachSafetyLoopsSampledFunc}{Status}{All}{}{Score}{0}%
\StoreBenchExecResult{Avr}{IcIIIsaReachSafetyLoopsSampledFunc}{Status}{All}{}{Count}{150}%
\StoreBenchExecResult{Avr}{IcIIIsaReachSafetyLoopsSampledFunc}{Status}{Correct}{}{Count}{42}%
\StoreBenchExecResult{Avr}{IcIIIsaReachSafetyLoopsSampledFunc}{Status}{Correct}{True}{Count}{26}%
\StoreBenchExecResult{Avr}{IcIIIsaReachSafetyLoopsSampledFunc}{Status}{Correct}{False}{Count}{16}%
\StoreBenchExecResult{Avr}{IcIIIsaReachSafetyLoopsSampledFunc}{Status}{Wrong}{}{Count}{0}%
\StoreBenchExecResult{Avr}{IcIIIsaReachSafetyLoopsSampledFunc}{Status}{Wrong}{True}{Count}{0}%
\StoreBenchExecResult{Avr}{IcIIIsaReachSafetyLoopsSampledFunc}{Status}{Wrong}{False}{Count}{0}%
\providecommand\StoreBenchExecResult[7]{\expandafter\newcommand\csname#1#2#3#4#5#6\endcsname{#7}}%
\StoreBenchExecResult{Avr}{IcIIIsaReachSafetyProductLinesSampledFunc}{Status}{All}{}{Score}{0}%
\StoreBenchExecResult{Avr}{IcIIIsaReachSafetyProductLinesSampledFunc}{Status}{All}{}{Count}{150}%
\StoreBenchExecResult{Avr}{IcIIIsaReachSafetyProductLinesSampledFunc}{Status}{Correct}{}{Count}{149}%
\StoreBenchExecResult{Avr}{IcIIIsaReachSafetyProductLinesSampledFunc}{Status}{Correct}{True}{Count}{74}%
\StoreBenchExecResult{Avr}{IcIIIsaReachSafetyProductLinesSampledFunc}{Status}{Correct}{False}{Count}{75}%
\StoreBenchExecResult{Avr}{IcIIIsaReachSafetyProductLinesSampledFunc}{Status}{Wrong}{}{Count}{0}%
\StoreBenchExecResult{Avr}{IcIIIsaReachSafetyProductLinesSampledFunc}{Status}{Wrong}{True}{Count}{0}%
\StoreBenchExecResult{Avr}{IcIIIsaReachSafetyProductLinesSampledFunc}{Status}{Wrong}{False}{Count}{0}%
\providecommand\StoreBenchExecResult[7]{\expandafter\newcommand\csname#1#2#3#4#5#6\endcsname{#7}}%
\StoreBenchExecResult{Avr}{IcIIIsaReachSafetySequentializedSampledFunc}{Status}{All}{}{Score}{0}%
\StoreBenchExecResult{Avr}{IcIIIsaReachSafetySequentializedSampledFunc}{Status}{All}{}{Count}{150}%
\StoreBenchExecResult{Avr}{IcIIIsaReachSafetySequentializedSampledFunc}{Status}{Correct}{}{Count}{103}%
\StoreBenchExecResult{Avr}{IcIIIsaReachSafetySequentializedSampledFunc}{Status}{Correct}{True}{Count}{24}%
\StoreBenchExecResult{Avr}{IcIIIsaReachSafetySequentializedSampledFunc}{Status}{Correct}{False}{Count}{79}%
\StoreBenchExecResult{Avr}{IcIIIsaReachSafetySequentializedSampledFunc}{Status}{Wrong}{}{Count}{0}%
\StoreBenchExecResult{Avr}{IcIIIsaReachSafetySequentializedSampledFunc}{Status}{Wrong}{True}{Count}{0}%
\StoreBenchExecResult{Avr}{IcIIIsaReachSafetySequentializedSampledFunc}{Status}{Wrong}{False}{Count}{0}%
\providecommand\StoreBenchExecResult[7]{\expandafter\newcommand\csname#1#2#3#4#5#6\endcsname{#7}}%
\StoreBenchExecResult{Avr}{IcIIIsaReachSafetyXCSPSampledFunc}{Status}{All}{}{Score}{0}%
\StoreBenchExecResult{Avr}{IcIIIsaReachSafetyXCSPSampledFunc}{Status}{All}{}{Count}{98}%
\StoreBenchExecResult{Avr}{IcIIIsaReachSafetyXCSPSampledFunc}{Status}{Correct}{}{Count}{88}%
\StoreBenchExecResult{Avr}{IcIIIsaReachSafetyXCSPSampledFunc}{Status}{Correct}{True}{Count}{51}%
\StoreBenchExecResult{Avr}{IcIIIsaReachSafetyXCSPSampledFunc}{Status}{Correct}{False}{Count}{37}%
\StoreBenchExecResult{Avr}{IcIIIsaReachSafetyXCSPSampledFunc}{Status}{Wrong}{}{Count}{0}%
\StoreBenchExecResult{Avr}{IcIIIsaReachSafetyXCSPSampledFunc}{Status}{Wrong}{True}{Count}{0}%
\StoreBenchExecResult{Avr}{IcIIIsaReachSafetyXCSPSampledFunc}{Status}{Wrong}{False}{Count}{0}%
\providecommand\StoreBenchExecResult[7]{\expandafter\newcommand\csname#1#2#3#4#5#6\endcsname{#7}}%
\StoreBenchExecResult{Avr}{IcIIIsaTerminationBitVectorsSampledFunc}{Status}{All}{}{Score}{0}%
\StoreBenchExecResult{Avr}{IcIIIsaTerminationBitVectorsSampledFunc}{Status}{All}{}{Count}{32}%
\StoreBenchExecResult{Avr}{IcIIIsaTerminationBitVectorsSampledFunc}{Status}{Correct}{}{Count}{16}%
\StoreBenchExecResult{Avr}{IcIIIsaTerminationBitVectorsSampledFunc}{Status}{Correct}{True}{Count}{5}%
\StoreBenchExecResult{Avr}{IcIIIsaTerminationBitVectorsSampledFunc}{Status}{Correct}{False}{Count}{11}%
\StoreBenchExecResult{Avr}{IcIIIsaTerminationBitVectorsSampledFunc}{Status}{Wrong}{}{Count}{0}%
\StoreBenchExecResult{Avr}{IcIIIsaTerminationBitVectorsSampledFunc}{Status}{Wrong}{True}{Count}{0}%
\StoreBenchExecResult{Avr}{IcIIIsaTerminationBitVectorsSampledFunc}{Status}{Wrong}{False}{Count}{0}%
\providecommand\StoreBenchExecResult[7]{\expandafter\newcommand\csname#1#2#3#4#5#6\endcsname{#7}}%
\StoreBenchExecResult{Avr}{IcIIIsaTerminationMainControlFlowSampledFunc}{Status}{All}{}{Score}{0}%
\StoreBenchExecResult{Avr}{IcIIIsaTerminationMainControlFlowSampledFunc}{Status}{All}{}{Count}{242}%
\StoreBenchExecResult{Avr}{IcIIIsaTerminationMainControlFlowSampledFunc}{Status}{Correct}{}{Count}{62}%
\StoreBenchExecResult{Avr}{IcIIIsaTerminationMainControlFlowSampledFunc}{Status}{Correct}{True}{Count}{18}%
\StoreBenchExecResult{Avr}{IcIIIsaTerminationMainControlFlowSampledFunc}{Status}{Correct}{False}{Count}{44}%
\StoreBenchExecResult{Avr}{IcIIIsaTerminationMainControlFlowSampledFunc}{Status}{Wrong}{}{Count}{0}%
\StoreBenchExecResult{Avr}{IcIIIsaTerminationMainControlFlowSampledFunc}{Status}{Wrong}{True}{Count}{0}%
\StoreBenchExecResult{Avr}{IcIIIsaTerminationMainControlFlowSampledFunc}{Status}{Wrong}{False}{Count}{0}%
\providecommand\StoreBenchExecResult[7]{\expandafter\newcommand\csname#1#2#3#4#5#6\endcsname{#7}}%
\StoreBenchExecResult{Avr}{IcIIIsaTerminationOtherSampledFunc}{Status}{All}{}{Score}{0}%
\StoreBenchExecResult{Avr}{IcIIIsaTerminationOtherSampledFunc}{Status}{All}{}{Count}{1080}%
\StoreBenchExecResult{Avr}{IcIIIsaTerminationOtherSampledFunc}{Status}{Correct}{}{Count}{825}%
\StoreBenchExecResult{Avr}{IcIIIsaTerminationOtherSampledFunc}{Status}{Correct}{True}{Count}{195}%
\StoreBenchExecResult{Avr}{IcIIIsaTerminationOtherSampledFunc}{Status}{Correct}{False}{Count}{630}%
\StoreBenchExecResult{Avr}{IcIIIsaTerminationOtherSampledFunc}{Status}{Wrong}{}{Count}{0}%
\StoreBenchExecResult{Avr}{IcIIIsaTerminationOtherSampledFunc}{Status}{Wrong}{True}{Count}{0}%
\StoreBenchExecResult{Avr}{IcIIIsaTerminationOtherSampledFunc}{Status}{Wrong}{False}{Count}{0}%
\edef\AvrIcIIIsaReachSafetySampledFuncStatusAllCount{\the\numexpr\AvrIcIIIsaReachSafetyArraysSampledFuncStatusAllCount+\AvrIcIIIsaReachSafetyBitVectorsSampledFuncStatusAllCount+\AvrIcIIIsaReachSafetyCombinationsSampledFuncStatusAllCount+\AvrIcIIIsaReachSafetyControlFlowSampledFuncStatusAllCount+\AvrIcIIIsaReachSafetyECASampledFuncStatusAllCount+\AvrIcIIIsaReachSafetyFloatsSampledFuncStatusAllCount+\AvrIcIIIsaReachSafetyHardnessSampledFuncStatusAllCount+\AvrIcIIIsaReachSafetyHardwareSampledFuncStatusAllCount+\AvrIcIIIsaReachSafetyHeapSampledFuncStatusAllCount+\AvrIcIIIsaReachSafetyLoopsSampledFuncStatusAllCount+\AvrIcIIIsaReachSafetyProductLinesSampledFuncStatusAllCount+\AvrIcIIIsaReachSafetySequentializedSampledFuncStatusAllCount+\AvrIcIIIsaReachSafetyXCSPSampledFuncStatusAllCount}
\edef\AvrIcIIIsaReachSafetySampledFuncStatusCorrectCount{\the\numexpr\AvrIcIIIsaReachSafetyArraysSampledFuncStatusCorrectCount+\AvrIcIIIsaReachSafetyBitVectorsSampledFuncStatusCorrectCount+\AvrIcIIIsaReachSafetyCombinationsSampledFuncStatusCorrectCount+\AvrIcIIIsaReachSafetyControlFlowSampledFuncStatusCorrectCount+\AvrIcIIIsaReachSafetyECASampledFuncStatusCorrectCount+\AvrIcIIIsaReachSafetyFloatsSampledFuncStatusCorrectCount+\AvrIcIIIsaReachSafetyHardnessSampledFuncStatusCorrectCount+\AvrIcIIIsaReachSafetyHardwareSampledFuncStatusCorrectCount+\AvrIcIIIsaReachSafetyHeapSampledFuncStatusCorrectCount+\AvrIcIIIsaReachSafetyLoopsSampledFuncStatusCorrectCount+\AvrIcIIIsaReachSafetyProductLinesSampledFuncStatusCorrectCount+\AvrIcIIIsaReachSafetySequentializedSampledFuncStatusCorrectCount+\AvrIcIIIsaReachSafetyXCSPSampledFuncStatusCorrectCount}
\edef\AvrIcIIIsaReachSafetySampledFuncStatusCorrectTrueCount{\the\numexpr\AvrIcIIIsaReachSafetyArraysSampledFuncStatusCorrectTrueCount+\AvrIcIIIsaReachSafetyBitVectorsSampledFuncStatusCorrectTrueCount+\AvrIcIIIsaReachSafetyCombinationsSampledFuncStatusCorrectTrueCount+\AvrIcIIIsaReachSafetyControlFlowSampledFuncStatusCorrectTrueCount+\AvrIcIIIsaReachSafetyECASampledFuncStatusCorrectTrueCount+\AvrIcIIIsaReachSafetyFloatsSampledFuncStatusCorrectTrueCount+\AvrIcIIIsaReachSafetyHardnessSampledFuncStatusCorrectTrueCount+\AvrIcIIIsaReachSafetyHardwareSampledFuncStatusCorrectTrueCount+\AvrIcIIIsaReachSafetyHeapSampledFuncStatusCorrectTrueCount+\AvrIcIIIsaReachSafetyLoopsSampledFuncStatusCorrectTrueCount+\AvrIcIIIsaReachSafetyProductLinesSampledFuncStatusCorrectTrueCount+\AvrIcIIIsaReachSafetySequentializedSampledFuncStatusCorrectTrueCount+\AvrIcIIIsaReachSafetyXCSPSampledFuncStatusCorrectTrueCount}
\edef\AvrIcIIIsaReachSafetySampledFuncStatusCorrectFalseCount{\the\numexpr\AvrIcIIIsaReachSafetyArraysSampledFuncStatusCorrectFalseCount+\AvrIcIIIsaReachSafetyBitVectorsSampledFuncStatusCorrectFalseCount+\AvrIcIIIsaReachSafetyCombinationsSampledFuncStatusCorrectFalseCount+\AvrIcIIIsaReachSafetyControlFlowSampledFuncStatusCorrectFalseCount+\AvrIcIIIsaReachSafetyECASampledFuncStatusCorrectFalseCount+\AvrIcIIIsaReachSafetyFloatsSampledFuncStatusCorrectFalseCount+\AvrIcIIIsaReachSafetyHardnessSampledFuncStatusCorrectFalseCount+\AvrIcIIIsaReachSafetyHardwareSampledFuncStatusCorrectFalseCount+\AvrIcIIIsaReachSafetyHeapSampledFuncStatusCorrectFalseCount+\AvrIcIIIsaReachSafetyLoopsSampledFuncStatusCorrectFalseCount+\AvrIcIIIsaReachSafetyProductLinesSampledFuncStatusCorrectFalseCount+\AvrIcIIIsaReachSafetySequentializedSampledFuncStatusCorrectFalseCount+\AvrIcIIIsaReachSafetyXCSPSampledFuncStatusCorrectFalseCount}
\edef\AvrIcIIIsaReachSafetySampledFuncStatusWrongCount{\the\numexpr\AvrIcIIIsaReachSafetyArraysSampledFuncStatusWrongCount+\AvrIcIIIsaReachSafetyBitVectorsSampledFuncStatusWrongCount+\AvrIcIIIsaReachSafetyCombinationsSampledFuncStatusWrongCount+\AvrIcIIIsaReachSafetyControlFlowSampledFuncStatusWrongCount+\AvrIcIIIsaReachSafetyECASampledFuncStatusWrongCount+\AvrIcIIIsaReachSafetyFloatsSampledFuncStatusWrongCount+\AvrIcIIIsaReachSafetyHardnessSampledFuncStatusWrongCount+\AvrIcIIIsaReachSafetyHardwareSampledFuncStatusWrongCount+\AvrIcIIIsaReachSafetyHeapSampledFuncStatusWrongCount+\AvrIcIIIsaReachSafetyLoopsSampledFuncStatusWrongCount+\AvrIcIIIsaReachSafetyProductLinesSampledFuncStatusWrongCount+\AvrIcIIIsaReachSafetySequentializedSampledFuncStatusWrongCount+\AvrIcIIIsaReachSafetyXCSPSampledFuncStatusWrongCount}
\edef\AvrIcIIIsaTerminationSampledFuncStatusAllCount{\the\numexpr\AvrIcIIIsaTerminationBitVectorsSampledFuncStatusAllCount+\AvrIcIIIsaTerminationMainControlFlowSampledFuncStatusAllCount+\AvrIcIIIsaTerminationOtherSampledFuncStatusAllCount}
\edef\AvrIcIIIsaTerminationSampledFuncStatusCorrectCount{\the\numexpr\AvrIcIIIsaTerminationBitVectorsSampledFuncStatusCorrectCount+\AvrIcIIIsaTerminationMainControlFlowSampledFuncStatusCorrectCount+\AvrIcIIIsaTerminationOtherSampledFuncStatusCorrectCount}
\edef\AvrIcIIIsaTerminationSampledFuncStatusCorrectTrueCount{\the\numexpr\AvrIcIIIsaTerminationBitVectorsSampledFuncStatusCorrectTrueCount+\AvrIcIIIsaTerminationMainControlFlowSampledFuncStatusCorrectTrueCount+\AvrIcIIIsaTerminationOtherSampledFuncStatusCorrectTrueCount}
\edef\AvrIcIIIsaTerminationSampledFuncStatusCorrectFalseCount{\the\numexpr\AvrIcIIIsaTerminationBitVectorsSampledFuncStatusCorrectFalseCount+\AvrIcIIIsaTerminationMainControlFlowSampledFuncStatusCorrectFalseCount+\AvrIcIIIsaTerminationOtherSampledFuncStatusCorrectFalseCount}
\edef\AvrIcIIIsaTerminationSampledFuncStatusWrongCount{\the\numexpr\AvrIcIIIsaTerminationBitVectorsSampledFuncStatusWrongCount+\AvrIcIIIsaTerminationMainControlFlowSampledFuncStatusWrongCount+\AvrIcIIIsaTerminationOtherSampledFuncStatusWrongCount}
\providecommand\StoreBenchExecResult[7]{\expandafter\newcommand\csname#1#2#3#4#5#6\endcsname{#7}}%
\StoreBenchExecResult{Avr}{KindReachSafetyArraysSampledFunc}{Status}{All}{}{Score}{0}%
\StoreBenchExecResult{Avr}{KindReachSafetyArraysSampledFunc}{Status}{All}{}{Count}{150}%
\StoreBenchExecResult{Avr}{KindReachSafetyArraysSampledFunc}{Status}{Correct}{}{Count}{50}%
\StoreBenchExecResult{Avr}{KindReachSafetyArraysSampledFunc}{Status}{Correct}{True}{Count}{0}%
\StoreBenchExecResult{Avr}{KindReachSafetyArraysSampledFunc}{Status}{Correct}{False}{Count}{50}%
\StoreBenchExecResult{Avr}{KindReachSafetyArraysSampledFunc}{Status}{Wrong}{}{Count}{0}%
\StoreBenchExecResult{Avr}{KindReachSafetyArraysSampledFunc}{Status}{Wrong}{True}{Count}{0}%
\StoreBenchExecResult{Avr}{KindReachSafetyArraysSampledFunc}{Status}{Wrong}{False}{Count}{0}%
\providecommand\StoreBenchExecResult[7]{\expandafter\newcommand\csname#1#2#3#4#5#6\endcsname{#7}}%
\StoreBenchExecResult{Avr}{KindReachSafetyBitVectorsSampledFunc}{Status}{All}{}{Score}{0}%
\StoreBenchExecResult{Avr}{KindReachSafetyBitVectorsSampledFunc}{Status}{All}{}{Count}{48}%
\StoreBenchExecResult{Avr}{KindReachSafetyBitVectorsSampledFunc}{Status}{Correct}{}{Count}{30}%
\StoreBenchExecResult{Avr}{KindReachSafetyBitVectorsSampledFunc}{Status}{Correct}{True}{Count}{18}%
\StoreBenchExecResult{Avr}{KindReachSafetyBitVectorsSampledFunc}{Status}{Correct}{False}{Count}{12}%
\StoreBenchExecResult{Avr}{KindReachSafetyBitVectorsSampledFunc}{Status}{Wrong}{}{Count}{0}%
\StoreBenchExecResult{Avr}{KindReachSafetyBitVectorsSampledFunc}{Status}{Wrong}{True}{Count}{0}%
\StoreBenchExecResult{Avr}{KindReachSafetyBitVectorsSampledFunc}{Status}{Wrong}{False}{Count}{0}%
\providecommand\StoreBenchExecResult[7]{\expandafter\newcommand\csname#1#2#3#4#5#6\endcsname{#7}}%
\StoreBenchExecResult{Avr}{KindReachSafetyCombinationsSampledFunc}{Status}{All}{}{Score}{0}%
\StoreBenchExecResult{Avr}{KindReachSafetyCombinationsSampledFunc}{Status}{All}{}{Count}{110}%
\StoreBenchExecResult{Avr}{KindReachSafetyCombinationsSampledFunc}{Status}{Correct}{}{Count}{71}%
\StoreBenchExecResult{Avr}{KindReachSafetyCombinationsSampledFunc}{Status}{Correct}{True}{Count}{0}%
\StoreBenchExecResult{Avr}{KindReachSafetyCombinationsSampledFunc}{Status}{Correct}{False}{Count}{71}%
\StoreBenchExecResult{Avr}{KindReachSafetyCombinationsSampledFunc}{Status}{Wrong}{}{Count}{0}%
\StoreBenchExecResult{Avr}{KindReachSafetyCombinationsSampledFunc}{Status}{Wrong}{True}{Count}{0}%
\StoreBenchExecResult{Avr}{KindReachSafetyCombinationsSampledFunc}{Status}{Wrong}{False}{Count}{0}%
\providecommand\StoreBenchExecResult[7]{\expandafter\newcommand\csname#1#2#3#4#5#6\endcsname{#7}}%
\StoreBenchExecResult{Avr}{KindReachSafetyControlFlowSampledFunc}{Status}{All}{}{Score}{0}%
\StoreBenchExecResult{Avr}{KindReachSafetyControlFlowSampledFunc}{Status}{All}{}{Count}{33}%
\StoreBenchExecResult{Avr}{KindReachSafetyControlFlowSampledFunc}{Status}{Correct}{}{Count}{25}%
\StoreBenchExecResult{Avr}{KindReachSafetyControlFlowSampledFunc}{Status}{Correct}{True}{Count}{21}%
\StoreBenchExecResult{Avr}{KindReachSafetyControlFlowSampledFunc}{Status}{Correct}{False}{Count}{4}%
\StoreBenchExecResult{Avr}{KindReachSafetyControlFlowSampledFunc}{Status}{Wrong}{}{Count}{0}%
\StoreBenchExecResult{Avr}{KindReachSafetyControlFlowSampledFunc}{Status}{Wrong}{True}{Count}{0}%
\StoreBenchExecResult{Avr}{KindReachSafetyControlFlowSampledFunc}{Status}{Wrong}{False}{Count}{0}%
\providecommand\StoreBenchExecResult[7]{\expandafter\newcommand\csname#1#2#3#4#5#6\endcsname{#7}}%
\StoreBenchExecResult{Avr}{KindReachSafetyECASampledFunc}{Status}{All}{}{Score}{0}%
\StoreBenchExecResult{Avr}{KindReachSafetyECASampledFunc}{Status}{All}{}{Count}{150}%
\StoreBenchExecResult{Avr}{KindReachSafetyECASampledFunc}{Status}{Correct}{}{Count}{100}%
\StoreBenchExecResult{Avr}{KindReachSafetyECASampledFunc}{Status}{Correct}{True}{Count}{50}%
\StoreBenchExecResult{Avr}{KindReachSafetyECASampledFunc}{Status}{Correct}{False}{Count}{50}%
\StoreBenchExecResult{Avr}{KindReachSafetyECASampledFunc}{Status}{Wrong}{}{Count}{0}%
\StoreBenchExecResult{Avr}{KindReachSafetyECASampledFunc}{Status}{Wrong}{True}{Count}{0}%
\StoreBenchExecResult{Avr}{KindReachSafetyECASampledFunc}{Status}{Wrong}{False}{Count}{0}%
\providecommand\StoreBenchExecResult[7]{\expandafter\newcommand\csname#1#2#3#4#5#6\endcsname{#7}}%
\StoreBenchExecResult{Avr}{KindReachSafetyFloatsSampledFunc}{Status}{All}{}{Score}{0}%
\StoreBenchExecResult{Avr}{KindReachSafetyFloatsSampledFunc}{Status}{All}{}{Count}{10}%
\StoreBenchExecResult{Avr}{KindReachSafetyFloatsSampledFunc}{Status}{Correct}{}{Count}{10}%
\StoreBenchExecResult{Avr}{KindReachSafetyFloatsSampledFunc}{Status}{Correct}{True}{Count}{10}%
\StoreBenchExecResult{Avr}{KindReachSafetyFloatsSampledFunc}{Status}{Correct}{False}{Count}{0}%
\StoreBenchExecResult{Avr}{KindReachSafetyFloatsSampledFunc}{Status}{Wrong}{}{Count}{0}%
\StoreBenchExecResult{Avr}{KindReachSafetyFloatsSampledFunc}{Status}{Wrong}{True}{Count}{0}%
\StoreBenchExecResult{Avr}{KindReachSafetyFloatsSampledFunc}{Status}{Wrong}{False}{Count}{0}%
\providecommand\StoreBenchExecResult[7]{\expandafter\newcommand\csname#1#2#3#4#5#6\endcsname{#7}}%
\StoreBenchExecResult{Avr}{KindReachSafetyHardnessSampledFunc}{Status}{All}{}{Score}{0}%
\StoreBenchExecResult{Avr}{KindReachSafetyHardnessSampledFunc}{Status}{All}{}{Count}{150}%
\StoreBenchExecResult{Avr}{KindReachSafetyHardnessSampledFunc}{Status}{Correct}{}{Count}{132}%
\StoreBenchExecResult{Avr}{KindReachSafetyHardnessSampledFunc}{Status}{Correct}{True}{Count}{132}%
\StoreBenchExecResult{Avr}{KindReachSafetyHardnessSampledFunc}{Status}{Correct}{False}{Count}{0}%
\StoreBenchExecResult{Avr}{KindReachSafetyHardnessSampledFunc}{Status}{Wrong}{}{Count}{0}%
\StoreBenchExecResult{Avr}{KindReachSafetyHardnessSampledFunc}{Status}{Wrong}{True}{Count}{0}%
\StoreBenchExecResult{Avr}{KindReachSafetyHardnessSampledFunc}{Status}{Wrong}{False}{Count}{0}%
\providecommand\StoreBenchExecResult[7]{\expandafter\newcommand\csname#1#2#3#4#5#6\endcsname{#7}}%
\StoreBenchExecResult{Avr}{KindReachSafetyHardwareSampledFunc}{Status}{All}{}{Score}{0}%
\StoreBenchExecResult{Avr}{KindReachSafetyHardwareSampledFunc}{Status}{All}{}{Count}{150}%
\StoreBenchExecResult{Avr}{KindReachSafetyHardwareSampledFunc}{Status}{Correct}{}{Count}{58}%
\StoreBenchExecResult{Avr}{KindReachSafetyHardwareSampledFunc}{Status}{Correct}{True}{Count}{3}%
\StoreBenchExecResult{Avr}{KindReachSafetyHardwareSampledFunc}{Status}{Correct}{False}{Count}{55}%
\StoreBenchExecResult{Avr}{KindReachSafetyHardwareSampledFunc}{Status}{Wrong}{}{Count}{0}%
\StoreBenchExecResult{Avr}{KindReachSafetyHardwareSampledFunc}{Status}{Wrong}{True}{Count}{0}%
\StoreBenchExecResult{Avr}{KindReachSafetyHardwareSampledFunc}{Status}{Wrong}{False}{Count}{0}%
\providecommand\StoreBenchExecResult[7]{\expandafter\newcommand\csname#1#2#3#4#5#6\endcsname{#7}}%
\StoreBenchExecResult{Avr}{KindReachSafetyHeapSampledFunc}{Status}{All}{}{Score}{0}%
\StoreBenchExecResult{Avr}{KindReachSafetyHeapSampledFunc}{Status}{All}{}{Count}{53}%
\StoreBenchExecResult{Avr}{KindReachSafetyHeapSampledFunc}{Status}{Correct}{}{Count}{48}%
\StoreBenchExecResult{Avr}{KindReachSafetyHeapSampledFunc}{Status}{Correct}{True}{Count}{31}%
\StoreBenchExecResult{Avr}{KindReachSafetyHeapSampledFunc}{Status}{Correct}{False}{Count}{17}%
\StoreBenchExecResult{Avr}{KindReachSafetyHeapSampledFunc}{Status}{Wrong}{}{Count}{0}%
\StoreBenchExecResult{Avr}{KindReachSafetyHeapSampledFunc}{Status}{Wrong}{True}{Count}{0}%
\StoreBenchExecResult{Avr}{KindReachSafetyHeapSampledFunc}{Status}{Wrong}{False}{Count}{0}%
\providecommand\StoreBenchExecResult[7]{\expandafter\newcommand\csname#1#2#3#4#5#6\endcsname{#7}}%
\StoreBenchExecResult{Avr}{KindReachSafetyLoopsSampledFunc}{Status}{All}{}{Score}{0}%
\StoreBenchExecResult{Avr}{KindReachSafetyLoopsSampledFunc}{Status}{All}{}{Count}{150}%
\StoreBenchExecResult{Avr}{KindReachSafetyLoopsSampledFunc}{Status}{Correct}{}{Count}{55}%
\StoreBenchExecResult{Avr}{KindReachSafetyLoopsSampledFunc}{Status}{Correct}{True}{Count}{15}%
\StoreBenchExecResult{Avr}{KindReachSafetyLoopsSampledFunc}{Status}{Correct}{False}{Count}{40}%
\StoreBenchExecResult{Avr}{KindReachSafetyLoopsSampledFunc}{Status}{Wrong}{}{Count}{0}%
\StoreBenchExecResult{Avr}{KindReachSafetyLoopsSampledFunc}{Status}{Wrong}{True}{Count}{0}%
\StoreBenchExecResult{Avr}{KindReachSafetyLoopsSampledFunc}{Status}{Wrong}{False}{Count}{0}%
\providecommand\StoreBenchExecResult[7]{\expandafter\newcommand\csname#1#2#3#4#5#6\endcsname{#7}}%
\StoreBenchExecResult{Avr}{KindReachSafetyProductLinesSampledFunc}{Status}{All}{}{Score}{0}%
\StoreBenchExecResult{Avr}{KindReachSafetyProductLinesSampledFunc}{Status}{All}{}{Count}{150}%
\StoreBenchExecResult{Avr}{KindReachSafetyProductLinesSampledFunc}{Status}{Correct}{}{Count}{91}%
\StoreBenchExecResult{Avr}{KindReachSafetyProductLinesSampledFunc}{Status}{Correct}{True}{Count}{16}%
\StoreBenchExecResult{Avr}{KindReachSafetyProductLinesSampledFunc}{Status}{Correct}{False}{Count}{75}%
\StoreBenchExecResult{Avr}{KindReachSafetyProductLinesSampledFunc}{Status}{Wrong}{}{Count}{0}%
\StoreBenchExecResult{Avr}{KindReachSafetyProductLinesSampledFunc}{Status}{Wrong}{True}{Count}{0}%
\StoreBenchExecResult{Avr}{KindReachSafetyProductLinesSampledFunc}{Status}{Wrong}{False}{Count}{0}%
\providecommand\StoreBenchExecResult[7]{\expandafter\newcommand\csname#1#2#3#4#5#6\endcsname{#7}}%
\StoreBenchExecResult{Avr}{KindReachSafetySequentializedSampledFunc}{Status}{All}{}{Score}{0}%
\StoreBenchExecResult{Avr}{KindReachSafetySequentializedSampledFunc}{Status}{All}{}{Count}{150}%
\StoreBenchExecResult{Avr}{KindReachSafetySequentializedSampledFunc}{Status}{Correct}{}{Count}{102}%
\StoreBenchExecResult{Avr}{KindReachSafetySequentializedSampledFunc}{Status}{Correct}{True}{Count}{14}%
\StoreBenchExecResult{Avr}{KindReachSafetySequentializedSampledFunc}{Status}{Correct}{False}{Count}{88}%
\StoreBenchExecResult{Avr}{KindReachSafetySequentializedSampledFunc}{Status}{Wrong}{}{Count}{0}%
\StoreBenchExecResult{Avr}{KindReachSafetySequentializedSampledFunc}{Status}{Wrong}{True}{Count}{0}%
\StoreBenchExecResult{Avr}{KindReachSafetySequentializedSampledFunc}{Status}{Wrong}{False}{Count}{0}%
\providecommand\StoreBenchExecResult[7]{\expandafter\newcommand\csname#1#2#3#4#5#6\endcsname{#7}}%
\StoreBenchExecResult{Avr}{KindReachSafetyXCSPSampledFunc}{Status}{All}{}{Score}{0}%
\StoreBenchExecResult{Avr}{KindReachSafetyXCSPSampledFunc}{Status}{All}{}{Count}{98}%
\StoreBenchExecResult{Avr}{KindReachSafetyXCSPSampledFunc}{Status}{Correct}{}{Count}{96}%
\StoreBenchExecResult{Avr}{KindReachSafetyXCSPSampledFunc}{Status}{Correct}{True}{Count}{52}%
\StoreBenchExecResult{Avr}{KindReachSafetyXCSPSampledFunc}{Status}{Correct}{False}{Count}{44}%
\StoreBenchExecResult{Avr}{KindReachSafetyXCSPSampledFunc}{Status}{Wrong}{}{Count}{0}%
\StoreBenchExecResult{Avr}{KindReachSafetyXCSPSampledFunc}{Status}{Wrong}{True}{Count}{0}%
\StoreBenchExecResult{Avr}{KindReachSafetyXCSPSampledFunc}{Status}{Wrong}{False}{Count}{0}%
\providecommand\StoreBenchExecResult[7]{\expandafter\newcommand\csname#1#2#3#4#5#6\endcsname{#7}}%
\StoreBenchExecResult{Avr}{KindTerminationBitVectorsSampledFunc}{Status}{All}{}{Score}{0}%
\StoreBenchExecResult{Avr}{KindTerminationBitVectorsSampledFunc}{Status}{All}{}{Count}{32}%
\StoreBenchExecResult{Avr}{KindTerminationBitVectorsSampledFunc}{Status}{Correct}{}{Count}{21}%
\StoreBenchExecResult{Avr}{KindTerminationBitVectorsSampledFunc}{Status}{Correct}{True}{Count}{10}%
\StoreBenchExecResult{Avr}{KindTerminationBitVectorsSampledFunc}{Status}{Correct}{False}{Count}{11}%
\StoreBenchExecResult{Avr}{KindTerminationBitVectorsSampledFunc}{Status}{Wrong}{}{Count}{0}%
\StoreBenchExecResult{Avr}{KindTerminationBitVectorsSampledFunc}{Status}{Wrong}{True}{Count}{0}%
\StoreBenchExecResult{Avr}{KindTerminationBitVectorsSampledFunc}{Status}{Wrong}{False}{Count}{0}%
\providecommand\StoreBenchExecResult[7]{\expandafter\newcommand\csname#1#2#3#4#5#6\endcsname{#7}}%
\StoreBenchExecResult{Avr}{KindTerminationMainControlFlowSampledFunc}{Status}{All}{}{Score}{0}%
\StoreBenchExecResult{Avr}{KindTerminationMainControlFlowSampledFunc}{Status}{All}{}{Count}{242}%
\StoreBenchExecResult{Avr}{KindTerminationMainControlFlowSampledFunc}{Status}{Correct}{}{Count}{62}%
\StoreBenchExecResult{Avr}{KindTerminationMainControlFlowSampledFunc}{Status}{Correct}{True}{Count}{10}%
\StoreBenchExecResult{Avr}{KindTerminationMainControlFlowSampledFunc}{Status}{Correct}{False}{Count}{52}%
\StoreBenchExecResult{Avr}{KindTerminationMainControlFlowSampledFunc}{Status}{Wrong}{}{Count}{0}%
\StoreBenchExecResult{Avr}{KindTerminationMainControlFlowSampledFunc}{Status}{Wrong}{True}{Count}{0}%
\StoreBenchExecResult{Avr}{KindTerminationMainControlFlowSampledFunc}{Status}{Wrong}{False}{Count}{0}%
\providecommand\StoreBenchExecResult[7]{\expandafter\newcommand\csname#1#2#3#4#5#6\endcsname{#7}}%
\StoreBenchExecResult{Avr}{KindTerminationOtherSampledFunc}{Status}{All}{}{Score}{0}%
\StoreBenchExecResult{Avr}{KindTerminationOtherSampledFunc}{Status}{All}{}{Count}{1080}%
\StoreBenchExecResult{Avr}{KindTerminationOtherSampledFunc}{Status}{Correct}{}{Count}{759}%
\StoreBenchExecResult{Avr}{KindTerminationOtherSampledFunc}{Status}{Correct}{True}{Count}{92}%
\StoreBenchExecResult{Avr}{KindTerminationOtherSampledFunc}{Status}{Correct}{False}{Count}{667}%
\StoreBenchExecResult{Avr}{KindTerminationOtherSampledFunc}{Status}{Wrong}{}{Count}{0}%
\StoreBenchExecResult{Avr}{KindTerminationOtherSampledFunc}{Status}{Wrong}{True}{Count}{0}%
\StoreBenchExecResult{Avr}{KindTerminationOtherSampledFunc}{Status}{Wrong}{False}{Count}{0}%
\edef\AvrKindReachSafetySampledFuncStatusAllCount{\the\numexpr\AvrKindReachSafetyArraysSampledFuncStatusAllCount+\AvrKindReachSafetyBitVectorsSampledFuncStatusAllCount+\AvrKindReachSafetyCombinationsSampledFuncStatusAllCount+\AvrKindReachSafetyControlFlowSampledFuncStatusAllCount+\AvrKindReachSafetyECASampledFuncStatusAllCount+\AvrKindReachSafetyFloatsSampledFuncStatusAllCount+\AvrKindReachSafetyHardnessSampledFuncStatusAllCount+\AvrKindReachSafetyHardwareSampledFuncStatusAllCount+\AvrKindReachSafetyHeapSampledFuncStatusAllCount+\AvrKindReachSafetyLoopsSampledFuncStatusAllCount+\AvrKindReachSafetyProductLinesSampledFuncStatusAllCount+\AvrKindReachSafetySequentializedSampledFuncStatusAllCount+\AvrKindReachSafetyXCSPSampledFuncStatusAllCount}
\edef\AvrKindReachSafetySampledFuncStatusCorrectCount{\the\numexpr\AvrKindReachSafetyArraysSampledFuncStatusCorrectCount+\AvrKindReachSafetyBitVectorsSampledFuncStatusCorrectCount+\AvrKindReachSafetyCombinationsSampledFuncStatusCorrectCount+\AvrKindReachSafetyControlFlowSampledFuncStatusCorrectCount+\AvrKindReachSafetyECASampledFuncStatusCorrectCount+\AvrKindReachSafetyFloatsSampledFuncStatusCorrectCount+\AvrKindReachSafetyHardnessSampledFuncStatusCorrectCount+\AvrKindReachSafetyHardwareSampledFuncStatusCorrectCount+\AvrKindReachSafetyHeapSampledFuncStatusCorrectCount+\AvrKindReachSafetyLoopsSampledFuncStatusCorrectCount+\AvrKindReachSafetyProductLinesSampledFuncStatusCorrectCount+\AvrKindReachSafetySequentializedSampledFuncStatusCorrectCount+\AvrKindReachSafetyXCSPSampledFuncStatusCorrectCount}
\edef\AvrKindReachSafetySampledFuncStatusCorrectTrueCount{\the\numexpr\AvrKindReachSafetyArraysSampledFuncStatusCorrectTrueCount+\AvrKindReachSafetyBitVectorsSampledFuncStatusCorrectTrueCount+\AvrKindReachSafetyCombinationsSampledFuncStatusCorrectTrueCount+\AvrKindReachSafetyControlFlowSampledFuncStatusCorrectTrueCount+\AvrKindReachSafetyECASampledFuncStatusCorrectTrueCount+\AvrKindReachSafetyFloatsSampledFuncStatusCorrectTrueCount+\AvrKindReachSafetyHardnessSampledFuncStatusCorrectTrueCount+\AvrKindReachSafetyHardwareSampledFuncStatusCorrectTrueCount+\AvrKindReachSafetyHeapSampledFuncStatusCorrectTrueCount+\AvrKindReachSafetyLoopsSampledFuncStatusCorrectTrueCount+\AvrKindReachSafetyProductLinesSampledFuncStatusCorrectTrueCount+\AvrKindReachSafetySequentializedSampledFuncStatusCorrectTrueCount+\AvrKindReachSafetyXCSPSampledFuncStatusCorrectTrueCount}
\edef\AvrKindReachSafetySampledFuncStatusCorrectFalseCount{\the\numexpr\AvrKindReachSafetyArraysSampledFuncStatusCorrectFalseCount+\AvrKindReachSafetyBitVectorsSampledFuncStatusCorrectFalseCount+\AvrKindReachSafetyCombinationsSampledFuncStatusCorrectFalseCount+\AvrKindReachSafetyControlFlowSampledFuncStatusCorrectFalseCount+\AvrKindReachSafetyECASampledFuncStatusCorrectFalseCount+\AvrKindReachSafetyFloatsSampledFuncStatusCorrectFalseCount+\AvrKindReachSafetyHardnessSampledFuncStatusCorrectFalseCount+\AvrKindReachSafetyHardwareSampledFuncStatusCorrectFalseCount+\AvrKindReachSafetyHeapSampledFuncStatusCorrectFalseCount+\AvrKindReachSafetyLoopsSampledFuncStatusCorrectFalseCount+\AvrKindReachSafetyProductLinesSampledFuncStatusCorrectFalseCount+\AvrKindReachSafetySequentializedSampledFuncStatusCorrectFalseCount+\AvrKindReachSafetyXCSPSampledFuncStatusCorrectFalseCount}
\edef\AvrKindReachSafetySampledFuncStatusWrongCount{\the\numexpr\AvrKindReachSafetyArraysSampledFuncStatusWrongCount+\AvrKindReachSafetyBitVectorsSampledFuncStatusWrongCount+\AvrKindReachSafetyCombinationsSampledFuncStatusWrongCount+\AvrKindReachSafetyControlFlowSampledFuncStatusWrongCount+\AvrKindReachSafetyECASampledFuncStatusWrongCount+\AvrKindReachSafetyFloatsSampledFuncStatusWrongCount+\AvrKindReachSafetyHardnessSampledFuncStatusWrongCount+\AvrKindReachSafetyHardwareSampledFuncStatusWrongCount+\AvrKindReachSafetyHeapSampledFuncStatusWrongCount+\AvrKindReachSafetyLoopsSampledFuncStatusWrongCount+\AvrKindReachSafetyProductLinesSampledFuncStatusWrongCount+\AvrKindReachSafetySequentializedSampledFuncStatusWrongCount+\AvrKindReachSafetyXCSPSampledFuncStatusWrongCount}
\edef\AvrKindTerminationSampledFuncStatusAllCount{\the\numexpr\AvrKindTerminationBitVectorsSampledFuncStatusAllCount+\AvrKindTerminationMainControlFlowSampledFuncStatusAllCount+\AvrKindTerminationOtherSampledFuncStatusAllCount}
\edef\AvrKindTerminationSampledFuncStatusCorrectCount{\the\numexpr\AvrKindTerminationBitVectorsSampledFuncStatusCorrectCount+\AvrKindTerminationMainControlFlowSampledFuncStatusCorrectCount+\AvrKindTerminationOtherSampledFuncStatusCorrectCount}
\edef\AvrKindTerminationSampledFuncStatusCorrectTrueCount{\the\numexpr\AvrKindTerminationBitVectorsSampledFuncStatusCorrectTrueCount+\AvrKindTerminationMainControlFlowSampledFuncStatusCorrectTrueCount+\AvrKindTerminationOtherSampledFuncStatusCorrectTrueCount}
\edef\AvrKindTerminationSampledFuncStatusCorrectFalseCount{\the\numexpr\AvrKindTerminationBitVectorsSampledFuncStatusCorrectFalseCount+\AvrKindTerminationMainControlFlowSampledFuncStatusCorrectFalseCount+\AvrKindTerminationOtherSampledFuncStatusCorrectFalseCount}
\edef\AvrKindTerminationSampledFuncStatusWrongCount{\the\numexpr\AvrKindTerminationBitVectorsSampledFuncStatusWrongCount+\AvrKindTerminationMainControlFlowSampledFuncStatusWrongCount+\AvrKindTerminationOtherSampledFuncStatusWrongCount}
\providecommand\StoreBenchExecResult[7]{\expandafter\newcommand\csname#1#2#3#4#5#6\endcsname{#7}}%
\StoreBenchExecResult{RicIII}{IcIIIReachSafetyBitVectorsSampledBvFunc}{Status}{All}{}{Score}{0}%
\StoreBenchExecResult{RicIII}{IcIIIReachSafetyBitVectorsSampledBvFunc}{Status}{All}{}{Count}{47}%
\StoreBenchExecResult{RicIII}{IcIIIReachSafetyBitVectorsSampledBvFunc}{Status}{Correct}{}{Count}{43}%
\StoreBenchExecResult{RicIII}{IcIIIReachSafetyBitVectorsSampledBvFunc}{Status}{Correct}{True}{Count}{32}%
\StoreBenchExecResult{RicIII}{IcIIIReachSafetyBitVectorsSampledBvFunc}{Status}{Correct}{False}{Count}{11}%
\StoreBenchExecResult{RicIII}{IcIIIReachSafetyBitVectorsSampledBvFunc}{Status}{Wrong}{}{Count}{0}%
\StoreBenchExecResult{RicIII}{IcIIIReachSafetyBitVectorsSampledBvFunc}{Status}{Wrong}{True}{Count}{0}%
\StoreBenchExecResult{RicIII}{IcIIIReachSafetyBitVectorsSampledBvFunc}{Status}{Wrong}{False}{Count}{0}%
\providecommand\StoreBenchExecResult[7]{\expandafter\newcommand\csname#1#2#3#4#5#6\endcsname{#7}}%
\StoreBenchExecResult{RicIII}{IcIIIReachSafetyCombinationsSampledBvFunc}{Status}{All}{}{Score}{0}%
\StoreBenchExecResult{RicIII}{IcIIIReachSafetyCombinationsSampledBvFunc}{Status}{All}{}{Count}{110}%
\StoreBenchExecResult{RicIII}{IcIIIReachSafetyCombinationsSampledBvFunc}{Status}{Correct}{}{Count}{83}%
\StoreBenchExecResult{RicIII}{IcIIIReachSafetyCombinationsSampledBvFunc}{Status}{Correct}{True}{Count}{12}%
\StoreBenchExecResult{RicIII}{IcIIIReachSafetyCombinationsSampledBvFunc}{Status}{Correct}{False}{Count}{71}%
\StoreBenchExecResult{RicIII}{IcIIIReachSafetyCombinationsSampledBvFunc}{Status}{Wrong}{}{Count}{0}%
\StoreBenchExecResult{RicIII}{IcIIIReachSafetyCombinationsSampledBvFunc}{Status}{Wrong}{True}{Count}{0}%
\StoreBenchExecResult{RicIII}{IcIIIReachSafetyCombinationsSampledBvFunc}{Status}{Wrong}{False}{Count}{0}%
\providecommand\StoreBenchExecResult[7]{\expandafter\newcommand\csname#1#2#3#4#5#6\endcsname{#7}}%
\StoreBenchExecResult{RicIII}{IcIIIReachSafetyControlFlowSampledBvFunc}{Status}{All}{}{Score}{0}%
\StoreBenchExecResult{RicIII}{IcIIIReachSafetyControlFlowSampledBvFunc}{Status}{All}{}{Count}{29}%
\StoreBenchExecResult{RicIII}{IcIIIReachSafetyControlFlowSampledBvFunc}{Status}{Correct}{}{Count}{27}%
\StoreBenchExecResult{RicIII}{IcIIIReachSafetyControlFlowSampledBvFunc}{Status}{Correct}{True}{Count}{25}%
\StoreBenchExecResult{RicIII}{IcIIIReachSafetyControlFlowSampledBvFunc}{Status}{Correct}{False}{Count}{2}%
\StoreBenchExecResult{RicIII}{IcIIIReachSafetyControlFlowSampledBvFunc}{Status}{Wrong}{}{Count}{0}%
\StoreBenchExecResult{RicIII}{IcIIIReachSafetyControlFlowSampledBvFunc}{Status}{Wrong}{True}{Count}{0}%
\StoreBenchExecResult{RicIII}{IcIIIReachSafetyControlFlowSampledBvFunc}{Status}{Wrong}{False}{Count}{0}%
\providecommand\StoreBenchExecResult[7]{\expandafter\newcommand\csname#1#2#3#4#5#6\endcsname{#7}}%
\StoreBenchExecResult{RicIII}{IcIIIReachSafetyECASampledBvFunc}{Status}{All}{}{Score}{0}%
\StoreBenchExecResult{RicIII}{IcIIIReachSafetyECASampledBvFunc}{Status}{All}{}{Count}{150}%
\StoreBenchExecResult{RicIII}{IcIIIReachSafetyECASampledBvFunc}{Status}{Correct}{}{Count}{73}%
\StoreBenchExecResult{RicIII}{IcIIIReachSafetyECASampledBvFunc}{Status}{Correct}{True}{Count}{45}%
\StoreBenchExecResult{RicIII}{IcIIIReachSafetyECASampledBvFunc}{Status}{Correct}{False}{Count}{28}%
\StoreBenchExecResult{RicIII}{IcIIIReachSafetyECASampledBvFunc}{Status}{Wrong}{}{Count}{0}%
\StoreBenchExecResult{RicIII}{IcIIIReachSafetyECASampledBvFunc}{Status}{Wrong}{True}{Count}{0}%
\StoreBenchExecResult{RicIII}{IcIIIReachSafetyECASampledBvFunc}{Status}{Wrong}{False}{Count}{0}%
\providecommand\StoreBenchExecResult[7]{\expandafter\newcommand\csname#1#2#3#4#5#6\endcsname{#7}}%
\StoreBenchExecResult{RicIII}{IcIIIReachSafetyFloatsSampledBvFunc}{Status}{All}{}{Score}{0}%
\StoreBenchExecResult{RicIII}{IcIIIReachSafetyFloatsSampledBvFunc}{Status}{All}{}{Count}{10}%
\StoreBenchExecResult{RicIII}{IcIIIReachSafetyFloatsSampledBvFunc}{Status}{Correct}{}{Count}{10}%
\StoreBenchExecResult{RicIII}{IcIIIReachSafetyFloatsSampledBvFunc}{Status}{Correct}{True}{Count}{10}%
\StoreBenchExecResult{RicIII}{IcIIIReachSafetyFloatsSampledBvFunc}{Status}{Correct}{False}{Count}{0}%
\StoreBenchExecResult{RicIII}{IcIIIReachSafetyFloatsSampledBvFunc}{Status}{Wrong}{}{Count}{0}%
\StoreBenchExecResult{RicIII}{IcIIIReachSafetyFloatsSampledBvFunc}{Status}{Wrong}{True}{Count}{0}%
\StoreBenchExecResult{RicIII}{IcIIIReachSafetyFloatsSampledBvFunc}{Status}{Wrong}{False}{Count}{0}%
\providecommand\StoreBenchExecResult[7]{\expandafter\newcommand\csname#1#2#3#4#5#6\endcsname{#7}}%
\StoreBenchExecResult{RicIII}{IcIIIReachSafetyHardnessSampledBvFunc}{Status}{All}{}{Score}{0}%
\StoreBenchExecResult{RicIII}{IcIIIReachSafetyHardnessSampledBvFunc}{Status}{All}{}{Count}{132}%
\StoreBenchExecResult{RicIII}{IcIIIReachSafetyHardnessSampledBvFunc}{Status}{Correct}{}{Count}{132}%
\StoreBenchExecResult{RicIII}{IcIIIReachSafetyHardnessSampledBvFunc}{Status}{Correct}{True}{Count}{132}%
\StoreBenchExecResult{RicIII}{IcIIIReachSafetyHardnessSampledBvFunc}{Status}{Correct}{False}{Count}{0}%
\StoreBenchExecResult{RicIII}{IcIIIReachSafetyHardnessSampledBvFunc}{Status}{Wrong}{}{Count}{0}%
\StoreBenchExecResult{RicIII}{IcIIIReachSafetyHardnessSampledBvFunc}{Status}{Wrong}{True}{Count}{0}%
\StoreBenchExecResult{RicIII}{IcIIIReachSafetyHardnessSampledBvFunc}{Status}{Wrong}{False}{Count}{0}%
\providecommand\StoreBenchExecResult[7]{\expandafter\newcommand\csname#1#2#3#4#5#6\endcsname{#7}}%
\StoreBenchExecResult{RicIII}{IcIIIReachSafetyHardwareSampledBvFunc}{Status}{All}{}{Score}{0}%
\StoreBenchExecResult{RicIII}{IcIIIReachSafetyHardwareSampledBvFunc}{Status}{All}{}{Count}{148}%
\StoreBenchExecResult{RicIII}{IcIIIReachSafetyHardwareSampledBvFunc}{Status}{Correct}{}{Count}{114}%
\StoreBenchExecResult{RicIII}{IcIIIReachSafetyHardwareSampledBvFunc}{Status}{Correct}{True}{Count}{53}%
\StoreBenchExecResult{RicIII}{IcIIIReachSafetyHardwareSampledBvFunc}{Status}{Correct}{False}{Count}{61}%
\StoreBenchExecResult{RicIII}{IcIIIReachSafetyHardwareSampledBvFunc}{Status}{Wrong}{}{Count}{0}%
\StoreBenchExecResult{RicIII}{IcIIIReachSafetyHardwareSampledBvFunc}{Status}{Wrong}{True}{Count}{0}%
\StoreBenchExecResult{RicIII}{IcIIIReachSafetyHardwareSampledBvFunc}{Status}{Wrong}{False}{Count}{0}%
\providecommand\StoreBenchExecResult[7]{\expandafter\newcommand\csname#1#2#3#4#5#6\endcsname{#7}}%
\StoreBenchExecResult{RicIII}{IcIIIReachSafetyHeapSampledBvFunc}{Status}{All}{}{Score}{0}%
\StoreBenchExecResult{RicIII}{IcIIIReachSafetyHeapSampledBvFunc}{Status}{All}{}{Count}{6}%
\StoreBenchExecResult{RicIII}{IcIIIReachSafetyHeapSampledBvFunc}{Status}{Correct}{}{Count}{6}%
\StoreBenchExecResult{RicIII}{IcIIIReachSafetyHeapSampledBvFunc}{Status}{Correct}{True}{Count}{4}%
\StoreBenchExecResult{RicIII}{IcIIIReachSafetyHeapSampledBvFunc}{Status}{Correct}{False}{Count}{2}%
\StoreBenchExecResult{RicIII}{IcIIIReachSafetyHeapSampledBvFunc}{Status}{Wrong}{}{Count}{0}%
\StoreBenchExecResult{RicIII}{IcIIIReachSafetyHeapSampledBvFunc}{Status}{Wrong}{True}{Count}{0}%
\StoreBenchExecResult{RicIII}{IcIIIReachSafetyHeapSampledBvFunc}{Status}{Wrong}{False}{Count}{0}%
\providecommand\StoreBenchExecResult[7]{\expandafter\newcommand\csname#1#2#3#4#5#6\endcsname{#7}}%
\StoreBenchExecResult{RicIII}{IcIIIReachSafetyLoopsSampledBvFunc}{Status}{All}{}{Score}{0}%
\StoreBenchExecResult{RicIII}{IcIIIReachSafetyLoopsSampledBvFunc}{Status}{All}{}{Count}{137}%
\StoreBenchExecResult{RicIII}{IcIIIReachSafetyLoopsSampledBvFunc}{Status}{Correct}{}{Count}{64}%
\StoreBenchExecResult{RicIII}{IcIIIReachSafetyLoopsSampledBvFunc}{Status}{Correct}{True}{Count}{32}%
\StoreBenchExecResult{RicIII}{IcIIIReachSafetyLoopsSampledBvFunc}{Status}{Correct}{False}{Count}{32}%
\StoreBenchExecResult{RicIII}{IcIIIReachSafetyLoopsSampledBvFunc}{Status}{Wrong}{}{Count}{0}%
\StoreBenchExecResult{RicIII}{IcIIIReachSafetyLoopsSampledBvFunc}{Status}{Wrong}{True}{Count}{0}%
\StoreBenchExecResult{RicIII}{IcIIIReachSafetyLoopsSampledBvFunc}{Status}{Wrong}{False}{Count}{0}%
\providecommand\StoreBenchExecResult[7]{\expandafter\newcommand\csname#1#2#3#4#5#6\endcsname{#7}}%
\StoreBenchExecResult{RicIII}{IcIIIReachSafetyProductLinesSampledBvFunc}{Status}{All}{}{Score}{0}%
\StoreBenchExecResult{RicIII}{IcIIIReachSafetyProductLinesSampledBvFunc}{Status}{All}{}{Count}{150}%
\StoreBenchExecResult{RicIII}{IcIIIReachSafetyProductLinesSampledBvFunc}{Status}{Correct}{}{Count}{147}%
\StoreBenchExecResult{RicIII}{IcIIIReachSafetyProductLinesSampledBvFunc}{Status}{Correct}{True}{Count}{72}%
\StoreBenchExecResult{RicIII}{IcIIIReachSafetyProductLinesSampledBvFunc}{Status}{Correct}{False}{Count}{75}%
\StoreBenchExecResult{RicIII}{IcIIIReachSafetyProductLinesSampledBvFunc}{Status}{Wrong}{}{Count}{0}%
\StoreBenchExecResult{RicIII}{IcIIIReachSafetyProductLinesSampledBvFunc}{Status}{Wrong}{True}{Count}{0}%
\StoreBenchExecResult{RicIII}{IcIIIReachSafetyProductLinesSampledBvFunc}{Status}{Wrong}{False}{Count}{0}%
\providecommand\StoreBenchExecResult[7]{\expandafter\newcommand\csname#1#2#3#4#5#6\endcsname{#7}}%
\StoreBenchExecResult{RicIII}{IcIIIReachSafetySequentializedSampledBvFunc}{Status}{All}{}{Score}{0}%
\StoreBenchExecResult{RicIII}{IcIIIReachSafetySequentializedSampledBvFunc}{Status}{All}{}{Count}{150}%
\StoreBenchExecResult{RicIII}{IcIIIReachSafetySequentializedSampledBvFunc}{Status}{Correct}{}{Count}{101}%
\StoreBenchExecResult{RicIII}{IcIIIReachSafetySequentializedSampledBvFunc}{Status}{Correct}{True}{Count}{24}%
\StoreBenchExecResult{RicIII}{IcIIIReachSafetySequentializedSampledBvFunc}{Status}{Correct}{False}{Count}{77}%
\StoreBenchExecResult{RicIII}{IcIIIReachSafetySequentializedSampledBvFunc}{Status}{Wrong}{}{Count}{0}%
\StoreBenchExecResult{RicIII}{IcIIIReachSafetySequentializedSampledBvFunc}{Status}{Wrong}{True}{Count}{0}%
\StoreBenchExecResult{RicIII}{IcIIIReachSafetySequentializedSampledBvFunc}{Status}{Wrong}{False}{Count}{0}%
\providecommand\StoreBenchExecResult[7]{\expandafter\newcommand\csname#1#2#3#4#5#6\endcsname{#7}}%
\StoreBenchExecResult{RicIII}{IcIIIReachSafetyXCSPSampledBvFunc}{Status}{All}{}{Score}{0}%
\StoreBenchExecResult{RicIII}{IcIIIReachSafetyXCSPSampledBvFunc}{Status}{All}{}{Count}{98}%
\StoreBenchExecResult{RicIII}{IcIIIReachSafetyXCSPSampledBvFunc}{Status}{Correct}{}{Count}{92}%
\StoreBenchExecResult{RicIII}{IcIIIReachSafetyXCSPSampledBvFunc}{Status}{Correct}{True}{Count}{51}%
\StoreBenchExecResult{RicIII}{IcIIIReachSafetyXCSPSampledBvFunc}{Status}{Correct}{False}{Count}{41}%
\StoreBenchExecResult{RicIII}{IcIIIReachSafetyXCSPSampledBvFunc}{Status}{Wrong}{}{Count}{0}%
\StoreBenchExecResult{RicIII}{IcIIIReachSafetyXCSPSampledBvFunc}{Status}{Wrong}{True}{Count}{0}%
\StoreBenchExecResult{RicIII}{IcIIIReachSafetyXCSPSampledBvFunc}{Status}{Wrong}{False}{Count}{0}%
\providecommand\StoreBenchExecResult[7]{\expandafter\newcommand\csname#1#2#3#4#5#6\endcsname{#7}}%
\StoreBenchExecResult{RicIII}{IcIIITerminationBitVectorsSampledBvFunc}{Status}{All}{}{Score}{0}%
\StoreBenchExecResult{RicIII}{IcIIITerminationBitVectorsSampledBvFunc}{Status}{All}{}{Count}{32}%
\StoreBenchExecResult{RicIII}{IcIIITerminationBitVectorsSampledBvFunc}{Status}{Correct}{}{Count}{24}%
\StoreBenchExecResult{RicIII}{IcIIITerminationBitVectorsSampledBvFunc}{Status}{Correct}{True}{Count}{13}%
\StoreBenchExecResult{RicIII}{IcIIITerminationBitVectorsSampledBvFunc}{Status}{Correct}{False}{Count}{11}%
\StoreBenchExecResult{RicIII}{IcIIITerminationBitVectorsSampledBvFunc}{Status}{Wrong}{}{Count}{0}%
\StoreBenchExecResult{RicIII}{IcIIITerminationBitVectorsSampledBvFunc}{Status}{Wrong}{True}{Count}{0}%
\StoreBenchExecResult{RicIII}{IcIIITerminationBitVectorsSampledBvFunc}{Status}{Wrong}{False}{Count}{0}%
\providecommand\StoreBenchExecResult[7]{\expandafter\newcommand\csname#1#2#3#4#5#6\endcsname{#7}}%
\StoreBenchExecResult{RicIII}{IcIIITerminationMainControlFlowSampledBvFunc}{Status}{All}{}{Score}{0}%
\StoreBenchExecResult{RicIII}{IcIIITerminationMainControlFlowSampledBvFunc}{Status}{All}{}{Count}{236}%
\StoreBenchExecResult{RicIII}{IcIIITerminationMainControlFlowSampledBvFunc}{Status}{Correct}{}{Count}{70}%
\StoreBenchExecResult{RicIII}{IcIIITerminationMainControlFlowSampledBvFunc}{Status}{Correct}{True}{Count}{22}%
\StoreBenchExecResult{RicIII}{IcIIITerminationMainControlFlowSampledBvFunc}{Status}{Correct}{False}{Count}{48}%
\StoreBenchExecResult{RicIII}{IcIIITerminationMainControlFlowSampledBvFunc}{Status}{Wrong}{}{Count}{0}%
\StoreBenchExecResult{RicIII}{IcIIITerminationMainControlFlowSampledBvFunc}{Status}{Wrong}{True}{Count}{0}%
\StoreBenchExecResult{RicIII}{IcIIITerminationMainControlFlowSampledBvFunc}{Status}{Wrong}{False}{Count}{0}%
\providecommand\StoreBenchExecResult[7]{\expandafter\newcommand\csname#1#2#3#4#5#6\endcsname{#7}}%
\StoreBenchExecResult{RicIII}{IcIIITerminationOtherSampledBvFunc}{Status}{All}{}{Score}{0}%
\StoreBenchExecResult{RicIII}{IcIIITerminationOtherSampledBvFunc}{Status}{All}{}{Count}{973}%
\StoreBenchExecResult{RicIII}{IcIIITerminationOtherSampledBvFunc}{Status}{Correct}{}{Count}{841}%
\StoreBenchExecResult{RicIII}{IcIIITerminationOtherSampledBvFunc}{Status}{Correct}{True}{Count}{206}%
\StoreBenchExecResult{RicIII}{IcIIITerminationOtherSampledBvFunc}{Status}{Correct}{False}{Count}{635}%
\StoreBenchExecResult{RicIII}{IcIIITerminationOtherSampledBvFunc}{Status}{Wrong}{}{Count}{0}%
\StoreBenchExecResult{RicIII}{IcIIITerminationOtherSampledBvFunc}{Status}{Wrong}{True}{Count}{0}%
\StoreBenchExecResult{RicIII}{IcIIITerminationOtherSampledBvFunc}{Status}{Wrong}{False}{Count}{0}%
\edef\RicIIIIcIIIReachSafetySampledBvFuncStatusAllCount{\the\numexpr\RicIIIIcIIIReachSafetyBitVectorsSampledBvFuncStatusAllCount+\RicIIIIcIIIReachSafetyCombinationsSampledBvFuncStatusAllCount+\RicIIIIcIIIReachSafetyControlFlowSampledBvFuncStatusAllCount+\RicIIIIcIIIReachSafetyECASampledBvFuncStatusAllCount+\RicIIIIcIIIReachSafetyFloatsSampledBvFuncStatusAllCount+\RicIIIIcIIIReachSafetyHardnessSampledBvFuncStatusAllCount+\RicIIIIcIIIReachSafetyHardwareSampledBvFuncStatusAllCount+\RicIIIIcIIIReachSafetyHeapSampledBvFuncStatusAllCount+\RicIIIIcIIIReachSafetyLoopsSampledBvFuncStatusAllCount+\RicIIIIcIIIReachSafetyProductLinesSampledBvFuncStatusAllCount+\RicIIIIcIIIReachSafetySequentializedSampledBvFuncStatusAllCount+\RicIIIIcIIIReachSafetyXCSPSampledBvFuncStatusAllCount}
\edef\RicIIIIcIIIReachSafetySampledBvFuncStatusCorrectCount{\the\numexpr\RicIIIIcIIIReachSafetyBitVectorsSampledBvFuncStatusCorrectCount+\RicIIIIcIIIReachSafetyCombinationsSampledBvFuncStatusCorrectCount+\RicIIIIcIIIReachSafetyControlFlowSampledBvFuncStatusCorrectCount+\RicIIIIcIIIReachSafetyECASampledBvFuncStatusCorrectCount+\RicIIIIcIIIReachSafetyFloatsSampledBvFuncStatusCorrectCount+\RicIIIIcIIIReachSafetyHardnessSampledBvFuncStatusCorrectCount+\RicIIIIcIIIReachSafetyHardwareSampledBvFuncStatusCorrectCount+\RicIIIIcIIIReachSafetyHeapSampledBvFuncStatusCorrectCount+\RicIIIIcIIIReachSafetyLoopsSampledBvFuncStatusCorrectCount+\RicIIIIcIIIReachSafetyProductLinesSampledBvFuncStatusCorrectCount+\RicIIIIcIIIReachSafetySequentializedSampledBvFuncStatusCorrectCount+\RicIIIIcIIIReachSafetyXCSPSampledBvFuncStatusCorrectCount}
\edef\RicIIIIcIIIReachSafetySampledBvFuncStatusCorrectTrueCount{\the\numexpr\RicIIIIcIIIReachSafetyBitVectorsSampledBvFuncStatusCorrectTrueCount+\RicIIIIcIIIReachSafetyCombinationsSampledBvFuncStatusCorrectTrueCount+\RicIIIIcIIIReachSafetyControlFlowSampledBvFuncStatusCorrectTrueCount+\RicIIIIcIIIReachSafetyECASampledBvFuncStatusCorrectTrueCount+\RicIIIIcIIIReachSafetyFloatsSampledBvFuncStatusCorrectTrueCount+\RicIIIIcIIIReachSafetyHardnessSampledBvFuncStatusCorrectTrueCount+\RicIIIIcIIIReachSafetyHardwareSampledBvFuncStatusCorrectTrueCount+\RicIIIIcIIIReachSafetyHeapSampledBvFuncStatusCorrectTrueCount+\RicIIIIcIIIReachSafetyLoopsSampledBvFuncStatusCorrectTrueCount+\RicIIIIcIIIReachSafetyProductLinesSampledBvFuncStatusCorrectTrueCount+\RicIIIIcIIIReachSafetySequentializedSampledBvFuncStatusCorrectTrueCount+\RicIIIIcIIIReachSafetyXCSPSampledBvFuncStatusCorrectTrueCount}
\edef\RicIIIIcIIIReachSafetySampledBvFuncStatusCorrectFalseCount{\the\numexpr\RicIIIIcIIIReachSafetyBitVectorsSampledBvFuncStatusCorrectFalseCount+\RicIIIIcIIIReachSafetyCombinationsSampledBvFuncStatusCorrectFalseCount+\RicIIIIcIIIReachSafetyControlFlowSampledBvFuncStatusCorrectFalseCount+\RicIIIIcIIIReachSafetyECASampledBvFuncStatusCorrectFalseCount+\RicIIIIcIIIReachSafetyFloatsSampledBvFuncStatusCorrectFalseCount+\RicIIIIcIIIReachSafetyHardnessSampledBvFuncStatusCorrectFalseCount+\RicIIIIcIIIReachSafetyHardwareSampledBvFuncStatusCorrectFalseCount+\RicIIIIcIIIReachSafetyHeapSampledBvFuncStatusCorrectFalseCount+\RicIIIIcIIIReachSafetyLoopsSampledBvFuncStatusCorrectFalseCount+\RicIIIIcIIIReachSafetyProductLinesSampledBvFuncStatusCorrectFalseCount+\RicIIIIcIIIReachSafetySequentializedSampledBvFuncStatusCorrectFalseCount+\RicIIIIcIIIReachSafetyXCSPSampledBvFuncStatusCorrectFalseCount}
\edef\RicIIIIcIIIReachSafetySampledBvFuncStatusWrongCount{\the\numexpr\RicIIIIcIIIReachSafetyBitVectorsSampledBvFuncStatusWrongCount+\RicIIIIcIIIReachSafetyCombinationsSampledBvFuncStatusWrongCount+\RicIIIIcIIIReachSafetyControlFlowSampledBvFuncStatusWrongCount+\RicIIIIcIIIReachSafetyECASampledBvFuncStatusWrongCount+\RicIIIIcIIIReachSafetyFloatsSampledBvFuncStatusWrongCount+\RicIIIIcIIIReachSafetyHardnessSampledBvFuncStatusWrongCount+\RicIIIIcIIIReachSafetyHardwareSampledBvFuncStatusWrongCount+\RicIIIIcIIIReachSafetyHeapSampledBvFuncStatusWrongCount+\RicIIIIcIIIReachSafetyLoopsSampledBvFuncStatusWrongCount+\RicIIIIcIIIReachSafetyProductLinesSampledBvFuncStatusWrongCount+\RicIIIIcIIIReachSafetySequentializedSampledBvFuncStatusWrongCount+\RicIIIIcIIIReachSafetyXCSPSampledBvFuncStatusWrongCount}
\edef\RicIIIIcIIITerminationSampledBvFuncStatusAllCount{\the\numexpr\RicIIIIcIIITerminationBitVectorsSampledBvFuncStatusAllCount+\RicIIIIcIIITerminationMainControlFlowSampledBvFuncStatusAllCount+\RicIIIIcIIITerminationOtherSampledBvFuncStatusAllCount}
\edef\RicIIIIcIIITerminationSampledBvFuncStatusCorrectCount{\the\numexpr\RicIIIIcIIITerminationBitVectorsSampledBvFuncStatusCorrectCount+\RicIIIIcIIITerminationMainControlFlowSampledBvFuncStatusCorrectCount+\RicIIIIcIIITerminationOtherSampledBvFuncStatusCorrectCount}
\edef\RicIIIIcIIITerminationSampledBvFuncStatusCorrectTrueCount{\the\numexpr\RicIIIIcIIITerminationBitVectorsSampledBvFuncStatusCorrectTrueCount+\RicIIIIcIIITerminationMainControlFlowSampledBvFuncStatusCorrectTrueCount+\RicIIIIcIIITerminationOtherSampledBvFuncStatusCorrectTrueCount}
\edef\RicIIIIcIIITerminationSampledBvFuncStatusCorrectFalseCount{\the\numexpr\RicIIIIcIIITerminationBitVectorsSampledBvFuncStatusCorrectFalseCount+\RicIIIIcIIITerminationMainControlFlowSampledBvFuncStatusCorrectFalseCount+\RicIIIIcIIITerminationOtherSampledBvFuncStatusCorrectFalseCount}
\edef\RicIIIIcIIITerminationSampledBvFuncStatusWrongCount{\the\numexpr\RicIIIIcIIITerminationBitVectorsSampledBvFuncStatusWrongCount+\RicIIIIcIIITerminationMainControlFlowSampledBvFuncStatusWrongCount+\RicIIIIcIIITerminationOtherSampledBvFuncStatusWrongCount}
\providecommand\StoreBenchExecResult[7]{\expandafter\newcommand\csname#1#2#3#4#5#6\endcsname{#7}}%
\StoreBenchExecResult{RicIII}{KindReachSafetyBitVectorsSampledBvFunc}{Status}{All}{}{Score}{0}%
\StoreBenchExecResult{RicIII}{KindReachSafetyBitVectorsSampledBvFunc}{Status}{All}{}{Count}{47}%
\StoreBenchExecResult{RicIII}{KindReachSafetyBitVectorsSampledBvFunc}{Status}{Correct}{}{Count}{33}%
\StoreBenchExecResult{RicIII}{KindReachSafetyBitVectorsSampledBvFunc}{Status}{Correct}{True}{Count}{22}%
\StoreBenchExecResult{RicIII}{KindReachSafetyBitVectorsSampledBvFunc}{Status}{Correct}{False}{Count}{11}%
\StoreBenchExecResult{RicIII}{KindReachSafetyBitVectorsSampledBvFunc}{Status}{Wrong}{}{Count}{0}%
\StoreBenchExecResult{RicIII}{KindReachSafetyBitVectorsSampledBvFunc}{Status}{Wrong}{True}{Count}{0}%
\StoreBenchExecResult{RicIII}{KindReachSafetyBitVectorsSampledBvFunc}{Status}{Wrong}{False}{Count}{0}%
\providecommand\StoreBenchExecResult[7]{\expandafter\newcommand\csname#1#2#3#4#5#6\endcsname{#7}}%
\StoreBenchExecResult{RicIII}{KindReachSafetyCombinationsSampledBvFunc}{Status}{All}{}{Score}{0}%
\StoreBenchExecResult{RicIII}{KindReachSafetyCombinationsSampledBvFunc}{Status}{All}{}{Count}{110}%
\StoreBenchExecResult{RicIII}{KindReachSafetyCombinationsSampledBvFunc}{Status}{Correct}{}{Count}{71}%
\StoreBenchExecResult{RicIII}{KindReachSafetyCombinationsSampledBvFunc}{Status}{Correct}{True}{Count}{0}%
\StoreBenchExecResult{RicIII}{KindReachSafetyCombinationsSampledBvFunc}{Status}{Correct}{False}{Count}{71}%
\StoreBenchExecResult{RicIII}{KindReachSafetyCombinationsSampledBvFunc}{Status}{Wrong}{}{Count}{0}%
\StoreBenchExecResult{RicIII}{KindReachSafetyCombinationsSampledBvFunc}{Status}{Wrong}{True}{Count}{0}%
\StoreBenchExecResult{RicIII}{KindReachSafetyCombinationsSampledBvFunc}{Status}{Wrong}{False}{Count}{0}%
\providecommand\StoreBenchExecResult[7]{\expandafter\newcommand\csname#1#2#3#4#5#6\endcsname{#7}}%
\StoreBenchExecResult{RicIII}{KindReachSafetyControlFlowSampledBvFunc}{Status}{All}{}{Score}{0}%
\StoreBenchExecResult{RicIII}{KindReachSafetyControlFlowSampledBvFunc}{Status}{All}{}{Count}{29}%
\StoreBenchExecResult{RicIII}{KindReachSafetyControlFlowSampledBvFunc}{Status}{Correct}{}{Count}{26}%
\StoreBenchExecResult{RicIII}{KindReachSafetyControlFlowSampledBvFunc}{Status}{Correct}{True}{Count}{24}%
\StoreBenchExecResult{RicIII}{KindReachSafetyControlFlowSampledBvFunc}{Status}{Correct}{False}{Count}{2}%
\StoreBenchExecResult{RicIII}{KindReachSafetyControlFlowSampledBvFunc}{Status}{Wrong}{}{Count}{0}%
\StoreBenchExecResult{RicIII}{KindReachSafetyControlFlowSampledBvFunc}{Status}{Wrong}{True}{Count}{0}%
\StoreBenchExecResult{RicIII}{KindReachSafetyControlFlowSampledBvFunc}{Status}{Wrong}{False}{Count}{0}%
\providecommand\StoreBenchExecResult[7]{\expandafter\newcommand\csname#1#2#3#4#5#6\endcsname{#7}}%
\StoreBenchExecResult{RicIII}{KindReachSafetyECASampledBvFunc}{Status}{All}{}{Score}{0}%
\StoreBenchExecResult{RicIII}{KindReachSafetyECASampledBvFunc}{Status}{All}{}{Count}{150}%
\StoreBenchExecResult{RicIII}{KindReachSafetyECASampledBvFunc}{Status}{Correct}{}{Count}{86}%
\StoreBenchExecResult{RicIII}{KindReachSafetyECASampledBvFunc}{Status}{Correct}{True}{Count}{49}%
\StoreBenchExecResult{RicIII}{KindReachSafetyECASampledBvFunc}{Status}{Correct}{False}{Count}{37}%
\StoreBenchExecResult{RicIII}{KindReachSafetyECASampledBvFunc}{Status}{Wrong}{}{Count}{0}%
\StoreBenchExecResult{RicIII}{KindReachSafetyECASampledBvFunc}{Status}{Wrong}{True}{Count}{0}%
\StoreBenchExecResult{RicIII}{KindReachSafetyECASampledBvFunc}{Status}{Wrong}{False}{Count}{0}%
\providecommand\StoreBenchExecResult[7]{\expandafter\newcommand\csname#1#2#3#4#5#6\endcsname{#7}}%
\StoreBenchExecResult{RicIII}{KindReachSafetyFloatsSampledBvFunc}{Status}{All}{}{Score}{0}%
\StoreBenchExecResult{RicIII}{KindReachSafetyFloatsSampledBvFunc}{Status}{All}{}{Count}{10}%
\StoreBenchExecResult{RicIII}{KindReachSafetyFloatsSampledBvFunc}{Status}{Correct}{}{Count}{10}%
\StoreBenchExecResult{RicIII}{KindReachSafetyFloatsSampledBvFunc}{Status}{Correct}{True}{Count}{10}%
\StoreBenchExecResult{RicIII}{KindReachSafetyFloatsSampledBvFunc}{Status}{Correct}{False}{Count}{0}%
\StoreBenchExecResult{RicIII}{KindReachSafetyFloatsSampledBvFunc}{Status}{Wrong}{}{Count}{0}%
\StoreBenchExecResult{RicIII}{KindReachSafetyFloatsSampledBvFunc}{Status}{Wrong}{True}{Count}{0}%
\StoreBenchExecResult{RicIII}{KindReachSafetyFloatsSampledBvFunc}{Status}{Wrong}{False}{Count}{0}%
\providecommand\StoreBenchExecResult[7]{\expandafter\newcommand\csname#1#2#3#4#5#6\endcsname{#7}}%
\StoreBenchExecResult{RicIII}{KindReachSafetyHardnessSampledBvFunc}{Status}{All}{}{Score}{0}%
\StoreBenchExecResult{RicIII}{KindReachSafetyHardnessSampledBvFunc}{Status}{All}{}{Count}{132}%
\StoreBenchExecResult{RicIII}{KindReachSafetyHardnessSampledBvFunc}{Status}{Correct}{}{Count}{132}%
\StoreBenchExecResult{RicIII}{KindReachSafetyHardnessSampledBvFunc}{Status}{Correct}{True}{Count}{132}%
\StoreBenchExecResult{RicIII}{KindReachSafetyHardnessSampledBvFunc}{Status}{Correct}{False}{Count}{0}%
\StoreBenchExecResult{RicIII}{KindReachSafetyHardnessSampledBvFunc}{Status}{Wrong}{}{Count}{0}%
\StoreBenchExecResult{RicIII}{KindReachSafetyHardnessSampledBvFunc}{Status}{Wrong}{True}{Count}{0}%
\StoreBenchExecResult{RicIII}{KindReachSafetyHardnessSampledBvFunc}{Status}{Wrong}{False}{Count}{0}%
\providecommand\StoreBenchExecResult[7]{\expandafter\newcommand\csname#1#2#3#4#5#6\endcsname{#7}}%
\StoreBenchExecResult{RicIII}{KindReachSafetyHardwareSampledBvFunc}{Status}{All}{}{Score}{0}%
\StoreBenchExecResult{RicIII}{KindReachSafetyHardwareSampledBvFunc}{Status}{All}{}{Count}{148}%
\StoreBenchExecResult{RicIII}{KindReachSafetyHardwareSampledBvFunc}{Status}{Correct}{}{Count}{76}%
\StoreBenchExecResult{RicIII}{KindReachSafetyHardwareSampledBvFunc}{Status}{Correct}{True}{Count}{15}%
\StoreBenchExecResult{RicIII}{KindReachSafetyHardwareSampledBvFunc}{Status}{Correct}{False}{Count}{61}%
\StoreBenchExecResult{RicIII}{KindReachSafetyHardwareSampledBvFunc}{Status}{Wrong}{}{Count}{0}%
\StoreBenchExecResult{RicIII}{KindReachSafetyHardwareSampledBvFunc}{Status}{Wrong}{True}{Count}{0}%
\StoreBenchExecResult{RicIII}{KindReachSafetyHardwareSampledBvFunc}{Status}{Wrong}{False}{Count}{0}%
\providecommand\StoreBenchExecResult[7]{\expandafter\newcommand\csname#1#2#3#4#5#6\endcsname{#7}}%
\StoreBenchExecResult{RicIII}{KindReachSafetyHeapSampledBvFunc}{Status}{All}{}{Score}{0}%
\StoreBenchExecResult{RicIII}{KindReachSafetyHeapSampledBvFunc}{Status}{All}{}{Count}{6}%
\StoreBenchExecResult{RicIII}{KindReachSafetyHeapSampledBvFunc}{Status}{Correct}{}{Count}{6}%
\StoreBenchExecResult{RicIII}{KindReachSafetyHeapSampledBvFunc}{Status}{Correct}{True}{Count}{4}%
\StoreBenchExecResult{RicIII}{KindReachSafetyHeapSampledBvFunc}{Status}{Correct}{False}{Count}{2}%
\StoreBenchExecResult{RicIII}{KindReachSafetyHeapSampledBvFunc}{Status}{Wrong}{}{Count}{0}%
\StoreBenchExecResult{RicIII}{KindReachSafetyHeapSampledBvFunc}{Status}{Wrong}{True}{Count}{0}%
\StoreBenchExecResult{RicIII}{KindReachSafetyHeapSampledBvFunc}{Status}{Wrong}{False}{Count}{0}%
\providecommand\StoreBenchExecResult[7]{\expandafter\newcommand\csname#1#2#3#4#5#6\endcsname{#7}}%
\StoreBenchExecResult{RicIII}{KindReachSafetyLoopsSampledBvFunc}{Status}{All}{}{Score}{0}%
\StoreBenchExecResult{RicIII}{KindReachSafetyLoopsSampledBvFunc}{Status}{All}{}{Count}{137}%
\StoreBenchExecResult{RicIII}{KindReachSafetyLoopsSampledBvFunc}{Status}{Correct}{}{Count}{60}%
\StoreBenchExecResult{RicIII}{KindReachSafetyLoopsSampledBvFunc}{Status}{Correct}{True}{Count}{23}%
\StoreBenchExecResult{RicIII}{KindReachSafetyLoopsSampledBvFunc}{Status}{Correct}{False}{Count}{37}%
\StoreBenchExecResult{RicIII}{KindReachSafetyLoopsSampledBvFunc}{Status}{Wrong}{}{Count}{0}%
\StoreBenchExecResult{RicIII}{KindReachSafetyLoopsSampledBvFunc}{Status}{Wrong}{True}{Count}{0}%
\StoreBenchExecResult{RicIII}{KindReachSafetyLoopsSampledBvFunc}{Status}{Wrong}{False}{Count}{0}%
\providecommand\StoreBenchExecResult[7]{\expandafter\newcommand\csname#1#2#3#4#5#6\endcsname{#7}}%
\StoreBenchExecResult{RicIII}{KindReachSafetyProductLinesSampledBvFunc}{Status}{All}{}{Score}{0}%
\StoreBenchExecResult{RicIII}{KindReachSafetyProductLinesSampledBvFunc}{Status}{All}{}{Count}{150}%
\StoreBenchExecResult{RicIII}{KindReachSafetyProductLinesSampledBvFunc}{Status}{Correct}{}{Count}{122}%
\StoreBenchExecResult{RicIII}{KindReachSafetyProductLinesSampledBvFunc}{Status}{Correct}{True}{Count}{47}%
\StoreBenchExecResult{RicIII}{KindReachSafetyProductLinesSampledBvFunc}{Status}{Correct}{False}{Count}{75}%
\StoreBenchExecResult{RicIII}{KindReachSafetyProductLinesSampledBvFunc}{Status}{Wrong}{}{Count}{0}%
\StoreBenchExecResult{RicIII}{KindReachSafetyProductLinesSampledBvFunc}{Status}{Wrong}{True}{Count}{0}%
\StoreBenchExecResult{RicIII}{KindReachSafetyProductLinesSampledBvFunc}{Status}{Wrong}{False}{Count}{0}%
\providecommand\StoreBenchExecResult[7]{\expandafter\newcommand\csname#1#2#3#4#5#6\endcsname{#7}}%
\StoreBenchExecResult{RicIII}{KindReachSafetySequentializedSampledBvFunc}{Status}{All}{}{Score}{0}%
\StoreBenchExecResult{RicIII}{KindReachSafetySequentializedSampledBvFunc}{Status}{All}{}{Count}{150}%
\StoreBenchExecResult{RicIII}{KindReachSafetySequentializedSampledBvFunc}{Status}{Correct}{}{Count}{118}%
\StoreBenchExecResult{RicIII}{KindReachSafetySequentializedSampledBvFunc}{Status}{Correct}{True}{Count}{28}%
\StoreBenchExecResult{RicIII}{KindReachSafetySequentializedSampledBvFunc}{Status}{Correct}{False}{Count}{90}%
\StoreBenchExecResult{RicIII}{KindReachSafetySequentializedSampledBvFunc}{Status}{Wrong}{}{Count}{0}%
\StoreBenchExecResult{RicIII}{KindReachSafetySequentializedSampledBvFunc}{Status}{Wrong}{True}{Count}{0}%
\StoreBenchExecResult{RicIII}{KindReachSafetySequentializedSampledBvFunc}{Status}{Wrong}{False}{Count}{0}%
\providecommand\StoreBenchExecResult[7]{\expandafter\newcommand\csname#1#2#3#4#5#6\endcsname{#7}}%
\StoreBenchExecResult{RicIII}{KindReachSafetyXCSPSampledBvFunc}{Status}{All}{}{Score}{0}%
\StoreBenchExecResult{RicIII}{KindReachSafetyXCSPSampledBvFunc}{Status}{All}{}{Count}{98}%
\StoreBenchExecResult{RicIII}{KindReachSafetyXCSPSampledBvFunc}{Status}{Correct}{}{Count}{93}%
\StoreBenchExecResult{RicIII}{KindReachSafetyXCSPSampledBvFunc}{Status}{Correct}{True}{Count}{50}%
\StoreBenchExecResult{RicIII}{KindReachSafetyXCSPSampledBvFunc}{Status}{Correct}{False}{Count}{43}%
\StoreBenchExecResult{RicIII}{KindReachSafetyXCSPSampledBvFunc}{Status}{Wrong}{}{Count}{0}%
\StoreBenchExecResult{RicIII}{KindReachSafetyXCSPSampledBvFunc}{Status}{Wrong}{True}{Count}{0}%
\StoreBenchExecResult{RicIII}{KindReachSafetyXCSPSampledBvFunc}{Status}{Wrong}{False}{Count}{0}%
\providecommand\StoreBenchExecResult[7]{\expandafter\newcommand\csname#1#2#3#4#5#6\endcsname{#7}}%
\StoreBenchExecResult{RicIII}{KindTerminationBitVectorsSampledBvFunc}{Status}{All}{}{Score}{0}%
\StoreBenchExecResult{RicIII}{KindTerminationBitVectorsSampledBvFunc}{Status}{All}{}{Count}{32}%
\StoreBenchExecResult{RicIII}{KindTerminationBitVectorsSampledBvFunc}{Status}{Correct}{}{Count}{21}%
\StoreBenchExecResult{RicIII}{KindTerminationBitVectorsSampledBvFunc}{Status}{Correct}{True}{Count}{10}%
\StoreBenchExecResult{RicIII}{KindTerminationBitVectorsSampledBvFunc}{Status}{Correct}{False}{Count}{11}%
\StoreBenchExecResult{RicIII}{KindTerminationBitVectorsSampledBvFunc}{Status}{Wrong}{}{Count}{0}%
\StoreBenchExecResult{RicIII}{KindTerminationBitVectorsSampledBvFunc}{Status}{Wrong}{True}{Count}{0}%
\StoreBenchExecResult{RicIII}{KindTerminationBitVectorsSampledBvFunc}{Status}{Wrong}{False}{Count}{0}%
\providecommand\StoreBenchExecResult[7]{\expandafter\newcommand\csname#1#2#3#4#5#6\endcsname{#7}}%
\StoreBenchExecResult{RicIII}{KindTerminationMainControlFlowSampledBvFunc}{Status}{All}{}{Score}{0}%
\StoreBenchExecResult{RicIII}{KindTerminationMainControlFlowSampledBvFunc}{Status}{All}{}{Count}{236}%
\StoreBenchExecResult{RicIII}{KindTerminationMainControlFlowSampledBvFunc}{Status}{Correct}{}{Count}{67}%
\StoreBenchExecResult{RicIII}{KindTerminationMainControlFlowSampledBvFunc}{Status}{Correct}{True}{Count}{18}%
\StoreBenchExecResult{RicIII}{KindTerminationMainControlFlowSampledBvFunc}{Status}{Correct}{False}{Count}{49}%
\StoreBenchExecResult{RicIII}{KindTerminationMainControlFlowSampledBvFunc}{Status}{Wrong}{}{Count}{0}%
\StoreBenchExecResult{RicIII}{KindTerminationMainControlFlowSampledBvFunc}{Status}{Wrong}{True}{Count}{0}%
\StoreBenchExecResult{RicIII}{KindTerminationMainControlFlowSampledBvFunc}{Status}{Wrong}{False}{Count}{0}%
\providecommand\StoreBenchExecResult[7]{\expandafter\newcommand\csname#1#2#3#4#5#6\endcsname{#7}}%
\StoreBenchExecResult{RicIII}{KindTerminationOtherSampledBvFunc}{Status}{All}{}{Score}{0}%
\StoreBenchExecResult{RicIII}{KindTerminationOtherSampledBvFunc}{Status}{All}{}{Count}{973}%
\StoreBenchExecResult{RicIII}{KindTerminationOtherSampledBvFunc}{Status}{Correct}{}{Count}{856}%
\StoreBenchExecResult{RicIII}{KindTerminationOtherSampledBvFunc}{Status}{Correct}{True}{Count}{189}%
\StoreBenchExecResult{RicIII}{KindTerminationOtherSampledBvFunc}{Status}{Correct}{False}{Count}{667}%
\StoreBenchExecResult{RicIII}{KindTerminationOtherSampledBvFunc}{Status}{Wrong}{}{Count}{0}%
\StoreBenchExecResult{RicIII}{KindTerminationOtherSampledBvFunc}{Status}{Wrong}{True}{Count}{0}%
\StoreBenchExecResult{RicIII}{KindTerminationOtherSampledBvFunc}{Status}{Wrong}{False}{Count}{0}%
\edef\RicIIIKindReachSafetySampledBvFuncStatusAllCount{\the\numexpr\RicIIIKindReachSafetyBitVectorsSampledBvFuncStatusAllCount+\RicIIIKindReachSafetyCombinationsSampledBvFuncStatusAllCount+\RicIIIKindReachSafetyControlFlowSampledBvFuncStatusAllCount+\RicIIIKindReachSafetyECASampledBvFuncStatusAllCount+\RicIIIKindReachSafetyFloatsSampledBvFuncStatusAllCount+\RicIIIKindReachSafetyHardnessSampledBvFuncStatusAllCount+\RicIIIKindReachSafetyHardwareSampledBvFuncStatusAllCount+\RicIIIKindReachSafetyHeapSampledBvFuncStatusAllCount+\RicIIIKindReachSafetyLoopsSampledBvFuncStatusAllCount+\RicIIIKindReachSafetyProductLinesSampledBvFuncStatusAllCount+\RicIIIKindReachSafetySequentializedSampledBvFuncStatusAllCount+\RicIIIKindReachSafetyXCSPSampledBvFuncStatusAllCount}
\edef\RicIIIKindReachSafetySampledBvFuncStatusCorrectCount{\the\numexpr\RicIIIKindReachSafetyBitVectorsSampledBvFuncStatusCorrectCount+\RicIIIKindReachSafetyCombinationsSampledBvFuncStatusCorrectCount+\RicIIIKindReachSafetyControlFlowSampledBvFuncStatusCorrectCount+\RicIIIKindReachSafetyECASampledBvFuncStatusCorrectCount+\RicIIIKindReachSafetyFloatsSampledBvFuncStatusCorrectCount+\RicIIIKindReachSafetyHardnessSampledBvFuncStatusCorrectCount+\RicIIIKindReachSafetyHardwareSampledBvFuncStatusCorrectCount+\RicIIIKindReachSafetyHeapSampledBvFuncStatusCorrectCount+\RicIIIKindReachSafetyLoopsSampledBvFuncStatusCorrectCount+\RicIIIKindReachSafetyProductLinesSampledBvFuncStatusCorrectCount+\RicIIIKindReachSafetySequentializedSampledBvFuncStatusCorrectCount+\RicIIIKindReachSafetyXCSPSampledBvFuncStatusCorrectCount}
\edef\RicIIIKindReachSafetySampledBvFuncStatusCorrectTrueCount{\the\numexpr\RicIIIKindReachSafetyBitVectorsSampledBvFuncStatusCorrectTrueCount+\RicIIIKindReachSafetyCombinationsSampledBvFuncStatusCorrectTrueCount+\RicIIIKindReachSafetyControlFlowSampledBvFuncStatusCorrectTrueCount+\RicIIIKindReachSafetyECASampledBvFuncStatusCorrectTrueCount+\RicIIIKindReachSafetyFloatsSampledBvFuncStatusCorrectTrueCount+\RicIIIKindReachSafetyHardnessSampledBvFuncStatusCorrectTrueCount+\RicIIIKindReachSafetyHardwareSampledBvFuncStatusCorrectTrueCount+\RicIIIKindReachSafetyHeapSampledBvFuncStatusCorrectTrueCount+\RicIIIKindReachSafetyLoopsSampledBvFuncStatusCorrectTrueCount+\RicIIIKindReachSafetyProductLinesSampledBvFuncStatusCorrectTrueCount+\RicIIIKindReachSafetySequentializedSampledBvFuncStatusCorrectTrueCount+\RicIIIKindReachSafetyXCSPSampledBvFuncStatusCorrectTrueCount}
\edef\RicIIIKindReachSafetySampledBvFuncStatusCorrectFalseCount{\the\numexpr\RicIIIKindReachSafetyBitVectorsSampledBvFuncStatusCorrectFalseCount+\RicIIIKindReachSafetyCombinationsSampledBvFuncStatusCorrectFalseCount+\RicIIIKindReachSafetyControlFlowSampledBvFuncStatusCorrectFalseCount+\RicIIIKindReachSafetyECASampledBvFuncStatusCorrectFalseCount+\RicIIIKindReachSafetyFloatsSampledBvFuncStatusCorrectFalseCount+\RicIIIKindReachSafetyHardnessSampledBvFuncStatusCorrectFalseCount+\RicIIIKindReachSafetyHardwareSampledBvFuncStatusCorrectFalseCount+\RicIIIKindReachSafetyHeapSampledBvFuncStatusCorrectFalseCount+\RicIIIKindReachSafetyLoopsSampledBvFuncStatusCorrectFalseCount+\RicIIIKindReachSafetyProductLinesSampledBvFuncStatusCorrectFalseCount+\RicIIIKindReachSafetySequentializedSampledBvFuncStatusCorrectFalseCount+\RicIIIKindReachSafetyXCSPSampledBvFuncStatusCorrectFalseCount}
\edef\RicIIIKindReachSafetySampledBvFuncStatusWrongCount{\the\numexpr\RicIIIKindReachSafetyBitVectorsSampledBvFuncStatusWrongCount+\RicIIIKindReachSafetyCombinationsSampledBvFuncStatusWrongCount+\RicIIIKindReachSafetyControlFlowSampledBvFuncStatusWrongCount+\RicIIIKindReachSafetyECASampledBvFuncStatusWrongCount+\RicIIIKindReachSafetyFloatsSampledBvFuncStatusWrongCount+\RicIIIKindReachSafetyHardnessSampledBvFuncStatusWrongCount+\RicIIIKindReachSafetyHardwareSampledBvFuncStatusWrongCount+\RicIIIKindReachSafetyHeapSampledBvFuncStatusWrongCount+\RicIIIKindReachSafetyLoopsSampledBvFuncStatusWrongCount+\RicIIIKindReachSafetyProductLinesSampledBvFuncStatusWrongCount+\RicIIIKindReachSafetySequentializedSampledBvFuncStatusWrongCount+\RicIIIKindReachSafetyXCSPSampledBvFuncStatusWrongCount}
\edef\RicIIIKindTerminationSampledBvFuncStatusAllCount{\the\numexpr\RicIIIKindTerminationBitVectorsSampledBvFuncStatusAllCount+\RicIIIKindTerminationMainControlFlowSampledBvFuncStatusAllCount+\RicIIIKindTerminationOtherSampledBvFuncStatusAllCount}
\edef\RicIIIKindTerminationSampledBvFuncStatusCorrectCount{\the\numexpr\RicIIIKindTerminationBitVectorsSampledBvFuncStatusCorrectCount+\RicIIIKindTerminationMainControlFlowSampledBvFuncStatusCorrectCount+\RicIIIKindTerminationOtherSampledBvFuncStatusCorrectCount}
\edef\RicIIIKindTerminationSampledBvFuncStatusCorrectTrueCount{\the\numexpr\RicIIIKindTerminationBitVectorsSampledBvFuncStatusCorrectTrueCount+\RicIIIKindTerminationMainControlFlowSampledBvFuncStatusCorrectTrueCount+\RicIIIKindTerminationOtherSampledBvFuncStatusCorrectTrueCount}
\edef\RicIIIKindTerminationSampledBvFuncStatusCorrectFalseCount{\the\numexpr\RicIIIKindTerminationBitVectorsSampledBvFuncStatusCorrectFalseCount+\RicIIIKindTerminationMainControlFlowSampledBvFuncStatusCorrectFalseCount+\RicIIIKindTerminationOtherSampledBvFuncStatusCorrectFalseCount}
\edef\RicIIIKindTerminationSampledBvFuncStatusWrongCount{\the\numexpr\RicIIIKindTerminationBitVectorsSampledBvFuncStatusWrongCount+\RicIIIKindTerminationMainControlFlowSampledBvFuncStatusWrongCount+\RicIIIKindTerminationOtherSampledBvFuncStatusWrongCount}
\providecommand\StoreBenchExecResult[7]{\expandafter\newcommand\csname#1#2#3#4#5#6\endcsname{#7}}%
\StoreBenchExecResult{Pono}{IcIIIiaMsatReachSafetyArraysSampledFunc}{Status}{All}{}{Score}{0}%
\StoreBenchExecResult{Pono}{IcIIIiaMsatReachSafetyArraysSampledFunc}{Status}{All}{}{Count}{150}%
\StoreBenchExecResult{Pono}{IcIIIiaMsatReachSafetyArraysSampledFunc}{Status}{Correct}{}{Count}{45}%
\StoreBenchExecResult{Pono}{IcIIIiaMsatReachSafetyArraysSampledFunc}{Status}{Correct}{True}{Count}{0}%
\StoreBenchExecResult{Pono}{IcIIIiaMsatReachSafetyArraysSampledFunc}{Status}{Correct}{False}{Count}{45}%
\StoreBenchExecResult{Pono}{IcIIIiaMsatReachSafetyArraysSampledFunc}{Status}{Wrong}{}{Count}{0}%
\StoreBenchExecResult{Pono}{IcIIIiaMsatReachSafetyArraysSampledFunc}{Status}{Wrong}{True}{Count}{0}%
\StoreBenchExecResult{Pono}{IcIIIiaMsatReachSafetyArraysSampledFunc}{Status}{Wrong}{False}{Count}{0}%
\providecommand\StoreBenchExecResult[7]{\expandafter\newcommand\csname#1#2#3#4#5#6\endcsname{#7}}%
\StoreBenchExecResult{Pono}{IcIIIiaMsatReachSafetyBitVectorsSampledFunc}{Status}{All}{}{Score}{0}%
\StoreBenchExecResult{Pono}{IcIIIiaMsatReachSafetyBitVectorsSampledFunc}{Status}{All}{}{Count}{48}%
\StoreBenchExecResult{Pono}{IcIIIiaMsatReachSafetyBitVectorsSampledFunc}{Status}{Correct}{}{Count}{30}%
\StoreBenchExecResult{Pono}{IcIIIiaMsatReachSafetyBitVectorsSampledFunc}{Status}{Correct}{True}{Count}{20}%
\StoreBenchExecResult{Pono}{IcIIIiaMsatReachSafetyBitVectorsSampledFunc}{Status}{Correct}{False}{Count}{10}%
\StoreBenchExecResult{Pono}{IcIIIiaMsatReachSafetyBitVectorsSampledFunc}{Status}{Wrong}{}{Count}{0}%
\StoreBenchExecResult{Pono}{IcIIIiaMsatReachSafetyBitVectorsSampledFunc}{Status}{Wrong}{True}{Count}{0}%
\StoreBenchExecResult{Pono}{IcIIIiaMsatReachSafetyBitVectorsSampledFunc}{Status}{Wrong}{False}{Count}{0}%
\providecommand\StoreBenchExecResult[7]{\expandafter\newcommand\csname#1#2#3#4#5#6\endcsname{#7}}%
\StoreBenchExecResult{Pono}{IcIIIiaMsatReachSafetyCombinationsSampledFunc}{Status}{All}{}{Score}{0}%
\StoreBenchExecResult{Pono}{IcIIIiaMsatReachSafetyCombinationsSampledFunc}{Status}{All}{}{Count}{110}%
\StoreBenchExecResult{Pono}{IcIIIiaMsatReachSafetyCombinationsSampledFunc}{Status}{Correct}{}{Count}{45}%
\StoreBenchExecResult{Pono}{IcIIIiaMsatReachSafetyCombinationsSampledFunc}{Status}{Correct}{True}{Count}{0}%
\StoreBenchExecResult{Pono}{IcIIIiaMsatReachSafetyCombinationsSampledFunc}{Status}{Correct}{False}{Count}{45}%
\StoreBenchExecResult{Pono}{IcIIIiaMsatReachSafetyCombinationsSampledFunc}{Status}{Wrong}{}{Count}{0}%
\StoreBenchExecResult{Pono}{IcIIIiaMsatReachSafetyCombinationsSampledFunc}{Status}{Wrong}{True}{Count}{0}%
\StoreBenchExecResult{Pono}{IcIIIiaMsatReachSafetyCombinationsSampledFunc}{Status}{Wrong}{False}{Count}{0}%
\providecommand\StoreBenchExecResult[7]{\expandafter\newcommand\csname#1#2#3#4#5#6\endcsname{#7}}%
\StoreBenchExecResult{Pono}{IcIIIiaMsatReachSafetyControlFlowSampledFunc}{Status}{All}{}{Score}{0}%
\StoreBenchExecResult{Pono}{IcIIIiaMsatReachSafetyControlFlowSampledFunc}{Status}{All}{}{Count}{33}%
\StoreBenchExecResult{Pono}{IcIIIiaMsatReachSafetyControlFlowSampledFunc}{Status}{Correct}{}{Count}{26}%
\StoreBenchExecResult{Pono}{IcIIIiaMsatReachSafetyControlFlowSampledFunc}{Status}{Correct}{True}{Count}{23}%
\StoreBenchExecResult{Pono}{IcIIIiaMsatReachSafetyControlFlowSampledFunc}{Status}{Correct}{False}{Count}{3}%
\StoreBenchExecResult{Pono}{IcIIIiaMsatReachSafetyControlFlowSampledFunc}{Status}{Wrong}{}{Count}{0}%
\StoreBenchExecResult{Pono}{IcIIIiaMsatReachSafetyControlFlowSampledFunc}{Status}{Wrong}{True}{Count}{0}%
\StoreBenchExecResult{Pono}{IcIIIiaMsatReachSafetyControlFlowSampledFunc}{Status}{Wrong}{False}{Count}{0}%
\providecommand\StoreBenchExecResult[7]{\expandafter\newcommand\csname#1#2#3#4#5#6\endcsname{#7}}%
\StoreBenchExecResult{Pono}{IcIIIiaMsatReachSafetyECASampledFunc}{Status}{All}{}{Score}{0}%
\StoreBenchExecResult{Pono}{IcIIIiaMsatReachSafetyECASampledFunc}{Status}{All}{}{Count}{150}%
\StoreBenchExecResult{Pono}{IcIIIiaMsatReachSafetyECASampledFunc}{Status}{Correct}{}{Count}{39}%
\StoreBenchExecResult{Pono}{IcIIIiaMsatReachSafetyECASampledFunc}{Status}{Correct}{True}{Count}{31}%
\StoreBenchExecResult{Pono}{IcIIIiaMsatReachSafetyECASampledFunc}{Status}{Correct}{False}{Count}{8}%
\StoreBenchExecResult{Pono}{IcIIIiaMsatReachSafetyECASampledFunc}{Status}{Wrong}{}{Count}{0}%
\StoreBenchExecResult{Pono}{IcIIIiaMsatReachSafetyECASampledFunc}{Status}{Wrong}{True}{Count}{0}%
\StoreBenchExecResult{Pono}{IcIIIiaMsatReachSafetyECASampledFunc}{Status}{Wrong}{False}{Count}{0}%
\providecommand\StoreBenchExecResult[7]{\expandafter\newcommand\csname#1#2#3#4#5#6\endcsname{#7}}%
\StoreBenchExecResult{Pono}{IcIIIiaMsatReachSafetyFloatsSampledFunc}{Status}{All}{}{Score}{0}%
\StoreBenchExecResult{Pono}{IcIIIiaMsatReachSafetyFloatsSampledFunc}{Status}{All}{}{Count}{10}%
\StoreBenchExecResult{Pono}{IcIIIiaMsatReachSafetyFloatsSampledFunc}{Status}{Correct}{}{Count}{10}%
\StoreBenchExecResult{Pono}{IcIIIiaMsatReachSafetyFloatsSampledFunc}{Status}{Correct}{True}{Count}{10}%
\StoreBenchExecResult{Pono}{IcIIIiaMsatReachSafetyFloatsSampledFunc}{Status}{Correct}{False}{Count}{0}%
\StoreBenchExecResult{Pono}{IcIIIiaMsatReachSafetyFloatsSampledFunc}{Status}{Wrong}{}{Count}{0}%
\StoreBenchExecResult{Pono}{IcIIIiaMsatReachSafetyFloatsSampledFunc}{Status}{Wrong}{True}{Count}{0}%
\StoreBenchExecResult{Pono}{IcIIIiaMsatReachSafetyFloatsSampledFunc}{Status}{Wrong}{False}{Count}{0}%
\providecommand\StoreBenchExecResult[7]{\expandafter\newcommand\csname#1#2#3#4#5#6\endcsname{#7}}%
\StoreBenchExecResult{Pono}{IcIIIiaMsatReachSafetyHardnessSampledFunc}{Status}{All}{}{Score}{0}%
\StoreBenchExecResult{Pono}{IcIIIiaMsatReachSafetyHardnessSampledFunc}{Status}{All}{}{Count}{150}%
\StoreBenchExecResult{Pono}{IcIIIiaMsatReachSafetyHardnessSampledFunc}{Status}{Correct}{}{Count}{138}%
\StoreBenchExecResult{Pono}{IcIIIiaMsatReachSafetyHardnessSampledFunc}{Status}{Correct}{True}{Count}{138}%
\StoreBenchExecResult{Pono}{IcIIIiaMsatReachSafetyHardnessSampledFunc}{Status}{Correct}{False}{Count}{0}%
\StoreBenchExecResult{Pono}{IcIIIiaMsatReachSafetyHardnessSampledFunc}{Status}{Wrong}{}{Count}{0}%
\StoreBenchExecResult{Pono}{IcIIIiaMsatReachSafetyHardnessSampledFunc}{Status}{Wrong}{True}{Count}{0}%
\StoreBenchExecResult{Pono}{IcIIIiaMsatReachSafetyHardnessSampledFunc}{Status}{Wrong}{False}{Count}{0}%
\providecommand\StoreBenchExecResult[7]{\expandafter\newcommand\csname#1#2#3#4#5#6\endcsname{#7}}%
\StoreBenchExecResult{Pono}{IcIIIiaMsatReachSafetyHardwareSampledFunc}{Status}{All}{}{Score}{0}%
\StoreBenchExecResult{Pono}{IcIIIiaMsatReachSafetyHardwareSampledFunc}{Status}{All}{}{Count}{150}%
\StoreBenchExecResult{Pono}{IcIIIiaMsatReachSafetyHardwareSampledFunc}{Status}{Correct}{}{Count}{20}%
\StoreBenchExecResult{Pono}{IcIIIiaMsatReachSafetyHardwareSampledFunc}{Status}{Correct}{True}{Count}{7}%
\StoreBenchExecResult{Pono}{IcIIIiaMsatReachSafetyHardwareSampledFunc}{Status}{Correct}{False}{Count}{13}%
\StoreBenchExecResult{Pono}{IcIIIiaMsatReachSafetyHardwareSampledFunc}{Status}{Wrong}{}{Count}{0}%
\StoreBenchExecResult{Pono}{IcIIIiaMsatReachSafetyHardwareSampledFunc}{Status}{Wrong}{True}{Count}{0}%
\StoreBenchExecResult{Pono}{IcIIIiaMsatReachSafetyHardwareSampledFunc}{Status}{Wrong}{False}{Count}{0}%
\providecommand\StoreBenchExecResult[7]{\expandafter\newcommand\csname#1#2#3#4#5#6\endcsname{#7}}%
\StoreBenchExecResult{Pono}{IcIIIiaMsatReachSafetyHeapSampledFunc}{Status}{All}{}{Score}{0}%
\StoreBenchExecResult{Pono}{IcIIIiaMsatReachSafetyHeapSampledFunc}{Status}{All}{}{Count}{53}%
\StoreBenchExecResult{Pono}{IcIIIiaMsatReachSafetyHeapSampledFunc}{Status}{Correct}{}{Count}{46}%
\StoreBenchExecResult{Pono}{IcIIIiaMsatReachSafetyHeapSampledFunc}{Status}{Correct}{True}{Count}{32}%
\StoreBenchExecResult{Pono}{IcIIIiaMsatReachSafetyHeapSampledFunc}{Status}{Correct}{False}{Count}{14}%
\StoreBenchExecResult{Pono}{IcIIIiaMsatReachSafetyHeapSampledFunc}{Status}{Wrong}{}{Count}{0}%
\StoreBenchExecResult{Pono}{IcIIIiaMsatReachSafetyHeapSampledFunc}{Status}{Wrong}{True}{Count}{0}%
\StoreBenchExecResult{Pono}{IcIIIiaMsatReachSafetyHeapSampledFunc}{Status}{Wrong}{False}{Count}{0}%
\providecommand\StoreBenchExecResult[7]{\expandafter\newcommand\csname#1#2#3#4#5#6\endcsname{#7}}%
\StoreBenchExecResult{Pono}{IcIIIiaMsatReachSafetyLoopsSampledFunc}{Status}{All}{}{Score}{0}%
\StoreBenchExecResult{Pono}{IcIIIiaMsatReachSafetyLoopsSampledFunc}{Status}{All}{}{Count}{150}%
\StoreBenchExecResult{Pono}{IcIIIiaMsatReachSafetyLoopsSampledFunc}{Status}{Correct}{}{Count}{65}%
\StoreBenchExecResult{Pono}{IcIIIiaMsatReachSafetyLoopsSampledFunc}{Status}{Correct}{True}{Count}{32}%
\StoreBenchExecResult{Pono}{IcIIIiaMsatReachSafetyLoopsSampledFunc}{Status}{Correct}{False}{Count}{33}%
\StoreBenchExecResult{Pono}{IcIIIiaMsatReachSafetyLoopsSampledFunc}{Status}{Wrong}{}{Count}{0}%
\StoreBenchExecResult{Pono}{IcIIIiaMsatReachSafetyLoopsSampledFunc}{Status}{Wrong}{True}{Count}{0}%
\StoreBenchExecResult{Pono}{IcIIIiaMsatReachSafetyLoopsSampledFunc}{Status}{Wrong}{False}{Count}{0}%
\providecommand\StoreBenchExecResult[7]{\expandafter\newcommand\csname#1#2#3#4#5#6\endcsname{#7}}%
\StoreBenchExecResult{Pono}{IcIIIiaMsatReachSafetyProductLinesSampledFunc}{Status}{All}{}{Score}{0}%
\StoreBenchExecResult{Pono}{IcIIIiaMsatReachSafetyProductLinesSampledFunc}{Status}{All}{}{Count}{150}%
\StoreBenchExecResult{Pono}{IcIIIiaMsatReachSafetyProductLinesSampledFunc}{Status}{Correct}{}{Count}{146}%
\StoreBenchExecResult{Pono}{IcIIIiaMsatReachSafetyProductLinesSampledFunc}{Status}{Correct}{True}{Count}{72}%
\StoreBenchExecResult{Pono}{IcIIIiaMsatReachSafetyProductLinesSampledFunc}{Status}{Correct}{False}{Count}{74}%
\StoreBenchExecResult{Pono}{IcIIIiaMsatReachSafetyProductLinesSampledFunc}{Status}{Wrong}{}{Count}{0}%
\StoreBenchExecResult{Pono}{IcIIIiaMsatReachSafetyProductLinesSampledFunc}{Status}{Wrong}{True}{Count}{0}%
\StoreBenchExecResult{Pono}{IcIIIiaMsatReachSafetyProductLinesSampledFunc}{Status}{Wrong}{False}{Count}{0}%
\providecommand\StoreBenchExecResult[7]{\expandafter\newcommand\csname#1#2#3#4#5#6\endcsname{#7}}%
\StoreBenchExecResult{Pono}{IcIIIiaMsatReachSafetySequentializedSampledFunc}{Status}{All}{}{Score}{0}%
\StoreBenchExecResult{Pono}{IcIIIiaMsatReachSafetySequentializedSampledFunc}{Status}{All}{}{Count}{150}%
\StoreBenchExecResult{Pono}{IcIIIiaMsatReachSafetySequentializedSampledFunc}{Status}{Correct}{}{Count}{50}%
\StoreBenchExecResult{Pono}{IcIIIiaMsatReachSafetySequentializedSampledFunc}{Status}{Correct}{True}{Count}{7}%
\StoreBenchExecResult{Pono}{IcIIIiaMsatReachSafetySequentializedSampledFunc}{Status}{Correct}{False}{Count}{43}%
\StoreBenchExecResult{Pono}{IcIIIiaMsatReachSafetySequentializedSampledFunc}{Status}{Wrong}{}{Count}{0}%
\StoreBenchExecResult{Pono}{IcIIIiaMsatReachSafetySequentializedSampledFunc}{Status}{Wrong}{True}{Count}{0}%
\StoreBenchExecResult{Pono}{IcIIIiaMsatReachSafetySequentializedSampledFunc}{Status}{Wrong}{False}{Count}{0}%
\providecommand\StoreBenchExecResult[7]{\expandafter\newcommand\csname#1#2#3#4#5#6\endcsname{#7}}%
\StoreBenchExecResult{Pono}{IcIIIiaMsatReachSafetyXCSPSampledFunc}{Status}{All}{}{Score}{0}%
\StoreBenchExecResult{Pono}{IcIIIiaMsatReachSafetyXCSPSampledFunc}{Status}{All}{}{Count}{98}%
\StoreBenchExecResult{Pono}{IcIIIiaMsatReachSafetyXCSPSampledFunc}{Status}{Correct}{}{Count}{90}%
\StoreBenchExecResult{Pono}{IcIIIiaMsatReachSafetyXCSPSampledFunc}{Status}{Correct}{True}{Count}{52}%
\StoreBenchExecResult{Pono}{IcIIIiaMsatReachSafetyXCSPSampledFunc}{Status}{Correct}{False}{Count}{38}%
\StoreBenchExecResult{Pono}{IcIIIiaMsatReachSafetyXCSPSampledFunc}{Status}{Wrong}{}{Count}{0}%
\StoreBenchExecResult{Pono}{IcIIIiaMsatReachSafetyXCSPSampledFunc}{Status}{Wrong}{True}{Count}{0}%
\StoreBenchExecResult{Pono}{IcIIIiaMsatReachSafetyXCSPSampledFunc}{Status}{Wrong}{False}{Count}{0}%
\providecommand\StoreBenchExecResult[7]{\expandafter\newcommand\csname#1#2#3#4#5#6\endcsname{#7}}%
\StoreBenchExecResult{Pono}{IcIIIiaMsatTerminationBitVectorsSampledFunc}{Status}{All}{}{Score}{0}%
\StoreBenchExecResult{Pono}{IcIIIiaMsatTerminationBitVectorsSampledFunc}{Status}{All}{}{Count}{32}%
\StoreBenchExecResult{Pono}{IcIIIiaMsatTerminationBitVectorsSampledFunc}{Status}{Correct}{}{Count}{24}%
\StoreBenchExecResult{Pono}{IcIIIiaMsatTerminationBitVectorsSampledFunc}{Status}{Correct}{True}{Count}{13}%
\StoreBenchExecResult{Pono}{IcIIIiaMsatTerminationBitVectorsSampledFunc}{Status}{Correct}{False}{Count}{11}%
\StoreBenchExecResult{Pono}{IcIIIiaMsatTerminationBitVectorsSampledFunc}{Status}{Wrong}{}{Count}{0}%
\StoreBenchExecResult{Pono}{IcIIIiaMsatTerminationBitVectorsSampledFunc}{Status}{Wrong}{True}{Count}{0}%
\StoreBenchExecResult{Pono}{IcIIIiaMsatTerminationBitVectorsSampledFunc}{Status}{Wrong}{False}{Count}{0}%
\providecommand\StoreBenchExecResult[7]{\expandafter\newcommand\csname#1#2#3#4#5#6\endcsname{#7}}%
\StoreBenchExecResult{Pono}{IcIIIiaMsatTerminationMainControlFlowSampledFunc}{Status}{All}{}{Score}{0}%
\StoreBenchExecResult{Pono}{IcIIIiaMsatTerminationMainControlFlowSampledFunc}{Status}{All}{}{Count}{242}%
\StoreBenchExecResult{Pono}{IcIIIiaMsatTerminationMainControlFlowSampledFunc}{Status}{Correct}{}{Count}{72}%
\StoreBenchExecResult{Pono}{IcIIIiaMsatTerminationMainControlFlowSampledFunc}{Status}{Correct}{True}{Count}{22}%
\StoreBenchExecResult{Pono}{IcIIIiaMsatTerminationMainControlFlowSampledFunc}{Status}{Correct}{False}{Count}{50}%
\StoreBenchExecResult{Pono}{IcIIIiaMsatTerminationMainControlFlowSampledFunc}{Status}{Wrong}{}{Count}{0}%
\StoreBenchExecResult{Pono}{IcIIIiaMsatTerminationMainControlFlowSampledFunc}{Status}{Wrong}{True}{Count}{0}%
\StoreBenchExecResult{Pono}{IcIIIiaMsatTerminationMainControlFlowSampledFunc}{Status}{Wrong}{False}{Count}{0}%
\providecommand\StoreBenchExecResult[7]{\expandafter\newcommand\csname#1#2#3#4#5#6\endcsname{#7}}%
\StoreBenchExecResult{Pono}{IcIIIiaMsatTerminationOtherSampledFunc}{Status}{All}{}{Score}{0}%
\StoreBenchExecResult{Pono}{IcIIIiaMsatTerminationOtherSampledFunc}{Status}{All}{}{Count}{1080}%
\StoreBenchExecResult{Pono}{IcIIIiaMsatTerminationOtherSampledFunc}{Status}{Correct}{}{Count}{671}%
\StoreBenchExecResult{Pono}{IcIIIiaMsatTerminationOtherSampledFunc}{Status}{Correct}{True}{Count}{123}%
\StoreBenchExecResult{Pono}{IcIIIiaMsatTerminationOtherSampledFunc}{Status}{Correct}{False}{Count}{548}%
\StoreBenchExecResult{Pono}{IcIIIiaMsatTerminationOtherSampledFunc}{Status}{Wrong}{}{Count}{0}%
\StoreBenchExecResult{Pono}{IcIIIiaMsatTerminationOtherSampledFunc}{Status}{Wrong}{True}{Count}{0}%
\StoreBenchExecResult{Pono}{IcIIIiaMsatTerminationOtherSampledFunc}{Status}{Wrong}{False}{Count}{0}%
\edef\PonoIcIIIiaMsatReachSafetySampledFuncStatusAllCount{\the\numexpr\PonoIcIIIiaMsatReachSafetyArraysSampledFuncStatusAllCount+\PonoIcIIIiaMsatReachSafetyBitVectorsSampledFuncStatusAllCount+\PonoIcIIIiaMsatReachSafetyCombinationsSampledFuncStatusAllCount+\PonoIcIIIiaMsatReachSafetyControlFlowSampledFuncStatusAllCount+\PonoIcIIIiaMsatReachSafetyECASampledFuncStatusAllCount+\PonoIcIIIiaMsatReachSafetyFloatsSampledFuncStatusAllCount+\PonoIcIIIiaMsatReachSafetyHardnessSampledFuncStatusAllCount+\PonoIcIIIiaMsatReachSafetyHardwareSampledFuncStatusAllCount+\PonoIcIIIiaMsatReachSafetyHeapSampledFuncStatusAllCount+\PonoIcIIIiaMsatReachSafetyLoopsSampledFuncStatusAllCount+\PonoIcIIIiaMsatReachSafetyProductLinesSampledFuncStatusAllCount+\PonoIcIIIiaMsatReachSafetySequentializedSampledFuncStatusAllCount+\PonoIcIIIiaMsatReachSafetyXCSPSampledFuncStatusAllCount}
\edef\PonoIcIIIiaMsatReachSafetySampledFuncStatusCorrectCount{\the\numexpr\PonoIcIIIiaMsatReachSafetyArraysSampledFuncStatusCorrectCount+\PonoIcIIIiaMsatReachSafetyBitVectorsSampledFuncStatusCorrectCount+\PonoIcIIIiaMsatReachSafetyCombinationsSampledFuncStatusCorrectCount+\PonoIcIIIiaMsatReachSafetyControlFlowSampledFuncStatusCorrectCount+\PonoIcIIIiaMsatReachSafetyECASampledFuncStatusCorrectCount+\PonoIcIIIiaMsatReachSafetyFloatsSampledFuncStatusCorrectCount+\PonoIcIIIiaMsatReachSafetyHardnessSampledFuncStatusCorrectCount+\PonoIcIIIiaMsatReachSafetyHardwareSampledFuncStatusCorrectCount+\PonoIcIIIiaMsatReachSafetyHeapSampledFuncStatusCorrectCount+\PonoIcIIIiaMsatReachSafetyLoopsSampledFuncStatusCorrectCount+\PonoIcIIIiaMsatReachSafetyProductLinesSampledFuncStatusCorrectCount+\PonoIcIIIiaMsatReachSafetySequentializedSampledFuncStatusCorrectCount+\PonoIcIIIiaMsatReachSafetyXCSPSampledFuncStatusCorrectCount}
\edef\PonoIcIIIiaMsatReachSafetySampledFuncStatusCorrectTrueCount{\the\numexpr\PonoIcIIIiaMsatReachSafetyArraysSampledFuncStatusCorrectTrueCount+\PonoIcIIIiaMsatReachSafetyBitVectorsSampledFuncStatusCorrectTrueCount+\PonoIcIIIiaMsatReachSafetyCombinationsSampledFuncStatusCorrectTrueCount+\PonoIcIIIiaMsatReachSafetyControlFlowSampledFuncStatusCorrectTrueCount+\PonoIcIIIiaMsatReachSafetyECASampledFuncStatusCorrectTrueCount+\PonoIcIIIiaMsatReachSafetyFloatsSampledFuncStatusCorrectTrueCount+\PonoIcIIIiaMsatReachSafetyHardnessSampledFuncStatusCorrectTrueCount+\PonoIcIIIiaMsatReachSafetyHardwareSampledFuncStatusCorrectTrueCount+\PonoIcIIIiaMsatReachSafetyHeapSampledFuncStatusCorrectTrueCount+\PonoIcIIIiaMsatReachSafetyLoopsSampledFuncStatusCorrectTrueCount+\PonoIcIIIiaMsatReachSafetyProductLinesSampledFuncStatusCorrectTrueCount+\PonoIcIIIiaMsatReachSafetySequentializedSampledFuncStatusCorrectTrueCount+\PonoIcIIIiaMsatReachSafetyXCSPSampledFuncStatusCorrectTrueCount}
\edef\PonoIcIIIiaMsatReachSafetySampledFuncStatusCorrectFalseCount{\the\numexpr\PonoIcIIIiaMsatReachSafetyArraysSampledFuncStatusCorrectFalseCount+\PonoIcIIIiaMsatReachSafetyBitVectorsSampledFuncStatusCorrectFalseCount+\PonoIcIIIiaMsatReachSafetyCombinationsSampledFuncStatusCorrectFalseCount+\PonoIcIIIiaMsatReachSafetyControlFlowSampledFuncStatusCorrectFalseCount+\PonoIcIIIiaMsatReachSafetyECASampledFuncStatusCorrectFalseCount+\PonoIcIIIiaMsatReachSafetyFloatsSampledFuncStatusCorrectFalseCount+\PonoIcIIIiaMsatReachSafetyHardnessSampledFuncStatusCorrectFalseCount+\PonoIcIIIiaMsatReachSafetyHardwareSampledFuncStatusCorrectFalseCount+\PonoIcIIIiaMsatReachSafetyHeapSampledFuncStatusCorrectFalseCount+\PonoIcIIIiaMsatReachSafetyLoopsSampledFuncStatusCorrectFalseCount+\PonoIcIIIiaMsatReachSafetyProductLinesSampledFuncStatusCorrectFalseCount+\PonoIcIIIiaMsatReachSafetySequentializedSampledFuncStatusCorrectFalseCount+\PonoIcIIIiaMsatReachSafetyXCSPSampledFuncStatusCorrectFalseCount}
\edef\PonoIcIIIiaMsatReachSafetySampledFuncStatusWrongCount{\the\numexpr\PonoIcIIIiaMsatReachSafetyArraysSampledFuncStatusWrongCount+\PonoIcIIIiaMsatReachSafetyBitVectorsSampledFuncStatusWrongCount+\PonoIcIIIiaMsatReachSafetyCombinationsSampledFuncStatusWrongCount+\PonoIcIIIiaMsatReachSafetyControlFlowSampledFuncStatusWrongCount+\PonoIcIIIiaMsatReachSafetyECASampledFuncStatusWrongCount+\PonoIcIIIiaMsatReachSafetyFloatsSampledFuncStatusWrongCount+\PonoIcIIIiaMsatReachSafetyHardnessSampledFuncStatusWrongCount+\PonoIcIIIiaMsatReachSafetyHardwareSampledFuncStatusWrongCount+\PonoIcIIIiaMsatReachSafetyHeapSampledFuncStatusWrongCount+\PonoIcIIIiaMsatReachSafetyLoopsSampledFuncStatusWrongCount+\PonoIcIIIiaMsatReachSafetyProductLinesSampledFuncStatusWrongCount+\PonoIcIIIiaMsatReachSafetySequentializedSampledFuncStatusWrongCount+\PonoIcIIIiaMsatReachSafetyXCSPSampledFuncStatusWrongCount}
\edef\PonoIcIIIiaMsatTerminationSampledFuncStatusAllCount{\the\numexpr\PonoIcIIIiaMsatTerminationBitVectorsSampledFuncStatusAllCount+\PonoIcIIIiaMsatTerminationMainControlFlowSampledFuncStatusAllCount+\PonoIcIIIiaMsatTerminationOtherSampledFuncStatusAllCount}
\edef\PonoIcIIIiaMsatTerminationSampledFuncStatusCorrectCount{\the\numexpr\PonoIcIIIiaMsatTerminationBitVectorsSampledFuncStatusCorrectCount+\PonoIcIIIiaMsatTerminationMainControlFlowSampledFuncStatusCorrectCount+\PonoIcIIIiaMsatTerminationOtherSampledFuncStatusCorrectCount}
\edef\PonoIcIIIiaMsatTerminationSampledFuncStatusCorrectTrueCount{\the\numexpr\PonoIcIIIiaMsatTerminationBitVectorsSampledFuncStatusCorrectTrueCount+\PonoIcIIIiaMsatTerminationMainControlFlowSampledFuncStatusCorrectTrueCount+\PonoIcIIIiaMsatTerminationOtherSampledFuncStatusCorrectTrueCount}
\edef\PonoIcIIIiaMsatTerminationSampledFuncStatusCorrectFalseCount{\the\numexpr\PonoIcIIIiaMsatTerminationBitVectorsSampledFuncStatusCorrectFalseCount+\PonoIcIIIiaMsatTerminationMainControlFlowSampledFuncStatusCorrectFalseCount+\PonoIcIIIiaMsatTerminationOtherSampledFuncStatusCorrectFalseCount}
\edef\PonoIcIIIiaMsatTerminationSampledFuncStatusWrongCount{\the\numexpr\PonoIcIIIiaMsatTerminationBitVectorsSampledFuncStatusWrongCount+\PonoIcIIIiaMsatTerminationMainControlFlowSampledFuncStatusWrongCount+\PonoIcIIIiaMsatTerminationOtherSampledFuncStatusWrongCount}
\providecommand\StoreBenchExecResult[7]{\expandafter\newcommand\csname#1#2#3#4#5#6\endcsname{#7}}%
\StoreBenchExecResult{Pono}{ImcBzlaReachSafetyBitVectorsSampledBvFunc}{Status}{All}{}{Score}{0}%
\StoreBenchExecResult{Pono}{ImcBzlaReachSafetyBitVectorsSampledBvFunc}{Status}{All}{}{Count}{47}%
\StoreBenchExecResult{Pono}{ImcBzlaReachSafetyBitVectorsSampledBvFunc}{Status}{Correct}{}{Count}{38}%
\StoreBenchExecResult{Pono}{ImcBzlaReachSafetyBitVectorsSampledBvFunc}{Status}{Correct}{True}{Count}{27}%
\StoreBenchExecResult{Pono}{ImcBzlaReachSafetyBitVectorsSampledBvFunc}{Status}{Correct}{False}{Count}{11}%
\StoreBenchExecResult{Pono}{ImcBzlaReachSafetyBitVectorsSampledBvFunc}{Status}{Wrong}{}{Count}{0}%
\StoreBenchExecResult{Pono}{ImcBzlaReachSafetyBitVectorsSampledBvFunc}{Status}{Wrong}{True}{Count}{0}%
\StoreBenchExecResult{Pono}{ImcBzlaReachSafetyBitVectorsSampledBvFunc}{Status}{Wrong}{False}{Count}{0}%
\providecommand\StoreBenchExecResult[7]{\expandafter\newcommand\csname#1#2#3#4#5#6\endcsname{#7}}%
\StoreBenchExecResult{Pono}{ImcBzlaReachSafetyCombinationsSampledBvFunc}{Status}{All}{}{Score}{0}%
\StoreBenchExecResult{Pono}{ImcBzlaReachSafetyCombinationsSampledBvFunc}{Status}{All}{}{Count}{110}%
\StoreBenchExecResult{Pono}{ImcBzlaReachSafetyCombinationsSampledBvFunc}{Status}{Correct}{}{Count}{72}%
\StoreBenchExecResult{Pono}{ImcBzlaReachSafetyCombinationsSampledBvFunc}{Status}{Correct}{True}{Count}{1}%
\StoreBenchExecResult{Pono}{ImcBzlaReachSafetyCombinationsSampledBvFunc}{Status}{Correct}{False}{Count}{71}%
\StoreBenchExecResult{Pono}{ImcBzlaReachSafetyCombinationsSampledBvFunc}{Status}{Wrong}{}{Count}{0}%
\StoreBenchExecResult{Pono}{ImcBzlaReachSafetyCombinationsSampledBvFunc}{Status}{Wrong}{True}{Count}{0}%
\StoreBenchExecResult{Pono}{ImcBzlaReachSafetyCombinationsSampledBvFunc}{Status}{Wrong}{False}{Count}{0}%
\providecommand\StoreBenchExecResult[7]{\expandafter\newcommand\csname#1#2#3#4#5#6\endcsname{#7}}%
\StoreBenchExecResult{Pono}{ImcBzlaReachSafetyControlFlowSampledBvFunc}{Status}{All}{}{Score}{0}%
\StoreBenchExecResult{Pono}{ImcBzlaReachSafetyControlFlowSampledBvFunc}{Status}{All}{}{Count}{29}%
\StoreBenchExecResult{Pono}{ImcBzlaReachSafetyControlFlowSampledBvFunc}{Status}{Correct}{}{Count}{26}%
\StoreBenchExecResult{Pono}{ImcBzlaReachSafetyControlFlowSampledBvFunc}{Status}{Correct}{True}{Count}{24}%
\StoreBenchExecResult{Pono}{ImcBzlaReachSafetyControlFlowSampledBvFunc}{Status}{Correct}{False}{Count}{2}%
\StoreBenchExecResult{Pono}{ImcBzlaReachSafetyControlFlowSampledBvFunc}{Status}{Wrong}{}{Count}{0}%
\StoreBenchExecResult{Pono}{ImcBzlaReachSafetyControlFlowSampledBvFunc}{Status}{Wrong}{True}{Count}{0}%
\StoreBenchExecResult{Pono}{ImcBzlaReachSafetyControlFlowSampledBvFunc}{Status}{Wrong}{False}{Count}{0}%
\providecommand\StoreBenchExecResult[7]{\expandafter\newcommand\csname#1#2#3#4#5#6\endcsname{#7}}%
\StoreBenchExecResult{Pono}{ImcBzlaReachSafetyECASampledBvFunc}{Status}{All}{}{Score}{0}%
\StoreBenchExecResult{Pono}{ImcBzlaReachSafetyECASampledBvFunc}{Status}{All}{}{Count}{150}%
\StoreBenchExecResult{Pono}{ImcBzlaReachSafetyECASampledBvFunc}{Status}{Correct}{}{Count}{64}%
\StoreBenchExecResult{Pono}{ImcBzlaReachSafetyECASampledBvFunc}{Status}{Correct}{True}{Count}{51}%
\StoreBenchExecResult{Pono}{ImcBzlaReachSafetyECASampledBvFunc}{Status}{Correct}{False}{Count}{13}%
\StoreBenchExecResult{Pono}{ImcBzlaReachSafetyECASampledBvFunc}{Status}{Wrong}{}{Count}{0}%
\StoreBenchExecResult{Pono}{ImcBzlaReachSafetyECASampledBvFunc}{Status}{Wrong}{True}{Count}{0}%
\StoreBenchExecResult{Pono}{ImcBzlaReachSafetyECASampledBvFunc}{Status}{Wrong}{False}{Count}{0}%
\providecommand\StoreBenchExecResult[7]{\expandafter\newcommand\csname#1#2#3#4#5#6\endcsname{#7}}%
\StoreBenchExecResult{Pono}{ImcBzlaReachSafetyFloatsSampledBvFunc}{Status}{All}{}{Score}{0}%
\StoreBenchExecResult{Pono}{ImcBzlaReachSafetyFloatsSampledBvFunc}{Status}{All}{}{Count}{10}%
\StoreBenchExecResult{Pono}{ImcBzlaReachSafetyFloatsSampledBvFunc}{Status}{Correct}{}{Count}{10}%
\StoreBenchExecResult{Pono}{ImcBzlaReachSafetyFloatsSampledBvFunc}{Status}{Correct}{True}{Count}{10}%
\StoreBenchExecResult{Pono}{ImcBzlaReachSafetyFloatsSampledBvFunc}{Status}{Correct}{False}{Count}{0}%
\StoreBenchExecResult{Pono}{ImcBzlaReachSafetyFloatsSampledBvFunc}{Status}{Wrong}{}{Count}{0}%
\StoreBenchExecResult{Pono}{ImcBzlaReachSafetyFloatsSampledBvFunc}{Status}{Wrong}{True}{Count}{0}%
\StoreBenchExecResult{Pono}{ImcBzlaReachSafetyFloatsSampledBvFunc}{Status}{Wrong}{False}{Count}{0}%
\providecommand\StoreBenchExecResult[7]{\expandafter\newcommand\csname#1#2#3#4#5#6\endcsname{#7}}%
\StoreBenchExecResult{Pono}{ImcBzlaReachSafetyHardnessSampledBvFunc}{Status}{All}{}{Score}{0}%
\StoreBenchExecResult{Pono}{ImcBzlaReachSafetyHardnessSampledBvFunc}{Status}{All}{}{Count}{132}%
\StoreBenchExecResult{Pono}{ImcBzlaReachSafetyHardnessSampledBvFunc}{Status}{Correct}{}{Count}{132}%
\StoreBenchExecResult{Pono}{ImcBzlaReachSafetyHardnessSampledBvFunc}{Status}{Correct}{True}{Count}{132}%
\StoreBenchExecResult{Pono}{ImcBzlaReachSafetyHardnessSampledBvFunc}{Status}{Correct}{False}{Count}{0}%
\StoreBenchExecResult{Pono}{ImcBzlaReachSafetyHardnessSampledBvFunc}{Status}{Wrong}{}{Count}{0}%
\StoreBenchExecResult{Pono}{ImcBzlaReachSafetyHardnessSampledBvFunc}{Status}{Wrong}{True}{Count}{0}%
\StoreBenchExecResult{Pono}{ImcBzlaReachSafetyHardnessSampledBvFunc}{Status}{Wrong}{False}{Count}{0}%
\providecommand\StoreBenchExecResult[7]{\expandafter\newcommand\csname#1#2#3#4#5#6\endcsname{#7}}%
\StoreBenchExecResult{Pono}{ImcBzlaReachSafetyHardwareSampledBvFunc}{Status}{All}{}{Score}{0}%
\StoreBenchExecResult{Pono}{ImcBzlaReachSafetyHardwareSampledBvFunc}{Status}{All}{}{Count}{148}%
\StoreBenchExecResult{Pono}{ImcBzlaReachSafetyHardwareSampledBvFunc}{Status}{Correct}{}{Count}{59}%
\StoreBenchExecResult{Pono}{ImcBzlaReachSafetyHardwareSampledBvFunc}{Status}{Correct}{True}{Count}{19}%
\StoreBenchExecResult{Pono}{ImcBzlaReachSafetyHardwareSampledBvFunc}{Status}{Correct}{False}{Count}{40}%
\StoreBenchExecResult{Pono}{ImcBzlaReachSafetyHardwareSampledBvFunc}{Status}{Wrong}{}{Count}{0}%
\StoreBenchExecResult{Pono}{ImcBzlaReachSafetyHardwareSampledBvFunc}{Status}{Wrong}{True}{Count}{0}%
\StoreBenchExecResult{Pono}{ImcBzlaReachSafetyHardwareSampledBvFunc}{Status}{Wrong}{False}{Count}{0}%
\providecommand\StoreBenchExecResult[7]{\expandafter\newcommand\csname#1#2#3#4#5#6\endcsname{#7}}%
\StoreBenchExecResult{Pono}{ImcBzlaReachSafetyHeapSampledBvFunc}{Status}{All}{}{Score}{0}%
\StoreBenchExecResult{Pono}{ImcBzlaReachSafetyHeapSampledBvFunc}{Status}{All}{}{Count}{6}%
\StoreBenchExecResult{Pono}{ImcBzlaReachSafetyHeapSampledBvFunc}{Status}{Correct}{}{Count}{6}%
\StoreBenchExecResult{Pono}{ImcBzlaReachSafetyHeapSampledBvFunc}{Status}{Correct}{True}{Count}{4}%
\StoreBenchExecResult{Pono}{ImcBzlaReachSafetyHeapSampledBvFunc}{Status}{Correct}{False}{Count}{2}%
\StoreBenchExecResult{Pono}{ImcBzlaReachSafetyHeapSampledBvFunc}{Status}{Wrong}{}{Count}{0}%
\StoreBenchExecResult{Pono}{ImcBzlaReachSafetyHeapSampledBvFunc}{Status}{Wrong}{True}{Count}{0}%
\StoreBenchExecResult{Pono}{ImcBzlaReachSafetyHeapSampledBvFunc}{Status}{Wrong}{False}{Count}{0}%
\providecommand\StoreBenchExecResult[7]{\expandafter\newcommand\csname#1#2#3#4#5#6\endcsname{#7}}%
\StoreBenchExecResult{Pono}{ImcBzlaReachSafetyLoopsSampledBvFunc}{Status}{All}{}{Score}{0}%
\StoreBenchExecResult{Pono}{ImcBzlaReachSafetyLoopsSampledBvFunc}{Status}{All}{}{Count}{137}%
\StoreBenchExecResult{Pono}{ImcBzlaReachSafetyLoopsSampledBvFunc}{Status}{Correct}{}{Count}{45}%
\StoreBenchExecResult{Pono}{ImcBzlaReachSafetyLoopsSampledBvFunc}{Status}{Correct}{True}{Count}{20}%
\StoreBenchExecResult{Pono}{ImcBzlaReachSafetyLoopsSampledBvFunc}{Status}{Correct}{False}{Count}{25}%
\StoreBenchExecResult{Pono}{ImcBzlaReachSafetyLoopsSampledBvFunc}{Status}{Wrong}{}{Count}{0}%
\StoreBenchExecResult{Pono}{ImcBzlaReachSafetyLoopsSampledBvFunc}{Status}{Wrong}{True}{Count}{0}%
\StoreBenchExecResult{Pono}{ImcBzlaReachSafetyLoopsSampledBvFunc}{Status}{Wrong}{False}{Count}{0}%
\providecommand\StoreBenchExecResult[7]{\expandafter\newcommand\csname#1#2#3#4#5#6\endcsname{#7}}%
\StoreBenchExecResult{Pono}{ImcBzlaReachSafetyProductLinesSampledBvFunc}{Status}{All}{}{Score}{0}%
\StoreBenchExecResult{Pono}{ImcBzlaReachSafetyProductLinesSampledBvFunc}{Status}{All}{}{Count}{150}%
\StoreBenchExecResult{Pono}{ImcBzlaReachSafetyProductLinesSampledBvFunc}{Status}{Correct}{}{Count}{148}%
\StoreBenchExecResult{Pono}{ImcBzlaReachSafetyProductLinesSampledBvFunc}{Status}{Correct}{True}{Count}{73}%
\StoreBenchExecResult{Pono}{ImcBzlaReachSafetyProductLinesSampledBvFunc}{Status}{Correct}{False}{Count}{75}%
\StoreBenchExecResult{Pono}{ImcBzlaReachSafetyProductLinesSampledBvFunc}{Status}{Wrong}{}{Count}{0}%
\StoreBenchExecResult{Pono}{ImcBzlaReachSafetyProductLinesSampledBvFunc}{Status}{Wrong}{True}{Count}{0}%
\StoreBenchExecResult{Pono}{ImcBzlaReachSafetyProductLinesSampledBvFunc}{Status}{Wrong}{False}{Count}{0}%
\providecommand\StoreBenchExecResult[7]{\expandafter\newcommand\csname#1#2#3#4#5#6\endcsname{#7}}%
\StoreBenchExecResult{Pono}{ImcBzlaReachSafetySequentializedSampledBvFunc}{Status}{All}{}{Score}{0}%
\StoreBenchExecResult{Pono}{ImcBzlaReachSafetySequentializedSampledBvFunc}{Status}{All}{}{Count}{150}%
\StoreBenchExecResult{Pono}{ImcBzlaReachSafetySequentializedSampledBvFunc}{Status}{Correct}{}{Count}{106}%
\StoreBenchExecResult{Pono}{ImcBzlaReachSafetySequentializedSampledBvFunc}{Status}{Correct}{True}{Count}{20}%
\StoreBenchExecResult{Pono}{ImcBzlaReachSafetySequentializedSampledBvFunc}{Status}{Correct}{False}{Count}{86}%
\StoreBenchExecResult{Pono}{ImcBzlaReachSafetySequentializedSampledBvFunc}{Status}{Wrong}{}{Count}{0}%
\StoreBenchExecResult{Pono}{ImcBzlaReachSafetySequentializedSampledBvFunc}{Status}{Wrong}{True}{Count}{0}%
\StoreBenchExecResult{Pono}{ImcBzlaReachSafetySequentializedSampledBvFunc}{Status}{Wrong}{False}{Count}{0}%
\providecommand\StoreBenchExecResult[7]{\expandafter\newcommand\csname#1#2#3#4#5#6\endcsname{#7}}%
\StoreBenchExecResult{Pono}{ImcBzlaReachSafetyXCSPSampledBvFunc}{Status}{All}{}{Score}{0}%
\StoreBenchExecResult{Pono}{ImcBzlaReachSafetyXCSPSampledBvFunc}{Status}{All}{}{Count}{98}%
\StoreBenchExecResult{Pono}{ImcBzlaReachSafetyXCSPSampledBvFunc}{Status}{Correct}{}{Count}{94}%
\StoreBenchExecResult{Pono}{ImcBzlaReachSafetyXCSPSampledBvFunc}{Status}{Correct}{True}{Count}{51}%
\StoreBenchExecResult{Pono}{ImcBzlaReachSafetyXCSPSampledBvFunc}{Status}{Correct}{False}{Count}{43}%
\StoreBenchExecResult{Pono}{ImcBzlaReachSafetyXCSPSampledBvFunc}{Status}{Wrong}{}{Count}{0}%
\StoreBenchExecResult{Pono}{ImcBzlaReachSafetyXCSPSampledBvFunc}{Status}{Wrong}{True}{Count}{0}%
\StoreBenchExecResult{Pono}{ImcBzlaReachSafetyXCSPSampledBvFunc}{Status}{Wrong}{False}{Count}{0}%
\providecommand\StoreBenchExecResult[7]{\expandafter\newcommand\csname#1#2#3#4#5#6\endcsname{#7}}%
\StoreBenchExecResult{Pono}{ImcBzlaTerminationBitVectorsSampledBvFunc}{Status}{All}{}{Score}{0}%
\StoreBenchExecResult{Pono}{ImcBzlaTerminationBitVectorsSampledBvFunc}{Status}{All}{}{Count}{32}%
\StoreBenchExecResult{Pono}{ImcBzlaTerminationBitVectorsSampledBvFunc}{Status}{Correct}{}{Count}{24}%
\StoreBenchExecResult{Pono}{ImcBzlaTerminationBitVectorsSampledBvFunc}{Status}{Correct}{True}{Count}{13}%
\StoreBenchExecResult{Pono}{ImcBzlaTerminationBitVectorsSampledBvFunc}{Status}{Correct}{False}{Count}{11}%
\StoreBenchExecResult{Pono}{ImcBzlaTerminationBitVectorsSampledBvFunc}{Status}{Wrong}{}{Count}{0}%
\StoreBenchExecResult{Pono}{ImcBzlaTerminationBitVectorsSampledBvFunc}{Status}{Wrong}{True}{Count}{0}%
\StoreBenchExecResult{Pono}{ImcBzlaTerminationBitVectorsSampledBvFunc}{Status}{Wrong}{False}{Count}{0}%
\providecommand\StoreBenchExecResult[7]{\expandafter\newcommand\csname#1#2#3#4#5#6\endcsname{#7}}%
\StoreBenchExecResult{Pono}{ImcBzlaTerminationMainControlFlowSampledBvFunc}{Status}{All}{}{Score}{0}%
\StoreBenchExecResult{Pono}{ImcBzlaTerminationMainControlFlowSampledBvFunc}{Status}{All}{}{Count}{236}%
\StoreBenchExecResult{Pono}{ImcBzlaTerminationMainControlFlowSampledBvFunc}{Status}{Correct}{}{Count}{70}%
\StoreBenchExecResult{Pono}{ImcBzlaTerminationMainControlFlowSampledBvFunc}{Status}{Correct}{True}{Count}{22}%
\StoreBenchExecResult{Pono}{ImcBzlaTerminationMainControlFlowSampledBvFunc}{Status}{Correct}{False}{Count}{48}%
\StoreBenchExecResult{Pono}{ImcBzlaTerminationMainControlFlowSampledBvFunc}{Status}{Wrong}{}{Count}{0}%
\StoreBenchExecResult{Pono}{ImcBzlaTerminationMainControlFlowSampledBvFunc}{Status}{Wrong}{True}{Count}{0}%
\StoreBenchExecResult{Pono}{ImcBzlaTerminationMainControlFlowSampledBvFunc}{Status}{Wrong}{False}{Count}{0}%
\providecommand\StoreBenchExecResult[7]{\expandafter\newcommand\csname#1#2#3#4#5#6\endcsname{#7}}%
\StoreBenchExecResult{Pono}{ImcBzlaTerminationOtherSampledBvFunc}{Status}{All}{}{Score}{0}%
\StoreBenchExecResult{Pono}{ImcBzlaTerminationOtherSampledBvFunc}{Status}{All}{}{Count}{973}%
\StoreBenchExecResult{Pono}{ImcBzlaTerminationOtherSampledBvFunc}{Status}{Correct}{}{Count}{759}%
\StoreBenchExecResult{Pono}{ImcBzlaTerminationOtherSampledBvFunc}{Status}{Correct}{True}{Count}{189}%
\StoreBenchExecResult{Pono}{ImcBzlaTerminationOtherSampledBvFunc}{Status}{Correct}{False}{Count}{570}%
\StoreBenchExecResult{Pono}{ImcBzlaTerminationOtherSampledBvFunc}{Status}{Wrong}{}{Count}{0}%
\StoreBenchExecResult{Pono}{ImcBzlaTerminationOtherSampledBvFunc}{Status}{Wrong}{True}{Count}{0}%
\StoreBenchExecResult{Pono}{ImcBzlaTerminationOtherSampledBvFunc}{Status}{Wrong}{False}{Count}{0}%
\edef\PonoImcBzlaReachSafetySampledBvFuncStatusAllCount{\the\numexpr\PonoImcBzlaReachSafetyBitVectorsSampledBvFuncStatusAllCount+\PonoImcBzlaReachSafetyCombinationsSampledBvFuncStatusAllCount+\PonoImcBzlaReachSafetyControlFlowSampledBvFuncStatusAllCount+\PonoImcBzlaReachSafetyECASampledBvFuncStatusAllCount+\PonoImcBzlaReachSafetyFloatsSampledBvFuncStatusAllCount+\PonoImcBzlaReachSafetyHardnessSampledBvFuncStatusAllCount+\PonoImcBzlaReachSafetyHardwareSampledBvFuncStatusAllCount+\PonoImcBzlaReachSafetyHeapSampledBvFuncStatusAllCount+\PonoImcBzlaReachSafetyLoopsSampledBvFuncStatusAllCount+\PonoImcBzlaReachSafetyProductLinesSampledBvFuncStatusAllCount+\PonoImcBzlaReachSafetySequentializedSampledBvFuncStatusAllCount+\PonoImcBzlaReachSafetyXCSPSampledBvFuncStatusAllCount}
\edef\PonoImcBzlaReachSafetySampledBvFuncStatusCorrectCount{\the\numexpr\PonoImcBzlaReachSafetyBitVectorsSampledBvFuncStatusCorrectCount+\PonoImcBzlaReachSafetyCombinationsSampledBvFuncStatusCorrectCount+\PonoImcBzlaReachSafetyControlFlowSampledBvFuncStatusCorrectCount+\PonoImcBzlaReachSafetyECASampledBvFuncStatusCorrectCount+\PonoImcBzlaReachSafetyFloatsSampledBvFuncStatusCorrectCount+\PonoImcBzlaReachSafetyHardnessSampledBvFuncStatusCorrectCount+\PonoImcBzlaReachSafetyHardwareSampledBvFuncStatusCorrectCount+\PonoImcBzlaReachSafetyHeapSampledBvFuncStatusCorrectCount+\PonoImcBzlaReachSafetyLoopsSampledBvFuncStatusCorrectCount+\PonoImcBzlaReachSafetyProductLinesSampledBvFuncStatusCorrectCount+\PonoImcBzlaReachSafetySequentializedSampledBvFuncStatusCorrectCount+\PonoImcBzlaReachSafetyXCSPSampledBvFuncStatusCorrectCount}
\edef\PonoImcBzlaReachSafetySampledBvFuncStatusCorrectTrueCount{\the\numexpr\PonoImcBzlaReachSafetyBitVectorsSampledBvFuncStatusCorrectTrueCount+\PonoImcBzlaReachSafetyCombinationsSampledBvFuncStatusCorrectTrueCount+\PonoImcBzlaReachSafetyControlFlowSampledBvFuncStatusCorrectTrueCount+\PonoImcBzlaReachSafetyECASampledBvFuncStatusCorrectTrueCount+\PonoImcBzlaReachSafetyFloatsSampledBvFuncStatusCorrectTrueCount+\PonoImcBzlaReachSafetyHardnessSampledBvFuncStatusCorrectTrueCount+\PonoImcBzlaReachSafetyHardwareSampledBvFuncStatusCorrectTrueCount+\PonoImcBzlaReachSafetyHeapSampledBvFuncStatusCorrectTrueCount+\PonoImcBzlaReachSafetyLoopsSampledBvFuncStatusCorrectTrueCount+\PonoImcBzlaReachSafetyProductLinesSampledBvFuncStatusCorrectTrueCount+\PonoImcBzlaReachSafetySequentializedSampledBvFuncStatusCorrectTrueCount+\PonoImcBzlaReachSafetyXCSPSampledBvFuncStatusCorrectTrueCount}
\edef\PonoImcBzlaReachSafetySampledBvFuncStatusCorrectFalseCount{\the\numexpr\PonoImcBzlaReachSafetyBitVectorsSampledBvFuncStatusCorrectFalseCount+\PonoImcBzlaReachSafetyCombinationsSampledBvFuncStatusCorrectFalseCount+\PonoImcBzlaReachSafetyControlFlowSampledBvFuncStatusCorrectFalseCount+\PonoImcBzlaReachSafetyECASampledBvFuncStatusCorrectFalseCount+\PonoImcBzlaReachSafetyFloatsSampledBvFuncStatusCorrectFalseCount+\PonoImcBzlaReachSafetyHardnessSampledBvFuncStatusCorrectFalseCount+\PonoImcBzlaReachSafetyHardwareSampledBvFuncStatusCorrectFalseCount+\PonoImcBzlaReachSafetyHeapSampledBvFuncStatusCorrectFalseCount+\PonoImcBzlaReachSafetyLoopsSampledBvFuncStatusCorrectFalseCount+\PonoImcBzlaReachSafetyProductLinesSampledBvFuncStatusCorrectFalseCount+\PonoImcBzlaReachSafetySequentializedSampledBvFuncStatusCorrectFalseCount+\PonoImcBzlaReachSafetyXCSPSampledBvFuncStatusCorrectFalseCount}
\edef\PonoImcBzlaReachSafetySampledBvFuncStatusWrongCount{\the\numexpr\PonoImcBzlaReachSafetyBitVectorsSampledBvFuncStatusWrongCount+\PonoImcBzlaReachSafetyCombinationsSampledBvFuncStatusWrongCount+\PonoImcBzlaReachSafetyControlFlowSampledBvFuncStatusWrongCount+\PonoImcBzlaReachSafetyECASampledBvFuncStatusWrongCount+\PonoImcBzlaReachSafetyFloatsSampledBvFuncStatusWrongCount+\PonoImcBzlaReachSafetyHardnessSampledBvFuncStatusWrongCount+\PonoImcBzlaReachSafetyHardwareSampledBvFuncStatusWrongCount+\PonoImcBzlaReachSafetyHeapSampledBvFuncStatusWrongCount+\PonoImcBzlaReachSafetyLoopsSampledBvFuncStatusWrongCount+\PonoImcBzlaReachSafetyProductLinesSampledBvFuncStatusWrongCount+\PonoImcBzlaReachSafetySequentializedSampledBvFuncStatusWrongCount+\PonoImcBzlaReachSafetyXCSPSampledBvFuncStatusWrongCount}
\edef\PonoImcBzlaTerminationSampledBvFuncStatusAllCount{\the\numexpr\PonoImcBzlaTerminationBitVectorsSampledBvFuncStatusAllCount+\PonoImcBzlaTerminationMainControlFlowSampledBvFuncStatusAllCount+\PonoImcBzlaTerminationOtherSampledBvFuncStatusAllCount}
\edef\PonoImcBzlaTerminationSampledBvFuncStatusCorrectCount{\the\numexpr\PonoImcBzlaTerminationBitVectorsSampledBvFuncStatusCorrectCount+\PonoImcBzlaTerminationMainControlFlowSampledBvFuncStatusCorrectCount+\PonoImcBzlaTerminationOtherSampledBvFuncStatusCorrectCount}
\edef\PonoImcBzlaTerminationSampledBvFuncStatusCorrectTrueCount{\the\numexpr\PonoImcBzlaTerminationBitVectorsSampledBvFuncStatusCorrectTrueCount+\PonoImcBzlaTerminationMainControlFlowSampledBvFuncStatusCorrectTrueCount+\PonoImcBzlaTerminationOtherSampledBvFuncStatusCorrectTrueCount}
\edef\PonoImcBzlaTerminationSampledBvFuncStatusCorrectFalseCount{\the\numexpr\PonoImcBzlaTerminationBitVectorsSampledBvFuncStatusCorrectFalseCount+\PonoImcBzlaTerminationMainControlFlowSampledBvFuncStatusCorrectFalseCount+\PonoImcBzlaTerminationOtherSampledBvFuncStatusCorrectFalseCount}
\edef\PonoImcBzlaTerminationSampledBvFuncStatusWrongCount{\the\numexpr\PonoImcBzlaTerminationBitVectorsSampledBvFuncStatusWrongCount+\PonoImcBzlaTerminationMainControlFlowSampledBvFuncStatusWrongCount+\PonoImcBzlaTerminationOtherSampledBvFuncStatusWrongCount}
\providecommand\StoreBenchExecResult[7]{\expandafter\newcommand\csname#1#2#3#4#5#6\endcsname{#7}}%
\StoreBenchExecResult{Pono}{KindReachSafetyArraysSampledFunc}{Status}{All}{}{Score}{0}%
\StoreBenchExecResult{Pono}{KindReachSafetyArraysSampledFunc}{Status}{All}{}{Count}{150}%
\StoreBenchExecResult{Pono}{KindReachSafetyArraysSampledFunc}{Status}{Correct}{}{Count}{51}%
\StoreBenchExecResult{Pono}{KindReachSafetyArraysSampledFunc}{Status}{Correct}{True}{Count}{1}%
\StoreBenchExecResult{Pono}{KindReachSafetyArraysSampledFunc}{Status}{Correct}{False}{Count}{50}%
\StoreBenchExecResult{Pono}{KindReachSafetyArraysSampledFunc}{Status}{Wrong}{}{Count}{0}%
\StoreBenchExecResult{Pono}{KindReachSafetyArraysSampledFunc}{Status}{Wrong}{True}{Count}{0}%
\StoreBenchExecResult{Pono}{KindReachSafetyArraysSampledFunc}{Status}{Wrong}{False}{Count}{0}%
\providecommand\StoreBenchExecResult[7]{\expandafter\newcommand\csname#1#2#3#4#5#6\endcsname{#7}}%
\StoreBenchExecResult{Pono}{KindReachSafetyBitVectorsSampledFunc}{Status}{All}{}{Score}{0}%
\StoreBenchExecResult{Pono}{KindReachSafetyBitVectorsSampledFunc}{Status}{All}{}{Count}{48}%
\StoreBenchExecResult{Pono}{KindReachSafetyBitVectorsSampledFunc}{Status}{Correct}{}{Count}{34}%
\StoreBenchExecResult{Pono}{KindReachSafetyBitVectorsSampledFunc}{Status}{Correct}{True}{Count}{22}%
\StoreBenchExecResult{Pono}{KindReachSafetyBitVectorsSampledFunc}{Status}{Correct}{False}{Count}{12}%
\StoreBenchExecResult{Pono}{KindReachSafetyBitVectorsSampledFunc}{Status}{Wrong}{}{Count}{0}%
\StoreBenchExecResult{Pono}{KindReachSafetyBitVectorsSampledFunc}{Status}{Wrong}{True}{Count}{0}%
\StoreBenchExecResult{Pono}{KindReachSafetyBitVectorsSampledFunc}{Status}{Wrong}{False}{Count}{0}%
\providecommand\StoreBenchExecResult[7]{\expandafter\newcommand\csname#1#2#3#4#5#6\endcsname{#7}}%
\StoreBenchExecResult{Pono}{KindReachSafetyCombinationsSampledFunc}{Status}{All}{}{Score}{0}%
\StoreBenchExecResult{Pono}{KindReachSafetyCombinationsSampledFunc}{Status}{All}{}{Count}{110}%
\StoreBenchExecResult{Pono}{KindReachSafetyCombinationsSampledFunc}{Status}{Correct}{}{Count}{71}%
\StoreBenchExecResult{Pono}{KindReachSafetyCombinationsSampledFunc}{Status}{Correct}{True}{Count}{0}%
\StoreBenchExecResult{Pono}{KindReachSafetyCombinationsSampledFunc}{Status}{Correct}{False}{Count}{71}%
\StoreBenchExecResult{Pono}{KindReachSafetyCombinationsSampledFunc}{Status}{Wrong}{}{Count}{0}%
\StoreBenchExecResult{Pono}{KindReachSafetyCombinationsSampledFunc}{Status}{Wrong}{True}{Count}{0}%
\StoreBenchExecResult{Pono}{KindReachSafetyCombinationsSampledFunc}{Status}{Wrong}{False}{Count}{0}%
\providecommand\StoreBenchExecResult[7]{\expandafter\newcommand\csname#1#2#3#4#5#6\endcsname{#7}}%
\StoreBenchExecResult{Pono}{KindReachSafetyControlFlowSampledFunc}{Status}{All}{}{Score}{0}%
\StoreBenchExecResult{Pono}{KindReachSafetyControlFlowSampledFunc}{Status}{All}{}{Count}{33}%
\StoreBenchExecResult{Pono}{KindReachSafetyControlFlowSampledFunc}{Status}{Correct}{}{Count}{25}%
\StoreBenchExecResult{Pono}{KindReachSafetyControlFlowSampledFunc}{Status}{Correct}{True}{Count}{21}%
\StoreBenchExecResult{Pono}{KindReachSafetyControlFlowSampledFunc}{Status}{Correct}{False}{Count}{4}%
\StoreBenchExecResult{Pono}{KindReachSafetyControlFlowSampledFunc}{Status}{Wrong}{}{Count}{0}%
\StoreBenchExecResult{Pono}{KindReachSafetyControlFlowSampledFunc}{Status}{Wrong}{True}{Count}{0}%
\StoreBenchExecResult{Pono}{KindReachSafetyControlFlowSampledFunc}{Status}{Wrong}{False}{Count}{0}%
\providecommand\StoreBenchExecResult[7]{\expandafter\newcommand\csname#1#2#3#4#5#6\endcsname{#7}}%
\StoreBenchExecResult{Pono}{KindReachSafetyECASampledFunc}{Status}{All}{}{Score}{0}%
\StoreBenchExecResult{Pono}{KindReachSafetyECASampledFunc}{Status}{All}{}{Count}{150}%
\StoreBenchExecResult{Pono}{KindReachSafetyECASampledFunc}{Status}{Correct}{}{Count}{99}%
\StoreBenchExecResult{Pono}{KindReachSafetyECASampledFunc}{Status}{Correct}{True}{Count}{51}%
\StoreBenchExecResult{Pono}{KindReachSafetyECASampledFunc}{Status}{Correct}{False}{Count}{48}%
\StoreBenchExecResult{Pono}{KindReachSafetyECASampledFunc}{Status}{Wrong}{}{Count}{0}%
\StoreBenchExecResult{Pono}{KindReachSafetyECASampledFunc}{Status}{Wrong}{True}{Count}{0}%
\StoreBenchExecResult{Pono}{KindReachSafetyECASampledFunc}{Status}{Wrong}{False}{Count}{0}%
\providecommand\StoreBenchExecResult[7]{\expandafter\newcommand\csname#1#2#3#4#5#6\endcsname{#7}}%
\StoreBenchExecResult{Pono}{KindReachSafetyFloatsSampledFunc}{Status}{All}{}{Score}{0}%
\StoreBenchExecResult{Pono}{KindReachSafetyFloatsSampledFunc}{Status}{All}{}{Count}{10}%
\StoreBenchExecResult{Pono}{KindReachSafetyFloatsSampledFunc}{Status}{Correct}{}{Count}{10}%
\StoreBenchExecResult{Pono}{KindReachSafetyFloatsSampledFunc}{Status}{Correct}{True}{Count}{10}%
\StoreBenchExecResult{Pono}{KindReachSafetyFloatsSampledFunc}{Status}{Correct}{False}{Count}{0}%
\StoreBenchExecResult{Pono}{KindReachSafetyFloatsSampledFunc}{Status}{Wrong}{}{Count}{0}%
\StoreBenchExecResult{Pono}{KindReachSafetyFloatsSampledFunc}{Status}{Wrong}{True}{Count}{0}%
\StoreBenchExecResult{Pono}{KindReachSafetyFloatsSampledFunc}{Status}{Wrong}{False}{Count}{0}%
\providecommand\StoreBenchExecResult[7]{\expandafter\newcommand\csname#1#2#3#4#5#6\endcsname{#7}}%
\StoreBenchExecResult{Pono}{KindReachSafetyHardnessSampledFunc}{Status}{All}{}{Score}{0}%
\StoreBenchExecResult{Pono}{KindReachSafetyHardnessSampledFunc}{Status}{All}{}{Count}{150}%
\StoreBenchExecResult{Pono}{KindReachSafetyHardnessSampledFunc}{Status}{Correct}{}{Count}{134}%
\StoreBenchExecResult{Pono}{KindReachSafetyHardnessSampledFunc}{Status}{Correct}{True}{Count}{134}%
\StoreBenchExecResult{Pono}{KindReachSafetyHardnessSampledFunc}{Status}{Correct}{False}{Count}{0}%
\StoreBenchExecResult{Pono}{KindReachSafetyHardnessSampledFunc}{Status}{Wrong}{}{Count}{0}%
\StoreBenchExecResult{Pono}{KindReachSafetyHardnessSampledFunc}{Status}{Wrong}{True}{Count}{0}%
\StoreBenchExecResult{Pono}{KindReachSafetyHardnessSampledFunc}{Status}{Wrong}{False}{Count}{0}%
\providecommand\StoreBenchExecResult[7]{\expandafter\newcommand\csname#1#2#3#4#5#6\endcsname{#7}}%
\StoreBenchExecResult{Pono}{KindReachSafetyHardwareSampledFunc}{Status}{All}{}{Score}{0}%
\StoreBenchExecResult{Pono}{KindReachSafetyHardwareSampledFunc}{Status}{All}{}{Count}{150}%
\StoreBenchExecResult{Pono}{KindReachSafetyHardwareSampledFunc}{Status}{Correct}{}{Count}{58}%
\StoreBenchExecResult{Pono}{KindReachSafetyHardwareSampledFunc}{Status}{Correct}{True}{Count}{3}%
\StoreBenchExecResult{Pono}{KindReachSafetyHardwareSampledFunc}{Status}{Correct}{False}{Count}{55}%
\StoreBenchExecResult{Pono}{KindReachSafetyHardwareSampledFunc}{Status}{Wrong}{}{Count}{0}%
\StoreBenchExecResult{Pono}{KindReachSafetyHardwareSampledFunc}{Status}{Wrong}{True}{Count}{0}%
\StoreBenchExecResult{Pono}{KindReachSafetyHardwareSampledFunc}{Status}{Wrong}{False}{Count}{0}%
\providecommand\StoreBenchExecResult[7]{\expandafter\newcommand\csname#1#2#3#4#5#6\endcsname{#7}}%
\StoreBenchExecResult{Pono}{KindReachSafetyHeapSampledFunc}{Status}{All}{}{Score}{0}%
\StoreBenchExecResult{Pono}{KindReachSafetyHeapSampledFunc}{Status}{All}{}{Count}{53}%
\StoreBenchExecResult{Pono}{KindReachSafetyHeapSampledFunc}{Status}{Correct}{}{Count}{53}%
\StoreBenchExecResult{Pono}{KindReachSafetyHeapSampledFunc}{Status}{Correct}{True}{Count}{36}%
\StoreBenchExecResult{Pono}{KindReachSafetyHeapSampledFunc}{Status}{Correct}{False}{Count}{17}%
\StoreBenchExecResult{Pono}{KindReachSafetyHeapSampledFunc}{Status}{Wrong}{}{Count}{0}%
\StoreBenchExecResult{Pono}{KindReachSafetyHeapSampledFunc}{Status}{Wrong}{True}{Count}{0}%
\StoreBenchExecResult{Pono}{KindReachSafetyHeapSampledFunc}{Status}{Wrong}{False}{Count}{0}%
\providecommand\StoreBenchExecResult[7]{\expandafter\newcommand\csname#1#2#3#4#5#6\endcsname{#7}}%
\StoreBenchExecResult{Pono}{KindReachSafetyLoopsSampledFunc}{Status}{All}{}{Score}{0}%
\StoreBenchExecResult{Pono}{KindReachSafetyLoopsSampledFunc}{Status}{All}{}{Count}{150}%
\StoreBenchExecResult{Pono}{KindReachSafetyLoopsSampledFunc}{Status}{Correct}{}{Count}{73}%
\StoreBenchExecResult{Pono}{KindReachSafetyLoopsSampledFunc}{Status}{Correct}{True}{Count}{28}%
\StoreBenchExecResult{Pono}{KindReachSafetyLoopsSampledFunc}{Status}{Correct}{False}{Count}{45}%
\StoreBenchExecResult{Pono}{KindReachSafetyLoopsSampledFunc}{Status}{Wrong}{}{Count}{0}%
\StoreBenchExecResult{Pono}{KindReachSafetyLoopsSampledFunc}{Status}{Wrong}{True}{Count}{0}%
\StoreBenchExecResult{Pono}{KindReachSafetyLoopsSampledFunc}{Status}{Wrong}{False}{Count}{0}%
\providecommand\StoreBenchExecResult[7]{\expandafter\newcommand\csname#1#2#3#4#5#6\endcsname{#7}}%
\StoreBenchExecResult{Pono}{KindReachSafetyProductLinesSampledFunc}{Status}{All}{}{Score}{0}%
\StoreBenchExecResult{Pono}{KindReachSafetyProductLinesSampledFunc}{Status}{All}{}{Count}{150}%
\StoreBenchExecResult{Pono}{KindReachSafetyProductLinesSampledFunc}{Status}{Correct}{}{Count}{95}%
\StoreBenchExecResult{Pono}{KindReachSafetyProductLinesSampledFunc}{Status}{Correct}{True}{Count}{20}%
\StoreBenchExecResult{Pono}{KindReachSafetyProductLinesSampledFunc}{Status}{Correct}{False}{Count}{75}%
\StoreBenchExecResult{Pono}{KindReachSafetyProductLinesSampledFunc}{Status}{Wrong}{}{Count}{0}%
\StoreBenchExecResult{Pono}{KindReachSafetyProductLinesSampledFunc}{Status}{Wrong}{True}{Count}{0}%
\StoreBenchExecResult{Pono}{KindReachSafetyProductLinesSampledFunc}{Status}{Wrong}{False}{Count}{0}%
\providecommand\StoreBenchExecResult[7]{\expandafter\newcommand\csname#1#2#3#4#5#6\endcsname{#7}}%
\StoreBenchExecResult{Pono}{KindReachSafetySequentializedSampledFunc}{Status}{All}{}{Score}{0}%
\StoreBenchExecResult{Pono}{KindReachSafetySequentializedSampledFunc}{Status}{All}{}{Count}{150}%
\StoreBenchExecResult{Pono}{KindReachSafetySequentializedSampledFunc}{Status}{Correct}{}{Count}{114}%
\StoreBenchExecResult{Pono}{KindReachSafetySequentializedSampledFunc}{Status}{Correct}{True}{Count}{27}%
\StoreBenchExecResult{Pono}{KindReachSafetySequentializedSampledFunc}{Status}{Correct}{False}{Count}{87}%
\StoreBenchExecResult{Pono}{KindReachSafetySequentializedSampledFunc}{Status}{Wrong}{}{Count}{0}%
\StoreBenchExecResult{Pono}{KindReachSafetySequentializedSampledFunc}{Status}{Wrong}{True}{Count}{0}%
\StoreBenchExecResult{Pono}{KindReachSafetySequentializedSampledFunc}{Status}{Wrong}{False}{Count}{0}%
\providecommand\StoreBenchExecResult[7]{\expandafter\newcommand\csname#1#2#3#4#5#6\endcsname{#7}}%
\StoreBenchExecResult{Pono}{KindReachSafetyXCSPSampledFunc}{Status}{All}{}{Score}{0}%
\StoreBenchExecResult{Pono}{KindReachSafetyXCSPSampledFunc}{Status}{All}{}{Count}{98}%
\StoreBenchExecResult{Pono}{KindReachSafetyXCSPSampledFunc}{Status}{Correct}{}{Count}{97}%
\StoreBenchExecResult{Pono}{KindReachSafetyXCSPSampledFunc}{Status}{Correct}{True}{Count}{52}%
\StoreBenchExecResult{Pono}{KindReachSafetyXCSPSampledFunc}{Status}{Correct}{False}{Count}{45}%
\StoreBenchExecResult{Pono}{KindReachSafetyXCSPSampledFunc}{Status}{Wrong}{}{Count}{0}%
\StoreBenchExecResult{Pono}{KindReachSafetyXCSPSampledFunc}{Status}{Wrong}{True}{Count}{0}%
\StoreBenchExecResult{Pono}{KindReachSafetyXCSPSampledFunc}{Status}{Wrong}{False}{Count}{0}%
\providecommand\StoreBenchExecResult[7]{\expandafter\newcommand\csname#1#2#3#4#5#6\endcsname{#7}}%
\StoreBenchExecResult{Pono}{KindTerminationBitVectorsSampledFunc}{Status}{All}{}{Score}{0}%
\StoreBenchExecResult{Pono}{KindTerminationBitVectorsSampledFunc}{Status}{All}{}{Count}{32}%
\StoreBenchExecResult{Pono}{KindTerminationBitVectorsSampledFunc}{Status}{Correct}{}{Count}{24}%
\StoreBenchExecResult{Pono}{KindTerminationBitVectorsSampledFunc}{Status}{Correct}{True}{Count}{13}%
\StoreBenchExecResult{Pono}{KindTerminationBitVectorsSampledFunc}{Status}{Correct}{False}{Count}{11}%
\StoreBenchExecResult{Pono}{KindTerminationBitVectorsSampledFunc}{Status}{Wrong}{}{Count}{0}%
\StoreBenchExecResult{Pono}{KindTerminationBitVectorsSampledFunc}{Status}{Wrong}{True}{Count}{0}%
\StoreBenchExecResult{Pono}{KindTerminationBitVectorsSampledFunc}{Status}{Wrong}{False}{Count}{0}%
\providecommand\StoreBenchExecResult[7]{\expandafter\newcommand\csname#1#2#3#4#5#6\endcsname{#7}}%
\StoreBenchExecResult{Pono}{KindTerminationMainControlFlowSampledFunc}{Status}{All}{}{Score}{0}%
\StoreBenchExecResult{Pono}{KindTerminationMainControlFlowSampledFunc}{Status}{All}{}{Count}{242}%
\StoreBenchExecResult{Pono}{KindTerminationMainControlFlowSampledFunc}{Status}{Correct}{}{Count}{80}%
\StoreBenchExecResult{Pono}{KindTerminationMainControlFlowSampledFunc}{Status}{Correct}{True}{Count}{28}%
\StoreBenchExecResult{Pono}{KindTerminationMainControlFlowSampledFunc}{Status}{Correct}{False}{Count}{52}%
\StoreBenchExecResult{Pono}{KindTerminationMainControlFlowSampledFunc}{Status}{Wrong}{}{Count}{0}%
\StoreBenchExecResult{Pono}{KindTerminationMainControlFlowSampledFunc}{Status}{Wrong}{True}{Count}{0}%
\StoreBenchExecResult{Pono}{KindTerminationMainControlFlowSampledFunc}{Status}{Wrong}{False}{Count}{0}%
\providecommand\StoreBenchExecResult[7]{\expandafter\newcommand\csname#1#2#3#4#5#6\endcsname{#7}}%
\StoreBenchExecResult{Pono}{KindTerminationOtherSampledFunc}{Status}{All}{}{Score}{0}%
\StoreBenchExecResult{Pono}{KindTerminationOtherSampledFunc}{Status}{All}{}{Count}{1080}%
\StoreBenchExecResult{Pono}{KindTerminationOtherSampledFunc}{Status}{Correct}{}{Count}{910}%
\StoreBenchExecResult{Pono}{KindTerminationOtherSampledFunc}{Status}{Correct}{True}{Count}{276}%
\StoreBenchExecResult{Pono}{KindTerminationOtherSampledFunc}{Status}{Correct}{False}{Count}{634}%
\StoreBenchExecResult{Pono}{KindTerminationOtherSampledFunc}{Status}{Wrong}{}{Count}{0}%
\StoreBenchExecResult{Pono}{KindTerminationOtherSampledFunc}{Status}{Wrong}{True}{Count}{0}%
\StoreBenchExecResult{Pono}{KindTerminationOtherSampledFunc}{Status}{Wrong}{False}{Count}{0}%
\edef\PonoKindReachSafetySampledFuncStatusAllCount{\the\numexpr\PonoKindReachSafetyArraysSampledFuncStatusAllCount+\PonoKindReachSafetyBitVectorsSampledFuncStatusAllCount+\PonoKindReachSafetyCombinationsSampledFuncStatusAllCount+\PonoKindReachSafetyControlFlowSampledFuncStatusAllCount+\PonoKindReachSafetyECASampledFuncStatusAllCount+\PonoKindReachSafetyFloatsSampledFuncStatusAllCount+\PonoKindReachSafetyHardnessSampledFuncStatusAllCount+\PonoKindReachSafetyHardwareSampledFuncStatusAllCount+\PonoKindReachSafetyHeapSampledFuncStatusAllCount+\PonoKindReachSafetyLoopsSampledFuncStatusAllCount+\PonoKindReachSafetyProductLinesSampledFuncStatusAllCount+\PonoKindReachSafetySequentializedSampledFuncStatusAllCount+\PonoKindReachSafetyXCSPSampledFuncStatusAllCount}
\edef\PonoKindReachSafetySampledFuncStatusCorrectCount{\the\numexpr\PonoKindReachSafetyArraysSampledFuncStatusCorrectCount+\PonoKindReachSafetyBitVectorsSampledFuncStatusCorrectCount+\PonoKindReachSafetyCombinationsSampledFuncStatusCorrectCount+\PonoKindReachSafetyControlFlowSampledFuncStatusCorrectCount+\PonoKindReachSafetyECASampledFuncStatusCorrectCount+\PonoKindReachSafetyFloatsSampledFuncStatusCorrectCount+\PonoKindReachSafetyHardnessSampledFuncStatusCorrectCount+\PonoKindReachSafetyHardwareSampledFuncStatusCorrectCount+\PonoKindReachSafetyHeapSampledFuncStatusCorrectCount+\PonoKindReachSafetyLoopsSampledFuncStatusCorrectCount+\PonoKindReachSafetyProductLinesSampledFuncStatusCorrectCount+\PonoKindReachSafetySequentializedSampledFuncStatusCorrectCount+\PonoKindReachSafetyXCSPSampledFuncStatusCorrectCount}
\edef\PonoKindReachSafetySampledFuncStatusCorrectTrueCount{\the\numexpr\PonoKindReachSafetyArraysSampledFuncStatusCorrectTrueCount+\PonoKindReachSafetyBitVectorsSampledFuncStatusCorrectTrueCount+\PonoKindReachSafetyCombinationsSampledFuncStatusCorrectTrueCount+\PonoKindReachSafetyControlFlowSampledFuncStatusCorrectTrueCount+\PonoKindReachSafetyECASampledFuncStatusCorrectTrueCount+\PonoKindReachSafetyFloatsSampledFuncStatusCorrectTrueCount+\PonoKindReachSafetyHardnessSampledFuncStatusCorrectTrueCount+\PonoKindReachSafetyHardwareSampledFuncStatusCorrectTrueCount+\PonoKindReachSafetyHeapSampledFuncStatusCorrectTrueCount+\PonoKindReachSafetyLoopsSampledFuncStatusCorrectTrueCount+\PonoKindReachSafetyProductLinesSampledFuncStatusCorrectTrueCount+\PonoKindReachSafetySequentializedSampledFuncStatusCorrectTrueCount+\PonoKindReachSafetyXCSPSampledFuncStatusCorrectTrueCount}
\edef\PonoKindReachSafetySampledFuncStatusCorrectFalseCount{\the\numexpr\PonoKindReachSafetyArraysSampledFuncStatusCorrectFalseCount+\PonoKindReachSafetyBitVectorsSampledFuncStatusCorrectFalseCount+\PonoKindReachSafetyCombinationsSampledFuncStatusCorrectFalseCount+\PonoKindReachSafetyControlFlowSampledFuncStatusCorrectFalseCount+\PonoKindReachSafetyECASampledFuncStatusCorrectFalseCount+\PonoKindReachSafetyFloatsSampledFuncStatusCorrectFalseCount+\PonoKindReachSafetyHardnessSampledFuncStatusCorrectFalseCount+\PonoKindReachSafetyHardwareSampledFuncStatusCorrectFalseCount+\PonoKindReachSafetyHeapSampledFuncStatusCorrectFalseCount+\PonoKindReachSafetyLoopsSampledFuncStatusCorrectFalseCount+\PonoKindReachSafetyProductLinesSampledFuncStatusCorrectFalseCount+\PonoKindReachSafetySequentializedSampledFuncStatusCorrectFalseCount+\PonoKindReachSafetyXCSPSampledFuncStatusCorrectFalseCount}
\edef\PonoKindReachSafetySampledFuncStatusWrongCount{\the\numexpr\PonoKindReachSafetyArraysSampledFuncStatusWrongCount+\PonoKindReachSafetyBitVectorsSampledFuncStatusWrongCount+\PonoKindReachSafetyCombinationsSampledFuncStatusWrongCount+\PonoKindReachSafetyControlFlowSampledFuncStatusWrongCount+\PonoKindReachSafetyECASampledFuncStatusWrongCount+\PonoKindReachSafetyFloatsSampledFuncStatusWrongCount+\PonoKindReachSafetyHardnessSampledFuncStatusWrongCount+\PonoKindReachSafetyHardwareSampledFuncStatusWrongCount+\PonoKindReachSafetyHeapSampledFuncStatusWrongCount+\PonoKindReachSafetyLoopsSampledFuncStatusWrongCount+\PonoKindReachSafetyProductLinesSampledFuncStatusWrongCount+\PonoKindReachSafetySequentializedSampledFuncStatusWrongCount+\PonoKindReachSafetyXCSPSampledFuncStatusWrongCount}
\edef\PonoKindTerminationSampledFuncStatusAllCount{\the\numexpr\PonoKindTerminationBitVectorsSampledFuncStatusAllCount+\PonoKindTerminationMainControlFlowSampledFuncStatusAllCount+\PonoKindTerminationOtherSampledFuncStatusAllCount}
\edef\PonoKindTerminationSampledFuncStatusCorrectCount{\the\numexpr\PonoKindTerminationBitVectorsSampledFuncStatusCorrectCount+\PonoKindTerminationMainControlFlowSampledFuncStatusCorrectCount+\PonoKindTerminationOtherSampledFuncStatusCorrectCount}
\edef\PonoKindTerminationSampledFuncStatusCorrectTrueCount{\the\numexpr\PonoKindTerminationBitVectorsSampledFuncStatusCorrectTrueCount+\PonoKindTerminationMainControlFlowSampledFuncStatusCorrectTrueCount+\PonoKindTerminationOtherSampledFuncStatusCorrectTrueCount}
\edef\PonoKindTerminationSampledFuncStatusCorrectFalseCount{\the\numexpr\PonoKindTerminationBitVectorsSampledFuncStatusCorrectFalseCount+\PonoKindTerminationMainControlFlowSampledFuncStatusCorrectFalseCount+\PonoKindTerminationOtherSampledFuncStatusCorrectFalseCount}
\edef\PonoKindTerminationSampledFuncStatusWrongCount{\the\numexpr\PonoKindTerminationBitVectorsSampledFuncStatusWrongCount+\PonoKindTerminationMainControlFlowSampledFuncStatusWrongCount+\PonoKindTerminationOtherSampledFuncStatusWrongCount}
\providecommand\StoreBenchExecResult[7]{\expandafter\newcommand\csname#1#2#3#4#5#6\endcsname{#7}}%
\StoreBenchExecResult{Abc}{ImcReachSafetyBitVectorsSampledBvRel}{Status}{All}{}{Score}{0}%
\StoreBenchExecResult{Abc}{ImcReachSafetyBitVectorsSampledBvRel}{Status}{All}{}{Count}{47}%
\StoreBenchExecResult{Abc}{ImcReachSafetyBitVectorsSampledBvRel}{Status}{Correct}{}{Count}{39}%
\StoreBenchExecResult{Abc}{ImcReachSafetyBitVectorsSampledBvRel}{Status}{Correct}{True}{Count}{28}%
\StoreBenchExecResult{Abc}{ImcReachSafetyBitVectorsSampledBvRel}{Status}{Correct}{False}{Count}{11}%
\StoreBenchExecResult{Abc}{ImcReachSafetyBitVectorsSampledBvRel}{Status}{Wrong}{}{Count}{0}%
\StoreBenchExecResult{Abc}{ImcReachSafetyBitVectorsSampledBvRel}{Status}{Wrong}{True}{Count}{0}%
\StoreBenchExecResult{Abc}{ImcReachSafetyBitVectorsSampledBvRel}{Status}{Wrong}{False}{Count}{0}%
\providecommand\StoreBenchExecResult[7]{\expandafter\newcommand\csname#1#2#3#4#5#6\endcsname{#7}}%
\StoreBenchExecResult{Abc}{ImcReachSafetyCombinationsSampledBvRel}{Status}{All}{}{Score}{0}%
\StoreBenchExecResult{Abc}{ImcReachSafetyCombinationsSampledBvRel}{Status}{All}{}{Count}{110}%
\StoreBenchExecResult{Abc}{ImcReachSafetyCombinationsSampledBvRel}{Status}{Correct}{}{Count}{73}%
\StoreBenchExecResult{Abc}{ImcReachSafetyCombinationsSampledBvRel}{Status}{Correct}{True}{Count}{2}%
\StoreBenchExecResult{Abc}{ImcReachSafetyCombinationsSampledBvRel}{Status}{Correct}{False}{Count}{71}%
\StoreBenchExecResult{Abc}{ImcReachSafetyCombinationsSampledBvRel}{Status}{Wrong}{}{Count}{0}%
\StoreBenchExecResult{Abc}{ImcReachSafetyCombinationsSampledBvRel}{Status}{Wrong}{True}{Count}{0}%
\StoreBenchExecResult{Abc}{ImcReachSafetyCombinationsSampledBvRel}{Status}{Wrong}{False}{Count}{0}%
\providecommand\StoreBenchExecResult[7]{\expandafter\newcommand\csname#1#2#3#4#5#6\endcsname{#7}}%
\StoreBenchExecResult{Abc}{ImcReachSafetyControlFlowSampledBvRel}{Status}{All}{}{Score}{0}%
\StoreBenchExecResult{Abc}{ImcReachSafetyControlFlowSampledBvRel}{Status}{All}{}{Count}{29}%
\StoreBenchExecResult{Abc}{ImcReachSafetyControlFlowSampledBvRel}{Status}{Correct}{}{Count}{26}%
\StoreBenchExecResult{Abc}{ImcReachSafetyControlFlowSampledBvRel}{Status}{Correct}{True}{Count}{24}%
\StoreBenchExecResult{Abc}{ImcReachSafetyControlFlowSampledBvRel}{Status}{Correct}{False}{Count}{2}%
\StoreBenchExecResult{Abc}{ImcReachSafetyControlFlowSampledBvRel}{Status}{Wrong}{}{Count}{0}%
\StoreBenchExecResult{Abc}{ImcReachSafetyControlFlowSampledBvRel}{Status}{Wrong}{True}{Count}{0}%
\StoreBenchExecResult{Abc}{ImcReachSafetyControlFlowSampledBvRel}{Status}{Wrong}{False}{Count}{0}%
\providecommand\StoreBenchExecResult[7]{\expandafter\newcommand\csname#1#2#3#4#5#6\endcsname{#7}}%
\StoreBenchExecResult{Abc}{ImcReachSafetyECASampledBvRel}{Status}{All}{}{Score}{0}%
\StoreBenchExecResult{Abc}{ImcReachSafetyECASampledBvRel}{Status}{All}{}{Count}{150}%
\StoreBenchExecResult{Abc}{ImcReachSafetyECASampledBvRel}{Status}{Correct}{}{Count}{113}%
\StoreBenchExecResult{Abc}{ImcReachSafetyECASampledBvRel}{Status}{Correct}{True}{Count}{50}%
\StoreBenchExecResult{Abc}{ImcReachSafetyECASampledBvRel}{Status}{Correct}{False}{Count}{63}%
\StoreBenchExecResult{Abc}{ImcReachSafetyECASampledBvRel}{Status}{Wrong}{}{Count}{0}%
\StoreBenchExecResult{Abc}{ImcReachSafetyECASampledBvRel}{Status}{Wrong}{True}{Count}{0}%
\StoreBenchExecResult{Abc}{ImcReachSafetyECASampledBvRel}{Status}{Wrong}{False}{Count}{0}%
\providecommand\StoreBenchExecResult[7]{\expandafter\newcommand\csname#1#2#3#4#5#6\endcsname{#7}}%
\StoreBenchExecResult{Abc}{ImcReachSafetyFloatsSampledBvRel}{Status}{All}{}{Score}{0}%
\StoreBenchExecResult{Abc}{ImcReachSafetyFloatsSampledBvRel}{Status}{All}{}{Count}{10}%
\StoreBenchExecResult{Abc}{ImcReachSafetyFloatsSampledBvRel}{Status}{Correct}{}{Count}{10}%
\StoreBenchExecResult{Abc}{ImcReachSafetyFloatsSampledBvRel}{Status}{Correct}{True}{Count}{10}%
\StoreBenchExecResult{Abc}{ImcReachSafetyFloatsSampledBvRel}{Status}{Correct}{False}{Count}{0}%
\StoreBenchExecResult{Abc}{ImcReachSafetyFloatsSampledBvRel}{Status}{Wrong}{}{Count}{0}%
\StoreBenchExecResult{Abc}{ImcReachSafetyFloatsSampledBvRel}{Status}{Wrong}{True}{Count}{0}%
\StoreBenchExecResult{Abc}{ImcReachSafetyFloatsSampledBvRel}{Status}{Wrong}{False}{Count}{0}%
\providecommand\StoreBenchExecResult[7]{\expandafter\newcommand\csname#1#2#3#4#5#6\endcsname{#7}}%
\StoreBenchExecResult{Abc}{ImcReachSafetyHardnessSampledBvRel}{Status}{All}{}{Score}{0}%
\StoreBenchExecResult{Abc}{ImcReachSafetyHardnessSampledBvRel}{Status}{All}{}{Count}{132}%
\StoreBenchExecResult{Abc}{ImcReachSafetyHardnessSampledBvRel}{Status}{Correct}{}{Count}{132}%
\StoreBenchExecResult{Abc}{ImcReachSafetyHardnessSampledBvRel}{Status}{Correct}{True}{Count}{132}%
\StoreBenchExecResult{Abc}{ImcReachSafetyHardnessSampledBvRel}{Status}{Correct}{False}{Count}{0}%
\StoreBenchExecResult{Abc}{ImcReachSafetyHardnessSampledBvRel}{Status}{Wrong}{}{Count}{0}%
\StoreBenchExecResult{Abc}{ImcReachSafetyHardnessSampledBvRel}{Status}{Wrong}{True}{Count}{0}%
\StoreBenchExecResult{Abc}{ImcReachSafetyHardnessSampledBvRel}{Status}{Wrong}{False}{Count}{0}%
\providecommand\StoreBenchExecResult[7]{\expandafter\newcommand\csname#1#2#3#4#5#6\endcsname{#7}}%
\StoreBenchExecResult{Abc}{ImcReachSafetyHardwareSampledBvRel}{Status}{All}{}{Score}{0}%
\StoreBenchExecResult{Abc}{ImcReachSafetyHardwareSampledBvRel}{Status}{All}{}{Count}{148}%
\StoreBenchExecResult{Abc}{ImcReachSafetyHardwareSampledBvRel}{Status}{Correct}{}{Count}{62}%
\StoreBenchExecResult{Abc}{ImcReachSafetyHardwareSampledBvRel}{Status}{Correct}{True}{Count}{18}%
\StoreBenchExecResult{Abc}{ImcReachSafetyHardwareSampledBvRel}{Status}{Correct}{False}{Count}{44}%
\StoreBenchExecResult{Abc}{ImcReachSafetyHardwareSampledBvRel}{Status}{Wrong}{}{Count}{0}%
\StoreBenchExecResult{Abc}{ImcReachSafetyHardwareSampledBvRel}{Status}{Wrong}{True}{Count}{0}%
\StoreBenchExecResult{Abc}{ImcReachSafetyHardwareSampledBvRel}{Status}{Wrong}{False}{Count}{0}%
\providecommand\StoreBenchExecResult[7]{\expandafter\newcommand\csname#1#2#3#4#5#6\endcsname{#7}}%
\StoreBenchExecResult{Abc}{ImcReachSafetyHeapSampledBvRel}{Status}{All}{}{Score}{0}%
\StoreBenchExecResult{Abc}{ImcReachSafetyHeapSampledBvRel}{Status}{All}{}{Count}{6}%
\StoreBenchExecResult{Abc}{ImcReachSafetyHeapSampledBvRel}{Status}{Correct}{}{Count}{6}%
\StoreBenchExecResult{Abc}{ImcReachSafetyHeapSampledBvRel}{Status}{Correct}{True}{Count}{4}%
\StoreBenchExecResult{Abc}{ImcReachSafetyHeapSampledBvRel}{Status}{Correct}{False}{Count}{2}%
\StoreBenchExecResult{Abc}{ImcReachSafetyHeapSampledBvRel}{Status}{Wrong}{}{Count}{0}%
\StoreBenchExecResult{Abc}{ImcReachSafetyHeapSampledBvRel}{Status}{Wrong}{True}{Count}{0}%
\StoreBenchExecResult{Abc}{ImcReachSafetyHeapSampledBvRel}{Status}{Wrong}{False}{Count}{0}%
\providecommand\StoreBenchExecResult[7]{\expandafter\newcommand\csname#1#2#3#4#5#6\endcsname{#7}}%
\StoreBenchExecResult{Abc}{ImcReachSafetyLoopsSampledBvRel}{Status}{All}{}{Score}{0}%
\StoreBenchExecResult{Abc}{ImcReachSafetyLoopsSampledBvRel}{Status}{All}{}{Count}{137}%
\StoreBenchExecResult{Abc}{ImcReachSafetyLoopsSampledBvRel}{Status}{Correct}{}{Count}{56}%
\StoreBenchExecResult{Abc}{ImcReachSafetyLoopsSampledBvRel}{Status}{Correct}{True}{Count}{29}%
\StoreBenchExecResult{Abc}{ImcReachSafetyLoopsSampledBvRel}{Status}{Correct}{False}{Count}{27}%
\StoreBenchExecResult{Abc}{ImcReachSafetyLoopsSampledBvRel}{Status}{Wrong}{}{Count}{0}%
\StoreBenchExecResult{Abc}{ImcReachSafetyLoopsSampledBvRel}{Status}{Wrong}{True}{Count}{0}%
\StoreBenchExecResult{Abc}{ImcReachSafetyLoopsSampledBvRel}{Status}{Wrong}{False}{Count}{0}%
\providecommand\StoreBenchExecResult[7]{\expandafter\newcommand\csname#1#2#3#4#5#6\endcsname{#7}}%
\StoreBenchExecResult{Abc}{ImcReachSafetyProductLinesSampledBvRel}{Status}{All}{}{Score}{0}%
\StoreBenchExecResult{Abc}{ImcReachSafetyProductLinesSampledBvRel}{Status}{All}{}{Count}{150}%
\StoreBenchExecResult{Abc}{ImcReachSafetyProductLinesSampledBvRel}{Status}{Correct}{}{Count}{150}%
\StoreBenchExecResult{Abc}{ImcReachSafetyProductLinesSampledBvRel}{Status}{Correct}{True}{Count}{75}%
\StoreBenchExecResult{Abc}{ImcReachSafetyProductLinesSampledBvRel}{Status}{Correct}{False}{Count}{75}%
\StoreBenchExecResult{Abc}{ImcReachSafetyProductLinesSampledBvRel}{Status}{Wrong}{}{Count}{0}%
\StoreBenchExecResult{Abc}{ImcReachSafetyProductLinesSampledBvRel}{Status}{Wrong}{True}{Count}{0}%
\StoreBenchExecResult{Abc}{ImcReachSafetyProductLinesSampledBvRel}{Status}{Wrong}{False}{Count}{0}%
\providecommand\StoreBenchExecResult[7]{\expandafter\newcommand\csname#1#2#3#4#5#6\endcsname{#7}}%
\StoreBenchExecResult{Abc}{ImcReachSafetySequentializedSampledBvRel}{Status}{All}{}{Score}{0}%
\StoreBenchExecResult{Abc}{ImcReachSafetySequentializedSampledBvRel}{Status}{All}{}{Count}{150}%
\StoreBenchExecResult{Abc}{ImcReachSafetySequentializedSampledBvRel}{Status}{Correct}{}{Count}{108}%
\StoreBenchExecResult{Abc}{ImcReachSafetySequentializedSampledBvRel}{Status}{Correct}{True}{Count}{21}%
\StoreBenchExecResult{Abc}{ImcReachSafetySequentializedSampledBvRel}{Status}{Correct}{False}{Count}{87}%
\StoreBenchExecResult{Abc}{ImcReachSafetySequentializedSampledBvRel}{Status}{Wrong}{}{Count}{0}%
\StoreBenchExecResult{Abc}{ImcReachSafetySequentializedSampledBvRel}{Status}{Wrong}{True}{Count}{0}%
\StoreBenchExecResult{Abc}{ImcReachSafetySequentializedSampledBvRel}{Status}{Wrong}{False}{Count}{0}%
\providecommand\StoreBenchExecResult[7]{\expandafter\newcommand\csname#1#2#3#4#5#6\endcsname{#7}}%
\StoreBenchExecResult{Abc}{ImcReachSafetyXCSPSampledBvRel}{Status}{All}{}{Score}{0}%
\StoreBenchExecResult{Abc}{ImcReachSafetyXCSPSampledBvRel}{Status}{All}{}{Count}{98}%
\StoreBenchExecResult{Abc}{ImcReachSafetyXCSPSampledBvRel}{Status}{Correct}{}{Count}{93}%
\StoreBenchExecResult{Abc}{ImcReachSafetyXCSPSampledBvRel}{Status}{Correct}{True}{Count}{51}%
\StoreBenchExecResult{Abc}{ImcReachSafetyXCSPSampledBvRel}{Status}{Correct}{False}{Count}{42}%
\StoreBenchExecResult{Abc}{ImcReachSafetyXCSPSampledBvRel}{Status}{Wrong}{}{Count}{0}%
\StoreBenchExecResult{Abc}{ImcReachSafetyXCSPSampledBvRel}{Status}{Wrong}{True}{Count}{0}%
\StoreBenchExecResult{Abc}{ImcReachSafetyXCSPSampledBvRel}{Status}{Wrong}{False}{Count}{0}%
\providecommand\StoreBenchExecResult[7]{\expandafter\newcommand\csname#1#2#3#4#5#6\endcsname{#7}}%
\StoreBenchExecResult{Abc}{ImcTerminationBitVectorsSampledBvRel}{Status}{All}{}{Score}{0}%
\StoreBenchExecResult{Abc}{ImcTerminationBitVectorsSampledBvRel}{Status}{All}{}{Count}{32}%
\StoreBenchExecResult{Abc}{ImcTerminationBitVectorsSampledBvRel}{Status}{Correct}{}{Count}{24}%
\StoreBenchExecResult{Abc}{ImcTerminationBitVectorsSampledBvRel}{Status}{Correct}{True}{Count}{13}%
\StoreBenchExecResult{Abc}{ImcTerminationBitVectorsSampledBvRel}{Status}{Correct}{False}{Count}{11}%
\StoreBenchExecResult{Abc}{ImcTerminationBitVectorsSampledBvRel}{Status}{Wrong}{}{Count}{0}%
\StoreBenchExecResult{Abc}{ImcTerminationBitVectorsSampledBvRel}{Status}{Wrong}{True}{Count}{0}%
\StoreBenchExecResult{Abc}{ImcTerminationBitVectorsSampledBvRel}{Status}{Wrong}{False}{Count}{0}%
\providecommand\StoreBenchExecResult[7]{\expandafter\newcommand\csname#1#2#3#4#5#6\endcsname{#7}}%
\StoreBenchExecResult{Abc}{ImcTerminationMainControlFlowSampledBvRel}{Status}{All}{}{Score}{0}%
\StoreBenchExecResult{Abc}{ImcTerminationMainControlFlowSampledBvRel}{Status}{All}{}{Count}{236}%
\StoreBenchExecResult{Abc}{ImcTerminationMainControlFlowSampledBvRel}{Status}{Correct}{}{Count}{75}%
\StoreBenchExecResult{Abc}{ImcTerminationMainControlFlowSampledBvRel}{Status}{Correct}{True}{Count}{27}%
\StoreBenchExecResult{Abc}{ImcTerminationMainControlFlowSampledBvRel}{Status}{Correct}{False}{Count}{48}%
\StoreBenchExecResult{Abc}{ImcTerminationMainControlFlowSampledBvRel}{Status}{Wrong}{}{Count}{0}%
\StoreBenchExecResult{Abc}{ImcTerminationMainControlFlowSampledBvRel}{Status}{Wrong}{True}{Count}{0}%
\StoreBenchExecResult{Abc}{ImcTerminationMainControlFlowSampledBvRel}{Status}{Wrong}{False}{Count}{0}%
\providecommand\StoreBenchExecResult[7]{\expandafter\newcommand\csname#1#2#3#4#5#6\endcsname{#7}}%
\StoreBenchExecResult{Abc}{ImcTerminationOtherSampledBvRel}{Status}{All}{}{Score}{0}%
\StoreBenchExecResult{Abc}{ImcTerminationOtherSampledBvRel}{Status}{All}{}{Count}{973}%
\StoreBenchExecResult{Abc}{ImcTerminationOtherSampledBvRel}{Status}{Correct}{}{Count}{880}%
\StoreBenchExecResult{Abc}{ImcTerminationOtherSampledBvRel}{Status}{Correct}{True}{Count}{209}%
\StoreBenchExecResult{Abc}{ImcTerminationOtherSampledBvRel}{Status}{Correct}{False}{Count}{671}%
\StoreBenchExecResult{Abc}{ImcTerminationOtherSampledBvRel}{Status}{Wrong}{}{Count}{0}%
\StoreBenchExecResult{Abc}{ImcTerminationOtherSampledBvRel}{Status}{Wrong}{True}{Count}{0}%
\StoreBenchExecResult{Abc}{ImcTerminationOtherSampledBvRel}{Status}{Wrong}{False}{Count}{0}%
\edef\AbcImcReachSafetySampledBvRelStatusAllCount{\the\numexpr\AbcImcReachSafetyBitVectorsSampledBvRelStatusAllCount+\AbcImcReachSafetyCombinationsSampledBvRelStatusAllCount+\AbcImcReachSafetyControlFlowSampledBvRelStatusAllCount+\AbcImcReachSafetyECASampledBvRelStatusAllCount+\AbcImcReachSafetyFloatsSampledBvRelStatusAllCount+\AbcImcReachSafetyHardnessSampledBvRelStatusAllCount+\AbcImcReachSafetyHardwareSampledBvRelStatusAllCount+\AbcImcReachSafetyHeapSampledBvRelStatusAllCount+\AbcImcReachSafetyLoopsSampledBvRelStatusAllCount+\AbcImcReachSafetyProductLinesSampledBvRelStatusAllCount+\AbcImcReachSafetySequentializedSampledBvRelStatusAllCount+\AbcImcReachSafetyXCSPSampledBvRelStatusAllCount}
\edef\AbcImcReachSafetySampledBvRelStatusCorrectCount{\the\numexpr\AbcImcReachSafetyBitVectorsSampledBvRelStatusCorrectCount+\AbcImcReachSafetyCombinationsSampledBvRelStatusCorrectCount+\AbcImcReachSafetyControlFlowSampledBvRelStatusCorrectCount+\AbcImcReachSafetyECASampledBvRelStatusCorrectCount+\AbcImcReachSafetyFloatsSampledBvRelStatusCorrectCount+\AbcImcReachSafetyHardnessSampledBvRelStatusCorrectCount+\AbcImcReachSafetyHardwareSampledBvRelStatusCorrectCount+\AbcImcReachSafetyHeapSampledBvRelStatusCorrectCount+\AbcImcReachSafetyLoopsSampledBvRelStatusCorrectCount+\AbcImcReachSafetyProductLinesSampledBvRelStatusCorrectCount+\AbcImcReachSafetySequentializedSampledBvRelStatusCorrectCount+\AbcImcReachSafetyXCSPSampledBvRelStatusCorrectCount}
\edef\AbcImcReachSafetySampledBvRelStatusCorrectTrueCount{\the\numexpr\AbcImcReachSafetyBitVectorsSampledBvRelStatusCorrectTrueCount+\AbcImcReachSafetyCombinationsSampledBvRelStatusCorrectTrueCount+\AbcImcReachSafetyControlFlowSampledBvRelStatusCorrectTrueCount+\AbcImcReachSafetyECASampledBvRelStatusCorrectTrueCount+\AbcImcReachSafetyFloatsSampledBvRelStatusCorrectTrueCount+\AbcImcReachSafetyHardnessSampledBvRelStatusCorrectTrueCount+\AbcImcReachSafetyHardwareSampledBvRelStatusCorrectTrueCount+\AbcImcReachSafetyHeapSampledBvRelStatusCorrectTrueCount+\AbcImcReachSafetyLoopsSampledBvRelStatusCorrectTrueCount+\AbcImcReachSafetyProductLinesSampledBvRelStatusCorrectTrueCount+\AbcImcReachSafetySequentializedSampledBvRelStatusCorrectTrueCount+\AbcImcReachSafetyXCSPSampledBvRelStatusCorrectTrueCount}
\edef\AbcImcReachSafetySampledBvRelStatusCorrectFalseCount{\the\numexpr\AbcImcReachSafetyBitVectorsSampledBvRelStatusCorrectFalseCount+\AbcImcReachSafetyCombinationsSampledBvRelStatusCorrectFalseCount+\AbcImcReachSafetyControlFlowSampledBvRelStatusCorrectFalseCount+\AbcImcReachSafetyECASampledBvRelStatusCorrectFalseCount+\AbcImcReachSafetyFloatsSampledBvRelStatusCorrectFalseCount+\AbcImcReachSafetyHardnessSampledBvRelStatusCorrectFalseCount+\AbcImcReachSafetyHardwareSampledBvRelStatusCorrectFalseCount+\AbcImcReachSafetyHeapSampledBvRelStatusCorrectFalseCount+\AbcImcReachSafetyLoopsSampledBvRelStatusCorrectFalseCount+\AbcImcReachSafetyProductLinesSampledBvRelStatusCorrectFalseCount+\AbcImcReachSafetySequentializedSampledBvRelStatusCorrectFalseCount+\AbcImcReachSafetyXCSPSampledBvRelStatusCorrectFalseCount}
\edef\AbcImcReachSafetySampledBvRelStatusWrongCount{\the\numexpr\AbcImcReachSafetyBitVectorsSampledBvRelStatusWrongCount+\AbcImcReachSafetyCombinationsSampledBvRelStatusWrongCount+\AbcImcReachSafetyControlFlowSampledBvRelStatusWrongCount+\AbcImcReachSafetyECASampledBvRelStatusWrongCount+\AbcImcReachSafetyFloatsSampledBvRelStatusWrongCount+\AbcImcReachSafetyHardnessSampledBvRelStatusWrongCount+\AbcImcReachSafetyHardwareSampledBvRelStatusWrongCount+\AbcImcReachSafetyHeapSampledBvRelStatusWrongCount+\AbcImcReachSafetyLoopsSampledBvRelStatusWrongCount+\AbcImcReachSafetyProductLinesSampledBvRelStatusWrongCount+\AbcImcReachSafetySequentializedSampledBvRelStatusWrongCount+\AbcImcReachSafetyXCSPSampledBvRelStatusWrongCount}
\edef\AbcImcTerminationSampledBvRelStatusAllCount{\the\numexpr\AbcImcTerminationBitVectorsSampledBvRelStatusAllCount+\AbcImcTerminationMainControlFlowSampledBvRelStatusAllCount+\AbcImcTerminationOtherSampledBvRelStatusAllCount}
\edef\AbcImcTerminationSampledBvRelStatusCorrectCount{\the\numexpr\AbcImcTerminationBitVectorsSampledBvRelStatusCorrectCount+\AbcImcTerminationMainControlFlowSampledBvRelStatusCorrectCount+\AbcImcTerminationOtherSampledBvRelStatusCorrectCount}
\edef\AbcImcTerminationSampledBvRelStatusCorrectTrueCount{\the\numexpr\AbcImcTerminationBitVectorsSampledBvRelStatusCorrectTrueCount+\AbcImcTerminationMainControlFlowSampledBvRelStatusCorrectTrueCount+\AbcImcTerminationOtherSampledBvRelStatusCorrectTrueCount}
\edef\AbcImcTerminationSampledBvRelStatusCorrectFalseCount{\the\numexpr\AbcImcTerminationBitVectorsSampledBvRelStatusCorrectFalseCount+\AbcImcTerminationMainControlFlowSampledBvRelStatusCorrectFalseCount+\AbcImcTerminationOtherSampledBvRelStatusCorrectFalseCount}
\edef\AbcImcTerminationSampledBvRelStatusWrongCount{\the\numexpr\AbcImcTerminationBitVectorsSampledBvRelStatusWrongCount+\AbcImcTerminationMainControlFlowSampledBvRelStatusWrongCount+\AbcImcTerminationOtherSampledBvRelStatusWrongCount}
\providecommand\StoreBenchExecResult[7]{\expandafter\newcommand\csname#1#2#3#4#5#6\endcsname{#7}}%
\StoreBenchExecResult{Abc}{PdrReachSafetyBitVectorsSampledBvRel}{Status}{All}{}{Score}{0}%
\StoreBenchExecResult{Abc}{PdrReachSafetyBitVectorsSampledBvRel}{Status}{All}{}{Count}{47}%
\StoreBenchExecResult{Abc}{PdrReachSafetyBitVectorsSampledBvRel}{Status}{Correct}{}{Count}{40}%
\StoreBenchExecResult{Abc}{PdrReachSafetyBitVectorsSampledBvRel}{Status}{Correct}{True}{Count}{29}%
\StoreBenchExecResult{Abc}{PdrReachSafetyBitVectorsSampledBvRel}{Status}{Correct}{False}{Count}{11}%
\StoreBenchExecResult{Abc}{PdrReachSafetyBitVectorsSampledBvRel}{Status}{Wrong}{}{Count}{0}%
\StoreBenchExecResult{Abc}{PdrReachSafetyBitVectorsSampledBvRel}{Status}{Wrong}{True}{Count}{0}%
\StoreBenchExecResult{Abc}{PdrReachSafetyBitVectorsSampledBvRel}{Status}{Wrong}{False}{Count}{0}%
\providecommand\StoreBenchExecResult[7]{\expandafter\newcommand\csname#1#2#3#4#5#6\endcsname{#7}}%
\StoreBenchExecResult{Abc}{PdrReachSafetyCombinationsSampledBvRel}{Status}{All}{}{Score}{0}%
\StoreBenchExecResult{Abc}{PdrReachSafetyCombinationsSampledBvRel}{Status}{All}{}{Count}{110}%
\StoreBenchExecResult{Abc}{PdrReachSafetyCombinationsSampledBvRel}{Status}{Correct}{}{Count}{51}%
\StoreBenchExecResult{Abc}{PdrReachSafetyCombinationsSampledBvRel}{Status}{Correct}{True}{Count}{4}%
\StoreBenchExecResult{Abc}{PdrReachSafetyCombinationsSampledBvRel}{Status}{Correct}{False}{Count}{47}%
\StoreBenchExecResult{Abc}{PdrReachSafetyCombinationsSampledBvRel}{Status}{Wrong}{}{Count}{0}%
\StoreBenchExecResult{Abc}{PdrReachSafetyCombinationsSampledBvRel}{Status}{Wrong}{True}{Count}{0}%
\StoreBenchExecResult{Abc}{PdrReachSafetyCombinationsSampledBvRel}{Status}{Wrong}{False}{Count}{0}%
\providecommand\StoreBenchExecResult[7]{\expandafter\newcommand\csname#1#2#3#4#5#6\endcsname{#7}}%
\StoreBenchExecResult{Abc}{PdrReachSafetyControlFlowSampledBvRel}{Status}{All}{}{Score}{0}%
\StoreBenchExecResult{Abc}{PdrReachSafetyControlFlowSampledBvRel}{Status}{All}{}{Count}{29}%
\StoreBenchExecResult{Abc}{PdrReachSafetyControlFlowSampledBvRel}{Status}{Correct}{}{Count}{26}%
\StoreBenchExecResult{Abc}{PdrReachSafetyControlFlowSampledBvRel}{Status}{Correct}{True}{Count}{24}%
\StoreBenchExecResult{Abc}{PdrReachSafetyControlFlowSampledBvRel}{Status}{Correct}{False}{Count}{2}%
\StoreBenchExecResult{Abc}{PdrReachSafetyControlFlowSampledBvRel}{Status}{Wrong}{}{Count}{0}%
\StoreBenchExecResult{Abc}{PdrReachSafetyControlFlowSampledBvRel}{Status}{Wrong}{True}{Count}{0}%
\StoreBenchExecResult{Abc}{PdrReachSafetyControlFlowSampledBvRel}{Status}{Wrong}{False}{Count}{0}%
\providecommand\StoreBenchExecResult[7]{\expandafter\newcommand\csname#1#2#3#4#5#6\endcsname{#7}}%
\StoreBenchExecResult{Abc}{PdrReachSafetyECASampledBvRel}{Status}{All}{}{Score}{0}%
\StoreBenchExecResult{Abc}{PdrReachSafetyECASampledBvRel}{Status}{All}{}{Count}{150}%
\StoreBenchExecResult{Abc}{PdrReachSafetyECASampledBvRel}{Status}{Correct}{}{Count}{56}%
\StoreBenchExecResult{Abc}{PdrReachSafetyECASampledBvRel}{Status}{Correct}{True}{Count}{40}%
\StoreBenchExecResult{Abc}{PdrReachSafetyECASampledBvRel}{Status}{Correct}{False}{Count}{16}%
\StoreBenchExecResult{Abc}{PdrReachSafetyECASampledBvRel}{Status}{Wrong}{}{Count}{0}%
\StoreBenchExecResult{Abc}{PdrReachSafetyECASampledBvRel}{Status}{Wrong}{True}{Count}{0}%
\StoreBenchExecResult{Abc}{PdrReachSafetyECASampledBvRel}{Status}{Wrong}{False}{Count}{0}%
\providecommand\StoreBenchExecResult[7]{\expandafter\newcommand\csname#1#2#3#4#5#6\endcsname{#7}}%
\StoreBenchExecResult{Abc}{PdrReachSafetyFloatsSampledBvRel}{Status}{All}{}{Score}{0}%
\StoreBenchExecResult{Abc}{PdrReachSafetyFloatsSampledBvRel}{Status}{All}{}{Count}{10}%
\StoreBenchExecResult{Abc}{PdrReachSafetyFloatsSampledBvRel}{Status}{Correct}{}{Count}{10}%
\StoreBenchExecResult{Abc}{PdrReachSafetyFloatsSampledBvRel}{Status}{Correct}{True}{Count}{10}%
\StoreBenchExecResult{Abc}{PdrReachSafetyFloatsSampledBvRel}{Status}{Correct}{False}{Count}{0}%
\StoreBenchExecResult{Abc}{PdrReachSafetyFloatsSampledBvRel}{Status}{Wrong}{}{Count}{0}%
\StoreBenchExecResult{Abc}{PdrReachSafetyFloatsSampledBvRel}{Status}{Wrong}{True}{Count}{0}%
\StoreBenchExecResult{Abc}{PdrReachSafetyFloatsSampledBvRel}{Status}{Wrong}{False}{Count}{0}%
\providecommand\StoreBenchExecResult[7]{\expandafter\newcommand\csname#1#2#3#4#5#6\endcsname{#7}}%
\StoreBenchExecResult{Abc}{PdrReachSafetyHardnessSampledBvRel}{Status}{All}{}{Score}{0}%
\StoreBenchExecResult{Abc}{PdrReachSafetyHardnessSampledBvRel}{Status}{All}{}{Count}{132}%
\StoreBenchExecResult{Abc}{PdrReachSafetyHardnessSampledBvRel}{Status}{Correct}{}{Count}{132}%
\StoreBenchExecResult{Abc}{PdrReachSafetyHardnessSampledBvRel}{Status}{Correct}{True}{Count}{132}%
\StoreBenchExecResult{Abc}{PdrReachSafetyHardnessSampledBvRel}{Status}{Correct}{False}{Count}{0}%
\StoreBenchExecResult{Abc}{PdrReachSafetyHardnessSampledBvRel}{Status}{Wrong}{}{Count}{0}%
\StoreBenchExecResult{Abc}{PdrReachSafetyHardnessSampledBvRel}{Status}{Wrong}{True}{Count}{0}%
\StoreBenchExecResult{Abc}{PdrReachSafetyHardnessSampledBvRel}{Status}{Wrong}{False}{Count}{0}%
\providecommand\StoreBenchExecResult[7]{\expandafter\newcommand\csname#1#2#3#4#5#6\endcsname{#7}}%
\StoreBenchExecResult{Abc}{PdrReachSafetyHardwareSampledBvRel}{Status}{All}{}{Score}{0}%
\StoreBenchExecResult{Abc}{PdrReachSafetyHardwareSampledBvRel}{Status}{All}{}{Count}{148}%
\StoreBenchExecResult{Abc}{PdrReachSafetyHardwareSampledBvRel}{Status}{Correct}{}{Count}{31}%
\StoreBenchExecResult{Abc}{PdrReachSafetyHardwareSampledBvRel}{Status}{Correct}{True}{Count}{17}%
\StoreBenchExecResult{Abc}{PdrReachSafetyHardwareSampledBvRel}{Status}{Correct}{False}{Count}{14}%
\StoreBenchExecResult{Abc}{PdrReachSafetyHardwareSampledBvRel}{Status}{Wrong}{}{Count}{0}%
\StoreBenchExecResult{Abc}{PdrReachSafetyHardwareSampledBvRel}{Status}{Wrong}{True}{Count}{0}%
\StoreBenchExecResult{Abc}{PdrReachSafetyHardwareSampledBvRel}{Status}{Wrong}{False}{Count}{0}%
\providecommand\StoreBenchExecResult[7]{\expandafter\newcommand\csname#1#2#3#4#5#6\endcsname{#7}}%
\StoreBenchExecResult{Abc}{PdrReachSafetyHeapSampledBvRel}{Status}{All}{}{Score}{0}%
\StoreBenchExecResult{Abc}{PdrReachSafetyHeapSampledBvRel}{Status}{All}{}{Count}{6}%
\StoreBenchExecResult{Abc}{PdrReachSafetyHeapSampledBvRel}{Status}{Correct}{}{Count}{6}%
\StoreBenchExecResult{Abc}{PdrReachSafetyHeapSampledBvRel}{Status}{Correct}{True}{Count}{4}%
\StoreBenchExecResult{Abc}{PdrReachSafetyHeapSampledBvRel}{Status}{Correct}{False}{Count}{2}%
\StoreBenchExecResult{Abc}{PdrReachSafetyHeapSampledBvRel}{Status}{Wrong}{}{Count}{0}%
\StoreBenchExecResult{Abc}{PdrReachSafetyHeapSampledBvRel}{Status}{Wrong}{True}{Count}{0}%
\StoreBenchExecResult{Abc}{PdrReachSafetyHeapSampledBvRel}{Status}{Wrong}{False}{Count}{0}%
\providecommand\StoreBenchExecResult[7]{\expandafter\newcommand\csname#1#2#3#4#5#6\endcsname{#7}}%
\StoreBenchExecResult{Abc}{PdrReachSafetyLoopsSampledBvRel}{Status}{All}{}{Score}{0}%
\StoreBenchExecResult{Abc}{PdrReachSafetyLoopsSampledBvRel}{Status}{All}{}{Count}{137}%
\StoreBenchExecResult{Abc}{PdrReachSafetyLoopsSampledBvRel}{Status}{Correct}{}{Count}{57}%
\StoreBenchExecResult{Abc}{PdrReachSafetyLoopsSampledBvRel}{Status}{Correct}{True}{Count}{26}%
\StoreBenchExecResult{Abc}{PdrReachSafetyLoopsSampledBvRel}{Status}{Correct}{False}{Count}{31}%
\StoreBenchExecResult{Abc}{PdrReachSafetyLoopsSampledBvRel}{Status}{Wrong}{}{Count}{0}%
\StoreBenchExecResult{Abc}{PdrReachSafetyLoopsSampledBvRel}{Status}{Wrong}{True}{Count}{0}%
\StoreBenchExecResult{Abc}{PdrReachSafetyLoopsSampledBvRel}{Status}{Wrong}{False}{Count}{0}%
\providecommand\StoreBenchExecResult[7]{\expandafter\newcommand\csname#1#2#3#4#5#6\endcsname{#7}}%
\StoreBenchExecResult{Abc}{PdrReachSafetyProductLinesSampledBvRel}{Status}{All}{}{Score}{0}%
\StoreBenchExecResult{Abc}{PdrReachSafetyProductLinesSampledBvRel}{Status}{All}{}{Count}{150}%
\StoreBenchExecResult{Abc}{PdrReachSafetyProductLinesSampledBvRel}{Status}{Correct}{}{Count}{138}%
\StoreBenchExecResult{Abc}{PdrReachSafetyProductLinesSampledBvRel}{Status}{Correct}{True}{Count}{68}%
\StoreBenchExecResult{Abc}{PdrReachSafetyProductLinesSampledBvRel}{Status}{Correct}{False}{Count}{70}%
\StoreBenchExecResult{Abc}{PdrReachSafetyProductLinesSampledBvRel}{Status}{Wrong}{}{Count}{0}%
\StoreBenchExecResult{Abc}{PdrReachSafetyProductLinesSampledBvRel}{Status}{Wrong}{True}{Count}{0}%
\StoreBenchExecResult{Abc}{PdrReachSafetyProductLinesSampledBvRel}{Status}{Wrong}{False}{Count}{0}%
\providecommand\StoreBenchExecResult[7]{\expandafter\newcommand\csname#1#2#3#4#5#6\endcsname{#7}}%
\StoreBenchExecResult{Abc}{PdrReachSafetySequentializedSampledBvRel}{Status}{All}{}{Score}{0}%
\StoreBenchExecResult{Abc}{PdrReachSafetySequentializedSampledBvRel}{Status}{All}{}{Count}{150}%
\StoreBenchExecResult{Abc}{PdrReachSafetySequentializedSampledBvRel}{Status}{Correct}{}{Count}{56}%
\StoreBenchExecResult{Abc}{PdrReachSafetySequentializedSampledBvRel}{Status}{Correct}{True}{Count}{7}%
\StoreBenchExecResult{Abc}{PdrReachSafetySequentializedSampledBvRel}{Status}{Correct}{False}{Count}{49}%
\StoreBenchExecResult{Abc}{PdrReachSafetySequentializedSampledBvRel}{Status}{Wrong}{}{Count}{0}%
\StoreBenchExecResult{Abc}{PdrReachSafetySequentializedSampledBvRel}{Status}{Wrong}{True}{Count}{0}%
\StoreBenchExecResult{Abc}{PdrReachSafetySequentializedSampledBvRel}{Status}{Wrong}{False}{Count}{0}%
\providecommand\StoreBenchExecResult[7]{\expandafter\newcommand\csname#1#2#3#4#5#6\endcsname{#7}}%
\StoreBenchExecResult{Abc}{PdrReachSafetyXCSPSampledBvRel}{Status}{All}{}{Score}{0}%
\StoreBenchExecResult{Abc}{PdrReachSafetyXCSPSampledBvRel}{Status}{All}{}{Count}{98}%
\StoreBenchExecResult{Abc}{PdrReachSafetyXCSPSampledBvRel}{Status}{Correct}{}{Count}{93}%
\StoreBenchExecResult{Abc}{PdrReachSafetyXCSPSampledBvRel}{Status}{Correct}{True}{Count}{51}%
\StoreBenchExecResult{Abc}{PdrReachSafetyXCSPSampledBvRel}{Status}{Correct}{False}{Count}{42}%
\StoreBenchExecResult{Abc}{PdrReachSafetyXCSPSampledBvRel}{Status}{Wrong}{}{Count}{0}%
\StoreBenchExecResult{Abc}{PdrReachSafetyXCSPSampledBvRel}{Status}{Wrong}{True}{Count}{0}%
\StoreBenchExecResult{Abc}{PdrReachSafetyXCSPSampledBvRel}{Status}{Wrong}{False}{Count}{0}%
\providecommand\StoreBenchExecResult[7]{\expandafter\newcommand\csname#1#2#3#4#5#6\endcsname{#7}}%
\StoreBenchExecResult{Abc}{PdrTerminationBitVectorsSampledBvRel}{Status}{All}{}{Score}{0}%
\StoreBenchExecResult{Abc}{PdrTerminationBitVectorsSampledBvRel}{Status}{All}{}{Count}{32}%
\StoreBenchExecResult{Abc}{PdrTerminationBitVectorsSampledBvRel}{Status}{Correct}{}{Count}{20}%
\StoreBenchExecResult{Abc}{PdrTerminationBitVectorsSampledBvRel}{Status}{Correct}{True}{Count}{9}%
\StoreBenchExecResult{Abc}{PdrTerminationBitVectorsSampledBvRel}{Status}{Correct}{False}{Count}{11}%
\StoreBenchExecResult{Abc}{PdrTerminationBitVectorsSampledBvRel}{Status}{Wrong}{}{Count}{0}%
\StoreBenchExecResult{Abc}{PdrTerminationBitVectorsSampledBvRel}{Status}{Wrong}{True}{Count}{0}%
\StoreBenchExecResult{Abc}{PdrTerminationBitVectorsSampledBvRel}{Status}{Wrong}{False}{Count}{0}%
\providecommand\StoreBenchExecResult[7]{\expandafter\newcommand\csname#1#2#3#4#5#6\endcsname{#7}}%
\StoreBenchExecResult{Abc}{PdrTerminationMainControlFlowSampledBvRel}{Status}{All}{}{Score}{0}%
\StoreBenchExecResult{Abc}{PdrTerminationMainControlFlowSampledBvRel}{Status}{All}{}{Count}{236}%
\StoreBenchExecResult{Abc}{PdrTerminationMainControlFlowSampledBvRel}{Status}{Correct}{}{Count}{51}%
\StoreBenchExecResult{Abc}{PdrTerminationMainControlFlowSampledBvRel}{Status}{Correct}{True}{Count}{6}%
\StoreBenchExecResult{Abc}{PdrTerminationMainControlFlowSampledBvRel}{Status}{Correct}{False}{Count}{45}%
\StoreBenchExecResult{Abc}{PdrTerminationMainControlFlowSampledBvRel}{Status}{Wrong}{}{Count}{0}%
\StoreBenchExecResult{Abc}{PdrTerminationMainControlFlowSampledBvRel}{Status}{Wrong}{True}{Count}{0}%
\StoreBenchExecResult{Abc}{PdrTerminationMainControlFlowSampledBvRel}{Status}{Wrong}{False}{Count}{0}%
\providecommand\StoreBenchExecResult[7]{\expandafter\newcommand\csname#1#2#3#4#5#6\endcsname{#7}}%
\StoreBenchExecResult{Abc}{PdrTerminationOtherSampledBvRel}{Status}{All}{}{Score}{0}%
\StoreBenchExecResult{Abc}{PdrTerminationOtherSampledBvRel}{Status}{All}{}{Count}{973}%
\StoreBenchExecResult{Abc}{PdrTerminationOtherSampledBvRel}{Status}{Correct}{}{Count}{540}%
\StoreBenchExecResult{Abc}{PdrTerminationOtherSampledBvRel}{Status}{Correct}{True}{Count}{50}%
\StoreBenchExecResult{Abc}{PdrTerminationOtherSampledBvRel}{Status}{Correct}{False}{Count}{490}%
\StoreBenchExecResult{Abc}{PdrTerminationOtherSampledBvRel}{Status}{Wrong}{}{Count}{0}%
\StoreBenchExecResult{Abc}{PdrTerminationOtherSampledBvRel}{Status}{Wrong}{True}{Count}{0}%
\StoreBenchExecResult{Abc}{PdrTerminationOtherSampledBvRel}{Status}{Wrong}{False}{Count}{0}%
\edef\AbcPdrReachSafetySampledBvRelStatusAllCount{\the\numexpr\AbcPdrReachSafetyBitVectorsSampledBvRelStatusAllCount+\AbcPdrReachSafetyCombinationsSampledBvRelStatusAllCount+\AbcPdrReachSafetyControlFlowSampledBvRelStatusAllCount+\AbcPdrReachSafetyECASampledBvRelStatusAllCount+\AbcPdrReachSafetyFloatsSampledBvRelStatusAllCount+\AbcPdrReachSafetyHardnessSampledBvRelStatusAllCount+\AbcPdrReachSafetyHardwareSampledBvRelStatusAllCount+\AbcPdrReachSafetyHeapSampledBvRelStatusAllCount+\AbcPdrReachSafetyLoopsSampledBvRelStatusAllCount+\AbcPdrReachSafetyProductLinesSampledBvRelStatusAllCount+\AbcPdrReachSafetySequentializedSampledBvRelStatusAllCount+\AbcPdrReachSafetyXCSPSampledBvRelStatusAllCount}
\edef\AbcPdrReachSafetySampledBvRelStatusCorrectCount{\the\numexpr\AbcPdrReachSafetyBitVectorsSampledBvRelStatusCorrectCount+\AbcPdrReachSafetyCombinationsSampledBvRelStatusCorrectCount+\AbcPdrReachSafetyControlFlowSampledBvRelStatusCorrectCount+\AbcPdrReachSafetyECASampledBvRelStatusCorrectCount+\AbcPdrReachSafetyFloatsSampledBvRelStatusCorrectCount+\AbcPdrReachSafetyHardnessSampledBvRelStatusCorrectCount+\AbcPdrReachSafetyHardwareSampledBvRelStatusCorrectCount+\AbcPdrReachSafetyHeapSampledBvRelStatusCorrectCount+\AbcPdrReachSafetyLoopsSampledBvRelStatusCorrectCount+\AbcPdrReachSafetyProductLinesSampledBvRelStatusCorrectCount+\AbcPdrReachSafetySequentializedSampledBvRelStatusCorrectCount+\AbcPdrReachSafetyXCSPSampledBvRelStatusCorrectCount}
\edef\AbcPdrReachSafetySampledBvRelStatusCorrectTrueCount{\the\numexpr\AbcPdrReachSafetyBitVectorsSampledBvRelStatusCorrectTrueCount+\AbcPdrReachSafetyCombinationsSampledBvRelStatusCorrectTrueCount+\AbcPdrReachSafetyControlFlowSampledBvRelStatusCorrectTrueCount+\AbcPdrReachSafetyECASampledBvRelStatusCorrectTrueCount+\AbcPdrReachSafetyFloatsSampledBvRelStatusCorrectTrueCount+\AbcPdrReachSafetyHardnessSampledBvRelStatusCorrectTrueCount+\AbcPdrReachSafetyHardwareSampledBvRelStatusCorrectTrueCount+\AbcPdrReachSafetyHeapSampledBvRelStatusCorrectTrueCount+\AbcPdrReachSafetyLoopsSampledBvRelStatusCorrectTrueCount+\AbcPdrReachSafetyProductLinesSampledBvRelStatusCorrectTrueCount+\AbcPdrReachSafetySequentializedSampledBvRelStatusCorrectTrueCount+\AbcPdrReachSafetyXCSPSampledBvRelStatusCorrectTrueCount}
\edef\AbcPdrReachSafetySampledBvRelStatusCorrectFalseCount{\the\numexpr\AbcPdrReachSafetyBitVectorsSampledBvRelStatusCorrectFalseCount+\AbcPdrReachSafetyCombinationsSampledBvRelStatusCorrectFalseCount+\AbcPdrReachSafetyControlFlowSampledBvRelStatusCorrectFalseCount+\AbcPdrReachSafetyECASampledBvRelStatusCorrectFalseCount+\AbcPdrReachSafetyFloatsSampledBvRelStatusCorrectFalseCount+\AbcPdrReachSafetyHardnessSampledBvRelStatusCorrectFalseCount+\AbcPdrReachSafetyHardwareSampledBvRelStatusCorrectFalseCount+\AbcPdrReachSafetyHeapSampledBvRelStatusCorrectFalseCount+\AbcPdrReachSafetyLoopsSampledBvRelStatusCorrectFalseCount+\AbcPdrReachSafetyProductLinesSampledBvRelStatusCorrectFalseCount+\AbcPdrReachSafetySequentializedSampledBvRelStatusCorrectFalseCount+\AbcPdrReachSafetyXCSPSampledBvRelStatusCorrectFalseCount}
\edef\AbcPdrReachSafetySampledBvRelStatusWrongCount{\the\numexpr\AbcPdrReachSafetyBitVectorsSampledBvRelStatusWrongCount+\AbcPdrReachSafetyCombinationsSampledBvRelStatusWrongCount+\AbcPdrReachSafetyControlFlowSampledBvRelStatusWrongCount+\AbcPdrReachSafetyECASampledBvRelStatusWrongCount+\AbcPdrReachSafetyFloatsSampledBvRelStatusWrongCount+\AbcPdrReachSafetyHardnessSampledBvRelStatusWrongCount+\AbcPdrReachSafetyHardwareSampledBvRelStatusWrongCount+\AbcPdrReachSafetyHeapSampledBvRelStatusWrongCount+\AbcPdrReachSafetyLoopsSampledBvRelStatusWrongCount+\AbcPdrReachSafetyProductLinesSampledBvRelStatusWrongCount+\AbcPdrReachSafetySequentializedSampledBvRelStatusWrongCount+\AbcPdrReachSafetyXCSPSampledBvRelStatusWrongCount}
\edef\AbcPdrTerminationSampledBvRelStatusAllCount{\the\numexpr\AbcPdrTerminationBitVectorsSampledBvRelStatusAllCount+\AbcPdrTerminationMainControlFlowSampledBvRelStatusAllCount+\AbcPdrTerminationOtherSampledBvRelStatusAllCount}
\edef\AbcPdrTerminationSampledBvRelStatusCorrectCount{\the\numexpr\AbcPdrTerminationBitVectorsSampledBvRelStatusCorrectCount+\AbcPdrTerminationMainControlFlowSampledBvRelStatusCorrectCount+\AbcPdrTerminationOtherSampledBvRelStatusCorrectCount}
\edef\AbcPdrTerminationSampledBvRelStatusCorrectTrueCount{\the\numexpr\AbcPdrTerminationBitVectorsSampledBvRelStatusCorrectTrueCount+\AbcPdrTerminationMainControlFlowSampledBvRelStatusCorrectTrueCount+\AbcPdrTerminationOtherSampledBvRelStatusCorrectTrueCount}
\edef\AbcPdrTerminationSampledBvRelStatusCorrectFalseCount{\the\numexpr\AbcPdrTerminationBitVectorsSampledBvRelStatusCorrectFalseCount+\AbcPdrTerminationMainControlFlowSampledBvRelStatusCorrectFalseCount+\AbcPdrTerminationOtherSampledBvRelStatusCorrectFalseCount}
\edef\AbcPdrTerminationSampledBvRelStatusWrongCount{\the\numexpr\AbcPdrTerminationBitVectorsSampledBvRelStatusWrongCount+\AbcPdrTerminationMainControlFlowSampledBvRelStatusWrongCount+\AbcPdrTerminationOtherSampledBvRelStatusWrongCount}
\providecommand\StoreBenchExecResult[7]{\expandafter\newcommand\csname#1#2#3#4#5#6\endcsname{#7}}%
\StoreBenchExecResult{Avr}{IcIIIsaReachSafetyArraysSampledRel}{Status}{All}{}{Score}{0}%
\StoreBenchExecResult{Avr}{IcIIIsaReachSafetyArraysSampledRel}{Status}{All}{}{Count}{150}%
\StoreBenchExecResult{Avr}{IcIIIsaReachSafetyArraysSampledRel}{Status}{Correct}{}{Count}{43}%
\StoreBenchExecResult{Avr}{IcIIIsaReachSafetyArraysSampledRel}{Status}{Correct}{True}{Count}{0}%
\StoreBenchExecResult{Avr}{IcIIIsaReachSafetyArraysSampledRel}{Status}{Correct}{False}{Count}{43}%
\StoreBenchExecResult{Avr}{IcIIIsaReachSafetyArraysSampledRel}{Status}{Wrong}{}{Count}{0}%
\StoreBenchExecResult{Avr}{IcIIIsaReachSafetyArraysSampledRel}{Status}{Wrong}{True}{Count}{0}%
\StoreBenchExecResult{Avr}{IcIIIsaReachSafetyArraysSampledRel}{Status}{Wrong}{False}{Count}{0}%
\providecommand\StoreBenchExecResult[7]{\expandafter\newcommand\csname#1#2#3#4#5#6\endcsname{#7}}%
\StoreBenchExecResult{Avr}{IcIIIsaReachSafetyBitVectorsSampledRel}{Status}{All}{}{Score}{0}%
\StoreBenchExecResult{Avr}{IcIIIsaReachSafetyBitVectorsSampledRel}{Status}{All}{}{Count}{48}%
\StoreBenchExecResult{Avr}{IcIIIsaReachSafetyBitVectorsSampledRel}{Status}{Correct}{}{Count}{32}%
\StoreBenchExecResult{Avr}{IcIIIsaReachSafetyBitVectorsSampledRel}{Status}{Correct}{True}{Count}{21}%
\StoreBenchExecResult{Avr}{IcIIIsaReachSafetyBitVectorsSampledRel}{Status}{Correct}{False}{Count}{11}%
\StoreBenchExecResult{Avr}{IcIIIsaReachSafetyBitVectorsSampledRel}{Status}{Wrong}{}{Count}{0}%
\StoreBenchExecResult{Avr}{IcIIIsaReachSafetyBitVectorsSampledRel}{Status}{Wrong}{True}{Count}{0}%
\StoreBenchExecResult{Avr}{IcIIIsaReachSafetyBitVectorsSampledRel}{Status}{Wrong}{False}{Count}{0}%
\providecommand\StoreBenchExecResult[7]{\expandafter\newcommand\csname#1#2#3#4#5#6\endcsname{#7}}%
\StoreBenchExecResult{Avr}{IcIIIsaReachSafetyCombinationsSampledRel}{Status}{All}{}{Score}{0}%
\StoreBenchExecResult{Avr}{IcIIIsaReachSafetyCombinationsSampledRel}{Status}{All}{}{Count}{110}%
\StoreBenchExecResult{Avr}{IcIIIsaReachSafetyCombinationsSampledRel}{Status}{Correct}{}{Count}{79}%
\StoreBenchExecResult{Avr}{IcIIIsaReachSafetyCombinationsSampledRel}{Status}{Correct}{True}{Count}{8}%
\StoreBenchExecResult{Avr}{IcIIIsaReachSafetyCombinationsSampledRel}{Status}{Correct}{False}{Count}{71}%
\StoreBenchExecResult{Avr}{IcIIIsaReachSafetyCombinationsSampledRel}{Status}{Wrong}{}{Count}{0}%
\StoreBenchExecResult{Avr}{IcIIIsaReachSafetyCombinationsSampledRel}{Status}{Wrong}{True}{Count}{0}%
\StoreBenchExecResult{Avr}{IcIIIsaReachSafetyCombinationsSampledRel}{Status}{Wrong}{False}{Count}{0}%
\providecommand\StoreBenchExecResult[7]{\expandafter\newcommand\csname#1#2#3#4#5#6\endcsname{#7}}%
\StoreBenchExecResult{Avr}{IcIIIsaReachSafetyControlFlowSampledRel}{Status}{All}{}{Score}{0}%
\StoreBenchExecResult{Avr}{IcIIIsaReachSafetyControlFlowSampledRel}{Status}{All}{}{Count}{33}%
\StoreBenchExecResult{Avr}{IcIIIsaReachSafetyControlFlowSampledRel}{Status}{Correct}{}{Count}{27}%
\StoreBenchExecResult{Avr}{IcIIIsaReachSafetyControlFlowSampledRel}{Status}{Correct}{True}{Count}{25}%
\StoreBenchExecResult{Avr}{IcIIIsaReachSafetyControlFlowSampledRel}{Status}{Correct}{False}{Count}{2}%
\StoreBenchExecResult{Avr}{IcIIIsaReachSafetyControlFlowSampledRel}{Status}{Wrong}{}{Count}{0}%
\StoreBenchExecResult{Avr}{IcIIIsaReachSafetyControlFlowSampledRel}{Status}{Wrong}{True}{Count}{0}%
\StoreBenchExecResult{Avr}{IcIIIsaReachSafetyControlFlowSampledRel}{Status}{Wrong}{False}{Count}{0}%
\providecommand\StoreBenchExecResult[7]{\expandafter\newcommand\csname#1#2#3#4#5#6\endcsname{#7}}%
\StoreBenchExecResult{Avr}{IcIIIsaReachSafetyECASampledRel}{Status}{All}{}{Score}{0}%
\StoreBenchExecResult{Avr}{IcIIIsaReachSafetyECASampledRel}{Status}{All}{}{Count}{150}%
\StoreBenchExecResult{Avr}{IcIIIsaReachSafetyECASampledRel}{Status}{Correct}{}{Count}{79}%
\StoreBenchExecResult{Avr}{IcIIIsaReachSafetyECASampledRel}{Status}{Correct}{True}{Count}{48}%
\StoreBenchExecResult{Avr}{IcIIIsaReachSafetyECASampledRel}{Status}{Correct}{False}{Count}{31}%
\StoreBenchExecResult{Avr}{IcIIIsaReachSafetyECASampledRel}{Status}{Wrong}{}{Count}{0}%
\StoreBenchExecResult{Avr}{IcIIIsaReachSafetyECASampledRel}{Status}{Wrong}{True}{Count}{0}%
\StoreBenchExecResult{Avr}{IcIIIsaReachSafetyECASampledRel}{Status}{Wrong}{False}{Count}{0}%
\providecommand\StoreBenchExecResult[7]{\expandafter\newcommand\csname#1#2#3#4#5#6\endcsname{#7}}%
\StoreBenchExecResult{Avr}{IcIIIsaReachSafetyFloatsSampledRel}{Status}{All}{}{Score}{0}%
\StoreBenchExecResult{Avr}{IcIIIsaReachSafetyFloatsSampledRel}{Status}{All}{}{Count}{10}%
\StoreBenchExecResult{Avr}{IcIIIsaReachSafetyFloatsSampledRel}{Status}{Correct}{}{Count}{10}%
\StoreBenchExecResult{Avr}{IcIIIsaReachSafetyFloatsSampledRel}{Status}{Correct}{True}{Count}{10}%
\StoreBenchExecResult{Avr}{IcIIIsaReachSafetyFloatsSampledRel}{Status}{Correct}{False}{Count}{0}%
\StoreBenchExecResult{Avr}{IcIIIsaReachSafetyFloatsSampledRel}{Status}{Wrong}{}{Count}{0}%
\StoreBenchExecResult{Avr}{IcIIIsaReachSafetyFloatsSampledRel}{Status}{Wrong}{True}{Count}{0}%
\StoreBenchExecResult{Avr}{IcIIIsaReachSafetyFloatsSampledRel}{Status}{Wrong}{False}{Count}{0}%
\providecommand\StoreBenchExecResult[7]{\expandafter\newcommand\csname#1#2#3#4#5#6\endcsname{#7}}%
\StoreBenchExecResult{Avr}{IcIIIsaReachSafetyHardnessSampledRel}{Status}{All}{}{Score}{0}%
\StoreBenchExecResult{Avr}{IcIIIsaReachSafetyHardnessSampledRel}{Status}{All}{}{Count}{150}%
\StoreBenchExecResult{Avr}{IcIIIsaReachSafetyHardnessSampledRel}{Status}{Correct}{}{Count}{147}%
\StoreBenchExecResult{Avr}{IcIIIsaReachSafetyHardnessSampledRel}{Status}{Correct}{True}{Count}{147}%
\StoreBenchExecResult{Avr}{IcIIIsaReachSafetyHardnessSampledRel}{Status}{Correct}{False}{Count}{0}%
\StoreBenchExecResult{Avr}{IcIIIsaReachSafetyHardnessSampledRel}{Status}{Wrong}{}{Count}{0}%
\StoreBenchExecResult{Avr}{IcIIIsaReachSafetyHardnessSampledRel}{Status}{Wrong}{True}{Count}{0}%
\StoreBenchExecResult{Avr}{IcIIIsaReachSafetyHardnessSampledRel}{Status}{Wrong}{False}{Count}{0}%
\providecommand\StoreBenchExecResult[7]{\expandafter\newcommand\csname#1#2#3#4#5#6\endcsname{#7}}%
\StoreBenchExecResult{Avr}{IcIIIsaReachSafetyHardwareSampledRel}{Status}{All}{}{Score}{0}%
\StoreBenchExecResult{Avr}{IcIIIsaReachSafetyHardwareSampledRel}{Status}{All}{}{Count}{150}%
\StoreBenchExecResult{Avr}{IcIIIsaReachSafetyHardwareSampledRel}{Status}{Correct}{}{Count}{24}%
\StoreBenchExecResult{Avr}{IcIIIsaReachSafetyHardwareSampledRel}{Status}{Correct}{True}{Count}{9}%
\StoreBenchExecResult{Avr}{IcIIIsaReachSafetyHardwareSampledRel}{Status}{Correct}{False}{Count}{15}%
\StoreBenchExecResult{Avr}{IcIIIsaReachSafetyHardwareSampledRel}{Status}{Wrong}{}{Count}{0}%
\StoreBenchExecResult{Avr}{IcIIIsaReachSafetyHardwareSampledRel}{Status}{Wrong}{True}{Count}{0}%
\StoreBenchExecResult{Avr}{IcIIIsaReachSafetyHardwareSampledRel}{Status}{Wrong}{False}{Count}{0}%
\providecommand\StoreBenchExecResult[7]{\expandafter\newcommand\csname#1#2#3#4#5#6\endcsname{#7}}%
\StoreBenchExecResult{Avr}{IcIIIsaReachSafetyHeapSampledRel}{Status}{All}{}{Score}{0}%
\StoreBenchExecResult{Avr}{IcIIIsaReachSafetyHeapSampledRel}{Status}{All}{}{Count}{53}%
\StoreBenchExecResult{Avr}{IcIIIsaReachSafetyHeapSampledRel}{Status}{Correct}{}{Count}{46}%
\StoreBenchExecResult{Avr}{IcIIIsaReachSafetyHeapSampledRel}{Status}{Correct}{True}{Count}{32}%
\StoreBenchExecResult{Avr}{IcIIIsaReachSafetyHeapSampledRel}{Status}{Correct}{False}{Count}{14}%
\StoreBenchExecResult{Avr}{IcIIIsaReachSafetyHeapSampledRel}{Status}{Wrong}{}{Count}{0}%
\StoreBenchExecResult{Avr}{IcIIIsaReachSafetyHeapSampledRel}{Status}{Wrong}{True}{Count}{0}%
\StoreBenchExecResult{Avr}{IcIIIsaReachSafetyHeapSampledRel}{Status}{Wrong}{False}{Count}{0}%
\providecommand\StoreBenchExecResult[7]{\expandafter\newcommand\csname#1#2#3#4#5#6\endcsname{#7}}%
\StoreBenchExecResult{Avr}{IcIIIsaReachSafetyLoopsSampledRel}{Status}{All}{}{Score}{0}%
\StoreBenchExecResult{Avr}{IcIIIsaReachSafetyLoopsSampledRel}{Status}{All}{}{Count}{150}%
\StoreBenchExecResult{Avr}{IcIIIsaReachSafetyLoopsSampledRel}{Status}{Correct}{}{Count}{44}%
\StoreBenchExecResult{Avr}{IcIIIsaReachSafetyLoopsSampledRel}{Status}{Correct}{True}{Count}{26}%
\StoreBenchExecResult{Avr}{IcIIIsaReachSafetyLoopsSampledRel}{Status}{Correct}{False}{Count}{18}%
\StoreBenchExecResult{Avr}{IcIIIsaReachSafetyLoopsSampledRel}{Status}{Wrong}{}{Count}{0}%
\StoreBenchExecResult{Avr}{IcIIIsaReachSafetyLoopsSampledRel}{Status}{Wrong}{True}{Count}{0}%
\StoreBenchExecResult{Avr}{IcIIIsaReachSafetyLoopsSampledRel}{Status}{Wrong}{False}{Count}{0}%
\providecommand\StoreBenchExecResult[7]{\expandafter\newcommand\csname#1#2#3#4#5#6\endcsname{#7}}%
\StoreBenchExecResult{Avr}{IcIIIsaReachSafetyProductLinesSampledRel}{Status}{All}{}{Score}{0}%
\StoreBenchExecResult{Avr}{IcIIIsaReachSafetyProductLinesSampledRel}{Status}{All}{}{Count}{150}%
\StoreBenchExecResult{Avr}{IcIIIsaReachSafetyProductLinesSampledRel}{Status}{Correct}{}{Count}{149}%
\StoreBenchExecResult{Avr}{IcIIIsaReachSafetyProductLinesSampledRel}{Status}{Correct}{True}{Count}{74}%
\StoreBenchExecResult{Avr}{IcIIIsaReachSafetyProductLinesSampledRel}{Status}{Correct}{False}{Count}{75}%
\StoreBenchExecResult{Avr}{IcIIIsaReachSafetyProductLinesSampledRel}{Status}{Wrong}{}{Count}{0}%
\StoreBenchExecResult{Avr}{IcIIIsaReachSafetyProductLinesSampledRel}{Status}{Wrong}{True}{Count}{0}%
\StoreBenchExecResult{Avr}{IcIIIsaReachSafetyProductLinesSampledRel}{Status}{Wrong}{False}{Count}{0}%
\providecommand\StoreBenchExecResult[7]{\expandafter\newcommand\csname#1#2#3#4#5#6\endcsname{#7}}%
\StoreBenchExecResult{Avr}{IcIIIsaReachSafetySequentializedSampledRel}{Status}{All}{}{Score}{0}%
\StoreBenchExecResult{Avr}{IcIIIsaReachSafetySequentializedSampledRel}{Status}{All}{}{Count}{150}%
\StoreBenchExecResult{Avr}{IcIIIsaReachSafetySequentializedSampledRel}{Status}{Correct}{}{Count}{88}%
\StoreBenchExecResult{Avr}{IcIIIsaReachSafetySequentializedSampledRel}{Status}{Correct}{True}{Count}{17}%
\StoreBenchExecResult{Avr}{IcIIIsaReachSafetySequentializedSampledRel}{Status}{Correct}{False}{Count}{71}%
\StoreBenchExecResult{Avr}{IcIIIsaReachSafetySequentializedSampledRel}{Status}{Wrong}{}{Count}{0}%
\StoreBenchExecResult{Avr}{IcIIIsaReachSafetySequentializedSampledRel}{Status}{Wrong}{True}{Count}{0}%
\StoreBenchExecResult{Avr}{IcIIIsaReachSafetySequentializedSampledRel}{Status}{Wrong}{False}{Count}{0}%
\providecommand\StoreBenchExecResult[7]{\expandafter\newcommand\csname#1#2#3#4#5#6\endcsname{#7}}%
\StoreBenchExecResult{Avr}{IcIIIsaReachSafetyXCSPSampledRel}{Status}{All}{}{Score}{0}%
\StoreBenchExecResult{Avr}{IcIIIsaReachSafetyXCSPSampledRel}{Status}{All}{}{Count}{98}%
\StoreBenchExecResult{Avr}{IcIIIsaReachSafetyXCSPSampledRel}{Status}{Correct}{}{Count}{89}%
\StoreBenchExecResult{Avr}{IcIIIsaReachSafetyXCSPSampledRel}{Status}{Correct}{True}{Count}{52}%
\StoreBenchExecResult{Avr}{IcIIIsaReachSafetyXCSPSampledRel}{Status}{Correct}{False}{Count}{37}%
\StoreBenchExecResult{Avr}{IcIIIsaReachSafetyXCSPSampledRel}{Status}{Wrong}{}{Count}{0}%
\StoreBenchExecResult{Avr}{IcIIIsaReachSafetyXCSPSampledRel}{Status}{Wrong}{True}{Count}{0}%
\StoreBenchExecResult{Avr}{IcIIIsaReachSafetyXCSPSampledRel}{Status}{Wrong}{False}{Count}{0}%
\providecommand\StoreBenchExecResult[7]{\expandafter\newcommand\csname#1#2#3#4#5#6\endcsname{#7}}%
\StoreBenchExecResult{Avr}{IcIIIsaTerminationBitVectorsSampledRel}{Status}{All}{}{Score}{0}%
\StoreBenchExecResult{Avr}{IcIIIsaTerminationBitVectorsSampledRel}{Status}{All}{}{Count}{32}%
\StoreBenchExecResult{Avr}{IcIIIsaTerminationBitVectorsSampledRel}{Status}{Correct}{}{Count}{16}%
\StoreBenchExecResult{Avr}{IcIIIsaTerminationBitVectorsSampledRel}{Status}{Correct}{True}{Count}{5}%
\StoreBenchExecResult{Avr}{IcIIIsaTerminationBitVectorsSampledRel}{Status}{Correct}{False}{Count}{11}%
\StoreBenchExecResult{Avr}{IcIIIsaTerminationBitVectorsSampledRel}{Status}{Wrong}{}{Count}{0}%
\StoreBenchExecResult{Avr}{IcIIIsaTerminationBitVectorsSampledRel}{Status}{Wrong}{True}{Count}{0}%
\StoreBenchExecResult{Avr}{IcIIIsaTerminationBitVectorsSampledRel}{Status}{Wrong}{False}{Count}{0}%
\providecommand\StoreBenchExecResult[7]{\expandafter\newcommand\csname#1#2#3#4#5#6\endcsname{#7}}%
\StoreBenchExecResult{Avr}{IcIIIsaTerminationMainControlFlowSampledRel}{Status}{All}{}{Score}{0}%
\StoreBenchExecResult{Avr}{IcIIIsaTerminationMainControlFlowSampledRel}{Status}{All}{}{Count}{242}%
\StoreBenchExecResult{Avr}{IcIIIsaTerminationMainControlFlowSampledRel}{Status}{Correct}{}{Count}{58}%
\StoreBenchExecResult{Avr}{IcIIIsaTerminationMainControlFlowSampledRel}{Status}{Correct}{True}{Count}{14}%
\StoreBenchExecResult{Avr}{IcIIIsaTerminationMainControlFlowSampledRel}{Status}{Correct}{False}{Count}{44}%
\StoreBenchExecResult{Avr}{IcIIIsaTerminationMainControlFlowSampledRel}{Status}{Wrong}{}{Count}{0}%
\StoreBenchExecResult{Avr}{IcIIIsaTerminationMainControlFlowSampledRel}{Status}{Wrong}{True}{Count}{0}%
\StoreBenchExecResult{Avr}{IcIIIsaTerminationMainControlFlowSampledRel}{Status}{Wrong}{False}{Count}{0}%
\providecommand\StoreBenchExecResult[7]{\expandafter\newcommand\csname#1#2#3#4#5#6\endcsname{#7}}%
\StoreBenchExecResult{Avr}{IcIIIsaTerminationOtherSampledRel}{Status}{All}{}{Score}{0}%
\StoreBenchExecResult{Avr}{IcIIIsaTerminationOtherSampledRel}{Status}{All}{}{Count}{1080}%
\StoreBenchExecResult{Avr}{IcIIIsaTerminationOtherSampledRel}{Status}{Correct}{}{Count}{785}%
\StoreBenchExecResult{Avr}{IcIIIsaTerminationOtherSampledRel}{Status}{Correct}{True}{Count}{197}%
\StoreBenchExecResult{Avr}{IcIIIsaTerminationOtherSampledRel}{Status}{Correct}{False}{Count}{588}%
\StoreBenchExecResult{Avr}{IcIIIsaTerminationOtherSampledRel}{Status}{Wrong}{}{Count}{0}%
\StoreBenchExecResult{Avr}{IcIIIsaTerminationOtherSampledRel}{Status}{Wrong}{True}{Count}{0}%
\StoreBenchExecResult{Avr}{IcIIIsaTerminationOtherSampledRel}{Status}{Wrong}{False}{Count}{0}%
\edef\AvrIcIIIsaReachSafetySampledRelStatusAllCount{\the\numexpr\AvrIcIIIsaReachSafetyArraysSampledRelStatusAllCount+\AvrIcIIIsaReachSafetyBitVectorsSampledRelStatusAllCount+\AvrIcIIIsaReachSafetyCombinationsSampledRelStatusAllCount+\AvrIcIIIsaReachSafetyControlFlowSampledRelStatusAllCount+\AvrIcIIIsaReachSafetyECASampledRelStatusAllCount+\AvrIcIIIsaReachSafetyFloatsSampledRelStatusAllCount+\AvrIcIIIsaReachSafetyHardnessSampledRelStatusAllCount+\AvrIcIIIsaReachSafetyHardwareSampledRelStatusAllCount+\AvrIcIIIsaReachSafetyHeapSampledRelStatusAllCount+\AvrIcIIIsaReachSafetyLoopsSampledRelStatusAllCount+\AvrIcIIIsaReachSafetyProductLinesSampledRelStatusAllCount+\AvrIcIIIsaReachSafetySequentializedSampledRelStatusAllCount+\AvrIcIIIsaReachSafetyXCSPSampledRelStatusAllCount}
\edef\AvrIcIIIsaReachSafetySampledRelStatusCorrectCount{\the\numexpr\AvrIcIIIsaReachSafetyArraysSampledRelStatusCorrectCount+\AvrIcIIIsaReachSafetyBitVectorsSampledRelStatusCorrectCount+\AvrIcIIIsaReachSafetyCombinationsSampledRelStatusCorrectCount+\AvrIcIIIsaReachSafetyControlFlowSampledRelStatusCorrectCount+\AvrIcIIIsaReachSafetyECASampledRelStatusCorrectCount+\AvrIcIIIsaReachSafetyFloatsSampledRelStatusCorrectCount+\AvrIcIIIsaReachSafetyHardnessSampledRelStatusCorrectCount+\AvrIcIIIsaReachSafetyHardwareSampledRelStatusCorrectCount+\AvrIcIIIsaReachSafetyHeapSampledRelStatusCorrectCount+\AvrIcIIIsaReachSafetyLoopsSampledRelStatusCorrectCount+\AvrIcIIIsaReachSafetyProductLinesSampledRelStatusCorrectCount+\AvrIcIIIsaReachSafetySequentializedSampledRelStatusCorrectCount+\AvrIcIIIsaReachSafetyXCSPSampledRelStatusCorrectCount}
\edef\AvrIcIIIsaReachSafetySampledRelStatusCorrectTrueCount{\the\numexpr\AvrIcIIIsaReachSafetyArraysSampledRelStatusCorrectTrueCount+\AvrIcIIIsaReachSafetyBitVectorsSampledRelStatusCorrectTrueCount+\AvrIcIIIsaReachSafetyCombinationsSampledRelStatusCorrectTrueCount+\AvrIcIIIsaReachSafetyControlFlowSampledRelStatusCorrectTrueCount+\AvrIcIIIsaReachSafetyECASampledRelStatusCorrectTrueCount+\AvrIcIIIsaReachSafetyFloatsSampledRelStatusCorrectTrueCount+\AvrIcIIIsaReachSafetyHardnessSampledRelStatusCorrectTrueCount+\AvrIcIIIsaReachSafetyHardwareSampledRelStatusCorrectTrueCount+\AvrIcIIIsaReachSafetyHeapSampledRelStatusCorrectTrueCount+\AvrIcIIIsaReachSafetyLoopsSampledRelStatusCorrectTrueCount+\AvrIcIIIsaReachSafetyProductLinesSampledRelStatusCorrectTrueCount+\AvrIcIIIsaReachSafetySequentializedSampledRelStatusCorrectTrueCount+\AvrIcIIIsaReachSafetyXCSPSampledRelStatusCorrectTrueCount}
\edef\AvrIcIIIsaReachSafetySampledRelStatusCorrectFalseCount{\the\numexpr\AvrIcIIIsaReachSafetyArraysSampledRelStatusCorrectFalseCount+\AvrIcIIIsaReachSafetyBitVectorsSampledRelStatusCorrectFalseCount+\AvrIcIIIsaReachSafetyCombinationsSampledRelStatusCorrectFalseCount+\AvrIcIIIsaReachSafetyControlFlowSampledRelStatusCorrectFalseCount+\AvrIcIIIsaReachSafetyECASampledRelStatusCorrectFalseCount+\AvrIcIIIsaReachSafetyFloatsSampledRelStatusCorrectFalseCount+\AvrIcIIIsaReachSafetyHardnessSampledRelStatusCorrectFalseCount+\AvrIcIIIsaReachSafetyHardwareSampledRelStatusCorrectFalseCount+\AvrIcIIIsaReachSafetyHeapSampledRelStatusCorrectFalseCount+\AvrIcIIIsaReachSafetyLoopsSampledRelStatusCorrectFalseCount+\AvrIcIIIsaReachSafetyProductLinesSampledRelStatusCorrectFalseCount+\AvrIcIIIsaReachSafetySequentializedSampledRelStatusCorrectFalseCount+\AvrIcIIIsaReachSafetyXCSPSampledRelStatusCorrectFalseCount}
\edef\AvrIcIIIsaReachSafetySampledRelStatusWrongCount{\the\numexpr\AvrIcIIIsaReachSafetyArraysSampledRelStatusWrongCount+\AvrIcIIIsaReachSafetyBitVectorsSampledRelStatusWrongCount+\AvrIcIIIsaReachSafetyCombinationsSampledRelStatusWrongCount+\AvrIcIIIsaReachSafetyControlFlowSampledRelStatusWrongCount+\AvrIcIIIsaReachSafetyECASampledRelStatusWrongCount+\AvrIcIIIsaReachSafetyFloatsSampledRelStatusWrongCount+\AvrIcIIIsaReachSafetyHardnessSampledRelStatusWrongCount+\AvrIcIIIsaReachSafetyHardwareSampledRelStatusWrongCount+\AvrIcIIIsaReachSafetyHeapSampledRelStatusWrongCount+\AvrIcIIIsaReachSafetyLoopsSampledRelStatusWrongCount+\AvrIcIIIsaReachSafetyProductLinesSampledRelStatusWrongCount+\AvrIcIIIsaReachSafetySequentializedSampledRelStatusWrongCount+\AvrIcIIIsaReachSafetyXCSPSampledRelStatusWrongCount}
\edef\AvrIcIIIsaTerminationSampledRelStatusAllCount{\the\numexpr\AvrIcIIIsaTerminationBitVectorsSampledRelStatusAllCount+\AvrIcIIIsaTerminationMainControlFlowSampledRelStatusAllCount+\AvrIcIIIsaTerminationOtherSampledRelStatusAllCount}
\edef\AvrIcIIIsaTerminationSampledRelStatusCorrectCount{\the\numexpr\AvrIcIIIsaTerminationBitVectorsSampledRelStatusCorrectCount+\AvrIcIIIsaTerminationMainControlFlowSampledRelStatusCorrectCount+\AvrIcIIIsaTerminationOtherSampledRelStatusCorrectCount}
\edef\AvrIcIIIsaTerminationSampledRelStatusCorrectTrueCount{\the\numexpr\AvrIcIIIsaTerminationBitVectorsSampledRelStatusCorrectTrueCount+\AvrIcIIIsaTerminationMainControlFlowSampledRelStatusCorrectTrueCount+\AvrIcIIIsaTerminationOtherSampledRelStatusCorrectTrueCount}
\edef\AvrIcIIIsaTerminationSampledRelStatusCorrectFalseCount{\the\numexpr\AvrIcIIIsaTerminationBitVectorsSampledRelStatusCorrectFalseCount+\AvrIcIIIsaTerminationMainControlFlowSampledRelStatusCorrectFalseCount+\AvrIcIIIsaTerminationOtherSampledRelStatusCorrectFalseCount}
\edef\AvrIcIIIsaTerminationSampledRelStatusWrongCount{\the\numexpr\AvrIcIIIsaTerminationBitVectorsSampledRelStatusWrongCount+\AvrIcIIIsaTerminationMainControlFlowSampledRelStatusWrongCount+\AvrIcIIIsaTerminationOtherSampledRelStatusWrongCount}
\providecommand\StoreBenchExecResult[7]{\expandafter\newcommand\csname#1#2#3#4#5#6\endcsname{#7}}%
\StoreBenchExecResult{Avr}{KindReachSafetyArraysSampledRel}{Status}{All}{}{Score}{0}%
\StoreBenchExecResult{Avr}{KindReachSafetyArraysSampledRel}{Status}{All}{}{Count}{150}%
\StoreBenchExecResult{Avr}{KindReachSafetyArraysSampledRel}{Status}{Correct}{}{Count}{46}%
\StoreBenchExecResult{Avr}{KindReachSafetyArraysSampledRel}{Status}{Correct}{True}{Count}{0}%
\StoreBenchExecResult{Avr}{KindReachSafetyArraysSampledRel}{Status}{Correct}{False}{Count}{46}%
\StoreBenchExecResult{Avr}{KindReachSafetyArraysSampledRel}{Status}{Wrong}{}{Count}{0}%
\StoreBenchExecResult{Avr}{KindReachSafetyArraysSampledRel}{Status}{Wrong}{True}{Count}{0}%
\StoreBenchExecResult{Avr}{KindReachSafetyArraysSampledRel}{Status}{Wrong}{False}{Count}{0}%
\providecommand\StoreBenchExecResult[7]{\expandafter\newcommand\csname#1#2#3#4#5#6\endcsname{#7}}%
\StoreBenchExecResult{Avr}{KindReachSafetyBitVectorsSampledRel}{Status}{All}{}{Score}{0}%
\StoreBenchExecResult{Avr}{KindReachSafetyBitVectorsSampledRel}{Status}{All}{}{Count}{48}%
\StoreBenchExecResult{Avr}{KindReachSafetyBitVectorsSampledRel}{Status}{Correct}{}{Count}{30}%
\StoreBenchExecResult{Avr}{KindReachSafetyBitVectorsSampledRel}{Status}{Correct}{True}{Count}{18}%
\StoreBenchExecResult{Avr}{KindReachSafetyBitVectorsSampledRel}{Status}{Correct}{False}{Count}{12}%
\StoreBenchExecResult{Avr}{KindReachSafetyBitVectorsSampledRel}{Status}{Wrong}{}{Count}{0}%
\StoreBenchExecResult{Avr}{KindReachSafetyBitVectorsSampledRel}{Status}{Wrong}{True}{Count}{0}%
\StoreBenchExecResult{Avr}{KindReachSafetyBitVectorsSampledRel}{Status}{Wrong}{False}{Count}{0}%
\providecommand\StoreBenchExecResult[7]{\expandafter\newcommand\csname#1#2#3#4#5#6\endcsname{#7}}%
\StoreBenchExecResult{Avr}{KindReachSafetyCombinationsSampledRel}{Status}{All}{}{Score}{0}%
\StoreBenchExecResult{Avr}{KindReachSafetyCombinationsSampledRel}{Status}{All}{}{Count}{110}%
\StoreBenchExecResult{Avr}{KindReachSafetyCombinationsSampledRel}{Status}{Correct}{}{Count}{71}%
\StoreBenchExecResult{Avr}{KindReachSafetyCombinationsSampledRel}{Status}{Correct}{True}{Count}{0}%
\StoreBenchExecResult{Avr}{KindReachSafetyCombinationsSampledRel}{Status}{Correct}{False}{Count}{71}%
\StoreBenchExecResult{Avr}{KindReachSafetyCombinationsSampledRel}{Status}{Wrong}{}{Count}{0}%
\StoreBenchExecResult{Avr}{KindReachSafetyCombinationsSampledRel}{Status}{Wrong}{True}{Count}{0}%
\StoreBenchExecResult{Avr}{KindReachSafetyCombinationsSampledRel}{Status}{Wrong}{False}{Count}{0}%
\providecommand\StoreBenchExecResult[7]{\expandafter\newcommand\csname#1#2#3#4#5#6\endcsname{#7}}%
\StoreBenchExecResult{Avr}{KindReachSafetyControlFlowSampledRel}{Status}{All}{}{Score}{0}%
\StoreBenchExecResult{Avr}{KindReachSafetyControlFlowSampledRel}{Status}{All}{}{Count}{33}%
\StoreBenchExecResult{Avr}{KindReachSafetyControlFlowSampledRel}{Status}{Correct}{}{Count}{24}%
\StoreBenchExecResult{Avr}{KindReachSafetyControlFlowSampledRel}{Status}{Correct}{True}{Count}{21}%
\StoreBenchExecResult{Avr}{KindReachSafetyControlFlowSampledRel}{Status}{Correct}{False}{Count}{3}%
\StoreBenchExecResult{Avr}{KindReachSafetyControlFlowSampledRel}{Status}{Wrong}{}{Count}{0}%
\StoreBenchExecResult{Avr}{KindReachSafetyControlFlowSampledRel}{Status}{Wrong}{True}{Count}{0}%
\StoreBenchExecResult{Avr}{KindReachSafetyControlFlowSampledRel}{Status}{Wrong}{False}{Count}{0}%
\providecommand\StoreBenchExecResult[7]{\expandafter\newcommand\csname#1#2#3#4#5#6\endcsname{#7}}%
\StoreBenchExecResult{Avr}{KindReachSafetyECASampledRel}{Status}{All}{}{Score}{0}%
\StoreBenchExecResult{Avr}{KindReachSafetyECASampledRel}{Status}{All}{}{Count}{150}%
\StoreBenchExecResult{Avr}{KindReachSafetyECASampledRel}{Status}{Correct}{}{Count}{88}%
\StoreBenchExecResult{Avr}{KindReachSafetyECASampledRel}{Status}{Correct}{True}{Count}{48}%
\StoreBenchExecResult{Avr}{KindReachSafetyECASampledRel}{Status}{Correct}{False}{Count}{40}%
\StoreBenchExecResult{Avr}{KindReachSafetyECASampledRel}{Status}{Wrong}{}{Count}{0}%
\StoreBenchExecResult{Avr}{KindReachSafetyECASampledRel}{Status}{Wrong}{True}{Count}{0}%
\StoreBenchExecResult{Avr}{KindReachSafetyECASampledRel}{Status}{Wrong}{False}{Count}{0}%
\providecommand\StoreBenchExecResult[7]{\expandafter\newcommand\csname#1#2#3#4#5#6\endcsname{#7}}%
\StoreBenchExecResult{Avr}{KindReachSafetyFloatsSampledRel}{Status}{All}{}{Score}{0}%
\StoreBenchExecResult{Avr}{KindReachSafetyFloatsSampledRel}{Status}{All}{}{Count}{10}%
\StoreBenchExecResult{Avr}{KindReachSafetyFloatsSampledRel}{Status}{Correct}{}{Count}{10}%
\StoreBenchExecResult{Avr}{KindReachSafetyFloatsSampledRel}{Status}{Correct}{True}{Count}{10}%
\StoreBenchExecResult{Avr}{KindReachSafetyFloatsSampledRel}{Status}{Correct}{False}{Count}{0}%
\StoreBenchExecResult{Avr}{KindReachSafetyFloatsSampledRel}{Status}{Wrong}{}{Count}{0}%
\StoreBenchExecResult{Avr}{KindReachSafetyFloatsSampledRel}{Status}{Wrong}{True}{Count}{0}%
\StoreBenchExecResult{Avr}{KindReachSafetyFloatsSampledRel}{Status}{Wrong}{False}{Count}{0}%
\providecommand\StoreBenchExecResult[7]{\expandafter\newcommand\csname#1#2#3#4#5#6\endcsname{#7}}%
\StoreBenchExecResult{Avr}{KindReachSafetyHardnessSampledRel}{Status}{All}{}{Score}{0}%
\StoreBenchExecResult{Avr}{KindReachSafetyHardnessSampledRel}{Status}{All}{}{Count}{150}%
\StoreBenchExecResult{Avr}{KindReachSafetyHardnessSampledRel}{Status}{Correct}{}{Count}{132}%
\StoreBenchExecResult{Avr}{KindReachSafetyHardnessSampledRel}{Status}{Correct}{True}{Count}{132}%
\StoreBenchExecResult{Avr}{KindReachSafetyHardnessSampledRel}{Status}{Correct}{False}{Count}{0}%
\StoreBenchExecResult{Avr}{KindReachSafetyHardnessSampledRel}{Status}{Wrong}{}{Count}{0}%
\StoreBenchExecResult{Avr}{KindReachSafetyHardnessSampledRel}{Status}{Wrong}{True}{Count}{0}%
\StoreBenchExecResult{Avr}{KindReachSafetyHardnessSampledRel}{Status}{Wrong}{False}{Count}{0}%
\providecommand\StoreBenchExecResult[7]{\expandafter\newcommand\csname#1#2#3#4#5#6\endcsname{#7}}%
\StoreBenchExecResult{Avr}{KindReachSafetyHardwareSampledRel}{Status}{All}{}{Score}{0}%
\StoreBenchExecResult{Avr}{KindReachSafetyHardwareSampledRel}{Status}{All}{}{Count}{150}%
\StoreBenchExecResult{Avr}{KindReachSafetyHardwareSampledRel}{Status}{Correct}{}{Count}{57}%
\StoreBenchExecResult{Avr}{KindReachSafetyHardwareSampledRel}{Status}{Correct}{True}{Count}{2}%
\StoreBenchExecResult{Avr}{KindReachSafetyHardwareSampledRel}{Status}{Correct}{False}{Count}{55}%
\StoreBenchExecResult{Avr}{KindReachSafetyHardwareSampledRel}{Status}{Wrong}{}{Count}{0}%
\StoreBenchExecResult{Avr}{KindReachSafetyHardwareSampledRel}{Status}{Wrong}{True}{Count}{0}%
\StoreBenchExecResult{Avr}{KindReachSafetyHardwareSampledRel}{Status}{Wrong}{False}{Count}{0}%
\providecommand\StoreBenchExecResult[7]{\expandafter\newcommand\csname#1#2#3#4#5#6\endcsname{#7}}%
\StoreBenchExecResult{Avr}{KindReachSafetyHeapSampledRel}{Status}{All}{}{Score}{0}%
\StoreBenchExecResult{Avr}{KindReachSafetyHeapSampledRel}{Status}{All}{}{Count}{53}%
\StoreBenchExecResult{Avr}{KindReachSafetyHeapSampledRel}{Status}{Correct}{}{Count}{46}%
\StoreBenchExecResult{Avr}{KindReachSafetyHeapSampledRel}{Status}{Correct}{True}{Count}{31}%
\StoreBenchExecResult{Avr}{KindReachSafetyHeapSampledRel}{Status}{Correct}{False}{Count}{15}%
\StoreBenchExecResult{Avr}{KindReachSafetyHeapSampledRel}{Status}{Wrong}{}{Count}{0}%
\StoreBenchExecResult{Avr}{KindReachSafetyHeapSampledRel}{Status}{Wrong}{True}{Count}{0}%
\StoreBenchExecResult{Avr}{KindReachSafetyHeapSampledRel}{Status}{Wrong}{False}{Count}{0}%
\providecommand\StoreBenchExecResult[7]{\expandafter\newcommand\csname#1#2#3#4#5#6\endcsname{#7}}%
\StoreBenchExecResult{Avr}{KindReachSafetyLoopsSampledRel}{Status}{All}{}{Score}{0}%
\StoreBenchExecResult{Avr}{KindReachSafetyLoopsSampledRel}{Status}{All}{}{Count}{150}%
\StoreBenchExecResult{Avr}{KindReachSafetyLoopsSampledRel}{Status}{Correct}{}{Count}{54}%
\StoreBenchExecResult{Avr}{KindReachSafetyLoopsSampledRel}{Status}{Correct}{True}{Count}{15}%
\StoreBenchExecResult{Avr}{KindReachSafetyLoopsSampledRel}{Status}{Correct}{False}{Count}{39}%
\StoreBenchExecResult{Avr}{KindReachSafetyLoopsSampledRel}{Status}{Wrong}{}{Count}{0}%
\StoreBenchExecResult{Avr}{KindReachSafetyLoopsSampledRel}{Status}{Wrong}{True}{Count}{0}%
\StoreBenchExecResult{Avr}{KindReachSafetyLoopsSampledRel}{Status}{Wrong}{False}{Count}{0}%
\providecommand\StoreBenchExecResult[7]{\expandafter\newcommand\csname#1#2#3#4#5#6\endcsname{#7}}%
\StoreBenchExecResult{Avr}{KindReachSafetyProductLinesSampledRel}{Status}{All}{}{Score}{0}%
\StoreBenchExecResult{Avr}{KindReachSafetyProductLinesSampledRel}{Status}{All}{}{Count}{150}%
\StoreBenchExecResult{Avr}{KindReachSafetyProductLinesSampledRel}{Status}{Correct}{}{Count}{91}%
\StoreBenchExecResult{Avr}{KindReachSafetyProductLinesSampledRel}{Status}{Correct}{True}{Count}{16}%
\StoreBenchExecResult{Avr}{KindReachSafetyProductLinesSampledRel}{Status}{Correct}{False}{Count}{75}%
\StoreBenchExecResult{Avr}{KindReachSafetyProductLinesSampledRel}{Status}{Wrong}{}{Count}{0}%
\StoreBenchExecResult{Avr}{KindReachSafetyProductLinesSampledRel}{Status}{Wrong}{True}{Count}{0}%
\StoreBenchExecResult{Avr}{KindReachSafetyProductLinesSampledRel}{Status}{Wrong}{False}{Count}{0}%
\providecommand\StoreBenchExecResult[7]{\expandafter\newcommand\csname#1#2#3#4#5#6\endcsname{#7}}%
\StoreBenchExecResult{Avr}{KindReachSafetySequentializedSampledRel}{Status}{All}{}{Score}{0}%
\StoreBenchExecResult{Avr}{KindReachSafetySequentializedSampledRel}{Status}{All}{}{Count}{150}%
\StoreBenchExecResult{Avr}{KindReachSafetySequentializedSampledRel}{Status}{Correct}{}{Count}{100}%
\StoreBenchExecResult{Avr}{KindReachSafetySequentializedSampledRel}{Status}{Correct}{True}{Count}{13}%
\StoreBenchExecResult{Avr}{KindReachSafetySequentializedSampledRel}{Status}{Correct}{False}{Count}{87}%
\StoreBenchExecResult{Avr}{KindReachSafetySequentializedSampledRel}{Status}{Wrong}{}{Count}{0}%
\StoreBenchExecResult{Avr}{KindReachSafetySequentializedSampledRel}{Status}{Wrong}{True}{Count}{0}%
\StoreBenchExecResult{Avr}{KindReachSafetySequentializedSampledRel}{Status}{Wrong}{False}{Count}{0}%
\providecommand\StoreBenchExecResult[7]{\expandafter\newcommand\csname#1#2#3#4#5#6\endcsname{#7}}%
\StoreBenchExecResult{Avr}{KindReachSafetyXCSPSampledRel}{Status}{All}{}{Score}{0}%
\StoreBenchExecResult{Avr}{KindReachSafetyXCSPSampledRel}{Status}{All}{}{Count}{98}%
\StoreBenchExecResult{Avr}{KindReachSafetyXCSPSampledRel}{Status}{Correct}{}{Count}{95}%
\StoreBenchExecResult{Avr}{KindReachSafetyXCSPSampledRel}{Status}{Correct}{True}{Count}{52}%
\StoreBenchExecResult{Avr}{KindReachSafetyXCSPSampledRel}{Status}{Correct}{False}{Count}{43}%
\StoreBenchExecResult{Avr}{KindReachSafetyXCSPSampledRel}{Status}{Wrong}{}{Count}{0}%
\StoreBenchExecResult{Avr}{KindReachSafetyXCSPSampledRel}{Status}{Wrong}{True}{Count}{0}%
\StoreBenchExecResult{Avr}{KindReachSafetyXCSPSampledRel}{Status}{Wrong}{False}{Count}{0}%
\providecommand\StoreBenchExecResult[7]{\expandafter\newcommand\csname#1#2#3#4#5#6\endcsname{#7}}%
\StoreBenchExecResult{Avr}{KindTerminationBitVectorsSampledRel}{Status}{All}{}{Score}{0}%
\StoreBenchExecResult{Avr}{KindTerminationBitVectorsSampledRel}{Status}{All}{}{Count}{32}%
\StoreBenchExecResult{Avr}{KindTerminationBitVectorsSampledRel}{Status}{Correct}{}{Count}{21}%
\StoreBenchExecResult{Avr}{KindTerminationBitVectorsSampledRel}{Status}{Correct}{True}{Count}{10}%
\StoreBenchExecResult{Avr}{KindTerminationBitVectorsSampledRel}{Status}{Correct}{False}{Count}{11}%
\StoreBenchExecResult{Avr}{KindTerminationBitVectorsSampledRel}{Status}{Wrong}{}{Count}{0}%
\StoreBenchExecResult{Avr}{KindTerminationBitVectorsSampledRel}{Status}{Wrong}{True}{Count}{0}%
\StoreBenchExecResult{Avr}{KindTerminationBitVectorsSampledRel}{Status}{Wrong}{False}{Count}{0}%
\providecommand\StoreBenchExecResult[7]{\expandafter\newcommand\csname#1#2#3#4#5#6\endcsname{#7}}%
\StoreBenchExecResult{Avr}{KindTerminationMainControlFlowSampledRel}{Status}{All}{}{Score}{0}%
\StoreBenchExecResult{Avr}{KindTerminationMainControlFlowSampledRel}{Status}{All}{}{Count}{242}%
\StoreBenchExecResult{Avr}{KindTerminationMainControlFlowSampledRel}{Status}{Correct}{}{Count}{62}%
\StoreBenchExecResult{Avr}{KindTerminationMainControlFlowSampledRel}{Status}{Correct}{True}{Count}{10}%
\StoreBenchExecResult{Avr}{KindTerminationMainControlFlowSampledRel}{Status}{Correct}{False}{Count}{52}%
\StoreBenchExecResult{Avr}{KindTerminationMainControlFlowSampledRel}{Status}{Wrong}{}{Count}{0}%
\StoreBenchExecResult{Avr}{KindTerminationMainControlFlowSampledRel}{Status}{Wrong}{True}{Count}{0}%
\StoreBenchExecResult{Avr}{KindTerminationMainControlFlowSampledRel}{Status}{Wrong}{False}{Count}{0}%
\providecommand\StoreBenchExecResult[7]{\expandafter\newcommand\csname#1#2#3#4#5#6\endcsname{#7}}%
\StoreBenchExecResult{Avr}{KindTerminationOtherSampledRel}{Status}{All}{}{Score}{0}%
\StoreBenchExecResult{Avr}{KindTerminationOtherSampledRel}{Status}{All}{}{Count}{1080}%
\StoreBenchExecResult{Avr}{KindTerminationOtherSampledRel}{Status}{Correct}{}{Count}{745}%
\StoreBenchExecResult{Avr}{KindTerminationOtherSampledRel}{Status}{Correct}{True}{Count}{90}%
\StoreBenchExecResult{Avr}{KindTerminationOtherSampledRel}{Status}{Correct}{False}{Count}{655}%
\StoreBenchExecResult{Avr}{KindTerminationOtherSampledRel}{Status}{Wrong}{}{Count}{0}%
\StoreBenchExecResult{Avr}{KindTerminationOtherSampledRel}{Status}{Wrong}{True}{Count}{0}%
\StoreBenchExecResult{Avr}{KindTerminationOtherSampledRel}{Status}{Wrong}{False}{Count}{0}%
\edef\AvrKindReachSafetySampledRelStatusAllCount{\the\numexpr\AvrKindReachSafetyArraysSampledRelStatusAllCount+\AvrKindReachSafetyBitVectorsSampledRelStatusAllCount+\AvrKindReachSafetyCombinationsSampledRelStatusAllCount+\AvrKindReachSafetyControlFlowSampledRelStatusAllCount+\AvrKindReachSafetyECASampledRelStatusAllCount+\AvrKindReachSafetyFloatsSampledRelStatusAllCount+\AvrKindReachSafetyHardnessSampledRelStatusAllCount+\AvrKindReachSafetyHardwareSampledRelStatusAllCount+\AvrKindReachSafetyHeapSampledRelStatusAllCount+\AvrKindReachSafetyLoopsSampledRelStatusAllCount+\AvrKindReachSafetyProductLinesSampledRelStatusAllCount+\AvrKindReachSafetySequentializedSampledRelStatusAllCount+\AvrKindReachSafetyXCSPSampledRelStatusAllCount}
\edef\AvrKindReachSafetySampledRelStatusCorrectCount{\the\numexpr\AvrKindReachSafetyArraysSampledRelStatusCorrectCount+\AvrKindReachSafetyBitVectorsSampledRelStatusCorrectCount+\AvrKindReachSafetyCombinationsSampledRelStatusCorrectCount+\AvrKindReachSafetyControlFlowSampledRelStatusCorrectCount+\AvrKindReachSafetyECASampledRelStatusCorrectCount+\AvrKindReachSafetyFloatsSampledRelStatusCorrectCount+\AvrKindReachSafetyHardnessSampledRelStatusCorrectCount+\AvrKindReachSafetyHardwareSampledRelStatusCorrectCount+\AvrKindReachSafetyHeapSampledRelStatusCorrectCount+\AvrKindReachSafetyLoopsSampledRelStatusCorrectCount+\AvrKindReachSafetyProductLinesSampledRelStatusCorrectCount+\AvrKindReachSafetySequentializedSampledRelStatusCorrectCount+\AvrKindReachSafetyXCSPSampledRelStatusCorrectCount}
\edef\AvrKindReachSafetySampledRelStatusCorrectTrueCount{\the\numexpr\AvrKindReachSafetyArraysSampledRelStatusCorrectTrueCount+\AvrKindReachSafetyBitVectorsSampledRelStatusCorrectTrueCount+\AvrKindReachSafetyCombinationsSampledRelStatusCorrectTrueCount+\AvrKindReachSafetyControlFlowSampledRelStatusCorrectTrueCount+\AvrKindReachSafetyECASampledRelStatusCorrectTrueCount+\AvrKindReachSafetyFloatsSampledRelStatusCorrectTrueCount+\AvrKindReachSafetyHardnessSampledRelStatusCorrectTrueCount+\AvrKindReachSafetyHardwareSampledRelStatusCorrectTrueCount+\AvrKindReachSafetyHeapSampledRelStatusCorrectTrueCount+\AvrKindReachSafetyLoopsSampledRelStatusCorrectTrueCount+\AvrKindReachSafetyProductLinesSampledRelStatusCorrectTrueCount+\AvrKindReachSafetySequentializedSampledRelStatusCorrectTrueCount+\AvrKindReachSafetyXCSPSampledRelStatusCorrectTrueCount}
\edef\AvrKindReachSafetySampledRelStatusCorrectFalseCount{\the\numexpr\AvrKindReachSafetyArraysSampledRelStatusCorrectFalseCount+\AvrKindReachSafetyBitVectorsSampledRelStatusCorrectFalseCount+\AvrKindReachSafetyCombinationsSampledRelStatusCorrectFalseCount+\AvrKindReachSafetyControlFlowSampledRelStatusCorrectFalseCount+\AvrKindReachSafetyECASampledRelStatusCorrectFalseCount+\AvrKindReachSafetyFloatsSampledRelStatusCorrectFalseCount+\AvrKindReachSafetyHardnessSampledRelStatusCorrectFalseCount+\AvrKindReachSafetyHardwareSampledRelStatusCorrectFalseCount+\AvrKindReachSafetyHeapSampledRelStatusCorrectFalseCount+\AvrKindReachSafetyLoopsSampledRelStatusCorrectFalseCount+\AvrKindReachSafetyProductLinesSampledRelStatusCorrectFalseCount+\AvrKindReachSafetySequentializedSampledRelStatusCorrectFalseCount+\AvrKindReachSafetyXCSPSampledRelStatusCorrectFalseCount}
\edef\AvrKindReachSafetySampledRelStatusWrongCount{\the\numexpr\AvrKindReachSafetyArraysSampledRelStatusWrongCount+\AvrKindReachSafetyBitVectorsSampledRelStatusWrongCount+\AvrKindReachSafetyCombinationsSampledRelStatusWrongCount+\AvrKindReachSafetyControlFlowSampledRelStatusWrongCount+\AvrKindReachSafetyECASampledRelStatusWrongCount+\AvrKindReachSafetyFloatsSampledRelStatusWrongCount+\AvrKindReachSafetyHardnessSampledRelStatusWrongCount+\AvrKindReachSafetyHardwareSampledRelStatusWrongCount+\AvrKindReachSafetyHeapSampledRelStatusWrongCount+\AvrKindReachSafetyLoopsSampledRelStatusWrongCount+\AvrKindReachSafetyProductLinesSampledRelStatusWrongCount+\AvrKindReachSafetySequentializedSampledRelStatusWrongCount+\AvrKindReachSafetyXCSPSampledRelStatusWrongCount}
\edef\AvrKindTerminationSampledRelStatusAllCount{\the\numexpr\AvrKindTerminationBitVectorsSampledRelStatusAllCount+\AvrKindTerminationMainControlFlowSampledRelStatusAllCount+\AvrKindTerminationOtherSampledRelStatusAllCount}
\edef\AvrKindTerminationSampledRelStatusCorrectCount{\the\numexpr\AvrKindTerminationBitVectorsSampledRelStatusCorrectCount+\AvrKindTerminationMainControlFlowSampledRelStatusCorrectCount+\AvrKindTerminationOtherSampledRelStatusCorrectCount}
\edef\AvrKindTerminationSampledRelStatusCorrectTrueCount{\the\numexpr\AvrKindTerminationBitVectorsSampledRelStatusCorrectTrueCount+\AvrKindTerminationMainControlFlowSampledRelStatusCorrectTrueCount+\AvrKindTerminationOtherSampledRelStatusCorrectTrueCount}
\edef\AvrKindTerminationSampledRelStatusCorrectFalseCount{\the\numexpr\AvrKindTerminationBitVectorsSampledRelStatusCorrectFalseCount+\AvrKindTerminationMainControlFlowSampledRelStatusCorrectFalseCount+\AvrKindTerminationOtherSampledRelStatusCorrectFalseCount}
\edef\AvrKindTerminationSampledRelStatusWrongCount{\the\numexpr\AvrKindTerminationBitVectorsSampledRelStatusWrongCount+\AvrKindTerminationMainControlFlowSampledRelStatusWrongCount+\AvrKindTerminationOtherSampledRelStatusWrongCount}
\providecommand\StoreBenchExecResult[7]{\expandafter\newcommand\csname#1#2#3#4#5#6\endcsname{#7}}%
\StoreBenchExecResult{RicIII}{IcIIIReachSafetyBitVectorsSampledBvRel}{Status}{All}{}{Score}{0}%
\StoreBenchExecResult{RicIII}{IcIIIReachSafetyBitVectorsSampledBvRel}{Status}{All}{}{Count}{47}%
\StoreBenchExecResult{RicIII}{IcIIIReachSafetyBitVectorsSampledBvRel}{Status}{Correct}{}{Count}{43}%
\StoreBenchExecResult{RicIII}{IcIIIReachSafetyBitVectorsSampledBvRel}{Status}{Correct}{True}{Count}{32}%
\StoreBenchExecResult{RicIII}{IcIIIReachSafetyBitVectorsSampledBvRel}{Status}{Correct}{False}{Count}{11}%
\StoreBenchExecResult{RicIII}{IcIIIReachSafetyBitVectorsSampledBvRel}{Status}{Wrong}{}{Count}{0}%
\StoreBenchExecResult{RicIII}{IcIIIReachSafetyBitVectorsSampledBvRel}{Status}{Wrong}{True}{Count}{0}%
\StoreBenchExecResult{RicIII}{IcIIIReachSafetyBitVectorsSampledBvRel}{Status}{Wrong}{False}{Count}{0}%
\providecommand\StoreBenchExecResult[7]{\expandafter\newcommand\csname#1#2#3#4#5#6\endcsname{#7}}%
\StoreBenchExecResult{RicIII}{IcIIIReachSafetyCombinationsSampledBvRel}{Status}{All}{}{Score}{0}%
\StoreBenchExecResult{RicIII}{IcIIIReachSafetyCombinationsSampledBvRel}{Status}{All}{}{Count}{110}%
\StoreBenchExecResult{RicIII}{IcIIIReachSafetyCombinationsSampledBvRel}{Status}{Correct}{}{Count}{75}%
\StoreBenchExecResult{RicIII}{IcIIIReachSafetyCombinationsSampledBvRel}{Status}{Correct}{True}{Count}{7}%
\StoreBenchExecResult{RicIII}{IcIIIReachSafetyCombinationsSampledBvRel}{Status}{Correct}{False}{Count}{68}%
\StoreBenchExecResult{RicIII}{IcIIIReachSafetyCombinationsSampledBvRel}{Status}{Wrong}{}{Count}{0}%
\StoreBenchExecResult{RicIII}{IcIIIReachSafetyCombinationsSampledBvRel}{Status}{Wrong}{True}{Count}{0}%
\StoreBenchExecResult{RicIII}{IcIIIReachSafetyCombinationsSampledBvRel}{Status}{Wrong}{False}{Count}{0}%
\providecommand\StoreBenchExecResult[7]{\expandafter\newcommand\csname#1#2#3#4#5#6\endcsname{#7}}%
\StoreBenchExecResult{RicIII}{IcIIIReachSafetyControlFlowSampledBvRel}{Status}{All}{}{Score}{0}%
\StoreBenchExecResult{RicIII}{IcIIIReachSafetyControlFlowSampledBvRel}{Status}{All}{}{Count}{29}%
\StoreBenchExecResult{RicIII}{IcIIIReachSafetyControlFlowSampledBvRel}{Status}{Correct}{}{Count}{27}%
\StoreBenchExecResult{RicIII}{IcIIIReachSafetyControlFlowSampledBvRel}{Status}{Correct}{True}{Count}{25}%
\StoreBenchExecResult{RicIII}{IcIIIReachSafetyControlFlowSampledBvRel}{Status}{Correct}{False}{Count}{2}%
\StoreBenchExecResult{RicIII}{IcIIIReachSafetyControlFlowSampledBvRel}{Status}{Wrong}{}{Count}{0}%
\StoreBenchExecResult{RicIII}{IcIIIReachSafetyControlFlowSampledBvRel}{Status}{Wrong}{True}{Count}{0}%
\StoreBenchExecResult{RicIII}{IcIIIReachSafetyControlFlowSampledBvRel}{Status}{Wrong}{False}{Count}{0}%
\providecommand\StoreBenchExecResult[7]{\expandafter\newcommand\csname#1#2#3#4#5#6\endcsname{#7}}%
\StoreBenchExecResult{RicIII}{IcIIIReachSafetyECASampledBvRel}{Status}{All}{}{Score}{0}%
\StoreBenchExecResult{RicIII}{IcIIIReachSafetyECASampledBvRel}{Status}{All}{}{Count}{150}%
\StoreBenchExecResult{RicIII}{IcIIIReachSafetyECASampledBvRel}{Status}{Correct}{}{Count}{70}%
\StoreBenchExecResult{RicIII}{IcIIIReachSafetyECASampledBvRel}{Status}{Correct}{True}{Count}{43}%
\StoreBenchExecResult{RicIII}{IcIIIReachSafetyECASampledBvRel}{Status}{Correct}{False}{Count}{27}%
\StoreBenchExecResult{RicIII}{IcIIIReachSafetyECASampledBvRel}{Status}{Wrong}{}{Count}{0}%
\StoreBenchExecResult{RicIII}{IcIIIReachSafetyECASampledBvRel}{Status}{Wrong}{True}{Count}{0}%
\StoreBenchExecResult{RicIII}{IcIIIReachSafetyECASampledBvRel}{Status}{Wrong}{False}{Count}{0}%
\providecommand\StoreBenchExecResult[7]{\expandafter\newcommand\csname#1#2#3#4#5#6\endcsname{#7}}%
\StoreBenchExecResult{RicIII}{IcIIIReachSafetyFloatsSampledBvRel}{Status}{All}{}{Score}{0}%
\StoreBenchExecResult{RicIII}{IcIIIReachSafetyFloatsSampledBvRel}{Status}{All}{}{Count}{10}%
\StoreBenchExecResult{RicIII}{IcIIIReachSafetyFloatsSampledBvRel}{Status}{Correct}{}{Count}{10}%
\StoreBenchExecResult{RicIII}{IcIIIReachSafetyFloatsSampledBvRel}{Status}{Correct}{True}{Count}{10}%
\StoreBenchExecResult{RicIII}{IcIIIReachSafetyFloatsSampledBvRel}{Status}{Correct}{False}{Count}{0}%
\StoreBenchExecResult{RicIII}{IcIIIReachSafetyFloatsSampledBvRel}{Status}{Wrong}{}{Count}{0}%
\StoreBenchExecResult{RicIII}{IcIIIReachSafetyFloatsSampledBvRel}{Status}{Wrong}{True}{Count}{0}%
\StoreBenchExecResult{RicIII}{IcIIIReachSafetyFloatsSampledBvRel}{Status}{Wrong}{False}{Count}{0}%
\providecommand\StoreBenchExecResult[7]{\expandafter\newcommand\csname#1#2#3#4#5#6\endcsname{#7}}%
\StoreBenchExecResult{RicIII}{IcIIIReachSafetyHardnessSampledBvRel}{Status}{All}{}{Score}{0}%
\StoreBenchExecResult{RicIII}{IcIIIReachSafetyHardnessSampledBvRel}{Status}{All}{}{Count}{132}%
\StoreBenchExecResult{RicIII}{IcIIIReachSafetyHardnessSampledBvRel}{Status}{Correct}{}{Count}{132}%
\StoreBenchExecResult{RicIII}{IcIIIReachSafetyHardnessSampledBvRel}{Status}{Correct}{True}{Count}{132}%
\StoreBenchExecResult{RicIII}{IcIIIReachSafetyHardnessSampledBvRel}{Status}{Correct}{False}{Count}{0}%
\StoreBenchExecResult{RicIII}{IcIIIReachSafetyHardnessSampledBvRel}{Status}{Wrong}{}{Count}{0}%
\StoreBenchExecResult{RicIII}{IcIIIReachSafetyHardnessSampledBvRel}{Status}{Wrong}{True}{Count}{0}%
\StoreBenchExecResult{RicIII}{IcIIIReachSafetyHardnessSampledBvRel}{Status}{Wrong}{False}{Count}{0}%
\providecommand\StoreBenchExecResult[7]{\expandafter\newcommand\csname#1#2#3#4#5#6\endcsname{#7}}%
\StoreBenchExecResult{RicIII}{IcIIIReachSafetyHardwareSampledBvRel}{Status}{All}{}{Score}{0}%
\StoreBenchExecResult{RicIII}{IcIIIReachSafetyHardwareSampledBvRel}{Status}{All}{}{Count}{148}%
\StoreBenchExecResult{RicIII}{IcIIIReachSafetyHardwareSampledBvRel}{Status}{Correct}{}{Count}{63}%
\StoreBenchExecResult{RicIII}{IcIIIReachSafetyHardwareSampledBvRel}{Status}{Correct}{True}{Count}{26}%
\StoreBenchExecResult{RicIII}{IcIIIReachSafetyHardwareSampledBvRel}{Status}{Correct}{False}{Count}{37}%
\StoreBenchExecResult{RicIII}{IcIIIReachSafetyHardwareSampledBvRel}{Status}{Wrong}{}{Count}{0}%
\StoreBenchExecResult{RicIII}{IcIIIReachSafetyHardwareSampledBvRel}{Status}{Wrong}{True}{Count}{0}%
\StoreBenchExecResult{RicIII}{IcIIIReachSafetyHardwareSampledBvRel}{Status}{Wrong}{False}{Count}{0}%
\providecommand\StoreBenchExecResult[7]{\expandafter\newcommand\csname#1#2#3#4#5#6\endcsname{#7}}%
\StoreBenchExecResult{RicIII}{IcIIIReachSafetyHeapSampledBvRel}{Status}{All}{}{Score}{0}%
\StoreBenchExecResult{RicIII}{IcIIIReachSafetyHeapSampledBvRel}{Status}{All}{}{Count}{6}%
\StoreBenchExecResult{RicIII}{IcIIIReachSafetyHeapSampledBvRel}{Status}{Correct}{}{Count}{6}%
\StoreBenchExecResult{RicIII}{IcIIIReachSafetyHeapSampledBvRel}{Status}{Correct}{True}{Count}{4}%
\StoreBenchExecResult{RicIII}{IcIIIReachSafetyHeapSampledBvRel}{Status}{Correct}{False}{Count}{2}%
\StoreBenchExecResult{RicIII}{IcIIIReachSafetyHeapSampledBvRel}{Status}{Wrong}{}{Count}{0}%
\StoreBenchExecResult{RicIII}{IcIIIReachSafetyHeapSampledBvRel}{Status}{Wrong}{True}{Count}{0}%
\StoreBenchExecResult{RicIII}{IcIIIReachSafetyHeapSampledBvRel}{Status}{Wrong}{False}{Count}{0}%
\providecommand\StoreBenchExecResult[7]{\expandafter\newcommand\csname#1#2#3#4#5#6\endcsname{#7}}%
\StoreBenchExecResult{RicIII}{IcIIIReachSafetyLoopsSampledBvRel}{Status}{All}{}{Score}{0}%
\StoreBenchExecResult{RicIII}{IcIIIReachSafetyLoopsSampledBvRel}{Status}{All}{}{Count}{137}%
\StoreBenchExecResult{RicIII}{IcIIIReachSafetyLoopsSampledBvRel}{Status}{Correct}{}{Count}{56}%
\StoreBenchExecResult{RicIII}{IcIIIReachSafetyLoopsSampledBvRel}{Status}{Correct}{True}{Count}{31}%
\StoreBenchExecResult{RicIII}{IcIIIReachSafetyLoopsSampledBvRel}{Status}{Correct}{False}{Count}{25}%
\StoreBenchExecResult{RicIII}{IcIIIReachSafetyLoopsSampledBvRel}{Status}{Wrong}{}{Count}{0}%
\StoreBenchExecResult{RicIII}{IcIIIReachSafetyLoopsSampledBvRel}{Status}{Wrong}{True}{Count}{0}%
\StoreBenchExecResult{RicIII}{IcIIIReachSafetyLoopsSampledBvRel}{Status}{Wrong}{False}{Count}{0}%
\providecommand\StoreBenchExecResult[7]{\expandafter\newcommand\csname#1#2#3#4#5#6\endcsname{#7}}%
\StoreBenchExecResult{RicIII}{IcIIIReachSafetyProductLinesSampledBvRel}{Status}{All}{}{Score}{0}%
\StoreBenchExecResult{RicIII}{IcIIIReachSafetyProductLinesSampledBvRel}{Status}{All}{}{Count}{150}%
\StoreBenchExecResult{RicIII}{IcIIIReachSafetyProductLinesSampledBvRel}{Status}{Correct}{}{Count}{146}%
\StoreBenchExecResult{RicIII}{IcIIIReachSafetyProductLinesSampledBvRel}{Status}{Correct}{True}{Count}{72}%
\StoreBenchExecResult{RicIII}{IcIIIReachSafetyProductLinesSampledBvRel}{Status}{Correct}{False}{Count}{74}%
\StoreBenchExecResult{RicIII}{IcIIIReachSafetyProductLinesSampledBvRel}{Status}{Wrong}{}{Count}{0}%
\StoreBenchExecResult{RicIII}{IcIIIReachSafetyProductLinesSampledBvRel}{Status}{Wrong}{True}{Count}{0}%
\StoreBenchExecResult{RicIII}{IcIIIReachSafetyProductLinesSampledBvRel}{Status}{Wrong}{False}{Count}{0}%
\providecommand\StoreBenchExecResult[7]{\expandafter\newcommand\csname#1#2#3#4#5#6\endcsname{#7}}%
\StoreBenchExecResult{RicIII}{IcIIIReachSafetySequentializedSampledBvRel}{Status}{All}{}{Score}{0}%
\StoreBenchExecResult{RicIII}{IcIIIReachSafetySequentializedSampledBvRel}{Status}{All}{}{Count}{150}%
\StoreBenchExecResult{RicIII}{IcIIIReachSafetySequentializedSampledBvRel}{Status}{Correct}{}{Count}{76}%
\StoreBenchExecResult{RicIII}{IcIIIReachSafetySequentializedSampledBvRel}{Status}{Correct}{True}{Count}{12}%
\StoreBenchExecResult{RicIII}{IcIIIReachSafetySequentializedSampledBvRel}{Status}{Correct}{False}{Count}{64}%
\StoreBenchExecResult{RicIII}{IcIIIReachSafetySequentializedSampledBvRel}{Status}{Wrong}{}{Count}{0}%
\StoreBenchExecResult{RicIII}{IcIIIReachSafetySequentializedSampledBvRel}{Status}{Wrong}{True}{Count}{0}%
\StoreBenchExecResult{RicIII}{IcIIIReachSafetySequentializedSampledBvRel}{Status}{Wrong}{False}{Count}{0}%
\providecommand\StoreBenchExecResult[7]{\expandafter\newcommand\csname#1#2#3#4#5#6\endcsname{#7}}%
\StoreBenchExecResult{RicIII}{IcIIIReachSafetyXCSPSampledBvRel}{Status}{All}{}{Score}{0}%
\StoreBenchExecResult{RicIII}{IcIIIReachSafetyXCSPSampledBvRel}{Status}{All}{}{Count}{98}%
\StoreBenchExecResult{RicIII}{IcIIIReachSafetyXCSPSampledBvRel}{Status}{Correct}{}{Count}{94}%
\StoreBenchExecResult{RicIII}{IcIIIReachSafetyXCSPSampledBvRel}{Status}{Correct}{True}{Count}{51}%
\StoreBenchExecResult{RicIII}{IcIIIReachSafetyXCSPSampledBvRel}{Status}{Correct}{False}{Count}{43}%
\StoreBenchExecResult{RicIII}{IcIIIReachSafetyXCSPSampledBvRel}{Status}{Wrong}{}{Count}{0}%
\StoreBenchExecResult{RicIII}{IcIIIReachSafetyXCSPSampledBvRel}{Status}{Wrong}{True}{Count}{0}%
\StoreBenchExecResult{RicIII}{IcIIIReachSafetyXCSPSampledBvRel}{Status}{Wrong}{False}{Count}{0}%
\providecommand\StoreBenchExecResult[7]{\expandafter\newcommand\csname#1#2#3#4#5#6\endcsname{#7}}%
\StoreBenchExecResult{RicIII}{IcIIITerminationBitVectorsSampledBvRel}{Status}{All}{}{Score}{0}%
\StoreBenchExecResult{RicIII}{IcIIITerminationBitVectorsSampledBvRel}{Status}{All}{}{Count}{32}%
\StoreBenchExecResult{RicIII}{IcIIITerminationBitVectorsSampledBvRel}{Status}{Correct}{}{Count}{24}%
\StoreBenchExecResult{RicIII}{IcIIITerminationBitVectorsSampledBvRel}{Status}{Correct}{True}{Count}{13}%
\StoreBenchExecResult{RicIII}{IcIIITerminationBitVectorsSampledBvRel}{Status}{Correct}{False}{Count}{11}%
\StoreBenchExecResult{RicIII}{IcIIITerminationBitVectorsSampledBvRel}{Status}{Wrong}{}{Count}{0}%
\StoreBenchExecResult{RicIII}{IcIIITerminationBitVectorsSampledBvRel}{Status}{Wrong}{True}{Count}{0}%
\StoreBenchExecResult{RicIII}{IcIIITerminationBitVectorsSampledBvRel}{Status}{Wrong}{False}{Count}{0}%
\providecommand\StoreBenchExecResult[7]{\expandafter\newcommand\csname#1#2#3#4#5#6\endcsname{#7}}%
\StoreBenchExecResult{RicIII}{IcIIITerminationMainControlFlowSampledBvRel}{Status}{All}{}{Score}{0}%
\StoreBenchExecResult{RicIII}{IcIIITerminationMainControlFlowSampledBvRel}{Status}{All}{}{Count}{236}%
\StoreBenchExecResult{RicIII}{IcIIITerminationMainControlFlowSampledBvRel}{Status}{Correct}{}{Count}{69}%
\StoreBenchExecResult{RicIII}{IcIIITerminationMainControlFlowSampledBvRel}{Status}{Correct}{True}{Count}{22}%
\StoreBenchExecResult{RicIII}{IcIIITerminationMainControlFlowSampledBvRel}{Status}{Correct}{False}{Count}{47}%
\StoreBenchExecResult{RicIII}{IcIIITerminationMainControlFlowSampledBvRel}{Status}{Wrong}{}{Count}{0}%
\StoreBenchExecResult{RicIII}{IcIIITerminationMainControlFlowSampledBvRel}{Status}{Wrong}{True}{Count}{0}%
\StoreBenchExecResult{RicIII}{IcIIITerminationMainControlFlowSampledBvRel}{Status}{Wrong}{False}{Count}{0}%
\providecommand\StoreBenchExecResult[7]{\expandafter\newcommand\csname#1#2#3#4#5#6\endcsname{#7}}%
\StoreBenchExecResult{RicIII}{IcIIITerminationOtherSampledBvRel}{Status}{All}{}{Score}{0}%
\StoreBenchExecResult{RicIII}{IcIIITerminationOtherSampledBvRel}{Status}{All}{}{Count}{973}%
\StoreBenchExecResult{RicIII}{IcIIITerminationOtherSampledBvRel}{Status}{Correct}{}{Count}{800}%
\StoreBenchExecResult{RicIII}{IcIIITerminationOtherSampledBvRel}{Status}{Correct}{True}{Count}{173}%
\StoreBenchExecResult{RicIII}{IcIIITerminationOtherSampledBvRel}{Status}{Correct}{False}{Count}{627}%
\StoreBenchExecResult{RicIII}{IcIIITerminationOtherSampledBvRel}{Status}{Wrong}{}{Count}{0}%
\StoreBenchExecResult{RicIII}{IcIIITerminationOtherSampledBvRel}{Status}{Wrong}{True}{Count}{0}%
\StoreBenchExecResult{RicIII}{IcIIITerminationOtherSampledBvRel}{Status}{Wrong}{False}{Count}{0}%
\edef\RicIIIIcIIIReachSafetySampledBvRelStatusAllCount{\the\numexpr\RicIIIIcIIIReachSafetyBitVectorsSampledBvRelStatusAllCount+\RicIIIIcIIIReachSafetyCombinationsSampledBvRelStatusAllCount+\RicIIIIcIIIReachSafetyControlFlowSampledBvRelStatusAllCount+\RicIIIIcIIIReachSafetyECASampledBvRelStatusAllCount+\RicIIIIcIIIReachSafetyFloatsSampledBvRelStatusAllCount+\RicIIIIcIIIReachSafetyHardnessSampledBvRelStatusAllCount+\RicIIIIcIIIReachSafetyHardwareSampledBvRelStatusAllCount+\RicIIIIcIIIReachSafetyHeapSampledBvRelStatusAllCount+\RicIIIIcIIIReachSafetyLoopsSampledBvRelStatusAllCount+\RicIIIIcIIIReachSafetyProductLinesSampledBvRelStatusAllCount+\RicIIIIcIIIReachSafetySequentializedSampledBvRelStatusAllCount+\RicIIIIcIIIReachSafetyXCSPSampledBvRelStatusAllCount}
\edef\RicIIIIcIIIReachSafetySampledBvRelStatusCorrectCount{\the\numexpr\RicIIIIcIIIReachSafetyBitVectorsSampledBvRelStatusCorrectCount+\RicIIIIcIIIReachSafetyCombinationsSampledBvRelStatusCorrectCount+\RicIIIIcIIIReachSafetyControlFlowSampledBvRelStatusCorrectCount+\RicIIIIcIIIReachSafetyECASampledBvRelStatusCorrectCount+\RicIIIIcIIIReachSafetyFloatsSampledBvRelStatusCorrectCount+\RicIIIIcIIIReachSafetyHardnessSampledBvRelStatusCorrectCount+\RicIIIIcIIIReachSafetyHardwareSampledBvRelStatusCorrectCount+\RicIIIIcIIIReachSafetyHeapSampledBvRelStatusCorrectCount+\RicIIIIcIIIReachSafetyLoopsSampledBvRelStatusCorrectCount+\RicIIIIcIIIReachSafetyProductLinesSampledBvRelStatusCorrectCount+\RicIIIIcIIIReachSafetySequentializedSampledBvRelStatusCorrectCount+\RicIIIIcIIIReachSafetyXCSPSampledBvRelStatusCorrectCount}
\edef\RicIIIIcIIIReachSafetySampledBvRelStatusCorrectTrueCount{\the\numexpr\RicIIIIcIIIReachSafetyBitVectorsSampledBvRelStatusCorrectTrueCount+\RicIIIIcIIIReachSafetyCombinationsSampledBvRelStatusCorrectTrueCount+\RicIIIIcIIIReachSafetyControlFlowSampledBvRelStatusCorrectTrueCount+\RicIIIIcIIIReachSafetyECASampledBvRelStatusCorrectTrueCount+\RicIIIIcIIIReachSafetyFloatsSampledBvRelStatusCorrectTrueCount+\RicIIIIcIIIReachSafetyHardnessSampledBvRelStatusCorrectTrueCount+\RicIIIIcIIIReachSafetyHardwareSampledBvRelStatusCorrectTrueCount+\RicIIIIcIIIReachSafetyHeapSampledBvRelStatusCorrectTrueCount+\RicIIIIcIIIReachSafetyLoopsSampledBvRelStatusCorrectTrueCount+\RicIIIIcIIIReachSafetyProductLinesSampledBvRelStatusCorrectTrueCount+\RicIIIIcIIIReachSafetySequentializedSampledBvRelStatusCorrectTrueCount+\RicIIIIcIIIReachSafetyXCSPSampledBvRelStatusCorrectTrueCount}
\edef\RicIIIIcIIIReachSafetySampledBvRelStatusCorrectFalseCount{\the\numexpr\RicIIIIcIIIReachSafetyBitVectorsSampledBvRelStatusCorrectFalseCount+\RicIIIIcIIIReachSafetyCombinationsSampledBvRelStatusCorrectFalseCount+\RicIIIIcIIIReachSafetyControlFlowSampledBvRelStatusCorrectFalseCount+\RicIIIIcIIIReachSafetyECASampledBvRelStatusCorrectFalseCount+\RicIIIIcIIIReachSafetyFloatsSampledBvRelStatusCorrectFalseCount+\RicIIIIcIIIReachSafetyHardnessSampledBvRelStatusCorrectFalseCount+\RicIIIIcIIIReachSafetyHardwareSampledBvRelStatusCorrectFalseCount+\RicIIIIcIIIReachSafetyHeapSampledBvRelStatusCorrectFalseCount+\RicIIIIcIIIReachSafetyLoopsSampledBvRelStatusCorrectFalseCount+\RicIIIIcIIIReachSafetyProductLinesSampledBvRelStatusCorrectFalseCount+\RicIIIIcIIIReachSafetySequentializedSampledBvRelStatusCorrectFalseCount+\RicIIIIcIIIReachSafetyXCSPSampledBvRelStatusCorrectFalseCount}
\edef\RicIIIIcIIIReachSafetySampledBvRelStatusWrongCount{\the\numexpr\RicIIIIcIIIReachSafetyBitVectorsSampledBvRelStatusWrongCount+\RicIIIIcIIIReachSafetyCombinationsSampledBvRelStatusWrongCount+\RicIIIIcIIIReachSafetyControlFlowSampledBvRelStatusWrongCount+\RicIIIIcIIIReachSafetyECASampledBvRelStatusWrongCount+\RicIIIIcIIIReachSafetyFloatsSampledBvRelStatusWrongCount+\RicIIIIcIIIReachSafetyHardnessSampledBvRelStatusWrongCount+\RicIIIIcIIIReachSafetyHardwareSampledBvRelStatusWrongCount+\RicIIIIcIIIReachSafetyHeapSampledBvRelStatusWrongCount+\RicIIIIcIIIReachSafetyLoopsSampledBvRelStatusWrongCount+\RicIIIIcIIIReachSafetyProductLinesSampledBvRelStatusWrongCount+\RicIIIIcIIIReachSafetySequentializedSampledBvRelStatusWrongCount+\RicIIIIcIIIReachSafetyXCSPSampledBvRelStatusWrongCount}
\edef\RicIIIIcIIITerminationSampledBvRelStatusAllCount{\the\numexpr\RicIIIIcIIITerminationBitVectorsSampledBvRelStatusAllCount+\RicIIIIcIIITerminationMainControlFlowSampledBvRelStatusAllCount+\RicIIIIcIIITerminationOtherSampledBvRelStatusAllCount}
\edef\RicIIIIcIIITerminationSampledBvRelStatusCorrectCount{\the\numexpr\RicIIIIcIIITerminationBitVectorsSampledBvRelStatusCorrectCount+\RicIIIIcIIITerminationMainControlFlowSampledBvRelStatusCorrectCount+\RicIIIIcIIITerminationOtherSampledBvRelStatusCorrectCount}
\edef\RicIIIIcIIITerminationSampledBvRelStatusCorrectTrueCount{\the\numexpr\RicIIIIcIIITerminationBitVectorsSampledBvRelStatusCorrectTrueCount+\RicIIIIcIIITerminationMainControlFlowSampledBvRelStatusCorrectTrueCount+\RicIIIIcIIITerminationOtherSampledBvRelStatusCorrectTrueCount}
\edef\RicIIIIcIIITerminationSampledBvRelStatusCorrectFalseCount{\the\numexpr\RicIIIIcIIITerminationBitVectorsSampledBvRelStatusCorrectFalseCount+\RicIIIIcIIITerminationMainControlFlowSampledBvRelStatusCorrectFalseCount+\RicIIIIcIIITerminationOtherSampledBvRelStatusCorrectFalseCount}
\edef\RicIIIIcIIITerminationSampledBvRelStatusWrongCount{\the\numexpr\RicIIIIcIIITerminationBitVectorsSampledBvRelStatusWrongCount+\RicIIIIcIIITerminationMainControlFlowSampledBvRelStatusWrongCount+\RicIIIIcIIITerminationOtherSampledBvRelStatusWrongCount}
\providecommand\StoreBenchExecResult[7]{\expandafter\newcommand\csname#1#2#3#4#5#6\endcsname{#7}}%
\StoreBenchExecResult{RicIII}{KindReachSafetyBitVectorsSampledBvRel}{Status}{All}{}{Score}{0}%
\StoreBenchExecResult{RicIII}{KindReachSafetyBitVectorsSampledBvRel}{Status}{All}{}{Count}{47}%
\StoreBenchExecResult{RicIII}{KindReachSafetyBitVectorsSampledBvRel}{Status}{Correct}{}{Count}{29}%
\StoreBenchExecResult{RicIII}{KindReachSafetyBitVectorsSampledBvRel}{Status}{Correct}{True}{Count}{18}%
\StoreBenchExecResult{RicIII}{KindReachSafetyBitVectorsSampledBvRel}{Status}{Correct}{False}{Count}{11}%
\StoreBenchExecResult{RicIII}{KindReachSafetyBitVectorsSampledBvRel}{Status}{Wrong}{}{Count}{0}%
\StoreBenchExecResult{RicIII}{KindReachSafetyBitVectorsSampledBvRel}{Status}{Wrong}{True}{Count}{0}%
\StoreBenchExecResult{RicIII}{KindReachSafetyBitVectorsSampledBvRel}{Status}{Wrong}{False}{Count}{0}%
\providecommand\StoreBenchExecResult[7]{\expandafter\newcommand\csname#1#2#3#4#5#6\endcsname{#7}}%
\StoreBenchExecResult{RicIII}{KindReachSafetyCombinationsSampledBvRel}{Status}{All}{}{Score}{0}%
\StoreBenchExecResult{RicIII}{KindReachSafetyCombinationsSampledBvRel}{Status}{All}{}{Count}{110}%
\StoreBenchExecResult{RicIII}{KindReachSafetyCombinationsSampledBvRel}{Status}{Correct}{}{Count}{71}%
\StoreBenchExecResult{RicIII}{KindReachSafetyCombinationsSampledBvRel}{Status}{Correct}{True}{Count}{0}%
\StoreBenchExecResult{RicIII}{KindReachSafetyCombinationsSampledBvRel}{Status}{Correct}{False}{Count}{71}%
\StoreBenchExecResult{RicIII}{KindReachSafetyCombinationsSampledBvRel}{Status}{Wrong}{}{Count}{0}%
\StoreBenchExecResult{RicIII}{KindReachSafetyCombinationsSampledBvRel}{Status}{Wrong}{True}{Count}{0}%
\StoreBenchExecResult{RicIII}{KindReachSafetyCombinationsSampledBvRel}{Status}{Wrong}{False}{Count}{0}%
\providecommand\StoreBenchExecResult[7]{\expandafter\newcommand\csname#1#2#3#4#5#6\endcsname{#7}}%
\StoreBenchExecResult{RicIII}{KindReachSafetyControlFlowSampledBvRel}{Status}{All}{}{Score}{0}%
\StoreBenchExecResult{RicIII}{KindReachSafetyControlFlowSampledBvRel}{Status}{All}{}{Count}{29}%
\StoreBenchExecResult{RicIII}{KindReachSafetyControlFlowSampledBvRel}{Status}{Correct}{}{Count}{23}%
\StoreBenchExecResult{RicIII}{KindReachSafetyControlFlowSampledBvRel}{Status}{Correct}{True}{Count}{21}%
\StoreBenchExecResult{RicIII}{KindReachSafetyControlFlowSampledBvRel}{Status}{Correct}{False}{Count}{2}%
\StoreBenchExecResult{RicIII}{KindReachSafetyControlFlowSampledBvRel}{Status}{Wrong}{}{Count}{0}%
\StoreBenchExecResult{RicIII}{KindReachSafetyControlFlowSampledBvRel}{Status}{Wrong}{True}{Count}{0}%
\StoreBenchExecResult{RicIII}{KindReachSafetyControlFlowSampledBvRel}{Status}{Wrong}{False}{Count}{0}%
\providecommand\StoreBenchExecResult[7]{\expandafter\newcommand\csname#1#2#3#4#5#6\endcsname{#7}}%
\StoreBenchExecResult{RicIII}{KindReachSafetyECASampledBvRel}{Status}{All}{}{Score}{0}%
\StoreBenchExecResult{RicIII}{KindReachSafetyECASampledBvRel}{Status}{All}{}{Count}{150}%
\StoreBenchExecResult{RicIII}{KindReachSafetyECASampledBvRel}{Status}{Correct}{}{Count}{85}%
\StoreBenchExecResult{RicIII}{KindReachSafetyECASampledBvRel}{Status}{Correct}{True}{Count}{50}%
\StoreBenchExecResult{RicIII}{KindReachSafetyECASampledBvRel}{Status}{Correct}{False}{Count}{35}%
\StoreBenchExecResult{RicIII}{KindReachSafetyECASampledBvRel}{Status}{Wrong}{}{Count}{0}%
\StoreBenchExecResult{RicIII}{KindReachSafetyECASampledBvRel}{Status}{Wrong}{True}{Count}{0}%
\StoreBenchExecResult{RicIII}{KindReachSafetyECASampledBvRel}{Status}{Wrong}{False}{Count}{0}%
\providecommand\StoreBenchExecResult[7]{\expandafter\newcommand\csname#1#2#3#4#5#6\endcsname{#7}}%
\StoreBenchExecResult{RicIII}{KindReachSafetyFloatsSampledBvRel}{Status}{All}{}{Score}{0}%
\StoreBenchExecResult{RicIII}{KindReachSafetyFloatsSampledBvRel}{Status}{All}{}{Count}{10}%
\StoreBenchExecResult{RicIII}{KindReachSafetyFloatsSampledBvRel}{Status}{Correct}{}{Count}{10}%
\StoreBenchExecResult{RicIII}{KindReachSafetyFloatsSampledBvRel}{Status}{Correct}{True}{Count}{10}%
\StoreBenchExecResult{RicIII}{KindReachSafetyFloatsSampledBvRel}{Status}{Correct}{False}{Count}{0}%
\StoreBenchExecResult{RicIII}{KindReachSafetyFloatsSampledBvRel}{Status}{Wrong}{}{Count}{0}%
\StoreBenchExecResult{RicIII}{KindReachSafetyFloatsSampledBvRel}{Status}{Wrong}{True}{Count}{0}%
\StoreBenchExecResult{RicIII}{KindReachSafetyFloatsSampledBvRel}{Status}{Wrong}{False}{Count}{0}%
\providecommand\StoreBenchExecResult[7]{\expandafter\newcommand\csname#1#2#3#4#5#6\endcsname{#7}}%
\StoreBenchExecResult{RicIII}{KindReachSafetyHardnessSampledBvRel}{Status}{All}{}{Score}{0}%
\StoreBenchExecResult{RicIII}{KindReachSafetyHardnessSampledBvRel}{Status}{All}{}{Count}{132}%
\StoreBenchExecResult{RicIII}{KindReachSafetyHardnessSampledBvRel}{Status}{Correct}{}{Count}{132}%
\StoreBenchExecResult{RicIII}{KindReachSafetyHardnessSampledBvRel}{Status}{Correct}{True}{Count}{132}%
\StoreBenchExecResult{RicIII}{KindReachSafetyHardnessSampledBvRel}{Status}{Correct}{False}{Count}{0}%
\StoreBenchExecResult{RicIII}{KindReachSafetyHardnessSampledBvRel}{Status}{Wrong}{}{Count}{0}%
\StoreBenchExecResult{RicIII}{KindReachSafetyHardnessSampledBvRel}{Status}{Wrong}{True}{Count}{0}%
\StoreBenchExecResult{RicIII}{KindReachSafetyHardnessSampledBvRel}{Status}{Wrong}{False}{Count}{0}%
\providecommand\StoreBenchExecResult[7]{\expandafter\newcommand\csname#1#2#3#4#5#6\endcsname{#7}}%
\StoreBenchExecResult{RicIII}{KindReachSafetyHardwareSampledBvRel}{Status}{All}{}{Score}{0}%
\StoreBenchExecResult{RicIII}{KindReachSafetyHardwareSampledBvRel}{Status}{All}{}{Count}{148}%
\StoreBenchExecResult{RicIII}{KindReachSafetyHardwareSampledBvRel}{Status}{Correct}{}{Count}{54}%
\StoreBenchExecResult{RicIII}{KindReachSafetyHardwareSampledBvRel}{Status}{Correct}{True}{Count}{2}%
\StoreBenchExecResult{RicIII}{KindReachSafetyHardwareSampledBvRel}{Status}{Correct}{False}{Count}{52}%
\StoreBenchExecResult{RicIII}{KindReachSafetyHardwareSampledBvRel}{Status}{Wrong}{}{Count}{0}%
\StoreBenchExecResult{RicIII}{KindReachSafetyHardwareSampledBvRel}{Status}{Wrong}{True}{Count}{0}%
\StoreBenchExecResult{RicIII}{KindReachSafetyHardwareSampledBvRel}{Status}{Wrong}{False}{Count}{0}%
\providecommand\StoreBenchExecResult[7]{\expandafter\newcommand\csname#1#2#3#4#5#6\endcsname{#7}}%
\StoreBenchExecResult{RicIII}{KindReachSafetyHeapSampledBvRel}{Status}{All}{}{Score}{0}%
\StoreBenchExecResult{RicIII}{KindReachSafetyHeapSampledBvRel}{Status}{All}{}{Count}{6}%
\StoreBenchExecResult{RicIII}{KindReachSafetyHeapSampledBvRel}{Status}{Correct}{}{Count}{6}%
\StoreBenchExecResult{RicIII}{KindReachSafetyHeapSampledBvRel}{Status}{Correct}{True}{Count}{4}%
\StoreBenchExecResult{RicIII}{KindReachSafetyHeapSampledBvRel}{Status}{Correct}{False}{Count}{2}%
\StoreBenchExecResult{RicIII}{KindReachSafetyHeapSampledBvRel}{Status}{Wrong}{}{Count}{0}%
\StoreBenchExecResult{RicIII}{KindReachSafetyHeapSampledBvRel}{Status}{Wrong}{True}{Count}{0}%
\StoreBenchExecResult{RicIII}{KindReachSafetyHeapSampledBvRel}{Status}{Wrong}{False}{Count}{0}%
\providecommand\StoreBenchExecResult[7]{\expandafter\newcommand\csname#1#2#3#4#5#6\endcsname{#7}}%
\StoreBenchExecResult{RicIII}{KindReachSafetyLoopsSampledBvRel}{Status}{All}{}{Score}{0}%
\StoreBenchExecResult{RicIII}{KindReachSafetyLoopsSampledBvRel}{Status}{All}{}{Count}{137}%
\StoreBenchExecResult{RicIII}{KindReachSafetyLoopsSampledBvRel}{Status}{Correct}{}{Count}{52}%
\StoreBenchExecResult{RicIII}{KindReachSafetyLoopsSampledBvRel}{Status}{Correct}{True}{Count}{15}%
\StoreBenchExecResult{RicIII}{KindReachSafetyLoopsSampledBvRel}{Status}{Correct}{False}{Count}{37}%
\StoreBenchExecResult{RicIII}{KindReachSafetyLoopsSampledBvRel}{Status}{Wrong}{}{Count}{0}%
\StoreBenchExecResult{RicIII}{KindReachSafetyLoopsSampledBvRel}{Status}{Wrong}{True}{Count}{0}%
\StoreBenchExecResult{RicIII}{KindReachSafetyLoopsSampledBvRel}{Status}{Wrong}{False}{Count}{0}%
\providecommand\StoreBenchExecResult[7]{\expandafter\newcommand\csname#1#2#3#4#5#6\endcsname{#7}}%
\StoreBenchExecResult{RicIII}{KindReachSafetyProductLinesSampledBvRel}{Status}{All}{}{Score}{0}%
\StoreBenchExecResult{RicIII}{KindReachSafetyProductLinesSampledBvRel}{Status}{All}{}{Count}{150}%
\StoreBenchExecResult{RicIII}{KindReachSafetyProductLinesSampledBvRel}{Status}{Correct}{}{Count}{91}%
\StoreBenchExecResult{RicIII}{KindReachSafetyProductLinesSampledBvRel}{Status}{Correct}{True}{Count}{16}%
\StoreBenchExecResult{RicIII}{KindReachSafetyProductLinesSampledBvRel}{Status}{Correct}{False}{Count}{75}%
\StoreBenchExecResult{RicIII}{KindReachSafetyProductLinesSampledBvRel}{Status}{Wrong}{}{Count}{0}%
\StoreBenchExecResult{RicIII}{KindReachSafetyProductLinesSampledBvRel}{Status}{Wrong}{True}{Count}{0}%
\StoreBenchExecResult{RicIII}{KindReachSafetyProductLinesSampledBvRel}{Status}{Wrong}{False}{Count}{0}%
\providecommand\StoreBenchExecResult[7]{\expandafter\newcommand\csname#1#2#3#4#5#6\endcsname{#7}}%
\StoreBenchExecResult{RicIII}{KindReachSafetySequentializedSampledBvRel}{Status}{All}{}{Score}{0}%
\StoreBenchExecResult{RicIII}{KindReachSafetySequentializedSampledBvRel}{Status}{All}{}{Count}{150}%
\StoreBenchExecResult{RicIII}{KindReachSafetySequentializedSampledBvRel}{Status}{Correct}{}{Count}{100}%
\StoreBenchExecResult{RicIII}{KindReachSafetySequentializedSampledBvRel}{Status}{Correct}{True}{Count}{13}%
\StoreBenchExecResult{RicIII}{KindReachSafetySequentializedSampledBvRel}{Status}{Correct}{False}{Count}{87}%
\StoreBenchExecResult{RicIII}{KindReachSafetySequentializedSampledBvRel}{Status}{Wrong}{}{Count}{0}%
\StoreBenchExecResult{RicIII}{KindReachSafetySequentializedSampledBvRel}{Status}{Wrong}{True}{Count}{0}%
\StoreBenchExecResult{RicIII}{KindReachSafetySequentializedSampledBvRel}{Status}{Wrong}{False}{Count}{0}%
\providecommand\StoreBenchExecResult[7]{\expandafter\newcommand\csname#1#2#3#4#5#6\endcsname{#7}}%
\StoreBenchExecResult{RicIII}{KindReachSafetyXCSPSampledBvRel}{Status}{All}{}{Score}{0}%
\StoreBenchExecResult{RicIII}{KindReachSafetyXCSPSampledBvRel}{Status}{All}{}{Count}{98}%
\StoreBenchExecResult{RicIII}{KindReachSafetyXCSPSampledBvRel}{Status}{Correct}{}{Count}{93}%
\StoreBenchExecResult{RicIII}{KindReachSafetyXCSPSampledBvRel}{Status}{Correct}{True}{Count}{51}%
\StoreBenchExecResult{RicIII}{KindReachSafetyXCSPSampledBvRel}{Status}{Correct}{False}{Count}{42}%
\StoreBenchExecResult{RicIII}{KindReachSafetyXCSPSampledBvRel}{Status}{Wrong}{}{Count}{0}%
\StoreBenchExecResult{RicIII}{KindReachSafetyXCSPSampledBvRel}{Status}{Wrong}{True}{Count}{0}%
\StoreBenchExecResult{RicIII}{KindReachSafetyXCSPSampledBvRel}{Status}{Wrong}{False}{Count}{0}%
\providecommand\StoreBenchExecResult[7]{\expandafter\newcommand\csname#1#2#3#4#5#6\endcsname{#7}}%
\StoreBenchExecResult{RicIII}{KindTerminationBitVectorsSampledBvRel}{Status}{All}{}{Score}{0}%
\StoreBenchExecResult{RicIII}{KindTerminationBitVectorsSampledBvRel}{Status}{All}{}{Count}{32}%
\StoreBenchExecResult{RicIII}{KindTerminationBitVectorsSampledBvRel}{Status}{Correct}{}{Count}{21}%
\StoreBenchExecResult{RicIII}{KindTerminationBitVectorsSampledBvRel}{Status}{Correct}{True}{Count}{10}%
\StoreBenchExecResult{RicIII}{KindTerminationBitVectorsSampledBvRel}{Status}{Correct}{False}{Count}{11}%
\StoreBenchExecResult{RicIII}{KindTerminationBitVectorsSampledBvRel}{Status}{Wrong}{}{Count}{0}%
\StoreBenchExecResult{RicIII}{KindTerminationBitVectorsSampledBvRel}{Status}{Wrong}{True}{Count}{0}%
\StoreBenchExecResult{RicIII}{KindTerminationBitVectorsSampledBvRel}{Status}{Wrong}{False}{Count}{0}%
\providecommand\StoreBenchExecResult[7]{\expandafter\newcommand\csname#1#2#3#4#5#6\endcsname{#7}}%
\StoreBenchExecResult{RicIII}{KindTerminationMainControlFlowSampledBvRel}{Status}{All}{}{Score}{0}%
\StoreBenchExecResult{RicIII}{KindTerminationMainControlFlowSampledBvRel}{Status}{All}{}{Count}{236}%
\StoreBenchExecResult{RicIII}{KindTerminationMainControlFlowSampledBvRel}{Status}{Correct}{}{Count}{59}%
\StoreBenchExecResult{RicIII}{KindTerminationMainControlFlowSampledBvRel}{Status}{Correct}{True}{Count}{10}%
\StoreBenchExecResult{RicIII}{KindTerminationMainControlFlowSampledBvRel}{Status}{Correct}{False}{Count}{49}%
\StoreBenchExecResult{RicIII}{KindTerminationMainControlFlowSampledBvRel}{Status}{Wrong}{}{Count}{0}%
\StoreBenchExecResult{RicIII}{KindTerminationMainControlFlowSampledBvRel}{Status}{Wrong}{True}{Count}{0}%
\StoreBenchExecResult{RicIII}{KindTerminationMainControlFlowSampledBvRel}{Status}{Wrong}{False}{Count}{0}%
\providecommand\StoreBenchExecResult[7]{\expandafter\newcommand\csname#1#2#3#4#5#6\endcsname{#7}}%
\StoreBenchExecResult{RicIII}{KindTerminationOtherSampledBvRel}{Status}{All}{}{Score}{0}%
\StoreBenchExecResult{RicIII}{KindTerminationOtherSampledBvRel}{Status}{All}{}{Count}{973}%
\StoreBenchExecResult{RicIII}{KindTerminationOtherSampledBvRel}{Status}{Correct}{}{Count}{716}%
\StoreBenchExecResult{RicIII}{KindTerminationOtherSampledBvRel}{Status}{Correct}{True}{Count}{48}%
\StoreBenchExecResult{RicIII}{KindTerminationOtherSampledBvRel}{Status}{Correct}{False}{Count}{668}%
\StoreBenchExecResult{RicIII}{KindTerminationOtherSampledBvRel}{Status}{Wrong}{}{Count}{0}%
\StoreBenchExecResult{RicIII}{KindTerminationOtherSampledBvRel}{Status}{Wrong}{True}{Count}{0}%
\StoreBenchExecResult{RicIII}{KindTerminationOtherSampledBvRel}{Status}{Wrong}{False}{Count}{0}%
\edef\RicIIIKindReachSafetySampledBvRelStatusAllCount{\the\numexpr\RicIIIKindReachSafetyBitVectorsSampledBvRelStatusAllCount+\RicIIIKindReachSafetyCombinationsSampledBvRelStatusAllCount+\RicIIIKindReachSafetyControlFlowSampledBvRelStatusAllCount+\RicIIIKindReachSafetyECASampledBvRelStatusAllCount+\RicIIIKindReachSafetyFloatsSampledBvRelStatusAllCount+\RicIIIKindReachSafetyHardnessSampledBvRelStatusAllCount+\RicIIIKindReachSafetyHardwareSampledBvRelStatusAllCount+\RicIIIKindReachSafetyHeapSampledBvRelStatusAllCount+\RicIIIKindReachSafetyLoopsSampledBvRelStatusAllCount+\RicIIIKindReachSafetyProductLinesSampledBvRelStatusAllCount+\RicIIIKindReachSafetySequentializedSampledBvRelStatusAllCount+\RicIIIKindReachSafetyXCSPSampledBvRelStatusAllCount}
\edef\RicIIIKindReachSafetySampledBvRelStatusCorrectCount{\the\numexpr\RicIIIKindReachSafetyBitVectorsSampledBvRelStatusCorrectCount+\RicIIIKindReachSafetyCombinationsSampledBvRelStatusCorrectCount+\RicIIIKindReachSafetyControlFlowSampledBvRelStatusCorrectCount+\RicIIIKindReachSafetyECASampledBvRelStatusCorrectCount+\RicIIIKindReachSafetyFloatsSampledBvRelStatusCorrectCount+\RicIIIKindReachSafetyHardnessSampledBvRelStatusCorrectCount+\RicIIIKindReachSafetyHardwareSampledBvRelStatusCorrectCount+\RicIIIKindReachSafetyHeapSampledBvRelStatusCorrectCount+\RicIIIKindReachSafetyLoopsSampledBvRelStatusCorrectCount+\RicIIIKindReachSafetyProductLinesSampledBvRelStatusCorrectCount+\RicIIIKindReachSafetySequentializedSampledBvRelStatusCorrectCount+\RicIIIKindReachSafetyXCSPSampledBvRelStatusCorrectCount}
\edef\RicIIIKindReachSafetySampledBvRelStatusCorrectTrueCount{\the\numexpr\RicIIIKindReachSafetyBitVectorsSampledBvRelStatusCorrectTrueCount+\RicIIIKindReachSafetyCombinationsSampledBvRelStatusCorrectTrueCount+\RicIIIKindReachSafetyControlFlowSampledBvRelStatusCorrectTrueCount+\RicIIIKindReachSafetyECASampledBvRelStatusCorrectTrueCount+\RicIIIKindReachSafetyFloatsSampledBvRelStatusCorrectTrueCount+\RicIIIKindReachSafetyHardnessSampledBvRelStatusCorrectTrueCount+\RicIIIKindReachSafetyHardwareSampledBvRelStatusCorrectTrueCount+\RicIIIKindReachSafetyHeapSampledBvRelStatusCorrectTrueCount+\RicIIIKindReachSafetyLoopsSampledBvRelStatusCorrectTrueCount+\RicIIIKindReachSafetyProductLinesSampledBvRelStatusCorrectTrueCount+\RicIIIKindReachSafetySequentializedSampledBvRelStatusCorrectTrueCount+\RicIIIKindReachSafetyXCSPSampledBvRelStatusCorrectTrueCount}
\edef\RicIIIKindReachSafetySampledBvRelStatusCorrectFalseCount{\the\numexpr\RicIIIKindReachSafetyBitVectorsSampledBvRelStatusCorrectFalseCount+\RicIIIKindReachSafetyCombinationsSampledBvRelStatusCorrectFalseCount+\RicIIIKindReachSafetyControlFlowSampledBvRelStatusCorrectFalseCount+\RicIIIKindReachSafetyECASampledBvRelStatusCorrectFalseCount+\RicIIIKindReachSafetyFloatsSampledBvRelStatusCorrectFalseCount+\RicIIIKindReachSafetyHardnessSampledBvRelStatusCorrectFalseCount+\RicIIIKindReachSafetyHardwareSampledBvRelStatusCorrectFalseCount+\RicIIIKindReachSafetyHeapSampledBvRelStatusCorrectFalseCount+\RicIIIKindReachSafetyLoopsSampledBvRelStatusCorrectFalseCount+\RicIIIKindReachSafetyProductLinesSampledBvRelStatusCorrectFalseCount+\RicIIIKindReachSafetySequentializedSampledBvRelStatusCorrectFalseCount+\RicIIIKindReachSafetyXCSPSampledBvRelStatusCorrectFalseCount}
\edef\RicIIIKindReachSafetySampledBvRelStatusWrongCount{\the\numexpr\RicIIIKindReachSafetyBitVectorsSampledBvRelStatusWrongCount+\RicIIIKindReachSafetyCombinationsSampledBvRelStatusWrongCount+\RicIIIKindReachSafetyControlFlowSampledBvRelStatusWrongCount+\RicIIIKindReachSafetyECASampledBvRelStatusWrongCount+\RicIIIKindReachSafetyFloatsSampledBvRelStatusWrongCount+\RicIIIKindReachSafetyHardnessSampledBvRelStatusWrongCount+\RicIIIKindReachSafetyHardwareSampledBvRelStatusWrongCount+\RicIIIKindReachSafetyHeapSampledBvRelStatusWrongCount+\RicIIIKindReachSafetyLoopsSampledBvRelStatusWrongCount+\RicIIIKindReachSafetyProductLinesSampledBvRelStatusWrongCount+\RicIIIKindReachSafetySequentializedSampledBvRelStatusWrongCount+\RicIIIKindReachSafetyXCSPSampledBvRelStatusWrongCount}
\edef\RicIIIKindTerminationSampledBvRelStatusAllCount{\the\numexpr\RicIIIKindTerminationBitVectorsSampledBvRelStatusAllCount+\RicIIIKindTerminationMainControlFlowSampledBvRelStatusAllCount+\RicIIIKindTerminationOtherSampledBvRelStatusAllCount}
\edef\RicIIIKindTerminationSampledBvRelStatusCorrectCount{\the\numexpr\RicIIIKindTerminationBitVectorsSampledBvRelStatusCorrectCount+\RicIIIKindTerminationMainControlFlowSampledBvRelStatusCorrectCount+\RicIIIKindTerminationOtherSampledBvRelStatusCorrectCount}
\edef\RicIIIKindTerminationSampledBvRelStatusCorrectTrueCount{\the\numexpr\RicIIIKindTerminationBitVectorsSampledBvRelStatusCorrectTrueCount+\RicIIIKindTerminationMainControlFlowSampledBvRelStatusCorrectTrueCount+\RicIIIKindTerminationOtherSampledBvRelStatusCorrectTrueCount}
\edef\RicIIIKindTerminationSampledBvRelStatusCorrectFalseCount{\the\numexpr\RicIIIKindTerminationBitVectorsSampledBvRelStatusCorrectFalseCount+\RicIIIKindTerminationMainControlFlowSampledBvRelStatusCorrectFalseCount+\RicIIIKindTerminationOtherSampledBvRelStatusCorrectFalseCount}
\edef\RicIIIKindTerminationSampledBvRelStatusWrongCount{\the\numexpr\RicIIIKindTerminationBitVectorsSampledBvRelStatusWrongCount+\RicIIIKindTerminationMainControlFlowSampledBvRelStatusWrongCount+\RicIIIKindTerminationOtherSampledBvRelStatusWrongCount}
\providecommand\StoreBenchExecResult[7]{\expandafter\newcommand\csname#1#2#3#4#5#6\endcsname{#7}}%
\StoreBenchExecResult{Pono}{IcIIIiaMsatReachSafetyArraysSampledRel}{Status}{All}{}{Score}{0}%
\StoreBenchExecResult{Pono}{IcIIIiaMsatReachSafetyArraysSampledRel}{Status}{All}{}{Count}{150}%
\StoreBenchExecResult{Pono}{IcIIIiaMsatReachSafetyArraysSampledRel}{Status}{Correct}{}{Count}{38}%
\StoreBenchExecResult{Pono}{IcIIIiaMsatReachSafetyArraysSampledRel}{Status}{Correct}{True}{Count}{0}%
\StoreBenchExecResult{Pono}{IcIIIiaMsatReachSafetyArraysSampledRel}{Status}{Correct}{False}{Count}{38}%
\StoreBenchExecResult{Pono}{IcIIIiaMsatReachSafetyArraysSampledRel}{Status}{Wrong}{}{Count}{0}%
\StoreBenchExecResult{Pono}{IcIIIiaMsatReachSafetyArraysSampledRel}{Status}{Wrong}{True}{Count}{0}%
\StoreBenchExecResult{Pono}{IcIIIiaMsatReachSafetyArraysSampledRel}{Status}{Wrong}{False}{Count}{0}%
\providecommand\StoreBenchExecResult[7]{\expandafter\newcommand\csname#1#2#3#4#5#6\endcsname{#7}}%
\StoreBenchExecResult{Pono}{IcIIIiaMsatReachSafetyBitVectorsSampledRel}{Status}{All}{}{Score}{0}%
\StoreBenchExecResult{Pono}{IcIIIiaMsatReachSafetyBitVectorsSampledRel}{Status}{All}{}{Count}{48}%
\StoreBenchExecResult{Pono}{IcIIIiaMsatReachSafetyBitVectorsSampledRel}{Status}{Correct}{}{Count}{25}%
\StoreBenchExecResult{Pono}{IcIIIiaMsatReachSafetyBitVectorsSampledRel}{Status}{Correct}{True}{Count}{16}%
\StoreBenchExecResult{Pono}{IcIIIiaMsatReachSafetyBitVectorsSampledRel}{Status}{Correct}{False}{Count}{9}%
\StoreBenchExecResult{Pono}{IcIIIiaMsatReachSafetyBitVectorsSampledRel}{Status}{Wrong}{}{Count}{0}%
\StoreBenchExecResult{Pono}{IcIIIiaMsatReachSafetyBitVectorsSampledRel}{Status}{Wrong}{True}{Count}{0}%
\StoreBenchExecResult{Pono}{IcIIIiaMsatReachSafetyBitVectorsSampledRel}{Status}{Wrong}{False}{Count}{0}%
\providecommand\StoreBenchExecResult[7]{\expandafter\newcommand\csname#1#2#3#4#5#6\endcsname{#7}}%
\StoreBenchExecResult{Pono}{IcIIIiaMsatReachSafetyCombinationsSampledRel}{Status}{All}{}{Score}{0}%
\StoreBenchExecResult{Pono}{IcIIIiaMsatReachSafetyCombinationsSampledRel}{Status}{All}{}{Count}{110}%
\StoreBenchExecResult{Pono}{IcIIIiaMsatReachSafetyCombinationsSampledRel}{Status}{Correct}{}{Count}{37}%
\StoreBenchExecResult{Pono}{IcIIIiaMsatReachSafetyCombinationsSampledRel}{Status}{Correct}{True}{Count}{1}%
\StoreBenchExecResult{Pono}{IcIIIiaMsatReachSafetyCombinationsSampledRel}{Status}{Correct}{False}{Count}{36}%
\StoreBenchExecResult{Pono}{IcIIIiaMsatReachSafetyCombinationsSampledRel}{Status}{Wrong}{}{Count}{0}%
\StoreBenchExecResult{Pono}{IcIIIiaMsatReachSafetyCombinationsSampledRel}{Status}{Wrong}{True}{Count}{0}%
\StoreBenchExecResult{Pono}{IcIIIiaMsatReachSafetyCombinationsSampledRel}{Status}{Wrong}{False}{Count}{0}%
\providecommand\StoreBenchExecResult[7]{\expandafter\newcommand\csname#1#2#3#4#5#6\endcsname{#7}}%
\StoreBenchExecResult{Pono}{IcIIIiaMsatReachSafetyControlFlowSampledRel}{Status}{All}{}{Score}{0}%
\StoreBenchExecResult{Pono}{IcIIIiaMsatReachSafetyControlFlowSampledRel}{Status}{All}{}{Count}{33}%
\StoreBenchExecResult{Pono}{IcIIIiaMsatReachSafetyControlFlowSampledRel}{Status}{Correct}{}{Count}{27}%
\StoreBenchExecResult{Pono}{IcIIIiaMsatReachSafetyControlFlowSampledRel}{Status}{Correct}{True}{Count}{25}%
\StoreBenchExecResult{Pono}{IcIIIiaMsatReachSafetyControlFlowSampledRel}{Status}{Correct}{False}{Count}{2}%
\StoreBenchExecResult{Pono}{IcIIIiaMsatReachSafetyControlFlowSampledRel}{Status}{Wrong}{}{Count}{0}%
\StoreBenchExecResult{Pono}{IcIIIiaMsatReachSafetyControlFlowSampledRel}{Status}{Wrong}{True}{Count}{0}%
\StoreBenchExecResult{Pono}{IcIIIiaMsatReachSafetyControlFlowSampledRel}{Status}{Wrong}{False}{Count}{0}%
\providecommand\StoreBenchExecResult[7]{\expandafter\newcommand\csname#1#2#3#4#5#6\endcsname{#7}}%
\StoreBenchExecResult{Pono}{IcIIIiaMsatReachSafetyECASampledRel}{Status}{All}{}{Score}{0}%
\StoreBenchExecResult{Pono}{IcIIIiaMsatReachSafetyECASampledRel}{Status}{All}{}{Count}{150}%
\StoreBenchExecResult{Pono}{IcIIIiaMsatReachSafetyECASampledRel}{Status}{Correct}{}{Count}{23}%
\StoreBenchExecResult{Pono}{IcIIIiaMsatReachSafetyECASampledRel}{Status}{Correct}{True}{Count}{20}%
\StoreBenchExecResult{Pono}{IcIIIiaMsatReachSafetyECASampledRel}{Status}{Correct}{False}{Count}{3}%
\StoreBenchExecResult{Pono}{IcIIIiaMsatReachSafetyECASampledRel}{Status}{Wrong}{}{Count}{0}%
\StoreBenchExecResult{Pono}{IcIIIiaMsatReachSafetyECASampledRel}{Status}{Wrong}{True}{Count}{0}%
\StoreBenchExecResult{Pono}{IcIIIiaMsatReachSafetyECASampledRel}{Status}{Wrong}{False}{Count}{0}%
\providecommand\StoreBenchExecResult[7]{\expandafter\newcommand\csname#1#2#3#4#5#6\endcsname{#7}}%
\StoreBenchExecResult{Pono}{IcIIIiaMsatReachSafetyFloatsSampledRel}{Status}{All}{}{Score}{0}%
\StoreBenchExecResult{Pono}{IcIIIiaMsatReachSafetyFloatsSampledRel}{Status}{All}{}{Count}{10}%
\StoreBenchExecResult{Pono}{IcIIIiaMsatReachSafetyFloatsSampledRel}{Status}{Correct}{}{Count}{10}%
\StoreBenchExecResult{Pono}{IcIIIiaMsatReachSafetyFloatsSampledRel}{Status}{Correct}{True}{Count}{10}%
\StoreBenchExecResult{Pono}{IcIIIiaMsatReachSafetyFloatsSampledRel}{Status}{Correct}{False}{Count}{0}%
\StoreBenchExecResult{Pono}{IcIIIiaMsatReachSafetyFloatsSampledRel}{Status}{Wrong}{}{Count}{0}%
\StoreBenchExecResult{Pono}{IcIIIiaMsatReachSafetyFloatsSampledRel}{Status}{Wrong}{True}{Count}{0}%
\StoreBenchExecResult{Pono}{IcIIIiaMsatReachSafetyFloatsSampledRel}{Status}{Wrong}{False}{Count}{0}%
\providecommand\StoreBenchExecResult[7]{\expandafter\newcommand\csname#1#2#3#4#5#6\endcsname{#7}}%
\StoreBenchExecResult{Pono}{IcIIIiaMsatReachSafetyHardnessSampledRel}{Status}{All}{}{Score}{0}%
\StoreBenchExecResult{Pono}{IcIIIiaMsatReachSafetyHardnessSampledRel}{Status}{All}{}{Count}{150}%
\StoreBenchExecResult{Pono}{IcIIIiaMsatReachSafetyHardnessSampledRel}{Status}{Correct}{}{Count}{142}%
\StoreBenchExecResult{Pono}{IcIIIiaMsatReachSafetyHardnessSampledRel}{Status}{Correct}{True}{Count}{142}%
\StoreBenchExecResult{Pono}{IcIIIiaMsatReachSafetyHardnessSampledRel}{Status}{Correct}{False}{Count}{0}%
\StoreBenchExecResult{Pono}{IcIIIiaMsatReachSafetyHardnessSampledRel}{Status}{Wrong}{}{Count}{0}%
\StoreBenchExecResult{Pono}{IcIIIiaMsatReachSafetyHardnessSampledRel}{Status}{Wrong}{True}{Count}{0}%
\StoreBenchExecResult{Pono}{IcIIIiaMsatReachSafetyHardnessSampledRel}{Status}{Wrong}{False}{Count}{0}%
\providecommand\StoreBenchExecResult[7]{\expandafter\newcommand\csname#1#2#3#4#5#6\endcsname{#7}}%
\StoreBenchExecResult{Pono}{IcIIIiaMsatReachSafetyHardwareSampledRel}{Status}{All}{}{Score}{0}%
\StoreBenchExecResult{Pono}{IcIIIiaMsatReachSafetyHardwareSampledRel}{Status}{All}{}{Count}{150}%
\StoreBenchExecResult{Pono}{IcIIIiaMsatReachSafetyHardwareSampledRel}{Status}{Correct}{}{Count}{24}%
\StoreBenchExecResult{Pono}{IcIIIiaMsatReachSafetyHardwareSampledRel}{Status}{Correct}{True}{Count}{7}%
\StoreBenchExecResult{Pono}{IcIIIiaMsatReachSafetyHardwareSampledRel}{Status}{Correct}{False}{Count}{17}%
\StoreBenchExecResult{Pono}{IcIIIiaMsatReachSafetyHardwareSampledRel}{Status}{Wrong}{}{Count}{0}%
\StoreBenchExecResult{Pono}{IcIIIiaMsatReachSafetyHardwareSampledRel}{Status}{Wrong}{True}{Count}{0}%
\StoreBenchExecResult{Pono}{IcIIIiaMsatReachSafetyHardwareSampledRel}{Status}{Wrong}{False}{Count}{0}%
\providecommand\StoreBenchExecResult[7]{\expandafter\newcommand\csname#1#2#3#4#5#6\endcsname{#7}}%
\StoreBenchExecResult{Pono}{IcIIIiaMsatReachSafetyHeapSampledRel}{Status}{All}{}{Score}{0}%
\StoreBenchExecResult{Pono}{IcIIIiaMsatReachSafetyHeapSampledRel}{Status}{All}{}{Count}{53}%
\StoreBenchExecResult{Pono}{IcIIIiaMsatReachSafetyHeapSampledRel}{Status}{Correct}{}{Count}{47}%
\StoreBenchExecResult{Pono}{IcIIIiaMsatReachSafetyHeapSampledRel}{Status}{Correct}{True}{Count}{32}%
\StoreBenchExecResult{Pono}{IcIIIiaMsatReachSafetyHeapSampledRel}{Status}{Correct}{False}{Count}{15}%
\StoreBenchExecResult{Pono}{IcIIIiaMsatReachSafetyHeapSampledRel}{Status}{Wrong}{}{Count}{0}%
\StoreBenchExecResult{Pono}{IcIIIiaMsatReachSafetyHeapSampledRel}{Status}{Wrong}{True}{Count}{0}%
\StoreBenchExecResult{Pono}{IcIIIiaMsatReachSafetyHeapSampledRel}{Status}{Wrong}{False}{Count}{0}%
\providecommand\StoreBenchExecResult[7]{\expandafter\newcommand\csname#1#2#3#4#5#6\endcsname{#7}}%
\StoreBenchExecResult{Pono}{IcIIIiaMsatReachSafetyLoopsSampledRel}{Status}{All}{}{Score}{0}%
\StoreBenchExecResult{Pono}{IcIIIiaMsatReachSafetyLoopsSampledRel}{Status}{All}{}{Count}{150}%
\StoreBenchExecResult{Pono}{IcIIIiaMsatReachSafetyLoopsSampledRel}{Status}{Correct}{}{Count}{34}%
\StoreBenchExecResult{Pono}{IcIIIiaMsatReachSafetyLoopsSampledRel}{Status}{Correct}{True}{Count}{11}%
\StoreBenchExecResult{Pono}{IcIIIiaMsatReachSafetyLoopsSampledRel}{Status}{Correct}{False}{Count}{23}%
\StoreBenchExecResult{Pono}{IcIIIiaMsatReachSafetyLoopsSampledRel}{Status}{Wrong}{}{Count}{0}%
\StoreBenchExecResult{Pono}{IcIIIiaMsatReachSafetyLoopsSampledRel}{Status}{Wrong}{True}{Count}{0}%
\StoreBenchExecResult{Pono}{IcIIIiaMsatReachSafetyLoopsSampledRel}{Status}{Wrong}{False}{Count}{0}%
\providecommand\StoreBenchExecResult[7]{\expandafter\newcommand\csname#1#2#3#4#5#6\endcsname{#7}}%
\StoreBenchExecResult{Pono}{IcIIIiaMsatReachSafetyProductLinesSampledRel}{Status}{All}{}{Score}{0}%
\StoreBenchExecResult{Pono}{IcIIIiaMsatReachSafetyProductLinesSampledRel}{Status}{All}{}{Count}{150}%
\StoreBenchExecResult{Pono}{IcIIIiaMsatReachSafetyProductLinesSampledRel}{Status}{Correct}{}{Count}{137}%
\StoreBenchExecResult{Pono}{IcIIIiaMsatReachSafetyProductLinesSampledRel}{Status}{Correct}{True}{Count}{69}%
\StoreBenchExecResult{Pono}{IcIIIiaMsatReachSafetyProductLinesSampledRel}{Status}{Correct}{False}{Count}{68}%
\StoreBenchExecResult{Pono}{IcIIIiaMsatReachSafetyProductLinesSampledRel}{Status}{Wrong}{}{Count}{0}%
\StoreBenchExecResult{Pono}{IcIIIiaMsatReachSafetyProductLinesSampledRel}{Status}{Wrong}{True}{Count}{0}%
\StoreBenchExecResult{Pono}{IcIIIiaMsatReachSafetyProductLinesSampledRel}{Status}{Wrong}{False}{Count}{0}%
\providecommand\StoreBenchExecResult[7]{\expandafter\newcommand\csname#1#2#3#4#5#6\endcsname{#7}}%
\StoreBenchExecResult{Pono}{IcIIIiaMsatReachSafetySequentializedSampledRel}{Status}{All}{}{Score}{0}%
\StoreBenchExecResult{Pono}{IcIIIiaMsatReachSafetySequentializedSampledRel}{Status}{All}{}{Count}{150}%
\StoreBenchExecResult{Pono}{IcIIIiaMsatReachSafetySequentializedSampledRel}{Status}{Correct}{}{Count}{44}%
\StoreBenchExecResult{Pono}{IcIIIiaMsatReachSafetySequentializedSampledRel}{Status}{Correct}{True}{Count}{3}%
\StoreBenchExecResult{Pono}{IcIIIiaMsatReachSafetySequentializedSampledRel}{Status}{Correct}{False}{Count}{41}%
\StoreBenchExecResult{Pono}{IcIIIiaMsatReachSafetySequentializedSampledRel}{Status}{Wrong}{}{Count}{0}%
\StoreBenchExecResult{Pono}{IcIIIiaMsatReachSafetySequentializedSampledRel}{Status}{Wrong}{True}{Count}{0}%
\StoreBenchExecResult{Pono}{IcIIIiaMsatReachSafetySequentializedSampledRel}{Status}{Wrong}{False}{Count}{0}%
\providecommand\StoreBenchExecResult[7]{\expandafter\newcommand\csname#1#2#3#4#5#6\endcsname{#7}}%
\StoreBenchExecResult{Pono}{IcIIIiaMsatReachSafetyXCSPSampledRel}{Status}{All}{}{Score}{0}%
\StoreBenchExecResult{Pono}{IcIIIiaMsatReachSafetyXCSPSampledRel}{Status}{All}{}{Count}{98}%
\StoreBenchExecResult{Pono}{IcIIIiaMsatReachSafetyXCSPSampledRel}{Status}{Correct}{}{Count}{90}%
\StoreBenchExecResult{Pono}{IcIIIiaMsatReachSafetyXCSPSampledRel}{Status}{Correct}{True}{Count}{52}%
\StoreBenchExecResult{Pono}{IcIIIiaMsatReachSafetyXCSPSampledRel}{Status}{Correct}{False}{Count}{38}%
\StoreBenchExecResult{Pono}{IcIIIiaMsatReachSafetyXCSPSampledRel}{Status}{Wrong}{}{Count}{0}%
\StoreBenchExecResult{Pono}{IcIIIiaMsatReachSafetyXCSPSampledRel}{Status}{Wrong}{True}{Count}{0}%
\StoreBenchExecResult{Pono}{IcIIIiaMsatReachSafetyXCSPSampledRel}{Status}{Wrong}{False}{Count}{0}%
\providecommand\StoreBenchExecResult[7]{\expandafter\newcommand\csname#1#2#3#4#5#6\endcsname{#7}}%
\StoreBenchExecResult{Pono}{IcIIIiaMsatTerminationBitVectorsSampledRel}{Status}{All}{}{Score}{0}%
\StoreBenchExecResult{Pono}{IcIIIiaMsatTerminationBitVectorsSampledRel}{Status}{All}{}{Count}{32}%
\StoreBenchExecResult{Pono}{IcIIIiaMsatTerminationBitVectorsSampledRel}{Status}{Correct}{}{Count}{19}%
\StoreBenchExecResult{Pono}{IcIIIiaMsatTerminationBitVectorsSampledRel}{Status}{Correct}{True}{Count}{8}%
\StoreBenchExecResult{Pono}{IcIIIiaMsatTerminationBitVectorsSampledRel}{Status}{Correct}{False}{Count}{11}%
\StoreBenchExecResult{Pono}{IcIIIiaMsatTerminationBitVectorsSampledRel}{Status}{Wrong}{}{Count}{0}%
\StoreBenchExecResult{Pono}{IcIIIiaMsatTerminationBitVectorsSampledRel}{Status}{Wrong}{True}{Count}{0}%
\StoreBenchExecResult{Pono}{IcIIIiaMsatTerminationBitVectorsSampledRel}{Status}{Wrong}{False}{Count}{0}%
\providecommand\StoreBenchExecResult[7]{\expandafter\newcommand\csname#1#2#3#4#5#6\endcsname{#7}}%
\StoreBenchExecResult{Pono}{IcIIIiaMsatTerminationMainControlFlowSampledRel}{Status}{All}{}{Score}{0}%
\StoreBenchExecResult{Pono}{IcIIIiaMsatTerminationMainControlFlowSampledRel}{Status}{All}{}{Count}{242}%
\StoreBenchExecResult{Pono}{IcIIIiaMsatTerminationMainControlFlowSampledRel}{Status}{Correct}{}{Count}{62}%
\StoreBenchExecResult{Pono}{IcIIIiaMsatTerminationMainControlFlowSampledRel}{Status}{Correct}{True}{Count}{16}%
\StoreBenchExecResult{Pono}{IcIIIiaMsatTerminationMainControlFlowSampledRel}{Status}{Correct}{False}{Count}{46}%
\StoreBenchExecResult{Pono}{IcIIIiaMsatTerminationMainControlFlowSampledRel}{Status}{Wrong}{}{Count}{0}%
\StoreBenchExecResult{Pono}{IcIIIiaMsatTerminationMainControlFlowSampledRel}{Status}{Wrong}{True}{Count}{0}%
\StoreBenchExecResult{Pono}{IcIIIiaMsatTerminationMainControlFlowSampledRel}{Status}{Wrong}{False}{Count}{0}%
\providecommand\StoreBenchExecResult[7]{\expandafter\newcommand\csname#1#2#3#4#5#6\endcsname{#7}}%
\StoreBenchExecResult{Pono}{IcIIIiaMsatTerminationOtherSampledRel}{Status}{All}{}{Score}{0}%
\StoreBenchExecResult{Pono}{IcIIIiaMsatTerminationOtherSampledRel}{Status}{All}{}{Count}{1080}%
\StoreBenchExecResult{Pono}{IcIIIiaMsatTerminationOtherSampledRel}{Status}{Correct}{}{Count}{674}%
\StoreBenchExecResult{Pono}{IcIIIiaMsatTerminationOtherSampledRel}{Status}{Correct}{True}{Count}{138}%
\StoreBenchExecResult{Pono}{IcIIIiaMsatTerminationOtherSampledRel}{Status}{Correct}{False}{Count}{536}%
\StoreBenchExecResult{Pono}{IcIIIiaMsatTerminationOtherSampledRel}{Status}{Wrong}{}{Count}{0}%
\StoreBenchExecResult{Pono}{IcIIIiaMsatTerminationOtherSampledRel}{Status}{Wrong}{True}{Count}{0}%
\StoreBenchExecResult{Pono}{IcIIIiaMsatTerminationOtherSampledRel}{Status}{Wrong}{False}{Count}{0}%
\edef\PonoIcIIIiaMsatReachSafetySampledRelStatusAllCount{\the\numexpr\PonoIcIIIiaMsatReachSafetyArraysSampledRelStatusAllCount+\PonoIcIIIiaMsatReachSafetyBitVectorsSampledRelStatusAllCount+\PonoIcIIIiaMsatReachSafetyCombinationsSampledRelStatusAllCount+\PonoIcIIIiaMsatReachSafetyControlFlowSampledRelStatusAllCount+\PonoIcIIIiaMsatReachSafetyECASampledRelStatusAllCount+\PonoIcIIIiaMsatReachSafetyFloatsSampledRelStatusAllCount+\PonoIcIIIiaMsatReachSafetyHardnessSampledRelStatusAllCount+\PonoIcIIIiaMsatReachSafetyHardwareSampledRelStatusAllCount+\PonoIcIIIiaMsatReachSafetyHeapSampledRelStatusAllCount+\PonoIcIIIiaMsatReachSafetyLoopsSampledRelStatusAllCount+\PonoIcIIIiaMsatReachSafetyProductLinesSampledRelStatusAllCount+\PonoIcIIIiaMsatReachSafetySequentializedSampledRelStatusAllCount+\PonoIcIIIiaMsatReachSafetyXCSPSampledRelStatusAllCount}
\edef\PonoIcIIIiaMsatReachSafetySampledRelStatusCorrectCount{\the\numexpr\PonoIcIIIiaMsatReachSafetyArraysSampledRelStatusCorrectCount+\PonoIcIIIiaMsatReachSafetyBitVectorsSampledRelStatusCorrectCount+\PonoIcIIIiaMsatReachSafetyCombinationsSampledRelStatusCorrectCount+\PonoIcIIIiaMsatReachSafetyControlFlowSampledRelStatusCorrectCount+\PonoIcIIIiaMsatReachSafetyECASampledRelStatusCorrectCount+\PonoIcIIIiaMsatReachSafetyFloatsSampledRelStatusCorrectCount+\PonoIcIIIiaMsatReachSafetyHardnessSampledRelStatusCorrectCount+\PonoIcIIIiaMsatReachSafetyHardwareSampledRelStatusCorrectCount+\PonoIcIIIiaMsatReachSafetyHeapSampledRelStatusCorrectCount+\PonoIcIIIiaMsatReachSafetyLoopsSampledRelStatusCorrectCount+\PonoIcIIIiaMsatReachSafetyProductLinesSampledRelStatusCorrectCount+\PonoIcIIIiaMsatReachSafetySequentializedSampledRelStatusCorrectCount+\PonoIcIIIiaMsatReachSafetyXCSPSampledRelStatusCorrectCount}
\edef\PonoIcIIIiaMsatReachSafetySampledRelStatusCorrectTrueCount{\the\numexpr\PonoIcIIIiaMsatReachSafetyArraysSampledRelStatusCorrectTrueCount+\PonoIcIIIiaMsatReachSafetyBitVectorsSampledRelStatusCorrectTrueCount+\PonoIcIIIiaMsatReachSafetyCombinationsSampledRelStatusCorrectTrueCount+\PonoIcIIIiaMsatReachSafetyControlFlowSampledRelStatusCorrectTrueCount+\PonoIcIIIiaMsatReachSafetyECASampledRelStatusCorrectTrueCount+\PonoIcIIIiaMsatReachSafetyFloatsSampledRelStatusCorrectTrueCount+\PonoIcIIIiaMsatReachSafetyHardnessSampledRelStatusCorrectTrueCount+\PonoIcIIIiaMsatReachSafetyHardwareSampledRelStatusCorrectTrueCount+\PonoIcIIIiaMsatReachSafetyHeapSampledRelStatusCorrectTrueCount+\PonoIcIIIiaMsatReachSafetyLoopsSampledRelStatusCorrectTrueCount+\PonoIcIIIiaMsatReachSafetyProductLinesSampledRelStatusCorrectTrueCount+\PonoIcIIIiaMsatReachSafetySequentializedSampledRelStatusCorrectTrueCount+\PonoIcIIIiaMsatReachSafetyXCSPSampledRelStatusCorrectTrueCount}
\edef\PonoIcIIIiaMsatReachSafetySampledRelStatusCorrectFalseCount{\the\numexpr\PonoIcIIIiaMsatReachSafetyArraysSampledRelStatusCorrectFalseCount+\PonoIcIIIiaMsatReachSafetyBitVectorsSampledRelStatusCorrectFalseCount+\PonoIcIIIiaMsatReachSafetyCombinationsSampledRelStatusCorrectFalseCount+\PonoIcIIIiaMsatReachSafetyControlFlowSampledRelStatusCorrectFalseCount+\PonoIcIIIiaMsatReachSafetyECASampledRelStatusCorrectFalseCount+\PonoIcIIIiaMsatReachSafetyFloatsSampledRelStatusCorrectFalseCount+\PonoIcIIIiaMsatReachSafetyHardnessSampledRelStatusCorrectFalseCount+\PonoIcIIIiaMsatReachSafetyHardwareSampledRelStatusCorrectFalseCount+\PonoIcIIIiaMsatReachSafetyHeapSampledRelStatusCorrectFalseCount+\PonoIcIIIiaMsatReachSafetyLoopsSampledRelStatusCorrectFalseCount+\PonoIcIIIiaMsatReachSafetyProductLinesSampledRelStatusCorrectFalseCount+\PonoIcIIIiaMsatReachSafetySequentializedSampledRelStatusCorrectFalseCount+\PonoIcIIIiaMsatReachSafetyXCSPSampledRelStatusCorrectFalseCount}
\edef\PonoIcIIIiaMsatReachSafetySampledRelStatusWrongCount{\the\numexpr\PonoIcIIIiaMsatReachSafetyArraysSampledRelStatusWrongCount+\PonoIcIIIiaMsatReachSafetyBitVectorsSampledRelStatusWrongCount+\PonoIcIIIiaMsatReachSafetyCombinationsSampledRelStatusWrongCount+\PonoIcIIIiaMsatReachSafetyControlFlowSampledRelStatusWrongCount+\PonoIcIIIiaMsatReachSafetyECASampledRelStatusWrongCount+\PonoIcIIIiaMsatReachSafetyFloatsSampledRelStatusWrongCount+\PonoIcIIIiaMsatReachSafetyHardnessSampledRelStatusWrongCount+\PonoIcIIIiaMsatReachSafetyHardwareSampledRelStatusWrongCount+\PonoIcIIIiaMsatReachSafetyHeapSampledRelStatusWrongCount+\PonoIcIIIiaMsatReachSafetyLoopsSampledRelStatusWrongCount+\PonoIcIIIiaMsatReachSafetyProductLinesSampledRelStatusWrongCount+\PonoIcIIIiaMsatReachSafetySequentializedSampledRelStatusWrongCount+\PonoIcIIIiaMsatReachSafetyXCSPSampledRelStatusWrongCount}
\edef\PonoIcIIIiaMsatTerminationSampledRelStatusAllCount{\the\numexpr\PonoIcIIIiaMsatTerminationBitVectorsSampledRelStatusAllCount+\PonoIcIIIiaMsatTerminationMainControlFlowSampledRelStatusAllCount+\PonoIcIIIiaMsatTerminationOtherSampledRelStatusAllCount}
\edef\PonoIcIIIiaMsatTerminationSampledRelStatusCorrectCount{\the\numexpr\PonoIcIIIiaMsatTerminationBitVectorsSampledRelStatusCorrectCount+\PonoIcIIIiaMsatTerminationMainControlFlowSampledRelStatusCorrectCount+\PonoIcIIIiaMsatTerminationOtherSampledRelStatusCorrectCount}
\edef\PonoIcIIIiaMsatTerminationSampledRelStatusCorrectTrueCount{\the\numexpr\PonoIcIIIiaMsatTerminationBitVectorsSampledRelStatusCorrectTrueCount+\PonoIcIIIiaMsatTerminationMainControlFlowSampledRelStatusCorrectTrueCount+\PonoIcIIIiaMsatTerminationOtherSampledRelStatusCorrectTrueCount}
\edef\PonoIcIIIiaMsatTerminationSampledRelStatusCorrectFalseCount{\the\numexpr\PonoIcIIIiaMsatTerminationBitVectorsSampledRelStatusCorrectFalseCount+\PonoIcIIIiaMsatTerminationMainControlFlowSampledRelStatusCorrectFalseCount+\PonoIcIIIiaMsatTerminationOtherSampledRelStatusCorrectFalseCount}
\edef\PonoIcIIIiaMsatTerminationSampledRelStatusWrongCount{\the\numexpr\PonoIcIIIiaMsatTerminationBitVectorsSampledRelStatusWrongCount+\PonoIcIIIiaMsatTerminationMainControlFlowSampledRelStatusWrongCount+\PonoIcIIIiaMsatTerminationOtherSampledRelStatusWrongCount}
\providecommand\StoreBenchExecResult[7]{\expandafter\newcommand\csname#1#2#3#4#5#6\endcsname{#7}}%
\StoreBenchExecResult{Pono}{ImcBzlaReachSafetyBitVectorsSampledBvRel}{Status}{All}{}{Score}{0}%
\StoreBenchExecResult{Pono}{ImcBzlaReachSafetyBitVectorsSampledBvRel}{Status}{All}{}{Count}{47}%
\StoreBenchExecResult{Pono}{ImcBzlaReachSafetyBitVectorsSampledBvRel}{Status}{Correct}{}{Count}{39}%
\StoreBenchExecResult{Pono}{ImcBzlaReachSafetyBitVectorsSampledBvRel}{Status}{Correct}{True}{Count}{28}%
\StoreBenchExecResult{Pono}{ImcBzlaReachSafetyBitVectorsSampledBvRel}{Status}{Correct}{False}{Count}{11}%
\StoreBenchExecResult{Pono}{ImcBzlaReachSafetyBitVectorsSampledBvRel}{Status}{Wrong}{}{Count}{0}%
\StoreBenchExecResult{Pono}{ImcBzlaReachSafetyBitVectorsSampledBvRel}{Status}{Wrong}{True}{Count}{0}%
\StoreBenchExecResult{Pono}{ImcBzlaReachSafetyBitVectorsSampledBvRel}{Status}{Wrong}{False}{Count}{0}%
\providecommand\StoreBenchExecResult[7]{\expandafter\newcommand\csname#1#2#3#4#5#6\endcsname{#7}}%
\StoreBenchExecResult{Pono}{ImcBzlaReachSafetyCombinationsSampledBvRel}{Status}{All}{}{Score}{0}%
\StoreBenchExecResult{Pono}{ImcBzlaReachSafetyCombinationsSampledBvRel}{Status}{All}{}{Count}{110}%
\StoreBenchExecResult{Pono}{ImcBzlaReachSafetyCombinationsSampledBvRel}{Status}{Correct}{}{Count}{72}%
\StoreBenchExecResult{Pono}{ImcBzlaReachSafetyCombinationsSampledBvRel}{Status}{Correct}{True}{Count}{1}%
\StoreBenchExecResult{Pono}{ImcBzlaReachSafetyCombinationsSampledBvRel}{Status}{Correct}{False}{Count}{71}%
\StoreBenchExecResult{Pono}{ImcBzlaReachSafetyCombinationsSampledBvRel}{Status}{Wrong}{}{Count}{0}%
\StoreBenchExecResult{Pono}{ImcBzlaReachSafetyCombinationsSampledBvRel}{Status}{Wrong}{True}{Count}{0}%
\StoreBenchExecResult{Pono}{ImcBzlaReachSafetyCombinationsSampledBvRel}{Status}{Wrong}{False}{Count}{0}%
\providecommand\StoreBenchExecResult[7]{\expandafter\newcommand\csname#1#2#3#4#5#6\endcsname{#7}}%
\StoreBenchExecResult{Pono}{ImcBzlaReachSafetyControlFlowSampledBvRel}{Status}{All}{}{Score}{0}%
\StoreBenchExecResult{Pono}{ImcBzlaReachSafetyControlFlowSampledBvRel}{Status}{All}{}{Count}{29}%
\StoreBenchExecResult{Pono}{ImcBzlaReachSafetyControlFlowSampledBvRel}{Status}{Correct}{}{Count}{27}%
\StoreBenchExecResult{Pono}{ImcBzlaReachSafetyControlFlowSampledBvRel}{Status}{Correct}{True}{Count}{25}%
\StoreBenchExecResult{Pono}{ImcBzlaReachSafetyControlFlowSampledBvRel}{Status}{Correct}{False}{Count}{2}%
\StoreBenchExecResult{Pono}{ImcBzlaReachSafetyControlFlowSampledBvRel}{Status}{Wrong}{}{Count}{0}%
\StoreBenchExecResult{Pono}{ImcBzlaReachSafetyControlFlowSampledBvRel}{Status}{Wrong}{True}{Count}{0}%
\StoreBenchExecResult{Pono}{ImcBzlaReachSafetyControlFlowSampledBvRel}{Status}{Wrong}{False}{Count}{0}%
\providecommand\StoreBenchExecResult[7]{\expandafter\newcommand\csname#1#2#3#4#5#6\endcsname{#7}}%
\StoreBenchExecResult{Pono}{ImcBzlaReachSafetyECASampledBvRel}{Status}{All}{}{Score}{0}%
\StoreBenchExecResult{Pono}{ImcBzlaReachSafetyECASampledBvRel}{Status}{All}{}{Count}{150}%
\StoreBenchExecResult{Pono}{ImcBzlaReachSafetyECASampledBvRel}{Status}{Correct}{}{Count}{64}%
\StoreBenchExecResult{Pono}{ImcBzlaReachSafetyECASampledBvRel}{Status}{Correct}{True}{Count}{50}%
\StoreBenchExecResult{Pono}{ImcBzlaReachSafetyECASampledBvRel}{Status}{Correct}{False}{Count}{14}%
\StoreBenchExecResult{Pono}{ImcBzlaReachSafetyECASampledBvRel}{Status}{Wrong}{}{Count}{0}%
\StoreBenchExecResult{Pono}{ImcBzlaReachSafetyECASampledBvRel}{Status}{Wrong}{True}{Count}{0}%
\StoreBenchExecResult{Pono}{ImcBzlaReachSafetyECASampledBvRel}{Status}{Wrong}{False}{Count}{0}%
\providecommand\StoreBenchExecResult[7]{\expandafter\newcommand\csname#1#2#3#4#5#6\endcsname{#7}}%
\StoreBenchExecResult{Pono}{ImcBzlaReachSafetyFloatsSampledBvRel}{Status}{All}{}{Score}{0}%
\StoreBenchExecResult{Pono}{ImcBzlaReachSafetyFloatsSampledBvRel}{Status}{All}{}{Count}{10}%
\StoreBenchExecResult{Pono}{ImcBzlaReachSafetyFloatsSampledBvRel}{Status}{Correct}{}{Count}{10}%
\StoreBenchExecResult{Pono}{ImcBzlaReachSafetyFloatsSampledBvRel}{Status}{Correct}{True}{Count}{10}%
\StoreBenchExecResult{Pono}{ImcBzlaReachSafetyFloatsSampledBvRel}{Status}{Correct}{False}{Count}{0}%
\StoreBenchExecResult{Pono}{ImcBzlaReachSafetyFloatsSampledBvRel}{Status}{Wrong}{}{Count}{0}%
\StoreBenchExecResult{Pono}{ImcBzlaReachSafetyFloatsSampledBvRel}{Status}{Wrong}{True}{Count}{0}%
\StoreBenchExecResult{Pono}{ImcBzlaReachSafetyFloatsSampledBvRel}{Status}{Wrong}{False}{Count}{0}%
\providecommand\StoreBenchExecResult[7]{\expandafter\newcommand\csname#1#2#3#4#5#6\endcsname{#7}}%
\StoreBenchExecResult{Pono}{ImcBzlaReachSafetyHardnessSampledBvRel}{Status}{All}{}{Score}{0}%
\StoreBenchExecResult{Pono}{ImcBzlaReachSafetyHardnessSampledBvRel}{Status}{All}{}{Count}{132}%
\StoreBenchExecResult{Pono}{ImcBzlaReachSafetyHardnessSampledBvRel}{Status}{Correct}{}{Count}{132}%
\StoreBenchExecResult{Pono}{ImcBzlaReachSafetyHardnessSampledBvRel}{Status}{Correct}{True}{Count}{132}%
\StoreBenchExecResult{Pono}{ImcBzlaReachSafetyHardnessSampledBvRel}{Status}{Correct}{False}{Count}{0}%
\StoreBenchExecResult{Pono}{ImcBzlaReachSafetyHardnessSampledBvRel}{Status}{Wrong}{}{Count}{0}%
\StoreBenchExecResult{Pono}{ImcBzlaReachSafetyHardnessSampledBvRel}{Status}{Wrong}{True}{Count}{0}%
\StoreBenchExecResult{Pono}{ImcBzlaReachSafetyHardnessSampledBvRel}{Status}{Wrong}{False}{Count}{0}%
\providecommand\StoreBenchExecResult[7]{\expandafter\newcommand\csname#1#2#3#4#5#6\endcsname{#7}}%
\StoreBenchExecResult{Pono}{ImcBzlaReachSafetyHardwareSampledBvRel}{Status}{All}{}{Score}{0}%
\StoreBenchExecResult{Pono}{ImcBzlaReachSafetyHardwareSampledBvRel}{Status}{All}{}{Count}{148}%
\StoreBenchExecResult{Pono}{ImcBzlaReachSafetyHardwareSampledBvRel}{Status}{Correct}{}{Count}{56}%
\StoreBenchExecResult{Pono}{ImcBzlaReachSafetyHardwareSampledBvRel}{Status}{Correct}{True}{Count}{18}%
\StoreBenchExecResult{Pono}{ImcBzlaReachSafetyHardwareSampledBvRel}{Status}{Correct}{False}{Count}{38}%
\StoreBenchExecResult{Pono}{ImcBzlaReachSafetyHardwareSampledBvRel}{Status}{Wrong}{}{Count}{0}%
\StoreBenchExecResult{Pono}{ImcBzlaReachSafetyHardwareSampledBvRel}{Status}{Wrong}{True}{Count}{0}%
\StoreBenchExecResult{Pono}{ImcBzlaReachSafetyHardwareSampledBvRel}{Status}{Wrong}{False}{Count}{0}%
\providecommand\StoreBenchExecResult[7]{\expandafter\newcommand\csname#1#2#3#4#5#6\endcsname{#7}}%
\StoreBenchExecResult{Pono}{ImcBzlaReachSafetyHeapSampledBvRel}{Status}{All}{}{Score}{0}%
\StoreBenchExecResult{Pono}{ImcBzlaReachSafetyHeapSampledBvRel}{Status}{All}{}{Count}{6}%
\StoreBenchExecResult{Pono}{ImcBzlaReachSafetyHeapSampledBvRel}{Status}{Correct}{}{Count}{6}%
\StoreBenchExecResult{Pono}{ImcBzlaReachSafetyHeapSampledBvRel}{Status}{Correct}{True}{Count}{4}%
\StoreBenchExecResult{Pono}{ImcBzlaReachSafetyHeapSampledBvRel}{Status}{Correct}{False}{Count}{2}%
\StoreBenchExecResult{Pono}{ImcBzlaReachSafetyHeapSampledBvRel}{Status}{Wrong}{}{Count}{0}%
\StoreBenchExecResult{Pono}{ImcBzlaReachSafetyHeapSampledBvRel}{Status}{Wrong}{True}{Count}{0}%
\StoreBenchExecResult{Pono}{ImcBzlaReachSafetyHeapSampledBvRel}{Status}{Wrong}{False}{Count}{0}%
\providecommand\StoreBenchExecResult[7]{\expandafter\newcommand\csname#1#2#3#4#5#6\endcsname{#7}}%
\StoreBenchExecResult{Pono}{ImcBzlaReachSafetyLoopsSampledBvRel}{Status}{All}{}{Score}{0}%
\StoreBenchExecResult{Pono}{ImcBzlaReachSafetyLoopsSampledBvRel}{Status}{All}{}{Count}{137}%
\StoreBenchExecResult{Pono}{ImcBzlaReachSafetyLoopsSampledBvRel}{Status}{Correct}{}{Count}{41}%
\StoreBenchExecResult{Pono}{ImcBzlaReachSafetyLoopsSampledBvRel}{Status}{Correct}{True}{Count}{16}%
\StoreBenchExecResult{Pono}{ImcBzlaReachSafetyLoopsSampledBvRel}{Status}{Correct}{False}{Count}{25}%
\StoreBenchExecResult{Pono}{ImcBzlaReachSafetyLoopsSampledBvRel}{Status}{Wrong}{}{Count}{0}%
\StoreBenchExecResult{Pono}{ImcBzlaReachSafetyLoopsSampledBvRel}{Status}{Wrong}{True}{Count}{0}%
\StoreBenchExecResult{Pono}{ImcBzlaReachSafetyLoopsSampledBvRel}{Status}{Wrong}{False}{Count}{0}%
\providecommand\StoreBenchExecResult[7]{\expandafter\newcommand\csname#1#2#3#4#5#6\endcsname{#7}}%
\StoreBenchExecResult{Pono}{ImcBzlaReachSafetyProductLinesSampledBvRel}{Status}{All}{}{Score}{0}%
\StoreBenchExecResult{Pono}{ImcBzlaReachSafetyProductLinesSampledBvRel}{Status}{All}{}{Count}{150}%
\StoreBenchExecResult{Pono}{ImcBzlaReachSafetyProductLinesSampledBvRel}{Status}{Correct}{}{Count}{150}%
\StoreBenchExecResult{Pono}{ImcBzlaReachSafetyProductLinesSampledBvRel}{Status}{Correct}{True}{Count}{75}%
\StoreBenchExecResult{Pono}{ImcBzlaReachSafetyProductLinesSampledBvRel}{Status}{Correct}{False}{Count}{75}%
\StoreBenchExecResult{Pono}{ImcBzlaReachSafetyProductLinesSampledBvRel}{Status}{Wrong}{}{Count}{0}%
\StoreBenchExecResult{Pono}{ImcBzlaReachSafetyProductLinesSampledBvRel}{Status}{Wrong}{True}{Count}{0}%
\StoreBenchExecResult{Pono}{ImcBzlaReachSafetyProductLinesSampledBvRel}{Status}{Wrong}{False}{Count}{0}%
\providecommand\StoreBenchExecResult[7]{\expandafter\newcommand\csname#1#2#3#4#5#6\endcsname{#7}}%
\StoreBenchExecResult{Pono}{ImcBzlaReachSafetySequentializedSampledBvRel}{Status}{All}{}{Score}{0}%
\StoreBenchExecResult{Pono}{ImcBzlaReachSafetySequentializedSampledBvRel}{Status}{All}{}{Count}{150}%
\StoreBenchExecResult{Pono}{ImcBzlaReachSafetySequentializedSampledBvRel}{Status}{Correct}{}{Count}{90}%
\StoreBenchExecResult{Pono}{ImcBzlaReachSafetySequentializedSampledBvRel}{Status}{Correct}{True}{Count}{6}%
\StoreBenchExecResult{Pono}{ImcBzlaReachSafetySequentializedSampledBvRel}{Status}{Correct}{False}{Count}{84}%
\StoreBenchExecResult{Pono}{ImcBzlaReachSafetySequentializedSampledBvRel}{Status}{Wrong}{}{Count}{0}%
\StoreBenchExecResult{Pono}{ImcBzlaReachSafetySequentializedSampledBvRel}{Status}{Wrong}{True}{Count}{0}%
\StoreBenchExecResult{Pono}{ImcBzlaReachSafetySequentializedSampledBvRel}{Status}{Wrong}{False}{Count}{0}%
\providecommand\StoreBenchExecResult[7]{\expandafter\newcommand\csname#1#2#3#4#5#6\endcsname{#7}}%
\StoreBenchExecResult{Pono}{ImcBzlaReachSafetyXCSPSampledBvRel}{Status}{All}{}{Score}{0}%
\StoreBenchExecResult{Pono}{ImcBzlaReachSafetyXCSPSampledBvRel}{Status}{All}{}{Count}{98}%
\StoreBenchExecResult{Pono}{ImcBzlaReachSafetyXCSPSampledBvRel}{Status}{Correct}{}{Count}{94}%
\StoreBenchExecResult{Pono}{ImcBzlaReachSafetyXCSPSampledBvRel}{Status}{Correct}{True}{Count}{52}%
\StoreBenchExecResult{Pono}{ImcBzlaReachSafetyXCSPSampledBvRel}{Status}{Correct}{False}{Count}{42}%
\StoreBenchExecResult{Pono}{ImcBzlaReachSafetyXCSPSampledBvRel}{Status}{Wrong}{}{Count}{0}%
\StoreBenchExecResult{Pono}{ImcBzlaReachSafetyXCSPSampledBvRel}{Status}{Wrong}{True}{Count}{0}%
\StoreBenchExecResult{Pono}{ImcBzlaReachSafetyXCSPSampledBvRel}{Status}{Wrong}{False}{Count}{0}%
\providecommand\StoreBenchExecResult[7]{\expandafter\newcommand\csname#1#2#3#4#5#6\endcsname{#7}}%
\StoreBenchExecResult{Pono}{ImcBzlaTerminationBitVectorsSampledBvRel}{Status}{All}{}{Score}{0}%
\StoreBenchExecResult{Pono}{ImcBzlaTerminationBitVectorsSampledBvRel}{Status}{All}{}{Count}{32}%
\StoreBenchExecResult{Pono}{ImcBzlaTerminationBitVectorsSampledBvRel}{Status}{Correct}{}{Count}{24}%
\StoreBenchExecResult{Pono}{ImcBzlaTerminationBitVectorsSampledBvRel}{Status}{Correct}{True}{Count}{13}%
\StoreBenchExecResult{Pono}{ImcBzlaTerminationBitVectorsSampledBvRel}{Status}{Correct}{False}{Count}{11}%
\StoreBenchExecResult{Pono}{ImcBzlaTerminationBitVectorsSampledBvRel}{Status}{Wrong}{}{Count}{0}%
\StoreBenchExecResult{Pono}{ImcBzlaTerminationBitVectorsSampledBvRel}{Status}{Wrong}{True}{Count}{0}%
\StoreBenchExecResult{Pono}{ImcBzlaTerminationBitVectorsSampledBvRel}{Status}{Wrong}{False}{Count}{0}%
\providecommand\StoreBenchExecResult[7]{\expandafter\newcommand\csname#1#2#3#4#5#6\endcsname{#7}}%
\StoreBenchExecResult{Pono}{ImcBzlaTerminationMainControlFlowSampledBvRel}{Status}{All}{}{Score}{0}%
\StoreBenchExecResult{Pono}{ImcBzlaTerminationMainControlFlowSampledBvRel}{Status}{All}{}{Count}{236}%
\StoreBenchExecResult{Pono}{ImcBzlaTerminationMainControlFlowSampledBvRel}{Status}{Correct}{}{Count}{70}%
\StoreBenchExecResult{Pono}{ImcBzlaTerminationMainControlFlowSampledBvRel}{Status}{Correct}{True}{Count}{22}%
\StoreBenchExecResult{Pono}{ImcBzlaTerminationMainControlFlowSampledBvRel}{Status}{Correct}{False}{Count}{48}%
\StoreBenchExecResult{Pono}{ImcBzlaTerminationMainControlFlowSampledBvRel}{Status}{Wrong}{}{Count}{0}%
\StoreBenchExecResult{Pono}{ImcBzlaTerminationMainControlFlowSampledBvRel}{Status}{Wrong}{True}{Count}{0}%
\StoreBenchExecResult{Pono}{ImcBzlaTerminationMainControlFlowSampledBvRel}{Status}{Wrong}{False}{Count}{0}%
\providecommand\StoreBenchExecResult[7]{\expandafter\newcommand\csname#1#2#3#4#5#6\endcsname{#7}}%
\StoreBenchExecResult{Pono}{ImcBzlaTerminationOtherSampledBvRel}{Status}{All}{}{Score}{0}%
\StoreBenchExecResult{Pono}{ImcBzlaTerminationOtherSampledBvRel}{Status}{All}{}{Count}{973}%
\StoreBenchExecResult{Pono}{ImcBzlaTerminationOtherSampledBvRel}{Status}{Correct}{}{Count}{761}%
\StoreBenchExecResult{Pono}{ImcBzlaTerminationOtherSampledBvRel}{Status}{Correct}{True}{Count}{188}%
\StoreBenchExecResult{Pono}{ImcBzlaTerminationOtherSampledBvRel}{Status}{Correct}{False}{Count}{573}%
\StoreBenchExecResult{Pono}{ImcBzlaTerminationOtherSampledBvRel}{Status}{Wrong}{}{Count}{0}%
\StoreBenchExecResult{Pono}{ImcBzlaTerminationOtherSampledBvRel}{Status}{Wrong}{True}{Count}{0}%
\StoreBenchExecResult{Pono}{ImcBzlaTerminationOtherSampledBvRel}{Status}{Wrong}{False}{Count}{0}%
\edef\PonoImcBzlaReachSafetySampledBvRelStatusAllCount{\the\numexpr\PonoImcBzlaReachSafetyBitVectorsSampledBvRelStatusAllCount+\PonoImcBzlaReachSafetyCombinationsSampledBvRelStatusAllCount+\PonoImcBzlaReachSafetyControlFlowSampledBvRelStatusAllCount+\PonoImcBzlaReachSafetyECASampledBvRelStatusAllCount+\PonoImcBzlaReachSafetyFloatsSampledBvRelStatusAllCount+\PonoImcBzlaReachSafetyHardnessSampledBvRelStatusAllCount+\PonoImcBzlaReachSafetyHardwareSampledBvRelStatusAllCount+\PonoImcBzlaReachSafetyHeapSampledBvRelStatusAllCount+\PonoImcBzlaReachSafetyLoopsSampledBvRelStatusAllCount+\PonoImcBzlaReachSafetyProductLinesSampledBvRelStatusAllCount+\PonoImcBzlaReachSafetySequentializedSampledBvRelStatusAllCount+\PonoImcBzlaReachSafetyXCSPSampledBvRelStatusAllCount}
\edef\PonoImcBzlaReachSafetySampledBvRelStatusCorrectCount{\the\numexpr\PonoImcBzlaReachSafetyBitVectorsSampledBvRelStatusCorrectCount+\PonoImcBzlaReachSafetyCombinationsSampledBvRelStatusCorrectCount+\PonoImcBzlaReachSafetyControlFlowSampledBvRelStatusCorrectCount+\PonoImcBzlaReachSafetyECASampledBvRelStatusCorrectCount+\PonoImcBzlaReachSafetyFloatsSampledBvRelStatusCorrectCount+\PonoImcBzlaReachSafetyHardnessSampledBvRelStatusCorrectCount+\PonoImcBzlaReachSafetyHardwareSampledBvRelStatusCorrectCount+\PonoImcBzlaReachSafetyHeapSampledBvRelStatusCorrectCount+\PonoImcBzlaReachSafetyLoopsSampledBvRelStatusCorrectCount+\PonoImcBzlaReachSafetyProductLinesSampledBvRelStatusCorrectCount+\PonoImcBzlaReachSafetySequentializedSampledBvRelStatusCorrectCount+\PonoImcBzlaReachSafetyXCSPSampledBvRelStatusCorrectCount}
\edef\PonoImcBzlaReachSafetySampledBvRelStatusCorrectTrueCount{\the\numexpr\PonoImcBzlaReachSafetyBitVectorsSampledBvRelStatusCorrectTrueCount+\PonoImcBzlaReachSafetyCombinationsSampledBvRelStatusCorrectTrueCount+\PonoImcBzlaReachSafetyControlFlowSampledBvRelStatusCorrectTrueCount+\PonoImcBzlaReachSafetyECASampledBvRelStatusCorrectTrueCount+\PonoImcBzlaReachSafetyFloatsSampledBvRelStatusCorrectTrueCount+\PonoImcBzlaReachSafetyHardnessSampledBvRelStatusCorrectTrueCount+\PonoImcBzlaReachSafetyHardwareSampledBvRelStatusCorrectTrueCount+\PonoImcBzlaReachSafetyHeapSampledBvRelStatusCorrectTrueCount+\PonoImcBzlaReachSafetyLoopsSampledBvRelStatusCorrectTrueCount+\PonoImcBzlaReachSafetyProductLinesSampledBvRelStatusCorrectTrueCount+\PonoImcBzlaReachSafetySequentializedSampledBvRelStatusCorrectTrueCount+\PonoImcBzlaReachSafetyXCSPSampledBvRelStatusCorrectTrueCount}
\edef\PonoImcBzlaReachSafetySampledBvRelStatusCorrectFalseCount{\the\numexpr\PonoImcBzlaReachSafetyBitVectorsSampledBvRelStatusCorrectFalseCount+\PonoImcBzlaReachSafetyCombinationsSampledBvRelStatusCorrectFalseCount+\PonoImcBzlaReachSafetyControlFlowSampledBvRelStatusCorrectFalseCount+\PonoImcBzlaReachSafetyECASampledBvRelStatusCorrectFalseCount+\PonoImcBzlaReachSafetyFloatsSampledBvRelStatusCorrectFalseCount+\PonoImcBzlaReachSafetyHardnessSampledBvRelStatusCorrectFalseCount+\PonoImcBzlaReachSafetyHardwareSampledBvRelStatusCorrectFalseCount+\PonoImcBzlaReachSafetyHeapSampledBvRelStatusCorrectFalseCount+\PonoImcBzlaReachSafetyLoopsSampledBvRelStatusCorrectFalseCount+\PonoImcBzlaReachSafetyProductLinesSampledBvRelStatusCorrectFalseCount+\PonoImcBzlaReachSafetySequentializedSampledBvRelStatusCorrectFalseCount+\PonoImcBzlaReachSafetyXCSPSampledBvRelStatusCorrectFalseCount}
\edef\PonoImcBzlaReachSafetySampledBvRelStatusWrongCount{\the\numexpr\PonoImcBzlaReachSafetyBitVectorsSampledBvRelStatusWrongCount+\PonoImcBzlaReachSafetyCombinationsSampledBvRelStatusWrongCount+\PonoImcBzlaReachSafetyControlFlowSampledBvRelStatusWrongCount+\PonoImcBzlaReachSafetyECASampledBvRelStatusWrongCount+\PonoImcBzlaReachSafetyFloatsSampledBvRelStatusWrongCount+\PonoImcBzlaReachSafetyHardnessSampledBvRelStatusWrongCount+\PonoImcBzlaReachSafetyHardwareSampledBvRelStatusWrongCount+\PonoImcBzlaReachSafetyHeapSampledBvRelStatusWrongCount+\PonoImcBzlaReachSafetyLoopsSampledBvRelStatusWrongCount+\PonoImcBzlaReachSafetyProductLinesSampledBvRelStatusWrongCount+\PonoImcBzlaReachSafetySequentializedSampledBvRelStatusWrongCount+\PonoImcBzlaReachSafetyXCSPSampledBvRelStatusWrongCount}
\edef\PonoImcBzlaTerminationSampledBvRelStatusAllCount{\the\numexpr\PonoImcBzlaTerminationBitVectorsSampledBvRelStatusAllCount+\PonoImcBzlaTerminationMainControlFlowSampledBvRelStatusAllCount+\PonoImcBzlaTerminationOtherSampledBvRelStatusAllCount}
\edef\PonoImcBzlaTerminationSampledBvRelStatusCorrectCount{\the\numexpr\PonoImcBzlaTerminationBitVectorsSampledBvRelStatusCorrectCount+\PonoImcBzlaTerminationMainControlFlowSampledBvRelStatusCorrectCount+\PonoImcBzlaTerminationOtherSampledBvRelStatusCorrectCount}
\edef\PonoImcBzlaTerminationSampledBvRelStatusCorrectTrueCount{\the\numexpr\PonoImcBzlaTerminationBitVectorsSampledBvRelStatusCorrectTrueCount+\PonoImcBzlaTerminationMainControlFlowSampledBvRelStatusCorrectTrueCount+\PonoImcBzlaTerminationOtherSampledBvRelStatusCorrectTrueCount}
\edef\PonoImcBzlaTerminationSampledBvRelStatusCorrectFalseCount{\the\numexpr\PonoImcBzlaTerminationBitVectorsSampledBvRelStatusCorrectFalseCount+\PonoImcBzlaTerminationMainControlFlowSampledBvRelStatusCorrectFalseCount+\PonoImcBzlaTerminationOtherSampledBvRelStatusCorrectFalseCount}
\edef\PonoImcBzlaTerminationSampledBvRelStatusWrongCount{\the\numexpr\PonoImcBzlaTerminationBitVectorsSampledBvRelStatusWrongCount+\PonoImcBzlaTerminationMainControlFlowSampledBvRelStatusWrongCount+\PonoImcBzlaTerminationOtherSampledBvRelStatusWrongCount}
\providecommand\StoreBenchExecResult[7]{\expandafter\newcommand\csname#1#2#3#4#5#6\endcsname{#7}}%
\StoreBenchExecResult{Pono}{KindReachSafetyArraysSampledRel}{Status}{All}{}{Score}{0}%
\StoreBenchExecResult{Pono}{KindReachSafetyArraysSampledRel}{Status}{All}{}{Count}{150}%
\StoreBenchExecResult{Pono}{KindReachSafetyArraysSampledRel}{Status}{Correct}{}{Count}{49}%
\StoreBenchExecResult{Pono}{KindReachSafetyArraysSampledRel}{Status}{Correct}{True}{Count}{0}%
\StoreBenchExecResult{Pono}{KindReachSafetyArraysSampledRel}{Status}{Correct}{False}{Count}{49}%
\StoreBenchExecResult{Pono}{KindReachSafetyArraysSampledRel}{Status}{Wrong}{}{Count}{0}%
\StoreBenchExecResult{Pono}{KindReachSafetyArraysSampledRel}{Status}{Wrong}{True}{Count}{0}%
\StoreBenchExecResult{Pono}{KindReachSafetyArraysSampledRel}{Status}{Wrong}{False}{Count}{0}%
\providecommand\StoreBenchExecResult[7]{\expandafter\newcommand\csname#1#2#3#4#5#6\endcsname{#7}}%
\StoreBenchExecResult{Pono}{KindReachSafetyBitVectorsSampledRel}{Status}{All}{}{Score}{0}%
\StoreBenchExecResult{Pono}{KindReachSafetyBitVectorsSampledRel}{Status}{All}{}{Count}{48}%
\StoreBenchExecResult{Pono}{KindReachSafetyBitVectorsSampledRel}{Status}{Correct}{}{Count}{30}%
\StoreBenchExecResult{Pono}{KindReachSafetyBitVectorsSampledRel}{Status}{Correct}{True}{Count}{18}%
\StoreBenchExecResult{Pono}{KindReachSafetyBitVectorsSampledRel}{Status}{Correct}{False}{Count}{12}%
\StoreBenchExecResult{Pono}{KindReachSafetyBitVectorsSampledRel}{Status}{Wrong}{}{Count}{0}%
\StoreBenchExecResult{Pono}{KindReachSafetyBitVectorsSampledRel}{Status}{Wrong}{True}{Count}{0}%
\StoreBenchExecResult{Pono}{KindReachSafetyBitVectorsSampledRel}{Status}{Wrong}{False}{Count}{0}%
\providecommand\StoreBenchExecResult[7]{\expandafter\newcommand\csname#1#2#3#4#5#6\endcsname{#7}}%
\StoreBenchExecResult{Pono}{KindReachSafetyCombinationsSampledRel}{Status}{All}{}{Score}{0}%
\StoreBenchExecResult{Pono}{KindReachSafetyCombinationsSampledRel}{Status}{All}{}{Count}{110}%
\StoreBenchExecResult{Pono}{KindReachSafetyCombinationsSampledRel}{Status}{Correct}{}{Count}{71}%
\StoreBenchExecResult{Pono}{KindReachSafetyCombinationsSampledRel}{Status}{Correct}{True}{Count}{0}%
\StoreBenchExecResult{Pono}{KindReachSafetyCombinationsSampledRel}{Status}{Correct}{False}{Count}{71}%
\StoreBenchExecResult{Pono}{KindReachSafetyCombinationsSampledRel}{Status}{Wrong}{}{Count}{0}%
\StoreBenchExecResult{Pono}{KindReachSafetyCombinationsSampledRel}{Status}{Wrong}{True}{Count}{0}%
\StoreBenchExecResult{Pono}{KindReachSafetyCombinationsSampledRel}{Status}{Wrong}{False}{Count}{0}%
\providecommand\StoreBenchExecResult[7]{\expandafter\newcommand\csname#1#2#3#4#5#6\endcsname{#7}}%
\StoreBenchExecResult{Pono}{KindReachSafetyControlFlowSampledRel}{Status}{All}{}{Score}{0}%
\StoreBenchExecResult{Pono}{KindReachSafetyControlFlowSampledRel}{Status}{All}{}{Count}{33}%
\StoreBenchExecResult{Pono}{KindReachSafetyControlFlowSampledRel}{Status}{Correct}{}{Count}{25}%
\StoreBenchExecResult{Pono}{KindReachSafetyControlFlowSampledRel}{Status}{Correct}{True}{Count}{21}%
\StoreBenchExecResult{Pono}{KindReachSafetyControlFlowSampledRel}{Status}{Correct}{False}{Count}{4}%
\StoreBenchExecResult{Pono}{KindReachSafetyControlFlowSampledRel}{Status}{Wrong}{}{Count}{0}%
\StoreBenchExecResult{Pono}{KindReachSafetyControlFlowSampledRel}{Status}{Wrong}{True}{Count}{0}%
\StoreBenchExecResult{Pono}{KindReachSafetyControlFlowSampledRel}{Status}{Wrong}{False}{Count}{0}%
\providecommand\StoreBenchExecResult[7]{\expandafter\newcommand\csname#1#2#3#4#5#6\endcsname{#7}}%
\StoreBenchExecResult{Pono}{KindReachSafetyECASampledRel}{Status}{All}{}{Score}{0}%
\StoreBenchExecResult{Pono}{KindReachSafetyECASampledRel}{Status}{All}{}{Count}{150}%
\StoreBenchExecResult{Pono}{KindReachSafetyECASampledRel}{Status}{Correct}{}{Count}{95}%
\StoreBenchExecResult{Pono}{KindReachSafetyECASampledRel}{Status}{Correct}{True}{Count}{51}%
\StoreBenchExecResult{Pono}{KindReachSafetyECASampledRel}{Status}{Correct}{False}{Count}{44}%
\StoreBenchExecResult{Pono}{KindReachSafetyECASampledRel}{Status}{Wrong}{}{Count}{0}%
\StoreBenchExecResult{Pono}{KindReachSafetyECASampledRel}{Status}{Wrong}{True}{Count}{0}%
\StoreBenchExecResult{Pono}{KindReachSafetyECASampledRel}{Status}{Wrong}{False}{Count}{0}%
\providecommand\StoreBenchExecResult[7]{\expandafter\newcommand\csname#1#2#3#4#5#6\endcsname{#7}}%
\StoreBenchExecResult{Pono}{KindReachSafetyFloatsSampledRel}{Status}{All}{}{Score}{0}%
\StoreBenchExecResult{Pono}{KindReachSafetyFloatsSampledRel}{Status}{All}{}{Count}{10}%
\StoreBenchExecResult{Pono}{KindReachSafetyFloatsSampledRel}{Status}{Correct}{}{Count}{10}%
\StoreBenchExecResult{Pono}{KindReachSafetyFloatsSampledRel}{Status}{Correct}{True}{Count}{10}%
\StoreBenchExecResult{Pono}{KindReachSafetyFloatsSampledRel}{Status}{Correct}{False}{Count}{0}%
\StoreBenchExecResult{Pono}{KindReachSafetyFloatsSampledRel}{Status}{Wrong}{}{Count}{0}%
\StoreBenchExecResult{Pono}{KindReachSafetyFloatsSampledRel}{Status}{Wrong}{True}{Count}{0}%
\StoreBenchExecResult{Pono}{KindReachSafetyFloatsSampledRel}{Status}{Wrong}{False}{Count}{0}%
\providecommand\StoreBenchExecResult[7]{\expandafter\newcommand\csname#1#2#3#4#5#6\endcsname{#7}}%
\StoreBenchExecResult{Pono}{KindReachSafetyHardnessSampledRel}{Status}{All}{}{Score}{0}%
\StoreBenchExecResult{Pono}{KindReachSafetyHardnessSampledRel}{Status}{All}{}{Count}{150}%
\StoreBenchExecResult{Pono}{KindReachSafetyHardnessSampledRel}{Status}{Correct}{}{Count}{134}%
\StoreBenchExecResult{Pono}{KindReachSafetyHardnessSampledRel}{Status}{Correct}{True}{Count}{134}%
\StoreBenchExecResult{Pono}{KindReachSafetyHardnessSampledRel}{Status}{Correct}{False}{Count}{0}%
\StoreBenchExecResult{Pono}{KindReachSafetyHardnessSampledRel}{Status}{Wrong}{}{Count}{0}%
\StoreBenchExecResult{Pono}{KindReachSafetyHardnessSampledRel}{Status}{Wrong}{True}{Count}{0}%
\StoreBenchExecResult{Pono}{KindReachSafetyHardnessSampledRel}{Status}{Wrong}{False}{Count}{0}%
\providecommand\StoreBenchExecResult[7]{\expandafter\newcommand\csname#1#2#3#4#5#6\endcsname{#7}}%
\StoreBenchExecResult{Pono}{KindReachSafetyHardwareSampledRel}{Status}{All}{}{Score}{0}%
\StoreBenchExecResult{Pono}{KindReachSafetyHardwareSampledRel}{Status}{All}{}{Count}{150}%
\StoreBenchExecResult{Pono}{KindReachSafetyHardwareSampledRel}{Status}{Correct}{}{Count}{56}%
\StoreBenchExecResult{Pono}{KindReachSafetyHardwareSampledRel}{Status}{Correct}{True}{Count}{2}%
\StoreBenchExecResult{Pono}{KindReachSafetyHardwareSampledRel}{Status}{Correct}{False}{Count}{54}%
\StoreBenchExecResult{Pono}{KindReachSafetyHardwareSampledRel}{Status}{Wrong}{}{Count}{0}%
\StoreBenchExecResult{Pono}{KindReachSafetyHardwareSampledRel}{Status}{Wrong}{True}{Count}{0}%
\StoreBenchExecResult{Pono}{KindReachSafetyHardwareSampledRel}{Status}{Wrong}{False}{Count}{0}%
\providecommand\StoreBenchExecResult[7]{\expandafter\newcommand\csname#1#2#3#4#5#6\endcsname{#7}}%
\StoreBenchExecResult{Pono}{KindReachSafetyHeapSampledRel}{Status}{All}{}{Score}{0}%
\StoreBenchExecResult{Pono}{KindReachSafetyHeapSampledRel}{Status}{All}{}{Count}{53}%
\StoreBenchExecResult{Pono}{KindReachSafetyHeapSampledRel}{Status}{Correct}{}{Count}{48}%
\StoreBenchExecResult{Pono}{KindReachSafetyHeapSampledRel}{Status}{Correct}{True}{Count}{31}%
\StoreBenchExecResult{Pono}{KindReachSafetyHeapSampledRel}{Status}{Correct}{False}{Count}{17}%
\StoreBenchExecResult{Pono}{KindReachSafetyHeapSampledRel}{Status}{Wrong}{}{Count}{0}%
\StoreBenchExecResult{Pono}{KindReachSafetyHeapSampledRel}{Status}{Wrong}{True}{Count}{0}%
\StoreBenchExecResult{Pono}{KindReachSafetyHeapSampledRel}{Status}{Wrong}{False}{Count}{0}%
\providecommand\StoreBenchExecResult[7]{\expandafter\newcommand\csname#1#2#3#4#5#6\endcsname{#7}}%
\StoreBenchExecResult{Pono}{KindReachSafetyLoopsSampledRel}{Status}{All}{}{Score}{0}%
\StoreBenchExecResult{Pono}{KindReachSafetyLoopsSampledRel}{Status}{All}{}{Count}{150}%
\StoreBenchExecResult{Pono}{KindReachSafetyLoopsSampledRel}{Status}{Correct}{}{Count}{57}%
\StoreBenchExecResult{Pono}{KindReachSafetyLoopsSampledRel}{Status}{Correct}{True}{Count}{15}%
\StoreBenchExecResult{Pono}{KindReachSafetyLoopsSampledRel}{Status}{Correct}{False}{Count}{42}%
\StoreBenchExecResult{Pono}{KindReachSafetyLoopsSampledRel}{Status}{Wrong}{}{Count}{0}%
\StoreBenchExecResult{Pono}{KindReachSafetyLoopsSampledRel}{Status}{Wrong}{True}{Count}{0}%
\StoreBenchExecResult{Pono}{KindReachSafetyLoopsSampledRel}{Status}{Wrong}{False}{Count}{0}%
\providecommand\StoreBenchExecResult[7]{\expandafter\newcommand\csname#1#2#3#4#5#6\endcsname{#7}}%
\StoreBenchExecResult{Pono}{KindReachSafetyProductLinesSampledRel}{Status}{All}{}{Score}{0}%
\StoreBenchExecResult{Pono}{KindReachSafetyProductLinesSampledRel}{Status}{All}{}{Count}{150}%
\StoreBenchExecResult{Pono}{KindReachSafetyProductLinesSampledRel}{Status}{Correct}{}{Count}{91}%
\StoreBenchExecResult{Pono}{KindReachSafetyProductLinesSampledRel}{Status}{Correct}{True}{Count}{16}%
\StoreBenchExecResult{Pono}{KindReachSafetyProductLinesSampledRel}{Status}{Correct}{False}{Count}{75}%
\StoreBenchExecResult{Pono}{KindReachSafetyProductLinesSampledRel}{Status}{Wrong}{}{Count}{0}%
\StoreBenchExecResult{Pono}{KindReachSafetyProductLinesSampledRel}{Status}{Wrong}{True}{Count}{0}%
\StoreBenchExecResult{Pono}{KindReachSafetyProductLinesSampledRel}{Status}{Wrong}{False}{Count}{0}%
\providecommand\StoreBenchExecResult[7]{\expandafter\newcommand\csname#1#2#3#4#5#6\endcsname{#7}}%
\StoreBenchExecResult{Pono}{KindReachSafetySequentializedSampledRel}{Status}{All}{}{Score}{0}%
\StoreBenchExecResult{Pono}{KindReachSafetySequentializedSampledRel}{Status}{All}{}{Count}{150}%
\StoreBenchExecResult{Pono}{KindReachSafetySequentializedSampledRel}{Status}{Correct}{}{Count}{100}%
\StoreBenchExecResult{Pono}{KindReachSafetySequentializedSampledRel}{Status}{Correct}{True}{Count}{13}%
\StoreBenchExecResult{Pono}{KindReachSafetySequentializedSampledRel}{Status}{Correct}{False}{Count}{87}%
\StoreBenchExecResult{Pono}{KindReachSafetySequentializedSampledRel}{Status}{Wrong}{}{Count}{0}%
\StoreBenchExecResult{Pono}{KindReachSafetySequentializedSampledRel}{Status}{Wrong}{True}{Count}{0}%
\StoreBenchExecResult{Pono}{KindReachSafetySequentializedSampledRel}{Status}{Wrong}{False}{Count}{0}%
\providecommand\StoreBenchExecResult[7]{\expandafter\newcommand\csname#1#2#3#4#5#6\endcsname{#7}}%
\StoreBenchExecResult{Pono}{KindReachSafetyXCSPSampledRel}{Status}{All}{}{Score}{0}%
\StoreBenchExecResult{Pono}{KindReachSafetyXCSPSampledRel}{Status}{All}{}{Count}{98}%
\StoreBenchExecResult{Pono}{KindReachSafetyXCSPSampledRel}{Status}{Correct}{}{Count}{95}%
\StoreBenchExecResult{Pono}{KindReachSafetyXCSPSampledRel}{Status}{Correct}{True}{Count}{52}%
\StoreBenchExecResult{Pono}{KindReachSafetyXCSPSampledRel}{Status}{Correct}{False}{Count}{43}%
\StoreBenchExecResult{Pono}{KindReachSafetyXCSPSampledRel}{Status}{Wrong}{}{Count}{0}%
\StoreBenchExecResult{Pono}{KindReachSafetyXCSPSampledRel}{Status}{Wrong}{True}{Count}{0}%
\StoreBenchExecResult{Pono}{KindReachSafetyXCSPSampledRel}{Status}{Wrong}{False}{Count}{0}%
\providecommand\StoreBenchExecResult[7]{\expandafter\newcommand\csname#1#2#3#4#5#6\endcsname{#7}}%
\StoreBenchExecResult{Pono}{KindTerminationBitVectorsSampledRel}{Status}{All}{}{Score}{0}%
\StoreBenchExecResult{Pono}{KindTerminationBitVectorsSampledRel}{Status}{All}{}{Count}{32}%
\StoreBenchExecResult{Pono}{KindTerminationBitVectorsSampledRel}{Status}{Correct}{}{Count}{24}%
\StoreBenchExecResult{Pono}{KindTerminationBitVectorsSampledRel}{Status}{Correct}{True}{Count}{13}%
\StoreBenchExecResult{Pono}{KindTerminationBitVectorsSampledRel}{Status}{Correct}{False}{Count}{11}%
\StoreBenchExecResult{Pono}{KindTerminationBitVectorsSampledRel}{Status}{Wrong}{}{Count}{0}%
\StoreBenchExecResult{Pono}{KindTerminationBitVectorsSampledRel}{Status}{Wrong}{True}{Count}{0}%
\StoreBenchExecResult{Pono}{KindTerminationBitVectorsSampledRel}{Status}{Wrong}{False}{Count}{0}%
\providecommand\StoreBenchExecResult[7]{\expandafter\newcommand\csname#1#2#3#4#5#6\endcsname{#7}}%
\StoreBenchExecResult{Pono}{KindTerminationMainControlFlowSampledRel}{Status}{All}{}{Score}{0}%
\StoreBenchExecResult{Pono}{KindTerminationMainControlFlowSampledRel}{Status}{All}{}{Count}{242}%
\StoreBenchExecResult{Pono}{KindTerminationMainControlFlowSampledRel}{Status}{Correct}{}{Count}{80}%
\StoreBenchExecResult{Pono}{KindTerminationMainControlFlowSampledRel}{Status}{Correct}{True}{Count}{28}%
\StoreBenchExecResult{Pono}{KindTerminationMainControlFlowSampledRel}{Status}{Correct}{False}{Count}{52}%
\StoreBenchExecResult{Pono}{KindTerminationMainControlFlowSampledRel}{Status}{Wrong}{}{Count}{0}%
\StoreBenchExecResult{Pono}{KindTerminationMainControlFlowSampledRel}{Status}{Wrong}{True}{Count}{0}%
\StoreBenchExecResult{Pono}{KindTerminationMainControlFlowSampledRel}{Status}{Wrong}{False}{Count}{0}%
\providecommand\StoreBenchExecResult[7]{\expandafter\newcommand\csname#1#2#3#4#5#6\endcsname{#7}}%
\StoreBenchExecResult{Pono}{KindTerminationOtherSampledRel}{Status}{All}{}{Score}{0}%
\StoreBenchExecResult{Pono}{KindTerminationOtherSampledRel}{Status}{All}{}{Count}{1080}%
\StoreBenchExecResult{Pono}{KindTerminationOtherSampledRel}{Status}{Correct}{}{Count}{911}%
\StoreBenchExecResult{Pono}{KindTerminationOtherSampledRel}{Status}{Correct}{True}{Count}{271}%
\StoreBenchExecResult{Pono}{KindTerminationOtherSampledRel}{Status}{Correct}{False}{Count}{640}%
\StoreBenchExecResult{Pono}{KindTerminationOtherSampledRel}{Status}{Wrong}{}{Count}{0}%
\StoreBenchExecResult{Pono}{KindTerminationOtherSampledRel}{Status}{Wrong}{True}{Count}{0}%
\StoreBenchExecResult{Pono}{KindTerminationOtherSampledRel}{Status}{Wrong}{False}{Count}{0}%
\edef\PonoKindReachSafetySampledRelStatusAllCount{\the\numexpr\PonoKindReachSafetyArraysSampledRelStatusAllCount+\PonoKindReachSafetyBitVectorsSampledRelStatusAllCount+\PonoKindReachSafetyCombinationsSampledRelStatusAllCount+\PonoKindReachSafetyControlFlowSampledRelStatusAllCount+\PonoKindReachSafetyECASampledRelStatusAllCount+\PonoKindReachSafetyFloatsSampledRelStatusAllCount+\PonoKindReachSafetyHardnessSampledRelStatusAllCount+\PonoKindReachSafetyHardwareSampledRelStatusAllCount+\PonoKindReachSafetyHeapSampledRelStatusAllCount+\PonoKindReachSafetyLoopsSampledRelStatusAllCount+\PonoKindReachSafetyProductLinesSampledRelStatusAllCount+\PonoKindReachSafetySequentializedSampledRelStatusAllCount+\PonoKindReachSafetyXCSPSampledRelStatusAllCount}
\edef\PonoKindReachSafetySampledRelStatusCorrectCount{\the\numexpr\PonoKindReachSafetyArraysSampledRelStatusCorrectCount+\PonoKindReachSafetyBitVectorsSampledRelStatusCorrectCount+\PonoKindReachSafetyCombinationsSampledRelStatusCorrectCount+\PonoKindReachSafetyControlFlowSampledRelStatusCorrectCount+\PonoKindReachSafetyECASampledRelStatusCorrectCount+\PonoKindReachSafetyFloatsSampledRelStatusCorrectCount+\PonoKindReachSafetyHardnessSampledRelStatusCorrectCount+\PonoKindReachSafetyHardwareSampledRelStatusCorrectCount+\PonoKindReachSafetyHeapSampledRelStatusCorrectCount+\PonoKindReachSafetyLoopsSampledRelStatusCorrectCount+\PonoKindReachSafetyProductLinesSampledRelStatusCorrectCount+\PonoKindReachSafetySequentializedSampledRelStatusCorrectCount+\PonoKindReachSafetyXCSPSampledRelStatusCorrectCount}
\edef\PonoKindReachSafetySampledRelStatusCorrectTrueCount{\the\numexpr\PonoKindReachSafetyArraysSampledRelStatusCorrectTrueCount+\PonoKindReachSafetyBitVectorsSampledRelStatusCorrectTrueCount+\PonoKindReachSafetyCombinationsSampledRelStatusCorrectTrueCount+\PonoKindReachSafetyControlFlowSampledRelStatusCorrectTrueCount+\PonoKindReachSafetyECASampledRelStatusCorrectTrueCount+\PonoKindReachSafetyFloatsSampledRelStatusCorrectTrueCount+\PonoKindReachSafetyHardnessSampledRelStatusCorrectTrueCount+\PonoKindReachSafetyHardwareSampledRelStatusCorrectTrueCount+\PonoKindReachSafetyHeapSampledRelStatusCorrectTrueCount+\PonoKindReachSafetyLoopsSampledRelStatusCorrectTrueCount+\PonoKindReachSafetyProductLinesSampledRelStatusCorrectTrueCount+\PonoKindReachSafetySequentializedSampledRelStatusCorrectTrueCount+\PonoKindReachSafetyXCSPSampledRelStatusCorrectTrueCount}
\edef\PonoKindReachSafetySampledRelStatusCorrectFalseCount{\the\numexpr\PonoKindReachSafetyArraysSampledRelStatusCorrectFalseCount+\PonoKindReachSafetyBitVectorsSampledRelStatusCorrectFalseCount+\PonoKindReachSafetyCombinationsSampledRelStatusCorrectFalseCount+\PonoKindReachSafetyControlFlowSampledRelStatusCorrectFalseCount+\PonoKindReachSafetyECASampledRelStatusCorrectFalseCount+\PonoKindReachSafetyFloatsSampledRelStatusCorrectFalseCount+\PonoKindReachSafetyHardnessSampledRelStatusCorrectFalseCount+\PonoKindReachSafetyHardwareSampledRelStatusCorrectFalseCount+\PonoKindReachSafetyHeapSampledRelStatusCorrectFalseCount+\PonoKindReachSafetyLoopsSampledRelStatusCorrectFalseCount+\PonoKindReachSafetyProductLinesSampledRelStatusCorrectFalseCount+\PonoKindReachSafetySequentializedSampledRelStatusCorrectFalseCount+\PonoKindReachSafetyXCSPSampledRelStatusCorrectFalseCount}
\edef\PonoKindReachSafetySampledRelStatusWrongCount{\the\numexpr\PonoKindReachSafetyArraysSampledRelStatusWrongCount+\PonoKindReachSafetyBitVectorsSampledRelStatusWrongCount+\PonoKindReachSafetyCombinationsSampledRelStatusWrongCount+\PonoKindReachSafetyControlFlowSampledRelStatusWrongCount+\PonoKindReachSafetyECASampledRelStatusWrongCount+\PonoKindReachSafetyFloatsSampledRelStatusWrongCount+\PonoKindReachSafetyHardnessSampledRelStatusWrongCount+\PonoKindReachSafetyHardwareSampledRelStatusWrongCount+\PonoKindReachSafetyHeapSampledRelStatusWrongCount+\PonoKindReachSafetyLoopsSampledRelStatusWrongCount+\PonoKindReachSafetyProductLinesSampledRelStatusWrongCount+\PonoKindReachSafetySequentializedSampledRelStatusWrongCount+\PonoKindReachSafetyXCSPSampledRelStatusWrongCount}
\edef\PonoKindTerminationSampledRelStatusAllCount{\the\numexpr\PonoKindTerminationBitVectorsSampledRelStatusAllCount+\PonoKindTerminationMainControlFlowSampledRelStatusAllCount+\PonoKindTerminationOtherSampledRelStatusAllCount}
\edef\PonoKindTerminationSampledRelStatusCorrectCount{\the\numexpr\PonoKindTerminationBitVectorsSampledRelStatusCorrectCount+\PonoKindTerminationMainControlFlowSampledRelStatusCorrectCount+\PonoKindTerminationOtherSampledRelStatusCorrectCount}
\edef\PonoKindTerminationSampledRelStatusCorrectTrueCount{\the\numexpr\PonoKindTerminationBitVectorsSampledRelStatusCorrectTrueCount+\PonoKindTerminationMainControlFlowSampledRelStatusCorrectTrueCount+\PonoKindTerminationOtherSampledRelStatusCorrectTrueCount}
\edef\PonoKindTerminationSampledRelStatusCorrectFalseCount{\the\numexpr\PonoKindTerminationBitVectorsSampledRelStatusCorrectFalseCount+\PonoKindTerminationMainControlFlowSampledRelStatusCorrectFalseCount+\PonoKindTerminationOtherSampledRelStatusCorrectFalseCount}
\edef\PonoKindTerminationSampledRelStatusWrongCount{\the\numexpr\PonoKindTerminationBitVectorsSampledRelStatusWrongCount+\PonoKindTerminationMainControlFlowSampledRelStatusWrongCount+\PonoKindTerminationOtherSampledRelStatusWrongCount}

%% file: eval-results/tex/data-commands.gates.tex
\providecommand\StoreBenchExecResult[7]{\expandafter\newcommand\csname#1#2#3#4#5#6\endcsname{#7}}%
\StoreBenchExecResult{Abc}{PdrReachSafetySampledBvFunc}{Gate}{All}{}{Sum}{151472009}%
\StoreBenchExecResult{Abc}{PdrReachSafetySampledBvFunc}{Gate}{All}{}{Min}{5}%
\StoreBenchExecResult{Abc}{PdrReachSafetySampledBvFunc}{Gate}{All}{}{Max}{2369075}%
\StoreBenchExecResult{Abc}{PdrReachSafetySampledBvFunc}{Gate}{All}{}{Avg}{129796.0659811482433590402742}%
\StoreBenchExecResult{Abc}{PdrReachSafetySampledBvFunc}{Gate}{All}{}{Median}{13758}%
\StoreBenchExecResult{Abc}{PdrReachSafetySampledBvFunc}{Gate}{All}{}{Stdev}{355987.1087239512947275353446}%
\StoreBenchExecResult{Abc}{PdrReachSafetySampledBvFunc}{Gate}{Correct}{}{Sum}{29398389}%
\StoreBenchExecResult{Abc}{PdrReachSafetySampledBvFunc}{Gate}{Correct}{}{Min}{5}%
\StoreBenchExecResult{Abc}{PdrReachSafetySampledBvFunc}{Gate}{Correct}{}{Max}{1062557}%
\StoreBenchExecResult{Abc}{PdrReachSafetySampledBvFunc}{Gate}{Correct}{}{Avg}{37025.67884130982367758186398}%
\StoreBenchExecResult{Abc}{PdrReachSafetySampledBvFunc}{Gate}{Correct}{}{Median}{7744.5}%
\StoreBenchExecResult{Abc}{PdrReachSafetySampledBvFunc}{Gate}{Correct}{}{Stdev}{85852.13695518692187695324434}%
\StoreBenchExecResult{Abc}{PdrReachSafetySampledBvFunc}{Gate}{Correct}{True}{Sum}{11702277}%
\StoreBenchExecResult{Abc}{PdrReachSafetySampledBvFunc}{Gate}{Correct}{True}{Min}{7}%
\StoreBenchExecResult{Abc}{PdrReachSafetySampledBvFunc}{Gate}{Correct}{True}{Max}{889022}%
\StoreBenchExecResult{Abc}{PdrReachSafetySampledBvFunc}{Gate}{Correct}{True}{Avg}{27026.04387990762124711316397}%
\StoreBenchExecResult{Abc}{PdrReachSafetySampledBvFunc}{Gate}{Correct}{True}{Median}{6234}%
\StoreBenchExecResult{Abc}{PdrReachSafetySampledBvFunc}{Gate}{Correct}{True}{Stdev}{80198.54870785162538347918603}%
\StoreBenchExecResult{Abc}{PdrReachSafetySampledBvFunc}{Gate}{Correct}{False}{Sum}{17696112}%
\StoreBenchExecResult{Abc}{PdrReachSafetySampledBvFunc}{Gate}{Correct}{False}{Min}{5}%
\StoreBenchExecResult{Abc}{PdrReachSafetySampledBvFunc}{Gate}{Correct}{False}{Max}{1062557}%
\StoreBenchExecResult{Abc}{PdrReachSafetySampledBvFunc}{Gate}{Correct}{False}{Avg}{49019.70083102493074792243767}%
\StoreBenchExecResult{Abc}{PdrReachSafetySampledBvFunc}{Gate}{Correct}{False}{Median}{13755}%
\StoreBenchExecResult{Abc}{PdrReachSafetySampledBvFunc}{Gate}{Correct}{False}{Stdev}{90734.86408009596844544419877}%
\providecommand\StoreBenchExecResult[7]{\expandafter\newcommand\csname#1#2#3#4#5#6\endcsname{#7}}%
\StoreBenchExecResult{Abc}{PdrReachSafetySampledBvRel}{Gate}{All}{}{Sum}{159430469}%
\StoreBenchExecResult{Abc}{PdrReachSafetySampledBvRel}{Gate}{All}{}{Min}{24}%
\StoreBenchExecResult{Abc}{PdrReachSafetySampledBvRel}{Gate}{All}{}{Max}{2371113}%
\StoreBenchExecResult{Abc}{PdrReachSafetySampledBvRel}{Gate}{All}{}{Avg}{136615.6546700942587832047986}%
\StoreBenchExecResult{Abc}{PdrReachSafetySampledBvRel}{Gate}{All}{}{Median}{21897}%
\StoreBenchExecResult{Abc}{PdrReachSafetySampledBvRel}{Gate}{All}{}{Stdev}{354429.4663319480135547588305}%
\StoreBenchExecResult{Abc}{PdrReachSafetySampledBvRel}{Gate}{Correct}{}{Sum}{26148356}%
\StoreBenchExecResult{Abc}{PdrReachSafetySampledBvRel}{Gate}{Correct}{}{Min}{24}%
\StoreBenchExecResult{Abc}{PdrReachSafetySampledBvRel}{Gate}{Correct}{}{Max}{1064048}%
\StoreBenchExecResult{Abc}{PdrReachSafetySampledBvRel}{Gate}{Correct}{}{Avg}{37569.47701149425287356321839}%
\StoreBenchExecResult{Abc}{PdrReachSafetySampledBvRel}{Gate}{Correct}{}{Median}{10454}%
\StoreBenchExecResult{Abc}{PdrReachSafetySampledBvRel}{Gate}{Correct}{}{Stdev}{86072.59239068450298226874840}%
\StoreBenchExecResult{Abc}{PdrReachSafetySampledBvRel}{Gate}{Correct}{True}{Sum}{12413779}%
\StoreBenchExecResult{Abc}{PdrReachSafetySampledBvRel}{Gate}{Correct}{True}{Min}{24}%
\StoreBenchExecResult{Abc}{PdrReachSafetySampledBvRel}{Gate}{Correct}{True}{Max}{898007}%
\StoreBenchExecResult{Abc}{PdrReachSafetySampledBvRel}{Gate}{Correct}{True}{Avg}{30130.53155339805825242718447}%
\StoreBenchExecResult{Abc}{PdrReachSafetySampledBvRel}{Gate}{Correct}{True}{Median}{8352}%
\StoreBenchExecResult{Abc}{PdrReachSafetySampledBvRel}{Gate}{Correct}{True}{Stdev}{82198.79211589493411141520533}%
\StoreBenchExecResult{Abc}{PdrReachSafetySampledBvRel}{Gate}{Correct}{False}{Sum}{13734577}%
\StoreBenchExecResult{Abc}{PdrReachSafetySampledBvRel}{Gate}{Correct}{False}{Min}{24}%
\StoreBenchExecResult{Abc}{PdrReachSafetySampledBvRel}{Gate}{Correct}{False}{Max}{1064048}%
\StoreBenchExecResult{Abc}{PdrReachSafetySampledBvRel}{Gate}{Correct}{False}{Avg}{48361.18661971830985915492958}%
\StoreBenchExecResult{Abc}{PdrReachSafetySampledBvRel}{Gate}{Correct}{False}{Median}{14883}%
\StoreBenchExecResult{Abc}{PdrReachSafetySampledBvRel}{Gate}{Correct}{False}{Stdev}{90318.28381370629902051001265}%
\newcommand{\AbcPdrGateReachSafetySampledBvSignTestGroupOneMedian}{21897.0}
\newcommand{\AbcPdrGateReachSafetySampledBvSignTestGroupTwoMedian}{13758.0}
\newcommand{\AbcPdrGateReachSafetySampledBvSignTestPlusCount}{1070}
\newcommand{\AbcPdrGateReachSafetySampledBvSignTestMinusCount}{97}
\newcommand{\AbcPdrGateReachSafetySampledBvSignTestAllCount}{1167}
\newcommand{\AbcPdrGateReachSafetySampledBvSignTestPVal}{6.030978747083162e-208}
\newcommand{\AbcImcCputimeReachSafetySampledBvSignTestGroupOneMedian}{0.342352}
\newcommand{\AbcImcCputimeReachSafetySampledBvSignTestGroupTwoMedian}{0.379281}
\newcommand{\AbcImcCputimeReachSafetySampledBvSignTestPlusCount}{457}
\newcommand{\AbcImcCputimeReachSafetySampledBvSignTestMinusCount}{400}
\newcommand{\AbcImcCputimeReachSafetySampledBvSignTestAllCount}{857}
\newcommand{\AbcImcCputimeReachSafetySampledBvSignTestPVal}{0.055694464833851036}
\newcommand{\AbcPdrCputimeReachSafetySampledBvSignTestGroupOneMedian}{0.661462}
\newcommand{\AbcPdrCputimeReachSafetySampledBvSignTestGroupTwoMedian}{0.232518}
\newcommand{\AbcPdrCputimeReachSafetySampledBvSignTestPlusCount}{436}
\newcommand{\AbcPdrCputimeReachSafetySampledBvSignTestMinusCount}{245}
\newcommand{\AbcPdrCputimeReachSafetySampledBvSignTestAllCount}{681}
\newcommand{\AbcPdrCputimeReachSafetySampledBvSignTestPVal}{2.3272876196738313e-13}
\providecommand\StoreBenchExecResult[7]{\expandafter\newcommand\csname#1#2#3#4#5#6\endcsname{#7}}%
\StoreBenchExecResult{Abc}{PdrTerminationSampledBvFunc}{Gate}{All}{}{Sum}{66669719}%
\StoreBenchExecResult{Abc}{PdrTerminationSampledBvFunc}{Gate}{All}{}{Min}{31}%
\StoreBenchExecResult{Abc}{PdrTerminationSampledBvFunc}{Gate}{All}{}{Max}{7471445}%
\StoreBenchExecResult{Abc}{PdrTerminationSampledBvFunc}{Gate}{All}{}{Avg}{53722.57775987107171635777599}%
\StoreBenchExecResult{Abc}{PdrTerminationSampledBvFunc}{Gate}{All}{}{Median}{8396}%
\StoreBenchExecResult{Abc}{PdrTerminationSampledBvFunc}{Gate}{All}{}{Stdev}{277867.1246349561671044716013}%
\StoreBenchExecResult{Abc}{PdrTerminationSampledBvFunc}{Gate}{Correct}{}{Sum}{13007541}%
\StoreBenchExecResult{Abc}{PdrTerminationSampledBvFunc}{Gate}{Correct}{}{Min}{31}%
\StoreBenchExecResult{Abc}{PdrTerminationSampledBvFunc}{Gate}{Correct}{}{Max}{273102}%
\StoreBenchExecResult{Abc}{PdrTerminationSampledBvFunc}{Gate}{Correct}{}{Avg}{19299.02225519287833827893175}%
\StoreBenchExecResult{Abc}{PdrTerminationSampledBvFunc}{Gate}{Correct}{}{Median}{8613}%
\StoreBenchExecResult{Abc}{PdrTerminationSampledBvFunc}{Gate}{Correct}{}{Stdev}{37631.79829270490959334818893}%
\StoreBenchExecResult{Abc}{PdrTerminationSampledBvFunc}{Gate}{Correct}{True}{Sum}{2685032}%
\StoreBenchExecResult{Abc}{PdrTerminationSampledBvFunc}{Gate}{Correct}{True}{Min}{31}%
\StoreBenchExecResult{Abc}{PdrTerminationSampledBvFunc}{Gate}{Correct}{True}{Max}{273102}%
\StoreBenchExecResult{Abc}{PdrTerminationSampledBvFunc}{Gate}{Correct}{True}{Avg}{24633.32110091743119266055046}%
\StoreBenchExecResult{Abc}{PdrTerminationSampledBvFunc}{Gate}{Correct}{True}{Median}{1196}%
\StoreBenchExecResult{Abc}{PdrTerminationSampledBvFunc}{Gate}{Correct}{True}{Stdev}{60195.31260087514228693126372}%
\StoreBenchExecResult{Abc}{PdrTerminationSampledBvFunc}{Gate}{Correct}{False}{Sum}{10322509}%
\StoreBenchExecResult{Abc}{PdrTerminationSampledBvFunc}{Gate}{Correct}{False}{Min}{35}%
\StoreBenchExecResult{Abc}{PdrTerminationSampledBvFunc}{Gate}{Correct}{False}{Max}{179438}%
\StoreBenchExecResult{Abc}{PdrTerminationSampledBvFunc}{Gate}{Correct}{False}{Avg}{18269.92743362831858407079646}%
\StoreBenchExecResult{Abc}{PdrTerminationSampledBvFunc}{Gate}{Correct}{False}{Median}{8883}%
\StoreBenchExecResult{Abc}{PdrTerminationSampledBvFunc}{Gate}{Correct}{False}{Stdev}{31365.04571066176946747190012}%
\providecommand\StoreBenchExecResult[7]{\expandafter\newcommand\csname#1#2#3#4#5#6\endcsname{#7}}%
\StoreBenchExecResult{Abc}{PdrTerminationSampledBvRel}{Gate}{All}{}{Sum}{58849426}%
\StoreBenchExecResult{Abc}{PdrTerminationSampledBvRel}{Gate}{All}{}{Min}{58}%
\StoreBenchExecResult{Abc}{PdrTerminationSampledBvRel}{Gate}{All}{}{Max}{3845121}%
\StoreBenchExecResult{Abc}{PdrTerminationSampledBvRel}{Gate}{All}{}{Avg}{47420.97179693795326349717969}%
\StoreBenchExecResult{Abc}{PdrTerminationSampledBvRel}{Gate}{All}{}{Median}{10801}%
\StoreBenchExecResult{Abc}{PdrTerminationSampledBvRel}{Gate}{All}{}{Stdev}{156656.1581104743377898133206}%
\StoreBenchExecResult{Abc}{PdrTerminationSampledBvRel}{Gate}{Correct}{}{Sum}{11294575}%
\StoreBenchExecResult{Abc}{PdrTerminationSampledBvRel}{Gate}{Correct}{}{Min}{58}%
\StoreBenchExecResult{Abc}{PdrTerminationSampledBvRel}{Gate}{Correct}{}{Max}{182772}%
\StoreBenchExecResult{Abc}{PdrTerminationSampledBvRel}{Gate}{Correct}{}{Avg}{18485.39279869067103109656301}%
\StoreBenchExecResult{Abc}{PdrTerminationSampledBvRel}{Gate}{Correct}{}{Median}{10830}%
\StoreBenchExecResult{Abc}{PdrTerminationSampledBvRel}{Gate}{Correct}{}{Stdev}{30277.24533679538785447214020}%
\StoreBenchExecResult{Abc}{PdrTerminationSampledBvRel}{Gate}{Correct}{True}{Sum}{399522}%
\StoreBenchExecResult{Abc}{PdrTerminationSampledBvRel}{Gate}{Correct}{True}{Min}{58}%
\StoreBenchExecResult{Abc}{PdrTerminationSampledBvRel}{Gate}{Correct}{True}{Max}{163862}%
\StoreBenchExecResult{Abc}{PdrTerminationSampledBvRel}{Gate}{Correct}{True}{Avg}{6146.492307692307692307692308}%
\StoreBenchExecResult{Abc}{PdrTerminationSampledBvRel}{Gate}{Correct}{True}{Median}{1153}%
\StoreBenchExecResult{Abc}{PdrTerminationSampledBvRel}{Gate}{Correct}{True}{Stdev}{24795.90735034440931288052198}%
\StoreBenchExecResult{Abc}{PdrTerminationSampledBvRel}{Gate}{Correct}{False}{Sum}{10895053}%
\StoreBenchExecResult{Abc}{PdrTerminationSampledBvRel}{Gate}{Correct}{False}{Min}{194}%
\StoreBenchExecResult{Abc}{PdrTerminationSampledBvRel}{Gate}{Correct}{False}{Max}{182772}%
\StoreBenchExecResult{Abc}{PdrTerminationSampledBvRel}{Gate}{Correct}{False}{Avg}{19954.30952380952380952380952}%
\StoreBenchExecResult{Abc}{PdrTerminationSampledBvRel}{Gate}{Correct}{False}{Median}{11080}%
\StoreBenchExecResult{Abc}{PdrTerminationSampledBvRel}{Gate}{Correct}{False}{Stdev}{30534.67700200297228931010202}%
\newcommand{\AbcPdrGateTerminationSampledBvSignTestGroupOneMedian}{10801.0}
\newcommand{\AbcPdrGateTerminationSampledBvSignTestGroupTwoMedian}{8396.0}
\newcommand{\AbcPdrGateTerminationSampledBvSignTestPlusCount}{1212}
\newcommand{\AbcPdrGateTerminationSampledBvSignTestMinusCount}{28}
\newcommand{\AbcPdrGateTerminationSampledBvSignTestAllCount}{1241}
\newcommand{\AbcPdrGateTerminationSampledBvSignTestPVal}{0.0}
\newcommand{\AbcImcCputimeTerminationSampledBvSignTestGroupOneMedian}{0.346793}
\newcommand{\AbcImcCputimeTerminationSampledBvSignTestGroupTwoMedian}{0.222718}
\newcommand{\AbcImcCputimeTerminationSampledBvSignTestPlusCount}{729}
\newcommand{\AbcImcCputimeTerminationSampledBvSignTestMinusCount}{218}
\newcommand{\AbcImcCputimeTerminationSampledBvSignTestAllCount}{947}
\newcommand{\AbcImcCputimeTerminationSampledBvSignTestPVal}{5.742152314832602e-65}
\newcommand{\AbcPdrCputimeTerminationSampledBvSignTestGroupOneMedian}{1.933537}
\newcommand{\AbcPdrCputimeTerminationSampledBvSignTestGroupTwoMedian}{0.377057}
\newcommand{\AbcPdrCputimeTerminationSampledBvSignTestPlusCount}{477}
\newcommand{\AbcPdrCputimeTerminationSampledBvSignTestMinusCount}{82}
\newcommand{\AbcPdrCputimeTerminationSampledBvSignTestAllCount}{559}
\newcommand{\AbcPdrCputimeTerminationSampledBvSignTestPVal}{1.0044471307983104e-68}

%% file: eval-results/tex/data-commands.scorr.tex
\providecommand\StoreBenchExecResult[7]{\expandafter\newcommand\csname#1#2#3#4#5#6\endcsname{#7}}%
\StoreBenchExecResult{Abc}{ScPdrReachSafetyBitVectorsSampledBvFunc}{Status}{All}{}{Score}{0}%
\StoreBenchExecResult{Abc}{ScPdrReachSafetyBitVectorsSampledBvFunc}{Status}{All}{}{Count}{47}%
\StoreBenchExecResult{Abc}{ScPdrReachSafetyBitVectorsSampledBvFunc}{Status}{Correct}{}{Count}{42}%
\StoreBenchExecResult{Abc}{ScPdrReachSafetyBitVectorsSampledBvFunc}{Status}{Correct}{True}{Count}{31}%
\StoreBenchExecResult{Abc}{ScPdrReachSafetyBitVectorsSampledBvFunc}{Status}{Correct}{False}{Count}{11}%
\StoreBenchExecResult{Abc}{ScPdrReachSafetyBitVectorsSampledBvFunc}{Status}{Wrong}{}{Count}{0}%
\StoreBenchExecResult{Abc}{ScPdrReachSafetyBitVectorsSampledBvFunc}{Status}{Wrong}{True}{Count}{0}%
\StoreBenchExecResult{Abc}{ScPdrReachSafetyBitVectorsSampledBvFunc}{Status}{Wrong}{False}{Count}{0}%
\providecommand\StoreBenchExecResult[7]{\expandafter\newcommand\csname#1#2#3#4#5#6\endcsname{#7}}%
\StoreBenchExecResult{Abc}{ScPdrReachSafetyCombinationsSampledBvFunc}{Status}{All}{}{Score}{0}%
\StoreBenchExecResult{Abc}{ScPdrReachSafetyCombinationsSampledBvFunc}{Status}{All}{}{Count}{110}%
\StoreBenchExecResult{Abc}{ScPdrReachSafetyCombinationsSampledBvFunc}{Status}{Correct}{}{Count}{83}%
\StoreBenchExecResult{Abc}{ScPdrReachSafetyCombinationsSampledBvFunc}{Status}{Correct}{True}{Count}{13}%
\StoreBenchExecResult{Abc}{ScPdrReachSafetyCombinationsSampledBvFunc}{Status}{Correct}{False}{Count}{70}%
\StoreBenchExecResult{Abc}{ScPdrReachSafetyCombinationsSampledBvFunc}{Status}{Wrong}{}{Count}{0}%
\StoreBenchExecResult{Abc}{ScPdrReachSafetyCombinationsSampledBvFunc}{Status}{Wrong}{True}{Count}{0}%
\StoreBenchExecResult{Abc}{ScPdrReachSafetyCombinationsSampledBvFunc}{Status}{Wrong}{False}{Count}{0}%
\providecommand\StoreBenchExecResult[7]{\expandafter\newcommand\csname#1#2#3#4#5#6\endcsname{#7}}%
\StoreBenchExecResult{Abc}{ScPdrReachSafetyControlFlowSampledBvFunc}{Status}{All}{}{Score}{0}%
\StoreBenchExecResult{Abc}{ScPdrReachSafetyControlFlowSampledBvFunc}{Status}{All}{}{Count}{29}%
\StoreBenchExecResult{Abc}{ScPdrReachSafetyControlFlowSampledBvFunc}{Status}{Correct}{}{Count}{27}%
\StoreBenchExecResult{Abc}{ScPdrReachSafetyControlFlowSampledBvFunc}{Status}{Correct}{True}{Count}{25}%
\StoreBenchExecResult{Abc}{ScPdrReachSafetyControlFlowSampledBvFunc}{Status}{Correct}{False}{Count}{2}%
\StoreBenchExecResult{Abc}{ScPdrReachSafetyControlFlowSampledBvFunc}{Status}{Wrong}{}{Count}{0}%
\StoreBenchExecResult{Abc}{ScPdrReachSafetyControlFlowSampledBvFunc}{Status}{Wrong}{True}{Count}{0}%
\StoreBenchExecResult{Abc}{ScPdrReachSafetyControlFlowSampledBvFunc}{Status}{Wrong}{False}{Count}{0}%
\providecommand\StoreBenchExecResult[7]{\expandafter\newcommand\csname#1#2#3#4#5#6\endcsname{#7}}%
\StoreBenchExecResult{Abc}{ScPdrReachSafetyECASampledBvFunc}{Status}{All}{}{Score}{0}%
\StoreBenchExecResult{Abc}{ScPdrReachSafetyECASampledBvFunc}{Status}{All}{}{Count}{150}%
\StoreBenchExecResult{Abc}{ScPdrReachSafetyECASampledBvFunc}{Status}{Correct}{}{Count}{48}%
\StoreBenchExecResult{Abc}{ScPdrReachSafetyECASampledBvFunc}{Status}{Correct}{True}{Count}{27}%
\StoreBenchExecResult{Abc}{ScPdrReachSafetyECASampledBvFunc}{Status}{Correct}{False}{Count}{21}%
\StoreBenchExecResult{Abc}{ScPdrReachSafetyECASampledBvFunc}{Status}{Wrong}{}{Count}{0}%
\StoreBenchExecResult{Abc}{ScPdrReachSafetyECASampledBvFunc}{Status}{Wrong}{True}{Count}{0}%
\StoreBenchExecResult{Abc}{ScPdrReachSafetyECASampledBvFunc}{Status}{Wrong}{False}{Count}{0}%
\providecommand\StoreBenchExecResult[7]{\expandafter\newcommand\csname#1#2#3#4#5#6\endcsname{#7}}%
\StoreBenchExecResult{Abc}{ScPdrReachSafetyFloatsSampledBvFunc}{Status}{All}{}{Score}{0}%
\StoreBenchExecResult{Abc}{ScPdrReachSafetyFloatsSampledBvFunc}{Status}{All}{}{Count}{10}%
\StoreBenchExecResult{Abc}{ScPdrReachSafetyFloatsSampledBvFunc}{Status}{Correct}{}{Count}{10}%
\StoreBenchExecResult{Abc}{ScPdrReachSafetyFloatsSampledBvFunc}{Status}{Correct}{True}{Count}{10}%
\StoreBenchExecResult{Abc}{ScPdrReachSafetyFloatsSampledBvFunc}{Status}{Correct}{False}{Count}{0}%
\StoreBenchExecResult{Abc}{ScPdrReachSafetyFloatsSampledBvFunc}{Status}{Wrong}{}{Count}{0}%
\StoreBenchExecResult{Abc}{ScPdrReachSafetyFloatsSampledBvFunc}{Status}{Wrong}{True}{Count}{0}%
\StoreBenchExecResult{Abc}{ScPdrReachSafetyFloatsSampledBvFunc}{Status}{Wrong}{False}{Count}{0}%
\providecommand\StoreBenchExecResult[7]{\expandafter\newcommand\csname#1#2#3#4#5#6\endcsname{#7}}%
\StoreBenchExecResult{Abc}{ScPdrReachSafetyHardnessSampledBvFunc}{Status}{All}{}{Score}{0}%
\StoreBenchExecResult{Abc}{ScPdrReachSafetyHardnessSampledBvFunc}{Status}{All}{}{Count}{132}%
\StoreBenchExecResult{Abc}{ScPdrReachSafetyHardnessSampledBvFunc}{Status}{Correct}{}{Count}{132}%
\StoreBenchExecResult{Abc}{ScPdrReachSafetyHardnessSampledBvFunc}{Status}{Correct}{True}{Count}{132}%
\StoreBenchExecResult{Abc}{ScPdrReachSafetyHardnessSampledBvFunc}{Status}{Correct}{False}{Count}{0}%
\StoreBenchExecResult{Abc}{ScPdrReachSafetyHardnessSampledBvFunc}{Status}{Wrong}{}{Count}{0}%
\StoreBenchExecResult{Abc}{ScPdrReachSafetyHardnessSampledBvFunc}{Status}{Wrong}{True}{Count}{0}%
\StoreBenchExecResult{Abc}{ScPdrReachSafetyHardnessSampledBvFunc}{Status}{Wrong}{False}{Count}{0}%
\providecommand\StoreBenchExecResult[7]{\expandafter\newcommand\csname#1#2#3#4#5#6\endcsname{#7}}%
\StoreBenchExecResult{Abc}{ScPdrReachSafetyHardwareSampledBvFunc}{Status}{All}{}{Score}{0}%
\StoreBenchExecResult{Abc}{ScPdrReachSafetyHardwareSampledBvFunc}{Status}{All}{}{Count}{148}%
\StoreBenchExecResult{Abc}{ScPdrReachSafetyHardwareSampledBvFunc}{Status}{Correct}{}{Count}{109}%
\StoreBenchExecResult{Abc}{ScPdrReachSafetyHardwareSampledBvFunc}{Status}{Correct}{True}{Count}{50}%
\StoreBenchExecResult{Abc}{ScPdrReachSafetyHardwareSampledBvFunc}{Status}{Correct}{False}{Count}{59}%
\StoreBenchExecResult{Abc}{ScPdrReachSafetyHardwareSampledBvFunc}{Status}{Wrong}{}{Count}{0}%
\StoreBenchExecResult{Abc}{ScPdrReachSafetyHardwareSampledBvFunc}{Status}{Wrong}{True}{Count}{0}%
\StoreBenchExecResult{Abc}{ScPdrReachSafetyHardwareSampledBvFunc}{Status}{Wrong}{False}{Count}{0}%
\providecommand\StoreBenchExecResult[7]{\expandafter\newcommand\csname#1#2#3#4#5#6\endcsname{#7}}%
\StoreBenchExecResult{Abc}{ScPdrReachSafetyHeapSampledBvFunc}{Status}{All}{}{Score}{0}%
\StoreBenchExecResult{Abc}{ScPdrReachSafetyHeapSampledBvFunc}{Status}{All}{}{Count}{6}%
\StoreBenchExecResult{Abc}{ScPdrReachSafetyHeapSampledBvFunc}{Status}{Correct}{}{Count}{6}%
\StoreBenchExecResult{Abc}{ScPdrReachSafetyHeapSampledBvFunc}{Status}{Correct}{True}{Count}{4}%
\StoreBenchExecResult{Abc}{ScPdrReachSafetyHeapSampledBvFunc}{Status}{Correct}{False}{Count}{2}%
\StoreBenchExecResult{Abc}{ScPdrReachSafetyHeapSampledBvFunc}{Status}{Wrong}{}{Count}{0}%
\StoreBenchExecResult{Abc}{ScPdrReachSafetyHeapSampledBvFunc}{Status}{Wrong}{True}{Count}{0}%
\StoreBenchExecResult{Abc}{ScPdrReachSafetyHeapSampledBvFunc}{Status}{Wrong}{False}{Count}{0}%
\providecommand\StoreBenchExecResult[7]{\expandafter\newcommand\csname#1#2#3#4#5#6\endcsname{#7}}%
\StoreBenchExecResult{Abc}{ScPdrReachSafetyLoopsSampledBvFunc}{Status}{All}{}{Score}{0}%
\StoreBenchExecResult{Abc}{ScPdrReachSafetyLoopsSampledBvFunc}{Status}{All}{}{Count}{137}%
\StoreBenchExecResult{Abc}{ScPdrReachSafetyLoopsSampledBvFunc}{Status}{Correct}{}{Count}{70}%
\StoreBenchExecResult{Abc}{ScPdrReachSafetyLoopsSampledBvFunc}{Status}{Correct}{True}{Count}{41}%
\StoreBenchExecResult{Abc}{ScPdrReachSafetyLoopsSampledBvFunc}{Status}{Correct}{False}{Count}{29}%
\StoreBenchExecResult{Abc}{ScPdrReachSafetyLoopsSampledBvFunc}{Status}{Wrong}{}{Count}{0}%
\StoreBenchExecResult{Abc}{ScPdrReachSafetyLoopsSampledBvFunc}{Status}{Wrong}{True}{Count}{0}%
\StoreBenchExecResult{Abc}{ScPdrReachSafetyLoopsSampledBvFunc}{Status}{Wrong}{False}{Count}{0}%
\providecommand\StoreBenchExecResult[7]{\expandafter\newcommand\csname#1#2#3#4#5#6\endcsname{#7}}%
\StoreBenchExecResult{Abc}{ScPdrReachSafetyProductLinesSampledBvFunc}{Status}{All}{}{Score}{0}%
\StoreBenchExecResult{Abc}{ScPdrReachSafetyProductLinesSampledBvFunc}{Status}{All}{}{Count}{150}%
\StoreBenchExecResult{Abc}{ScPdrReachSafetyProductLinesSampledBvFunc}{Status}{Correct}{}{Count}{146}%
\StoreBenchExecResult{Abc}{ScPdrReachSafetyProductLinesSampledBvFunc}{Status}{Correct}{True}{Count}{74}%
\StoreBenchExecResult{Abc}{ScPdrReachSafetyProductLinesSampledBvFunc}{Status}{Correct}{False}{Count}{72}%
\StoreBenchExecResult{Abc}{ScPdrReachSafetyProductLinesSampledBvFunc}{Status}{Wrong}{}{Count}{0}%
\StoreBenchExecResult{Abc}{ScPdrReachSafetyProductLinesSampledBvFunc}{Status}{Wrong}{True}{Count}{0}%
\StoreBenchExecResult{Abc}{ScPdrReachSafetyProductLinesSampledBvFunc}{Status}{Wrong}{False}{Count}{0}%
\providecommand\StoreBenchExecResult[7]{\expandafter\newcommand\csname#1#2#3#4#5#6\endcsname{#7}}%
\StoreBenchExecResult{Abc}{ScPdrReachSafetySequentializedSampledBvFunc}{Status}{All}{}{Score}{0}%
\StoreBenchExecResult{Abc}{ScPdrReachSafetySequentializedSampledBvFunc}{Status}{All}{}{Count}{150}%
\StoreBenchExecResult{Abc}{ScPdrReachSafetySequentializedSampledBvFunc}{Status}{Correct}{}{Count}{93}%
\StoreBenchExecResult{Abc}{ScPdrReachSafetySequentializedSampledBvFunc}{Status}{Correct}{True}{Count}{21}%
\StoreBenchExecResult{Abc}{ScPdrReachSafetySequentializedSampledBvFunc}{Status}{Correct}{False}{Count}{72}%
\StoreBenchExecResult{Abc}{ScPdrReachSafetySequentializedSampledBvFunc}{Status}{Wrong}{}{Count}{0}%
\StoreBenchExecResult{Abc}{ScPdrReachSafetySequentializedSampledBvFunc}{Status}{Wrong}{True}{Count}{0}%
\StoreBenchExecResult{Abc}{ScPdrReachSafetySequentializedSampledBvFunc}{Status}{Wrong}{False}{Count}{0}%
\providecommand\StoreBenchExecResult[7]{\expandafter\newcommand\csname#1#2#3#4#5#6\endcsname{#7}}%
\StoreBenchExecResult{Abc}{ScPdrReachSafetyXCSPSampledBvFunc}{Status}{All}{}{Score}{0}%
\StoreBenchExecResult{Abc}{ScPdrReachSafetyXCSPSampledBvFunc}{Status}{All}{}{Count}{98}%
\StoreBenchExecResult{Abc}{ScPdrReachSafetyXCSPSampledBvFunc}{Status}{Correct}{}{Count}{90}%
\StoreBenchExecResult{Abc}{ScPdrReachSafetyXCSPSampledBvFunc}{Status}{Correct}{True}{Count}{47}%
\StoreBenchExecResult{Abc}{ScPdrReachSafetyXCSPSampledBvFunc}{Status}{Correct}{False}{Count}{43}%
\StoreBenchExecResult{Abc}{ScPdrReachSafetyXCSPSampledBvFunc}{Status}{Wrong}{}{Count}{0}%
\StoreBenchExecResult{Abc}{ScPdrReachSafetyXCSPSampledBvFunc}{Status}{Wrong}{True}{Count}{0}%
\StoreBenchExecResult{Abc}{ScPdrReachSafetyXCSPSampledBvFunc}{Status}{Wrong}{False}{Count}{0}%
\providecommand\StoreBenchExecResult[7]{\expandafter\newcommand\csname#1#2#3#4#5#6\endcsname{#7}}%
\StoreBenchExecResult{Abc}{ScPdrTerminationBitVectorsSampledBvFunc}{Status}{All}{}{Score}{0}%
\StoreBenchExecResult{Abc}{ScPdrTerminationBitVectorsSampledBvFunc}{Status}{All}{}{Count}{32}%
\StoreBenchExecResult{Abc}{ScPdrTerminationBitVectorsSampledBvFunc}{Status}{Correct}{}{Count}{21}%
\StoreBenchExecResult{Abc}{ScPdrTerminationBitVectorsSampledBvFunc}{Status}{Correct}{True}{Count}{11}%
\StoreBenchExecResult{Abc}{ScPdrTerminationBitVectorsSampledBvFunc}{Status}{Correct}{False}{Count}{10}%
\StoreBenchExecResult{Abc}{ScPdrTerminationBitVectorsSampledBvFunc}{Status}{Wrong}{}{Count}{0}%
\StoreBenchExecResult{Abc}{ScPdrTerminationBitVectorsSampledBvFunc}{Status}{Wrong}{True}{Count}{0}%
\StoreBenchExecResult{Abc}{ScPdrTerminationBitVectorsSampledBvFunc}{Status}{Wrong}{False}{Count}{0}%
\providecommand\StoreBenchExecResult[7]{\expandafter\newcommand\csname#1#2#3#4#5#6\endcsname{#7}}%
\StoreBenchExecResult{Abc}{ScPdrTerminationMainControlFlowSampledBvFunc}{Status}{All}{}{Score}{0}%
\StoreBenchExecResult{Abc}{ScPdrTerminationMainControlFlowSampledBvFunc}{Status}{All}{}{Count}{236}%
\StoreBenchExecResult{Abc}{ScPdrTerminationMainControlFlowSampledBvFunc}{Status}{Correct}{}{Count}{66}%
\StoreBenchExecResult{Abc}{ScPdrTerminationMainControlFlowSampledBvFunc}{Status}{Correct}{True}{Count}{18}%
\StoreBenchExecResult{Abc}{ScPdrTerminationMainControlFlowSampledBvFunc}{Status}{Correct}{False}{Count}{48}%
\StoreBenchExecResult{Abc}{ScPdrTerminationMainControlFlowSampledBvFunc}{Status}{Wrong}{}{Count}{0}%
\StoreBenchExecResult{Abc}{ScPdrTerminationMainControlFlowSampledBvFunc}{Status}{Wrong}{True}{Count}{0}%
\StoreBenchExecResult{Abc}{ScPdrTerminationMainControlFlowSampledBvFunc}{Status}{Wrong}{False}{Count}{0}%
\providecommand\StoreBenchExecResult[7]{\expandafter\newcommand\csname#1#2#3#4#5#6\endcsname{#7}}%
\StoreBenchExecResult{Abc}{ScPdrTerminationOtherSampledBvFunc}{Status}{All}{}{Score}{0}%
\StoreBenchExecResult{Abc}{ScPdrTerminationOtherSampledBvFunc}{Status}{All}{}{Count}{973}%
\StoreBenchExecResult{Abc}{ScPdrTerminationOtherSampledBvFunc}{Status}{Correct}{}{Count}{751}%
\StoreBenchExecResult{Abc}{ScPdrTerminationOtherSampledBvFunc}{Status}{Correct}{True}{Count}{183}%
\StoreBenchExecResult{Abc}{ScPdrTerminationOtherSampledBvFunc}{Status}{Correct}{False}{Count}{568}%
\StoreBenchExecResult{Abc}{ScPdrTerminationOtherSampledBvFunc}{Status}{Wrong}{}{Count}{0}%
\StoreBenchExecResult{Abc}{ScPdrTerminationOtherSampledBvFunc}{Status}{Wrong}{True}{Count}{0}%
\StoreBenchExecResult{Abc}{ScPdrTerminationOtherSampledBvFunc}{Status}{Wrong}{False}{Count}{0}%
\edef\AbcScPdrReachSafetySampledBvFuncStatusAllCount{\the\numexpr\AbcScPdrReachSafetyBitVectorsSampledBvFuncStatusAllCount+\AbcScPdrReachSafetyCombinationsSampledBvFuncStatusAllCount+\AbcScPdrReachSafetyControlFlowSampledBvFuncStatusAllCount+\AbcScPdrReachSafetyECASampledBvFuncStatusAllCount+\AbcScPdrReachSafetyFloatsSampledBvFuncStatusAllCount+\AbcScPdrReachSafetyHardnessSampledBvFuncStatusAllCount+\AbcScPdrReachSafetyHardwareSampledBvFuncStatusAllCount+\AbcScPdrReachSafetyHeapSampledBvFuncStatusAllCount+\AbcScPdrReachSafetyLoopsSampledBvFuncStatusAllCount+\AbcScPdrReachSafetyProductLinesSampledBvFuncStatusAllCount+\AbcScPdrReachSafetySequentializedSampledBvFuncStatusAllCount+\AbcScPdrReachSafetyXCSPSampledBvFuncStatusAllCount}
\edef\AbcScPdrReachSafetySampledBvFuncStatusCorrectCount{\the\numexpr\AbcScPdrReachSafetyBitVectorsSampledBvFuncStatusCorrectCount+\AbcScPdrReachSafetyCombinationsSampledBvFuncStatusCorrectCount+\AbcScPdrReachSafetyControlFlowSampledBvFuncStatusCorrectCount+\AbcScPdrReachSafetyECASampledBvFuncStatusCorrectCount+\AbcScPdrReachSafetyFloatsSampledBvFuncStatusCorrectCount+\AbcScPdrReachSafetyHardnessSampledBvFuncStatusCorrectCount+\AbcScPdrReachSafetyHardwareSampledBvFuncStatusCorrectCount+\AbcScPdrReachSafetyHeapSampledBvFuncStatusCorrectCount+\AbcScPdrReachSafetyLoopsSampledBvFuncStatusCorrectCount+\AbcScPdrReachSafetyProductLinesSampledBvFuncStatusCorrectCount+\AbcScPdrReachSafetySequentializedSampledBvFuncStatusCorrectCount+\AbcScPdrReachSafetyXCSPSampledBvFuncStatusCorrectCount}
\edef\AbcScPdrReachSafetySampledBvFuncStatusCorrectTrueCount{\the\numexpr\AbcScPdrReachSafetyBitVectorsSampledBvFuncStatusCorrectTrueCount+\AbcScPdrReachSafetyCombinationsSampledBvFuncStatusCorrectTrueCount+\AbcScPdrReachSafetyControlFlowSampledBvFuncStatusCorrectTrueCount+\AbcScPdrReachSafetyECASampledBvFuncStatusCorrectTrueCount+\AbcScPdrReachSafetyFloatsSampledBvFuncStatusCorrectTrueCount+\AbcScPdrReachSafetyHardnessSampledBvFuncStatusCorrectTrueCount+\AbcScPdrReachSafetyHardwareSampledBvFuncStatusCorrectTrueCount+\AbcScPdrReachSafetyHeapSampledBvFuncStatusCorrectTrueCount+\AbcScPdrReachSafetyLoopsSampledBvFuncStatusCorrectTrueCount+\AbcScPdrReachSafetyProductLinesSampledBvFuncStatusCorrectTrueCount+\AbcScPdrReachSafetySequentializedSampledBvFuncStatusCorrectTrueCount+\AbcScPdrReachSafetyXCSPSampledBvFuncStatusCorrectTrueCount}
\edef\AbcScPdrReachSafetySampledBvFuncStatusCorrectFalseCount{\the\numexpr\AbcScPdrReachSafetyBitVectorsSampledBvFuncStatusCorrectFalseCount+\AbcScPdrReachSafetyCombinationsSampledBvFuncStatusCorrectFalseCount+\AbcScPdrReachSafetyControlFlowSampledBvFuncStatusCorrectFalseCount+\AbcScPdrReachSafetyECASampledBvFuncStatusCorrectFalseCount+\AbcScPdrReachSafetyFloatsSampledBvFuncStatusCorrectFalseCount+\AbcScPdrReachSafetyHardnessSampledBvFuncStatusCorrectFalseCount+\AbcScPdrReachSafetyHardwareSampledBvFuncStatusCorrectFalseCount+\AbcScPdrReachSafetyHeapSampledBvFuncStatusCorrectFalseCount+\AbcScPdrReachSafetyLoopsSampledBvFuncStatusCorrectFalseCount+\AbcScPdrReachSafetyProductLinesSampledBvFuncStatusCorrectFalseCount+\AbcScPdrReachSafetySequentializedSampledBvFuncStatusCorrectFalseCount+\AbcScPdrReachSafetyXCSPSampledBvFuncStatusCorrectFalseCount}
\edef\AbcScPdrReachSafetySampledBvFuncStatusWrongCount{\the\numexpr\AbcScPdrReachSafetyBitVectorsSampledBvFuncStatusWrongCount+\AbcScPdrReachSafetyCombinationsSampledBvFuncStatusWrongCount+\AbcScPdrReachSafetyControlFlowSampledBvFuncStatusWrongCount+\AbcScPdrReachSafetyECASampledBvFuncStatusWrongCount+\AbcScPdrReachSafetyFloatsSampledBvFuncStatusWrongCount+\AbcScPdrReachSafetyHardnessSampledBvFuncStatusWrongCount+\AbcScPdrReachSafetyHardwareSampledBvFuncStatusWrongCount+\AbcScPdrReachSafetyHeapSampledBvFuncStatusWrongCount+\AbcScPdrReachSafetyLoopsSampledBvFuncStatusWrongCount+\AbcScPdrReachSafetyProductLinesSampledBvFuncStatusWrongCount+\AbcScPdrReachSafetySequentializedSampledBvFuncStatusWrongCount+\AbcScPdrReachSafetyXCSPSampledBvFuncStatusWrongCount}
\edef\AbcScPdrTerminationSampledBvFuncStatusAllCount{\the\numexpr\AbcScPdrTerminationBitVectorsSampledBvFuncStatusAllCount+\AbcScPdrTerminationMainControlFlowSampledBvFuncStatusAllCount+\AbcScPdrTerminationOtherSampledBvFuncStatusAllCount}
\edef\AbcScPdrTerminationSampledBvFuncStatusCorrectCount{\the\numexpr\AbcScPdrTerminationBitVectorsSampledBvFuncStatusCorrectCount+\AbcScPdrTerminationMainControlFlowSampledBvFuncStatusCorrectCount+\AbcScPdrTerminationOtherSampledBvFuncStatusCorrectCount}
\edef\AbcScPdrTerminationSampledBvFuncStatusCorrectTrueCount{\the\numexpr\AbcScPdrTerminationBitVectorsSampledBvFuncStatusCorrectTrueCount+\AbcScPdrTerminationMainControlFlowSampledBvFuncStatusCorrectTrueCount+\AbcScPdrTerminationOtherSampledBvFuncStatusCorrectTrueCount}
\edef\AbcScPdrTerminationSampledBvFuncStatusCorrectFalseCount{\the\numexpr\AbcScPdrTerminationBitVectorsSampledBvFuncStatusCorrectFalseCount+\AbcScPdrTerminationMainControlFlowSampledBvFuncStatusCorrectFalseCount+\AbcScPdrTerminationOtherSampledBvFuncStatusCorrectFalseCount}
\edef\AbcScPdrTerminationSampledBvFuncStatusWrongCount{\the\numexpr\AbcScPdrTerminationBitVectorsSampledBvFuncStatusWrongCount+\AbcScPdrTerminationMainControlFlowSampledBvFuncStatusWrongCount+\AbcScPdrTerminationOtherSampledBvFuncStatusWrongCount}
\providecommand\StoreBenchExecResult[7]{\expandafter\newcommand\csname#1#2#3#4#5#6\endcsname{#7}}%
\StoreBenchExecResult{Abc}{ScPdrReachSafetyBitVectorsSampledBvRel}{Status}{All}{}{Score}{0}%
\StoreBenchExecResult{Abc}{ScPdrReachSafetyBitVectorsSampledBvRel}{Status}{All}{}{Count}{47}%
\StoreBenchExecResult{Abc}{ScPdrReachSafetyBitVectorsSampledBvRel}{Status}{Correct}{}{Count}{38}%
\StoreBenchExecResult{Abc}{ScPdrReachSafetyBitVectorsSampledBvRel}{Status}{Correct}{True}{Count}{28}%
\StoreBenchExecResult{Abc}{ScPdrReachSafetyBitVectorsSampledBvRel}{Status}{Correct}{False}{Count}{10}%
\StoreBenchExecResult{Abc}{ScPdrReachSafetyBitVectorsSampledBvRel}{Status}{Wrong}{}{Count}{0}%
\StoreBenchExecResult{Abc}{ScPdrReachSafetyBitVectorsSampledBvRel}{Status}{Wrong}{True}{Count}{0}%
\StoreBenchExecResult{Abc}{ScPdrReachSafetyBitVectorsSampledBvRel}{Status}{Wrong}{False}{Count}{0}%
\providecommand\StoreBenchExecResult[7]{\expandafter\newcommand\csname#1#2#3#4#5#6\endcsname{#7}}%
\StoreBenchExecResult{Abc}{ScPdrReachSafetyCombinationsSampledBvRel}{Status}{All}{}{Score}{0}%
\StoreBenchExecResult{Abc}{ScPdrReachSafetyCombinationsSampledBvRel}{Status}{All}{}{Count}{110}%
\StoreBenchExecResult{Abc}{ScPdrReachSafetyCombinationsSampledBvRel}{Status}{Correct}{}{Count}{51}%
\StoreBenchExecResult{Abc}{ScPdrReachSafetyCombinationsSampledBvRel}{Status}{Correct}{True}{Count}{3}%
\StoreBenchExecResult{Abc}{ScPdrReachSafetyCombinationsSampledBvRel}{Status}{Correct}{False}{Count}{48}%
\StoreBenchExecResult{Abc}{ScPdrReachSafetyCombinationsSampledBvRel}{Status}{Wrong}{}{Count}{0}%
\StoreBenchExecResult{Abc}{ScPdrReachSafetyCombinationsSampledBvRel}{Status}{Wrong}{True}{Count}{0}%
\StoreBenchExecResult{Abc}{ScPdrReachSafetyCombinationsSampledBvRel}{Status}{Wrong}{False}{Count}{0}%
\providecommand\StoreBenchExecResult[7]{\expandafter\newcommand\csname#1#2#3#4#5#6\endcsname{#7}}%
\StoreBenchExecResult{Abc}{ScPdrReachSafetyControlFlowSampledBvRel}{Status}{All}{}{Score}{0}%
\StoreBenchExecResult{Abc}{ScPdrReachSafetyControlFlowSampledBvRel}{Status}{All}{}{Count}{29}%
\StoreBenchExecResult{Abc}{ScPdrReachSafetyControlFlowSampledBvRel}{Status}{Correct}{}{Count}{26}%
\StoreBenchExecResult{Abc}{ScPdrReachSafetyControlFlowSampledBvRel}{Status}{Correct}{True}{Count}{24}%
\StoreBenchExecResult{Abc}{ScPdrReachSafetyControlFlowSampledBvRel}{Status}{Correct}{False}{Count}{2}%
\StoreBenchExecResult{Abc}{ScPdrReachSafetyControlFlowSampledBvRel}{Status}{Wrong}{}{Count}{0}%
\StoreBenchExecResult{Abc}{ScPdrReachSafetyControlFlowSampledBvRel}{Status}{Wrong}{True}{Count}{0}%
\StoreBenchExecResult{Abc}{ScPdrReachSafetyControlFlowSampledBvRel}{Status}{Wrong}{False}{Count}{0}%
\providecommand\StoreBenchExecResult[7]{\expandafter\newcommand\csname#1#2#3#4#5#6\endcsname{#7}}%
\StoreBenchExecResult{Abc}{ScPdrReachSafetyECASampledBvRel}{Status}{All}{}{Score}{0}%
\StoreBenchExecResult{Abc}{ScPdrReachSafetyECASampledBvRel}{Status}{All}{}{Count}{150}%
\StoreBenchExecResult{Abc}{ScPdrReachSafetyECASampledBvRel}{Status}{Correct}{}{Count}{53}%
\StoreBenchExecResult{Abc}{ScPdrReachSafetyECASampledBvRel}{Status}{Correct}{True}{Count}{36}%
\StoreBenchExecResult{Abc}{ScPdrReachSafetyECASampledBvRel}{Status}{Correct}{False}{Count}{17}%
\StoreBenchExecResult{Abc}{ScPdrReachSafetyECASampledBvRel}{Status}{Wrong}{}{Count}{0}%
\StoreBenchExecResult{Abc}{ScPdrReachSafetyECASampledBvRel}{Status}{Wrong}{True}{Count}{0}%
\StoreBenchExecResult{Abc}{ScPdrReachSafetyECASampledBvRel}{Status}{Wrong}{False}{Count}{0}%
\providecommand\StoreBenchExecResult[7]{\expandafter\newcommand\csname#1#2#3#4#5#6\endcsname{#7}}%
\StoreBenchExecResult{Abc}{ScPdrReachSafetyFloatsSampledBvRel}{Status}{All}{}{Score}{0}%
\StoreBenchExecResult{Abc}{ScPdrReachSafetyFloatsSampledBvRel}{Status}{All}{}{Count}{10}%
\StoreBenchExecResult{Abc}{ScPdrReachSafetyFloatsSampledBvRel}{Status}{Correct}{}{Count}{10}%
\StoreBenchExecResult{Abc}{ScPdrReachSafetyFloatsSampledBvRel}{Status}{Correct}{True}{Count}{10}%
\StoreBenchExecResult{Abc}{ScPdrReachSafetyFloatsSampledBvRel}{Status}{Correct}{False}{Count}{0}%
\StoreBenchExecResult{Abc}{ScPdrReachSafetyFloatsSampledBvRel}{Status}{Wrong}{}{Count}{0}%
\StoreBenchExecResult{Abc}{ScPdrReachSafetyFloatsSampledBvRel}{Status}{Wrong}{True}{Count}{0}%
\StoreBenchExecResult{Abc}{ScPdrReachSafetyFloatsSampledBvRel}{Status}{Wrong}{False}{Count}{0}%
\providecommand\StoreBenchExecResult[7]{\expandafter\newcommand\csname#1#2#3#4#5#6\endcsname{#7}}%
\StoreBenchExecResult{Abc}{ScPdrReachSafetyHardnessSampledBvRel}{Status}{All}{}{Score}{0}%
\StoreBenchExecResult{Abc}{ScPdrReachSafetyHardnessSampledBvRel}{Status}{All}{}{Count}{132}%
\StoreBenchExecResult{Abc}{ScPdrReachSafetyHardnessSampledBvRel}{Status}{Correct}{}{Count}{132}%
\StoreBenchExecResult{Abc}{ScPdrReachSafetyHardnessSampledBvRel}{Status}{Correct}{True}{Count}{132}%
\StoreBenchExecResult{Abc}{ScPdrReachSafetyHardnessSampledBvRel}{Status}{Correct}{False}{Count}{0}%
\StoreBenchExecResult{Abc}{ScPdrReachSafetyHardnessSampledBvRel}{Status}{Wrong}{}{Count}{0}%
\StoreBenchExecResult{Abc}{ScPdrReachSafetyHardnessSampledBvRel}{Status}{Wrong}{True}{Count}{0}%
\StoreBenchExecResult{Abc}{ScPdrReachSafetyHardnessSampledBvRel}{Status}{Wrong}{False}{Count}{0}%
\providecommand\StoreBenchExecResult[7]{\expandafter\newcommand\csname#1#2#3#4#5#6\endcsname{#7}}%
\StoreBenchExecResult{Abc}{ScPdrReachSafetyHardwareSampledBvRel}{Status}{All}{}{Score}{0}%
\StoreBenchExecResult{Abc}{ScPdrReachSafetyHardwareSampledBvRel}{Status}{All}{}{Count}{148}%
\StoreBenchExecResult{Abc}{ScPdrReachSafetyHardwareSampledBvRel}{Status}{Correct}{}{Count}{30}%
\StoreBenchExecResult{Abc}{ScPdrReachSafetyHardwareSampledBvRel}{Status}{Correct}{True}{Count}{17}%
\StoreBenchExecResult{Abc}{ScPdrReachSafetyHardwareSampledBvRel}{Status}{Correct}{False}{Count}{13}%
\StoreBenchExecResult{Abc}{ScPdrReachSafetyHardwareSampledBvRel}{Status}{Wrong}{}{Count}{0}%
\StoreBenchExecResult{Abc}{ScPdrReachSafetyHardwareSampledBvRel}{Status}{Wrong}{True}{Count}{0}%
\StoreBenchExecResult{Abc}{ScPdrReachSafetyHardwareSampledBvRel}{Status}{Wrong}{False}{Count}{0}%
\providecommand\StoreBenchExecResult[7]{\expandafter\newcommand\csname#1#2#3#4#5#6\endcsname{#7}}%
\StoreBenchExecResult{Abc}{ScPdrReachSafetyHeapSampledBvRel}{Status}{All}{}{Score}{0}%
\StoreBenchExecResult{Abc}{ScPdrReachSafetyHeapSampledBvRel}{Status}{All}{}{Count}{6}%
\StoreBenchExecResult{Abc}{ScPdrReachSafetyHeapSampledBvRel}{Status}{Correct}{}{Count}{6}%
\StoreBenchExecResult{Abc}{ScPdrReachSafetyHeapSampledBvRel}{Status}{Correct}{True}{Count}{4}%
\StoreBenchExecResult{Abc}{ScPdrReachSafetyHeapSampledBvRel}{Status}{Correct}{False}{Count}{2}%
\StoreBenchExecResult{Abc}{ScPdrReachSafetyHeapSampledBvRel}{Status}{Wrong}{}{Count}{0}%
\StoreBenchExecResult{Abc}{ScPdrReachSafetyHeapSampledBvRel}{Status}{Wrong}{True}{Count}{0}%
\StoreBenchExecResult{Abc}{ScPdrReachSafetyHeapSampledBvRel}{Status}{Wrong}{False}{Count}{0}%
\providecommand\StoreBenchExecResult[7]{\expandafter\newcommand\csname#1#2#3#4#5#6\endcsname{#7}}%
\StoreBenchExecResult{Abc}{ScPdrReachSafetyLoopsSampledBvRel}{Status}{All}{}{Score}{0}%
\StoreBenchExecResult{Abc}{ScPdrReachSafetyLoopsSampledBvRel}{Status}{All}{}{Count}{137}%
\StoreBenchExecResult{Abc}{ScPdrReachSafetyLoopsSampledBvRel}{Status}{Correct}{}{Count}{57}%
\StoreBenchExecResult{Abc}{ScPdrReachSafetyLoopsSampledBvRel}{Status}{Correct}{True}{Count}{27}%
\StoreBenchExecResult{Abc}{ScPdrReachSafetyLoopsSampledBvRel}{Status}{Correct}{False}{Count}{30}%
\StoreBenchExecResult{Abc}{ScPdrReachSafetyLoopsSampledBvRel}{Status}{Wrong}{}{Count}{0}%
\StoreBenchExecResult{Abc}{ScPdrReachSafetyLoopsSampledBvRel}{Status}{Wrong}{True}{Count}{0}%
\StoreBenchExecResult{Abc}{ScPdrReachSafetyLoopsSampledBvRel}{Status}{Wrong}{False}{Count}{0}%
\providecommand\StoreBenchExecResult[7]{\expandafter\newcommand\csname#1#2#3#4#5#6\endcsname{#7}}%
\StoreBenchExecResult{Abc}{ScPdrReachSafetyProductLinesSampledBvRel}{Status}{All}{}{Score}{0}%
\StoreBenchExecResult{Abc}{ScPdrReachSafetyProductLinesSampledBvRel}{Status}{All}{}{Count}{150}%
\StoreBenchExecResult{Abc}{ScPdrReachSafetyProductLinesSampledBvRel}{Status}{Correct}{}{Count}{134}%
\StoreBenchExecResult{Abc}{ScPdrReachSafetyProductLinesSampledBvRel}{Status}{Correct}{True}{Count}{68}%
\StoreBenchExecResult{Abc}{ScPdrReachSafetyProductLinesSampledBvRel}{Status}{Correct}{False}{Count}{66}%
\StoreBenchExecResult{Abc}{ScPdrReachSafetyProductLinesSampledBvRel}{Status}{Wrong}{}{Count}{0}%
\StoreBenchExecResult{Abc}{ScPdrReachSafetyProductLinesSampledBvRel}{Status}{Wrong}{True}{Count}{0}%
\StoreBenchExecResult{Abc}{ScPdrReachSafetyProductLinesSampledBvRel}{Status}{Wrong}{False}{Count}{0}%
\providecommand\StoreBenchExecResult[7]{\expandafter\newcommand\csname#1#2#3#4#5#6\endcsname{#7}}%
\StoreBenchExecResult{Abc}{ScPdrReachSafetySequentializedSampledBvRel}{Status}{All}{}{Score}{0}%
\StoreBenchExecResult{Abc}{ScPdrReachSafetySequentializedSampledBvRel}{Status}{All}{}{Count}{150}%
\StoreBenchExecResult{Abc}{ScPdrReachSafetySequentializedSampledBvRel}{Status}{Correct}{}{Count}{56}%
\StoreBenchExecResult{Abc}{ScPdrReachSafetySequentializedSampledBvRel}{Status}{Correct}{True}{Count}{7}%
\StoreBenchExecResult{Abc}{ScPdrReachSafetySequentializedSampledBvRel}{Status}{Correct}{False}{Count}{49}%
\StoreBenchExecResult{Abc}{ScPdrReachSafetySequentializedSampledBvRel}{Status}{Wrong}{}{Count}{0}%
\StoreBenchExecResult{Abc}{ScPdrReachSafetySequentializedSampledBvRel}{Status}{Wrong}{True}{Count}{0}%
\StoreBenchExecResult{Abc}{ScPdrReachSafetySequentializedSampledBvRel}{Status}{Wrong}{False}{Count}{0}%
\providecommand\StoreBenchExecResult[7]{\expandafter\newcommand\csname#1#2#3#4#5#6\endcsname{#7}}%
\StoreBenchExecResult{Abc}{ScPdrReachSafetyXCSPSampledBvRel}{Status}{All}{}{Score}{0}%
\StoreBenchExecResult{Abc}{ScPdrReachSafetyXCSPSampledBvRel}{Status}{All}{}{Count}{98}%
\StoreBenchExecResult{Abc}{ScPdrReachSafetyXCSPSampledBvRel}{Status}{Correct}{}{Count}{89}%
\StoreBenchExecResult{Abc}{ScPdrReachSafetyXCSPSampledBvRel}{Status}{Correct}{True}{Count}{47}%
\StoreBenchExecResult{Abc}{ScPdrReachSafetyXCSPSampledBvRel}{Status}{Correct}{False}{Count}{42}%
\StoreBenchExecResult{Abc}{ScPdrReachSafetyXCSPSampledBvRel}{Status}{Wrong}{}{Count}{0}%
\StoreBenchExecResult{Abc}{ScPdrReachSafetyXCSPSampledBvRel}{Status}{Wrong}{True}{Count}{0}%
\StoreBenchExecResult{Abc}{ScPdrReachSafetyXCSPSampledBvRel}{Status}{Wrong}{False}{Count}{0}%
\providecommand\StoreBenchExecResult[7]{\expandafter\newcommand\csname#1#2#3#4#5#6\endcsname{#7}}%
\StoreBenchExecResult{Abc}{ScPdrTerminationBitVectorsSampledBvRel}{Status}{All}{}{Score}{0}%
\StoreBenchExecResult{Abc}{ScPdrTerminationBitVectorsSampledBvRel}{Status}{All}{}{Count}{32}%
\StoreBenchExecResult{Abc}{ScPdrTerminationBitVectorsSampledBvRel}{Status}{Correct}{}{Count}{22}%
\StoreBenchExecResult{Abc}{ScPdrTerminationBitVectorsSampledBvRel}{Status}{Correct}{True}{Count}{11}%
\StoreBenchExecResult{Abc}{ScPdrTerminationBitVectorsSampledBvRel}{Status}{Correct}{False}{Count}{11}%
\StoreBenchExecResult{Abc}{ScPdrTerminationBitVectorsSampledBvRel}{Status}{Wrong}{}{Count}{0}%
\StoreBenchExecResult{Abc}{ScPdrTerminationBitVectorsSampledBvRel}{Status}{Wrong}{True}{Count}{0}%
\StoreBenchExecResult{Abc}{ScPdrTerminationBitVectorsSampledBvRel}{Status}{Wrong}{False}{Count}{0}%
\providecommand\StoreBenchExecResult[7]{\expandafter\newcommand\csname#1#2#3#4#5#6\endcsname{#7}}%
\StoreBenchExecResult{Abc}{ScPdrTerminationMainControlFlowSampledBvRel}{Status}{All}{}{Score}{0}%
\StoreBenchExecResult{Abc}{ScPdrTerminationMainControlFlowSampledBvRel}{Status}{All}{}{Count}{236}%
\StoreBenchExecResult{Abc}{ScPdrTerminationMainControlFlowSampledBvRel}{Status}{Correct}{}{Count}{51}%
\StoreBenchExecResult{Abc}{ScPdrTerminationMainControlFlowSampledBvRel}{Status}{Correct}{True}{Count}{6}%
\StoreBenchExecResult{Abc}{ScPdrTerminationMainControlFlowSampledBvRel}{Status}{Correct}{False}{Count}{45}%
\StoreBenchExecResult{Abc}{ScPdrTerminationMainControlFlowSampledBvRel}{Status}{Wrong}{}{Count}{0}%
\StoreBenchExecResult{Abc}{ScPdrTerminationMainControlFlowSampledBvRel}{Status}{Wrong}{True}{Count}{0}%
\StoreBenchExecResult{Abc}{ScPdrTerminationMainControlFlowSampledBvRel}{Status}{Wrong}{False}{Count}{0}%
\providecommand\StoreBenchExecResult[7]{\expandafter\newcommand\csname#1#2#3#4#5#6\endcsname{#7}}%
\StoreBenchExecResult{Abc}{ScPdrTerminationOtherSampledBvRel}{Status}{All}{}{Score}{0}%
\StoreBenchExecResult{Abc}{ScPdrTerminationOtherSampledBvRel}{Status}{All}{}{Count}{973}%
\StoreBenchExecResult{Abc}{ScPdrTerminationOtherSampledBvRel}{Status}{Correct}{}{Count}{559}%
\StoreBenchExecResult{Abc}{ScPdrTerminationOtherSampledBvRel}{Status}{Correct}{True}{Count}{68}%
\StoreBenchExecResult{Abc}{ScPdrTerminationOtherSampledBvRel}{Status}{Correct}{False}{Count}{491}%
\StoreBenchExecResult{Abc}{ScPdrTerminationOtherSampledBvRel}{Status}{Wrong}{}{Count}{0}%
\StoreBenchExecResult{Abc}{ScPdrTerminationOtherSampledBvRel}{Status}{Wrong}{True}{Count}{0}%
\StoreBenchExecResult{Abc}{ScPdrTerminationOtherSampledBvRel}{Status}{Wrong}{False}{Count}{0}%
\edef\AbcScPdrReachSafetySampledBvRelStatusAllCount{\the\numexpr\AbcScPdrReachSafetyBitVectorsSampledBvRelStatusAllCount+\AbcScPdrReachSafetyCombinationsSampledBvRelStatusAllCount+\AbcScPdrReachSafetyControlFlowSampledBvRelStatusAllCount+\AbcScPdrReachSafetyECASampledBvRelStatusAllCount+\AbcScPdrReachSafetyFloatsSampledBvRelStatusAllCount+\AbcScPdrReachSafetyHardnessSampledBvRelStatusAllCount+\AbcScPdrReachSafetyHardwareSampledBvRelStatusAllCount+\AbcScPdrReachSafetyHeapSampledBvRelStatusAllCount+\AbcScPdrReachSafetyLoopsSampledBvRelStatusAllCount+\AbcScPdrReachSafetyProductLinesSampledBvRelStatusAllCount+\AbcScPdrReachSafetySequentializedSampledBvRelStatusAllCount+\AbcScPdrReachSafetyXCSPSampledBvRelStatusAllCount}
\edef\AbcScPdrReachSafetySampledBvRelStatusCorrectCount{\the\numexpr\AbcScPdrReachSafetyBitVectorsSampledBvRelStatusCorrectCount+\AbcScPdrReachSafetyCombinationsSampledBvRelStatusCorrectCount+\AbcScPdrReachSafetyControlFlowSampledBvRelStatusCorrectCount+\AbcScPdrReachSafetyECASampledBvRelStatusCorrectCount+\AbcScPdrReachSafetyFloatsSampledBvRelStatusCorrectCount+\AbcScPdrReachSafetyHardnessSampledBvRelStatusCorrectCount+\AbcScPdrReachSafetyHardwareSampledBvRelStatusCorrectCount+\AbcScPdrReachSafetyHeapSampledBvRelStatusCorrectCount+\AbcScPdrReachSafetyLoopsSampledBvRelStatusCorrectCount+\AbcScPdrReachSafetyProductLinesSampledBvRelStatusCorrectCount+\AbcScPdrReachSafetySequentializedSampledBvRelStatusCorrectCount+\AbcScPdrReachSafetyXCSPSampledBvRelStatusCorrectCount}
\edef\AbcScPdrReachSafetySampledBvRelStatusCorrectTrueCount{\the\numexpr\AbcScPdrReachSafetyBitVectorsSampledBvRelStatusCorrectTrueCount+\AbcScPdrReachSafetyCombinationsSampledBvRelStatusCorrectTrueCount+\AbcScPdrReachSafetyControlFlowSampledBvRelStatusCorrectTrueCount+\AbcScPdrReachSafetyECASampledBvRelStatusCorrectTrueCount+\AbcScPdrReachSafetyFloatsSampledBvRelStatusCorrectTrueCount+\AbcScPdrReachSafetyHardnessSampledBvRelStatusCorrectTrueCount+\AbcScPdrReachSafetyHardwareSampledBvRelStatusCorrectTrueCount+\AbcScPdrReachSafetyHeapSampledBvRelStatusCorrectTrueCount+\AbcScPdrReachSafetyLoopsSampledBvRelStatusCorrectTrueCount+\AbcScPdrReachSafetyProductLinesSampledBvRelStatusCorrectTrueCount+\AbcScPdrReachSafetySequentializedSampledBvRelStatusCorrectTrueCount+\AbcScPdrReachSafetyXCSPSampledBvRelStatusCorrectTrueCount}
\edef\AbcScPdrReachSafetySampledBvRelStatusCorrectFalseCount{\the\numexpr\AbcScPdrReachSafetyBitVectorsSampledBvRelStatusCorrectFalseCount+\AbcScPdrReachSafetyCombinationsSampledBvRelStatusCorrectFalseCount+\AbcScPdrReachSafetyControlFlowSampledBvRelStatusCorrectFalseCount+\AbcScPdrReachSafetyECASampledBvRelStatusCorrectFalseCount+\AbcScPdrReachSafetyFloatsSampledBvRelStatusCorrectFalseCount+\AbcScPdrReachSafetyHardnessSampledBvRelStatusCorrectFalseCount+\AbcScPdrReachSafetyHardwareSampledBvRelStatusCorrectFalseCount+\AbcScPdrReachSafetyHeapSampledBvRelStatusCorrectFalseCount+\AbcScPdrReachSafetyLoopsSampledBvRelStatusCorrectFalseCount+\AbcScPdrReachSafetyProductLinesSampledBvRelStatusCorrectFalseCount+\AbcScPdrReachSafetySequentializedSampledBvRelStatusCorrectFalseCount+\AbcScPdrReachSafetyXCSPSampledBvRelStatusCorrectFalseCount}
\edef\AbcScPdrReachSafetySampledBvRelStatusWrongCount{\the\numexpr\AbcScPdrReachSafetyBitVectorsSampledBvRelStatusWrongCount+\AbcScPdrReachSafetyCombinationsSampledBvRelStatusWrongCount+\AbcScPdrReachSafetyControlFlowSampledBvRelStatusWrongCount+\AbcScPdrReachSafetyECASampledBvRelStatusWrongCount+\AbcScPdrReachSafetyFloatsSampledBvRelStatusWrongCount+\AbcScPdrReachSafetyHardnessSampledBvRelStatusWrongCount+\AbcScPdrReachSafetyHardwareSampledBvRelStatusWrongCount+\AbcScPdrReachSafetyHeapSampledBvRelStatusWrongCount+\AbcScPdrReachSafetyLoopsSampledBvRelStatusWrongCount+\AbcScPdrReachSafetyProductLinesSampledBvRelStatusWrongCount+\AbcScPdrReachSafetySequentializedSampledBvRelStatusWrongCount+\AbcScPdrReachSafetyXCSPSampledBvRelStatusWrongCount}
\edef\AbcScPdrTerminationSampledBvRelStatusAllCount{\the\numexpr\AbcScPdrTerminationBitVectorsSampledBvRelStatusAllCount+\AbcScPdrTerminationMainControlFlowSampledBvRelStatusAllCount+\AbcScPdrTerminationOtherSampledBvRelStatusAllCount}
\edef\AbcScPdrTerminationSampledBvRelStatusCorrectCount{\the\numexpr\AbcScPdrTerminationBitVectorsSampledBvRelStatusCorrectCount+\AbcScPdrTerminationMainControlFlowSampledBvRelStatusCorrectCount+\AbcScPdrTerminationOtherSampledBvRelStatusCorrectCount}
\edef\AbcScPdrTerminationSampledBvRelStatusCorrectTrueCount{\the\numexpr\AbcScPdrTerminationBitVectorsSampledBvRelStatusCorrectTrueCount+\AbcScPdrTerminationMainControlFlowSampledBvRelStatusCorrectTrueCount+\AbcScPdrTerminationOtherSampledBvRelStatusCorrectTrueCount}
\edef\AbcScPdrTerminationSampledBvRelStatusCorrectFalseCount{\the\numexpr\AbcScPdrTerminationBitVectorsSampledBvRelStatusCorrectFalseCount+\AbcScPdrTerminationMainControlFlowSampledBvRelStatusCorrectFalseCount+\AbcScPdrTerminationOtherSampledBvRelStatusCorrectFalseCount}
\edef\AbcScPdrTerminationSampledBvRelStatusWrongCount{\the\numexpr\AbcScPdrTerminationBitVectorsSampledBvRelStatusWrongCount+\AbcScPdrTerminationMainControlFlowSampledBvRelStatusWrongCount+\AbcScPdrTerminationOtherSampledBvRelStatusWrongCount}
\providecommand\StoreBenchExecResult[7]{\expandafter\newcommand\csname#1#2#3#4#5#6\endcsname{#7}}%
\StoreBenchExecResult{Abc}{PdrScPdrReachSafetyFuncCommon}{Status}{All}{}{Score}{0}%
\StoreBenchExecResult{Abc}{PdrScPdrReachSafetyFuncCommon}{Status}{All}{}{Count}{647}%
\StoreBenchExecResult{Abc}{PdrScPdrReachSafetyFuncCommon}{Status}{Correct}{}{Count}{647}%
\StoreBenchExecResult{Abc}{PdrScPdrReachSafetyFuncCommon}{Status}{Correct}{True}{Count}{380}%
\StoreBenchExecResult{Abc}{PdrScPdrReachSafetyFuncCommon}{Status}{Correct}{False}{Count}{267}%
\StoreBenchExecResult{Abc}{PdrScPdrReachSafetyFuncCommon}{Status}{Wrong}{}{Count}{0}%
\StoreBenchExecResult{Abc}{PdrScPdrReachSafetyFuncCommon}{Status}{Wrong}{True}{Count}{0}%
\StoreBenchExecResult{Abc}{PdrScPdrReachSafetyFuncCommon}{Status}{Wrong}{False}{Count}{0}%
\StoreBenchExecResult{Abc}{PdrScPdrReachSafetyFuncCommon}{Cputime}{All}{}{Sum}{8024.722471}%
\StoreBenchExecResult{Abc}{PdrScPdrReachSafetyFuncCommon}{Cputime}{All}{}{Min}{0.046402}%
\StoreBenchExecResult{Abc}{PdrScPdrReachSafetyFuncCommon}{Cputime}{All}{}{Max}{836.431817}%
\StoreBenchExecResult{Abc}{PdrScPdrReachSafetyFuncCommon}{Cputime}{All}{}{Avg}{12.40297136166924265842349304}%
\StoreBenchExecResult{Abc}{PdrScPdrReachSafetyFuncCommon}{Cputime}{All}{}{Median}{0.212988}%
\StoreBenchExecResult{Abc}{PdrScPdrReachSafetyFuncCommon}{Cputime}{All}{}{Stdev}{59.78637218452247679428575330}%
\StoreBenchExecResult{Abc}{PdrScPdrReachSafetyFuncCommon}{Cputime}{All}{}{Unit}{s}%
\StoreBenchExecResult{Abc}{PdrScPdrReachSafetyFuncCommon}{Cputime}{Correct}{}{Sum}{8024.722471}%
\StoreBenchExecResult{Abc}{PdrScPdrReachSafetyFuncCommon}{Cputime}{Correct}{}{Min}{0.046402}%
\StoreBenchExecResult{Abc}{PdrScPdrReachSafetyFuncCommon}{Cputime}{Correct}{}{Max}{836.431817}%
\StoreBenchExecResult{Abc}{PdrScPdrReachSafetyFuncCommon}{Cputime}{Correct}{}{Avg}{12.40297136166924265842349304}%
\StoreBenchExecResult{Abc}{PdrScPdrReachSafetyFuncCommon}{Cputime}{Correct}{}{Median}{0.212988}%
\StoreBenchExecResult{Abc}{PdrScPdrReachSafetyFuncCommon}{Cputime}{Correct}{}{Stdev}{59.78637218452247679428575330}%
\StoreBenchExecResult{Abc}{PdrScPdrReachSafetyFuncCommon}{Cputime}{Correct}{}{Unit}{s}%
\StoreBenchExecResult{Abc}{PdrScPdrReachSafetyFuncCommon}{Cputime}{Correct}{True}{Sum}{2543.226838}%
\StoreBenchExecResult{Abc}{PdrScPdrReachSafetyFuncCommon}{Cputime}{Correct}{True}{Min}{0.048955}%
\StoreBenchExecResult{Abc}{PdrScPdrReachSafetyFuncCommon}{Cputime}{Correct}{True}{Max}{549.633321}%
\StoreBenchExecResult{Abc}{PdrScPdrReachSafetyFuncCommon}{Cputime}{Correct}{True}{Avg}{6.692702205263157894736842105}%
\StoreBenchExecResult{Abc}{PdrScPdrReachSafetyFuncCommon}{Cputime}{Correct}{True}{Median}{0.1441405}%
\StoreBenchExecResult{Abc}{PdrScPdrReachSafetyFuncCommon}{Cputime}{Correct}{True}{Stdev}{39.50140426639080464286956873}%
\StoreBenchExecResult{Abc}{PdrScPdrReachSafetyFuncCommon}{Cputime}{Correct}{True}{Unit}{s}%
\StoreBenchExecResult{Abc}{PdrScPdrReachSafetyFuncCommon}{Cputime}{Correct}{False}{Sum}{5481.495633}%
\StoreBenchExecResult{Abc}{PdrScPdrReachSafetyFuncCommon}{Cputime}{Correct}{False}{Min}{0.046402}%
\StoreBenchExecResult{Abc}{PdrScPdrReachSafetyFuncCommon}{Cputime}{Correct}{False}{Max}{836.431817}%
\StoreBenchExecResult{Abc}{PdrScPdrReachSafetyFuncCommon}{Cputime}{Correct}{False}{Avg}{20.52994619101123595505617978}%
\StoreBenchExecResult{Abc}{PdrScPdrReachSafetyFuncCommon}{Cputime}{Correct}{False}{Median}{0.53767}%
\StoreBenchExecResult{Abc}{PdrScPdrReachSafetyFuncCommon}{Cputime}{Correct}{False}{Stdev}{79.55119519371678871895394459}%
\StoreBenchExecResult{Abc}{PdrScPdrReachSafetyFuncCommon}{Cputime}{Correct}{False}{Unit}{s}%
\StoreBenchExecResult{Abc}{PdrScPdrReachSafetyFuncCommon}{Walltime}{All}{}{Sum}{8027.484721661545261067}%
\StoreBenchExecResult{Abc}{PdrScPdrReachSafetyFuncCommon}{Walltime}{All}{}{Min}{0.04655615147203207}%
\StoreBenchExecResult{Abc}{PdrScPdrReachSafetyFuncCommon}{Walltime}{All}{}{Max}{836.78276585415}%
\StoreBenchExecResult{Abc}{PdrScPdrReachSafetyFuncCommon}{Walltime}{All}{}{Avg}{12.40724068262989993982534776}%
\StoreBenchExecResult{Abc}{PdrScPdrReachSafetyFuncCommon}{Walltime}{All}{}{Median}{0.2131260670721531}%
\StoreBenchExecResult{Abc}{PdrScPdrReachSafetyFuncCommon}{Walltime}{All}{}{Stdev}{59.80738969848060476762251281}%
\StoreBenchExecResult{Abc}{PdrScPdrReachSafetyFuncCommon}{Walltime}{All}{}{Unit}{s}%
\StoreBenchExecResult{Abc}{PdrScPdrReachSafetyFuncCommon}{Walltime}{Correct}{}{Sum}{8027.484721661545261067}%
\StoreBenchExecResult{Abc}{PdrScPdrReachSafetyFuncCommon}{Walltime}{Correct}{}{Min}{0.04655615147203207}%
\StoreBenchExecResult{Abc}{PdrScPdrReachSafetyFuncCommon}{Walltime}{Correct}{}{Max}{836.78276585415}%
\StoreBenchExecResult{Abc}{PdrScPdrReachSafetyFuncCommon}{Walltime}{Correct}{}{Avg}{12.40724068262989993982534776}%
\StoreBenchExecResult{Abc}{PdrScPdrReachSafetyFuncCommon}{Walltime}{Correct}{}{Median}{0.2131260670721531}%
\StoreBenchExecResult{Abc}{PdrScPdrReachSafetyFuncCommon}{Walltime}{Correct}{}{Stdev}{59.80738969848060476762251281}%
\StoreBenchExecResult{Abc}{PdrScPdrReachSafetyFuncCommon}{Walltime}{Correct}{}{Unit}{s}%
\StoreBenchExecResult{Abc}{PdrScPdrReachSafetyFuncCommon}{Walltime}{Correct}{True}{Sum}{2544.042003061622363994}%
\StoreBenchExecResult{Abc}{PdrScPdrReachSafetyFuncCommon}{Walltime}{Correct}{True}{Min}{0.04914864897727966}%
\StoreBenchExecResult{Abc}{PdrScPdrReachSafetyFuncCommon}{Walltime}{Correct}{True}{Max}{549.7329334067181}%
\StoreBenchExecResult{Abc}{PdrScPdrReachSafetyFuncCommon}{Walltime}{Correct}{True}{Avg}{6.694847376477953589457894737}%
\StoreBenchExecResult{Abc}{PdrScPdrReachSafetyFuncCommon}{Walltime}{Correct}{True}{Median}{0.144380588550120585}%
\StoreBenchExecResult{Abc}{PdrScPdrReachSafetyFuncCommon}{Walltime}{Correct}{True}{Stdev}{39.51133913063002163151937389}%
\StoreBenchExecResult{Abc}{PdrScPdrReachSafetyFuncCommon}{Walltime}{Correct}{True}{Unit}{s}%
\StoreBenchExecResult{Abc}{PdrScPdrReachSafetyFuncCommon}{Walltime}{Correct}{False}{Sum}{5483.442718599922897073}%
\StoreBenchExecResult{Abc}{PdrScPdrReachSafetyFuncCommon}{Walltime}{Correct}{False}{Min}{0.04655615147203207}%
\StoreBenchExecResult{Abc}{PdrScPdrReachSafetyFuncCommon}{Walltime}{Correct}{False}{Max}{836.78276585415}%
\StoreBenchExecResult{Abc}{PdrScPdrReachSafetyFuncCommon}{Walltime}{Correct}{False}{Avg}{20.53723864644165879053558052}%
\StoreBenchExecResult{Abc}{PdrScPdrReachSafetyFuncCommon}{Walltime}{Correct}{False}{Median}{0.537820640951395}%
\StoreBenchExecResult{Abc}{PdrScPdrReachSafetyFuncCommon}{Walltime}{Correct}{False}{Stdev}{79.58192443121352883979678786}%
\StoreBenchExecResult{Abc}{PdrScPdrReachSafetyFuncCommon}{Walltime}{Correct}{False}{Unit}{s}%
\StoreBenchExecResult{Abc}{PdrScPdrReachSafetyFuncCommon}{Memory}{All}{}{Sum}{26346.520576}%
\StoreBenchExecResult{Abc}{PdrScPdrReachSafetyFuncCommon}{Memory}{All}{}{Min}{7.086080}%
\StoreBenchExecResult{Abc}{PdrScPdrReachSafetyFuncCommon}{Memory}{All}{}{Max}{564.817920}%
\StoreBenchExecResult{Abc}{PdrScPdrReachSafetyFuncCommon}{Memory}{All}{}{Avg}{40.72105189489953632148377125}%
\StoreBenchExecResult{Abc}{PdrScPdrReachSafetyFuncCommon}{Memory}{All}{}{Median}{19.607552}%
\StoreBenchExecResult{Abc}{PdrScPdrReachSafetyFuncCommon}{Memory}{All}{}{Stdev}{58.39203058346753083596849452}%
\StoreBenchExecResult{Abc}{PdrScPdrReachSafetyFuncCommon}{Memory}{All}{}{Unit}{MB}%
\StoreBenchExecResult{Abc}{PdrScPdrReachSafetyFuncCommon}{Memory}{Correct}{}{Sum}{26346.520576}%
\StoreBenchExecResult{Abc}{PdrScPdrReachSafetyFuncCommon}{Memory}{Correct}{}{Min}{7.086080}%
\StoreBenchExecResult{Abc}{PdrScPdrReachSafetyFuncCommon}{Memory}{Correct}{}{Max}{564.817920}%
\StoreBenchExecResult{Abc}{PdrScPdrReachSafetyFuncCommon}{Memory}{Correct}{}{Avg}{40.72105189489953632148377125}%
\StoreBenchExecResult{Abc}{PdrScPdrReachSafetyFuncCommon}{Memory}{Correct}{}{Median}{19.607552}%
\StoreBenchExecResult{Abc}{PdrScPdrReachSafetyFuncCommon}{Memory}{Correct}{}{Stdev}{58.39203058346753083596849452}%
\StoreBenchExecResult{Abc}{PdrScPdrReachSafetyFuncCommon}{Memory}{Correct}{}{Unit}{MB}%
\StoreBenchExecResult{Abc}{PdrScPdrReachSafetyFuncCommon}{Memory}{Correct}{True}{Sum}{12808.675328}%
\StoreBenchExecResult{Abc}{PdrScPdrReachSafetyFuncCommon}{Memory}{Correct}{True}{Min}{10.776576}%
\StoreBenchExecResult{Abc}{PdrScPdrReachSafetyFuncCommon}{Memory}{Correct}{True}{Max}{510.070784}%
\StoreBenchExecResult{Abc}{PdrScPdrReachSafetyFuncCommon}{Memory}{Correct}{True}{Avg}{33.70704033684210526315789474}%
\StoreBenchExecResult{Abc}{PdrScPdrReachSafetyFuncCommon}{Memory}{Correct}{True}{Median}{18.690048}%
\StoreBenchExecResult{Abc}{PdrScPdrReachSafetyFuncCommon}{Memory}{Correct}{True}{Stdev}{50.01029605694186662693227633}%
\StoreBenchExecResult{Abc}{PdrScPdrReachSafetyFuncCommon}{Memory}{Correct}{True}{Unit}{MB}%
\StoreBenchExecResult{Abc}{PdrScPdrReachSafetyFuncCommon}{Memory}{Correct}{False}{Sum}{13537.845248}%
\StoreBenchExecResult{Abc}{PdrScPdrReachSafetyFuncCommon}{Memory}{Correct}{False}{Min}{7.086080}%
\StoreBenchExecResult{Abc}{PdrScPdrReachSafetyFuncCommon}{Memory}{Correct}{False}{Max}{564.817920}%
\StoreBenchExecResult{Abc}{PdrScPdrReachSafetyFuncCommon}{Memory}{Correct}{False}{Avg}{50.70354025468164794007490637}%
\StoreBenchExecResult{Abc}{PdrScPdrReachSafetyFuncCommon}{Memory}{Correct}{False}{Median}{20.611072}%
\StoreBenchExecResult{Abc}{PdrScPdrReachSafetyFuncCommon}{Memory}{Correct}{False}{Stdev}{67.32829845124025302570856497}%
\StoreBenchExecResult{Abc}{PdrScPdrReachSafetyFuncCommon}{Memory}{Correct}{False}{Unit}{MB}%
\StoreBenchExecResult{Abc}{PdrScPdrReachSafetyFuncCommon}{Gate}{All}{}{Sum}{14126879}%
\StoreBenchExecResult{Abc}{PdrScPdrReachSafetyFuncCommon}{Gate}{All}{}{Min}{5}%
\StoreBenchExecResult{Abc}{PdrScPdrReachSafetyFuncCommon}{Gate}{All}{}{Max}{249491}%
\StoreBenchExecResult{Abc}{PdrScPdrReachSafetyFuncCommon}{Gate}{All}{}{Avg}{21834.43431221020092735703246}%
\StoreBenchExecResult{Abc}{PdrScPdrReachSafetyFuncCommon}{Gate}{All}{}{Median}{6784}%
\StoreBenchExecResult{Abc}{PdrScPdrReachSafetyFuncCommon}{Gate}{All}{}{Stdev}{37968.95863873109635420422428}%
\StoreBenchExecResult{Abc}{PdrScPdrReachSafetyFuncCommon}{Gate}{Correct}{}{Sum}{14126879}%
\StoreBenchExecResult{Abc}{PdrScPdrReachSafetyFuncCommon}{Gate}{Correct}{}{Min}{5}%
\StoreBenchExecResult{Abc}{PdrScPdrReachSafetyFuncCommon}{Gate}{Correct}{}{Max}{249491}%
\StoreBenchExecResult{Abc}{PdrScPdrReachSafetyFuncCommon}{Gate}{Correct}{}{Avg}{21834.43431221020092735703246}%
\StoreBenchExecResult{Abc}{PdrScPdrReachSafetyFuncCommon}{Gate}{Correct}{}{Median}{6784}%
\StoreBenchExecResult{Abc}{PdrScPdrReachSafetyFuncCommon}{Gate}{Correct}{}{Stdev}{37968.95863873109635420422428}%
\StoreBenchExecResult{Abc}{PdrScPdrReachSafetyFuncCommon}{Gate}{Correct}{True}{Sum}{4277161}%
\StoreBenchExecResult{Abc}{PdrScPdrReachSafetyFuncCommon}{Gate}{Correct}{True}{Min}{7}%
\StoreBenchExecResult{Abc}{PdrScPdrReachSafetyFuncCommon}{Gate}{Correct}{True}{Max}{163181}%
\StoreBenchExecResult{Abc}{PdrScPdrReachSafetyFuncCommon}{Gate}{Correct}{True}{Avg}{11255.68684210526315789473684}%
\StoreBenchExecResult{Abc}{PdrScPdrReachSafetyFuncCommon}{Gate}{Correct}{True}{Median}{5669.5}%
\StoreBenchExecResult{Abc}{PdrScPdrReachSafetyFuncCommon}{Gate}{Correct}{True}{Stdev}{21442.92588807333953973767185}%
\StoreBenchExecResult{Abc}{PdrScPdrReachSafetyFuncCommon}{Gate}{Correct}{False}{Sum}{9849718}%
\StoreBenchExecResult{Abc}{PdrScPdrReachSafetyFuncCommon}{Gate}{Correct}{False}{Min}{5}%
\StoreBenchExecResult{Abc}{PdrScPdrReachSafetyFuncCommon}{Gate}{Correct}{False}{Max}{249491}%
\StoreBenchExecResult{Abc}{PdrScPdrReachSafetyFuncCommon}{Gate}{Correct}{False}{Avg}{36890.32958801498127340823970}%
\StoreBenchExecResult{Abc}{PdrScPdrReachSafetyFuncCommon}{Gate}{Correct}{False}{Median}{8600}%
\StoreBenchExecResult{Abc}{PdrScPdrReachSafetyFuncCommon}{Gate}{Correct}{False}{Stdev}{49528.46169117680152970007347}%
\StoreBenchExecResult{Abc}{PdrScPdrReachSafetyFuncCommon}{VerifTime}{All}{}{Sum}{7880.51}%
\StoreBenchExecResult{Abc}{PdrScPdrReachSafetyFuncCommon}{VerifTime}{All}{}{Min}{0.01}%
\StoreBenchExecResult{Abc}{PdrScPdrReachSafetyFuncCommon}{VerifTime}{All}{}{Max}{836.49}%
\StoreBenchExecResult{Abc}{PdrScPdrReachSafetyFuncCommon}{VerifTime}{All}{}{Avg}{12.18007727975270479134466770}%
\StoreBenchExecResult{Abc}{PdrScPdrReachSafetyFuncCommon}{VerifTime}{All}{}{Median}{0.1}%
\StoreBenchExecResult{Abc}{PdrScPdrReachSafetyFuncCommon}{VerifTime}{All}{}{Stdev}{59.76643781308823704196597370}%
\StoreBenchExecResult{Abc}{PdrScPdrReachSafetyFuncCommon}{VerifTime}{Correct}{}{Sum}{7880.51}%
\StoreBenchExecResult{Abc}{PdrScPdrReachSafetyFuncCommon}{VerifTime}{Correct}{}{Min}{0.01}%
\StoreBenchExecResult{Abc}{PdrScPdrReachSafetyFuncCommon}{VerifTime}{Correct}{}{Max}{836.49}%
\StoreBenchExecResult{Abc}{PdrScPdrReachSafetyFuncCommon}{VerifTime}{Correct}{}{Avg}{12.18007727975270479134466770}%
\StoreBenchExecResult{Abc}{PdrScPdrReachSafetyFuncCommon}{VerifTime}{Correct}{}{Median}{0.1}%
\StoreBenchExecResult{Abc}{PdrScPdrReachSafetyFuncCommon}{VerifTime}{Correct}{}{Stdev}{59.76643781308823704196597370}%
\StoreBenchExecResult{Abc}{PdrScPdrReachSafetyFuncCommon}{VerifTime}{Correct}{True}{Sum}{2492.11}%
\StoreBenchExecResult{Abc}{PdrScPdrReachSafetyFuncCommon}{VerifTime}{Correct}{True}{Min}{0.01}%
\StoreBenchExecResult{Abc}{PdrScPdrReachSafetyFuncCommon}{VerifTime}{Correct}{True}{Max}{549.49}%
\StoreBenchExecResult{Abc}{PdrScPdrReachSafetyFuncCommon}{VerifTime}{Correct}{True}{Avg}{6.558184210526315789473684211}%
\StoreBenchExecResult{Abc}{PdrScPdrReachSafetyFuncCommon}{VerifTime}{Correct}{True}{Median}{0.05}%
\StoreBenchExecResult{Abc}{PdrScPdrReachSafetyFuncCommon}{VerifTime}{Correct}{True}{Stdev}{39.49919053957759215486054784}%
\StoreBenchExecResult{Abc}{PdrScPdrReachSafetyFuncCommon}{VerifTime}{Correct}{False}{Sum}{5388.40}%
\StoreBenchExecResult{Abc}{PdrScPdrReachSafetyFuncCommon}{VerifTime}{Correct}{False}{Min}{0.01}%
\StoreBenchExecResult{Abc}{PdrScPdrReachSafetyFuncCommon}{VerifTime}{Correct}{False}{Max}{836.49}%
\StoreBenchExecResult{Abc}{PdrScPdrReachSafetyFuncCommon}{VerifTime}{Correct}{False}{Avg}{20.18127340823970037453183521}%
\StoreBenchExecResult{Abc}{PdrScPdrReachSafetyFuncCommon}{VerifTime}{Correct}{False}{Median}{0.28}%
\StoreBenchExecResult{Abc}{PdrScPdrReachSafetyFuncCommon}{VerifTime}{Correct}{False}{Stdev}{79.53816969278841776545426691}%
\StoreBenchExecResult{Abc}{ScPdrScPdrReachSafetyFuncCommon}{Status}{All}{}{Score}{0}%
\StoreBenchExecResult{Abc}{ScPdrScPdrReachSafetyFuncCommon}{Status}{All}{}{Count}{647}%
\StoreBenchExecResult{Abc}{ScPdrScPdrReachSafetyFuncCommon}{Status}{Correct}{}{Count}{647}%
\StoreBenchExecResult{Abc}{ScPdrScPdrReachSafetyFuncCommon}{Status}{Correct}{True}{Count}{380}%
\StoreBenchExecResult{Abc}{ScPdrScPdrReachSafetyFuncCommon}{Status}{Correct}{False}{Count}{267}%
\StoreBenchExecResult{Abc}{ScPdrScPdrReachSafetyFuncCommon}{Status}{Wrong}{}{Count}{0}%
\StoreBenchExecResult{Abc}{ScPdrScPdrReachSafetyFuncCommon}{Status}{Wrong}{True}{Count}{0}%
\StoreBenchExecResult{Abc}{ScPdrScPdrReachSafetyFuncCommon}{Status}{Wrong}{False}{Count}{0}%
\StoreBenchExecResult{Abc}{ScPdrScPdrReachSafetyFuncCommon}{Cputime}{All}{}{Sum}{19140.009918}%
\StoreBenchExecResult{Abc}{ScPdrScPdrReachSafetyFuncCommon}{Cputime}{All}{}{Min}{0.048323}%
\StoreBenchExecResult{Abc}{ScPdrScPdrReachSafetyFuncCommon}{Cputime}{All}{}{Max}{698.512336}%
\StoreBenchExecResult{Abc}{ScPdrScPdrReachSafetyFuncCommon}{Cputime}{All}{}{Avg}{29.58270466460587326120556414}%
\StoreBenchExecResult{Abc}{ScPdrScPdrReachSafetyFuncCommon}{Cputime}{All}{}{Median}{0.384011}%
\StoreBenchExecResult{Abc}{ScPdrScPdrReachSafetyFuncCommon}{Cputime}{All}{}{Stdev}{87.36607720544816697572750199}%
\StoreBenchExecResult{Abc}{ScPdrScPdrReachSafetyFuncCommon}{Cputime}{All}{}{Unit}{s}%
\StoreBenchExecResult{Abc}{ScPdrScPdrReachSafetyFuncCommon}{Cputime}{Correct}{}{Sum}{19140.009918}%
\StoreBenchExecResult{Abc}{ScPdrScPdrReachSafetyFuncCommon}{Cputime}{Correct}{}{Min}{0.048323}%
\StoreBenchExecResult{Abc}{ScPdrScPdrReachSafetyFuncCommon}{Cputime}{Correct}{}{Max}{698.512336}%
\StoreBenchExecResult{Abc}{ScPdrScPdrReachSafetyFuncCommon}{Cputime}{Correct}{}{Avg}{29.58270466460587326120556414}%
\StoreBenchExecResult{Abc}{ScPdrScPdrReachSafetyFuncCommon}{Cputime}{Correct}{}{Median}{0.384011}%
\StoreBenchExecResult{Abc}{ScPdrScPdrReachSafetyFuncCommon}{Cputime}{Correct}{}{Stdev}{87.36607720544816697572750199}%
\StoreBenchExecResult{Abc}{ScPdrScPdrReachSafetyFuncCommon}{Cputime}{Correct}{}{Unit}{s}%
\StoreBenchExecResult{Abc}{ScPdrScPdrReachSafetyFuncCommon}{Cputime}{Correct}{True}{Sum}{5094.735284}%
\StoreBenchExecResult{Abc}{ScPdrScPdrReachSafetyFuncCommon}{Cputime}{Correct}{True}{Min}{0.051299}%
\StoreBenchExecResult{Abc}{ScPdrScPdrReachSafetyFuncCommon}{Cputime}{Correct}{True}{Max}{399.273418}%
\StoreBenchExecResult{Abc}{ScPdrScPdrReachSafetyFuncCommon}{Cputime}{Correct}{True}{Avg}{13.40719811578947368421052632}%
\StoreBenchExecResult{Abc}{ScPdrScPdrReachSafetyFuncCommon}{Cputime}{Correct}{True}{Median}{0.183816}%
\StoreBenchExecResult{Abc}{ScPdrScPdrReachSafetyFuncCommon}{Cputime}{Correct}{True}{Stdev}{56.23604598951302072581965252}%
\StoreBenchExecResult{Abc}{ScPdrScPdrReachSafetyFuncCommon}{Cputime}{Correct}{True}{Unit}{s}%
\StoreBenchExecResult{Abc}{ScPdrScPdrReachSafetyFuncCommon}{Cputime}{Correct}{False}{Sum}{14045.274634}%
\StoreBenchExecResult{Abc}{ScPdrScPdrReachSafetyFuncCommon}{Cputime}{Correct}{False}{Min}{0.048323}%
\StoreBenchExecResult{Abc}{ScPdrScPdrReachSafetyFuncCommon}{Cputime}{Correct}{False}{Max}{698.512336}%
\StoreBenchExecResult{Abc}{ScPdrScPdrReachSafetyFuncCommon}{Cputime}{Correct}{False}{Avg}{52.60402484644194756554307116}%
\StoreBenchExecResult{Abc}{ScPdrScPdrReachSafetyFuncCommon}{Cputime}{Correct}{False}{Median}{2.950293}%
\StoreBenchExecResult{Abc}{ScPdrScPdrReachSafetyFuncCommon}{Cputime}{Correct}{False}{Stdev}{114.4235481521652678406087617}%
\StoreBenchExecResult{Abc}{ScPdrScPdrReachSafetyFuncCommon}{Cputime}{Correct}{False}{Unit}{s}%
\StoreBenchExecResult{Abc}{ScPdrScPdrReachSafetyFuncCommon}{Walltime}{All}{}{Sum}{19146.225942060351354531}%
\StoreBenchExecResult{Abc}{ScPdrScPdrReachSafetyFuncCommon}{Walltime}{All}{}{Min}{0.04849046561866999}%
\StoreBenchExecResult{Abc}{ScPdrScPdrReachSafetyFuncCommon}{Walltime}{All}{}{Max}{698.5888067018241}%
\StoreBenchExecResult{Abc}{ScPdrScPdrReachSafetyFuncCommon}{Walltime}{All}{}{Avg}{29.59231212064969297454559505}%
\StoreBenchExecResult{Abc}{ScPdrScPdrReachSafetyFuncCommon}{Walltime}{All}{}{Median}{0.38553504180163145}%
\StoreBenchExecResult{Abc}{ScPdrScPdrReachSafetyFuncCommon}{Walltime}{All}{}{Stdev}{87.39258312373416290549377329}%
\StoreBenchExecResult{Abc}{ScPdrScPdrReachSafetyFuncCommon}{Walltime}{All}{}{Unit}{s}%
\StoreBenchExecResult{Abc}{ScPdrScPdrReachSafetyFuncCommon}{Walltime}{Correct}{}{Sum}{19146.225942060351354531}%
\StoreBenchExecResult{Abc}{ScPdrScPdrReachSafetyFuncCommon}{Walltime}{Correct}{}{Min}{0.04849046561866999}%
\StoreBenchExecResult{Abc}{ScPdrScPdrReachSafetyFuncCommon}{Walltime}{Correct}{}{Max}{698.5888067018241}%
\StoreBenchExecResult{Abc}{ScPdrScPdrReachSafetyFuncCommon}{Walltime}{Correct}{}{Avg}{29.59231212064969297454559505}%
\StoreBenchExecResult{Abc}{ScPdrScPdrReachSafetyFuncCommon}{Walltime}{Correct}{}{Median}{0.38553504180163145}%
\StoreBenchExecResult{Abc}{ScPdrScPdrReachSafetyFuncCommon}{Walltime}{Correct}{}{Stdev}{87.39258312373416290549377329}%
\StoreBenchExecResult{Abc}{ScPdrScPdrReachSafetyFuncCommon}{Walltime}{Correct}{}{Unit}{s}%
\StoreBenchExecResult{Abc}{ScPdrScPdrReachSafetyFuncCommon}{Walltime}{Correct}{True}{Sum}{5096.384113317355479230}%
\StoreBenchExecResult{Abc}{ScPdrScPdrReachSafetyFuncCommon}{Walltime}{Correct}{True}{Min}{0.0514822518453002}%
\StoreBenchExecResult{Abc}{ScPdrScPdrReachSafetyFuncCommon}{Walltime}{Correct}{True}{Max}{399.4103391766548}%
\StoreBenchExecResult{Abc}{ScPdrScPdrReachSafetyFuncCommon}{Walltime}{Correct}{True}{Avg}{13.4115371403088302085}%
\StoreBenchExecResult{Abc}{ScPdrScPdrReachSafetyFuncCommon}{Walltime}{Correct}{True}{Median}{0.18397414218634367}%
\StoreBenchExecResult{Abc}{ScPdrScPdrReachSafetyFuncCommon}{Walltime}{Correct}{True}{Stdev}{56.25351981589441117672550918}%
\StoreBenchExecResult{Abc}{ScPdrScPdrReachSafetyFuncCommon}{Walltime}{Correct}{True}{Unit}{s}%
\StoreBenchExecResult{Abc}{ScPdrScPdrReachSafetyFuncCommon}{Walltime}{Correct}{False}{Sum}{14049.841828742995875301}%
\StoreBenchExecResult{Abc}{ScPdrScPdrReachSafetyFuncCommon}{Walltime}{Correct}{False}{Min}{0.04849046561866999}%
\StoreBenchExecResult{Abc}{ScPdrScPdrReachSafetyFuncCommon}{Walltime}{Correct}{False}{Max}{698.5888067018241}%
\StoreBenchExecResult{Abc}{ScPdrScPdrReachSafetyFuncCommon}{Walltime}{Correct}{False}{Avg}{52.62113044473032163034082397}%
\StoreBenchExecResult{Abc}{ScPdrScPdrReachSafetyFuncCommon}{Walltime}{Correct}{False}{Median}{2.950880047865212}%
\StoreBenchExecResult{Abc}{ScPdrScPdrReachSafetyFuncCommon}{Walltime}{Correct}{False}{Stdev}{114.4577985502932466007260524}%
\StoreBenchExecResult{Abc}{ScPdrScPdrReachSafetyFuncCommon}{Walltime}{Correct}{False}{Unit}{s}%
\StoreBenchExecResult{Abc}{ScPdrScPdrReachSafetyFuncCommon}{Memory}{All}{}{Sum}{18029.125632}%
\StoreBenchExecResult{Abc}{ScPdrScPdrReachSafetyFuncCommon}{Memory}{All}{}{Min}{7.557120}%
\StoreBenchExecResult{Abc}{ScPdrScPdrReachSafetyFuncCommon}{Memory}{All}{}{Max}{193.683456}%
\StoreBenchExecResult{Abc}{ScPdrScPdrReachSafetyFuncCommon}{Memory}{All}{}{Avg}{27.86572740649149922720247295}%
\StoreBenchExecResult{Abc}{ScPdrScPdrReachSafetyFuncCommon}{Memory}{All}{}{Median}{15.417344}%
\StoreBenchExecResult{Abc}{ScPdrScPdrReachSafetyFuncCommon}{Memory}{All}{}{Stdev}{28.59736824119984951603040078}%
\StoreBenchExecResult{Abc}{ScPdrScPdrReachSafetyFuncCommon}{Memory}{All}{}{Unit}{MB}%
\StoreBenchExecResult{Abc}{ScPdrScPdrReachSafetyFuncCommon}{Memory}{Correct}{}{Sum}{18029.125632}%
\StoreBenchExecResult{Abc}{ScPdrScPdrReachSafetyFuncCommon}{Memory}{Correct}{}{Min}{7.557120}%
\StoreBenchExecResult{Abc}{ScPdrScPdrReachSafetyFuncCommon}{Memory}{Correct}{}{Max}{193.683456}%
\StoreBenchExecResult{Abc}{ScPdrScPdrReachSafetyFuncCommon}{Memory}{Correct}{}{Avg}{27.86572740649149922720247295}%
\StoreBenchExecResult{Abc}{ScPdrScPdrReachSafetyFuncCommon}{Memory}{Correct}{}{Median}{15.417344}%
\StoreBenchExecResult{Abc}{ScPdrScPdrReachSafetyFuncCommon}{Memory}{Correct}{}{Stdev}{28.59736824119984951603040078}%
\StoreBenchExecResult{Abc}{ScPdrScPdrReachSafetyFuncCommon}{Memory}{Correct}{}{Unit}{MB}%
\StoreBenchExecResult{Abc}{ScPdrScPdrReachSafetyFuncCommon}{Memory}{Correct}{True}{Sum}{8126.582784}%
\StoreBenchExecResult{Abc}{ScPdrScPdrReachSafetyFuncCommon}{Memory}{Correct}{True}{Min}{10.780672}%
\StoreBenchExecResult{Abc}{ScPdrScPdrReachSafetyFuncCommon}{Memory}{Correct}{True}{Max}{193.683456}%
\StoreBenchExecResult{Abc}{ScPdrScPdrReachSafetyFuncCommon}{Memory}{Correct}{True}{Avg}{21.38574416842105263157894737}%
\StoreBenchExecResult{Abc}{ScPdrScPdrReachSafetyFuncCommon}{Memory}{Correct}{True}{Median}{14.032896}%
\StoreBenchExecResult{Abc}{ScPdrScPdrReachSafetyFuncCommon}{Memory}{Correct}{True}{Stdev}{22.36990020244313291564888574}%
\StoreBenchExecResult{Abc}{ScPdrScPdrReachSafetyFuncCommon}{Memory}{Correct}{True}{Unit}{MB}%
\StoreBenchExecResult{Abc}{ScPdrScPdrReachSafetyFuncCommon}{Memory}{Correct}{False}{Sum}{9902.542848}%
\StoreBenchExecResult{Abc}{ScPdrScPdrReachSafetyFuncCommon}{Memory}{Correct}{False}{Min}{7.557120}%
\StoreBenchExecResult{Abc}{ScPdrScPdrReachSafetyFuncCommon}{Memory}{Correct}{False}{Max}{168.542208}%
\StoreBenchExecResult{Abc}{ScPdrScPdrReachSafetyFuncCommon}{Memory}{Correct}{False}{Avg}{37.08817546067415730337078652}%
\StoreBenchExecResult{Abc}{ScPdrScPdrReachSafetyFuncCommon}{Memory}{Correct}{False}{Median}{20.226048}%
\StoreBenchExecResult{Abc}{ScPdrScPdrReachSafetyFuncCommon}{Memory}{Correct}{False}{Stdev}{33.53685630761678501500104534}%
\StoreBenchExecResult{Abc}{ScPdrScPdrReachSafetyFuncCommon}{Memory}{Correct}{False}{Unit}{MB}%
\StoreBenchExecResult{Abc}{ScPdrScPdrReachSafetyFuncCommon}{Gate}{All}{}{Sum}{3242888}%
\StoreBenchExecResult{Abc}{ScPdrScPdrReachSafetyFuncCommon}{Gate}{All}{}{Min}{0}%
\StoreBenchExecResult{Abc}{ScPdrScPdrReachSafetyFuncCommon}{Gate}{All}{}{Max}{110127}%
\StoreBenchExecResult{Abc}{ScPdrScPdrReachSafetyFuncCommon}{Gate}{All}{}{Avg}{5012.191653786707882534775889}%
\StoreBenchExecResult{Abc}{ScPdrScPdrReachSafetyFuncCommon}{Gate}{All}{}{Median}{0}%
\StoreBenchExecResult{Abc}{ScPdrScPdrReachSafetyFuncCommon}{Gate}{All}{}{Stdev}{13380.78775784355093296184195}%
\StoreBenchExecResult{Abc}{ScPdrScPdrReachSafetyFuncCommon}{Gate}{Correct}{}{Sum}{3242888}%
\StoreBenchExecResult{Abc}{ScPdrScPdrReachSafetyFuncCommon}{Gate}{Correct}{}{Min}{0}%
\StoreBenchExecResult{Abc}{ScPdrScPdrReachSafetyFuncCommon}{Gate}{Correct}{}{Max}{110127}%
\StoreBenchExecResult{Abc}{ScPdrScPdrReachSafetyFuncCommon}{Gate}{Correct}{}{Avg}{5012.191653786707882534775889}%
\StoreBenchExecResult{Abc}{ScPdrScPdrReachSafetyFuncCommon}{Gate}{Correct}{}{Median}{0}%
\StoreBenchExecResult{Abc}{ScPdrScPdrReachSafetyFuncCommon}{Gate}{Correct}{}{Stdev}{13380.78775784355093296184195}%
\StoreBenchExecResult{Abc}{ScPdrScPdrReachSafetyFuncCommon}{Gate}{Correct}{True}{Sum}{178113}%
\StoreBenchExecResult{Abc}{ScPdrScPdrReachSafetyFuncCommon}{Gate}{Correct}{True}{Min}{0}%
\StoreBenchExecResult{Abc}{ScPdrScPdrReachSafetyFuncCommon}{Gate}{Correct}{True}{Max}{16656}%
\StoreBenchExecResult{Abc}{ScPdrScPdrReachSafetyFuncCommon}{Gate}{Correct}{True}{Avg}{468.7184210526315789473684211}%
\StoreBenchExecResult{Abc}{ScPdrScPdrReachSafetyFuncCommon}{Gate}{Correct}{True}{Median}{0}%
\StoreBenchExecResult{Abc}{ScPdrScPdrReachSafetyFuncCommon}{Gate}{Correct}{True}{Stdev}{2319.652986111598178852850775}%
\StoreBenchExecResult{Abc}{ScPdrScPdrReachSafetyFuncCommon}{Gate}{Correct}{False}{Sum}{3064775}%
\StoreBenchExecResult{Abc}{ScPdrScPdrReachSafetyFuncCommon}{Gate}{Correct}{False}{Min}{0}%
\StoreBenchExecResult{Abc}{ScPdrScPdrReachSafetyFuncCommon}{Gate}{Correct}{False}{Max}{110127}%
\StoreBenchExecResult{Abc}{ScPdrScPdrReachSafetyFuncCommon}{Gate}{Correct}{False}{Avg}{11478.55805243445692883895131}%
\StoreBenchExecResult{Abc}{ScPdrScPdrReachSafetyFuncCommon}{Gate}{Correct}{False}{Median}{3273}%
\StoreBenchExecResult{Abc}{ScPdrScPdrReachSafetyFuncCommon}{Gate}{Correct}{False}{Stdev}{18841.84318673723085681326539}%
\StoreBenchExecResult{Abc}{ScPdrScPdrReachSafetyFuncCommon}{ScTime}{All}{}{Sum}{15991.51}%
\StoreBenchExecResult{Abc}{ScPdrScPdrReachSafetyFuncCommon}{ScTime}{All}{}{Min}{0.00}%
\StoreBenchExecResult{Abc}{ScPdrScPdrReachSafetyFuncCommon}{ScTime}{All}{}{Max}{506.33}%
\StoreBenchExecResult{Abc}{ScPdrScPdrReachSafetyFuncCommon}{ScTime}{All}{}{Avg}{24.71639876352395672333848532}%
\StoreBenchExecResult{Abc}{ScPdrScPdrReachSafetyFuncCommon}{ScTime}{All}{}{Median}{0.23}%
\StoreBenchExecResult{Abc}{ScPdrScPdrReachSafetyFuncCommon}{ScTime}{All}{}{Stdev}{79.28013436712325276321226363}%
\StoreBenchExecResult{Abc}{ScPdrScPdrReachSafetyFuncCommon}{ScTime}{Correct}{}{Sum}{15991.51}%
\StoreBenchExecResult{Abc}{ScPdrScPdrReachSafetyFuncCommon}{ScTime}{Correct}{}{Min}{0.00}%
\StoreBenchExecResult{Abc}{ScPdrScPdrReachSafetyFuncCommon}{ScTime}{Correct}{}{Max}{506.33}%
\StoreBenchExecResult{Abc}{ScPdrScPdrReachSafetyFuncCommon}{ScTime}{Correct}{}{Avg}{24.71639876352395672333848532}%
\StoreBenchExecResult{Abc}{ScPdrScPdrReachSafetyFuncCommon}{ScTime}{Correct}{}{Median}{0.23}%
\StoreBenchExecResult{Abc}{ScPdrScPdrReachSafetyFuncCommon}{ScTime}{Correct}{}{Stdev}{79.28013436712325276321226363}%
\StoreBenchExecResult{Abc}{ScPdrScPdrReachSafetyFuncCommon}{ScTime}{Correct}{True}{Sum}{3924.79}%
\StoreBenchExecResult{Abc}{ScPdrScPdrReachSafetyFuncCommon}{ScTime}{Correct}{True}{Min}{0.00}%
\StoreBenchExecResult{Abc}{ScPdrScPdrReachSafetyFuncCommon}{ScTime}{Correct}{True}{Max}{398.61}%
\StoreBenchExecResult{Abc}{ScPdrScPdrReachSafetyFuncCommon}{ScTime}{Correct}{True}{Avg}{10.32839473684210526315789474}%
\StoreBenchExecResult{Abc}{ScPdrScPdrReachSafetyFuncCommon}{ScTime}{Correct}{True}{Median}{0.10}%
\StoreBenchExecResult{Abc}{ScPdrScPdrReachSafetyFuncCommon}{ScTime}{Correct}{True}{Stdev}{49.85040057171027687156306057}%
\StoreBenchExecResult{Abc}{ScPdrScPdrReachSafetyFuncCommon}{ScTime}{Correct}{False}{Sum}{12066.72}%
\StoreBenchExecResult{Abc}{ScPdrScPdrReachSafetyFuncCommon}{ScTime}{Correct}{False}{Min}{0.00}%
\StoreBenchExecResult{Abc}{ScPdrScPdrReachSafetyFuncCommon}{ScTime}{Correct}{False}{Max}{506.33}%
\StoreBenchExecResult{Abc}{ScPdrScPdrReachSafetyFuncCommon}{ScTime}{Correct}{False}{Avg}{45.19370786516853932584269663}%
\StoreBenchExecResult{Abc}{ScPdrScPdrReachSafetyFuncCommon}{ScTime}{Correct}{False}{Median}{2.21}%
\StoreBenchExecResult{Abc}{ScPdrScPdrReachSafetyFuncCommon}{ScTime}{Correct}{False}{Stdev}{104.7856203187134765580609730}%
\StoreBenchExecResult{Abc}{ScPdrScPdrReachSafetyFuncCommon}{SolveTime}{All}{}{Sum}{19000.069999999999819055}%
\StoreBenchExecResult{Abc}{ScPdrScPdrReachSafetyFuncCommon}{SolveTime}{All}{}{Min}{0.01}%
\StoreBenchExecResult{Abc}{ScPdrScPdrReachSafetyFuncCommon}{SolveTime}{All}{}{Max}{698.52}%
\StoreBenchExecResult{Abc}{ScPdrScPdrReachSafetyFuncCommon}{SolveTime}{All}{}{Avg}{29.36641421947449740193972179}%
\StoreBenchExecResult{Abc}{ScPdrScPdrReachSafetyFuncCommon}{SolveTime}{All}{}{Median}{0.29000000000000004}%
\StoreBenchExecResult{Abc}{ScPdrScPdrReachSafetyFuncCommon}{SolveTime}{All}{}{Stdev}{87.17153475316669409950326350}%
\StoreBenchExecResult{Abc}{ScPdrScPdrReachSafetyFuncCommon}{SolveTime}{Correct}{}{Sum}{19000.069999999999819055}%
\StoreBenchExecResult{Abc}{ScPdrScPdrReachSafetyFuncCommon}{SolveTime}{Correct}{}{Min}{0.01}%
\StoreBenchExecResult{Abc}{ScPdrScPdrReachSafetyFuncCommon}{SolveTime}{Correct}{}{Max}{698.52}%
\StoreBenchExecResult{Abc}{ScPdrScPdrReachSafetyFuncCommon}{SolveTime}{Correct}{}{Avg}{29.36641421947449740193972179}%
\StoreBenchExecResult{Abc}{ScPdrScPdrReachSafetyFuncCommon}{SolveTime}{Correct}{}{Median}{0.29000000000000004}%
\StoreBenchExecResult{Abc}{ScPdrScPdrReachSafetyFuncCommon}{SolveTime}{Correct}{}{Stdev}{87.17153475316669409950326350}%
\StoreBenchExecResult{Abc}{ScPdrScPdrReachSafetyFuncCommon}{SolveTime}{Correct}{True}{Sum}{5043.450000000000015175}%
\StoreBenchExecResult{Abc}{ScPdrScPdrReachSafetyFuncCommon}{SolveTime}{Correct}{True}{Min}{0.01}%
\StoreBenchExecResult{Abc}{ScPdrScPdrReachSafetyFuncCommon}{SolveTime}{Correct}{True}{Max}{398.62}%
\StoreBenchExecResult{Abc}{ScPdrScPdrReachSafetyFuncCommon}{SolveTime}{Correct}{True}{Avg}{13.27223684210526319782894737}%
\StoreBenchExecResult{Abc}{ScPdrScPdrReachSafetyFuncCommon}{SolveTime}{Correct}{True}{Median}{0.11}%
\StoreBenchExecResult{Abc}{ScPdrScPdrReachSafetyFuncCommon}{SolveTime}{Correct}{True}{Stdev}{56.14569091584491332580991291}%
\StoreBenchExecResult{Abc}{ScPdrScPdrReachSafetyFuncCommon}{SolveTime}{Correct}{False}{Sum}{13956.61999999999980388}%
\StoreBenchExecResult{Abc}{ScPdrScPdrReachSafetyFuncCommon}{SolveTime}{Correct}{False}{Min}{0.01}%
\StoreBenchExecResult{Abc}{ScPdrScPdrReachSafetyFuncCommon}{SolveTime}{Correct}{False}{Max}{698.52}%
\StoreBenchExecResult{Abc}{ScPdrScPdrReachSafetyFuncCommon}{SolveTime}{Correct}{False}{Avg}{52.27198501872659102576779026}%
\StoreBenchExecResult{Abc}{ScPdrScPdrReachSafetyFuncCommon}{SolveTime}{Correct}{False}{Median}{2.82}%
\StoreBenchExecResult{Abc}{ScPdrScPdrReachSafetyFuncCommon}{SolveTime}{Correct}{False}{Stdev}{114.1664176773239358969118219}%
\StoreBenchExecResult{Abc}{ScPdrScPdrReachSafetyFuncCommon}{VerifTime}{All}{}{Sum}{3008.56}%
\StoreBenchExecResult{Abc}{ScPdrScPdrReachSafetyFuncCommon}{VerifTime}{All}{}{Min}{0.01}%
\StoreBenchExecResult{Abc}{ScPdrScPdrReachSafetyFuncCommon}{VerifTime}{All}{}{Max}{636.14}%
\StoreBenchExecResult{Abc}{ScPdrScPdrReachSafetyFuncCommon}{VerifTime}{All}{}{Avg}{4.650015455950540958268933539}%
\StoreBenchExecResult{Abc}{ScPdrScPdrReachSafetyFuncCommon}{VerifTime}{All}{}{Median}{0.01}%
\StoreBenchExecResult{Abc}{ScPdrScPdrReachSafetyFuncCommon}{VerifTime}{All}{}{Stdev}{36.32110929433053434499219804}%
\StoreBenchExecResult{Abc}{ScPdrScPdrReachSafetyFuncCommon}{VerifTime}{Correct}{}{Sum}{3008.56}%
\StoreBenchExecResult{Abc}{ScPdrScPdrReachSafetyFuncCommon}{VerifTime}{Correct}{}{Min}{0.01}%
\StoreBenchExecResult{Abc}{ScPdrScPdrReachSafetyFuncCommon}{VerifTime}{Correct}{}{Max}{636.14}%
\StoreBenchExecResult{Abc}{ScPdrScPdrReachSafetyFuncCommon}{VerifTime}{Correct}{}{Avg}{4.650015455950540958268933539}%
\StoreBenchExecResult{Abc}{ScPdrScPdrReachSafetyFuncCommon}{VerifTime}{Correct}{}{Median}{0.01}%
\StoreBenchExecResult{Abc}{ScPdrScPdrReachSafetyFuncCommon}{VerifTime}{Correct}{}{Stdev}{36.32110929433053434499219804}%
\StoreBenchExecResult{Abc}{ScPdrScPdrReachSafetyFuncCommon}{VerifTime}{Correct}{True}{Sum}{1118.66}%
\StoreBenchExecResult{Abc}{ScPdrScPdrReachSafetyFuncCommon}{VerifTime}{Correct}{True}{Min}{0.01}%
\StoreBenchExecResult{Abc}{ScPdrScPdrReachSafetyFuncCommon}{VerifTime}{Correct}{True}{Max}{369.31}%
\StoreBenchExecResult{Abc}{ScPdrScPdrReachSafetyFuncCommon}{VerifTime}{Correct}{True}{Avg}{2.943842105263157894736842105}%
\StoreBenchExecResult{Abc}{ScPdrScPdrReachSafetyFuncCommon}{VerifTime}{Correct}{True}{Median}{0.01}%
\StoreBenchExecResult{Abc}{ScPdrScPdrReachSafetyFuncCommon}{VerifTime}{Correct}{True}{Stdev}{26.34266234549487335860685984}%
\StoreBenchExecResult{Abc}{ScPdrScPdrReachSafetyFuncCommon}{VerifTime}{Correct}{False}{Sum}{1889.90}%
\StoreBenchExecResult{Abc}{ScPdrScPdrReachSafetyFuncCommon}{VerifTime}{Correct}{False}{Min}{0.01}%
\StoreBenchExecResult{Abc}{ScPdrScPdrReachSafetyFuncCommon}{VerifTime}{Correct}{False}{Max}{636.14}%
\StoreBenchExecResult{Abc}{ScPdrScPdrReachSafetyFuncCommon}{VerifTime}{Correct}{False}{Avg}{7.078277153558052434456928839}%
\StoreBenchExecResult{Abc}{ScPdrScPdrReachSafetyFuncCommon}{VerifTime}{Correct}{False}{Median}{0.03}%
\StoreBenchExecResult{Abc}{ScPdrScPdrReachSafetyFuncCommon}{VerifTime}{Correct}{False}{Stdev}{46.89461840375766751591403889}%
\newcommand{\AbcPdrScPdrReachSafetyFuncCommonVerifTimeAllGeoMean}{0.20065062804637795}
\newcommand{\AbcScPdrScPdrReachSafetyFuncCommonVerifTimeAllGeoMean}{0.03325419211328067}
\newcommand{\AbcScPdrScPdrReachSafetyFuncCommonSolveTimeAllGeoMean}{0.6927201252358874}
\newcommand{\ScPdrReachSafetyFuncReducedToZeroCount}{356}
\providecommand\StoreBenchExecResult[7]{\expandafter\newcommand\csname#1#2#3#4#5#6\endcsname{#7}}%
\StoreBenchExecResult{Abc}{PdrScPdrReachSafetyRelCommon}{Status}{All}{}{Score}{0}%
\StoreBenchExecResult{Abc}{PdrScPdrReachSafetyRelCommon}{Status}{All}{}{Count}{647}%
\StoreBenchExecResult{Abc}{PdrScPdrReachSafetyRelCommon}{Status}{Correct}{}{Count}{647}%
\StoreBenchExecResult{Abc}{PdrScPdrReachSafetyRelCommon}{Status}{Correct}{True}{Count}{380}%
\StoreBenchExecResult{Abc}{PdrScPdrReachSafetyRelCommon}{Status}{Correct}{False}{Count}{267}%
\StoreBenchExecResult{Abc}{PdrScPdrReachSafetyRelCommon}{Status}{Wrong}{}{Count}{0}%
\StoreBenchExecResult{Abc}{PdrScPdrReachSafetyRelCommon}{Status}{Wrong}{True}{Count}{0}%
\StoreBenchExecResult{Abc}{PdrScPdrReachSafetyRelCommon}{Status}{Wrong}{False}{Count}{0}%
\StoreBenchExecResult{Abc}{PdrScPdrReachSafetyRelCommon}{Cputime}{All}{}{Sum}{38012.754737}%
\StoreBenchExecResult{Abc}{PdrScPdrReachSafetyRelCommon}{Cputime}{All}{}{Min}{0.047499}%
\StoreBenchExecResult{Abc}{PdrScPdrReachSafetyRelCommon}{Cputime}{All}{}{Max}{896.415424}%
\StoreBenchExecResult{Abc}{PdrScPdrReachSafetyRelCommon}{Cputime}{All}{}{Avg}{58.75232571406491499227202473}%
\StoreBenchExecResult{Abc}{PdrScPdrReachSafetyRelCommon}{Cputime}{All}{}{Median}{0.57586}%
\StoreBenchExecResult{Abc}{PdrScPdrReachSafetyRelCommon}{Cputime}{All}{}{Stdev}{152.8739700534984141613743076}%
\StoreBenchExecResult{Abc}{PdrScPdrReachSafetyRelCommon}{Cputime}{All}{}{Unit}{s}%
\StoreBenchExecResult{Abc}{PdrScPdrReachSafetyRelCommon}{Cputime}{Correct}{}{Sum}{38012.754737}%
\StoreBenchExecResult{Abc}{PdrScPdrReachSafetyRelCommon}{Cputime}{Correct}{}{Min}{0.047499}%
\StoreBenchExecResult{Abc}{PdrScPdrReachSafetyRelCommon}{Cputime}{Correct}{}{Max}{896.415424}%
\StoreBenchExecResult{Abc}{PdrScPdrReachSafetyRelCommon}{Cputime}{Correct}{}{Avg}{58.75232571406491499227202473}%
\StoreBenchExecResult{Abc}{PdrScPdrReachSafetyRelCommon}{Cputime}{Correct}{}{Median}{0.57586}%
\StoreBenchExecResult{Abc}{PdrScPdrReachSafetyRelCommon}{Cputime}{Correct}{}{Stdev}{152.8739700534984141613743076}%
\StoreBenchExecResult{Abc}{PdrScPdrReachSafetyRelCommon}{Cputime}{Correct}{}{Unit}{s}%
\StoreBenchExecResult{Abc}{PdrScPdrReachSafetyRelCommon}{Cputime}{Correct}{True}{Sum}{10863.374674}%
\StoreBenchExecResult{Abc}{PdrScPdrReachSafetyRelCommon}{Cputime}{Correct}{True}{Min}{0.048702}%
\StoreBenchExecResult{Abc}{PdrScPdrReachSafetyRelCommon}{Cputime}{Correct}{True}{Max}{896.415424}%
\StoreBenchExecResult{Abc}{PdrScPdrReachSafetyRelCommon}{Cputime}{Correct}{True}{Avg}{28.58782808947368421052631579}%
\StoreBenchExecResult{Abc}{PdrScPdrReachSafetyRelCommon}{Cputime}{Correct}{True}{Median}{0.137511}%
\StoreBenchExecResult{Abc}{PdrScPdrReachSafetyRelCommon}{Cputime}{Correct}{True}{Stdev}{114.2919176824740614578721097}%
\StoreBenchExecResult{Abc}{PdrScPdrReachSafetyRelCommon}{Cputime}{Correct}{True}{Unit}{s}%
\StoreBenchExecResult{Abc}{PdrScPdrReachSafetyRelCommon}{Cputime}{Correct}{False}{Sum}{27149.380063}%
\StoreBenchExecResult{Abc}{PdrScPdrReachSafetyRelCommon}{Cputime}{Correct}{False}{Min}{0.047499}%
\StoreBenchExecResult{Abc}{PdrScPdrReachSafetyRelCommon}{Cputime}{Correct}{False}{Max}{830.044536}%
\StoreBenchExecResult{Abc}{PdrScPdrReachSafetyRelCommon}{Cputime}{Correct}{False}{Avg}{101.6830713970037453183520599}%
\StoreBenchExecResult{Abc}{PdrScPdrReachSafetyRelCommon}{Cputime}{Correct}{False}{Median}{3.41436}%
\StoreBenchExecResult{Abc}{PdrScPdrReachSafetyRelCommon}{Cputime}{Correct}{False}{Stdev}{186.8226601508320980909527242}%
\StoreBenchExecResult{Abc}{PdrScPdrReachSafetyRelCommon}{Cputime}{Correct}{False}{Unit}{s}%
\StoreBenchExecResult{Abc}{PdrScPdrReachSafetyRelCommon}{Walltime}{All}{}{Sum}{38025.589361446909848739}%
\StoreBenchExecResult{Abc}{PdrScPdrReachSafetyRelCommon}{Walltime}{All}{}{Min}{0.04767234530299902}%
\StoreBenchExecResult{Abc}{PdrScPdrReachSafetyRelCommon}{Walltime}{All}{}{Max}{896.7487301426008}%
\StoreBenchExecResult{Abc}{PdrScPdrReachSafetyRelCommon}{Walltime}{All}{}{Avg}{58.77216284613123624225502318}%
\StoreBenchExecResult{Abc}{PdrScPdrReachSafetyRelCommon}{Walltime}{All}{}{Median}{0.5760467201471329}%
\StoreBenchExecResult{Abc}{PdrScPdrReachSafetyRelCommon}{Walltime}{All}{}{Stdev}{152.9225019173356293279237595}%
\StoreBenchExecResult{Abc}{PdrScPdrReachSafetyRelCommon}{Walltime}{All}{}{Unit}{s}%
\StoreBenchExecResult{Abc}{PdrScPdrReachSafetyRelCommon}{Walltime}{Correct}{}{Sum}{38025.589361446909848739}%
\StoreBenchExecResult{Abc}{PdrScPdrReachSafetyRelCommon}{Walltime}{Correct}{}{Min}{0.04767234530299902}%
\StoreBenchExecResult{Abc}{PdrScPdrReachSafetyRelCommon}{Walltime}{Correct}{}{Max}{896.7487301426008}%
\StoreBenchExecResult{Abc}{PdrScPdrReachSafetyRelCommon}{Walltime}{Correct}{}{Avg}{58.77216284613123624225502318}%
\StoreBenchExecResult{Abc}{PdrScPdrReachSafetyRelCommon}{Walltime}{Correct}{}{Median}{0.5760467201471329}%
\StoreBenchExecResult{Abc}{PdrScPdrReachSafetyRelCommon}{Walltime}{Correct}{}{Stdev}{152.9225019173356293279237595}%
\StoreBenchExecResult{Abc}{PdrScPdrReachSafetyRelCommon}{Walltime}{Correct}{}{Unit}{s}%
\StoreBenchExecResult{Abc}{PdrScPdrReachSafetyRelCommon}{Walltime}{Correct}{True}{Sum}{10867.079992641694913525}%
\StoreBenchExecResult{Abc}{PdrScPdrReachSafetyRelCommon}{Walltime}{Correct}{True}{Min}{0.04889687988907099}%
\StoreBenchExecResult{Abc}{PdrScPdrReachSafetyRelCommon}{Walltime}{Correct}{True}{Max}{896.7487301426008}%
\StoreBenchExecResult{Abc}{PdrScPdrReachSafetyRelCommon}{Walltime}{Correct}{True}{Avg}{28.59757892800446029875}%
\StoreBenchExecResult{Abc}{PdrScPdrReachSafetyRelCommon}{Walltime}{Correct}{True}{Median}{0.14085948560386896}%
\StoreBenchExecResult{Abc}{PdrScPdrReachSafetyRelCommon}{Walltime}{Correct}{True}{Stdev}{114.3295515965162574531910900}%
\StoreBenchExecResult{Abc}{PdrScPdrReachSafetyRelCommon}{Walltime}{Correct}{True}{Unit}{s}%
\StoreBenchExecResult{Abc}{PdrScPdrReachSafetyRelCommon}{Walltime}{Correct}{False}{Sum}{27158.509368805214935214}%
\StoreBenchExecResult{Abc}{PdrScPdrReachSafetyRelCommon}{Walltime}{Correct}{False}{Min}{0.04767234530299902}%
\StoreBenchExecResult{Abc}{PdrScPdrReachSafetyRelCommon}{Walltime}{Correct}{False}{Max}{830.3341630129144}%
\StoreBenchExecResult{Abc}{PdrScPdrReachSafetyRelCommon}{Walltime}{Correct}{False}{Avg}{101.7172635535775840270187266}%
\StoreBenchExecResult{Abc}{PdrScPdrReachSafetyRelCommon}{Walltime}{Correct}{False}{Median}{3.415569666773081}%
\StoreBenchExecResult{Abc}{PdrScPdrReachSafetyRelCommon}{Walltime}{Correct}{False}{Stdev}{186.8805095793070480794933387}%
\StoreBenchExecResult{Abc}{PdrScPdrReachSafetyRelCommon}{Walltime}{Correct}{False}{Unit}{s}%
\StoreBenchExecResult{Abc}{PdrScPdrReachSafetyRelCommon}{Memory}{All}{}{Sum}{47635.767296}%
\StoreBenchExecResult{Abc}{PdrScPdrReachSafetyRelCommon}{Memory}{All}{}{Min}{8.908800}%
\StoreBenchExecResult{Abc}{PdrScPdrReachSafetyRelCommon}{Memory}{All}{}{Max}{706.969600}%
\StoreBenchExecResult{Abc}{PdrScPdrReachSafetyRelCommon}{Memory}{All}{}{Avg}{73.62560633075734157650695518}%
\StoreBenchExecResult{Abc}{PdrScPdrReachSafetyRelCommon}{Memory}{All}{}{Median}{25.980928}%
\StoreBenchExecResult{Abc}{PdrScPdrReachSafetyRelCommon}{Memory}{All}{}{Stdev}{107.2842785713000455416321006}%
\StoreBenchExecResult{Abc}{PdrScPdrReachSafetyRelCommon}{Memory}{All}{}{Unit}{MB}%
\StoreBenchExecResult{Abc}{PdrScPdrReachSafetyRelCommon}{Memory}{Correct}{}{Sum}{47635.767296}%
\StoreBenchExecResult{Abc}{PdrScPdrReachSafetyRelCommon}{Memory}{Correct}{}{Min}{8.908800}%
\StoreBenchExecResult{Abc}{PdrScPdrReachSafetyRelCommon}{Memory}{Correct}{}{Max}{706.969600}%
\StoreBenchExecResult{Abc}{PdrScPdrReachSafetyRelCommon}{Memory}{Correct}{}{Avg}{73.62560633075734157650695518}%
\StoreBenchExecResult{Abc}{PdrScPdrReachSafetyRelCommon}{Memory}{Correct}{}{Median}{25.980928}%
\StoreBenchExecResult{Abc}{PdrScPdrReachSafetyRelCommon}{Memory}{Correct}{}{Stdev}{107.2842785713000455416321006}%
\StoreBenchExecResult{Abc}{PdrScPdrReachSafetyRelCommon}{Memory}{Correct}{}{Unit}{MB}%
\StoreBenchExecResult{Abc}{PdrScPdrReachSafetyRelCommon}{Memory}{Correct}{True}{Sum}{18351.521792}%
\StoreBenchExecResult{Abc}{PdrScPdrReachSafetyRelCommon}{Memory}{Correct}{True}{Min}{10.600448}%
\StoreBenchExecResult{Abc}{PdrScPdrReachSafetyRelCommon}{Memory}{Correct}{True}{Max}{606.322688}%
\StoreBenchExecResult{Abc}{PdrScPdrReachSafetyRelCommon}{Memory}{Correct}{True}{Avg}{48.2934784}%
\StoreBenchExecResult{Abc}{PdrScPdrReachSafetyRelCommon}{Memory}{Correct}{True}{Median}{23.121920}%
\StoreBenchExecResult{Abc}{PdrScPdrReachSafetyRelCommon}{Memory}{Correct}{True}{Stdev}{74.04175032348996434683298927}%
\StoreBenchExecResult{Abc}{PdrScPdrReachSafetyRelCommon}{Memory}{Correct}{True}{Unit}{MB}%
\StoreBenchExecResult{Abc}{PdrScPdrReachSafetyRelCommon}{Memory}{Correct}{False}{Sum}{29284.245504}%
\StoreBenchExecResult{Abc}{PdrScPdrReachSafetyRelCommon}{Memory}{Correct}{False}{Min}{8.908800}%
\StoreBenchExecResult{Abc}{PdrScPdrReachSafetyRelCommon}{Memory}{Correct}{False}{Max}{706.969600}%
\StoreBenchExecResult{Abc}{PdrScPdrReachSafetyRelCommon}{Memory}{Correct}{False}{Avg}{109.6788221123595505617977528}%
\StoreBenchExecResult{Abc}{PdrScPdrReachSafetyRelCommon}{Memory}{Correct}{False}{Median}{38.092800}%
\StoreBenchExecResult{Abc}{PdrScPdrReachSafetyRelCommon}{Memory}{Correct}{False}{Stdev}{133.6995780461648941706200551}%
\StoreBenchExecResult{Abc}{PdrScPdrReachSafetyRelCommon}{Memory}{Correct}{False}{Unit}{MB}%
\StoreBenchExecResult{Abc}{PdrScPdrReachSafetyRelCommon}{Gate}{All}{}{Sum}{16550161}%
\StoreBenchExecResult{Abc}{PdrScPdrReachSafetyRelCommon}{Gate}{All}{}{Min}{24}%
\StoreBenchExecResult{Abc}{PdrScPdrReachSafetyRelCommon}{Gate}{All}{}{Max}{260466}%
\StoreBenchExecResult{Abc}{PdrScPdrReachSafetyRelCommon}{Gate}{All}{}{Avg}{25579.84698608964451313755796}%
\StoreBenchExecResult{Abc}{PdrScPdrReachSafetyRelCommon}{Gate}{All}{}{Median}{9871}%
\StoreBenchExecResult{Abc}{PdrScPdrReachSafetyRelCommon}{Gate}{All}{}{Stdev}{38402.17372473767866319879372}%
\StoreBenchExecResult{Abc}{PdrScPdrReachSafetyRelCommon}{Gate}{Correct}{}{Sum}{16550161}%
\StoreBenchExecResult{Abc}{PdrScPdrReachSafetyRelCommon}{Gate}{Correct}{}{Min}{24}%
\StoreBenchExecResult{Abc}{PdrScPdrReachSafetyRelCommon}{Gate}{Correct}{}{Max}{260466}%
\StoreBenchExecResult{Abc}{PdrScPdrReachSafetyRelCommon}{Gate}{Correct}{}{Avg}{25579.84698608964451313755796}%
\StoreBenchExecResult{Abc}{PdrScPdrReachSafetyRelCommon}{Gate}{Correct}{}{Median}{9871}%
\StoreBenchExecResult{Abc}{PdrScPdrReachSafetyRelCommon}{Gate}{Correct}{}{Stdev}{38402.17372473767866319879372}%
\StoreBenchExecResult{Abc}{PdrScPdrReachSafetyRelCommon}{Gate}{Correct}{True}{Sum}{5883658}%
\StoreBenchExecResult{Abc}{PdrScPdrReachSafetyRelCommon}{Gate}{Correct}{True}{Min}{24}%
\StoreBenchExecResult{Abc}{PdrScPdrReachSafetyRelCommon}{Gate}{Correct}{True}{Max}{166334}%
\StoreBenchExecResult{Abc}{PdrScPdrReachSafetyRelCommon}{Gate}{Correct}{True}{Avg}{15483.31052631578947368421053}%
\StoreBenchExecResult{Abc}{PdrScPdrReachSafetyRelCommon}{Gate}{Correct}{True}{Median}{8066.5}%
\StoreBenchExecResult{Abc}{PdrScPdrReachSafetyRelCommon}{Gate}{Correct}{True}{Stdev}{23100.05859136781580949428482}%
\StoreBenchExecResult{Abc}{PdrScPdrReachSafetyRelCommon}{Gate}{Correct}{False}{Sum}{10666503}%
\StoreBenchExecResult{Abc}{PdrScPdrReachSafetyRelCommon}{Gate}{Correct}{False}{Min}{24}%
\StoreBenchExecResult{Abc}{PdrScPdrReachSafetyRelCommon}{Gate}{Correct}{False}{Max}{260466}%
\StoreBenchExecResult{Abc}{PdrScPdrReachSafetyRelCommon}{Gate}{Correct}{False}{Avg}{39949.44943820224719101123596}%
\StoreBenchExecResult{Abc}{PdrScPdrReachSafetyRelCommon}{Gate}{Correct}{False}{Median}{14425}%
\StoreBenchExecResult{Abc}{PdrScPdrReachSafetyRelCommon}{Gate}{Correct}{False}{Stdev}{49624.30636268662831998193658}%
\StoreBenchExecResult{Abc}{PdrScPdrReachSafetyRelCommon}{VerifTime}{All}{}{Sum}{37964.22}%
\StoreBenchExecResult{Abc}{PdrScPdrReachSafetyRelCommon}{VerifTime}{All}{}{Min}{0.01}%
\StoreBenchExecResult{Abc}{PdrScPdrReachSafetyRelCommon}{VerifTime}{All}{}{Max}{896.61}%
\StoreBenchExecResult{Abc}{PdrScPdrReachSafetyRelCommon}{VerifTime}{All}{}{Avg}{58.67731066460587326120556414}%
\StoreBenchExecResult{Abc}{PdrScPdrReachSafetyRelCommon}{VerifTime}{All}{}{Median}{0.52}%
\StoreBenchExecResult{Abc}{PdrScPdrReachSafetyRelCommon}{VerifTime}{All}{}{Stdev}{152.8819451835044192614877782}%
\StoreBenchExecResult{Abc}{PdrScPdrReachSafetyRelCommon}{VerifTime}{Correct}{}{Sum}{37964.22}%
\StoreBenchExecResult{Abc}{PdrScPdrReachSafetyRelCommon}{VerifTime}{Correct}{}{Min}{0.01}%
\StoreBenchExecResult{Abc}{PdrScPdrReachSafetyRelCommon}{VerifTime}{Correct}{}{Max}{896.61}%
\StoreBenchExecResult{Abc}{PdrScPdrReachSafetyRelCommon}{VerifTime}{Correct}{}{Avg}{58.67731066460587326120556414}%
\StoreBenchExecResult{Abc}{PdrScPdrReachSafetyRelCommon}{VerifTime}{Correct}{}{Median}{0.52}%
\StoreBenchExecResult{Abc}{PdrScPdrReachSafetyRelCommon}{VerifTime}{Correct}{}{Stdev}{152.8819451835044192614877782}%
\StoreBenchExecResult{Abc}{PdrScPdrReachSafetyRelCommon}{VerifTime}{Correct}{True}{Sum}{10839.95}%
\StoreBenchExecResult{Abc}{PdrScPdrReachSafetyRelCommon}{VerifTime}{Correct}{True}{Min}{0.01}%
\StoreBenchExecResult{Abc}{PdrScPdrReachSafetyRelCommon}{VerifTime}{Correct}{True}{Max}{896.61}%
\StoreBenchExecResult{Abc}{PdrScPdrReachSafetyRelCommon}{VerifTime}{Correct}{True}{Avg}{28.52618421052631578947368421}%
\StoreBenchExecResult{Abc}{PdrScPdrReachSafetyRelCommon}{VerifTime}{Correct}{True}{Median}{0.07}%
\StoreBenchExecResult{Abc}{PdrScPdrReachSafetyRelCommon}{VerifTime}{Correct}{True}{Stdev}{114.3113941823536456470570175}%
\StoreBenchExecResult{Abc}{PdrScPdrReachSafetyRelCommon}{VerifTime}{Correct}{False}{Sum}{27124.27}%
\StoreBenchExecResult{Abc}{PdrScPdrReachSafetyRelCommon}{VerifTime}{Correct}{False}{Min}{0.01}%
\StoreBenchExecResult{Abc}{PdrScPdrReachSafetyRelCommon}{VerifTime}{Correct}{False}{Max}{830.12}%
\StoreBenchExecResult{Abc}{PdrScPdrReachSafetyRelCommon}{VerifTime}{Correct}{False}{Avg}{101.5890262172284644194756554}%
\StoreBenchExecResult{Abc}{PdrScPdrReachSafetyRelCommon}{VerifTime}{Correct}{False}{Median}{3.37}%
\StoreBenchExecResult{Abc}{PdrScPdrReachSafetyRelCommon}{VerifTime}{Correct}{False}{Stdev}{186.8289589482141014274469400}%
\StoreBenchExecResult{Abc}{ScPdrScPdrReachSafetyRelCommon}{Status}{All}{}{Score}{0}%
\StoreBenchExecResult{Abc}{ScPdrScPdrReachSafetyRelCommon}{Status}{All}{}{Count}{647}%
\StoreBenchExecResult{Abc}{ScPdrScPdrReachSafetyRelCommon}{Status}{Correct}{}{Count}{647}%
\StoreBenchExecResult{Abc}{ScPdrScPdrReachSafetyRelCommon}{Status}{Correct}{True}{Count}{380}%
\StoreBenchExecResult{Abc}{ScPdrScPdrReachSafetyRelCommon}{Status}{Correct}{False}{Count}{267}%
\StoreBenchExecResult{Abc}{ScPdrScPdrReachSafetyRelCommon}{Status}{Wrong}{}{Count}{0}%
\StoreBenchExecResult{Abc}{ScPdrScPdrReachSafetyRelCommon}{Status}{Wrong}{True}{Count}{0}%
\StoreBenchExecResult{Abc}{ScPdrScPdrReachSafetyRelCommon}{Status}{Wrong}{False}{Count}{0}%
\StoreBenchExecResult{Abc}{ScPdrScPdrReachSafetyRelCommon}{Cputime}{All}{}{Sum}{51374.950380}%
\StoreBenchExecResult{Abc}{ScPdrScPdrReachSafetyRelCommon}{Cputime}{All}{}{Min}{0.050359}%
\StoreBenchExecResult{Abc}{ScPdrScPdrReachSafetyRelCommon}{Cputime}{All}{}{Max}{884.855196}%
\StoreBenchExecResult{Abc}{ScPdrScPdrReachSafetyRelCommon}{Cputime}{All}{}{Avg}{79.40486921174652241112828439}%
\StoreBenchExecResult{Abc}{ScPdrScPdrReachSafetyRelCommon}{Cputime}{All}{}{Median}{2.19159}%
\StoreBenchExecResult{Abc}{ScPdrScPdrReachSafetyRelCommon}{Cputime}{All}{}{Stdev}{166.5166376636680289262511738}%
\StoreBenchExecResult{Abc}{ScPdrScPdrReachSafetyRelCommon}{Cputime}{All}{}{Unit}{s}%
\StoreBenchExecResult{Abc}{ScPdrScPdrReachSafetyRelCommon}{Cputime}{Correct}{}{Sum}{51374.950380}%
\StoreBenchExecResult{Abc}{ScPdrScPdrReachSafetyRelCommon}{Cputime}{Correct}{}{Min}{0.050359}%
\StoreBenchExecResult{Abc}{ScPdrScPdrReachSafetyRelCommon}{Cputime}{Correct}{}{Max}{884.855196}%
\StoreBenchExecResult{Abc}{ScPdrScPdrReachSafetyRelCommon}{Cputime}{Correct}{}{Avg}{79.40486921174652241112828439}%
\StoreBenchExecResult{Abc}{ScPdrScPdrReachSafetyRelCommon}{Cputime}{Correct}{}{Median}{2.19159}%
\StoreBenchExecResult{Abc}{ScPdrScPdrReachSafetyRelCommon}{Cputime}{Correct}{}{Stdev}{166.5166376636680289262511738}%
\StoreBenchExecResult{Abc}{ScPdrScPdrReachSafetyRelCommon}{Cputime}{Correct}{}{Unit}{s}%
\StoreBenchExecResult{Abc}{ScPdrScPdrReachSafetyRelCommon}{Cputime}{Correct}{True}{Sum}{14379.853970}%
\StoreBenchExecResult{Abc}{ScPdrScPdrReachSafetyRelCommon}{Cputime}{Correct}{True}{Min}{0.051163}%
\StoreBenchExecResult{Abc}{ScPdrScPdrReachSafetyRelCommon}{Cputime}{Correct}{True}{Max}{884.855196}%
\StoreBenchExecResult{Abc}{ScPdrScPdrReachSafetyRelCommon}{Cputime}{Correct}{True}{Avg}{37.84172097368421052631578947}%
\StoreBenchExecResult{Abc}{ScPdrScPdrReachSafetyRelCommon}{Cputime}{Correct}{True}{Median}{0.985041}%
\StoreBenchExecResult{Abc}{ScPdrScPdrReachSafetyRelCommon}{Cputime}{Correct}{True}{Stdev}{110.8263033846615317284088758}%
\StoreBenchExecResult{Abc}{ScPdrScPdrReachSafetyRelCommon}{Cputime}{Correct}{True}{Unit}{s}%
\StoreBenchExecResult{Abc}{ScPdrScPdrReachSafetyRelCommon}{Cputime}{Correct}{False}{Sum}{36995.096410}%
\StoreBenchExecResult{Abc}{ScPdrScPdrReachSafetyRelCommon}{Cputime}{Correct}{False}{Min}{0.050359}%
\StoreBenchExecResult{Abc}{ScPdrScPdrReachSafetyRelCommon}{Cputime}{Correct}{False}{Max}{816.958714}%
\StoreBenchExecResult{Abc}{ScPdrScPdrReachSafetyRelCommon}{Cputime}{Correct}{False}{Avg}{138.5584135205992509363295880}%
\StoreBenchExecResult{Abc}{ScPdrScPdrReachSafetyRelCommon}{Cputime}{Correct}{False}{Median}{18.654254}%
\StoreBenchExecResult{Abc}{ScPdrScPdrReachSafetyRelCommon}{Cputime}{Correct}{False}{Stdev}{209.1701370069875534856799785}%
\StoreBenchExecResult{Abc}{ScPdrScPdrReachSafetyRelCommon}{Cputime}{Correct}{False}{Unit}{s}%
\StoreBenchExecResult{Abc}{ScPdrScPdrReachSafetyRelCommon}{Walltime}{All}{}{Sum}{51391.403616611845753337}%
\StoreBenchExecResult{Abc}{ScPdrScPdrReachSafetyRelCommon}{Walltime}{All}{}{Min}{0.05049137119203806}%
\StoreBenchExecResult{Abc}{ScPdrScPdrReachSafetyRelCommon}{Walltime}{All}{}{Max}{885.2453976608813}%
\StoreBenchExecResult{Abc}{ScPdrScPdrReachSafetyRelCommon}{Walltime}{All}{}{Avg}{79.43029925287765958784698609}%
\StoreBenchExecResult{Abc}{ScPdrScPdrReachSafetyRelCommon}{Walltime}{All}{}{Median}{2.1929925605654716}%
\StoreBenchExecResult{Abc}{ScPdrScPdrReachSafetyRelCommon}{Walltime}{All}{}{Stdev}{166.5666315691870144501225339}%
\StoreBenchExecResult{Abc}{ScPdrScPdrReachSafetyRelCommon}{Walltime}{All}{}{Unit}{s}%
\StoreBenchExecResult{Abc}{ScPdrScPdrReachSafetyRelCommon}{Walltime}{Correct}{}{Sum}{51391.403616611845753337}%
\StoreBenchExecResult{Abc}{ScPdrScPdrReachSafetyRelCommon}{Walltime}{Correct}{}{Min}{0.05049137119203806}%
\StoreBenchExecResult{Abc}{ScPdrScPdrReachSafetyRelCommon}{Walltime}{Correct}{}{Max}{885.2453976608813}%
\StoreBenchExecResult{Abc}{ScPdrScPdrReachSafetyRelCommon}{Walltime}{Correct}{}{Avg}{79.43029925287765958784698609}%
\StoreBenchExecResult{Abc}{ScPdrScPdrReachSafetyRelCommon}{Walltime}{Correct}{}{Median}{2.1929925605654716}%
\StoreBenchExecResult{Abc}{ScPdrScPdrReachSafetyRelCommon}{Walltime}{Correct}{}{Stdev}{166.5666315691870144501225339}%
\StoreBenchExecResult{Abc}{ScPdrScPdrReachSafetyRelCommon}{Walltime}{Correct}{}{Unit}{s}%
\StoreBenchExecResult{Abc}{ScPdrScPdrReachSafetyRelCommon}{Walltime}{Correct}{True}{Sum}{14384.633698186837114201}%
\StoreBenchExecResult{Abc}{ScPdrScPdrReachSafetyRelCommon}{Walltime}{Correct}{True}{Min}{0.05129480082541704}%
\StoreBenchExecResult{Abc}{ScPdrScPdrReachSafetyRelCommon}{Walltime}{Correct}{True}{Max}{885.2453976608813}%
\StoreBenchExecResult{Abc}{ScPdrScPdrReachSafetyRelCommon}{Walltime}{Correct}{True}{Avg}{37.85429920575483451105526316}%
\StoreBenchExecResult{Abc}{ScPdrScPdrReachSafetyRelCommon}{Walltime}{Correct}{True}{Median}{0.9857011125423014}%
\StoreBenchExecResult{Abc}{ScPdrScPdrReachSafetyRelCommon}{Walltime}{Correct}{True}{Stdev}{110.8592085198322419575683040}%
\StoreBenchExecResult{Abc}{ScPdrScPdrReachSafetyRelCommon}{Walltime}{Correct}{True}{Unit}{s}%
\StoreBenchExecResult{Abc}{ScPdrScPdrReachSafetyRelCommon}{Walltime}{Correct}{False}{Sum}{37006.769918425008639136}%
\StoreBenchExecResult{Abc}{ScPdrScPdrReachSafetyRelCommon}{Walltime}{Correct}{False}{Min}{0.05049137119203806}%
\StoreBenchExecResult{Abc}{ScPdrScPdrReachSafetyRelCommon}{Walltime}{Correct}{False}{Max}{817.2782592717558}%
\StoreBenchExecResult{Abc}{ScPdrScPdrReachSafetyRelCommon}{Walltime}{Correct}{False}{Avg}{138.6021345259363619443295880}%
\StoreBenchExecResult{Abc}{ScPdrScPdrReachSafetyRelCommon}{Walltime}{Correct}{False}{Median}{18.65882681682706}%
\StoreBenchExecResult{Abc}{ScPdrScPdrReachSafetyRelCommon}{Walltime}{Correct}{False}{Stdev}{209.2329592063307849195077853}%
\StoreBenchExecResult{Abc}{ScPdrScPdrReachSafetyRelCommon}{Walltime}{Correct}{False}{Unit}{s}%
\StoreBenchExecResult{Abc}{ScPdrScPdrReachSafetyRelCommon}{Memory}{All}{}{Sum}{46413.856768}%
\StoreBenchExecResult{Abc}{ScPdrScPdrReachSafetyRelCommon}{Memory}{All}{}{Min}{9.359360}%
\StoreBenchExecResult{Abc}{ScPdrScPdrReachSafetyRelCommon}{Memory}{All}{}{Max}{742.928384}%
\StoreBenchExecResult{Abc}{ScPdrScPdrReachSafetyRelCommon}{Memory}{All}{}{Avg}{71.73702746213292117465224111}%
\StoreBenchExecResult{Abc}{ScPdrScPdrReachSafetyRelCommon}{Memory}{All}{}{Median}{25.522176}%
\StoreBenchExecResult{Abc}{ScPdrScPdrReachSafetyRelCommon}{Memory}{All}{}{Stdev}{101.8313866365918739712831505}%
\StoreBenchExecResult{Abc}{ScPdrScPdrReachSafetyRelCommon}{Memory}{All}{}{Unit}{MB}%
\StoreBenchExecResult{Abc}{ScPdrScPdrReachSafetyRelCommon}{Memory}{Correct}{}{Sum}{46413.856768}%
\StoreBenchExecResult{Abc}{ScPdrScPdrReachSafetyRelCommon}{Memory}{Correct}{}{Min}{9.359360}%
\StoreBenchExecResult{Abc}{ScPdrScPdrReachSafetyRelCommon}{Memory}{Correct}{}{Max}{742.928384}%
\StoreBenchExecResult{Abc}{ScPdrScPdrReachSafetyRelCommon}{Memory}{Correct}{}{Avg}{71.73702746213292117465224111}%
\StoreBenchExecResult{Abc}{ScPdrScPdrReachSafetyRelCommon}{Memory}{Correct}{}{Median}{25.522176}%
\StoreBenchExecResult{Abc}{ScPdrScPdrReachSafetyRelCommon}{Memory}{Correct}{}{Stdev}{101.8313866365918739712831505}%
\StoreBenchExecResult{Abc}{ScPdrScPdrReachSafetyRelCommon}{Memory}{Correct}{}{Unit}{MB}%
\StoreBenchExecResult{Abc}{ScPdrScPdrReachSafetyRelCommon}{Memory}{Correct}{True}{Sum}{17248.366592}%
\StoreBenchExecResult{Abc}{ScPdrScPdrReachSafetyRelCommon}{Memory}{Correct}{True}{Min}{10.821632}%
\StoreBenchExecResult{Abc}{ScPdrScPdrReachSafetyRelCommon}{Memory}{Correct}{True}{Max}{395.120640}%
\StoreBenchExecResult{Abc}{ScPdrScPdrReachSafetyRelCommon}{Memory}{Correct}{True}{Avg}{45.3904384}%
\StoreBenchExecResult{Abc}{ScPdrScPdrReachSafetyRelCommon}{Memory}{Correct}{True}{Median}{21.403648}%
\StoreBenchExecResult{Abc}{ScPdrScPdrReachSafetyRelCommon}{Memory}{Correct}{True}{Stdev}{64.50390325564293940896432556}%
\StoreBenchExecResult{Abc}{ScPdrScPdrReachSafetyRelCommon}{Memory}{Correct}{True}{Unit}{MB}%
\StoreBenchExecResult{Abc}{ScPdrScPdrReachSafetyRelCommon}{Memory}{Correct}{False}{Sum}{29165.490176}%
\StoreBenchExecResult{Abc}{ScPdrScPdrReachSafetyRelCommon}{Memory}{Correct}{False}{Min}{9.359360}%
\StoreBenchExecResult{Abc}{ScPdrScPdrReachSafetyRelCommon}{Memory}{Correct}{False}{Max}{742.928384}%
\StoreBenchExecResult{Abc}{ScPdrScPdrReachSafetyRelCommon}{Memory}{Correct}{False}{Avg}{109.2340456029962546816479401}%
\StoreBenchExecResult{Abc}{ScPdrScPdrReachSafetyRelCommon}{Memory}{Correct}{False}{Median}{45.432832}%
\StoreBenchExecResult{Abc}{ScPdrScPdrReachSafetyRelCommon}{Memory}{Correct}{False}{Stdev}{129.6622206794410864086561823}%
\StoreBenchExecResult{Abc}{ScPdrScPdrReachSafetyRelCommon}{Memory}{Correct}{False}{Unit}{MB}%
\StoreBenchExecResult{Abc}{ScPdrScPdrReachSafetyRelCommon}{Gate}{All}{}{Sum}{12021759}%
\StoreBenchExecResult{Abc}{ScPdrScPdrReachSafetyRelCommon}{Gate}{All}{}{Min}{0}%
\StoreBenchExecResult{Abc}{ScPdrScPdrReachSafetyRelCommon}{Gate}{All}{}{Max}{250198}%
\StoreBenchExecResult{Abc}{ScPdrScPdrReachSafetyRelCommon}{Gate}{All}{}{Avg}{18580.77125193199381761978362}%
\StoreBenchExecResult{Abc}{ScPdrScPdrReachSafetyRelCommon}{Gate}{All}{}{Median}{6835}%
\StoreBenchExecResult{Abc}{ScPdrScPdrReachSafetyRelCommon}{Gate}{All}{}{Stdev}{35446.66605220592337331040394}%
\StoreBenchExecResult{Abc}{ScPdrScPdrReachSafetyRelCommon}{Gate}{Correct}{}{Sum}{12021759}%
\StoreBenchExecResult{Abc}{ScPdrScPdrReachSafetyRelCommon}{Gate}{Correct}{}{Min}{0}%
\StoreBenchExecResult{Abc}{ScPdrScPdrReachSafetyRelCommon}{Gate}{Correct}{}{Max}{250198}%
\StoreBenchExecResult{Abc}{ScPdrScPdrReachSafetyRelCommon}{Gate}{Correct}{}{Avg}{18580.77125193199381761978362}%
\StoreBenchExecResult{Abc}{ScPdrScPdrReachSafetyRelCommon}{Gate}{Correct}{}{Median}{6835}%
\StoreBenchExecResult{Abc}{ScPdrScPdrReachSafetyRelCommon}{Gate}{Correct}{}{Stdev}{35446.66605220592337331040394}%
\StoreBenchExecResult{Abc}{ScPdrScPdrReachSafetyRelCommon}{Gate}{Correct}{True}{Sum}{2655124}%
\StoreBenchExecResult{Abc}{ScPdrScPdrReachSafetyRelCommon}{Gate}{Correct}{True}{Min}{0}%
\StoreBenchExecResult{Abc}{ScPdrScPdrReachSafetyRelCommon}{Gate}{Correct}{True}{Max}{163965}%
\StoreBenchExecResult{Abc}{ScPdrScPdrReachSafetyRelCommon}{Gate}{Correct}{True}{Avg}{6987.168421052631578947368421}%
\StoreBenchExecResult{Abc}{ScPdrScPdrReachSafetyRelCommon}{Gate}{Correct}{True}{Median}{0}%
\StoreBenchExecResult{Abc}{ScPdrScPdrReachSafetyRelCommon}{Gate}{Correct}{True}{Stdev}{16819.36590810074853612351160}%
\StoreBenchExecResult{Abc}{ScPdrScPdrReachSafetyRelCommon}{Gate}{Correct}{False}{Sum}{9366635}%
\StoreBenchExecResult{Abc}{ScPdrScPdrReachSafetyRelCommon}{Gate}{Correct}{False}{Min}{22}%
\StoreBenchExecResult{Abc}{ScPdrScPdrReachSafetyRelCommon}{Gate}{Correct}{False}{Max}{250198}%
\StoreBenchExecResult{Abc}{ScPdrScPdrReachSafetyRelCommon}{Gate}{Correct}{False}{Avg}{35081.02996254681647940074906}%
\StoreBenchExecResult{Abc}{ScPdrScPdrReachSafetyRelCommon}{Gate}{Correct}{False}{Median}{13076}%
\StoreBenchExecResult{Abc}{ScPdrScPdrReachSafetyRelCommon}{Gate}{Correct}{False}{Stdev}{46674.64624596921530899679088}%
\StoreBenchExecResult{Abc}{ScPdrScPdrReachSafetyRelCommon}{ScTime}{All}{}{Sum}{18864.33}%
\StoreBenchExecResult{Abc}{ScPdrScPdrReachSafetyRelCommon}{ScTime}{All}{}{Min}{0.00}%
\StoreBenchExecResult{Abc}{ScPdrScPdrReachSafetyRelCommon}{ScTime}{All}{}{Max}{700.54}%
\StoreBenchExecResult{Abc}{ScPdrScPdrReachSafetyRelCommon}{ScTime}{All}{}{Avg}{29.15661514683153013910355487}%
\StoreBenchExecResult{Abc}{ScPdrScPdrReachSafetyRelCommon}{ScTime}{All}{}{Median}{0.68}%
\StoreBenchExecResult{Abc}{ScPdrScPdrReachSafetyRelCommon}{ScTime}{All}{}{Stdev}{87.90164347341823912924696076}%
\StoreBenchExecResult{Abc}{ScPdrScPdrReachSafetyRelCommon}{ScTime}{Correct}{}{Sum}{18864.33}%
\StoreBenchExecResult{Abc}{ScPdrScPdrReachSafetyRelCommon}{ScTime}{Correct}{}{Min}{0.00}%
\StoreBenchExecResult{Abc}{ScPdrScPdrReachSafetyRelCommon}{ScTime}{Correct}{}{Max}{700.54}%
\StoreBenchExecResult{Abc}{ScPdrScPdrReachSafetyRelCommon}{ScTime}{Correct}{}{Avg}{29.15661514683153013910355487}%
\StoreBenchExecResult{Abc}{ScPdrScPdrReachSafetyRelCommon}{ScTime}{Correct}{}{Median}{0.68}%
\StoreBenchExecResult{Abc}{ScPdrScPdrReachSafetyRelCommon}{ScTime}{Correct}{}{Stdev}{87.90164347341823912924696076}%
\StoreBenchExecResult{Abc}{ScPdrScPdrReachSafetyRelCommon}{ScTime}{Correct}{True}{Sum}{5001.94}%
\StoreBenchExecResult{Abc}{ScPdrScPdrReachSafetyRelCommon}{ScTime}{Correct}{True}{Min}{0.00}%
\StoreBenchExecResult{Abc}{ScPdrScPdrReachSafetyRelCommon}{ScTime}{Correct}{True}{Max}{355.88}%
\StoreBenchExecResult{Abc}{ScPdrScPdrReachSafetyRelCommon}{ScTime}{Correct}{True}{Avg}{13.163}%
\StoreBenchExecResult{Abc}{ScPdrScPdrReachSafetyRelCommon}{ScTime}{Correct}{True}{Median}{0.51}%
\StoreBenchExecResult{Abc}{ScPdrScPdrReachSafetyRelCommon}{ScTime}{Correct}{True}{Stdev}{50.71624661782454894525145572}%
\StoreBenchExecResult{Abc}{ScPdrScPdrReachSafetyRelCommon}{ScTime}{Correct}{False}{Sum}{13862.39}%
\StoreBenchExecResult{Abc}{ScPdrScPdrReachSafetyRelCommon}{ScTime}{Correct}{False}{Min}{0.00}%
\StoreBenchExecResult{Abc}{ScPdrScPdrReachSafetyRelCommon}{ScTime}{Correct}{False}{Max}{700.54}%
\StoreBenchExecResult{Abc}{ScPdrScPdrReachSafetyRelCommon}{ScTime}{Correct}{False}{Avg}{51.91906367041198501872659176}%
\StoreBenchExecResult{Abc}{ScPdrScPdrReachSafetyRelCommon}{ScTime}{Correct}{False}{Median}{3.23}%
\StoreBenchExecResult{Abc}{ScPdrScPdrReachSafetyRelCommon}{ScTime}{Correct}{False}{Stdev}{119.0823064475202152101805858}%
\StoreBenchExecResult{Abc}{ScPdrScPdrReachSafetyRelCommon}{SolveTime}{All}{}{Sum}{51329.239999999999688325}%
\StoreBenchExecResult{Abc}{ScPdrScPdrReachSafetyRelCommon}{SolveTime}{All}{}{Min}{0.01}%
\StoreBenchExecResult{Abc}{ScPdrScPdrReachSafetyRelCommon}{SolveTime}{All}{}{Max}{885.0}%
\StoreBenchExecResult{Abc}{ScPdrScPdrReachSafetyRelCommon}{SolveTime}{All}{}{Avg}{79.33421947449768112569551777}%
\StoreBenchExecResult{Abc}{ScPdrScPdrReachSafetyRelCommon}{SolveTime}{All}{}{Median}{2.13}%
\StoreBenchExecResult{Abc}{ScPdrScPdrReachSafetyRelCommon}{SolveTime}{All}{}{Stdev}{166.4969073292420553197921729}%
\StoreBenchExecResult{Abc}{ScPdrScPdrReachSafetyRelCommon}{SolveTime}{Correct}{}{Sum}{51329.239999999999688325}%
\StoreBenchExecResult{Abc}{ScPdrScPdrReachSafetyRelCommon}{SolveTime}{Correct}{}{Min}{0.01}%
\StoreBenchExecResult{Abc}{ScPdrScPdrReachSafetyRelCommon}{SolveTime}{Correct}{}{Max}{885.0}%
\StoreBenchExecResult{Abc}{ScPdrScPdrReachSafetyRelCommon}{SolveTime}{Correct}{}{Avg}{79.33421947449768112569551777}%
\StoreBenchExecResult{Abc}{ScPdrScPdrReachSafetyRelCommon}{SolveTime}{Correct}{}{Median}{2.13}%
\StoreBenchExecResult{Abc}{ScPdrScPdrReachSafetyRelCommon}{SolveTime}{Correct}{}{Stdev}{166.4969073292420553197921729}%
\StoreBenchExecResult{Abc}{ScPdrScPdrReachSafetyRelCommon}{SolveTime}{Correct}{True}{Sum}{14356.730000000000065215}%
\StoreBenchExecResult{Abc}{ScPdrScPdrReachSafetyRelCommon}{SolveTime}{Correct}{True}{Min}{0.01}%
\StoreBenchExecResult{Abc}{ScPdrScPdrReachSafetyRelCommon}{SolveTime}{Correct}{True}{Max}{885.0}%
\StoreBenchExecResult{Abc}{ScPdrScPdrReachSafetyRelCommon}{SolveTime}{Correct}{True}{Avg}{37.78086842105263175056578947}%
\StoreBenchExecResult{Abc}{ScPdrScPdrReachSafetyRelCommon}{SolveTime}{Correct}{True}{Median}{0.935}%
\StoreBenchExecResult{Abc}{ScPdrScPdrReachSafetyRelCommon}{SolveTime}{Correct}{True}{Stdev}{110.8212485778386651757933374}%
\StoreBenchExecResult{Abc}{ScPdrScPdrReachSafetyRelCommon}{SolveTime}{Correct}{False}{Sum}{36972.50999999999962311}%
\StoreBenchExecResult{Abc}{ScPdrScPdrReachSafetyRelCommon}{SolveTime}{Correct}{False}{Min}{0.01}%
\StoreBenchExecResult{Abc}{ScPdrScPdrReachSafetyRelCommon}{SolveTime}{Correct}{False}{Max}{817.04}%
\StoreBenchExecResult{Abc}{ScPdrScPdrReachSafetyRelCommon}{SolveTime}{Correct}{False}{Avg}{138.4738202247190997120224719}%
\StoreBenchExecResult{Abc}{ScPdrScPdrReachSafetyRelCommon}{SolveTime}{Correct}{False}{Median}{18.6}%
\StoreBenchExecResult{Abc}{ScPdrScPdrReachSafetyRelCommon}{SolveTime}{Correct}{False}{Stdev}{209.1426007335321840796002145}%
\StoreBenchExecResult{Abc}{ScPdrScPdrReachSafetyRelCommon}{VerifTime}{All}{}{Sum}{32464.91}%
\StoreBenchExecResult{Abc}{ScPdrScPdrReachSafetyRelCommon}{VerifTime}{All}{}{Min}{0.01}%
\StoreBenchExecResult{Abc}{ScPdrScPdrReachSafetyRelCommon}{VerifTime}{All}{}{Max}{862.56}%
\StoreBenchExecResult{Abc}{ScPdrScPdrReachSafetyRelCommon}{VerifTime}{All}{}{Avg}{50.17760432766615146831530139}%
\StoreBenchExecResult{Abc}{ScPdrScPdrReachSafetyRelCommon}{VerifTime}{All}{}{Median}{0.33}%
\StoreBenchExecResult{Abc}{ScPdrScPdrReachSafetyRelCommon}{VerifTime}{All}{}{Stdev}{131.9035389362441147326830869}%
\StoreBenchExecResult{Abc}{ScPdrScPdrReachSafetyRelCommon}{VerifTime}{Correct}{}{Sum}{32464.91}%
\StoreBenchExecResult{Abc}{ScPdrScPdrReachSafetyRelCommon}{VerifTime}{Correct}{}{Min}{0.01}%
\StoreBenchExecResult{Abc}{ScPdrScPdrReachSafetyRelCommon}{VerifTime}{Correct}{}{Max}{862.56}%
\StoreBenchExecResult{Abc}{ScPdrScPdrReachSafetyRelCommon}{VerifTime}{Correct}{}{Avg}{50.17760432766615146831530139}%
\StoreBenchExecResult{Abc}{ScPdrScPdrReachSafetyRelCommon}{VerifTime}{Correct}{}{Median}{0.33}%
\StoreBenchExecResult{Abc}{ScPdrScPdrReachSafetyRelCommon}{VerifTime}{Correct}{}{Stdev}{131.9035389362441147326830869}%
\StoreBenchExecResult{Abc}{ScPdrScPdrReachSafetyRelCommon}{VerifTime}{Correct}{True}{Sum}{9354.79}%
\StoreBenchExecResult{Abc}{ScPdrScPdrReachSafetyRelCommon}{VerifTime}{Correct}{True}{Min}{0.01}%
\StoreBenchExecResult{Abc}{ScPdrScPdrReachSafetyRelCommon}{VerifTime}{Correct}{True}{Max}{862.56}%
\StoreBenchExecResult{Abc}{ScPdrScPdrReachSafetyRelCommon}{VerifTime}{Correct}{True}{Avg}{24.61786842105263157894736842}%
\StoreBenchExecResult{Abc}{ScPdrScPdrReachSafetyRelCommon}{VerifTime}{Correct}{True}{Median}{0.01}%
\StoreBenchExecResult{Abc}{ScPdrScPdrReachSafetyRelCommon}{VerifTime}{Correct}{True}{Stdev}{98.43296934266620978667070382}%
\StoreBenchExecResult{Abc}{ScPdrScPdrReachSafetyRelCommon}{VerifTime}{Correct}{False}{Sum}{23110.12}%
\StoreBenchExecResult{Abc}{ScPdrScPdrReachSafetyRelCommon}{VerifTime}{Correct}{False}{Min}{0.01}%
\StoreBenchExecResult{Abc}{ScPdrScPdrReachSafetyRelCommon}{VerifTime}{Correct}{False}{Max}{763.66}%
\StoreBenchExecResult{Abc}{ScPdrScPdrReachSafetyRelCommon}{VerifTime}{Correct}{False}{Avg}{86.55475655430711610486891386}%
\StoreBenchExecResult{Abc}{ScPdrScPdrReachSafetyRelCommon}{VerifTime}{Correct}{False}{Median}{3.56}%
\StoreBenchExecResult{Abc}{ScPdrScPdrReachSafetyRelCommon}{VerifTime}{Correct}{False}{Stdev}{161.6099307571250606113375433}%
\newcommand{\AbcPdrScPdrReachSafetyRelCommonVerifTimeAllGeoMean}{0.8295486995069015}
\newcommand{\AbcScPdrScPdrReachSafetyRelCommonVerifTimeAllGeoMean}{0.47120866920081395}
\newcommand{\AbcScPdrScPdrReachSafetyRelCommonSolveTimeAllGeoMean}{3.6505065632375513}
\newcommand{\ScPdrReachSafetyRelReducedToZeroCount}{214}
\providecommand\StoreBenchExecResult[7]{\expandafter\newcommand\csname#1#2#3#4#5#6\endcsname{#7}}%
\StoreBenchExecResult{Abc}{PdrScPdrTerminationFuncCommon}{Status}{All}{}{Score}{0}%
\StoreBenchExecResult{Abc}{PdrScPdrTerminationFuncCommon}{Status}{All}{}{Count}{542}%
\StoreBenchExecResult{Abc}{PdrScPdrTerminationFuncCommon}{Status}{Correct}{}{Count}{542}%
\StoreBenchExecResult{Abc}{PdrScPdrTerminationFuncCommon}{Status}{Correct}{True}{Count}{60}%
\StoreBenchExecResult{Abc}{PdrScPdrTerminationFuncCommon}{Status}{Correct}{False}{Count}{482}%
\StoreBenchExecResult{Abc}{PdrScPdrTerminationFuncCommon}{Status}{Wrong}{}{Count}{0}%
\StoreBenchExecResult{Abc}{PdrScPdrTerminationFuncCommon}{Status}{Wrong}{True}{Count}{0}%
\StoreBenchExecResult{Abc}{PdrScPdrTerminationFuncCommon}{Status}{Wrong}{False}{Count}{0}%
\StoreBenchExecResult{Abc}{PdrScPdrTerminationFuncCommon}{Cputime}{All}{}{Sum}{6805.690049}%
\StoreBenchExecResult{Abc}{PdrScPdrTerminationFuncCommon}{Cputime}{All}{}{Min}{0.048442}%
\StoreBenchExecResult{Abc}{PdrScPdrTerminationFuncCommon}{Cputime}{All}{}{Max}{847.098501}%
\StoreBenchExecResult{Abc}{PdrScPdrTerminationFuncCommon}{Cputime}{All}{}{Avg}{12.55662370664206642066420664}%
\StoreBenchExecResult{Abc}{PdrScPdrTerminationFuncCommon}{Cputime}{All}{}{Median}{0.3559295}%
\StoreBenchExecResult{Abc}{PdrScPdrTerminationFuncCommon}{Cputime}{All}{}{Stdev}{73.19808248878570653990947685}%
\StoreBenchExecResult{Abc}{PdrScPdrTerminationFuncCommon}{Cputime}{All}{}{Unit}{s}%
\StoreBenchExecResult{Abc}{PdrScPdrTerminationFuncCommon}{Cputime}{Correct}{}{Sum}{6805.690049}%
\StoreBenchExecResult{Abc}{PdrScPdrTerminationFuncCommon}{Cputime}{Correct}{}{Min}{0.048442}%
\StoreBenchExecResult{Abc}{PdrScPdrTerminationFuncCommon}{Cputime}{Correct}{}{Max}{847.098501}%
\StoreBenchExecResult{Abc}{PdrScPdrTerminationFuncCommon}{Cputime}{Correct}{}{Avg}{12.55662370664206642066420664}%
\StoreBenchExecResult{Abc}{PdrScPdrTerminationFuncCommon}{Cputime}{Correct}{}{Median}{0.3559295}%
\StoreBenchExecResult{Abc}{PdrScPdrTerminationFuncCommon}{Cputime}{Correct}{}{Stdev}{73.19808248878570653990947685}%
\StoreBenchExecResult{Abc}{PdrScPdrTerminationFuncCommon}{Cputime}{Correct}{}{Unit}{s}%
\StoreBenchExecResult{Abc}{PdrScPdrTerminationFuncCommon}{Cputime}{Correct}{True}{Sum}{283.900928}%
\StoreBenchExecResult{Abc}{PdrScPdrTerminationFuncCommon}{Cputime}{Correct}{True}{Min}{0.050897}%
\StoreBenchExecResult{Abc}{PdrScPdrTerminationFuncCommon}{Cputime}{Correct}{True}{Max}{96.419281}%
\StoreBenchExecResult{Abc}{PdrScPdrTerminationFuncCommon}{Cputime}{Correct}{True}{Avg}{4.731682133333333333333333333}%
\StoreBenchExecResult{Abc}{PdrScPdrTerminationFuncCommon}{Cputime}{Correct}{True}{Median}{0.0657725}%
\StoreBenchExecResult{Abc}{PdrScPdrTerminationFuncCommon}{Cputime}{Correct}{True}{Stdev}{17.07672195010170803660042839}%
\StoreBenchExecResult{Abc}{PdrScPdrTerminationFuncCommon}{Cputime}{Correct}{True}{Unit}{s}%
\StoreBenchExecResult{Abc}{PdrScPdrTerminationFuncCommon}{Cputime}{Correct}{False}{Sum}{6521.789121}%
\StoreBenchExecResult{Abc}{PdrScPdrTerminationFuncCommon}{Cputime}{Correct}{False}{Min}{0.048442}%
\StoreBenchExecResult{Abc}{PdrScPdrTerminationFuncCommon}{Cputime}{Correct}{False}{Max}{847.098501}%
\StoreBenchExecResult{Abc}{PdrScPdrTerminationFuncCommon}{Cputime}{Correct}{False}{Avg}{13.53068282365145228215767635}%
\StoreBenchExecResult{Abc}{PdrScPdrTerminationFuncCommon}{Cputime}{Correct}{False}{Median}{0.465653}%
\StoreBenchExecResult{Abc}{PdrScPdrTerminationFuncCommon}{Cputime}{Correct}{False}{Stdev}{77.33080800626976371670624773}%
\StoreBenchExecResult{Abc}{PdrScPdrTerminationFuncCommon}{Cputime}{Correct}{False}{Unit}{s}%
\StoreBenchExecResult{Abc}{PdrScPdrTerminationFuncCommon}{Walltime}{All}{}{Sum}{6808.458385964855609016}%
\StoreBenchExecResult{Abc}{PdrScPdrTerminationFuncCommon}{Walltime}{All}{}{Min}{0.04858383350074291}%
\StoreBenchExecResult{Abc}{PdrScPdrTerminationFuncCommon}{Walltime}{All}{}{Max}{847.3846403351054}%
\StoreBenchExecResult{Abc}{PdrScPdrTerminationFuncCommon}{Walltime}{All}{}{Avg}{12.56173133941855278416236162}%
\StoreBenchExecResult{Abc}{PdrScPdrTerminationFuncCommon}{Walltime}{All}{}{Median}{0.356116090435534715}%
\StoreBenchExecResult{Abc}{PdrScPdrTerminationFuncCommon}{Walltime}{All}{}{Stdev}{73.22461609258266799489522960}%
\StoreBenchExecResult{Abc}{PdrScPdrTerminationFuncCommon}{Walltime}{All}{}{Unit}{s}%
\StoreBenchExecResult{Abc}{PdrScPdrTerminationFuncCommon}{Walltime}{Correct}{}{Sum}{6808.458385964855609016}%
\StoreBenchExecResult{Abc}{PdrScPdrTerminationFuncCommon}{Walltime}{Correct}{}{Min}{0.04858383350074291}%
\StoreBenchExecResult{Abc}{PdrScPdrTerminationFuncCommon}{Walltime}{Correct}{}{Max}{847.3846403351054}%
\StoreBenchExecResult{Abc}{PdrScPdrTerminationFuncCommon}{Walltime}{Correct}{}{Avg}{12.56173133941855278416236162}%
\StoreBenchExecResult{Abc}{PdrScPdrTerminationFuncCommon}{Walltime}{Correct}{}{Median}{0.356116090435534715}%
\StoreBenchExecResult{Abc}{PdrScPdrTerminationFuncCommon}{Walltime}{Correct}{}{Stdev}{73.22461609258266799489522960}%
\StoreBenchExecResult{Abc}{PdrScPdrTerminationFuncCommon}{Walltime}{Correct}{}{Unit}{s}%
\StoreBenchExecResult{Abc}{PdrScPdrTerminationFuncCommon}{Walltime}{Correct}{True}{Sum}{284.044658583588900119}%
\StoreBenchExecResult{Abc}{PdrScPdrTerminationFuncCommon}{Walltime}{Correct}{True}{Min}{0.051056032069027424}%
\StoreBenchExecResult{Abc}{PdrScPdrTerminationFuncCommon}{Walltime}{Correct}{True}{Max}{96.45572651177645}%
\StoreBenchExecResult{Abc}{PdrScPdrTerminationFuncCommon}{Walltime}{Correct}{True}{Avg}{4.734077643059815001983333333}%
\StoreBenchExecResult{Abc}{PdrScPdrTerminationFuncCommon}{Walltime}{Correct}{True}{Median}{0.065953914541751145}%
\StoreBenchExecResult{Abc}{PdrScPdrTerminationFuncCommon}{Walltime}{Correct}{True}{Stdev}{17.08276392323480750546403794}%
\StoreBenchExecResult{Abc}{PdrScPdrTerminationFuncCommon}{Walltime}{Correct}{True}{Unit}{s}%
\StoreBenchExecResult{Abc}{PdrScPdrTerminationFuncCommon}{Walltime}{Correct}{False}{Sum}{6524.413727381266708897}%
\StoreBenchExecResult{Abc}{PdrScPdrTerminationFuncCommon}{Walltime}{Correct}{False}{Min}{0.04858383350074291}%
\StoreBenchExecResult{Abc}{PdrScPdrTerminationFuncCommon}{Walltime}{Correct}{False}{Max}{847.3846403351054}%
\StoreBenchExecResult{Abc}{PdrScPdrTerminationFuncCommon}{Walltime}{Correct}{False}{Avg}{13.53612806510636246659128631}%
\StoreBenchExecResult{Abc}{PdrScPdrTerminationFuncCommon}{Walltime}{Correct}{False}{Median}{0.4658575593493879}%
\StoreBenchExecResult{Abc}{PdrScPdrTerminationFuncCommon}{Walltime}{Correct}{False}{Stdev}{77.35884551780291487459038106}%
\StoreBenchExecResult{Abc}{PdrScPdrTerminationFuncCommon}{Walltime}{Correct}{False}{Unit}{s}%
\StoreBenchExecResult{Abc}{PdrScPdrTerminationFuncCommon}{Memory}{All}{}{Sum}{13493.473280}%
\StoreBenchExecResult{Abc}{PdrScPdrTerminationFuncCommon}{Memory}{All}{}{Min}{9.076736}%
\StoreBenchExecResult{Abc}{PdrScPdrTerminationFuncCommon}{Memory}{All}{}{Max}{215.097344}%
\StoreBenchExecResult{Abc}{PdrScPdrTerminationFuncCommon}{Memory}{All}{}{Avg}{24.89570715867158671586715867}%
\StoreBenchExecResult{Abc}{PdrScPdrTerminationFuncCommon}{Memory}{All}{}{Median}{16.990208}%
\StoreBenchExecResult{Abc}{PdrScPdrTerminationFuncCommon}{Memory}{All}{}{Stdev}{26.24351306395599187447456169}%
\StoreBenchExecResult{Abc}{PdrScPdrTerminationFuncCommon}{Memory}{All}{}{Unit}{MB}%
\StoreBenchExecResult{Abc}{PdrScPdrTerminationFuncCommon}{Memory}{Correct}{}{Sum}{13493.473280}%
\StoreBenchExecResult{Abc}{PdrScPdrTerminationFuncCommon}{Memory}{Correct}{}{Min}{9.076736}%
\StoreBenchExecResult{Abc}{PdrScPdrTerminationFuncCommon}{Memory}{Correct}{}{Max}{215.097344}%
\StoreBenchExecResult{Abc}{PdrScPdrTerminationFuncCommon}{Memory}{Correct}{}{Avg}{24.89570715867158671586715867}%
\StoreBenchExecResult{Abc}{PdrScPdrTerminationFuncCommon}{Memory}{Correct}{}{Median}{16.990208}%
\StoreBenchExecResult{Abc}{PdrScPdrTerminationFuncCommon}{Memory}{Correct}{}{Stdev}{26.24351306395599187447456169}%
\StoreBenchExecResult{Abc}{PdrScPdrTerminationFuncCommon}{Memory}{Correct}{}{Unit}{MB}%
\StoreBenchExecResult{Abc}{PdrScPdrTerminationFuncCommon}{Memory}{Correct}{True}{Sum}{2137.247744}%
\StoreBenchExecResult{Abc}{PdrScPdrTerminationFuncCommon}{Memory}{Correct}{True}{Min}{12.374016}%
\StoreBenchExecResult{Abc}{PdrScPdrTerminationFuncCommon}{Memory}{Correct}{True}{Max}{215.097344}%
\StoreBenchExecResult{Abc}{PdrScPdrTerminationFuncCommon}{Memory}{Correct}{True}{Avg}{35.62079573333333333333333333}%
\StoreBenchExecResult{Abc}{PdrScPdrTerminationFuncCommon}{Memory}{Correct}{True}{Median}{15.663104}%
\StoreBenchExecResult{Abc}{PdrScPdrTerminationFuncCommon}{Memory}{Correct}{True}{Stdev}{48.18640143982783512690331107}%
\StoreBenchExecResult{Abc}{PdrScPdrTerminationFuncCommon}{Memory}{Correct}{True}{Unit}{MB}%
\StoreBenchExecResult{Abc}{PdrScPdrTerminationFuncCommon}{Memory}{Correct}{False}{Sum}{11356.225536}%
\StoreBenchExecResult{Abc}{PdrScPdrTerminationFuncCommon}{Memory}{Correct}{False}{Min}{9.076736}%
\StoreBenchExecResult{Abc}{PdrScPdrTerminationFuncCommon}{Memory}{Correct}{False}{Max}{161.959936}%
\StoreBenchExecResult{Abc}{PdrScPdrTerminationFuncCommon}{Memory}{Correct}{False}{Avg}{23.56063389211618257261410788}%
\StoreBenchExecResult{Abc}{PdrScPdrTerminationFuncCommon}{Memory}{Correct}{False}{Median}{17.143808}%
\StoreBenchExecResult{Abc}{PdrScPdrTerminationFuncCommon}{Memory}{Correct}{False}{Stdev}{21.66372461687284077936998571}%
\StoreBenchExecResult{Abc}{PdrScPdrTerminationFuncCommon}{Memory}{Correct}{False}{Unit}{MB}%
\StoreBenchExecResult{Abc}{PdrScPdrTerminationFuncCommon}{Gate}{All}{}{Sum}{8196885}%
\StoreBenchExecResult{Abc}{PdrScPdrTerminationFuncCommon}{Gate}{All}{}{Min}{31}%
\StoreBenchExecResult{Abc}{PdrScPdrTerminationFuncCommon}{Gate}{All}{}{Max}{158370}%
\StoreBenchExecResult{Abc}{PdrScPdrTerminationFuncCommon}{Gate}{All}{}{Avg}{15123.40405904059040590405904}%
\StoreBenchExecResult{Abc}{PdrScPdrTerminationFuncCommon}{Gate}{All}{}{Median}{8606}%
\StoreBenchExecResult{Abc}{PdrScPdrTerminationFuncCommon}{Gate}{All}{}{Stdev}{28221.65985179842022933519719}%
\StoreBenchExecResult{Abc}{PdrScPdrTerminationFuncCommon}{Gate}{Correct}{}{Sum}{8196885}%
\StoreBenchExecResult{Abc}{PdrScPdrTerminationFuncCommon}{Gate}{Correct}{}{Min}{31}%
\StoreBenchExecResult{Abc}{PdrScPdrTerminationFuncCommon}{Gate}{Correct}{}{Max}{158370}%
\StoreBenchExecResult{Abc}{PdrScPdrTerminationFuncCommon}{Gate}{Correct}{}{Avg}{15123.40405904059040590405904}%
\StoreBenchExecResult{Abc}{PdrScPdrTerminationFuncCommon}{Gate}{Correct}{}{Median}{8606}%
\StoreBenchExecResult{Abc}{PdrScPdrTerminationFuncCommon}{Gate}{Correct}{}{Stdev}{28221.65985179842022933519719}%
\StoreBenchExecResult{Abc}{PdrScPdrTerminationFuncCommon}{Gate}{Correct}{True}{Sum}{326287}%
\StoreBenchExecResult{Abc}{PdrScPdrTerminationFuncCommon}{Gate}{Correct}{True}{Min}{31}%
\StoreBenchExecResult{Abc}{PdrScPdrTerminationFuncCommon}{Gate}{Correct}{True}{Max}{158370}%
\StoreBenchExecResult{Abc}{PdrScPdrTerminationFuncCommon}{Gate}{Correct}{True}{Avg}{5438.116666666666666666666667}%
\StoreBenchExecResult{Abc}{PdrScPdrTerminationFuncCommon}{Gate}{Correct}{True}{Median}{41}%
\StoreBenchExecResult{Abc}{PdrScPdrTerminationFuncCommon}{Gate}{Correct}{True}{Stdev}{25294.65571162656133894412513}%
\StoreBenchExecResult{Abc}{PdrScPdrTerminationFuncCommon}{Gate}{Correct}{False}{Sum}{7870598}%
\StoreBenchExecResult{Abc}{PdrScPdrTerminationFuncCommon}{Gate}{Correct}{False}{Min}{35}%
\StoreBenchExecResult{Abc}{PdrScPdrTerminationFuncCommon}{Gate}{Correct}{False}{Max}{154246}%
\StoreBenchExecResult{Abc}{PdrScPdrTerminationFuncCommon}{Gate}{Correct}{False}{Avg}{16329.04149377593360995850622}%
\StoreBenchExecResult{Abc}{PdrScPdrTerminationFuncCommon}{Gate}{Correct}{False}{Median}{8885}%
\StoreBenchExecResult{Abc}{PdrScPdrTerminationFuncCommon}{Gate}{Correct}{False}{Stdev}{28334.26621027315071549437319}%
\StoreBenchExecResult{Abc}{PdrScPdrTerminationFuncCommon}{VerifTime}{All}{}{Sum}{6737.35}%
\StoreBenchExecResult{Abc}{PdrScPdrTerminationFuncCommon}{VerifTime}{All}{}{Min}{0.01}%
\StoreBenchExecResult{Abc}{PdrScPdrTerminationFuncCommon}{VerifTime}{All}{}{Max}{847.32}%
\StoreBenchExecResult{Abc}{PdrScPdrTerminationFuncCommon}{VerifTime}{All}{}{Avg}{12.43053505535055350553505535}%
\StoreBenchExecResult{Abc}{PdrScPdrTerminationFuncCommon}{VerifTime}{All}{}{Median}{0.27}%
\StoreBenchExecResult{Abc}{PdrScPdrTerminationFuncCommon}{VerifTime}{All}{}{Stdev}{73.20494211011106374268228078}%
\StoreBenchExecResult{Abc}{PdrScPdrTerminationFuncCommon}{VerifTime}{Correct}{}{Sum}{6737.35}%
\StoreBenchExecResult{Abc}{PdrScPdrTerminationFuncCommon}{VerifTime}{Correct}{}{Min}{0.01}%
\StoreBenchExecResult{Abc}{PdrScPdrTerminationFuncCommon}{VerifTime}{Correct}{}{Max}{847.32}%
\StoreBenchExecResult{Abc}{PdrScPdrTerminationFuncCommon}{VerifTime}{Correct}{}{Avg}{12.43053505535055350553505535}%
\StoreBenchExecResult{Abc}{PdrScPdrTerminationFuncCommon}{VerifTime}{Correct}{}{Median}{0.27}%
\StoreBenchExecResult{Abc}{PdrScPdrTerminationFuncCommon}{VerifTime}{Correct}{}{Stdev}{73.20494211011106374268228078}%
\StoreBenchExecResult{Abc}{PdrScPdrTerminationFuncCommon}{VerifTime}{Correct}{True}{Sum}{278.93}%
\StoreBenchExecResult{Abc}{PdrScPdrTerminationFuncCommon}{VerifTime}{Correct}{True}{Min}{0.01}%
\StoreBenchExecResult{Abc}{PdrScPdrTerminationFuncCommon}{VerifTime}{Correct}{True}{Max}{96.4}%
\StoreBenchExecResult{Abc}{PdrScPdrTerminationFuncCommon}{VerifTime}{Correct}{True}{Avg}{4.648833333333333333333333333}%
\StoreBenchExecResult{Abc}{PdrScPdrTerminationFuncCommon}{VerifTime}{Correct}{True}{Median}{0.02}%
\StoreBenchExecResult{Abc}{PdrScPdrTerminationFuncCommon}{VerifTime}{Correct}{True}{Stdev}{17.03882410767310615535216612}%
\StoreBenchExecResult{Abc}{PdrScPdrTerminationFuncCommon}{VerifTime}{Correct}{False}{Sum}{6458.42}%
\StoreBenchExecResult{Abc}{PdrScPdrTerminationFuncCommon}{VerifTime}{Correct}{False}{Min}{0.01}%
\StoreBenchExecResult{Abc}{PdrScPdrTerminationFuncCommon}{VerifTime}{Correct}{False}{Max}{847.32}%
\StoreBenchExecResult{Abc}{PdrScPdrTerminationFuncCommon}{VerifTime}{Correct}{False}{Avg}{13.39921161825726141078838174}%
\StoreBenchExecResult{Abc}{PdrScPdrTerminationFuncCommon}{VerifTime}{Correct}{False}{Median}{0.375}%
\StoreBenchExecResult{Abc}{PdrScPdrTerminationFuncCommon}{VerifTime}{Correct}{False}{Stdev}{77.33976048628253712534410714}%
\StoreBenchExecResult{Abc}{ScPdrScPdrTerminationFuncCommon}{Status}{All}{}{Score}{0}%
\StoreBenchExecResult{Abc}{ScPdrScPdrTerminationFuncCommon}{Status}{All}{}{Count}{542}%
\StoreBenchExecResult{Abc}{ScPdrScPdrTerminationFuncCommon}{Status}{Correct}{}{Count}{542}%
\StoreBenchExecResult{Abc}{ScPdrScPdrTerminationFuncCommon}{Status}{Correct}{True}{Count}{60}%
\StoreBenchExecResult{Abc}{ScPdrScPdrTerminationFuncCommon}{Status}{Correct}{False}{Count}{482}%
\StoreBenchExecResult{Abc}{ScPdrScPdrTerminationFuncCommon}{Status}{Wrong}{}{Count}{0}%
\StoreBenchExecResult{Abc}{ScPdrScPdrTerminationFuncCommon}{Status}{Wrong}{True}{Count}{0}%
\StoreBenchExecResult{Abc}{ScPdrScPdrTerminationFuncCommon}{Status}{Wrong}{False}{Count}{0}%
\StoreBenchExecResult{Abc}{ScPdrScPdrTerminationFuncCommon}{Cputime}{All}{}{Sum}{13712.814326}%
\StoreBenchExecResult{Abc}{ScPdrScPdrTerminationFuncCommon}{Cputime}{All}{}{Min}{0.050012}%
\StoreBenchExecResult{Abc}{ScPdrScPdrTerminationFuncCommon}{Cputime}{All}{}{Max}{894.140251}%
\StoreBenchExecResult{Abc}{ScPdrScPdrTerminationFuncCommon}{Cputime}{All}{}{Avg}{25.30039543542435424354243542}%
\StoreBenchExecResult{Abc}{ScPdrScPdrTerminationFuncCommon}{Cputime}{All}{}{Median}{0.2466425}%
\StoreBenchExecResult{Abc}{ScPdrScPdrTerminationFuncCommon}{Cputime}{All}{}{Stdev}{120.9543063397083616984646391}%
\StoreBenchExecResult{Abc}{ScPdrScPdrTerminationFuncCommon}{Cputime}{All}{}{Unit}{s}%
\StoreBenchExecResult{Abc}{ScPdrScPdrTerminationFuncCommon}{Cputime}{Correct}{}{Sum}{13712.814326}%
\StoreBenchExecResult{Abc}{ScPdrScPdrTerminationFuncCommon}{Cputime}{Correct}{}{Min}{0.050012}%
\StoreBenchExecResult{Abc}{ScPdrScPdrTerminationFuncCommon}{Cputime}{Correct}{}{Max}{894.140251}%
\StoreBenchExecResult{Abc}{ScPdrScPdrTerminationFuncCommon}{Cputime}{Correct}{}{Avg}{25.30039543542435424354243542}%
\StoreBenchExecResult{Abc}{ScPdrScPdrTerminationFuncCommon}{Cputime}{Correct}{}{Median}{0.2466425}%
\StoreBenchExecResult{Abc}{ScPdrScPdrTerminationFuncCommon}{Cputime}{Correct}{}{Stdev}{120.9543063397083616984646391}%
\StoreBenchExecResult{Abc}{ScPdrScPdrTerminationFuncCommon}{Cputime}{Correct}{}{Unit}{s}%
\StoreBenchExecResult{Abc}{ScPdrScPdrTerminationFuncCommon}{Cputime}{Correct}{True}{Sum}{216.590145}%
\StoreBenchExecResult{Abc}{ScPdrScPdrTerminationFuncCommon}{Cputime}{Correct}{True}{Min}{0.051109}%
\StoreBenchExecResult{Abc}{ScPdrScPdrTerminationFuncCommon}{Cputime}{Correct}{True}{Max}{88.224353}%
\StoreBenchExecResult{Abc}{ScPdrScPdrTerminationFuncCommon}{Cputime}{Correct}{True}{Avg}{3.60983575}%
\StoreBenchExecResult{Abc}{ScPdrScPdrTerminationFuncCommon}{Cputime}{Correct}{True}{Median}{0.0658305}%
\StoreBenchExecResult{Abc}{ScPdrScPdrTerminationFuncCommon}{Cputime}{Correct}{True}{Stdev}{13.79433946068727693055461161}%
\StoreBenchExecResult{Abc}{ScPdrScPdrTerminationFuncCommon}{Cputime}{Correct}{True}{Unit}{s}%
\StoreBenchExecResult{Abc}{ScPdrScPdrTerminationFuncCommon}{Cputime}{Correct}{False}{Sum}{13496.224181}%
\StoreBenchExecResult{Abc}{ScPdrScPdrTerminationFuncCommon}{Cputime}{Correct}{False}{Min}{0.050012}%
\StoreBenchExecResult{Abc}{ScPdrScPdrTerminationFuncCommon}{Cputime}{Correct}{False}{Max}{894.140251}%
\StoreBenchExecResult{Abc}{ScPdrScPdrTerminationFuncCommon}{Cputime}{Correct}{False}{Avg}{28.00046510580912863070539419}%
\StoreBenchExecResult{Abc}{ScPdrScPdrTerminationFuncCommon}{Cputime}{Correct}{False}{Median}{0.2779435}%
\StoreBenchExecResult{Abc}{ScPdrScPdrTerminationFuncCommon}{Cputime}{Correct}{False}{Stdev}{127.9122977893423830336379080}%
\StoreBenchExecResult{Abc}{ScPdrScPdrTerminationFuncCommon}{Cputime}{Correct}{False}{Unit}{s}%
\StoreBenchExecResult{Abc}{ScPdrScPdrTerminationFuncCommon}{Walltime}{All}{}{Sum}{13717.263146388344358376}%
\StoreBenchExecResult{Abc}{ScPdrScPdrTerminationFuncCommon}{Walltime}{All}{}{Min}{0.05015954840928316}%
\StoreBenchExecResult{Abc}{ScPdrScPdrTerminationFuncCommon}{Walltime}{All}{}{Max}{894.4672521091998}%
\StoreBenchExecResult{Abc}{ScPdrScPdrTerminationFuncCommon}{Walltime}{All}{}{Avg}{25.30860359112240656526937269}%
\StoreBenchExecResult{Abc}{ScPdrScPdrTerminationFuncCommon}{Walltime}{All}{}{Median}{0.24772225273773075}%
\StoreBenchExecResult{Abc}{ScPdrScPdrTerminationFuncCommon}{Walltime}{All}{}{Stdev}{120.9904669051962562251259794}%
\StoreBenchExecResult{Abc}{ScPdrScPdrTerminationFuncCommon}{Walltime}{All}{}{Unit}{s}%
\StoreBenchExecResult{Abc}{ScPdrScPdrTerminationFuncCommon}{Walltime}{Correct}{}{Sum}{13717.263146388344358376}%
\StoreBenchExecResult{Abc}{ScPdrScPdrTerminationFuncCommon}{Walltime}{Correct}{}{Min}{0.05015954840928316}%
\StoreBenchExecResult{Abc}{ScPdrScPdrTerminationFuncCommon}{Walltime}{Correct}{}{Max}{894.4672521091998}%
\StoreBenchExecResult{Abc}{ScPdrScPdrTerminationFuncCommon}{Walltime}{Correct}{}{Avg}{25.30860359112240656526937269}%
\StoreBenchExecResult{Abc}{ScPdrScPdrTerminationFuncCommon}{Walltime}{Correct}{}{Median}{0.24772225273773075}%
\StoreBenchExecResult{Abc}{ScPdrScPdrTerminationFuncCommon}{Walltime}{Correct}{}{Stdev}{120.9904669051962562251259794}%
\StoreBenchExecResult{Abc}{ScPdrScPdrTerminationFuncCommon}{Walltime}{Correct}{}{Unit}{s}%
\StoreBenchExecResult{Abc}{ScPdrScPdrTerminationFuncCommon}{Walltime}{Correct}{True}{Sum}{216.707466729916632039}%
\StoreBenchExecResult{Abc}{ScPdrScPdrTerminationFuncCommon}{Walltime}{Correct}{True}{Min}{0.051295832730829716}%
\StoreBenchExecResult{Abc}{ScPdrScPdrTerminationFuncCommon}{Walltime}{Correct}{True}{Max}{88.25422892253846}%
\StoreBenchExecResult{Abc}{ScPdrScPdrTerminationFuncCommon}{Walltime}{Correct}{True}{Avg}{3.61179111216527720065}%
\StoreBenchExecResult{Abc}{ScPdrScPdrTerminationFuncCommon}{Walltime}{Correct}{True}{Median}{0.06668338505551219}%
\StoreBenchExecResult{Abc}{ScPdrScPdrTerminationFuncCommon}{Walltime}{Correct}{True}{Stdev}{13.79901139724956407444713081}%
\StoreBenchExecResult{Abc}{ScPdrScPdrTerminationFuncCommon}{Walltime}{Correct}{True}{Unit}{s}%
\StoreBenchExecResult{Abc}{ScPdrScPdrTerminationFuncCommon}{Walltime}{Correct}{False}{Sum}{13500.555679658427726337}%
\StoreBenchExecResult{Abc}{ScPdrScPdrTerminationFuncCommon}{Walltime}{Correct}{False}{Min}{0.05015954840928316}%
\StoreBenchExecResult{Abc}{ScPdrScPdrTerminationFuncCommon}{Walltime}{Correct}{False}{Max}{894.4672521091998}%
\StoreBenchExecResult{Abc}{ScPdrScPdrTerminationFuncCommon}{Walltime}{Correct}{False}{Avg}{28.00945161754860524136307054}%
\StoreBenchExecResult{Abc}{ScPdrScPdrTerminationFuncCommon}{Walltime}{Correct}{False}{Median}{0.278196970466524365}%
\StoreBenchExecResult{Abc}{ScPdrScPdrTerminationFuncCommon}{Walltime}{Correct}{False}{Stdev}{127.9505366661650244836643919}%
\StoreBenchExecResult{Abc}{ScPdrScPdrTerminationFuncCommon}{Walltime}{Correct}{False}{Unit}{s}%
\StoreBenchExecResult{Abc}{ScPdrScPdrTerminationFuncCommon}{Memory}{All}{}{Sum}{10766.733312}%
\StoreBenchExecResult{Abc}{ScPdrScPdrTerminationFuncCommon}{Memory}{All}{}{Min}{8.953856}%
\StoreBenchExecResult{Abc}{ScPdrScPdrTerminationFuncCommon}{Memory}{All}{}{Max}{115.695616}%
\StoreBenchExecResult{Abc}{ScPdrScPdrTerminationFuncCommon}{Memory}{All}{}{Avg}{19.86482160885608856088560886}%
\StoreBenchExecResult{Abc}{ScPdrScPdrTerminationFuncCommon}{Memory}{All}{}{Median}{12.994560}%
\StoreBenchExecResult{Abc}{ScPdrScPdrTerminationFuncCommon}{Memory}{All}{}{Stdev}{20.97193740501673262343549027}%
\StoreBenchExecResult{Abc}{ScPdrScPdrTerminationFuncCommon}{Memory}{All}{}{Unit}{MB}%
\StoreBenchExecResult{Abc}{ScPdrScPdrTerminationFuncCommon}{Memory}{Correct}{}{Sum}{10766.733312}%
\StoreBenchExecResult{Abc}{ScPdrScPdrTerminationFuncCommon}{Memory}{Correct}{}{Min}{8.953856}%
\StoreBenchExecResult{Abc}{ScPdrScPdrTerminationFuncCommon}{Memory}{Correct}{}{Max}{115.695616}%
\StoreBenchExecResult{Abc}{ScPdrScPdrTerminationFuncCommon}{Memory}{Correct}{}{Avg}{19.86482160885608856088560886}%
\StoreBenchExecResult{Abc}{ScPdrScPdrTerminationFuncCommon}{Memory}{Correct}{}{Median}{12.994560}%
\StoreBenchExecResult{Abc}{ScPdrScPdrTerminationFuncCommon}{Memory}{Correct}{}{Stdev}{20.97193740501673262343549027}%
\StoreBenchExecResult{Abc}{ScPdrScPdrTerminationFuncCommon}{Memory}{Correct}{}{Unit}{MB}%
\StoreBenchExecResult{Abc}{ScPdrScPdrTerminationFuncCommon}{Memory}{Correct}{True}{Sum}{1400.475648}%
\StoreBenchExecResult{Abc}{ScPdrScPdrTerminationFuncCommon}{Memory}{Correct}{True}{Min}{10.690560}%
\StoreBenchExecResult{Abc}{ScPdrScPdrTerminationFuncCommon}{Memory}{Correct}{True}{Max}{115.695616}%
\StoreBenchExecResult{Abc}{ScPdrScPdrTerminationFuncCommon}{Memory}{Correct}{True}{Avg}{23.3412608}%
\StoreBenchExecResult{Abc}{ScPdrScPdrTerminationFuncCommon}{Memory}{Correct}{True}{Median}{11.919360}%
\StoreBenchExecResult{Abc}{ScPdrScPdrTerminationFuncCommon}{Memory}{Correct}{True}{Stdev}{23.78220231596003445845601874}%
\StoreBenchExecResult{Abc}{ScPdrScPdrTerminationFuncCommon}{Memory}{Correct}{True}{Unit}{MB}%
\StoreBenchExecResult{Abc}{ScPdrScPdrTerminationFuncCommon}{Memory}{Correct}{False}{Sum}{9366.257664}%
\StoreBenchExecResult{Abc}{ScPdrScPdrTerminationFuncCommon}{Memory}{Correct}{False}{Min}{8.953856}%
\StoreBenchExecResult{Abc}{ScPdrScPdrTerminationFuncCommon}{Memory}{Correct}{False}{Max}{115.200000}%
\StoreBenchExecResult{Abc}{ScPdrScPdrTerminationFuncCommon}{Memory}{Correct}{False}{Avg}{19.43206984232365145228215768}%
\StoreBenchExecResult{Abc}{ScPdrScPdrTerminationFuncCommon}{Memory}{Correct}{False}{Median}{13.037568}%
\StoreBenchExecResult{Abc}{ScPdrScPdrTerminationFuncCommon}{Memory}{Correct}{False}{Stdev}{20.55417985673020881222502223}%
\StoreBenchExecResult{Abc}{ScPdrScPdrTerminationFuncCommon}{Memory}{Correct}{False}{Unit}{MB}%
\StoreBenchExecResult{Abc}{ScPdrScPdrTerminationFuncCommon}{Gate}{All}{}{Sum}{1486416}%
\StoreBenchExecResult{Abc}{ScPdrScPdrTerminationFuncCommon}{Gate}{All}{}{Min}{0}%
\StoreBenchExecResult{Abc}{ScPdrScPdrTerminationFuncCommon}{Gate}{All}{}{Max}{89562}%
\StoreBenchExecResult{Abc}{ScPdrScPdrTerminationFuncCommon}{Gate}{All}{}{Avg}{2742.464944649446494464944649}%
\StoreBenchExecResult{Abc}{ScPdrScPdrTerminationFuncCommon}{Gate}{All}{}{Median}{636.5}%
\StoreBenchExecResult{Abc}{ScPdrScPdrTerminationFuncCommon}{Gate}{All}{}{Stdev}{8667.848893718849210539144235}%
\StoreBenchExecResult{Abc}{ScPdrScPdrTerminationFuncCommon}{Gate}{Correct}{}{Sum}{1486416}%
\StoreBenchExecResult{Abc}{ScPdrScPdrTerminationFuncCommon}{Gate}{Correct}{}{Min}{0}%
\StoreBenchExecResult{Abc}{ScPdrScPdrTerminationFuncCommon}{Gate}{Correct}{}{Max}{89562}%
\StoreBenchExecResult{Abc}{ScPdrScPdrTerminationFuncCommon}{Gate}{Correct}{}{Avg}{2742.464944649446494464944649}%
\StoreBenchExecResult{Abc}{ScPdrScPdrTerminationFuncCommon}{Gate}{Correct}{}{Median}{636.5}%
\StoreBenchExecResult{Abc}{ScPdrScPdrTerminationFuncCommon}{Gate}{Correct}{}{Stdev}{8667.848893718849210539144235}%
\StoreBenchExecResult{Abc}{ScPdrScPdrTerminationFuncCommon}{Gate}{Correct}{True}{Sum}{13974}%
\StoreBenchExecResult{Abc}{ScPdrScPdrTerminationFuncCommon}{Gate}{Correct}{True}{Min}{0}%
\StoreBenchExecResult{Abc}{ScPdrScPdrTerminationFuncCommon}{Gate}{Correct}{True}{Max}{2177}%
\StoreBenchExecResult{Abc}{ScPdrScPdrTerminationFuncCommon}{Gate}{Correct}{True}{Avg}{232.9}%
\StoreBenchExecResult{Abc}{ScPdrScPdrTerminationFuncCommon}{Gate}{Correct}{True}{Median}{0}%
\StoreBenchExecResult{Abc}{ScPdrScPdrTerminationFuncCommon}{Gate}{Correct}{True}{Stdev}{425.1779509805276814549831028}%
\StoreBenchExecResult{Abc}{ScPdrScPdrTerminationFuncCommon}{Gate}{Correct}{False}{Sum}{1472442}%
\StoreBenchExecResult{Abc}{ScPdrScPdrTerminationFuncCommon}{Gate}{Correct}{False}{Min}{4}%
\StoreBenchExecResult{Abc}{ScPdrScPdrTerminationFuncCommon}{Gate}{Correct}{False}{Max}{89562}%
\StoreBenchExecResult{Abc}{ScPdrScPdrTerminationFuncCommon}{Gate}{Correct}{False}{Avg}{3054.858921161825726141078838}%
\StoreBenchExecResult{Abc}{ScPdrScPdrTerminationFuncCommon}{Gate}{Correct}{False}{Median}{786}%
\StoreBenchExecResult{Abc}{ScPdrScPdrTerminationFuncCommon}{Gate}{Correct}{False}{Stdev}{9142.210896923741457548227125}%
\StoreBenchExecResult{Abc}{ScPdrScPdrTerminationFuncCommon}{ScTime}{All}{}{Sum}{13425.97}%
\StoreBenchExecResult{Abc}{ScPdrScPdrTerminationFuncCommon}{ScTime}{All}{}{Min}{0.00}%
\StoreBenchExecResult{Abc}{ScPdrScPdrTerminationFuncCommon}{ScTime}{All}{}{Max}{892.68}%
\StoreBenchExecResult{Abc}{ScPdrScPdrTerminationFuncCommon}{ScTime}{All}{}{Avg}{24.77116236162361623616236162}%
\StoreBenchExecResult{Abc}{ScPdrScPdrTerminationFuncCommon}{ScTime}{All}{}{Median}{0.15}%
\StoreBenchExecResult{Abc}{ScPdrScPdrTerminationFuncCommon}{ScTime}{All}{}{Stdev}{120.4436709591258268224805883}%
\StoreBenchExecResult{Abc}{ScPdrScPdrTerminationFuncCommon}{ScTime}{Correct}{}{Sum}{13425.97}%
\StoreBenchExecResult{Abc}{ScPdrScPdrTerminationFuncCommon}{ScTime}{Correct}{}{Min}{0.00}%
\StoreBenchExecResult{Abc}{ScPdrScPdrTerminationFuncCommon}{ScTime}{Correct}{}{Max}{892.68}%
\StoreBenchExecResult{Abc}{ScPdrScPdrTerminationFuncCommon}{ScTime}{Correct}{}{Avg}{24.77116236162361623616236162}%
\StoreBenchExecResult{Abc}{ScPdrScPdrTerminationFuncCommon}{ScTime}{Correct}{}{Median}{0.15}%
\StoreBenchExecResult{Abc}{ScPdrScPdrTerminationFuncCommon}{ScTime}{Correct}{}{Stdev}{120.4436709591258268224805883}%
\StoreBenchExecResult{Abc}{ScPdrScPdrTerminationFuncCommon}{ScTime}{Correct}{True}{Sum}{134.05}%
\StoreBenchExecResult{Abc}{ScPdrScPdrTerminationFuncCommon}{ScTime}{Correct}{True}{Min}{0.00}%
\StoreBenchExecResult{Abc}{ScPdrScPdrTerminationFuncCommon}{ScTime}{Correct}{True}{Max}{86.81}%
\StoreBenchExecResult{Abc}{ScPdrScPdrTerminationFuncCommon}{ScTime}{Correct}{True}{Avg}{2.234166666666666666666666667}%
\StoreBenchExecResult{Abc}{ScPdrScPdrTerminationFuncCommon}{ScTime}{Correct}{True}{Median}{0.00}%
\StoreBenchExecResult{Abc}{ScPdrScPdrTerminationFuncCommon}{ScTime}{Correct}{True}{Stdev}{12.47902070031494454083202614}%
\StoreBenchExecResult{Abc}{ScPdrScPdrTerminationFuncCommon}{ScTime}{Correct}{False}{Sum}{13291.92}%
\StoreBenchExecResult{Abc}{ScPdrScPdrTerminationFuncCommon}{ScTime}{Correct}{False}{Min}{0.00}%
\StoreBenchExecResult{Abc}{ScPdrScPdrTerminationFuncCommon}{ScTime}{Correct}{False}{Max}{892.68}%
\StoreBenchExecResult{Abc}{ScPdrScPdrTerminationFuncCommon}{ScTime}{Correct}{False}{Avg}{27.57659751037344398340248963}%
\StoreBenchExecResult{Abc}{ScPdrScPdrTerminationFuncCommon}{ScTime}{Correct}{False}{Median}{0.18}%
\StoreBenchExecResult{Abc}{ScPdrScPdrTerminationFuncCommon}{ScTime}{Correct}{False}{Stdev}{127.3656424536644128588018660}%
\StoreBenchExecResult{Abc}{ScPdrScPdrTerminationFuncCommon}{SolveTime}{All}{}{Sum}{13645.060000000000407500}%
\StoreBenchExecResult{Abc}{ScPdrScPdrTerminationFuncCommon}{SolveTime}{All}{}{Min}{0.01}%
\StoreBenchExecResult{Abc}{ScPdrScPdrTerminationFuncCommon}{SolveTime}{All}{}{Max}{893.41}%
\StoreBenchExecResult{Abc}{ScPdrScPdrTerminationFuncCommon}{SolveTime}{All}{}{Avg}{25.17538745387453949723247232}%
\StoreBenchExecResult{Abc}{ScPdrScPdrTerminationFuncCommon}{SolveTime}{All}{}{Median}{0.17}%
\StoreBenchExecResult{Abc}{ScPdrScPdrTerminationFuncCommon}{SolveTime}{All}{}{Stdev}{120.7897981817010887496859452}%
\StoreBenchExecResult{Abc}{ScPdrScPdrTerminationFuncCommon}{SolveTime}{Correct}{}{Sum}{13645.060000000000407500}%
\StoreBenchExecResult{Abc}{ScPdrScPdrTerminationFuncCommon}{SolveTime}{Correct}{}{Min}{0.01}%
\StoreBenchExecResult{Abc}{ScPdrScPdrTerminationFuncCommon}{SolveTime}{Correct}{}{Max}{893.41}%
\StoreBenchExecResult{Abc}{ScPdrScPdrTerminationFuncCommon}{SolveTime}{Correct}{}{Avg}{25.17538745387453949723247232}%
\StoreBenchExecResult{Abc}{ScPdrScPdrTerminationFuncCommon}{SolveTime}{Correct}{}{Median}{0.17}%
\StoreBenchExecResult{Abc}{ScPdrScPdrTerminationFuncCommon}{SolveTime}{Correct}{}{Stdev}{120.7897981817010887496859452}%
\StoreBenchExecResult{Abc}{ScPdrScPdrTerminationFuncCommon}{SolveTime}{Correct}{True}{Sum}{211.460000000000010125}%
\StoreBenchExecResult{Abc}{ScPdrScPdrTerminationFuncCommon}{SolveTime}{Correct}{True}{Min}{0.01}%
\StoreBenchExecResult{Abc}{ScPdrScPdrTerminationFuncCommon}{SolveTime}{Correct}{True}{Max}{86.82000000000001}%
\StoreBenchExecResult{Abc}{ScPdrScPdrTerminationFuncCommon}{SolveTime}{Correct}{True}{Avg}{3.524333333333333502083333333}%
\StoreBenchExecResult{Abc}{ScPdrScPdrTerminationFuncCommon}{SolveTime}{Correct}{True}{Median}{0.02}%
\StoreBenchExecResult{Abc}{ScPdrScPdrTerminationFuncCommon}{SolveTime}{Correct}{True}{Stdev}{13.60060640396433144677016503}%
\StoreBenchExecResult{Abc}{ScPdrScPdrTerminationFuncCommon}{SolveTime}{Correct}{False}{Sum}{13433.600000000000397375}%
\StoreBenchExecResult{Abc}{ScPdrScPdrTerminationFuncCommon}{SolveTime}{Correct}{False}{Min}{0.01}%
\StoreBenchExecResult{Abc}{ScPdrScPdrTerminationFuncCommon}{SolveTime}{Correct}{False}{Max}{893.41}%
\StoreBenchExecResult{Abc}{ScPdrScPdrTerminationFuncCommon}{SolveTime}{Correct}{False}{Avg}{27.87053941908713775389004149}%
\StoreBenchExecResult{Abc}{ScPdrScPdrTerminationFuncCommon}{SolveTime}{Correct}{False}{Median}{0.2}%
\StoreBenchExecResult{Abc}{ScPdrScPdrTerminationFuncCommon}{SolveTime}{Correct}{False}{Stdev}{127.7408975219331549257025995}%
\StoreBenchExecResult{Abc}{ScPdrScPdrTerminationFuncCommon}{VerifTime}{All}{}{Sum}{219.09}%
\StoreBenchExecResult{Abc}{ScPdrScPdrTerminationFuncCommon}{VerifTime}{All}{}{Min}{0.01}%
\StoreBenchExecResult{Abc}{ScPdrScPdrTerminationFuncCommon}{VerifTime}{All}{}{Max}{35.82}%
\StoreBenchExecResult{Abc}{ScPdrScPdrTerminationFuncCommon}{VerifTime}{All}{}{Avg}{0.4042250922509225092250922509}%
\StoreBenchExecResult{Abc}{ScPdrScPdrTerminationFuncCommon}{VerifTime}{All}{}{Median}{0.02}%
\StoreBenchExecResult{Abc}{ScPdrScPdrTerminationFuncCommon}{VerifTime}{All}{}{Stdev}{2.643752652869568429638427437}%
\StoreBenchExecResult{Abc}{ScPdrScPdrTerminationFuncCommon}{VerifTime}{Correct}{}{Sum}{219.09}%
\StoreBenchExecResult{Abc}{ScPdrScPdrTerminationFuncCommon}{VerifTime}{Correct}{}{Min}{0.01}%
\StoreBenchExecResult{Abc}{ScPdrScPdrTerminationFuncCommon}{VerifTime}{Correct}{}{Max}{35.82}%
\StoreBenchExecResult{Abc}{ScPdrScPdrTerminationFuncCommon}{VerifTime}{Correct}{}{Avg}{0.4042250922509225092250922509}%
\StoreBenchExecResult{Abc}{ScPdrScPdrTerminationFuncCommon}{VerifTime}{Correct}{}{Median}{0.02}%
\StoreBenchExecResult{Abc}{ScPdrScPdrTerminationFuncCommon}{VerifTime}{Correct}{}{Stdev}{2.643752652869568429638427437}%
\StoreBenchExecResult{Abc}{ScPdrScPdrTerminationFuncCommon}{VerifTime}{Correct}{True}{Sum}{77.41}%
\StoreBenchExecResult{Abc}{ScPdrScPdrTerminationFuncCommon}{VerifTime}{Correct}{True}{Min}{0.01}%
\StoreBenchExecResult{Abc}{ScPdrScPdrTerminationFuncCommon}{VerifTime}{Correct}{True}{Max}{35.82}%
\StoreBenchExecResult{Abc}{ScPdrScPdrTerminationFuncCommon}{VerifTime}{Correct}{True}{Avg}{1.290166666666666666666666667}%
\StoreBenchExecResult{Abc}{ScPdrScPdrTerminationFuncCommon}{VerifTime}{Correct}{True}{Median}{0.015}%
\StoreBenchExecResult{Abc}{ScPdrScPdrTerminationFuncCommon}{VerifTime}{Correct}{True}{Stdev}{5.904989272264222797684173497}%
\StoreBenchExecResult{Abc}{ScPdrScPdrTerminationFuncCommon}{VerifTime}{Correct}{False}{Sum}{141.68}%
\StoreBenchExecResult{Abc}{ScPdrScPdrTerminationFuncCommon}{VerifTime}{Correct}{False}{Min}{0.01}%
\StoreBenchExecResult{Abc}{ScPdrScPdrTerminationFuncCommon}{VerifTime}{Correct}{False}{Max}{33.23}%
\StoreBenchExecResult{Abc}{ScPdrScPdrTerminationFuncCommon}{VerifTime}{Correct}{False}{Avg}{0.2939419087136929460580912863}%
\StoreBenchExecResult{Abc}{ScPdrScPdrTerminationFuncCommon}{VerifTime}{Correct}{False}{Median}{0.02}%
\StoreBenchExecResult{Abc}{ScPdrScPdrTerminationFuncCommon}{VerifTime}{Correct}{False}{Stdev}{1.846371517560257318762087115}%
\newcommand{\AbcPdrScPdrTerminationFuncCommonVerifTimeAllGeoMean}{0.36886109415133195}
\newcommand{\AbcScPdrScPdrTerminationFuncCommonVerifTimeAllGeoMean}{0.027855840475691826}
\newcommand{\AbcScPdrScPdrTerminationFuncCommonSolveTimeAllGeoMean}{0.3104732952552977}
\newcommand{\ScPdrTerminationFuncReducedToZeroCount}{37}
\providecommand\StoreBenchExecResult[7]{\expandafter\newcommand\csname#1#2#3#4#5#6\endcsname{#7}}%
\StoreBenchExecResult{Abc}{PdrScPdrTerminationRelCommon}{Status}{All}{}{Score}{0}%
\StoreBenchExecResult{Abc}{PdrScPdrTerminationRelCommon}{Status}{All}{}{Count}{542}%
\StoreBenchExecResult{Abc}{PdrScPdrTerminationRelCommon}{Status}{Correct}{}{Count}{542}%
\StoreBenchExecResult{Abc}{PdrScPdrTerminationRelCommon}{Status}{Correct}{True}{Count}{60}%
\StoreBenchExecResult{Abc}{PdrScPdrTerminationRelCommon}{Status}{Correct}{False}{Count}{482}%
\StoreBenchExecResult{Abc}{PdrScPdrTerminationRelCommon}{Status}{Wrong}{}{Count}{0}%
\StoreBenchExecResult{Abc}{PdrScPdrTerminationRelCommon}{Status}{Wrong}{True}{Count}{0}%
\StoreBenchExecResult{Abc}{PdrScPdrTerminationRelCommon}{Status}{Wrong}{False}{Count}{0}%
\StoreBenchExecResult{Abc}{PdrScPdrTerminationRelCommon}{Cputime}{All}{}{Sum}{13123.563978}%
\StoreBenchExecResult{Abc}{PdrScPdrTerminationRelCommon}{Cputime}{All}{}{Min}{0.052709}%
\StoreBenchExecResult{Abc}{PdrScPdrTerminationRelCommon}{Cputime}{All}{}{Max}{853.852519}%
\StoreBenchExecResult{Abc}{PdrScPdrTerminationRelCommon}{Cputime}{All}{}{Avg}{24.21321767158671586715867159}%
\StoreBenchExecResult{Abc}{PdrScPdrTerminationRelCommon}{Cputime}{All}{}{Median}{1.8822315}%
\StoreBenchExecResult{Abc}{PdrScPdrTerminationRelCommon}{Cputime}{All}{}{Stdev}{83.02828475086081493692684871}%
\StoreBenchExecResult{Abc}{PdrScPdrTerminationRelCommon}{Cputime}{All}{}{Unit}{s}%
\StoreBenchExecResult{Abc}{PdrScPdrTerminationRelCommon}{Cputime}{Correct}{}{Sum}{13123.563978}%
\StoreBenchExecResult{Abc}{PdrScPdrTerminationRelCommon}{Cputime}{Correct}{}{Min}{0.052709}%
\StoreBenchExecResult{Abc}{PdrScPdrTerminationRelCommon}{Cputime}{Correct}{}{Max}{853.852519}%
\StoreBenchExecResult{Abc}{PdrScPdrTerminationRelCommon}{Cputime}{Correct}{}{Avg}{24.21321767158671586715867159}%
\StoreBenchExecResult{Abc}{PdrScPdrTerminationRelCommon}{Cputime}{Correct}{}{Median}{1.8822315}%
\StoreBenchExecResult{Abc}{PdrScPdrTerminationRelCommon}{Cputime}{Correct}{}{Stdev}{83.02828475086081493692684871}%
\StoreBenchExecResult{Abc}{PdrScPdrTerminationRelCommon}{Cputime}{Correct}{}{Unit}{s}%
\StoreBenchExecResult{Abc}{PdrScPdrTerminationRelCommon}{Cputime}{Correct}{True}{Sum}{1924.866148}%
\StoreBenchExecResult{Abc}{PdrScPdrTerminationRelCommon}{Cputime}{Correct}{True}{Min}{0.052905}%
\StoreBenchExecResult{Abc}{PdrScPdrTerminationRelCommon}{Cputime}{Correct}{True}{Max}{853.852519}%
\StoreBenchExecResult{Abc}{PdrScPdrTerminationRelCommon}{Cputime}{Correct}{True}{Avg}{32.08110246666666666666666667}%
\StoreBenchExecResult{Abc}{PdrScPdrTerminationRelCommon}{Cputime}{Correct}{True}{Median}{0.068145}%
\StoreBenchExecResult{Abc}{PdrScPdrTerminationRelCommon}{Cputime}{Correct}{True}{Stdev}{134.7502600449686509455839862}%
\StoreBenchExecResult{Abc}{PdrScPdrTerminationRelCommon}{Cputime}{Correct}{True}{Unit}{s}%
\StoreBenchExecResult{Abc}{PdrScPdrTerminationRelCommon}{Cputime}{Correct}{False}{Sum}{11198.697830}%
\StoreBenchExecResult{Abc}{PdrScPdrTerminationRelCommon}{Cputime}{Correct}{False}{Min}{0.052709}%
\StoreBenchExecResult{Abc}{PdrScPdrTerminationRelCommon}{Cputime}{Correct}{False}{Max}{709.295928}%
\StoreBenchExecResult{Abc}{PdrScPdrTerminationRelCommon}{Cputime}{Correct}{False}{Avg}{23.23381292531120331950207469}%
\StoreBenchExecResult{Abc}{PdrScPdrTerminationRelCommon}{Cputime}{Correct}{False}{Median}{2.134092}%
\StoreBenchExecResult{Abc}{PdrScPdrTerminationRelCommon}{Cputime}{Correct}{False}{Stdev}{74.04648011779321981615363243}%
\StoreBenchExecResult{Abc}{PdrScPdrTerminationRelCommon}{Cputime}{Correct}{False}{Unit}{s}%
\StoreBenchExecResult{Abc}{PdrScPdrTerminationRelCommon}{Walltime}{All}{}{Sum}{13127.924281796440359369}%
\StoreBenchExecResult{Abc}{PdrScPdrTerminationRelCommon}{Walltime}{All}{}{Min}{0.052910907194018364}%
\StoreBenchExecResult{Abc}{PdrScPdrTerminationRelCommon}{Walltime}{All}{}{Max}{854.1706596203148}%
\StoreBenchExecResult{Abc}{PdrScPdrTerminationRelCommon}{Walltime}{All}{}{Avg}{24.22126251253955785861439114}%
\StoreBenchExecResult{Abc}{PdrScPdrTerminationRelCommon}{Walltime}{All}{}{Median}{1.88286541821435095}%
\StoreBenchExecResult{Abc}{PdrScPdrTerminationRelCommon}{Walltime}{All}{}{Stdev}{83.05623177234609441829427322}%
\StoreBenchExecResult{Abc}{PdrScPdrTerminationRelCommon}{Walltime}{All}{}{Unit}{s}%
\StoreBenchExecResult{Abc}{PdrScPdrTerminationRelCommon}{Walltime}{Correct}{}{Sum}{13127.924281796440359369}%
\StoreBenchExecResult{Abc}{PdrScPdrTerminationRelCommon}{Walltime}{Correct}{}{Min}{0.052910907194018364}%
\StoreBenchExecResult{Abc}{PdrScPdrTerminationRelCommon}{Walltime}{Correct}{}{Max}{854.1706596203148}%
\StoreBenchExecResult{Abc}{PdrScPdrTerminationRelCommon}{Walltime}{Correct}{}{Avg}{24.22126251253955785861439114}%
\StoreBenchExecResult{Abc}{PdrScPdrTerminationRelCommon}{Walltime}{Correct}{}{Median}{1.88286541821435095}%
\StoreBenchExecResult{Abc}{PdrScPdrTerminationRelCommon}{Walltime}{Correct}{}{Stdev}{83.05623177234609441829427322}%
\StoreBenchExecResult{Abc}{PdrScPdrTerminationRelCommon}{Walltime}{Correct}{}{Unit}{s}%
\StoreBenchExecResult{Abc}{PdrScPdrTerminationRelCommon}{Walltime}{Correct}{True}{Sum}{1925.544310632161724981}%
\StoreBenchExecResult{Abc}{PdrScPdrTerminationRelCommon}{Walltime}{Correct}{True}{Min}{0.0530650382861495}%
\StoreBenchExecResult{Abc}{PdrScPdrTerminationRelCommon}{Walltime}{Correct}{True}{Max}{854.1706596203148}%
\StoreBenchExecResult{Abc}{PdrScPdrTerminationRelCommon}{Walltime}{Correct}{True}{Avg}{32.09240517720269541635}%
\StoreBenchExecResult{Abc}{PdrScPdrTerminationRelCommon}{Walltime}{Correct}{True}{Median}{0.068295935634523625}%
\StoreBenchExecResult{Abc}{PdrScPdrTerminationRelCommon}{Walltime}{Correct}{True}{Stdev}{134.8000651780752696768801350}%
\StoreBenchExecResult{Abc}{PdrScPdrTerminationRelCommon}{Walltime}{Correct}{True}{Unit}{s}%
\StoreBenchExecResult{Abc}{PdrScPdrTerminationRelCommon}{Walltime}{Correct}{False}{Sum}{11202.379971164278634388}%
\StoreBenchExecResult{Abc}{PdrScPdrTerminationRelCommon}{Walltime}{Correct}{False}{Min}{0.052910907194018364}%
\StoreBenchExecResult{Abc}{PdrScPdrTerminationRelCommon}{Walltime}{Correct}{False}{Max}{709.5522584225982}%
\StoreBenchExecResult{Abc}{PdrScPdrTerminationRelCommon}{Walltime}{Correct}{False}{Avg}{23.24145222233252828711203320}%
\StoreBenchExecResult{Abc}{PdrScPdrTerminationRelCommon}{Walltime}{Correct}{False}{Median}{2.13485610112547865}%
\StoreBenchExecResult{Abc}{PdrScPdrTerminationRelCommon}{Walltime}{Correct}{False}{Stdev}{74.07038703762874369615210979}%
\StoreBenchExecResult{Abc}{PdrScPdrTerminationRelCommon}{Walltime}{Correct}{False}{Unit}{s}%
\StoreBenchExecResult{Abc}{PdrScPdrTerminationRelCommon}{Memory}{All}{}{Sum}{21002.530816}%
\StoreBenchExecResult{Abc}{PdrScPdrTerminationRelCommon}{Memory}{All}{}{Min}{9.789440}%
\StoreBenchExecResult{Abc}{PdrScPdrTerminationRelCommon}{Memory}{All}{}{Max}{885.702656}%
\StoreBenchExecResult{Abc}{PdrScPdrTerminationRelCommon}{Memory}{All}{}{Avg}{38.75005685608856088560885609}%
\StoreBenchExecResult{Abc}{PdrScPdrTerminationRelCommon}{Memory}{All}{}{Median}{23.453696}%
\StoreBenchExecResult{Abc}{PdrScPdrTerminationRelCommon}{Memory}{All}{}{Stdev}{56.66160448391649746101287648}%
\StoreBenchExecResult{Abc}{PdrScPdrTerminationRelCommon}{Memory}{All}{}{Unit}{MB}%
\StoreBenchExecResult{Abc}{PdrScPdrTerminationRelCommon}{Memory}{Correct}{}{Sum}{21002.530816}%
\StoreBenchExecResult{Abc}{PdrScPdrTerminationRelCommon}{Memory}{Correct}{}{Min}{9.789440}%
\StoreBenchExecResult{Abc}{PdrScPdrTerminationRelCommon}{Memory}{Correct}{}{Max}{885.702656}%
\StoreBenchExecResult{Abc}{PdrScPdrTerminationRelCommon}{Memory}{Correct}{}{Avg}{38.75005685608856088560885609}%
\StoreBenchExecResult{Abc}{PdrScPdrTerminationRelCommon}{Memory}{Correct}{}{Median}{23.453696}%
\StoreBenchExecResult{Abc}{PdrScPdrTerminationRelCommon}{Memory}{Correct}{}{Stdev}{56.66160448391649746101287648}%
\StoreBenchExecResult{Abc}{PdrScPdrTerminationRelCommon}{Memory}{Correct}{}{Unit}{MB}%
\StoreBenchExecResult{Abc}{PdrScPdrTerminationRelCommon}{Memory}{Correct}{True}{Sum}{3537.575936}%
\StoreBenchExecResult{Abc}{PdrScPdrTerminationRelCommon}{Memory}{Correct}{True}{Min}{14.180352}%
\StoreBenchExecResult{Abc}{PdrScPdrTerminationRelCommon}{Memory}{Correct}{True}{Max}{885.702656}%
\StoreBenchExecResult{Abc}{PdrScPdrTerminationRelCommon}{Memory}{Correct}{True}{Avg}{58.95959893333333333333333333}%
\StoreBenchExecResult{Abc}{PdrScPdrTerminationRelCommon}{Memory}{Correct}{True}{Median}{15.548416}%
\StoreBenchExecResult{Abc}{PdrScPdrTerminationRelCommon}{Memory}{Correct}{True}{Stdev}{128.3245202001183162637515225}%
\StoreBenchExecResult{Abc}{PdrScPdrTerminationRelCommon}{Memory}{Correct}{True}{Unit}{MB}%
\StoreBenchExecResult{Abc}{PdrScPdrTerminationRelCommon}{Memory}{Correct}{False}{Sum}{17464.954880}%
\StoreBenchExecResult{Abc}{PdrScPdrTerminationRelCommon}{Memory}{Correct}{False}{Min}{9.789440}%
\StoreBenchExecResult{Abc}{PdrScPdrTerminationRelCommon}{Memory}{Correct}{False}{Max}{260.206592}%
\StoreBenchExecResult{Abc}{PdrScPdrTerminationRelCommon}{Memory}{Correct}{False}{Avg}{36.23434622406639004149377593}%
\StoreBenchExecResult{Abc}{PdrScPdrTerminationRelCommon}{Memory}{Correct}{False}{Median}{24.000512}%
\StoreBenchExecResult{Abc}{PdrScPdrTerminationRelCommon}{Memory}{Correct}{False}{Stdev}{38.77063944979779406162677431}%
\StoreBenchExecResult{Abc}{PdrScPdrTerminationRelCommon}{Memory}{Correct}{False}{Unit}{MB}%
\StoreBenchExecResult{Abc}{PdrScPdrTerminationRelCommon}{Gate}{All}{}{Sum}{9596262}%
\StoreBenchExecResult{Abc}{PdrScPdrTerminationRelCommon}{Gate}{All}{}{Min}{58}%
\StoreBenchExecResult{Abc}{PdrScPdrTerminationRelCommon}{Gate}{All}{}{Max}{163862}%
\StoreBenchExecResult{Abc}{PdrScPdrTerminationRelCommon}{Gate}{All}{}{Avg}{17705.28044280442804428044280}%
\StoreBenchExecResult{Abc}{PdrScPdrTerminationRelCommon}{Gate}{All}{}{Median}{10849}%
\StoreBenchExecResult{Abc}{PdrScPdrTerminationRelCommon}{Gate}{All}{}{Stdev}{28586.24528240297033128225619}%
\StoreBenchExecResult{Abc}{PdrScPdrTerminationRelCommon}{Gate}{Correct}{}{Sum}{9596262}%
\StoreBenchExecResult{Abc}{PdrScPdrTerminationRelCommon}{Gate}{Correct}{}{Min}{58}%
\StoreBenchExecResult{Abc}{PdrScPdrTerminationRelCommon}{Gate}{Correct}{}{Max}{163862}%
\StoreBenchExecResult{Abc}{PdrScPdrTerminationRelCommon}{Gate}{Correct}{}{Avg}{17705.28044280442804428044280}%
\StoreBenchExecResult{Abc}{PdrScPdrTerminationRelCommon}{Gate}{Correct}{}{Median}{10849}%
\StoreBenchExecResult{Abc}{PdrScPdrTerminationRelCommon}{Gate}{Correct}{}{Stdev}{28586.24528240297033128225619}%
\StoreBenchExecResult{Abc}{PdrScPdrTerminationRelCommon}{Gate}{Correct}{True}{Sum}{375981}%
\StoreBenchExecResult{Abc}{PdrScPdrTerminationRelCommon}{Gate}{Correct}{True}{Min}{58}%
\StoreBenchExecResult{Abc}{PdrScPdrTerminationRelCommon}{Gate}{Correct}{True}{Max}{163862}%
\StoreBenchExecResult{Abc}{PdrScPdrTerminationRelCommon}{Gate}{Correct}{True}{Avg}{6266.35}%
\StoreBenchExecResult{Abc}{PdrScPdrTerminationRelCommon}{Gate}{Correct}{True}{Median}{849}%
\StoreBenchExecResult{Abc}{PdrScPdrTerminationRelCommon}{Gate}{Correct}{True}{Stdev}{25783.58836018563184973151543}%
\StoreBenchExecResult{Abc}{PdrScPdrTerminationRelCommon}{Gate}{Correct}{False}{Sum}{9220281}%
\StoreBenchExecResult{Abc}{PdrScPdrTerminationRelCommon}{Gate}{Correct}{False}{Min}{194}%
\StoreBenchExecResult{Abc}{PdrScPdrTerminationRelCommon}{Gate}{Correct}{False}{Max}{160247}%
\StoreBenchExecResult{Abc}{PdrScPdrTerminationRelCommon}{Gate}{Correct}{False}{Avg}{19129.21369294605809128630705}%
\StoreBenchExecResult{Abc}{PdrScPdrTerminationRelCommon}{Gate}{Correct}{False}{Median}{11183}%
\StoreBenchExecResult{Abc}{PdrScPdrTerminationRelCommon}{Gate}{Correct}{False}{Stdev}{28597.65765939079350748543799}%
\StoreBenchExecResult{Abc}{PdrScPdrTerminationRelCommon}{VerifTime}{All}{}{Sum}{13076.44}%
\StoreBenchExecResult{Abc}{PdrScPdrTerminationRelCommon}{VerifTime}{All}{}{Min}{0.01}%
\StoreBenchExecResult{Abc}{PdrScPdrTerminationRelCommon}{VerifTime}{All}{}{Max}{854.08}%
\StoreBenchExecResult{Abc}{PdrScPdrTerminationRelCommon}{VerifTime}{All}{}{Avg}{24.12627306273062730627306273}%
\StoreBenchExecResult{Abc}{PdrScPdrTerminationRelCommon}{VerifTime}{All}{}{Median}{1.805}%
\StoreBenchExecResult{Abc}{PdrScPdrTerminationRelCommon}{VerifTime}{All}{}{Stdev}{83.02173825355202657601142794}%
\StoreBenchExecResult{Abc}{PdrScPdrTerminationRelCommon}{VerifTime}{Correct}{}{Sum}{13076.44}%
\StoreBenchExecResult{Abc}{PdrScPdrTerminationRelCommon}{VerifTime}{Correct}{}{Min}{0.01}%
\StoreBenchExecResult{Abc}{PdrScPdrTerminationRelCommon}{VerifTime}{Correct}{}{Max}{854.08}%
\StoreBenchExecResult{Abc}{PdrScPdrTerminationRelCommon}{VerifTime}{Correct}{}{Avg}{24.12627306273062730627306273}%
\StoreBenchExecResult{Abc}{PdrScPdrTerminationRelCommon}{VerifTime}{Correct}{}{Median}{1.805}%
\StoreBenchExecResult{Abc}{PdrScPdrTerminationRelCommon}{VerifTime}{Correct}{}{Stdev}{83.02173825355202657601142794}%
\StoreBenchExecResult{Abc}{PdrScPdrTerminationRelCommon}{VerifTime}{Correct}{True}{Sum}{1922.13}%
\StoreBenchExecResult{Abc}{PdrScPdrTerminationRelCommon}{VerifTime}{Correct}{True}{Min}{0.02}%
\StoreBenchExecResult{Abc}{PdrScPdrTerminationRelCommon}{VerifTime}{Correct}{True}{Max}{854.08}%
\StoreBenchExecResult{Abc}{PdrScPdrTerminationRelCommon}{VerifTime}{Correct}{True}{Avg}{32.0355}%
\StoreBenchExecResult{Abc}{PdrScPdrTerminationRelCommon}{VerifTime}{Correct}{True}{Median}{0.02}%
\StoreBenchExecResult{Abc}{PdrScPdrTerminationRelCommon}{VerifTime}{Correct}{True}{Stdev}{134.7721918080655094010830933}%
\StoreBenchExecResult{Abc}{PdrScPdrTerminationRelCommon}{VerifTime}{Correct}{False}{Sum}{11154.31}%
\StoreBenchExecResult{Abc}{PdrScPdrTerminationRelCommon}{VerifTime}{Correct}{False}{Min}{0.01}%
\StoreBenchExecResult{Abc}{PdrScPdrTerminationRelCommon}{VerifTime}{Correct}{False}{Max}{709.49}%
\StoreBenchExecResult{Abc}{PdrScPdrTerminationRelCommon}{VerifTime}{Correct}{False}{Avg}{23.14172199170124481327800830}%
\StoreBenchExecResult{Abc}{PdrScPdrTerminationRelCommon}{VerifTime}{Correct}{False}{Median}{2.05}%
\StoreBenchExecResult{Abc}{PdrScPdrTerminationRelCommon}{VerifTime}{Correct}{False}{Stdev}{74.03263963740619917062288124}%
\StoreBenchExecResult{Abc}{ScPdrScPdrTerminationRelCommon}{Status}{All}{}{Score}{0}%
\StoreBenchExecResult{Abc}{ScPdrScPdrTerminationRelCommon}{Status}{All}{}{Count}{542}%
\StoreBenchExecResult{Abc}{ScPdrScPdrTerminationRelCommon}{Status}{Correct}{}{Count}{542}%
\StoreBenchExecResult{Abc}{ScPdrScPdrTerminationRelCommon}{Status}{Correct}{True}{Count}{60}%
\StoreBenchExecResult{Abc}{ScPdrScPdrTerminationRelCommon}{Status}{Correct}{False}{Count}{482}%
\StoreBenchExecResult{Abc}{ScPdrScPdrTerminationRelCommon}{Status}{Wrong}{}{Count}{0}%
\StoreBenchExecResult{Abc}{ScPdrScPdrTerminationRelCommon}{Status}{Wrong}{True}{Count}{0}%
\StoreBenchExecResult{Abc}{ScPdrScPdrTerminationRelCommon}{Status}{Wrong}{False}{Count}{0}%
\StoreBenchExecResult{Abc}{ScPdrScPdrTerminationRelCommon}{Cputime}{All}{}{Sum}{29783.074795}%
\StoreBenchExecResult{Abc}{ScPdrScPdrTerminationRelCommon}{Cputime}{All}{}{Min}{0.051302}%
\StoreBenchExecResult{Abc}{ScPdrScPdrTerminationRelCommon}{Cputime}{All}{}{Max}{897.503116}%
\StoreBenchExecResult{Abc}{ScPdrScPdrTerminationRelCommon}{Cputime}{All}{}{Avg}{54.9503225}%
\StoreBenchExecResult{Abc}{ScPdrScPdrTerminationRelCommon}{Cputime}{All}{}{Median}{3.5387445}%
\StoreBenchExecResult{Abc}{ScPdrScPdrTerminationRelCommon}{Cputime}{All}{}{Stdev}{165.6464110210388517247152018}%
\StoreBenchExecResult{Abc}{ScPdrScPdrTerminationRelCommon}{Cputime}{All}{}{Unit}{s}%
\StoreBenchExecResult{Abc}{ScPdrScPdrTerminationRelCommon}{Cputime}{Correct}{}{Sum}{29783.074795}%
\StoreBenchExecResult{Abc}{ScPdrScPdrTerminationRelCommon}{Cputime}{Correct}{}{Min}{0.051302}%
\StoreBenchExecResult{Abc}{ScPdrScPdrTerminationRelCommon}{Cputime}{Correct}{}{Max}{897.503116}%
\StoreBenchExecResult{Abc}{ScPdrScPdrTerminationRelCommon}{Cputime}{Correct}{}{Avg}{54.9503225}%
\StoreBenchExecResult{Abc}{ScPdrScPdrTerminationRelCommon}{Cputime}{Correct}{}{Median}{3.5387445}%
\StoreBenchExecResult{Abc}{ScPdrScPdrTerminationRelCommon}{Cputime}{Correct}{}{Stdev}{165.6464110210388517247152018}%
\StoreBenchExecResult{Abc}{ScPdrScPdrTerminationRelCommon}{Cputime}{Correct}{}{Unit}{s}%
\StoreBenchExecResult{Abc}{ScPdrScPdrTerminationRelCommon}{Cputime}{Correct}{True}{Sum}{1433.396930}%
\StoreBenchExecResult{Abc}{ScPdrScPdrTerminationRelCommon}{Cputime}{Correct}{True}{Min}{0.051302}%
\StoreBenchExecResult{Abc}{ScPdrScPdrTerminationRelCommon}{Cputime}{Correct}{True}{Max}{629.402551}%
\StoreBenchExecResult{Abc}{ScPdrScPdrTerminationRelCommon}{Cputime}{Correct}{True}{Avg}{23.88994883333333333333333333}%
\StoreBenchExecResult{Abc}{ScPdrScPdrTerminationRelCommon}{Cputime}{Correct}{True}{Median}{0.0659285}%
\StoreBenchExecResult{Abc}{ScPdrScPdrTerminationRelCommon}{Cputime}{Correct}{True}{Stdev}{92.63497686759486009996940906}%
\StoreBenchExecResult{Abc}{ScPdrScPdrTerminationRelCommon}{Cputime}{Correct}{True}{Unit}{s}%
\StoreBenchExecResult{Abc}{ScPdrScPdrTerminationRelCommon}{Cputime}{Correct}{False}{Sum}{28349.677865}%
\StoreBenchExecResult{Abc}{ScPdrScPdrTerminationRelCommon}{Cputime}{Correct}{False}{Min}{0.055736}%
\StoreBenchExecResult{Abc}{ScPdrScPdrTerminationRelCommon}{Cputime}{Correct}{False}{Max}{897.503116}%
\StoreBenchExecResult{Abc}{ScPdrScPdrTerminationRelCommon}{Cputime}{Correct}{False}{Avg}{58.81675905601659751037344398}%
\StoreBenchExecResult{Abc}{ScPdrScPdrTerminationRelCommon}{Cputime}{Correct}{False}{Median}{6.039934}%
\StoreBenchExecResult{Abc}{ScPdrScPdrTerminationRelCommon}{Cputime}{Correct}{False}{Stdev}{172.1949393805342231578462097}%
\StoreBenchExecResult{Abc}{ScPdrScPdrTerminationRelCommon}{Cputime}{Correct}{False}{Unit}{s}%
\StoreBenchExecResult{Abc}{ScPdrScPdrTerminationRelCommon}{Walltime}{All}{}{Sum}{29792.413600717671084270}%
\StoreBenchExecResult{Abc}{ScPdrScPdrTerminationRelCommon}{Walltime}{All}{}{Min}{0.05148467980325222}%
\StoreBenchExecResult{Abc}{ScPdrScPdrTerminationRelCommon}{Walltime}{All}{}{Max}{897.8748909356073}%
\StoreBenchExecResult{Abc}{ScPdrScPdrTerminationRelCommon}{Walltime}{All}{}{Avg}{54.96755276885179166839483395}%
\StoreBenchExecResult{Abc}{ScPdrScPdrTerminationRelCommon}{Walltime}{All}{}{Median}{3.540411017369479}%
\StoreBenchExecResult{Abc}{ScPdrScPdrTerminationRelCommon}{Walltime}{All}{}{Stdev}{165.6981803299836898398755418}%
\StoreBenchExecResult{Abc}{ScPdrScPdrTerminationRelCommon}{Walltime}{All}{}{Unit}{s}%
\StoreBenchExecResult{Abc}{ScPdrScPdrTerminationRelCommon}{Walltime}{Correct}{}{Sum}{29792.413600717671084270}%
\StoreBenchExecResult{Abc}{ScPdrScPdrTerminationRelCommon}{Walltime}{Correct}{}{Min}{0.05148467980325222}%
\StoreBenchExecResult{Abc}{ScPdrScPdrTerminationRelCommon}{Walltime}{Correct}{}{Max}{897.8748909356073}%
\StoreBenchExecResult{Abc}{ScPdrScPdrTerminationRelCommon}{Walltime}{Correct}{}{Avg}{54.96755276885179166839483395}%
\StoreBenchExecResult{Abc}{ScPdrScPdrTerminationRelCommon}{Walltime}{Correct}{}{Median}{3.540411017369479}%
\StoreBenchExecResult{Abc}{ScPdrScPdrTerminationRelCommon}{Walltime}{Correct}{}{Stdev}{165.6981803299836898398755418}%
\StoreBenchExecResult{Abc}{ScPdrScPdrTerminationRelCommon}{Walltime}{Correct}{}{Unit}{s}%
\StoreBenchExecResult{Abc}{ScPdrScPdrTerminationRelCommon}{Walltime}{Correct}{True}{Sum}{1433.903442555107211330}%
\StoreBenchExecResult{Abc}{ScPdrScPdrTerminationRelCommon}{Walltime}{Correct}{True}{Min}{0.05148467980325222}%
\StoreBenchExecResult{Abc}{ScPdrScPdrTerminationRelCommon}{Walltime}{Correct}{True}{Max}{629.6233206978068}%
\StoreBenchExecResult{Abc}{ScPdrScPdrTerminationRelCommon}{Walltime}{Correct}{True}{Avg}{23.8983907092517868555}%
\StoreBenchExecResult{Abc}{ScPdrScPdrTerminationRelCommon}{Walltime}{Correct}{True}{Median}{0.06613929430022836}%
\StoreBenchExecResult{Abc}{ScPdrScPdrTerminationRelCommon}{Walltime}{Correct}{True}{Stdev}{92.66803948799911860260476441}%
\StoreBenchExecResult{Abc}{ScPdrScPdrTerminationRelCommon}{Walltime}{Correct}{True}{Unit}{s}%
\StoreBenchExecResult{Abc}{ScPdrScPdrTerminationRelCommon}{Walltime}{Correct}{False}{Sum}{28358.51015816256387294}%
\StoreBenchExecResult{Abc}{ScPdrScPdrTerminationRelCommon}{Walltime}{Correct}{False}{Min}{0.05587849859148264}%
\StoreBenchExecResult{Abc}{ScPdrScPdrTerminationRelCommon}{Walltime}{Correct}{False}{Max}{897.8748909356073}%
\StoreBenchExecResult{Abc}{ScPdrScPdrTerminationRelCommon}{Walltime}{Correct}{False}{Avg}{58.83508331568996654136929461}%
\StoreBenchExecResult{Abc}{ScPdrScPdrTerminationRelCommon}{Walltime}{Correct}{False}{Median}{6.04167615389451375}%
\StoreBenchExecResult{Abc}{ScPdrScPdrTerminationRelCommon}{Walltime}{Correct}{False}{Stdev}{172.2485031548042325851387456}%
\StoreBenchExecResult{Abc}{ScPdrScPdrTerminationRelCommon}{Walltime}{Correct}{False}{Unit}{s}%
\StoreBenchExecResult{Abc}{ScPdrScPdrTerminationRelCommon}{Memory}{All}{}{Sum}{19925.471232}%
\StoreBenchExecResult{Abc}{ScPdrScPdrTerminationRelCommon}{Memory}{All}{}{Min}{10.031104}%
\StoreBenchExecResult{Abc}{ScPdrScPdrTerminationRelCommon}{Memory}{All}{}{Max}{445.108224}%
\StoreBenchExecResult{Abc}{ScPdrScPdrTerminationRelCommon}{Memory}{All}{}{Avg}{36.76286205166051660516605166}%
\StoreBenchExecResult{Abc}{ScPdrScPdrTerminationRelCommon}{Memory}{All}{}{Median}{24.295424}%
\StoreBenchExecResult{Abc}{ScPdrScPdrTerminationRelCommon}{Memory}{All}{}{Stdev}{42.23153758780649230195872686}%
\StoreBenchExecResult{Abc}{ScPdrScPdrTerminationRelCommon}{Memory}{All}{}{Unit}{MB}%
\StoreBenchExecResult{Abc}{ScPdrScPdrTerminationRelCommon}{Memory}{Correct}{}{Sum}{19925.471232}%
\StoreBenchExecResult{Abc}{ScPdrScPdrTerminationRelCommon}{Memory}{Correct}{}{Min}{10.031104}%
\StoreBenchExecResult{Abc}{ScPdrScPdrTerminationRelCommon}{Memory}{Correct}{}{Max}{445.108224}%
\StoreBenchExecResult{Abc}{ScPdrScPdrTerminationRelCommon}{Memory}{Correct}{}{Avg}{36.76286205166051660516605166}%
\StoreBenchExecResult{Abc}{ScPdrScPdrTerminationRelCommon}{Memory}{Correct}{}{Median}{24.295424}%
\StoreBenchExecResult{Abc}{ScPdrScPdrTerminationRelCommon}{Memory}{Correct}{}{Stdev}{42.23153758780649230195872686}%
\StoreBenchExecResult{Abc}{ScPdrScPdrTerminationRelCommon}{Memory}{Correct}{}{Unit}{MB}%
\StoreBenchExecResult{Abc}{ScPdrScPdrTerminationRelCommon}{Memory}{Correct}{True}{Sum}{2581.803008}%
\StoreBenchExecResult{Abc}{ScPdrScPdrTerminationRelCommon}{Memory}{Correct}{True}{Min}{10.723328}%
\StoreBenchExecResult{Abc}{ScPdrScPdrTerminationRelCommon}{Memory}{Correct}{True}{Max}{308.940800}%
\StoreBenchExecResult{Abc}{ScPdrScPdrTerminationRelCommon}{Memory}{Correct}{True}{Avg}{43.03005013333333333333333333}%
\StoreBenchExecResult{Abc}{ScPdrScPdrTerminationRelCommon}{Memory}{Correct}{True}{Median}{11.847680}%
\StoreBenchExecResult{Abc}{ScPdrScPdrTerminationRelCommon}{Memory}{Correct}{True}{Stdev}{62.52602252501157336741922203}%
\StoreBenchExecResult{Abc}{ScPdrScPdrTerminationRelCommon}{Memory}{Correct}{True}{Unit}{MB}%
\StoreBenchExecResult{Abc}{ScPdrScPdrTerminationRelCommon}{Memory}{Correct}{False}{Sum}{17343.668224}%
\StoreBenchExecResult{Abc}{ScPdrScPdrTerminationRelCommon}{Memory}{Correct}{False}{Min}{10.031104}%
\StoreBenchExecResult{Abc}{ScPdrScPdrTerminationRelCommon}{Memory}{Correct}{False}{Max}{445.108224}%
\StoreBenchExecResult{Abc}{ScPdrScPdrTerminationRelCommon}{Memory}{Correct}{False}{Avg}{35.98271415767634854771784232}%
\StoreBenchExecResult{Abc}{ScPdrScPdrTerminationRelCommon}{Memory}{Correct}{False}{Median}{24.807424}%
\StoreBenchExecResult{Abc}{ScPdrScPdrTerminationRelCommon}{Memory}{Correct}{False}{Stdev}{38.90189467701519967968408464}%
\StoreBenchExecResult{Abc}{ScPdrScPdrTerminationRelCommon}{Memory}{Correct}{False}{Unit}{MB}%
\StoreBenchExecResult{Abc}{ScPdrScPdrTerminationRelCommon}{Gate}{All}{}{Sum}{6116211}%
\StoreBenchExecResult{Abc}{ScPdrScPdrTerminationRelCommon}{Gate}{All}{}{Min}{0}%
\StoreBenchExecResult{Abc}{ScPdrScPdrTerminationRelCommon}{Gate}{All}{}{Max}{155351}%
\StoreBenchExecResult{Abc}{ScPdrScPdrTerminationRelCommon}{Gate}{All}{}{Avg}{11284.52214022140221402214022}%
\StoreBenchExecResult{Abc}{ScPdrScPdrTerminationRelCommon}{Gate}{All}{}{Median}{9521}%
\StoreBenchExecResult{Abc}{ScPdrScPdrTerminationRelCommon}{Gate}{All}{}{Stdev}{16649.53568223485491247394813}%
\StoreBenchExecResult{Abc}{ScPdrScPdrTerminationRelCommon}{Gate}{Correct}{}{Sum}{6116211}%
\StoreBenchExecResult{Abc}{ScPdrScPdrTerminationRelCommon}{Gate}{Correct}{}{Min}{0}%
\StoreBenchExecResult{Abc}{ScPdrScPdrTerminationRelCommon}{Gate}{Correct}{}{Max}{155351}%
\StoreBenchExecResult{Abc}{ScPdrScPdrTerminationRelCommon}{Gate}{Correct}{}{Avg}{11284.52214022140221402214022}%
\StoreBenchExecResult{Abc}{ScPdrScPdrTerminationRelCommon}{Gate}{Correct}{}{Median}{9521}%
\StoreBenchExecResult{Abc}{ScPdrScPdrTerminationRelCommon}{Gate}{Correct}{}{Stdev}{16649.53568223485491247394813}%
\StoreBenchExecResult{Abc}{ScPdrScPdrTerminationRelCommon}{Gate}{Correct}{True}{Sum}{341545}%
\StoreBenchExecResult{Abc}{ScPdrScPdrTerminationRelCommon}{Gate}{Correct}{True}{Min}{0}%
\StoreBenchExecResult{Abc}{ScPdrScPdrTerminationRelCommon}{Gate}{Correct}{True}{Max}{155351}%
\StoreBenchExecResult{Abc}{ScPdrScPdrTerminationRelCommon}{Gate}{Correct}{True}{Avg}{5692.416666666666666666666667}%
\StoreBenchExecResult{Abc}{ScPdrScPdrTerminationRelCommon}{Gate}{Correct}{True}{Median}{0}%
\StoreBenchExecResult{Abc}{ScPdrScPdrTerminationRelCommon}{Gate}{Correct}{True}{Stdev}{24502.53443849681077766832524}%
\StoreBenchExecResult{Abc}{ScPdrScPdrTerminationRelCommon}{Gate}{Correct}{False}{Sum}{5774666}%
\StoreBenchExecResult{Abc}{ScPdrScPdrTerminationRelCommon}{Gate}{Correct}{False}{Min}{178}%
\StoreBenchExecResult{Abc}{ScPdrScPdrTerminationRelCommon}{Gate}{Correct}{False}{Max}{154186}%
\StoreBenchExecResult{Abc}{ScPdrScPdrTerminationRelCommon}{Gate}{Correct}{False}{Avg}{11980.63485477178423236514523}%
\StoreBenchExecResult{Abc}{ScPdrScPdrTerminationRelCommon}{Gate}{Correct}{False}{Median}{9634.5}%
\StoreBenchExecResult{Abc}{ScPdrScPdrTerminationRelCommon}{Gate}{Correct}{False}{Stdev}{15251.27710017003910446367599}%
\StoreBenchExecResult{Abc}{ScPdrScPdrTerminationRelCommon}{ScTime}{All}{}{Sum}{15942.75}%
\StoreBenchExecResult{Abc}{ScPdrScPdrTerminationRelCommon}{ScTime}{All}{}{Min}{0.00}%
\StoreBenchExecResult{Abc}{ScPdrScPdrTerminationRelCommon}{ScTime}{All}{}{Max}{884.96}%
\StoreBenchExecResult{Abc}{ScPdrScPdrTerminationRelCommon}{ScTime}{All}{}{Avg}{29.41466789667896678966789668}%
\StoreBenchExecResult{Abc}{ScPdrScPdrTerminationRelCommon}{ScTime}{All}{}{Median}{0.76}%
\StoreBenchExecResult{Abc}{ScPdrScPdrTerminationRelCommon}{ScTime}{All}{}{Stdev}{140.3485082173588579235508270}%
\StoreBenchExecResult{Abc}{ScPdrScPdrTerminationRelCommon}{ScTime}{Correct}{}{Sum}{15942.75}%
\StoreBenchExecResult{Abc}{ScPdrScPdrTerminationRelCommon}{ScTime}{Correct}{}{Min}{0.00}%
\StoreBenchExecResult{Abc}{ScPdrScPdrTerminationRelCommon}{ScTime}{Correct}{}{Max}{884.96}%
\StoreBenchExecResult{Abc}{ScPdrScPdrTerminationRelCommon}{ScTime}{Correct}{}{Avg}{29.41466789667896678966789668}%
\StoreBenchExecResult{Abc}{ScPdrScPdrTerminationRelCommon}{ScTime}{Correct}{}{Median}{0.76}%
\StoreBenchExecResult{Abc}{ScPdrScPdrTerminationRelCommon}{ScTime}{Correct}{}{Stdev}{140.3485082173588579235508270}%
\StoreBenchExecResult{Abc}{ScPdrScPdrTerminationRelCommon}{ScTime}{Correct}{True}{Sum}{403.61}%
\StoreBenchExecResult{Abc}{ScPdrScPdrTerminationRelCommon}{ScTime}{Correct}{True}{Min}{0.00}%
\StoreBenchExecResult{Abc}{ScPdrScPdrTerminationRelCommon}{ScTime}{Correct}{True}{Max}{277.22}%
\StoreBenchExecResult{Abc}{ScPdrScPdrTerminationRelCommon}{ScTime}{Correct}{True}{Avg}{6.726833333333333333333333333}%
\StoreBenchExecResult{Abc}{ScPdrScPdrTerminationRelCommon}{ScTime}{Correct}{True}{Median}{0.01}%
\StoreBenchExecResult{Abc}{ScPdrScPdrTerminationRelCommon}{ScTime}{Correct}{True}{Stdev}{38.65851115824589308652445957}%
\StoreBenchExecResult{Abc}{ScPdrScPdrTerminationRelCommon}{ScTime}{Correct}{False}{Sum}{15539.14}%
\StoreBenchExecResult{Abc}{ScPdrScPdrTerminationRelCommon}{ScTime}{Correct}{False}{Min}{0.00}%
\StoreBenchExecResult{Abc}{ScPdrScPdrTerminationRelCommon}{ScTime}{Correct}{False}{Max}{884.96}%
\StoreBenchExecResult{Abc}{ScPdrScPdrTerminationRelCommon}{ScTime}{Correct}{False}{Avg}{32.23887966804979253112033195}%
\StoreBenchExecResult{Abc}{ScPdrScPdrTerminationRelCommon}{ScTime}{Correct}{False}{Median}{0.81}%
\StoreBenchExecResult{Abc}{ScPdrScPdrTerminationRelCommon}{ScTime}{Correct}{False}{Stdev}{147.9581486118746013227330333}%
\StoreBenchExecResult{Abc}{ScPdrScPdrTerminationRelCommon}{SolveTime}{All}{}{Sum}{29740.44000000000027654}%
\StoreBenchExecResult{Abc}{ScPdrScPdrTerminationRelCommon}{SolveTime}{All}{}{Min}{0.01}%
\StoreBenchExecResult{Abc}{ScPdrScPdrTerminationRelCommon}{SolveTime}{All}{}{Max}{897.0500000000001}%
\StoreBenchExecResult{Abc}{ScPdrScPdrTerminationRelCommon}{SolveTime}{All}{}{Avg}{54.87166051660516656188191882}%
\StoreBenchExecResult{Abc}{ScPdrScPdrTerminationRelCommon}{SolveTime}{All}{}{Median}{3.47}%
\StoreBenchExecResult{Abc}{ScPdrScPdrTerminationRelCommon}{SolveTime}{All}{}{Stdev}{165.5884346612185598519535322}%
\StoreBenchExecResult{Abc}{ScPdrScPdrTerminationRelCommon}{SolveTime}{Correct}{}{Sum}{29740.44000000000027654}%
\StoreBenchExecResult{Abc}{ScPdrScPdrTerminationRelCommon}{SolveTime}{Correct}{}{Min}{0.01}%
\StoreBenchExecResult{Abc}{ScPdrScPdrTerminationRelCommon}{SolveTime}{Correct}{}{Max}{897.0500000000001}%
\StoreBenchExecResult{Abc}{ScPdrScPdrTerminationRelCommon}{SolveTime}{Correct}{}{Avg}{54.87166051660516656188191882}%
\StoreBenchExecResult{Abc}{ScPdrScPdrTerminationRelCommon}{SolveTime}{Correct}{}{Median}{3.47}%
\StoreBenchExecResult{Abc}{ScPdrScPdrTerminationRelCommon}{SolveTime}{Correct}{}{Stdev}{165.5884346612185598519535322}%
\StoreBenchExecResult{Abc}{ScPdrScPdrTerminationRelCommon}{SolveTime}{Correct}{True}{Sum}{1430.19000000000005190}%
\StoreBenchExecResult{Abc}{ScPdrScPdrTerminationRelCommon}{SolveTime}{Correct}{True}{Min}{0.01}%
\StoreBenchExecResult{Abc}{ScPdrScPdrTerminationRelCommon}{SolveTime}{Correct}{True}{Max}{629.23}%
\StoreBenchExecResult{Abc}{ScPdrScPdrTerminationRelCommon}{SolveTime}{Correct}{True}{Avg}{23.836500000000000865}%
\StoreBenchExecResult{Abc}{ScPdrScPdrTerminationRelCommon}{SolveTime}{Correct}{True}{Median}{0.02}%
\StoreBenchExecResult{Abc}{ScPdrScPdrTerminationRelCommon}{SolveTime}{Correct}{True}{Stdev}{92.59923395696460682233831042}%
\StoreBenchExecResult{Abc}{ScPdrScPdrTerminationRelCommon}{SolveTime}{Correct}{False}{Sum}{28310.25000000000022464}%
\StoreBenchExecResult{Abc}{ScPdrScPdrTerminationRelCommon}{SolveTime}{Correct}{False}{Min}{0.01}%
\StoreBenchExecResult{Abc}{ScPdrScPdrTerminationRelCommon}{SolveTime}{Correct}{False}{Max}{897.0500000000001}%
\StoreBenchExecResult{Abc}{ScPdrScPdrTerminationRelCommon}{SolveTime}{Correct}{False}{Avg}{58.73495850622406685609958506}%
\StoreBenchExecResult{Abc}{ScPdrScPdrTerminationRelCommon}{SolveTime}{Correct}{False}{Median}{5.97000000000000025}%
\StoreBenchExecResult{Abc}{ScPdrScPdrTerminationRelCommon}{SolveTime}{Correct}{False}{Stdev}{172.1352554323419674579498607}%
\StoreBenchExecResult{Abc}{ScPdrScPdrTerminationRelCommon}{VerifTime}{All}{}{Sum}{13797.69}%
\StoreBenchExecResult{Abc}{ScPdrScPdrTerminationRelCommon}{VerifTime}{All}{}{Min}{0.01}%
\StoreBenchExecResult{Abc}{ScPdrScPdrTerminationRelCommon}{VerifTime}{All}{}{Max}{644.31}%
\StoreBenchExecResult{Abc}{ScPdrScPdrTerminationRelCommon}{VerifTime}{All}{}{Avg}{25.45699261992619926199261993}%
\StoreBenchExecResult{Abc}{ScPdrScPdrTerminationRelCommon}{VerifTime}{All}{}{Median}{1.335}%
\StoreBenchExecResult{Abc}{ScPdrScPdrTerminationRelCommon}{VerifTime}{All}{}{Stdev}{84.19334090291689739111938940}%
\StoreBenchExecResult{Abc}{ScPdrScPdrTerminationRelCommon}{VerifTime}{Correct}{}{Sum}{13797.69}%
\StoreBenchExecResult{Abc}{ScPdrScPdrTerminationRelCommon}{VerifTime}{Correct}{}{Min}{0.01}%
\StoreBenchExecResult{Abc}{ScPdrScPdrTerminationRelCommon}{VerifTime}{Correct}{}{Max}{644.31}%
\StoreBenchExecResult{Abc}{ScPdrScPdrTerminationRelCommon}{VerifTime}{Correct}{}{Avg}{25.45699261992619926199261993}%
\StoreBenchExecResult{Abc}{ScPdrScPdrTerminationRelCommon}{VerifTime}{Correct}{}{Median}{1.335}%
\StoreBenchExecResult{Abc}{ScPdrScPdrTerminationRelCommon}{VerifTime}{Correct}{}{Stdev}{84.19334090291689739111938940}%
\StoreBenchExecResult{Abc}{ScPdrScPdrTerminationRelCommon}{VerifTime}{Correct}{True}{Sum}{1026.58}%
\StoreBenchExecResult{Abc}{ScPdrScPdrTerminationRelCommon}{VerifTime}{Correct}{True}{Min}{0.01}%
\StoreBenchExecResult{Abc}{ScPdrScPdrTerminationRelCommon}{VerifTime}{Correct}{True}{Max}{504.6}%
\StoreBenchExecResult{Abc}{ScPdrScPdrTerminationRelCommon}{VerifTime}{Correct}{True}{Avg}{17.10966666666666666666666667}%
\StoreBenchExecResult{Abc}{ScPdrScPdrTerminationRelCommon}{VerifTime}{Correct}{True}{Median}{0.015}%
\StoreBenchExecResult{Abc}{ScPdrScPdrTerminationRelCommon}{VerifTime}{Correct}{True}{Stdev}{71.25207180535563595728926382}%
\StoreBenchExecResult{Abc}{ScPdrScPdrTerminationRelCommon}{VerifTime}{Correct}{False}{Sum}{12771.11}%
\StoreBenchExecResult{Abc}{ScPdrScPdrTerminationRelCommon}{VerifTime}{Correct}{False}{Min}{0.01}%
\StoreBenchExecResult{Abc}{ScPdrScPdrTerminationRelCommon}{VerifTime}{Correct}{False}{Max}{644.31}%
\StoreBenchExecResult{Abc}{ScPdrScPdrTerminationRelCommon}{VerifTime}{Correct}{False}{Avg}{26.49607883817427385892116183}%
\StoreBenchExecResult{Abc}{ScPdrScPdrTerminationRelCommon}{VerifTime}{Correct}{False}{Median}{1.575}%
\StoreBenchExecResult{Abc}{ScPdrScPdrTerminationRelCommon}{VerifTime}{Correct}{False}{Stdev}{85.61062774420749462336488221}%
\newcommand{\AbcPdrScPdrTerminationRelCommonVerifTimeAllGeoMean}{1.4844906438413679}
\newcommand{\AbcScPdrScPdrTerminationRelCommonVerifTimeAllGeoMean}{1.3002195853899066}
\newcommand{\AbcScPdrScPdrTerminationRelCommonSolveTimeAllGeoMean}{2.8723334411004893}
\newcommand{\ScPdrTerminationRelReducedToZeroCount}{33}

%% file: eval-results/tex/data-commands.l2s-simp.tex
\providecommand\StoreBenchExecResult[7]{\expandafter\newcommand\csname#1#2#3#4#5#6\endcsname{#7}}%
\StoreBenchExecResult{CpvLIIs}{PonoKindFunc}{Status}{All}{}{Score}{0}%
\StoreBenchExecResult{CpvLIIs}{PonoKindFunc}{Status}{All}{}{Count}{1993}%
\StoreBenchExecResult{CpvLIIs}{PonoKindFunc}{Status}{Correct}{}{Count}{1014}%
\StoreBenchExecResult{CpvLIIs}{PonoKindFunc}{Status}{Correct}{True}{Count}{317}%
\StoreBenchExecResult{CpvLIIs}{PonoKindFunc}{Status}{Correct}{False}{Count}{697}%
\StoreBenchExecResult{CpvLIIs}{PonoKindFunc}{Status}{Wrong}{}{Count}{0}%
\StoreBenchExecResult{CpvLIIs}{PonoKindFunc}{Status}{Wrong}{True}{Count}{0}%
\StoreBenchExecResult{CpvLIIs}{PonoKindFunc}{Status}{Wrong}{False}{Count}{0}%
\providecommand\StoreBenchExecResult[7]{\expandafter\newcommand\csname#1#2#3#4#5#6\endcsname{#7}}%
\StoreBenchExecResult{CpvLIIs}{PonoKindFuncNoSimp}{Status}{All}{}{Score}{0}%
\StoreBenchExecResult{CpvLIIs}{PonoKindFuncNoSimp}{Status}{All}{}{Count}{1993}%
\StoreBenchExecResult{CpvLIIs}{PonoKindFuncNoSimp}{Status}{Correct}{}{Count}{392}%
\StoreBenchExecResult{CpvLIIs}{PonoKindFuncNoSimp}{Status}{Correct}{True}{Count}{316}%
\StoreBenchExecResult{CpvLIIs}{PonoKindFuncNoSimp}{Status}{Correct}{False}{Count}{76}%
\StoreBenchExecResult{CpvLIIs}{PonoKindFuncNoSimp}{Status}{Wrong}{}{Count}{0}%
\StoreBenchExecResult{CpvLIIs}{PonoKindFuncNoSimp}{Status}{Wrong}{True}{Count}{0}%
\StoreBenchExecResult{CpvLIIs}{PonoKindFuncNoSimp}{Status}{Wrong}{False}{Count}{0}%
\providecommand\StoreBenchExecResult[7]{\expandafter\newcommand\csname#1#2#3#4#5#6\endcsname{#7}}%
\StoreBenchExecResult{CpvLIIs}{PonoKindRel}{Status}{All}{}{Score}{0}%
\StoreBenchExecResult{CpvLIIs}{PonoKindRel}{Status}{All}{}{Count}{1993}%
\StoreBenchExecResult{CpvLIIs}{PonoKindRel}{Status}{Correct}{}{Count}{1083}%
\StoreBenchExecResult{CpvLIIs}{PonoKindRel}{Status}{Correct}{True}{Count}{378}%
\StoreBenchExecResult{CpvLIIs}{PonoKindRel}{Status}{Correct}{False}{Count}{705}%
\StoreBenchExecResult{CpvLIIs}{PonoKindRel}{Status}{Wrong}{}{Count}{0}%
\StoreBenchExecResult{CpvLIIs}{PonoKindRel}{Status}{Wrong}{True}{Count}{0}%
\StoreBenchExecResult{CpvLIIs}{PonoKindRel}{Status}{Wrong}{False}{Count}{0}%
\providecommand\StoreBenchExecResult[7]{\expandafter\newcommand\csname#1#2#3#4#5#6\endcsname{#7}}%
\StoreBenchExecResult{CpvLIIs}{PonoKindRelNoSimp}{Status}{All}{}{Score}{0}%
\StoreBenchExecResult{CpvLIIs}{PonoKindRelNoSimp}{Status}{All}{}{Count}{1993}%
\StoreBenchExecResult{CpvLIIs}{PonoKindRelNoSimp}{Status}{Correct}{}{Count}{780}%
\StoreBenchExecResult{CpvLIIs}{PonoKindRelNoSimp}{Status}{Correct}{True}{Count}{378}%
\StoreBenchExecResult{CpvLIIs}{PonoKindRelNoSimp}{Status}{Correct}{False}{Count}{402}%
\StoreBenchExecResult{CpvLIIs}{PonoKindRelNoSimp}{Status}{Wrong}{}{Count}{0}%
\StoreBenchExecResult{CpvLIIs}{PonoKindRelNoSimp}{Status}{Wrong}{True}{Count}{0}%
\StoreBenchExecResult{CpvLIIs}{PonoKindRelNoSimp}{Status}{Wrong}{False}{Count}{0}%
\providecommand\StoreBenchExecResult[7]{\expandafter\newcommand\csname#1#2#3#4#5#6\endcsname{#7}}%
\StoreBenchExecResult{CpvLIIs}{RicIIIIcIIIFunc}{Status}{All}{}{Score}{0}%
\StoreBenchExecResult{CpvLIIs}{RicIIIIcIIIFunc}{Status}{All}{}{Count}{1993}%
\StoreBenchExecResult{CpvLIIs}{RicIIIIcIIIFunc}{Status}{Correct}{}{Count}{977}%
\StoreBenchExecResult{CpvLIIs}{RicIIIIcIIIFunc}{Status}{Correct}{True}{Count}{281}%
\StoreBenchExecResult{CpvLIIs}{RicIIIIcIIIFunc}{Status}{Correct}{False}{Count}{696}%
\StoreBenchExecResult{CpvLIIs}{RicIIIIcIIIFunc}{Status}{Wrong}{}{Count}{0}%
\StoreBenchExecResult{CpvLIIs}{RicIIIIcIIIFunc}{Status}{Wrong}{True}{Count}{0}%
\StoreBenchExecResult{CpvLIIs}{RicIIIIcIIIFunc}{Status}{Wrong}{False}{Count}{0}%
\providecommand\StoreBenchExecResult[7]{\expandafter\newcommand\csname#1#2#3#4#5#6\endcsname{#7}}%
\StoreBenchExecResult{CpvLIIs}{RicIIIIcIIIFuncNoSimp}{Status}{All}{}{Score}{0}%
\StoreBenchExecResult{CpvLIIs}{RicIIIIcIIIFuncNoSimp}{Status}{All}{}{Count}{1993}%
\StoreBenchExecResult{CpvLIIs}{RicIIIIcIIIFuncNoSimp}{Status}{Correct}{}{Count}{770}%
\StoreBenchExecResult{CpvLIIs}{RicIIIIcIIIFuncNoSimp}{Status}{Correct}{True}{Count}{237}%
\StoreBenchExecResult{CpvLIIs}{RicIIIIcIIIFuncNoSimp}{Status}{Correct}{False}{Count}{533}%
\StoreBenchExecResult{CpvLIIs}{RicIIIIcIIIFuncNoSimp}{Status}{Wrong}{}{Count}{0}%
\StoreBenchExecResult{CpvLIIs}{RicIIIIcIIIFuncNoSimp}{Status}{Wrong}{True}{Count}{0}%
\StoreBenchExecResult{CpvLIIs}{RicIIIIcIIIFuncNoSimp}{Status}{Wrong}{False}{Count}{0}%
\providecommand\StoreBenchExecResult[7]{\expandafter\newcommand\csname#1#2#3#4#5#6\endcsname{#7}}%
\StoreBenchExecResult{CpvLIIs}{RicIIIIcIIIRel}{Status}{All}{}{Score}{0}%
\StoreBenchExecResult{CpvLIIs}{RicIIIIcIIIRel}{Status}{All}{}{Count}{1993}%
\StoreBenchExecResult{CpvLIIs}{RicIIIIcIIIRel}{Status}{Correct}{}{Count}{932}%
\StoreBenchExecResult{CpvLIIs}{RicIIIIcIIIRel}{Status}{Correct}{True}{Count}{244}%
\StoreBenchExecResult{CpvLIIs}{RicIIIIcIIIRel}{Status}{Correct}{False}{Count}{688}%
\StoreBenchExecResult{CpvLIIs}{RicIIIIcIIIRel}{Status}{Wrong}{}{Count}{0}%
\StoreBenchExecResult{CpvLIIs}{RicIIIIcIIIRel}{Status}{Wrong}{True}{Count}{0}%
\StoreBenchExecResult{CpvLIIs}{RicIIIIcIIIRel}{Status}{Wrong}{False}{Count}{0}%
\providecommand\StoreBenchExecResult[7]{\expandafter\newcommand\csname#1#2#3#4#5#6\endcsname{#7}}%
\StoreBenchExecResult{CpvLIIs}{RicIIIIcIIIRelNoSimp}{Status}{All}{}{Score}{0}%
\StoreBenchExecResult{CpvLIIs}{RicIIIIcIIIRelNoSimp}{Status}{All}{}{Count}{1993}%
\StoreBenchExecResult{CpvLIIs}{RicIIIIcIIIRelNoSimp}{Status}{Correct}{}{Count}{575}%
\StoreBenchExecResult{CpvLIIs}{RicIIIIcIIIRelNoSimp}{Status}{Correct}{True}{Count}{262}%
\StoreBenchExecResult{CpvLIIs}{RicIIIIcIIIRelNoSimp}{Status}{Correct}{False}{Count}{313}%
\StoreBenchExecResult{CpvLIIs}{RicIIIIcIIIRelNoSimp}{Status}{Wrong}{}{Count}{0}%
\StoreBenchExecResult{CpvLIIs}{RicIIIIcIIIRelNoSimp}{Status}{Wrong}{True}{Count}{0}%
\StoreBenchExecResult{CpvLIIs}{RicIIIIcIIIRelNoSimp}{Status}{Wrong}{False}{Count}{0}%
\newcommand{\CpvLIIsPonoKindFuncReductRateMax}{99.22178988326849}
\newcommand{\CpvLIIsPonoKindFuncReductRateMin}{0.0}
\newcommand{\CpvLIIsPonoKindFuncReductRateAvg}{96.6770426394351}
\newcommand{\CpvLIIsPonoKindRelReductRateMax}{98.83720930232558}
\newcommand{\CpvLIIsPonoKindRelReductRateMin}{0.0}
\newcommand{\CpvLIIsPonoKindRelReductRateAvg}{98.35231166928418}

%% file: eval-results/tex/data-commands.transver.tex
\providecommand\StoreBenchExecResult[7]{\expandafter\newcommand\csname#1#2#3#4#5#6\endcsname{#7}}%
\StoreBenchExecResult{TransverCpv}{PonoKindFuncTerminationBitVectors}{Status}{All}{}{Score}{0}%
\StoreBenchExecResult{TransverCpv}{PonoKindFuncTerminationBitVectors}{Status}{All}{}{Count}{34}%
\StoreBenchExecResult{TransverCpv}{PonoKindFuncTerminationBitVectors}{Status}{Correct}{}{Count}{22}%
\StoreBenchExecResult{TransverCpv}{PonoKindFuncTerminationBitVectors}{Status}{Correct}{True}{Count}{11}%
\StoreBenchExecResult{TransverCpv}{PonoKindFuncTerminationBitVectors}{Status}{Correct}{False}{Count}{11}%
\StoreBenchExecResult{TransverCpv}{PonoKindFuncTerminationBitVectors}{Status}{Wrong}{}{Count}{0}%
\StoreBenchExecResult{TransverCpv}{PonoKindFuncTerminationBitVectors}{Status}{Wrong}{True}{Count}{0}%
\StoreBenchExecResult{TransverCpv}{PonoKindFuncTerminationBitVectors}{Status}{Wrong}{False}{Count}{0}%
\providecommand\StoreBenchExecResult[7]{\expandafter\newcommand\csname#1#2#3#4#5#6\endcsname{#7}}%
\StoreBenchExecResult{Cpv}{PonoKindFuncLIIsTerminationBitVectors}{Status}{All}{}{Score}{0}%
\StoreBenchExecResult{Cpv}{PonoKindFuncLIIsTerminationBitVectors}{Status}{All}{}{Count}{34}%
\StoreBenchExecResult{Cpv}{PonoKindFuncLIIsTerminationBitVectors}{Status}{Correct}{}{Count}{24}%
\StoreBenchExecResult{Cpv}{PonoKindFuncLIIsTerminationBitVectors}{Status}{Correct}{True}{Count}{13}%
\StoreBenchExecResult{Cpv}{PonoKindFuncLIIsTerminationBitVectors}{Status}{Correct}{False}{Count}{11}%
\StoreBenchExecResult{Cpv}{PonoKindFuncLIIsTerminationBitVectors}{Status}{Wrong}{}{Count}{0}%
\StoreBenchExecResult{Cpv}{PonoKindFuncLIIsTerminationBitVectors}{Status}{Wrong}{True}{Count}{0}%
\StoreBenchExecResult{Cpv}{PonoKindFuncLIIsTerminationBitVectors}{Status}{Wrong}{False}{Count}{0}%
\providecommand\StoreBenchExecResult[7]{\expandafter\newcommand\csname#1#2#3#4#5#6\endcsname{#7}}%
\StoreBenchExecResult{Cpv}{PonoKindFuncLIIsTransVerSucceededTerminationBitVectors}{Status}{All}{}{Score}{0}%
\StoreBenchExecResult{Cpv}{PonoKindFuncLIIsTransVerSucceededTerminationBitVectors}{Status}{All}{}{Count}{31}%
\StoreBenchExecResult{Cpv}{PonoKindFuncLIIsTransVerSucceededTerminationBitVectors}{Status}{Correct}{}{Count}{23}%
\StoreBenchExecResult{Cpv}{PonoKindFuncLIIsTransVerSucceededTerminationBitVectors}{Status}{Correct}{True}{Count}{12}%
\StoreBenchExecResult{Cpv}{PonoKindFuncLIIsTransVerSucceededTerminationBitVectors}{Status}{Correct}{False}{Count}{11}%
\StoreBenchExecResult{Cpv}{PonoKindFuncLIIsTransVerSucceededTerminationBitVectors}{Status}{Wrong}{}{Count}{0}%
\StoreBenchExecResult{Cpv}{PonoKindFuncLIIsTransVerSucceededTerminationBitVectors}{Status}{Wrong}{True}{Count}{0}%
\StoreBenchExecResult{Cpv}{PonoKindFuncLIIsTransVerSucceededTerminationBitVectors}{Status}{Wrong}{False}{Count}{0}%
\providecommand\StoreBenchExecResult[7]{\expandafter\newcommand\csname#1#2#3#4#5#6\endcsname{#7}}%
\StoreBenchExecResult{TransverCpv}{PonoKindFuncTerminationMainControlFlow}{Status}{All}{}{Score}{0}%
\StoreBenchExecResult{TransverCpv}{PonoKindFuncTerminationMainControlFlow}{Status}{All}{}{Count}{253}%
\StoreBenchExecResult{TransverCpv}{PonoKindFuncTerminationMainControlFlow}{Status}{Correct}{}{Count}{70}%
\StoreBenchExecResult{TransverCpv}{PonoKindFuncTerminationMainControlFlow}{Status}{Correct}{True}{Count}{27}%
\StoreBenchExecResult{TransverCpv}{PonoKindFuncTerminationMainControlFlow}{Status}{Correct}{False}{Count}{43}%
\StoreBenchExecResult{TransverCpv}{PonoKindFuncTerminationMainControlFlow}{Status}{Wrong}{}{Count}{0}%
\StoreBenchExecResult{TransverCpv}{PonoKindFuncTerminationMainControlFlow}{Status}{Wrong}{True}{Count}{0}%
\StoreBenchExecResult{TransverCpv}{PonoKindFuncTerminationMainControlFlow}{Status}{Wrong}{False}{Count}{0}%
\providecommand\StoreBenchExecResult[7]{\expandafter\newcommand\csname#1#2#3#4#5#6\endcsname{#7}}%
\StoreBenchExecResult{Cpv}{PonoKindFuncLIIsTerminationMainControlFlow}{Status}{All}{}{Score}{0}%
\StoreBenchExecResult{Cpv}{PonoKindFuncLIIsTerminationMainControlFlow}{Status}{All}{}{Count}{253}%
\StoreBenchExecResult{Cpv}{PonoKindFuncLIIsTerminationMainControlFlow}{Status}{Correct}{}{Count}{80}%
\StoreBenchExecResult{Cpv}{PonoKindFuncLIIsTerminationMainControlFlow}{Status}{Correct}{True}{Count}{28}%
\StoreBenchExecResult{Cpv}{PonoKindFuncLIIsTerminationMainControlFlow}{Status}{Correct}{False}{Count}{52}%
\StoreBenchExecResult{Cpv}{PonoKindFuncLIIsTerminationMainControlFlow}{Status}{Wrong}{}{Count}{0}%
\StoreBenchExecResult{Cpv}{PonoKindFuncLIIsTerminationMainControlFlow}{Status}{Wrong}{True}{Count}{0}%
\StoreBenchExecResult{Cpv}{PonoKindFuncLIIsTerminationMainControlFlow}{Status}{Wrong}{False}{Count}{0}%
\providecommand\StoreBenchExecResult[7]{\expandafter\newcommand\csname#1#2#3#4#5#6\endcsname{#7}}%
\StoreBenchExecResult{Cpv}{PonoKindFuncLIIsTransVerSucceededTerminationMainControlFlow}{Status}{All}{}{Score}{0}%
\StoreBenchExecResult{Cpv}{PonoKindFuncLIIsTransVerSucceededTerminationMainControlFlow}{Status}{All}{}{Count}{231}%
\StoreBenchExecResult{Cpv}{PonoKindFuncLIIsTransVerSucceededTerminationMainControlFlow}{Status}{Correct}{}{Count}{71}%
\StoreBenchExecResult{Cpv}{PonoKindFuncLIIsTransVerSucceededTerminationMainControlFlow}{Status}{Correct}{True}{Count}{27}%
\StoreBenchExecResult{Cpv}{PonoKindFuncLIIsTransVerSucceededTerminationMainControlFlow}{Status}{Correct}{False}{Count}{44}%
\StoreBenchExecResult{Cpv}{PonoKindFuncLIIsTransVerSucceededTerminationMainControlFlow}{Status}{Wrong}{}{Count}{0}%
\StoreBenchExecResult{Cpv}{PonoKindFuncLIIsTransVerSucceededTerminationMainControlFlow}{Status}{Wrong}{True}{Count}{0}%
\StoreBenchExecResult{Cpv}{PonoKindFuncLIIsTransVerSucceededTerminationMainControlFlow}{Status}{Wrong}{False}{Count}{0}%
\providecommand\StoreBenchExecResult[7]{\expandafter\newcommand\csname#1#2#3#4#5#6\endcsname{#7}}%
\StoreBenchExecResult{TransverCpv}{PonoKindFuncTerminationOther}{Status}{All}{}{Score}{0}%
\StoreBenchExecResult{TransverCpv}{PonoKindFuncTerminationOther}{Status}{All}{}{Count}{1526}%
\StoreBenchExecResult{TransverCpv}{PonoKindFuncTerminationOther}{Status}{Correct}{}{Count}{389}%
\StoreBenchExecResult{TransverCpv}{PonoKindFuncTerminationOther}{Status}{Correct}{True}{Count}{95}%
\StoreBenchExecResult{TransverCpv}{PonoKindFuncTerminationOther}{Status}{Correct}{False}{Count}{294}%
\StoreBenchExecResult{TransverCpv}{PonoKindFuncTerminationOther}{Status}{Wrong}{}{Count}{0}%
\StoreBenchExecResult{TransverCpv}{PonoKindFuncTerminationOther}{Status}{Wrong}{True}{Count}{0}%
\StoreBenchExecResult{TransverCpv}{PonoKindFuncTerminationOther}{Status}{Wrong}{False}{Count}{0}%
\providecommand\StoreBenchExecResult[7]{\expandafter\newcommand\csname#1#2#3#4#5#6\endcsname{#7}}%
\StoreBenchExecResult{Cpv}{PonoKindFuncLIIsTerminationOther}{Status}{All}{}{Score}{0}%
\StoreBenchExecResult{Cpv}{PonoKindFuncLIIsTerminationOther}{Status}{All}{}{Count}{1526}%
\StoreBenchExecResult{Cpv}{PonoKindFuncLIIsTerminationOther}{Status}{Correct}{}{Count}{910}%
\StoreBenchExecResult{Cpv}{PonoKindFuncLIIsTerminationOther}{Status}{Correct}{True}{Count}{276}%
\StoreBenchExecResult{Cpv}{PonoKindFuncLIIsTerminationOther}{Status}{Correct}{False}{Count}{634}%
\StoreBenchExecResult{Cpv}{PonoKindFuncLIIsTerminationOther}{Status}{Wrong}{}{Count}{0}%
\StoreBenchExecResult{Cpv}{PonoKindFuncLIIsTerminationOther}{Status}{Wrong}{True}{Count}{0}%
\StoreBenchExecResult{Cpv}{PonoKindFuncLIIsTerminationOther}{Status}{Wrong}{False}{Count}{0}%
\providecommand\StoreBenchExecResult[7]{\expandafter\newcommand\csname#1#2#3#4#5#6\endcsname{#7}}%
\StoreBenchExecResult{Cpv}{PonoKindFuncLIIsTransVerSucceededTerminationOther}{Status}{All}{}{Score}{0}%
\StoreBenchExecResult{Cpv}{PonoKindFuncLIIsTransVerSucceededTerminationOther}{Status}{All}{}{Count}{669}%
\StoreBenchExecResult{Cpv}{PonoKindFuncLIIsTransVerSucceededTerminationOther}{Status}{Correct}{}{Count}{448}%
\StoreBenchExecResult{Cpv}{PonoKindFuncLIIsTransVerSucceededTerminationOther}{Status}{Correct}{True}{Count}{143}%
\StoreBenchExecResult{Cpv}{PonoKindFuncLIIsTransVerSucceededTerminationOther}{Status}{Correct}{False}{Count}{305}%
\StoreBenchExecResult{Cpv}{PonoKindFuncLIIsTransVerSucceededTerminationOther}{Status}{Wrong}{}{Count}{0}%
\StoreBenchExecResult{Cpv}{PonoKindFuncLIIsTransVerSucceededTerminationOther}{Status}{Wrong}{True}{Count}{0}%
\StoreBenchExecResult{Cpv}{PonoKindFuncLIIsTransVerSucceededTerminationOther}{Status}{Wrong}{False}{Count}{0}%
\edef\TransverCpvPonoKindFuncTerminationStatusAllCount{\the\numexpr\TransverCpvPonoKindFuncTerminationBitVectorsStatusAllCount+\TransverCpvPonoKindFuncTerminationMainControlFlowStatusAllCount+\TransverCpvPonoKindFuncTerminationOtherStatusAllCount}
\edef\TransverCpvPonoKindFuncTerminationStatusCorrectCount{\the\numexpr\TransverCpvPonoKindFuncTerminationBitVectorsStatusCorrectCount+\TransverCpvPonoKindFuncTerminationMainControlFlowStatusCorrectCount+\TransverCpvPonoKindFuncTerminationOtherStatusCorrectCount}
\edef\TransverCpvPonoKindFuncTerminationStatusCorrectTrueCount{\the\numexpr\TransverCpvPonoKindFuncTerminationBitVectorsStatusCorrectTrueCount+\TransverCpvPonoKindFuncTerminationMainControlFlowStatusCorrectTrueCount+\TransverCpvPonoKindFuncTerminationOtherStatusCorrectTrueCount}
\edef\TransverCpvPonoKindFuncTerminationStatusCorrectFalseCount{\the\numexpr\TransverCpvPonoKindFuncTerminationBitVectorsStatusCorrectFalseCount+\TransverCpvPonoKindFuncTerminationMainControlFlowStatusCorrectFalseCount+\TransverCpvPonoKindFuncTerminationOtherStatusCorrectFalseCount}
\edef\TransverCpvPonoKindFuncTerminationStatusWrongCount{\the\numexpr\TransverCpvPonoKindFuncTerminationBitVectorsStatusWrongCount+\TransverCpvPonoKindFuncTerminationMainControlFlowStatusWrongCount+\TransverCpvPonoKindFuncTerminationOtherStatusWrongCount}
\edef\CpvPonoKindFuncLIIsTerminationStatusAllCount{\the\numexpr\CpvPonoKindFuncLIIsTerminationBitVectorsStatusAllCount+\CpvPonoKindFuncLIIsTerminationMainControlFlowStatusAllCount+\CpvPonoKindFuncLIIsTerminationOtherStatusAllCount}
\edef\CpvPonoKindFuncLIIsTerminationStatusCorrectCount{\the\numexpr\CpvPonoKindFuncLIIsTerminationBitVectorsStatusCorrectCount+\CpvPonoKindFuncLIIsTerminationMainControlFlowStatusCorrectCount+\CpvPonoKindFuncLIIsTerminationOtherStatusCorrectCount}
\edef\CpvPonoKindFuncLIIsTerminationStatusCorrectTrueCount{\the\numexpr\CpvPonoKindFuncLIIsTerminationBitVectorsStatusCorrectTrueCount+\CpvPonoKindFuncLIIsTerminationMainControlFlowStatusCorrectTrueCount+\CpvPonoKindFuncLIIsTerminationOtherStatusCorrectTrueCount}
\edef\CpvPonoKindFuncLIIsTerminationStatusCorrectFalseCount{\the\numexpr\CpvPonoKindFuncLIIsTerminationBitVectorsStatusCorrectFalseCount+\CpvPonoKindFuncLIIsTerminationMainControlFlowStatusCorrectFalseCount+\CpvPonoKindFuncLIIsTerminationOtherStatusCorrectFalseCount}
\edef\CpvPonoKindFuncLIIsTerminationStatusWrongCount{\the\numexpr\CpvPonoKindFuncLIIsTerminationBitVectorsStatusWrongCount+\CpvPonoKindFuncLIIsTerminationMainControlFlowStatusWrongCount+\CpvPonoKindFuncLIIsTerminationOtherStatusWrongCount}
\edef\CpvPonoKindFuncLIIsTransVerSucceededTerminationStatusAllCount{\the\numexpr\CpvPonoKindFuncLIIsTransVerSucceededTerminationBitVectorsStatusAllCount+\CpvPonoKindFuncLIIsTransVerSucceededTerminationMainControlFlowStatusAllCount+\CpvPonoKindFuncLIIsTransVerSucceededTerminationOtherStatusAllCount}
\edef\CpvPonoKindFuncLIIsTransVerSucceededTerminationStatusCorrectCount{\the\numexpr\CpvPonoKindFuncLIIsTransVerSucceededTerminationBitVectorsStatusCorrectCount+\CpvPonoKindFuncLIIsTransVerSucceededTerminationMainControlFlowStatusCorrectCount+\CpvPonoKindFuncLIIsTransVerSucceededTerminationOtherStatusCorrectCount}
\edef\CpvPonoKindFuncLIIsTransVerSucceededTerminationStatusCorrectTrueCount{\the\numexpr\CpvPonoKindFuncLIIsTransVerSucceededTerminationBitVectorsStatusCorrectTrueCount+\CpvPonoKindFuncLIIsTransVerSucceededTerminationMainControlFlowStatusCorrectTrueCount+\CpvPonoKindFuncLIIsTransVerSucceededTerminationOtherStatusCorrectTrueCount}
\edef\CpvPonoKindFuncLIIsTransVerSucceededTerminationStatusCorrectFalseCount{\the\numexpr\CpvPonoKindFuncLIIsTransVerSucceededTerminationBitVectorsStatusCorrectFalseCount+\CpvPonoKindFuncLIIsTransVerSucceededTerminationMainControlFlowStatusCorrectFalseCount+\CpvPonoKindFuncLIIsTransVerSucceededTerminationOtherStatusCorrectFalseCount}
\edef\CpvPonoKindFuncLIIsTransVerSucceededTerminationStatusWrongCount{\the\numexpr\CpvPonoKindFuncLIIsTransVerSucceededTerminationBitVectorsStatusWrongCount+\CpvPonoKindFuncLIIsTransVerSucceededTerminationMainControlFlowStatusWrongCount+\CpvPonoKindFuncLIIsTransVerSucceededTerminationOtherStatusWrongCount}

%% file: eval-results/tex/data-commands.wit-val.tex
\providecommand\StoreBenchExecResult[7]{\expandafter\newcommand\csname#1#2#3#4#5#6\endcsname{#7}}%
\StoreBenchExecResult{CpvWitgen}{PonoBmcFunc}{Status}{All}{}{Score}{0}%
\StoreBenchExecResult{CpvWitgen}{PonoBmcFunc}{Status}{All}{}{Count}{3000}%
\StoreBenchExecResult{CpvWitgen}{PonoBmcFunc}{Status}{Correct}{}{Count}{1083}%
\StoreBenchExecResult{CpvWitgen}{PonoBmcFunc}{Status}{Correct}{True}{Count}{0}%
\StoreBenchExecResult{CpvWitgen}{PonoBmcFunc}{Status}{Correct}{False}{Count}{1083}%
\StoreBenchExecResult{CpvWitgen}{PonoBmcFunc}{Status}{Wrong}{}{Count}{0}%
\StoreBenchExecResult{CpvWitgen}{PonoBmcFunc}{Status}{Wrong}{True}{Count}{0}%
\StoreBenchExecResult{CpvWitgen}{PonoBmcFunc}{Status}{Wrong}{False}{Count}{0}%
\providecommand\StoreBenchExecResult[7]{\expandafter\newcommand\csname#1#2#3#4#5#6\endcsname{#7}}%
\StoreBenchExecResult{CpvWitgen}{PonoBmcFuncLIIs}{Status}{All}{}{Score}{0}%
\StoreBenchExecResult{CpvWitgen}{PonoBmcFuncLIIs}{Status}{All}{}{Count}{784}%
\StoreBenchExecResult{CpvWitgen}{PonoBmcFuncLIIs}{Status}{Correct}{}{Count}{698}%
\StoreBenchExecResult{CpvWitgen}{PonoBmcFuncLIIs}{Status}{Correct}{True}{Count}{0}%
\StoreBenchExecResult{CpvWitgen}{PonoBmcFuncLIIs}{Status}{Correct}{False}{Count}{698}%
\StoreBenchExecResult{CpvWitgen}{PonoBmcFuncLIIs}{Status}{Wrong}{}{Count}{0}%
\StoreBenchExecResult{CpvWitgen}{PonoBmcFuncLIIs}{Status}{Wrong}{True}{Count}{0}%
\StoreBenchExecResult{CpvWitgen}{PonoBmcFuncLIIs}{Status}{Wrong}{False}{Count}{0}%
\providecommand\StoreBenchExecResult[7]{\expandafter\newcommand\csname#1#2#3#4#5#6\endcsname{#7}}%
\StoreBenchExecResult{CpaValidate}{ReachValVI}{Status}{All}{}{Score}{0}%
\StoreBenchExecResult{CpaValidate}{ReachValVI}{Status}{All}{}{Count}{3000}%
\StoreBenchExecResult{CpaValidate}{ReachValVI}{Status}{Correct}{}{Count}{931}%
\StoreBenchExecResult{CpaValidate}{ReachValVI}{Status}{Correct}{True}{Count}{0}%
\StoreBenchExecResult{CpaValidate}{ReachValVI}{Status}{Correct}{False}{Count}{931}%
\StoreBenchExecResult{CpaValidate}{ReachValVI}{Status}{Wrong}{}{Count}{61}%
\StoreBenchExecResult{CpaValidate}{ReachValVI}{Status}{Wrong}{True}{Count}{61}%
\StoreBenchExecResult{CpaValidate}{ReachValVI}{Status}{Wrong}{False}{Count}{0}%
\providecommand\StoreBenchExecResult[7]{\expandafter\newcommand\csname#1#2#3#4#5#6\endcsname{#7}}%
\StoreBenchExecResult{CpaValidate}{ReachValVII}{Status}{All}{}{Score}{0}%
\StoreBenchExecResult{CpaValidate}{ReachValVII}{Status}{All}{}{Count}{3000}%
\StoreBenchExecResult{CpaValidate}{ReachValVII}{Status}{Correct}{}{Count}{925}%
\StoreBenchExecResult{CpaValidate}{ReachValVII}{Status}{Correct}{True}{Count}{0}%
\StoreBenchExecResult{CpaValidate}{ReachValVII}{Status}{Correct}{False}{Count}{925}%
\StoreBenchExecResult{CpaValidate}{ReachValVII}{Status}{Wrong}{}{Count}{59}%
\StoreBenchExecResult{CpaValidate}{ReachValVII}{Status}{Wrong}{True}{Count}{59}%
\StoreBenchExecResult{CpaValidate}{ReachValVII}{Status}{Wrong}{False}{Count}{0}%
\providecommand\StoreBenchExecResult[7]{\expandafter\newcommand\csname#1#2#3#4#5#6\endcsname{#7}}%
\StoreBenchExecResult{CpaValidate}{TermValVI}{Status}{All}{}{Score}{0}%
\StoreBenchExecResult{CpaValidate}{TermValVI}{Status}{All}{}{Count}{784}%
\StoreBenchExecResult{CpaValidate}{TermValVI}{Status}{Correct}{}{Count}{113}%
\StoreBenchExecResult{CpaValidate}{TermValVI}{Status}{Correct}{True}{Count}{0}%
\StoreBenchExecResult{CpaValidate}{TermValVI}{Status}{Correct}{False}{Count}{113}%
\StoreBenchExecResult{CpaValidate}{TermValVI}{Status}{Wrong}{}{Count}{0}%
\StoreBenchExecResult{CpaValidate}{TermValVI}{Status}{Wrong}{True}{Count}{0}%
\StoreBenchExecResult{CpaValidate}{TermValVI}{Status}{Wrong}{False}{Count}{0}%
\providecommand\StoreBenchExecResult[7]{\expandafter\newcommand\csname#1#2#3#4#5#6\endcsname{#7}}%
\StoreBenchExecResult{CpaValidate}{TermValVII}{Status}{All}{}{Score}{0}%
\StoreBenchExecResult{CpaValidate}{TermValVII}{Status}{All}{}{Count}{784}%
\StoreBenchExecResult{CpaValidate}{TermValVII}{Status}{Correct}{}{Count}{496}%
\StoreBenchExecResult{CpaValidate}{TermValVII}{Status}{Correct}{True}{Count}{0}%
\StoreBenchExecResult{CpaValidate}{TermValVII}{Status}{Correct}{False}{Count}{496}%
\StoreBenchExecResult{CpaValidate}{TermValVII}{Status}{Wrong}{}{Count}{0}%
\StoreBenchExecResult{CpaValidate}{TermValVII}{Status}{Wrong}{True}{Count}{0}%
\StoreBenchExecResult{CpaValidate}{TermValVII}{Status}{Wrong}{False}{Count}{0}%
\providecommand\StoreBenchExecResult[7]{\expandafter\newcommand\csname#1#2#3#4#5#6\endcsname{#7}}%
\StoreBenchExecResult{WitchValidate}{ReachValVI}{Status}{All}{}{Score}{0}%
\StoreBenchExecResult{WitchValidate}{ReachValVI}{Status}{All}{}{Count}{3000}%
\StoreBenchExecResult{WitchValidate}{ReachValVI}{Status}{Correct}{}{Count}{1012}%
\StoreBenchExecResult{WitchValidate}{ReachValVI}{Status}{Correct}{True}{Count}{0}%
\StoreBenchExecResult{WitchValidate}{ReachValVI}{Status}{Correct}{False}{Count}{1012}%
\StoreBenchExecResult{WitchValidate}{ReachValVI}{Status}{Wrong}{}{Count}{3}%
\StoreBenchExecResult{WitchValidate}{ReachValVI}{Status}{Wrong}{True}{Count}{3}%
\StoreBenchExecResult{WitchValidate}{ReachValVI}{Status}{Wrong}{False}{Count}{0}%
\providecommand\StoreBenchExecResult[7]{\expandafter\newcommand\csname#1#2#3#4#5#6\endcsname{#7}}%
\StoreBenchExecResult{WitchValidate}{ReachValVII}{Status}{All}{}{Score}{0}%
\StoreBenchExecResult{WitchValidate}{ReachValVII}{Status}{All}{}{Count}{3000}%
\StoreBenchExecResult{WitchValidate}{ReachValVII}{Status}{Correct}{}{Count}{1011}%
\StoreBenchExecResult{WitchValidate}{ReachValVII}{Status}{Correct}{True}{Count}{0}%
\StoreBenchExecResult{WitchValidate}{ReachValVII}{Status}{Correct}{False}{Count}{1011}%
\StoreBenchExecResult{WitchValidate}{ReachValVII}{Status}{Wrong}{}{Count}{4}%
\StoreBenchExecResult{WitchValidate}{ReachValVII}{Status}{Wrong}{True}{Count}{4}%
\StoreBenchExecResult{WitchValidate}{ReachValVII}{Status}{Wrong}{False}{Count}{0}%
\providecommand\StoreBenchExecResult[7]{\expandafter\newcommand\csname#1#2#3#4#5#6\endcsname{#7}}%
\StoreBenchExecResult{WitchValidate}{TermValVI}{Status}{All}{}{Score}{0}%
\StoreBenchExecResult{WitchValidate}{TermValVI}{Status}{All}{}{Count}{784}%
\StoreBenchExecResult{WitchValidate}{TermValVI}{Status}{Correct}{}{Count}{0}%
\StoreBenchExecResult{WitchValidate}{TermValVI}{Status}{Correct}{True}{Count}{0}%
\StoreBenchExecResult{WitchValidate}{TermValVI}{Status}{Correct}{False}{Count}{0}%
\StoreBenchExecResult{WitchValidate}{TermValVI}{Status}{Wrong}{}{Count}{0}%
\StoreBenchExecResult{WitchValidate}{TermValVI}{Status}{Wrong}{True}{Count}{0}%
\StoreBenchExecResult{WitchValidate}{TermValVI}{Status}{Wrong}{False}{Count}{0}%
\providecommand\StoreBenchExecResult[7]{\expandafter\newcommand\csname#1#2#3#4#5#6\endcsname{#7}}%
\StoreBenchExecResult{WitchValidate}{TermValVII}{Status}{All}{}{Score}{0}%
\StoreBenchExecResult{WitchValidate}{TermValVII}{Status}{All}{}{Count}{784}%
\StoreBenchExecResult{WitchValidate}{TermValVII}{Status}{Correct}{}{Count}{603}%
\StoreBenchExecResult{WitchValidate}{TermValVII}{Status}{Correct}{True}{Count}{0}%
\StoreBenchExecResult{WitchValidate}{TermValVII}{Status}{Correct}{False}{Count}{603}%
\StoreBenchExecResult{WitchValidate}{TermValVII}{Status}{Wrong}{}{Count}{0}%
\StoreBenchExecResult{WitchValidate}{TermValVII}{Status}{Wrong}{True}{Count}{0}%
\StoreBenchExecResult{WitchValidate}{TermValVII}{Status}{Wrong}{False}{Count}{0}%
\providecommand\StoreBenchExecResult[7]{\expandafter\newcommand\csname#1#2#3#4#5#6\endcsname{#7}}%
\StoreBenchExecResult{Vb}{WitValReachSafety}{Status}{All}{}{Score}{0}%
\StoreBenchExecResult{Vb}{WitValReachSafety}{Status}{All}{}{Count}{3000}%
\StoreBenchExecResult{Vb}{WitValReachSafety}{Status}{Correct}{}{Count}{1018}%
\StoreBenchExecResult{Vb}{WitValReachSafety}{Status}{Correct}{True}{Count}{0}%
\StoreBenchExecResult{Vb}{WitValReachSafety}{Status}{Correct}{False}{Count}{1018}%
\StoreBenchExecResult{Vb}{WitValReachSafety}{Status}{Wrong}{}{Count}{0}%
\StoreBenchExecResult{Vb}{WitValReachSafety}{Status}{Wrong}{True}{Count}{0}%
\StoreBenchExecResult{Vb}{WitValReachSafety}{Status}{Wrong}{False}{Count}{0}%
\providecommand\StoreBenchExecResult[7]{\expandafter\newcommand\csname#1#2#3#4#5#6\endcsname{#7}}%
\StoreBenchExecResult{Vb}{WitValTermination}{Status}{All}{}{Score}{0}%
\StoreBenchExecResult{Vb}{WitValTermination}{Status}{All}{}{Count}{784}%
\StoreBenchExecResult{Vb}{WitValTermination}{Status}{Correct}{}{Count}{676}%
\StoreBenchExecResult{Vb}{WitValTermination}{Status}{Correct}{True}{Count}{0}%
\StoreBenchExecResult{Vb}{WitValTermination}{Status}{Correct}{False}{Count}{676}%
\StoreBenchExecResult{Vb}{WitValTermination}{Status}{Wrong}{}{Count}{0}%
\StoreBenchExecResult{Vb}{WitValTermination}{Status}{Wrong}{True}{Count}{0}%
\StoreBenchExecResult{Vb}{WitValTermination}{Status}{Wrong}{False}{Count}{0}%

%% file: eval-results/tex/data-commands.svcomp-reach.tex
\providecommand\StoreBenchExecResult[7]{\expandafter\newcommand\csname#1#2#3#4#5#6\endcsname{#7}}%
\StoreBenchExecResult{Cpv}{SvcompReachSafetyArrays}{Status}{All}{}{Score}{0}%
\StoreBenchExecResult{Cpv}{SvcompReachSafetyArrays}{Status}{All}{}{Count}{440}%
\StoreBenchExecResult{Cpv}{SvcompReachSafetyArrays}{Status}{Correct}{}{Count}{105}%
\StoreBenchExecResult{Cpv}{SvcompReachSafetyArrays}{Status}{Correct}{True}{Count}{32}%
\StoreBenchExecResult{Cpv}{SvcompReachSafetyArrays}{Status}{Correct}{False}{Count}{73}%
\StoreBenchExecResult{Cpv}{SvcompReachSafetyArrays}{Status}{Wrong}{}{Count}{0}%
\StoreBenchExecResult{Cpv}{SvcompReachSafetyArrays}{Status}{Wrong}{True}{Count}{0}%
\StoreBenchExecResult{Cpv}{SvcompReachSafetyArrays}{Status}{Wrong}{False}{Count}{0}%
\providecommand\StoreBenchExecResult[7]{\expandafter\newcommand\csname#1#2#3#4#5#6\endcsname{#7}}%
\StoreBenchExecResult{Cpachecker}{SvcompReachSafetyArrays}{Status}{All}{}{Score}{0}%
\StoreBenchExecResult{Cpachecker}{SvcompReachSafetyArrays}{Status}{All}{}{Count}{440}%
\StoreBenchExecResult{Cpachecker}{SvcompReachSafetyArrays}{Status}{Correct}{}{Count}{74}%
\StoreBenchExecResult{Cpachecker}{SvcompReachSafetyArrays}{Status}{Correct}{True}{Count}{4}%
\StoreBenchExecResult{Cpachecker}{SvcompReachSafetyArrays}{Status}{Correct}{False}{Count}{70}%
\StoreBenchExecResult{Cpachecker}{SvcompReachSafetyArrays}{Status}{Wrong}{}{Count}{0}%
\StoreBenchExecResult{Cpachecker}{SvcompReachSafetyArrays}{Status}{Wrong}{True}{Count}{0}%
\StoreBenchExecResult{Cpachecker}{SvcompReachSafetyArrays}{Status}{Wrong}{False}{Count}{0}%
\providecommand\StoreBenchExecResult[7]{\expandafter\newcommand\csname#1#2#3#4#5#6\endcsname{#7}}%
\StoreBenchExecResult{Esbmc}{KindReachSafetyArrays}{Status}{All}{}{Score}{0}%
\StoreBenchExecResult{Esbmc}{KindReachSafetyArrays}{Status}{All}{}{Count}{440}%
\StoreBenchExecResult{Esbmc}{KindReachSafetyArrays}{Status}{Correct}{}{Count}{99}%
\StoreBenchExecResult{Esbmc}{KindReachSafetyArrays}{Status}{Correct}{True}{Count}{23}%
\StoreBenchExecResult{Esbmc}{KindReachSafetyArrays}{Status}{Correct}{False}{Count}{76}%
\StoreBenchExecResult{Esbmc}{KindReachSafetyArrays}{Status}{Wrong}{}{Count}{0}%
\StoreBenchExecResult{Esbmc}{KindReachSafetyArrays}{Status}{Wrong}{True}{Count}{0}%
\StoreBenchExecResult{Esbmc}{KindReachSafetyArrays}{Status}{Wrong}{False}{Count}{0}%
\providecommand\StoreBenchExecResult[7]{\expandafter\newcommand\csname#1#2#3#4#5#6\endcsname{#7}}%
\StoreBenchExecResult{KratosII}{SvcompReachSafetyArrays}{Status}{All}{}{Score}{0}%
\StoreBenchExecResult{KratosII}{SvcompReachSafetyArrays}{Status}{All}{}{Count}{440}%
\StoreBenchExecResult{KratosII}{SvcompReachSafetyArrays}{Status}{Correct}{}{Count}{83}%
\StoreBenchExecResult{KratosII}{SvcompReachSafetyArrays}{Status}{Correct}{True}{Count}{7}%
\StoreBenchExecResult{KratosII}{SvcompReachSafetyArrays}{Status}{Correct}{False}{Count}{76}%
\StoreBenchExecResult{KratosII}{SvcompReachSafetyArrays}{Status}{Wrong}{}{Count}{0}%
\StoreBenchExecResult{KratosII}{SvcompReachSafetyArrays}{Status}{Wrong}{True}{Count}{0}%
\StoreBenchExecResult{KratosII}{SvcompReachSafetyArrays}{Status}{Wrong}{False}{Count}{0}%
\providecommand\StoreBenchExecResult[7]{\expandafter\newcommand\csname#1#2#3#4#5#6\endcsname{#7}}%
\StoreBenchExecResult{Symbiotic}{SvcompReachSafetyArrays}{Status}{All}{}{Score}{0}%
\StoreBenchExecResult{Symbiotic}{SvcompReachSafetyArrays}{Status}{All}{}{Count}{440}%
\StoreBenchExecResult{Symbiotic}{SvcompReachSafetyArrays}{Status}{Correct}{}{Count}{154}%
\StoreBenchExecResult{Symbiotic}{SvcompReachSafetyArrays}{Status}{Correct}{True}{Count}{68}%
\StoreBenchExecResult{Symbiotic}{SvcompReachSafetyArrays}{Status}{Correct}{False}{Count}{86}%
\StoreBenchExecResult{Symbiotic}{SvcompReachSafetyArrays}{Status}{Wrong}{}{Count}{0}%
\StoreBenchExecResult{Symbiotic}{SvcompReachSafetyArrays}{Status}{Wrong}{True}{Count}{0}%
\StoreBenchExecResult{Symbiotic}{SvcompReachSafetyArrays}{Status}{Wrong}{False}{Count}{0}%
\providecommand\StoreBenchExecResult[7]{\expandafter\newcommand\csname#1#2#3#4#5#6\endcsname{#7}}%
\StoreBenchExecResult{Uautomizer}{DefaultReachSafetyArrays}{Status}{All}{}{Score}{0}%
\StoreBenchExecResult{Uautomizer}{DefaultReachSafetyArrays}{Status}{All}{}{Count}{440}%
\StoreBenchExecResult{Uautomizer}{DefaultReachSafetyArrays}{Status}{Correct}{}{Count}{87}%
\StoreBenchExecResult{Uautomizer}{DefaultReachSafetyArrays}{Status}{Correct}{True}{Count}{11}%
\StoreBenchExecResult{Uautomizer}{DefaultReachSafetyArrays}{Status}{Correct}{False}{Count}{76}%
\StoreBenchExecResult{Uautomizer}{DefaultReachSafetyArrays}{Status}{Wrong}{}{Count}{0}%
\StoreBenchExecResult{Uautomizer}{DefaultReachSafetyArrays}{Status}{Wrong}{True}{Count}{0}%
\StoreBenchExecResult{Uautomizer}{DefaultReachSafetyArrays}{Status}{Wrong}{False}{Count}{0}%
\newcommand{\CpvSvcompReachSafetyArraysBestCount}{18}
\newcommand{\CpvSvcompReachSafetyArraysUniqCount}{18}
\newcommand{\CpacheckerSvcompReachSafetyArraysBestCount}{0}
\newcommand{\CpacheckerSvcompReachSafetyArraysUniqCount}{0}
\newcommand{\EsbmcKindReachSafetyArraysBestCount}{73}
\newcommand{\EsbmcKindReachSafetyArraysUniqCount}{1}
\newcommand{\KratosIISvcompReachSafetyArraysBestCount}{7}
\newcommand{\KratosIISvcompReachSafetyArraysUniqCount}{0}
\newcommand{\SymbioticSvcompReachSafetyArraysBestCount}{80}
\newcommand{\SymbioticSvcompReachSafetyArraysUniqCount}{59}
\newcommand{\UautomizerDefaultReachSafetyArraysBestCount}{5}
\newcommand{\UautomizerDefaultReachSafetyArraysUniqCount}{2}
\providecommand\StoreBenchExecResult[7]{\expandafter\newcommand\csname#1#2#3#4#5#6\endcsname{#7}}%
\StoreBenchExecResult{Vb}{SvcompReachSafetyArrays}{Status}{All}{}{Score}{0}%
\StoreBenchExecResult{Vb}{SvcompReachSafetyArrays}{Status}{All}{}{Count}{440}%
\StoreBenchExecResult{Vb}{SvcompReachSafetyArrays}{Status}{Correct}{}{Count}{183}%
\StoreBenchExecResult{Vb}{SvcompReachSafetyArrays}{Status}{Correct}{True}{Count}{94}%
\StoreBenchExecResult{Vb}{SvcompReachSafetyArrays}{Status}{Correct}{False}{Count}{89}%
\StoreBenchExecResult{Vb}{SvcompReachSafetyArrays}{Status}{Wrong}{}{Count}{0}%
\StoreBenchExecResult{Vb}{SvcompReachSafetyArrays}{Status}{Wrong}{True}{Count}{0}%
\StoreBenchExecResult{Vb}{SvcompReachSafetyArrays}{Status}{Wrong}{False}{Count}{0}%
\providecommand\StoreBenchExecResult[7]{\expandafter\newcommand\csname#1#2#3#4#5#6\endcsname{#7}}%
\StoreBenchExecResult{Cpv}{SvcompReachSafetyBitVectors}{Status}{All}{}{Score}{0}%
\StoreBenchExecResult{Cpv}{SvcompReachSafetyBitVectors}{Status}{All}{}{Count}{48}%
\StoreBenchExecResult{Cpv}{SvcompReachSafetyBitVectors}{Status}{Correct}{}{Count}{45}%
\StoreBenchExecResult{Cpv}{SvcompReachSafetyBitVectors}{Status}{Correct}{True}{Count}{33}%
\StoreBenchExecResult{Cpv}{SvcompReachSafetyBitVectors}{Status}{Correct}{False}{Count}{12}%
\StoreBenchExecResult{Cpv}{SvcompReachSafetyBitVectors}{Status}{Wrong}{}{Count}{0}%
\StoreBenchExecResult{Cpv}{SvcompReachSafetyBitVectors}{Status}{Wrong}{True}{Count}{0}%
\StoreBenchExecResult{Cpv}{SvcompReachSafetyBitVectors}{Status}{Wrong}{False}{Count}{0}%
\providecommand\StoreBenchExecResult[7]{\expandafter\newcommand\csname#1#2#3#4#5#6\endcsname{#7}}%
\StoreBenchExecResult{Cpachecker}{SvcompReachSafetyBitVectors}{Status}{All}{}{Score}{0}%
\StoreBenchExecResult{Cpachecker}{SvcompReachSafetyBitVectors}{Status}{All}{}{Count}{48}%
\StoreBenchExecResult{Cpachecker}{SvcompReachSafetyBitVectors}{Status}{Correct}{}{Count}{43}%
\StoreBenchExecResult{Cpachecker}{SvcompReachSafetyBitVectors}{Status}{Correct}{True}{Count}{31}%
\StoreBenchExecResult{Cpachecker}{SvcompReachSafetyBitVectors}{Status}{Correct}{False}{Count}{12}%
\StoreBenchExecResult{Cpachecker}{SvcompReachSafetyBitVectors}{Status}{Wrong}{}{Count}{0}%
\StoreBenchExecResult{Cpachecker}{SvcompReachSafetyBitVectors}{Status}{Wrong}{True}{Count}{0}%
\StoreBenchExecResult{Cpachecker}{SvcompReachSafetyBitVectors}{Status}{Wrong}{False}{Count}{0}%
\providecommand\StoreBenchExecResult[7]{\expandafter\newcommand\csname#1#2#3#4#5#6\endcsname{#7}}%
\StoreBenchExecResult{Esbmc}{KindReachSafetyBitVectors}{Status}{All}{}{Score}{0}%
\StoreBenchExecResult{Esbmc}{KindReachSafetyBitVectors}{Status}{All}{}{Count}{48}%
\StoreBenchExecResult{Esbmc}{KindReachSafetyBitVectors}{Status}{Correct}{}{Count}{34}%
\StoreBenchExecResult{Esbmc}{KindReachSafetyBitVectors}{Status}{Correct}{True}{Count}{22}%
\StoreBenchExecResult{Esbmc}{KindReachSafetyBitVectors}{Status}{Correct}{False}{Count}{12}%
\StoreBenchExecResult{Esbmc}{KindReachSafetyBitVectors}{Status}{Wrong}{}{Count}{0}%
\StoreBenchExecResult{Esbmc}{KindReachSafetyBitVectors}{Status}{Wrong}{True}{Count}{0}%
\StoreBenchExecResult{Esbmc}{KindReachSafetyBitVectors}{Status}{Wrong}{False}{Count}{0}%
\providecommand\StoreBenchExecResult[7]{\expandafter\newcommand\csname#1#2#3#4#5#6\endcsname{#7}}%
\StoreBenchExecResult{KratosII}{SvcompReachSafetyBitVectors}{Status}{All}{}{Score}{0}%
\StoreBenchExecResult{KratosII}{SvcompReachSafetyBitVectors}{Status}{All}{}{Count}{48}%
\StoreBenchExecResult{KratosII}{SvcompReachSafetyBitVectors}{Status}{Correct}{}{Count}{44}%
\StoreBenchExecResult{KratosII}{SvcompReachSafetyBitVectors}{Status}{Correct}{True}{Count}{32}%
\StoreBenchExecResult{KratosII}{SvcompReachSafetyBitVectors}{Status}{Correct}{False}{Count}{12}%
\StoreBenchExecResult{KratosII}{SvcompReachSafetyBitVectors}{Status}{Wrong}{}{Count}{0}%
\StoreBenchExecResult{KratosII}{SvcompReachSafetyBitVectors}{Status}{Wrong}{True}{Count}{0}%
\StoreBenchExecResult{KratosII}{SvcompReachSafetyBitVectors}{Status}{Wrong}{False}{Count}{0}%
\providecommand\StoreBenchExecResult[7]{\expandafter\newcommand\csname#1#2#3#4#5#6\endcsname{#7}}%
\StoreBenchExecResult{Symbiotic}{SvcompReachSafetyBitVectors}{Status}{All}{}{Score}{0}%
\StoreBenchExecResult{Symbiotic}{SvcompReachSafetyBitVectors}{Status}{All}{}{Count}{48}%
\StoreBenchExecResult{Symbiotic}{SvcompReachSafetyBitVectors}{Status}{Correct}{}{Count}{38}%
\StoreBenchExecResult{Symbiotic}{SvcompReachSafetyBitVectors}{Status}{Correct}{True}{Count}{26}%
\StoreBenchExecResult{Symbiotic}{SvcompReachSafetyBitVectors}{Status}{Correct}{False}{Count}{12}%
\StoreBenchExecResult{Symbiotic}{SvcompReachSafetyBitVectors}{Status}{Wrong}{}{Count}{0}%
\StoreBenchExecResult{Symbiotic}{SvcompReachSafetyBitVectors}{Status}{Wrong}{True}{Count}{0}%
\StoreBenchExecResult{Symbiotic}{SvcompReachSafetyBitVectors}{Status}{Wrong}{False}{Count}{0}%
\providecommand\StoreBenchExecResult[7]{\expandafter\newcommand\csname#1#2#3#4#5#6\endcsname{#7}}%
\StoreBenchExecResult{Uautomizer}{DefaultReachSafetyBitVectors}{Status}{All}{}{Score}{0}%
\StoreBenchExecResult{Uautomizer}{DefaultReachSafetyBitVectors}{Status}{All}{}{Count}{48}%
\StoreBenchExecResult{Uautomizer}{DefaultReachSafetyBitVectors}{Status}{Correct}{}{Count}{34}%
\StoreBenchExecResult{Uautomizer}{DefaultReachSafetyBitVectors}{Status}{Correct}{True}{Count}{24}%
\StoreBenchExecResult{Uautomizer}{DefaultReachSafetyBitVectors}{Status}{Correct}{False}{Count}{10}%
\StoreBenchExecResult{Uautomizer}{DefaultReachSafetyBitVectors}{Status}{Wrong}{}{Count}{0}%
\StoreBenchExecResult{Uautomizer}{DefaultReachSafetyBitVectors}{Status}{Wrong}{True}{Count}{0}%
\StoreBenchExecResult{Uautomizer}{DefaultReachSafetyBitVectors}{Status}{Wrong}{False}{Count}{0}%
\newcommand{\CpvSvcompReachSafetyBitVectorsBestCount}{14}
\newcommand{\CpvSvcompReachSafetyBitVectorsUniqCount}{1}
\newcommand{\CpacheckerSvcompReachSafetyBitVectorsBestCount}{0}
\newcommand{\CpacheckerSvcompReachSafetyBitVectorsUniqCount}{0}
\newcommand{\EsbmcKindReachSafetyBitVectorsBestCount}{19}
\newcommand{\EsbmcKindReachSafetyBitVectorsUniqCount}{0}
\newcommand{\KratosIISvcompReachSafetyBitVectorsBestCount}{8}
\newcommand{\KratosIISvcompReachSafetyBitVectorsUniqCount}{0}
\newcommand{\SymbioticSvcompReachSafetyBitVectorsBestCount}{4}
\newcommand{\SymbioticSvcompReachSafetyBitVectorsUniqCount}{0}
\newcommand{\UautomizerDefaultReachSafetyBitVectorsBestCount}{0}
\newcommand{\UautomizerDefaultReachSafetyBitVectorsUniqCount}{0}
\providecommand\StoreBenchExecResult[7]{\expandafter\newcommand\csname#1#2#3#4#5#6\endcsname{#7}}%
\StoreBenchExecResult{Vb}{SvcompReachSafetyBitVectors}{Status}{All}{}{Score}{0}%
\StoreBenchExecResult{Vb}{SvcompReachSafetyBitVectors}{Status}{All}{}{Count}{48}%
\StoreBenchExecResult{Vb}{SvcompReachSafetyBitVectors}{Status}{Correct}{}{Count}{45}%
\StoreBenchExecResult{Vb}{SvcompReachSafetyBitVectors}{Status}{Correct}{True}{Count}{33}%
\StoreBenchExecResult{Vb}{SvcompReachSafetyBitVectors}{Status}{Correct}{False}{Count}{12}%
\StoreBenchExecResult{Vb}{SvcompReachSafetyBitVectors}{Status}{Wrong}{}{Count}{0}%
\StoreBenchExecResult{Vb}{SvcompReachSafetyBitVectors}{Status}{Wrong}{True}{Count}{0}%
\StoreBenchExecResult{Vb}{SvcompReachSafetyBitVectors}{Status}{Wrong}{False}{Count}{0}%
\providecommand\StoreBenchExecResult[7]{\expandafter\newcommand\csname#1#2#3#4#5#6\endcsname{#7}}%
\StoreBenchExecResult{Cpv}{SvcompReachSafetyCombinations}{Status}{All}{}{Score}{0}%
\StoreBenchExecResult{Cpv}{SvcompReachSafetyCombinations}{Status}{All}{}{Count}{671}%
\StoreBenchExecResult{Cpv}{SvcompReachSafetyCombinations}{Status}{Correct}{}{Count}{238}%
\StoreBenchExecResult{Cpv}{SvcompReachSafetyCombinations}{Status}{Correct}{True}{Count}{32}%
\StoreBenchExecResult{Cpv}{SvcompReachSafetyCombinations}{Status}{Correct}{False}{Count}{206}%
\StoreBenchExecResult{Cpv}{SvcompReachSafetyCombinations}{Status}{Wrong}{}{Count}{0}%
\StoreBenchExecResult{Cpv}{SvcompReachSafetyCombinations}{Status}{Wrong}{True}{Count}{0}%
\StoreBenchExecResult{Cpv}{SvcompReachSafetyCombinations}{Status}{Wrong}{False}{Count}{0}%
\providecommand\StoreBenchExecResult[7]{\expandafter\newcommand\csname#1#2#3#4#5#6\endcsname{#7}}%
\StoreBenchExecResult{Cpachecker}{SvcompReachSafetyCombinations}{Status}{All}{}{Score}{0}%
\StoreBenchExecResult{Cpachecker}{SvcompReachSafetyCombinations}{Status}{All}{}{Count}{671}%
\StoreBenchExecResult{Cpachecker}{SvcompReachSafetyCombinations}{Status}{Correct}{}{Count}{337}%
\StoreBenchExecResult{Cpachecker}{SvcompReachSafetyCombinations}{Status}{Correct}{True}{Count}{31}%
\StoreBenchExecResult{Cpachecker}{SvcompReachSafetyCombinations}{Status}{Correct}{False}{Count}{306}%
\StoreBenchExecResult{Cpachecker}{SvcompReachSafetyCombinations}{Status}{Wrong}{}{Count}{0}%
\StoreBenchExecResult{Cpachecker}{SvcompReachSafetyCombinations}{Status}{Wrong}{True}{Count}{0}%
\StoreBenchExecResult{Cpachecker}{SvcompReachSafetyCombinations}{Status}{Wrong}{False}{Count}{0}%
\providecommand\StoreBenchExecResult[7]{\expandafter\newcommand\csname#1#2#3#4#5#6\endcsname{#7}}%
\StoreBenchExecResult{Esbmc}{KindReachSafetyCombinations}{Status}{All}{}{Score}{0}%
\StoreBenchExecResult{Esbmc}{KindReachSafetyCombinations}{Status}{All}{}{Count}{671}%
\StoreBenchExecResult{Esbmc}{KindReachSafetyCombinations}{Status}{Correct}{}{Count}{398}%
\StoreBenchExecResult{Esbmc}{KindReachSafetyCombinations}{Status}{Correct}{True}{Count}{47}%
\StoreBenchExecResult{Esbmc}{KindReachSafetyCombinations}{Status}{Correct}{False}{Count}{351}%
\StoreBenchExecResult{Esbmc}{KindReachSafetyCombinations}{Status}{Wrong}{}{Count}{0}%
\StoreBenchExecResult{Esbmc}{KindReachSafetyCombinations}{Status}{Wrong}{True}{Count}{0}%
\StoreBenchExecResult{Esbmc}{KindReachSafetyCombinations}{Status}{Wrong}{False}{Count}{0}%
\providecommand\StoreBenchExecResult[7]{\expandafter\newcommand\csname#1#2#3#4#5#6\endcsname{#7}}%
\StoreBenchExecResult{KratosII}{SvcompReachSafetyCombinations}{Status}{All}{}{Score}{0}%
\StoreBenchExecResult{KratosII}{SvcompReachSafetyCombinations}{Status}{All}{}{Count}{671}%
\StoreBenchExecResult{KratosII}{SvcompReachSafetyCombinations}{Status}{Correct}{}{Count}{80}%
\StoreBenchExecResult{KratosII}{SvcompReachSafetyCombinations}{Status}{Correct}{True}{Count}{15}%
\StoreBenchExecResult{KratosII}{SvcompReachSafetyCombinations}{Status}{Correct}{False}{Count}{65}%
\StoreBenchExecResult{KratosII}{SvcompReachSafetyCombinations}{Status}{Wrong}{}{Count}{0}%
\StoreBenchExecResult{KratosII}{SvcompReachSafetyCombinations}{Status}{Wrong}{True}{Count}{0}%
\StoreBenchExecResult{KratosII}{SvcompReachSafetyCombinations}{Status}{Wrong}{False}{Count}{0}%
\providecommand\StoreBenchExecResult[7]{\expandafter\newcommand\csname#1#2#3#4#5#6\endcsname{#7}}%
\StoreBenchExecResult{Symbiotic}{SvcompReachSafetyCombinations}{Status}{All}{}{Score}{0}%
\StoreBenchExecResult{Symbiotic}{SvcompReachSafetyCombinations}{Status}{All}{}{Count}{671}%
\StoreBenchExecResult{Symbiotic}{SvcompReachSafetyCombinations}{Status}{Correct}{}{Count}{223}%
\StoreBenchExecResult{Symbiotic}{SvcompReachSafetyCombinations}{Status}{Correct}{True}{Count}{0}%
\StoreBenchExecResult{Symbiotic}{SvcompReachSafetyCombinations}{Status}{Correct}{False}{Count}{223}%
\StoreBenchExecResult{Symbiotic}{SvcompReachSafetyCombinations}{Status}{Wrong}{}{Count}{0}%
\StoreBenchExecResult{Symbiotic}{SvcompReachSafetyCombinations}{Status}{Wrong}{True}{Count}{0}%
\StoreBenchExecResult{Symbiotic}{SvcompReachSafetyCombinations}{Status}{Wrong}{False}{Count}{0}%
\providecommand\StoreBenchExecResult[7]{\expandafter\newcommand\csname#1#2#3#4#5#6\endcsname{#7}}%
\StoreBenchExecResult{Uautomizer}{DefaultReachSafetyCombinations}{Status}{All}{}{Score}{0}%
\StoreBenchExecResult{Uautomizer}{DefaultReachSafetyCombinations}{Status}{All}{}{Count}{671}%
\StoreBenchExecResult{Uautomizer}{DefaultReachSafetyCombinations}{Status}{Correct}{}{Count}{108}%
\StoreBenchExecResult{Uautomizer}{DefaultReachSafetyCombinations}{Status}{Correct}{True}{Count}{13}%
\StoreBenchExecResult{Uautomizer}{DefaultReachSafetyCombinations}{Status}{Correct}{False}{Count}{95}%
\StoreBenchExecResult{Uautomizer}{DefaultReachSafetyCombinations}{Status}{Wrong}{}{Count}{0}%
\StoreBenchExecResult{Uautomizer}{DefaultReachSafetyCombinations}{Status}{Wrong}{True}{Count}{0}%
\StoreBenchExecResult{Uautomizer}{DefaultReachSafetyCombinations}{Status}{Wrong}{False}{Count}{0}%
\newcommand{\CpvSvcompReachSafetyCombinationsBestCount}{20}
\newcommand{\CpvSvcompReachSafetyCombinationsUniqCount}{5}
\newcommand{\CpacheckerSvcompReachSafetyCombinationsBestCount}{66}
\newcommand{\CpacheckerSvcompReachSafetyCombinationsUniqCount}{41}
\newcommand{\EsbmcKindReachSafetyCombinationsBestCount}{303}
\newcommand{\EsbmcKindReachSafetyCombinationsUniqCount}{26}
\newcommand{\KratosIISvcompReachSafetyCombinationsBestCount}{23}
\newcommand{\KratosIISvcompReachSafetyCombinationsUniqCount}{0}
\newcommand{\SymbioticSvcompReachSafetyCombinationsBestCount}{89}
\newcommand{\SymbioticSvcompReachSafetyCombinationsUniqCount}{35}
\newcommand{\UautomizerDefaultReachSafetyCombinationsBestCount}{0}
\newcommand{\UautomizerDefaultReachSafetyCombinationsUniqCount}{0}
\providecommand\StoreBenchExecResult[7]{\expandafter\newcommand\csname#1#2#3#4#5#6\endcsname{#7}}%
\StoreBenchExecResult{Vb}{SvcompReachSafetyCombinations}{Status}{All}{}{Score}{0}%
\StoreBenchExecResult{Vb}{SvcompReachSafetyCombinations}{Status}{All}{}{Count}{671}%
\StoreBenchExecResult{Vb}{SvcompReachSafetyCombinations}{Status}{Correct}{}{Count}{501}%
\StoreBenchExecResult{Vb}{SvcompReachSafetyCombinations}{Status}{Correct}{True}{Count}{81}%
\StoreBenchExecResult{Vb}{SvcompReachSafetyCombinations}{Status}{Correct}{False}{Count}{420}%
\StoreBenchExecResult{Vb}{SvcompReachSafetyCombinations}{Status}{Wrong}{}{Count}{0}%
\StoreBenchExecResult{Vb}{SvcompReachSafetyCombinations}{Status}{Wrong}{True}{Count}{0}%
\StoreBenchExecResult{Vb}{SvcompReachSafetyCombinations}{Status}{Wrong}{False}{Count}{0}%
\providecommand\StoreBenchExecResult[7]{\expandafter\newcommand\csname#1#2#3#4#5#6\endcsname{#7}}%
\StoreBenchExecResult{Cpv}{SvcompReachSafetyControlFlow}{Status}{All}{}{Score}{0}%
\StoreBenchExecResult{Cpv}{SvcompReachSafetyControlFlow}{Status}{All}{}{Count}{73}%
\StoreBenchExecResult{Cpv}{SvcompReachSafetyControlFlow}{Status}{Correct}{}{Count}{29}%
\StoreBenchExecResult{Cpv}{SvcompReachSafetyControlFlow}{Status}{Correct}{True}{Count}{25}%
\StoreBenchExecResult{Cpv}{SvcompReachSafetyControlFlow}{Status}{Correct}{False}{Count}{4}%
\StoreBenchExecResult{Cpv}{SvcompReachSafetyControlFlow}{Status}{Wrong}{}{Count}{0}%
\StoreBenchExecResult{Cpv}{SvcompReachSafetyControlFlow}{Status}{Wrong}{True}{Count}{0}%
\StoreBenchExecResult{Cpv}{SvcompReachSafetyControlFlow}{Status}{Wrong}{False}{Count}{0}%
\providecommand\StoreBenchExecResult[7]{\expandafter\newcommand\csname#1#2#3#4#5#6\endcsname{#7}}%
\StoreBenchExecResult{Cpachecker}{SvcompReachSafetyControlFlow}{Status}{All}{}{Score}{0}%
\StoreBenchExecResult{Cpachecker}{SvcompReachSafetyControlFlow}{Status}{All}{}{Count}{73}%
\StoreBenchExecResult{Cpachecker}{SvcompReachSafetyControlFlow}{Status}{Correct}{}{Count}{40}%
\StoreBenchExecResult{Cpachecker}{SvcompReachSafetyControlFlow}{Status}{Correct}{True}{Count}{27}%
\StoreBenchExecResult{Cpachecker}{SvcompReachSafetyControlFlow}{Status}{Correct}{False}{Count}{13}%
\StoreBenchExecResult{Cpachecker}{SvcompReachSafetyControlFlow}{Status}{Wrong}{}{Count}{0}%
\StoreBenchExecResult{Cpachecker}{SvcompReachSafetyControlFlow}{Status}{Wrong}{True}{Count}{0}%
\StoreBenchExecResult{Cpachecker}{SvcompReachSafetyControlFlow}{Status}{Wrong}{False}{Count}{0}%
\providecommand\StoreBenchExecResult[7]{\expandafter\newcommand\csname#1#2#3#4#5#6\endcsname{#7}}%
\StoreBenchExecResult{Esbmc}{KindReachSafetyControlFlow}{Status}{All}{}{Score}{0}%
\StoreBenchExecResult{Esbmc}{KindReachSafetyControlFlow}{Status}{All}{}{Count}{73}%
\StoreBenchExecResult{Esbmc}{KindReachSafetyControlFlow}{Status}{Correct}{}{Count}{25}%
\StoreBenchExecResult{Esbmc}{KindReachSafetyControlFlow}{Status}{Correct}{True}{Count}{17}%
\StoreBenchExecResult{Esbmc}{KindReachSafetyControlFlow}{Status}{Correct}{False}{Count}{8}%
\StoreBenchExecResult{Esbmc}{KindReachSafetyControlFlow}{Status}{Wrong}{}{Count}{0}%
\StoreBenchExecResult{Esbmc}{KindReachSafetyControlFlow}{Status}{Wrong}{True}{Count}{0}%
\StoreBenchExecResult{Esbmc}{KindReachSafetyControlFlow}{Status}{Wrong}{False}{Count}{0}%
\providecommand\StoreBenchExecResult[7]{\expandafter\newcommand\csname#1#2#3#4#5#6\endcsname{#7}}%
\StoreBenchExecResult{KratosII}{SvcompReachSafetyControlFlow}{Status}{All}{}{Score}{0}%
\StoreBenchExecResult{KratosII}{SvcompReachSafetyControlFlow}{Status}{All}{}{Count}{73}%
\StoreBenchExecResult{KratosII}{SvcompReachSafetyControlFlow}{Status}{Correct}{}{Count}{29}%
\StoreBenchExecResult{KratosII}{SvcompReachSafetyControlFlow}{Status}{Correct}{True}{Count}{24}%
\StoreBenchExecResult{KratosII}{SvcompReachSafetyControlFlow}{Status}{Correct}{False}{Count}{5}%
\StoreBenchExecResult{KratosII}{SvcompReachSafetyControlFlow}{Status}{Wrong}{}{Count}{0}%
\StoreBenchExecResult{KratosII}{SvcompReachSafetyControlFlow}{Status}{Wrong}{True}{Count}{0}%
\StoreBenchExecResult{KratosII}{SvcompReachSafetyControlFlow}{Status}{Wrong}{False}{Count}{0}%
\providecommand\StoreBenchExecResult[7]{\expandafter\newcommand\csname#1#2#3#4#5#6\endcsname{#7}}%
\StoreBenchExecResult{Symbiotic}{SvcompReachSafetyControlFlow}{Status}{All}{}{Score}{0}%
\StoreBenchExecResult{Symbiotic}{SvcompReachSafetyControlFlow}{Status}{All}{}{Count}{73}%
\StoreBenchExecResult{Symbiotic}{SvcompReachSafetyControlFlow}{Status}{Correct}{}{Count}{37}%
\StoreBenchExecResult{Symbiotic}{SvcompReachSafetyControlFlow}{Status}{Correct}{True}{Count}{21}%
\StoreBenchExecResult{Symbiotic}{SvcompReachSafetyControlFlow}{Status}{Correct}{False}{Count}{16}%
\StoreBenchExecResult{Symbiotic}{SvcompReachSafetyControlFlow}{Status}{Wrong}{}{Count}{0}%
\StoreBenchExecResult{Symbiotic}{SvcompReachSafetyControlFlow}{Status}{Wrong}{True}{Count}{0}%
\StoreBenchExecResult{Symbiotic}{SvcompReachSafetyControlFlow}{Status}{Wrong}{False}{Count}{0}%
\providecommand\StoreBenchExecResult[7]{\expandafter\newcommand\csname#1#2#3#4#5#6\endcsname{#7}}%
\StoreBenchExecResult{Uautomizer}{DefaultReachSafetyControlFlow}{Status}{All}{}{Score}{0}%
\StoreBenchExecResult{Uautomizer}{DefaultReachSafetyControlFlow}{Status}{All}{}{Count}{73}%
\StoreBenchExecResult{Uautomizer}{DefaultReachSafetyControlFlow}{Status}{Correct}{}{Count}{40}%
\StoreBenchExecResult{Uautomizer}{DefaultReachSafetyControlFlow}{Status}{Correct}{True}{Count}{27}%
\StoreBenchExecResult{Uautomizer}{DefaultReachSafetyControlFlow}{Status}{Correct}{False}{Count}{13}%
\StoreBenchExecResult{Uautomizer}{DefaultReachSafetyControlFlow}{Status}{Wrong}{}{Count}{0}%
\StoreBenchExecResult{Uautomizer}{DefaultReachSafetyControlFlow}{Status}{Wrong}{True}{Count}{0}%
\StoreBenchExecResult{Uautomizer}{DefaultReachSafetyControlFlow}{Status}{Wrong}{False}{Count}{0}%
\newcommand{\CpvSvcompReachSafetyControlFlowBestCount}{9}
\newcommand{\CpvSvcompReachSafetyControlFlowUniqCount}{0}
\newcommand{\CpacheckerSvcompReachSafetyControlFlowBestCount}{1}
\newcommand{\CpacheckerSvcompReachSafetyControlFlowUniqCount}{1}
\newcommand{\EsbmcKindReachSafetyControlFlowBestCount}{21}
\newcommand{\EsbmcKindReachSafetyControlFlowUniqCount}{2}
\newcommand{\KratosIISvcompReachSafetyControlFlowBestCount}{3}
\newcommand{\KratosIISvcompReachSafetyControlFlowUniqCount}{0}
\newcommand{\SymbioticSvcompReachSafetyControlFlowBestCount}{12}
\newcommand{\SymbioticSvcompReachSafetyControlFlowUniqCount}{0}
\newcommand{\UautomizerDefaultReachSafetyControlFlowBestCount}{1}
\newcommand{\UautomizerDefaultReachSafetyControlFlowUniqCount}{1}
\providecommand\StoreBenchExecResult[7]{\expandafter\newcommand\csname#1#2#3#4#5#6\endcsname{#7}}%
\StoreBenchExecResult{Vb}{SvcompReachSafetyControlFlow}{Status}{All}{}{Score}{0}%
\StoreBenchExecResult{Vb}{SvcompReachSafetyControlFlow}{Status}{All}{}{Count}{73}%
\StoreBenchExecResult{Vb}{SvcompReachSafetyControlFlow}{Status}{Correct}{}{Count}{47}%
\StoreBenchExecResult{Vb}{SvcompReachSafetyControlFlow}{Status}{Correct}{True}{Count}{29}%
\StoreBenchExecResult{Vb}{SvcompReachSafetyControlFlow}{Status}{Correct}{False}{Count}{18}%
\StoreBenchExecResult{Vb}{SvcompReachSafetyControlFlow}{Status}{Wrong}{}{Count}{0}%
\StoreBenchExecResult{Vb}{SvcompReachSafetyControlFlow}{Status}{Wrong}{True}{Count}{0}%
\StoreBenchExecResult{Vb}{SvcompReachSafetyControlFlow}{Status}{Wrong}{False}{Count}{0}%
\providecommand\StoreBenchExecResult[7]{\expandafter\newcommand\csname#1#2#3#4#5#6\endcsname{#7}}%
\StoreBenchExecResult{Cpv}{SvcompReachSafetyECA}{Status}{All}{}{Score}{0}%
\StoreBenchExecResult{Cpv}{SvcompReachSafetyECA}{Status}{All}{}{Count}{1263}%
\StoreBenchExecResult{Cpv}{SvcompReachSafetyECA}{Status}{Correct}{}{Count}{596}%
\StoreBenchExecResult{Cpv}{SvcompReachSafetyECA}{Status}{Correct}{True}{Count}{432}%
\StoreBenchExecResult{Cpv}{SvcompReachSafetyECA}{Status}{Correct}{False}{Count}{164}%
\StoreBenchExecResult{Cpv}{SvcompReachSafetyECA}{Status}{Wrong}{}{Count}{0}%
\StoreBenchExecResult{Cpv}{SvcompReachSafetyECA}{Status}{Wrong}{True}{Count}{0}%
\StoreBenchExecResult{Cpv}{SvcompReachSafetyECA}{Status}{Wrong}{False}{Count}{0}%
\providecommand\StoreBenchExecResult[7]{\expandafter\newcommand\csname#1#2#3#4#5#6\endcsname{#7}}%
\StoreBenchExecResult{Cpachecker}{SvcompReachSafetyECA}{Status}{All}{}{Score}{0}%
\StoreBenchExecResult{Cpachecker}{SvcompReachSafetyECA}{Status}{All}{}{Count}{1263}%
\StoreBenchExecResult{Cpachecker}{SvcompReachSafetyECA}{Status}{Correct}{}{Count}{729}%
\StoreBenchExecResult{Cpachecker}{SvcompReachSafetyECA}{Status}{Correct}{True}{Count}{478}%
\StoreBenchExecResult{Cpachecker}{SvcompReachSafetyECA}{Status}{Correct}{False}{Count}{251}%
\StoreBenchExecResult{Cpachecker}{SvcompReachSafetyECA}{Status}{Wrong}{}{Count}{0}%
\StoreBenchExecResult{Cpachecker}{SvcompReachSafetyECA}{Status}{Wrong}{True}{Count}{0}%
\StoreBenchExecResult{Cpachecker}{SvcompReachSafetyECA}{Status}{Wrong}{False}{Count}{0}%
\providecommand\StoreBenchExecResult[7]{\expandafter\newcommand\csname#1#2#3#4#5#6\endcsname{#7}}%
\StoreBenchExecResult{Esbmc}{KindReachSafetyECA}{Status}{All}{}{Score}{0}%
\StoreBenchExecResult{Esbmc}{KindReachSafetyECA}{Status}{All}{}{Count}{1263}%
\StoreBenchExecResult{Esbmc}{KindReachSafetyECA}{Status}{Correct}{}{Count}{312}%
\StoreBenchExecResult{Esbmc}{KindReachSafetyECA}{Status}{Correct}{True}{Count}{0}%
\StoreBenchExecResult{Esbmc}{KindReachSafetyECA}{Status}{Correct}{False}{Count}{312}%
\StoreBenchExecResult{Esbmc}{KindReachSafetyECA}{Status}{Wrong}{}{Count}{0}%
\StoreBenchExecResult{Esbmc}{KindReachSafetyECA}{Status}{Wrong}{True}{Count}{0}%
\StoreBenchExecResult{Esbmc}{KindReachSafetyECA}{Status}{Wrong}{False}{Count}{0}%
\providecommand\StoreBenchExecResult[7]{\expandafter\newcommand\csname#1#2#3#4#5#6\endcsname{#7}}%
\StoreBenchExecResult{KratosII}{SvcompReachSafetyECA}{Status}{All}{}{Score}{0}%
\StoreBenchExecResult{KratosII}{SvcompReachSafetyECA}{Status}{All}{}{Count}{1263}%
\StoreBenchExecResult{KratosII}{SvcompReachSafetyECA}{Status}{Correct}{}{Count}{662}%
\StoreBenchExecResult{KratosII}{SvcompReachSafetyECA}{Status}{Correct}{True}{Count}{312}%
\StoreBenchExecResult{KratosII}{SvcompReachSafetyECA}{Status}{Correct}{False}{Count}{350}%
\StoreBenchExecResult{KratosII}{SvcompReachSafetyECA}{Status}{Wrong}{}{Count}{0}%
\StoreBenchExecResult{KratosII}{SvcompReachSafetyECA}{Status}{Wrong}{True}{Count}{0}%
\StoreBenchExecResult{KratosII}{SvcompReachSafetyECA}{Status}{Wrong}{False}{Count}{0}%
\providecommand\StoreBenchExecResult[7]{\expandafter\newcommand\csname#1#2#3#4#5#6\endcsname{#7}}%
\StoreBenchExecResult{Symbiotic}{SvcompReachSafetyECA}{Status}{All}{}{Score}{0}%
\StoreBenchExecResult{Symbiotic}{SvcompReachSafetyECA}{Status}{All}{}{Count}{1263}%
\StoreBenchExecResult{Symbiotic}{SvcompReachSafetyECA}{Status}{Correct}{}{Count}{283}%
\StoreBenchExecResult{Symbiotic}{SvcompReachSafetyECA}{Status}{Correct}{True}{Count}{0}%
\StoreBenchExecResult{Symbiotic}{SvcompReachSafetyECA}{Status}{Correct}{False}{Count}{283}%
\StoreBenchExecResult{Symbiotic}{SvcompReachSafetyECA}{Status}{Wrong}{}{Count}{0}%
\StoreBenchExecResult{Symbiotic}{SvcompReachSafetyECA}{Status}{Wrong}{True}{Count}{0}%
\StoreBenchExecResult{Symbiotic}{SvcompReachSafetyECA}{Status}{Wrong}{False}{Count}{0}%
\providecommand\StoreBenchExecResult[7]{\expandafter\newcommand\csname#1#2#3#4#5#6\endcsname{#7}}%
\StoreBenchExecResult{Uautomizer}{DefaultReachSafetyECA}{Status}{All}{}{Score}{0}%
\StoreBenchExecResult{Uautomizer}{DefaultReachSafetyECA}{Status}{All}{}{Count}{1263}%
\StoreBenchExecResult{Uautomizer}{DefaultReachSafetyECA}{Status}{Correct}{}{Count}{613}%
\StoreBenchExecResult{Uautomizer}{DefaultReachSafetyECA}{Status}{Correct}{True}{Count}{421}%
\StoreBenchExecResult{Uautomizer}{DefaultReachSafetyECA}{Status}{Correct}{False}{Count}{192}%
\StoreBenchExecResult{Uautomizer}{DefaultReachSafetyECA}{Status}{Wrong}{}{Count}{0}%
\StoreBenchExecResult{Uautomizer}{DefaultReachSafetyECA}{Status}{Wrong}{True}{Count}{0}%
\StoreBenchExecResult{Uautomizer}{DefaultReachSafetyECA}{Status}{Wrong}{False}{Count}{0}%
\newcommand{\CpvSvcompReachSafetyECABestCount}{255}
\newcommand{\CpvSvcompReachSafetyECAUniqCount}{4}
\newcommand{\CpacheckerSvcompReachSafetyECABestCount}{215}
\newcommand{\CpacheckerSvcompReachSafetyECAUniqCount}{67}
\newcommand{\EsbmcKindReachSafetyECABestCount}{58}
\newcommand{\EsbmcKindReachSafetyECAUniqCount}{0}
\newcommand{\KratosIISvcompReachSafetyECABestCount}{77}
\newcommand{\KratosIISvcompReachSafetyECAUniqCount}{48}
\newcommand{\SymbioticSvcompReachSafetyECABestCount}{233}
\newcommand{\SymbioticSvcompReachSafetyECAUniqCount}{20}
\newcommand{\UautomizerDefaultReachSafetyECABestCount}{126}
\newcommand{\UautomizerDefaultReachSafetyECAUniqCount}{70}
\providecommand\StoreBenchExecResult[7]{\expandafter\newcommand\csname#1#2#3#4#5#6\endcsname{#7}}%
\StoreBenchExecResult{Vb}{SvcompReachSafetyECA}{Status}{All}{}{Score}{0}%
\StoreBenchExecResult{Vb}{SvcompReachSafetyECA}{Status}{All}{}{Count}{1263}%
\StoreBenchExecResult{Vb}{SvcompReachSafetyECA}{Status}{Correct}{}{Count}{964}%
\StoreBenchExecResult{Vb}{SvcompReachSafetyECA}{Status}{Correct}{True}{Count}{572}%
\StoreBenchExecResult{Vb}{SvcompReachSafetyECA}{Status}{Correct}{False}{Count}{392}%
\StoreBenchExecResult{Vb}{SvcompReachSafetyECA}{Status}{Wrong}{}{Count}{0}%
\StoreBenchExecResult{Vb}{SvcompReachSafetyECA}{Status}{Wrong}{True}{Count}{0}%
\StoreBenchExecResult{Vb}{SvcompReachSafetyECA}{Status}{Wrong}{False}{Count}{0}%
\providecommand\StoreBenchExecResult[7]{\expandafter\newcommand\csname#1#2#3#4#5#6\endcsname{#7}}%
\StoreBenchExecResult{Cpv}{SvcompReachSafetyFloats}{Status}{All}{}{Score}{0}%
\StoreBenchExecResult{Cpv}{SvcompReachSafetyFloats}{Status}{All}{}{Count}{1400}%
\StoreBenchExecResult{Cpv}{SvcompReachSafetyFloats}{Status}{Correct}{}{Count}{192}%
\StoreBenchExecResult{Cpv}{SvcompReachSafetyFloats}{Status}{Correct}{True}{Count}{159}%
\StoreBenchExecResult{Cpv}{SvcompReachSafetyFloats}{Status}{Correct}{False}{Count}{33}%
\StoreBenchExecResult{Cpv}{SvcompReachSafetyFloats}{Status}{Wrong}{}{Count}{0}%
\StoreBenchExecResult{Cpv}{SvcompReachSafetyFloats}{Status}{Wrong}{True}{Count}{0}%
\StoreBenchExecResult{Cpv}{SvcompReachSafetyFloats}{Status}{Wrong}{False}{Count}{0}%
\providecommand\StoreBenchExecResult[7]{\expandafter\newcommand\csname#1#2#3#4#5#6\endcsname{#7}}%
\StoreBenchExecResult{Cpachecker}{SvcompReachSafetyFloats}{Status}{All}{}{Score}{0}%
\StoreBenchExecResult{Cpachecker}{SvcompReachSafetyFloats}{Status}{All}{}{Count}{1400}%
\StoreBenchExecResult{Cpachecker}{SvcompReachSafetyFloats}{Status}{Correct}{}{Count}{446}%
\StoreBenchExecResult{Cpachecker}{SvcompReachSafetyFloats}{Status}{Correct}{True}{Count}{381}%
\StoreBenchExecResult{Cpachecker}{SvcompReachSafetyFloats}{Status}{Correct}{False}{Count}{65}%
\StoreBenchExecResult{Cpachecker}{SvcompReachSafetyFloats}{Status}{Wrong}{}{Count}{0}%
\StoreBenchExecResult{Cpachecker}{SvcompReachSafetyFloats}{Status}{Wrong}{True}{Count}{0}%
\StoreBenchExecResult{Cpachecker}{SvcompReachSafetyFloats}{Status}{Wrong}{False}{Count}{0}%
\providecommand\StoreBenchExecResult[7]{\expandafter\newcommand\csname#1#2#3#4#5#6\endcsname{#7}}%
\StoreBenchExecResult{Esbmc}{KindReachSafetyFloats}{Status}{All}{}{Score}{0}%
\StoreBenchExecResult{Esbmc}{KindReachSafetyFloats}{Status}{All}{}{Count}{1400}%
\StoreBenchExecResult{Esbmc}{KindReachSafetyFloats}{Status}{Correct}{}{Count}{567}%
\StoreBenchExecResult{Esbmc}{KindReachSafetyFloats}{Status}{Correct}{True}{Count}{452}%
\StoreBenchExecResult{Esbmc}{KindReachSafetyFloats}{Status}{Correct}{False}{Count}{115}%
\StoreBenchExecResult{Esbmc}{KindReachSafetyFloats}{Status}{Wrong}{}{Count}{0}%
\StoreBenchExecResult{Esbmc}{KindReachSafetyFloats}{Status}{Wrong}{True}{Count}{0}%
\StoreBenchExecResult{Esbmc}{KindReachSafetyFloats}{Status}{Wrong}{False}{Count}{0}%
\providecommand\StoreBenchExecResult[7]{\expandafter\newcommand\csname#1#2#3#4#5#6\endcsname{#7}}%
\StoreBenchExecResult{KratosII}{SvcompReachSafetyFloats}{Status}{All}{}{Score}{0}%
\StoreBenchExecResult{KratosII}{SvcompReachSafetyFloats}{Status}{All}{}{Count}{1400}%
\StoreBenchExecResult{KratosII}{SvcompReachSafetyFloats}{Status}{Correct}{}{Count}{268}%
\StoreBenchExecResult{KratosII}{SvcompReachSafetyFloats}{Status}{Correct}{True}{Count}{219}%
\StoreBenchExecResult{KratosII}{SvcompReachSafetyFloats}{Status}{Correct}{False}{Count}{49}%
\StoreBenchExecResult{KratosII}{SvcompReachSafetyFloats}{Status}{Wrong}{}{Count}{0}%
\StoreBenchExecResult{KratosII}{SvcompReachSafetyFloats}{Status}{Wrong}{True}{Count}{0}%
\StoreBenchExecResult{KratosII}{SvcompReachSafetyFloats}{Status}{Wrong}{False}{Count}{0}%
\providecommand\StoreBenchExecResult[7]{\expandafter\newcommand\csname#1#2#3#4#5#6\endcsname{#7}}%
\StoreBenchExecResult{Symbiotic}{SvcompReachSafetyFloats}{Status}{All}{}{Score}{0}%
\StoreBenchExecResult{Symbiotic}{SvcompReachSafetyFloats}{Status}{All}{}{Count}{1400}%
\StoreBenchExecResult{Symbiotic}{SvcompReachSafetyFloats}{Status}{Correct}{}{Count}{407}%
\StoreBenchExecResult{Symbiotic}{SvcompReachSafetyFloats}{Status}{Correct}{True}{Count}{347}%
\StoreBenchExecResult{Symbiotic}{SvcompReachSafetyFloats}{Status}{Correct}{False}{Count}{60}%
\StoreBenchExecResult{Symbiotic}{SvcompReachSafetyFloats}{Status}{Wrong}{}{Count}{1}%
\StoreBenchExecResult{Symbiotic}{SvcompReachSafetyFloats}{Status}{Wrong}{True}{Count}{0}%
\StoreBenchExecResult{Symbiotic}{SvcompReachSafetyFloats}{Status}{Wrong}{False}{Count}{1}%
\providecommand\StoreBenchExecResult[7]{\expandafter\newcommand\csname#1#2#3#4#5#6\endcsname{#7}}%
\StoreBenchExecResult{Uautomizer}{DefaultReachSafetyFloats}{Status}{All}{}{Score}{0}%
\StoreBenchExecResult{Uautomizer}{DefaultReachSafetyFloats}{Status}{All}{}{Count}{1400}%
\StoreBenchExecResult{Uautomizer}{DefaultReachSafetyFloats}{Status}{Correct}{}{Count}{432}%
\StoreBenchExecResult{Uautomizer}{DefaultReachSafetyFloats}{Status}{Correct}{True}{Count}{360}%
\StoreBenchExecResult{Uautomizer}{DefaultReachSafetyFloats}{Status}{Correct}{False}{Count}{72}%
\StoreBenchExecResult{Uautomizer}{DefaultReachSafetyFloats}{Status}{Wrong}{}{Count}{0}%
\StoreBenchExecResult{Uautomizer}{DefaultReachSafetyFloats}{Status}{Wrong}{True}{Count}{0}%
\StoreBenchExecResult{Uautomizer}{DefaultReachSafetyFloats}{Status}{Wrong}{False}{Count}{0}%
\newcommand{\CpvSvcompReachSafetyFloatsBestCount}{12}
\newcommand{\CpvSvcompReachSafetyFloatsUniqCount}{4}
\newcommand{\CpacheckerSvcompReachSafetyFloatsBestCount}{14}
\newcommand{\CpacheckerSvcompReachSafetyFloatsUniqCount}{1}
\newcommand{\EsbmcKindReachSafetyFloatsBestCount}{363}
\newcommand{\EsbmcKindReachSafetyFloatsUniqCount}{64}
\newcommand{\KratosIISvcompReachSafetyFloatsBestCount}{92}
\newcommand{\KratosIISvcompReachSafetyFloatsUniqCount}{1}
\newcommand{\SymbioticSvcompReachSafetyFloatsBestCount}{113}
\newcommand{\SymbioticSvcompReachSafetyFloatsUniqCount}{19}
\newcommand{\UautomizerDefaultReachSafetyFloatsBestCount}{22}
\newcommand{\UautomizerDefaultReachSafetyFloatsUniqCount}{8}
\providecommand\StoreBenchExecResult[7]{\expandafter\newcommand\csname#1#2#3#4#5#6\endcsname{#7}}%
\StoreBenchExecResult{Vb}{SvcompReachSafetyFloats}{Status}{All}{}{Score}{0}%
\StoreBenchExecResult{Vb}{SvcompReachSafetyFloats}{Status}{All}{}{Count}{1400}%
\StoreBenchExecResult{Vb}{SvcompReachSafetyFloats}{Status}{Correct}{}{Count}{616}%
\StoreBenchExecResult{Vb}{SvcompReachSafetyFloats}{Status}{Correct}{True}{Count}{477}%
\StoreBenchExecResult{Vb}{SvcompReachSafetyFloats}{Status}{Correct}{False}{Count}{139}%
\StoreBenchExecResult{Vb}{SvcompReachSafetyFloats}{Status}{Wrong}{}{Count}{0}%
\StoreBenchExecResult{Vb}{SvcompReachSafetyFloats}{Status}{Wrong}{True}{Count}{0}%
\StoreBenchExecResult{Vb}{SvcompReachSafetyFloats}{Status}{Wrong}{False}{Count}{0}%
\providecommand\StoreBenchExecResult[7]{\expandafter\newcommand\csname#1#2#3#4#5#6\endcsname{#7}}%
\StoreBenchExecResult{Cpv}{SvcompReachSafetyHardness}{Status}{All}{}{Score}{0}%
\StoreBenchExecResult{Cpv}{SvcompReachSafetyHardness}{Status}{All}{}{Count}{6789}%
\StoreBenchExecResult{Cpv}{SvcompReachSafetyHardness}{Status}{Correct}{}{Count}{6077}%
\StoreBenchExecResult{Cpv}{SvcompReachSafetyHardness}{Status}{Correct}{True}{Count}{6077}%
\StoreBenchExecResult{Cpv}{SvcompReachSafetyHardness}{Status}{Correct}{False}{Count}{0}%
\StoreBenchExecResult{Cpv}{SvcompReachSafetyHardness}{Status}{Wrong}{}{Count}{0}%
\StoreBenchExecResult{Cpv}{SvcompReachSafetyHardness}{Status}{Wrong}{True}{Count}{0}%
\StoreBenchExecResult{Cpv}{SvcompReachSafetyHardness}{Status}{Wrong}{False}{Count}{0}%
\providecommand\StoreBenchExecResult[7]{\expandafter\newcommand\csname#1#2#3#4#5#6\endcsname{#7}}%
\StoreBenchExecResult{Cpachecker}{SvcompReachSafetyHardness}{Status}{All}{}{Score}{0}%
\StoreBenchExecResult{Cpachecker}{SvcompReachSafetyHardness}{Status}{All}{}{Count}{6789}%
\StoreBenchExecResult{Cpachecker}{SvcompReachSafetyHardness}{Status}{Correct}{}{Count}{4565}%
\StoreBenchExecResult{Cpachecker}{SvcompReachSafetyHardness}{Status}{Correct}{True}{Count}{4565}%
\StoreBenchExecResult{Cpachecker}{SvcompReachSafetyHardness}{Status}{Correct}{False}{Count}{0}%
\StoreBenchExecResult{Cpachecker}{SvcompReachSafetyHardness}{Status}{Wrong}{}{Count}{0}%
\StoreBenchExecResult{Cpachecker}{SvcompReachSafetyHardness}{Status}{Wrong}{True}{Count}{0}%
\StoreBenchExecResult{Cpachecker}{SvcompReachSafetyHardness}{Status}{Wrong}{False}{Count}{0}%
\providecommand\StoreBenchExecResult[7]{\expandafter\newcommand\csname#1#2#3#4#5#6\endcsname{#7}}%
\StoreBenchExecResult{Esbmc}{KindReachSafetyHardness}{Status}{All}{}{Score}{0}%
\StoreBenchExecResult{Esbmc}{KindReachSafetyHardness}{Status}{All}{}{Count}{6789}%
\StoreBenchExecResult{Esbmc}{KindReachSafetyHardness}{Status}{Correct}{}{Count}{6371}%
\StoreBenchExecResult{Esbmc}{KindReachSafetyHardness}{Status}{Correct}{True}{Count}{6371}%
\StoreBenchExecResult{Esbmc}{KindReachSafetyHardness}{Status}{Correct}{False}{Count}{0}%
\StoreBenchExecResult{Esbmc}{KindReachSafetyHardness}{Status}{Wrong}{}{Count}{0}%
\StoreBenchExecResult{Esbmc}{KindReachSafetyHardness}{Status}{Wrong}{True}{Count}{0}%
\StoreBenchExecResult{Esbmc}{KindReachSafetyHardness}{Status}{Wrong}{False}{Count}{0}%
\providecommand\StoreBenchExecResult[7]{\expandafter\newcommand\csname#1#2#3#4#5#6\endcsname{#7}}%
\StoreBenchExecResult{KratosII}{SvcompReachSafetyHardness}{Status}{All}{}{Score}{0}%
\StoreBenchExecResult{KratosII}{SvcompReachSafetyHardness}{Status}{All}{}{Count}{6789}%
\StoreBenchExecResult{KratosII}{SvcompReachSafetyHardness}{Status}{Correct}{}{Count}{4469}%
\StoreBenchExecResult{KratosII}{SvcompReachSafetyHardness}{Status}{Correct}{True}{Count}{4469}%
\StoreBenchExecResult{KratosII}{SvcompReachSafetyHardness}{Status}{Correct}{False}{Count}{0}%
\StoreBenchExecResult{KratosII}{SvcompReachSafetyHardness}{Status}{Wrong}{}{Count}{0}%
\StoreBenchExecResult{KratosII}{SvcompReachSafetyHardness}{Status}{Wrong}{True}{Count}{0}%
\StoreBenchExecResult{KratosII}{SvcompReachSafetyHardness}{Status}{Wrong}{False}{Count}{0}%
\providecommand\StoreBenchExecResult[7]{\expandafter\newcommand\csname#1#2#3#4#5#6\endcsname{#7}}%
\StoreBenchExecResult{Symbiotic}{SvcompReachSafetyHardness}{Status}{All}{}{Score}{0}%
\StoreBenchExecResult{Symbiotic}{SvcompReachSafetyHardness}{Status}{All}{}{Count}{6789}%
\StoreBenchExecResult{Symbiotic}{SvcompReachSafetyHardness}{Status}{Correct}{}{Count}{1247}%
\StoreBenchExecResult{Symbiotic}{SvcompReachSafetyHardness}{Status}{Correct}{True}{Count}{1247}%
\StoreBenchExecResult{Symbiotic}{SvcompReachSafetyHardness}{Status}{Correct}{False}{Count}{0}%
\StoreBenchExecResult{Symbiotic}{SvcompReachSafetyHardness}{Status}{Wrong}{}{Count}{0}%
\StoreBenchExecResult{Symbiotic}{SvcompReachSafetyHardness}{Status}{Wrong}{True}{Count}{0}%
\StoreBenchExecResult{Symbiotic}{SvcompReachSafetyHardness}{Status}{Wrong}{False}{Count}{0}%
\providecommand\StoreBenchExecResult[7]{\expandafter\newcommand\csname#1#2#3#4#5#6\endcsname{#7}}%
\StoreBenchExecResult{Uautomizer}{DefaultReachSafetyHardness}{Status}{All}{}{Score}{0}%
\StoreBenchExecResult{Uautomizer}{DefaultReachSafetyHardness}{Status}{All}{}{Count}{6789}%
\StoreBenchExecResult{Uautomizer}{DefaultReachSafetyHardness}{Status}{Correct}{}{Count}{348}%
\StoreBenchExecResult{Uautomizer}{DefaultReachSafetyHardness}{Status}{Correct}{True}{Count}{348}%
\StoreBenchExecResult{Uautomizer}{DefaultReachSafetyHardness}{Status}{Correct}{False}{Count}{0}%
\StoreBenchExecResult{Uautomizer}{DefaultReachSafetyHardness}{Status}{Wrong}{}{Count}{0}%
\StoreBenchExecResult{Uautomizer}{DefaultReachSafetyHardness}{Status}{Wrong}{True}{Count}{0}%
\StoreBenchExecResult{Uautomizer}{DefaultReachSafetyHardness}{Status}{Wrong}{False}{Count}{0}%
\newcommand{\CpvSvcompReachSafetyHardnessBestCount}{3302}
\newcommand{\CpvSvcompReachSafetyHardnessUniqCount}{277}
\newcommand{\CpacheckerSvcompReachSafetyHardnessBestCount}{180}
\newcommand{\CpacheckerSvcompReachSafetyHardnessUniqCount}{11}
\newcommand{\EsbmcKindReachSafetyHardnessBestCount}{3133}
\newcommand{\EsbmcKindReachSafetyHardnessUniqCount}{191}
\newcommand{\KratosIISvcompReachSafetyHardnessBestCount}{61}
\newcommand{\KratosIISvcompReachSafetyHardnessUniqCount}{15}
\newcommand{\SymbioticSvcompReachSafetyHardnessBestCount}{25}
\newcommand{\SymbioticSvcompReachSafetyHardnessUniqCount}{2}
\newcommand{\UautomizerDefaultReachSafetyHardnessBestCount}{0}
\newcommand{\UautomizerDefaultReachSafetyHardnessUniqCount}{0}
\providecommand\StoreBenchExecResult[7]{\expandafter\newcommand\csname#1#2#3#4#5#6\endcsname{#7}}%
\StoreBenchExecResult{Vb}{SvcompReachSafetyHardness}{Status}{All}{}{Score}{0}%
\StoreBenchExecResult{Vb}{SvcompReachSafetyHardness}{Status}{All}{}{Count}{6789}%
\StoreBenchExecResult{Vb}{SvcompReachSafetyHardness}{Status}{Correct}{}{Count}{6701}%
\StoreBenchExecResult{Vb}{SvcompReachSafetyHardness}{Status}{Correct}{True}{Count}{6701}%
\StoreBenchExecResult{Vb}{SvcompReachSafetyHardness}{Status}{Correct}{False}{Count}{0}%
\StoreBenchExecResult{Vb}{SvcompReachSafetyHardness}{Status}{Wrong}{}{Count}{0}%
\StoreBenchExecResult{Vb}{SvcompReachSafetyHardness}{Status}{Wrong}{True}{Count}{0}%
\StoreBenchExecResult{Vb}{SvcompReachSafetyHardness}{Status}{Wrong}{False}{Count}{0}%
\providecommand\StoreBenchExecResult[7]{\expandafter\newcommand\csname#1#2#3#4#5#6\endcsname{#7}}%
\StoreBenchExecResult{Cpv}{SvcompReachSafetyHardware}{Status}{All}{}{Score}{0}%
\StoreBenchExecResult{Cpv}{SvcompReachSafetyHardware}{Status}{All}{}{Count}{1224}%
\StoreBenchExecResult{Cpv}{SvcompReachSafetyHardware}{Status}{Correct}{}{Count}{792}%
\StoreBenchExecResult{Cpv}{SvcompReachSafetyHardware}{Status}{Correct}{True}{Count}{449}%
\StoreBenchExecResult{Cpv}{SvcompReachSafetyHardware}{Status}{Correct}{False}{Count}{343}%
\StoreBenchExecResult{Cpv}{SvcompReachSafetyHardware}{Status}{Wrong}{}{Count}{0}%
\StoreBenchExecResult{Cpv}{SvcompReachSafetyHardware}{Status}{Wrong}{True}{Count}{0}%
\StoreBenchExecResult{Cpv}{SvcompReachSafetyHardware}{Status}{Wrong}{False}{Count}{0}%
\providecommand\StoreBenchExecResult[7]{\expandafter\newcommand\csname#1#2#3#4#5#6\endcsname{#7}}%
\StoreBenchExecResult{Cpachecker}{SvcompReachSafetyHardware}{Status}{All}{}{Score}{0}%
\StoreBenchExecResult{Cpachecker}{SvcompReachSafetyHardware}{Status}{All}{}{Count}{1224}%
\StoreBenchExecResult{Cpachecker}{SvcompReachSafetyHardware}{Status}{Correct}{}{Count}{251}%
\StoreBenchExecResult{Cpachecker}{SvcompReachSafetyHardware}{Status}{Correct}{True}{Count}{116}%
\StoreBenchExecResult{Cpachecker}{SvcompReachSafetyHardware}{Status}{Correct}{False}{Count}{135}%
\StoreBenchExecResult{Cpachecker}{SvcompReachSafetyHardware}{Status}{Wrong}{}{Count}{0}%
\StoreBenchExecResult{Cpachecker}{SvcompReachSafetyHardware}{Status}{Wrong}{True}{Count}{0}%
\StoreBenchExecResult{Cpachecker}{SvcompReachSafetyHardware}{Status}{Wrong}{False}{Count}{0}%
\providecommand\StoreBenchExecResult[7]{\expandafter\newcommand\csname#1#2#3#4#5#6\endcsname{#7}}%
\StoreBenchExecResult{Esbmc}{KindReachSafetyHardware}{Status}{All}{}{Score}{0}%
\StoreBenchExecResult{Esbmc}{KindReachSafetyHardware}{Status}{All}{}{Count}{1224}%
\StoreBenchExecResult{Esbmc}{KindReachSafetyHardware}{Status}{Correct}{}{Count}{359}%
\StoreBenchExecResult{Esbmc}{KindReachSafetyHardware}{Status}{Correct}{True}{Count}{5}%
\StoreBenchExecResult{Esbmc}{KindReachSafetyHardware}{Status}{Correct}{False}{Count}{354}%
\StoreBenchExecResult{Esbmc}{KindReachSafetyHardware}{Status}{Wrong}{}{Count}{0}%
\StoreBenchExecResult{Esbmc}{KindReachSafetyHardware}{Status}{Wrong}{True}{Count}{0}%
\StoreBenchExecResult{Esbmc}{KindReachSafetyHardware}{Status}{Wrong}{False}{Count}{0}%
\providecommand\StoreBenchExecResult[7]{\expandafter\newcommand\csname#1#2#3#4#5#6\endcsname{#7}}%
\StoreBenchExecResult{KratosII}{SvcompReachSafetyHardware}{Status}{All}{}{Score}{0}%
\StoreBenchExecResult{KratosII}{SvcompReachSafetyHardware}{Status}{All}{}{Count}{1224}%
\StoreBenchExecResult{KratosII}{SvcompReachSafetyHardware}{Status}{Correct}{}{Count}{459}%
\StoreBenchExecResult{KratosII}{SvcompReachSafetyHardware}{Status}{Correct}{True}{Count}{182}%
\StoreBenchExecResult{KratosII}{SvcompReachSafetyHardware}{Status}{Correct}{False}{Count}{277}%
\StoreBenchExecResult{KratosII}{SvcompReachSafetyHardware}{Status}{Wrong}{}{Count}{0}%
\StoreBenchExecResult{KratosII}{SvcompReachSafetyHardware}{Status}{Wrong}{True}{Count}{0}%
\StoreBenchExecResult{KratosII}{SvcompReachSafetyHardware}{Status}{Wrong}{False}{Count}{0}%
\providecommand\StoreBenchExecResult[7]{\expandafter\newcommand\csname#1#2#3#4#5#6\endcsname{#7}}%
\StoreBenchExecResult{Symbiotic}{SvcompReachSafetyHardware}{Status}{All}{}{Score}{0}%
\StoreBenchExecResult{Symbiotic}{SvcompReachSafetyHardware}{Status}{All}{}{Count}{1224}%
\StoreBenchExecResult{Symbiotic}{SvcompReachSafetyHardware}{Status}{Correct}{}{Count}{98}%
\StoreBenchExecResult{Symbiotic}{SvcompReachSafetyHardware}{Status}{Correct}{True}{Count}{48}%
\StoreBenchExecResult{Symbiotic}{SvcompReachSafetyHardware}{Status}{Correct}{False}{Count}{50}%
\StoreBenchExecResult{Symbiotic}{SvcompReachSafetyHardware}{Status}{Wrong}{}{Count}{0}%
\StoreBenchExecResult{Symbiotic}{SvcompReachSafetyHardware}{Status}{Wrong}{True}{Count}{0}%
\StoreBenchExecResult{Symbiotic}{SvcompReachSafetyHardware}{Status}{Wrong}{False}{Count}{0}%
\providecommand\StoreBenchExecResult[7]{\expandafter\newcommand\csname#1#2#3#4#5#6\endcsname{#7}}%
\StoreBenchExecResult{Uautomizer}{DefaultReachSafetyHardware}{Status}{All}{}{Score}{0}%
\StoreBenchExecResult{Uautomizer}{DefaultReachSafetyHardware}{Status}{All}{}{Count}{1224}%
\StoreBenchExecResult{Uautomizer}{DefaultReachSafetyHardware}{Status}{Correct}{}{Count}{83}%
\StoreBenchExecResult{Uautomizer}{DefaultReachSafetyHardware}{Status}{Correct}{True}{Count}{51}%
\StoreBenchExecResult{Uautomizer}{DefaultReachSafetyHardware}{Status}{Correct}{False}{Count}{32}%
\StoreBenchExecResult{Uautomizer}{DefaultReachSafetyHardware}{Status}{Wrong}{}{Count}{0}%
\StoreBenchExecResult{Uautomizer}{DefaultReachSafetyHardware}{Status}{Wrong}{True}{Count}{0}%
\StoreBenchExecResult{Uautomizer}{DefaultReachSafetyHardware}{Status}{Wrong}{False}{Count}{0}%
\newcommand{\CpvSvcompReachSafetyHardwareBestCount}{592}
\newcommand{\CpvSvcompReachSafetyHardwareUniqCount}{283}
\newcommand{\CpacheckerSvcompReachSafetyHardwareBestCount}{3}
\newcommand{\CpacheckerSvcompReachSafetyHardwareUniqCount}{0}
\newcommand{\EsbmcKindReachSafetyHardwareBestCount}{213}
\newcommand{\EsbmcKindReachSafetyHardwareUniqCount}{26}
\newcommand{\KratosIISvcompReachSafetyHardwareBestCount}{15}
\newcommand{\KratosIISvcompReachSafetyHardwareUniqCount}{0}
\newcommand{\SymbioticSvcompReachSafetyHardwareBestCount}{0}
\newcommand{\SymbioticSvcompReachSafetyHardwareUniqCount}{0}
\newcommand{\UautomizerDefaultReachSafetyHardwareBestCount}{0}
\newcommand{\UautomizerDefaultReachSafetyHardwareUniqCount}{0}
\providecommand\StoreBenchExecResult[7]{\expandafter\newcommand\csname#1#2#3#4#5#6\endcsname{#7}}%
\StoreBenchExecResult{Vb}{SvcompReachSafetyHardware}{Status}{All}{}{Score}{0}%
\StoreBenchExecResult{Vb}{SvcompReachSafetyHardware}{Status}{All}{}{Count}{1224}%
\StoreBenchExecResult{Vb}{SvcompReachSafetyHardware}{Status}{Correct}{}{Count}{823}%
\StoreBenchExecResult{Vb}{SvcompReachSafetyHardware}{Status}{Correct}{True}{Count}{450}%
\StoreBenchExecResult{Vb}{SvcompReachSafetyHardware}{Status}{Correct}{False}{Count}{373}%
\StoreBenchExecResult{Vb}{SvcompReachSafetyHardware}{Status}{Wrong}{}{Count}{0}%
\StoreBenchExecResult{Vb}{SvcompReachSafetyHardware}{Status}{Wrong}{True}{Count}{0}%
\StoreBenchExecResult{Vb}{SvcompReachSafetyHardware}{Status}{Wrong}{False}{Count}{0}%
\providecommand\StoreBenchExecResult[7]{\expandafter\newcommand\csname#1#2#3#4#5#6\endcsname{#7}}%
\StoreBenchExecResult{Cpv}{SvcompReachSafetyHeap}{Status}{All}{}{Score}{0}%
\StoreBenchExecResult{Cpv}{SvcompReachSafetyHeap}{Status}{All}{}{Count}{237}%
\StoreBenchExecResult{Cpv}{SvcompReachSafetyHeap}{Status}{Correct}{}{Count}{72}%
\StoreBenchExecResult{Cpv}{SvcompReachSafetyHeap}{Status}{Correct}{True}{Count}{50}%
\StoreBenchExecResult{Cpv}{SvcompReachSafetyHeap}{Status}{Correct}{False}{Count}{22}%
\StoreBenchExecResult{Cpv}{SvcompReachSafetyHeap}{Status}{Wrong}{}{Count}{0}%
\StoreBenchExecResult{Cpv}{SvcompReachSafetyHeap}{Status}{Wrong}{True}{Count}{0}%
\StoreBenchExecResult{Cpv}{SvcompReachSafetyHeap}{Status}{Wrong}{False}{Count}{0}%
\providecommand\StoreBenchExecResult[7]{\expandafter\newcommand\csname#1#2#3#4#5#6\endcsname{#7}}%
\StoreBenchExecResult{Cpachecker}{SvcompReachSafetyHeap}{Status}{All}{}{Score}{0}%
\StoreBenchExecResult{Cpachecker}{SvcompReachSafetyHeap}{Status}{All}{}{Count}{237}%
\StoreBenchExecResult{Cpachecker}{SvcompReachSafetyHeap}{Status}{Correct}{}{Count}{189}%
\StoreBenchExecResult{Cpachecker}{SvcompReachSafetyHeap}{Status}{Correct}{True}{Count}{118}%
\StoreBenchExecResult{Cpachecker}{SvcompReachSafetyHeap}{Status}{Correct}{False}{Count}{71}%
\StoreBenchExecResult{Cpachecker}{SvcompReachSafetyHeap}{Status}{Wrong}{}{Count}{0}%
\StoreBenchExecResult{Cpachecker}{SvcompReachSafetyHeap}{Status}{Wrong}{True}{Count}{0}%
\StoreBenchExecResult{Cpachecker}{SvcompReachSafetyHeap}{Status}{Wrong}{False}{Count}{0}%
\providecommand\StoreBenchExecResult[7]{\expandafter\newcommand\csname#1#2#3#4#5#6\endcsname{#7}}%
\StoreBenchExecResult{Esbmc}{KindReachSafetyHeap}{Status}{All}{}{Score}{0}%
\StoreBenchExecResult{Esbmc}{KindReachSafetyHeap}{Status}{All}{}{Count}{237}%
\StoreBenchExecResult{Esbmc}{KindReachSafetyHeap}{Status}{Correct}{}{Count}{212}%
\StoreBenchExecResult{Esbmc}{KindReachSafetyHeap}{Status}{Correct}{True}{Count}{141}%
\StoreBenchExecResult{Esbmc}{KindReachSafetyHeap}{Status}{Correct}{False}{Count}{71}%
\StoreBenchExecResult{Esbmc}{KindReachSafetyHeap}{Status}{Wrong}{}{Count}{0}%
\StoreBenchExecResult{Esbmc}{KindReachSafetyHeap}{Status}{Wrong}{True}{Count}{0}%
\StoreBenchExecResult{Esbmc}{KindReachSafetyHeap}{Status}{Wrong}{False}{Count}{0}%
\providecommand\StoreBenchExecResult[7]{\expandafter\newcommand\csname#1#2#3#4#5#6\endcsname{#7}}%
\StoreBenchExecResult{KratosII}{SvcompReachSafetyHeap}{Status}{All}{}{Score}{0}%
\StoreBenchExecResult{KratosII}{SvcompReachSafetyHeap}{Status}{All}{}{Count}{237}%
\StoreBenchExecResult{KratosII}{SvcompReachSafetyHeap}{Status}{Correct}{}{Count}{73}%
\StoreBenchExecResult{KratosII}{SvcompReachSafetyHeap}{Status}{Correct}{True}{Count}{48}%
\StoreBenchExecResult{KratosII}{SvcompReachSafetyHeap}{Status}{Correct}{False}{Count}{25}%
\StoreBenchExecResult{KratosII}{SvcompReachSafetyHeap}{Status}{Wrong}{}{Count}{0}%
\StoreBenchExecResult{KratosII}{SvcompReachSafetyHeap}{Status}{Wrong}{True}{Count}{0}%
\StoreBenchExecResult{KratosII}{SvcompReachSafetyHeap}{Status}{Wrong}{False}{Count}{0}%
\providecommand\StoreBenchExecResult[7]{\expandafter\newcommand\csname#1#2#3#4#5#6\endcsname{#7}}%
\StoreBenchExecResult{Symbiotic}{SvcompReachSafetyHeap}{Status}{All}{}{Score}{0}%
\StoreBenchExecResult{Symbiotic}{SvcompReachSafetyHeap}{Status}{All}{}{Count}{237}%
\StoreBenchExecResult{Symbiotic}{SvcompReachSafetyHeap}{Status}{Correct}{}{Count}{189}%
\StoreBenchExecResult{Symbiotic}{SvcompReachSafetyHeap}{Status}{Correct}{True}{Count}{120}%
\StoreBenchExecResult{Symbiotic}{SvcompReachSafetyHeap}{Status}{Correct}{False}{Count}{69}%
\StoreBenchExecResult{Symbiotic}{SvcompReachSafetyHeap}{Status}{Wrong}{}{Count}{0}%
\StoreBenchExecResult{Symbiotic}{SvcompReachSafetyHeap}{Status}{Wrong}{True}{Count}{0}%
\StoreBenchExecResult{Symbiotic}{SvcompReachSafetyHeap}{Status}{Wrong}{False}{Count}{0}%
\providecommand\StoreBenchExecResult[7]{\expandafter\newcommand\csname#1#2#3#4#5#6\endcsname{#7}}%
\StoreBenchExecResult{Uautomizer}{DefaultReachSafetyHeap}{Status}{All}{}{Score}{0}%
\StoreBenchExecResult{Uautomizer}{DefaultReachSafetyHeap}{Status}{All}{}{Count}{237}%
\StoreBenchExecResult{Uautomizer}{DefaultReachSafetyHeap}{Status}{Correct}{}{Count}{156}%
\StoreBenchExecResult{Uautomizer}{DefaultReachSafetyHeap}{Status}{Correct}{True}{Count}{94}%
\StoreBenchExecResult{Uautomizer}{DefaultReachSafetyHeap}{Status}{Correct}{False}{Count}{62}%
\StoreBenchExecResult{Uautomizer}{DefaultReachSafetyHeap}{Status}{Wrong}{}{Count}{0}%
\StoreBenchExecResult{Uautomizer}{DefaultReachSafetyHeap}{Status}{Wrong}{True}{Count}{0}%
\StoreBenchExecResult{Uautomizer}{DefaultReachSafetyHeap}{Status}{Wrong}{False}{Count}{0}%
\newcommand{\CpvSvcompReachSafetyHeapBestCount}{0}
\newcommand{\CpvSvcompReachSafetyHeapUniqCount}{0}
\newcommand{\CpacheckerSvcompReachSafetyHeapBestCount}{0}
\newcommand{\CpacheckerSvcompReachSafetyHeapUniqCount}{0}
\newcommand{\EsbmcKindReachSafetyHeapBestCount}{169}
\newcommand{\EsbmcKindReachSafetyHeapUniqCount}{31}
\newcommand{\KratosIISvcompReachSafetyHeapBestCount}{18}
\newcommand{\KratosIISvcompReachSafetyHeapUniqCount}{0}
\newcommand{\SymbioticSvcompReachSafetyHeapBestCount}{37}
\newcommand{\SymbioticSvcompReachSafetyHeapUniqCount}{1}
\newcommand{\UautomizerDefaultReachSafetyHeapBestCount}{1}
\newcommand{\UautomizerDefaultReachSafetyHeapUniqCount}{1}
\providecommand\StoreBenchExecResult[7]{\expandafter\newcommand\csname#1#2#3#4#5#6\endcsname{#7}}%
\StoreBenchExecResult{Vb}{SvcompReachSafetyHeap}{Status}{All}{}{Score}{0}%
\StoreBenchExecResult{Vb}{SvcompReachSafetyHeap}{Status}{All}{}{Count}{237}%
\StoreBenchExecResult{Vb}{SvcompReachSafetyHeap}{Status}{Correct}{}{Count}{225}%
\StoreBenchExecResult{Vb}{SvcompReachSafetyHeap}{Status}{Correct}{True}{Count}{153}%
\StoreBenchExecResult{Vb}{SvcompReachSafetyHeap}{Status}{Correct}{False}{Count}{72}%
\StoreBenchExecResult{Vb}{SvcompReachSafetyHeap}{Status}{Wrong}{}{Count}{0}%
\StoreBenchExecResult{Vb}{SvcompReachSafetyHeap}{Status}{Wrong}{True}{Count}{0}%
\StoreBenchExecResult{Vb}{SvcompReachSafetyHeap}{Status}{Wrong}{False}{Count}{0}%
\providecommand\StoreBenchExecResult[7]{\expandafter\newcommand\csname#1#2#3#4#5#6\endcsname{#7}}%
\StoreBenchExecResult{Cpv}{SvcompReachSafetyLoops}{Status}{All}{}{Score}{0}%
\StoreBenchExecResult{Cpv}{SvcompReachSafetyLoops}{Status}{All}{}{Count}{762}%
\StoreBenchExecResult{Cpv}{SvcompReachSafetyLoops}{Status}{Correct}{}{Count}{483}%
\StoreBenchExecResult{Cpv}{SvcompReachSafetyLoops}{Status}{Correct}{True}{Count}{358}%
\StoreBenchExecResult{Cpv}{SvcompReachSafetyLoops}{Status}{Correct}{False}{Count}{125}%
\StoreBenchExecResult{Cpv}{SvcompReachSafetyLoops}{Status}{Wrong}{}{Count}{0}%
\StoreBenchExecResult{Cpv}{SvcompReachSafetyLoops}{Status}{Wrong}{True}{Count}{0}%
\StoreBenchExecResult{Cpv}{SvcompReachSafetyLoops}{Status}{Wrong}{False}{Count}{0}%
\providecommand\StoreBenchExecResult[7]{\expandafter\newcommand\csname#1#2#3#4#5#6\endcsname{#7}}%
\StoreBenchExecResult{Cpachecker}{SvcompReachSafetyLoops}{Status}{All}{}{Score}{0}%
\StoreBenchExecResult{Cpachecker}{SvcompReachSafetyLoops}{Status}{All}{}{Count}{762}%
\StoreBenchExecResult{Cpachecker}{SvcompReachSafetyLoops}{Status}{Correct}{}{Count}{480}%
\StoreBenchExecResult{Cpachecker}{SvcompReachSafetyLoops}{Status}{Correct}{True}{Count}{348}%
\StoreBenchExecResult{Cpachecker}{SvcompReachSafetyLoops}{Status}{Correct}{False}{Count}{132}%
\StoreBenchExecResult{Cpachecker}{SvcompReachSafetyLoops}{Status}{Wrong}{}{Count}{1}%
\StoreBenchExecResult{Cpachecker}{SvcompReachSafetyLoops}{Status}{Wrong}{True}{Count}{0}%
\StoreBenchExecResult{Cpachecker}{SvcompReachSafetyLoops}{Status}{Wrong}{False}{Count}{1}%
\providecommand\StoreBenchExecResult[7]{\expandafter\newcommand\csname#1#2#3#4#5#6\endcsname{#7}}%
\StoreBenchExecResult{Esbmc}{KindReachSafetyLoops}{Status}{All}{}{Score}{0}%
\StoreBenchExecResult{Esbmc}{KindReachSafetyLoops}{Status}{All}{}{Count}{762}%
\StoreBenchExecResult{Esbmc}{KindReachSafetyLoops}{Status}{Correct}{}{Count}{456}%
\StoreBenchExecResult{Esbmc}{KindReachSafetyLoops}{Status}{Correct}{True}{Count}{315}%
\StoreBenchExecResult{Esbmc}{KindReachSafetyLoops}{Status}{Correct}{False}{Count}{141}%
\StoreBenchExecResult{Esbmc}{KindReachSafetyLoops}{Status}{Wrong}{}{Count}{0}%
\StoreBenchExecResult{Esbmc}{KindReachSafetyLoops}{Status}{Wrong}{True}{Count}{0}%
\StoreBenchExecResult{Esbmc}{KindReachSafetyLoops}{Status}{Wrong}{False}{Count}{0}%
\providecommand\StoreBenchExecResult[7]{\expandafter\newcommand\csname#1#2#3#4#5#6\endcsname{#7}}%
\StoreBenchExecResult{KratosII}{SvcompReachSafetyLoops}{Status}{All}{}{Score}{0}%
\StoreBenchExecResult{KratosII}{SvcompReachSafetyLoops}{Status}{All}{}{Count}{762}%
\StoreBenchExecResult{KratosII}{SvcompReachSafetyLoops}{Status}{Correct}{}{Count}{462}%
\StoreBenchExecResult{KratosII}{SvcompReachSafetyLoops}{Status}{Correct}{True}{Count}{322}%
\StoreBenchExecResult{KratosII}{SvcompReachSafetyLoops}{Status}{Correct}{False}{Count}{140}%
\StoreBenchExecResult{KratosII}{SvcompReachSafetyLoops}{Status}{Wrong}{}{Count}{0}%
\StoreBenchExecResult{KratosII}{SvcompReachSafetyLoops}{Status}{Wrong}{True}{Count}{0}%
\StoreBenchExecResult{KratosII}{SvcompReachSafetyLoops}{Status}{Wrong}{False}{Count}{0}%
\providecommand\StoreBenchExecResult[7]{\expandafter\newcommand\csname#1#2#3#4#5#6\endcsname{#7}}%
\StoreBenchExecResult{Symbiotic}{SvcompReachSafetyLoops}{Status}{All}{}{Score}{0}%
\StoreBenchExecResult{Symbiotic}{SvcompReachSafetyLoops}{Status}{All}{}{Count}{762}%
\StoreBenchExecResult{Symbiotic}{SvcompReachSafetyLoops}{Status}{Correct}{}{Count}{593}%
\StoreBenchExecResult{Symbiotic}{SvcompReachSafetyLoops}{Status}{Correct}{True}{Count}{426}%
\StoreBenchExecResult{Symbiotic}{SvcompReachSafetyLoops}{Status}{Correct}{False}{Count}{167}%
\StoreBenchExecResult{Symbiotic}{SvcompReachSafetyLoops}{Status}{Wrong}{}{Count}{0}%
\StoreBenchExecResult{Symbiotic}{SvcompReachSafetyLoops}{Status}{Wrong}{True}{Count}{0}%
\StoreBenchExecResult{Symbiotic}{SvcompReachSafetyLoops}{Status}{Wrong}{False}{Count}{0}%
\providecommand\StoreBenchExecResult[7]{\expandafter\newcommand\csname#1#2#3#4#5#6\endcsname{#7}}%
\StoreBenchExecResult{Uautomizer}{DefaultReachSafetyLoops}{Status}{All}{}{Score}{0}%
\StoreBenchExecResult{Uautomizer}{DefaultReachSafetyLoops}{Status}{All}{}{Count}{762}%
\StoreBenchExecResult{Uautomizer}{DefaultReachSafetyLoops}{Status}{Correct}{}{Count}{507}%
\StoreBenchExecResult{Uautomizer}{DefaultReachSafetyLoops}{Status}{Correct}{True}{Count}{378}%
\StoreBenchExecResult{Uautomizer}{DefaultReachSafetyLoops}{Status}{Correct}{False}{Count}{129}%
\StoreBenchExecResult{Uautomizer}{DefaultReachSafetyLoops}{Status}{Wrong}{}{Count}{0}%
\StoreBenchExecResult{Uautomizer}{DefaultReachSafetyLoops}{Status}{Wrong}{True}{Count}{0}%
\StoreBenchExecResult{Uautomizer}{DefaultReachSafetyLoops}{Status}{Wrong}{False}{Count}{0}%
\newcommand{\CpvSvcompReachSafetyLoopsBestCount}{40}
\newcommand{\CpvSvcompReachSafetyLoopsUniqCount}{4}
\newcommand{\CpacheckerSvcompReachSafetyLoopsBestCount}{26}
\newcommand{\CpacheckerSvcompReachSafetyLoopsUniqCount}{0}
\newcommand{\EsbmcKindReachSafetyLoopsBestCount}{189}
\newcommand{\EsbmcKindReachSafetyLoopsUniqCount}{4}
\newcommand{\KratosIISvcompReachSafetyLoopsBestCount}{202}
\newcommand{\KratosIISvcompReachSafetyLoopsUniqCount}{1}
\newcommand{\SymbioticSvcompReachSafetyLoopsBestCount}{131}
\newcommand{\SymbioticSvcompReachSafetyLoopsUniqCount}{49}
\newcommand{\UautomizerDefaultReachSafetyLoopsBestCount}{102}
\newcommand{\UautomizerDefaultReachSafetyLoopsUniqCount}{36}
\providecommand\StoreBenchExecResult[7]{\expandafter\newcommand\csname#1#2#3#4#5#6\endcsname{#7}}%
\StoreBenchExecResult{Vb}{SvcompReachSafetyLoops}{Status}{All}{}{Score}{0}%
\StoreBenchExecResult{Vb}{SvcompReachSafetyLoops}{Status}{All}{}{Count}{762}%
\StoreBenchExecResult{Vb}{SvcompReachSafetyLoops}{Status}{Correct}{}{Count}{690}%
\StoreBenchExecResult{Vb}{SvcompReachSafetyLoops}{Status}{Correct}{True}{Count}{507}%
\StoreBenchExecResult{Vb}{SvcompReachSafetyLoops}{Status}{Correct}{False}{Count}{183}%
\StoreBenchExecResult{Vb}{SvcompReachSafetyLoops}{Status}{Wrong}{}{Count}{0}%
\StoreBenchExecResult{Vb}{SvcompReachSafetyLoops}{Status}{Wrong}{True}{Count}{0}%
\StoreBenchExecResult{Vb}{SvcompReachSafetyLoops}{Status}{Wrong}{False}{Count}{0}%
\providecommand\StoreBenchExecResult[7]{\expandafter\newcommand\csname#1#2#3#4#5#6\endcsname{#7}}%
\StoreBenchExecResult{Cpv}{SvcompReachSafetyProductLines}{Status}{All}{}{Score}{0}%
\StoreBenchExecResult{Cpv}{SvcompReachSafetyProductLines}{Status}{All}{}{Count}{597}%
\StoreBenchExecResult{Cpv}{SvcompReachSafetyProductLines}{Status}{Correct}{}{Count}{584}%
\StoreBenchExecResult{Cpv}{SvcompReachSafetyProductLines}{Status}{Correct}{True}{Count}{323}%
\StoreBenchExecResult{Cpv}{SvcompReachSafetyProductLines}{Status}{Correct}{False}{Count}{261}%
\StoreBenchExecResult{Cpv}{SvcompReachSafetyProductLines}{Status}{Wrong}{}{Count}{0}%
\StoreBenchExecResult{Cpv}{SvcompReachSafetyProductLines}{Status}{Wrong}{True}{Count}{0}%
\StoreBenchExecResult{Cpv}{SvcompReachSafetyProductLines}{Status}{Wrong}{False}{Count}{0}%
\providecommand\StoreBenchExecResult[7]{\expandafter\newcommand\csname#1#2#3#4#5#6\endcsname{#7}}%
\StoreBenchExecResult{Cpachecker}{SvcompReachSafetyProductLines}{Status}{All}{}{Score}{0}%
\StoreBenchExecResult{Cpachecker}{SvcompReachSafetyProductLines}{Status}{All}{}{Count}{597}%
\StoreBenchExecResult{Cpachecker}{SvcompReachSafetyProductLines}{Status}{Correct}{}{Count}{597}%
\StoreBenchExecResult{Cpachecker}{SvcompReachSafetyProductLines}{Status}{Correct}{True}{Count}{332}%
\StoreBenchExecResult{Cpachecker}{SvcompReachSafetyProductLines}{Status}{Correct}{False}{Count}{265}%
\StoreBenchExecResult{Cpachecker}{SvcompReachSafetyProductLines}{Status}{Wrong}{}{Count}{0}%
\StoreBenchExecResult{Cpachecker}{SvcompReachSafetyProductLines}{Status}{Wrong}{True}{Count}{0}%
\StoreBenchExecResult{Cpachecker}{SvcompReachSafetyProductLines}{Status}{Wrong}{False}{Count}{0}%
\providecommand\StoreBenchExecResult[7]{\expandafter\newcommand\csname#1#2#3#4#5#6\endcsname{#7}}%
\StoreBenchExecResult{Esbmc}{KindReachSafetyProductLines}{Status}{All}{}{Score}{0}%
\StoreBenchExecResult{Esbmc}{KindReachSafetyProductLines}{Status}{All}{}{Count}{597}%
\StoreBenchExecResult{Esbmc}{KindReachSafetyProductLines}{Status}{Correct}{}{Count}{503}%
\StoreBenchExecResult{Esbmc}{KindReachSafetyProductLines}{Status}{Correct}{True}{Count}{238}%
\StoreBenchExecResult{Esbmc}{KindReachSafetyProductLines}{Status}{Correct}{False}{Count}{265}%
\StoreBenchExecResult{Esbmc}{KindReachSafetyProductLines}{Status}{Wrong}{}{Count}{0}%
\StoreBenchExecResult{Esbmc}{KindReachSafetyProductLines}{Status}{Wrong}{True}{Count}{0}%
\StoreBenchExecResult{Esbmc}{KindReachSafetyProductLines}{Status}{Wrong}{False}{Count}{0}%
\providecommand\StoreBenchExecResult[7]{\expandafter\newcommand\csname#1#2#3#4#5#6\endcsname{#7}}%
\StoreBenchExecResult{KratosII}{SvcompReachSafetyProductLines}{Status}{All}{}{Score}{0}%
\StoreBenchExecResult{KratosII}{SvcompReachSafetyProductLines}{Status}{All}{}{Count}{597}%
\StoreBenchExecResult{KratosII}{SvcompReachSafetyProductLines}{Status}{Correct}{}{Count}{579}%
\StoreBenchExecResult{KratosII}{SvcompReachSafetyProductLines}{Status}{Correct}{True}{Count}{314}%
\StoreBenchExecResult{KratosII}{SvcompReachSafetyProductLines}{Status}{Correct}{False}{Count}{265}%
\StoreBenchExecResult{KratosII}{SvcompReachSafetyProductLines}{Status}{Wrong}{}{Count}{0}%
\StoreBenchExecResult{KratosII}{SvcompReachSafetyProductLines}{Status}{Wrong}{True}{Count}{0}%
\StoreBenchExecResult{KratosII}{SvcompReachSafetyProductLines}{Status}{Wrong}{False}{Count}{0}%
\providecommand\StoreBenchExecResult[7]{\expandafter\newcommand\csname#1#2#3#4#5#6\endcsname{#7}}%
\StoreBenchExecResult{Symbiotic}{SvcompReachSafetyProductLines}{Status}{All}{}{Score}{0}%
\StoreBenchExecResult{Symbiotic}{SvcompReachSafetyProductLines}{Status}{All}{}{Count}{597}%
\StoreBenchExecResult{Symbiotic}{SvcompReachSafetyProductLines}{Status}{Correct}{}{Count}{532}%
\StoreBenchExecResult{Symbiotic}{SvcompReachSafetyProductLines}{Status}{Correct}{True}{Count}{267}%
\StoreBenchExecResult{Symbiotic}{SvcompReachSafetyProductLines}{Status}{Correct}{False}{Count}{265}%
\StoreBenchExecResult{Symbiotic}{SvcompReachSafetyProductLines}{Status}{Wrong}{}{Count}{0}%
\StoreBenchExecResult{Symbiotic}{SvcompReachSafetyProductLines}{Status}{Wrong}{True}{Count}{0}%
\StoreBenchExecResult{Symbiotic}{SvcompReachSafetyProductLines}{Status}{Wrong}{False}{Count}{0}%
\providecommand\StoreBenchExecResult[7]{\expandafter\newcommand\csname#1#2#3#4#5#6\endcsname{#7}}%
\StoreBenchExecResult{Uautomizer}{DefaultReachSafetyProductLines}{Status}{All}{}{Score}{0}%
\StoreBenchExecResult{Uautomizer}{DefaultReachSafetyProductLines}{Status}{All}{}{Count}{597}%
\StoreBenchExecResult{Uautomizer}{DefaultReachSafetyProductLines}{Status}{Correct}{}{Count}{356}%
\StoreBenchExecResult{Uautomizer}{DefaultReachSafetyProductLines}{Status}{Correct}{True}{Count}{256}%
\StoreBenchExecResult{Uautomizer}{DefaultReachSafetyProductLines}{Status}{Correct}{False}{Count}{100}%
\StoreBenchExecResult{Uautomizer}{DefaultReachSafetyProductLines}{Status}{Wrong}{}{Count}{0}%
\StoreBenchExecResult{Uautomizer}{DefaultReachSafetyProductLines}{Status}{Wrong}{True}{Count}{0}%
\StoreBenchExecResult{Uautomizer}{DefaultReachSafetyProductLines}{Status}{Wrong}{False}{Count}{0}%
\newcommand{\CpvSvcompReachSafetyProductLinesBestCount}{84}
\newcommand{\CpvSvcompReachSafetyProductLinesUniqCount}{0}
\newcommand{\CpacheckerSvcompReachSafetyProductLinesBestCount}{10}
\newcommand{\CpacheckerSvcompReachSafetyProductLinesUniqCount}{9}
\newcommand{\EsbmcKindReachSafetyProductLinesBestCount}{342}
\newcommand{\EsbmcKindReachSafetyProductLinesUniqCount}{0}
\newcommand{\KratosIISvcompReachSafetyProductLinesBestCount}{112}
\newcommand{\KratosIISvcompReachSafetyProductLinesUniqCount}{0}
\newcommand{\SymbioticSvcompReachSafetyProductLinesBestCount}{49}
\newcommand{\SymbioticSvcompReachSafetyProductLinesUniqCount}{0}
\newcommand{\UautomizerDefaultReachSafetyProductLinesBestCount}{0}
\newcommand{\UautomizerDefaultReachSafetyProductLinesUniqCount}{0}
\providecommand\StoreBenchExecResult[7]{\expandafter\newcommand\csname#1#2#3#4#5#6\endcsname{#7}}%
\StoreBenchExecResult{Vb}{SvcompReachSafetyProductLines}{Status}{All}{}{Score}{0}%
\StoreBenchExecResult{Vb}{SvcompReachSafetyProductLines}{Status}{All}{}{Count}{597}%
\StoreBenchExecResult{Vb}{SvcompReachSafetyProductLines}{Status}{Correct}{}{Count}{597}%
\StoreBenchExecResult{Vb}{SvcompReachSafetyProductLines}{Status}{Correct}{True}{Count}{332}%
\StoreBenchExecResult{Vb}{SvcompReachSafetyProductLines}{Status}{Correct}{False}{Count}{265}%
\StoreBenchExecResult{Vb}{SvcompReachSafetyProductLines}{Status}{Wrong}{}{Count}{0}%
\StoreBenchExecResult{Vb}{SvcompReachSafetyProductLines}{Status}{Wrong}{True}{Count}{0}%
\StoreBenchExecResult{Vb}{SvcompReachSafetyProductLines}{Status}{Wrong}{False}{Count}{0}%
\providecommand\StoreBenchExecResult[7]{\expandafter\newcommand\csname#1#2#3#4#5#6\endcsname{#7}}%
\StoreBenchExecResult{Cpv}{SvcompReachSafetySequentialized}{Status}{All}{}{Score}{0}%
\StoreBenchExecResult{Cpv}{SvcompReachSafetySequentialized}{Status}{All}{}{Count}{585}%
\StoreBenchExecResult{Cpv}{SvcompReachSafetySequentialized}{Status}{Correct}{}{Count}{165}%
\StoreBenchExecResult{Cpv}{SvcompReachSafetySequentialized}{Status}{Correct}{True}{Count}{42}%
\StoreBenchExecResult{Cpv}{SvcompReachSafetySequentialized}{Status}{Correct}{False}{Count}{123}%
\StoreBenchExecResult{Cpv}{SvcompReachSafetySequentialized}{Status}{Wrong}{}{Count}{0}%
\StoreBenchExecResult{Cpv}{SvcompReachSafetySequentialized}{Status}{Wrong}{True}{Count}{0}%
\StoreBenchExecResult{Cpv}{SvcompReachSafetySequentialized}{Status}{Wrong}{False}{Count}{0}%
\providecommand\StoreBenchExecResult[7]{\expandafter\newcommand\csname#1#2#3#4#5#6\endcsname{#7}}%
\StoreBenchExecResult{Cpachecker}{SvcompReachSafetySequentialized}{Status}{All}{}{Score}{0}%
\StoreBenchExecResult{Cpachecker}{SvcompReachSafetySequentialized}{Status}{All}{}{Count}{585}%
\StoreBenchExecResult{Cpachecker}{SvcompReachSafetySequentialized}{Status}{Correct}{}{Count}{455}%
\StoreBenchExecResult{Cpachecker}{SvcompReachSafetySequentialized}{Status}{Correct}{True}{Count}{107}%
\StoreBenchExecResult{Cpachecker}{SvcompReachSafetySequentialized}{Status}{Correct}{False}{Count}{348}%
\StoreBenchExecResult{Cpachecker}{SvcompReachSafetySequentialized}{Status}{Wrong}{}{Count}{0}%
\StoreBenchExecResult{Cpachecker}{SvcompReachSafetySequentialized}{Status}{Wrong}{True}{Count}{0}%
\StoreBenchExecResult{Cpachecker}{SvcompReachSafetySequentialized}{Status}{Wrong}{False}{Count}{0}%
\providecommand\StoreBenchExecResult[7]{\expandafter\newcommand\csname#1#2#3#4#5#6\endcsname{#7}}%
\StoreBenchExecResult{Esbmc}{KindReachSafetySequentialized}{Status}{All}{}{Score}{0}%
\StoreBenchExecResult{Esbmc}{KindReachSafetySequentialized}{Status}{All}{}{Count}{585}%
\StoreBenchExecResult{Esbmc}{KindReachSafetySequentialized}{Status}{Correct}{}{Count}{188}%
\StoreBenchExecResult{Esbmc}{KindReachSafetySequentialized}{Status}{Correct}{True}{Count}{40}%
\StoreBenchExecResult{Esbmc}{KindReachSafetySequentialized}{Status}{Correct}{False}{Count}{148}%
\StoreBenchExecResult{Esbmc}{KindReachSafetySequentialized}{Status}{Wrong}{}{Count}{0}%
\StoreBenchExecResult{Esbmc}{KindReachSafetySequentialized}{Status}{Wrong}{True}{Count}{0}%
\StoreBenchExecResult{Esbmc}{KindReachSafetySequentialized}{Status}{Wrong}{False}{Count}{0}%
\providecommand\StoreBenchExecResult[7]{\expandafter\newcommand\csname#1#2#3#4#5#6\endcsname{#7}}%
\StoreBenchExecResult{KratosII}{SvcompReachSafetySequentialized}{Status}{All}{}{Score}{0}%
\StoreBenchExecResult{KratosII}{SvcompReachSafetySequentialized}{Status}{All}{}{Count}{585}%
\StoreBenchExecResult{KratosII}{SvcompReachSafetySequentialized}{Status}{Correct}{}{Count}{15}%
\StoreBenchExecResult{KratosII}{SvcompReachSafetySequentialized}{Status}{Correct}{True}{Count}{8}%
\StoreBenchExecResult{KratosII}{SvcompReachSafetySequentialized}{Status}{Correct}{False}{Count}{7}%
\StoreBenchExecResult{KratosII}{SvcompReachSafetySequentialized}{Status}{Wrong}{}{Count}{0}%
\StoreBenchExecResult{KratosII}{SvcompReachSafetySequentialized}{Status}{Wrong}{True}{Count}{0}%
\StoreBenchExecResult{KratosII}{SvcompReachSafetySequentialized}{Status}{Wrong}{False}{Count}{0}%
\providecommand\StoreBenchExecResult[7]{\expandafter\newcommand\csname#1#2#3#4#5#6\endcsname{#7}}%
\StoreBenchExecResult{Symbiotic}{SvcompReachSafetySequentialized}{Status}{All}{}{Score}{0}%
\StoreBenchExecResult{Symbiotic}{SvcompReachSafetySequentialized}{Status}{All}{}{Count}{585}%
\StoreBenchExecResult{Symbiotic}{SvcompReachSafetySequentialized}{Status}{Correct}{}{Count}{296}%
\StoreBenchExecResult{Symbiotic}{SvcompReachSafetySequentialized}{Status}{Correct}{True}{Count}{51}%
\StoreBenchExecResult{Symbiotic}{SvcompReachSafetySequentialized}{Status}{Correct}{False}{Count}{245}%
\StoreBenchExecResult{Symbiotic}{SvcompReachSafetySequentialized}{Status}{Wrong}{}{Count}{0}%
\StoreBenchExecResult{Symbiotic}{SvcompReachSafetySequentialized}{Status}{Wrong}{True}{Count}{0}%
\StoreBenchExecResult{Symbiotic}{SvcompReachSafetySequentialized}{Status}{Wrong}{False}{Count}{0}%
\providecommand\StoreBenchExecResult[7]{\expandafter\newcommand\csname#1#2#3#4#5#6\endcsname{#7}}%
\StoreBenchExecResult{Uautomizer}{DefaultReachSafetySequentialized}{Status}{All}{}{Score}{0}%
\StoreBenchExecResult{Uautomizer}{DefaultReachSafetySequentialized}{Status}{All}{}{Count}{585}%
\StoreBenchExecResult{Uautomizer}{DefaultReachSafetySequentialized}{Status}{Correct}{}{Count}{56}%
\StoreBenchExecResult{Uautomizer}{DefaultReachSafetySequentialized}{Status}{Correct}{True}{Count}{6}%
\StoreBenchExecResult{Uautomizer}{DefaultReachSafetySequentialized}{Status}{Correct}{False}{Count}{50}%
\StoreBenchExecResult{Uautomizer}{DefaultReachSafetySequentialized}{Status}{Wrong}{}{Count}{0}%
\StoreBenchExecResult{Uautomizer}{DefaultReachSafetySequentialized}{Status}{Wrong}{True}{Count}{0}%
\StoreBenchExecResult{Uautomizer}{DefaultReachSafetySequentialized}{Status}{Wrong}{False}{Count}{0}%
\newcommand{\CpvSvcompReachSafetySequentializedBestCount}{17}
\newcommand{\CpvSvcompReachSafetySequentializedUniqCount}{8}
\newcommand{\CpacheckerSvcompReachSafetySequentializedBestCount}{166}
\newcommand{\CpacheckerSvcompReachSafetySequentializedUniqCount}{113}
\newcommand{\EsbmcKindReachSafetySequentializedBestCount}{183}
\newcommand{\EsbmcKindReachSafetySequentializedUniqCount}{43}
\newcommand{\KratosIISvcompReachSafetySequentializedBestCount}{3}
\newcommand{\KratosIISvcompReachSafetySequentializedUniqCount}{0}
\newcommand{\SymbioticSvcompReachSafetySequentializedBestCount}{153}
\newcommand{\SymbioticSvcompReachSafetySequentializedUniqCount}{5}
\newcommand{\UautomizerDefaultReachSafetySequentializedBestCount}{1}
\newcommand{\UautomizerDefaultReachSafetySequentializedUniqCount}{0}
\providecommand\StoreBenchExecResult[7]{\expandafter\newcommand\csname#1#2#3#4#5#6\endcsname{#7}}%
\StoreBenchExecResult{Vb}{SvcompReachSafetySequentialized}{Status}{All}{}{Score}{0}%
\StoreBenchExecResult{Vb}{SvcompReachSafetySequentialized}{Status}{All}{}{Count}{585}%
\StoreBenchExecResult{Vb}{SvcompReachSafetySequentialized}{Status}{Correct}{}{Count}{523}%
\StoreBenchExecResult{Vb}{SvcompReachSafetySequentialized}{Status}{Correct}{True}{Count}{143}%
\StoreBenchExecResult{Vb}{SvcompReachSafetySequentialized}{Status}{Correct}{False}{Count}{380}%
\StoreBenchExecResult{Vb}{SvcompReachSafetySequentialized}{Status}{Wrong}{}{Count}{0}%
\StoreBenchExecResult{Vb}{SvcompReachSafetySequentialized}{Status}{Wrong}{True}{Count}{0}%
\StoreBenchExecResult{Vb}{SvcompReachSafetySequentialized}{Status}{Wrong}{False}{Count}{0}%
\providecommand\StoreBenchExecResult[7]{\expandafter\newcommand\csname#1#2#3#4#5#6\endcsname{#7}}%
\StoreBenchExecResult{Cpv}{SvcompReachSafetyXCSP}{Status}{All}{}{Score}{0}%
\StoreBenchExecResult{Cpv}{SvcompReachSafetyXCSP}{Status}{All}{}{Count}{119}%
\StoreBenchExecResult{Cpv}{SvcompReachSafetyXCSP}{Status}{Correct}{}{Count}{105}%
\StoreBenchExecResult{Cpv}{SvcompReachSafetyXCSP}{Status}{Correct}{True}{Count}{52}%
\StoreBenchExecResult{Cpv}{SvcompReachSafetyXCSP}{Status}{Correct}{False}{Count}{53}%
\StoreBenchExecResult{Cpv}{SvcompReachSafetyXCSP}{Status}{Wrong}{}{Count}{0}%
\StoreBenchExecResult{Cpv}{SvcompReachSafetyXCSP}{Status}{Wrong}{True}{Count}{0}%
\StoreBenchExecResult{Cpv}{SvcompReachSafetyXCSP}{Status}{Wrong}{False}{Count}{0}%
\providecommand\StoreBenchExecResult[7]{\expandafter\newcommand\csname#1#2#3#4#5#6\endcsname{#7}}%
\StoreBenchExecResult{Cpachecker}{SvcompReachSafetyXCSP}{Status}{All}{}{Score}{0}%
\StoreBenchExecResult{Cpachecker}{SvcompReachSafetyXCSP}{Status}{All}{}{Count}{119}%
\StoreBenchExecResult{Cpachecker}{SvcompReachSafetyXCSP}{Status}{Correct}{}{Count}{101}%
\StoreBenchExecResult{Cpachecker}{SvcompReachSafetyXCSP}{Status}{Correct}{True}{Count}{52}%
\StoreBenchExecResult{Cpachecker}{SvcompReachSafetyXCSP}{Status}{Correct}{False}{Count}{49}%
\StoreBenchExecResult{Cpachecker}{SvcompReachSafetyXCSP}{Status}{Wrong}{}{Count}{0}%
\StoreBenchExecResult{Cpachecker}{SvcompReachSafetyXCSP}{Status}{Wrong}{True}{Count}{0}%
\StoreBenchExecResult{Cpachecker}{SvcompReachSafetyXCSP}{Status}{Wrong}{False}{Count}{0}%
\providecommand\StoreBenchExecResult[7]{\expandafter\newcommand\csname#1#2#3#4#5#6\endcsname{#7}}%
\StoreBenchExecResult{Esbmc}{KindReachSafetyXCSP}{Status}{All}{}{Score}{0}%
\StoreBenchExecResult{Esbmc}{KindReachSafetyXCSP}{Status}{All}{}{Count}{119}%
\StoreBenchExecResult{Esbmc}{KindReachSafetyXCSP}{Status}{Correct}{}{Count}{107}%
\StoreBenchExecResult{Esbmc}{KindReachSafetyXCSP}{Status}{Correct}{True}{Count}{53}%
\StoreBenchExecResult{Esbmc}{KindReachSafetyXCSP}{Status}{Correct}{False}{Count}{54}%
\StoreBenchExecResult{Esbmc}{KindReachSafetyXCSP}{Status}{Wrong}{}{Count}{0}%
\StoreBenchExecResult{Esbmc}{KindReachSafetyXCSP}{Status}{Wrong}{True}{Count}{0}%
\StoreBenchExecResult{Esbmc}{KindReachSafetyXCSP}{Status}{Wrong}{False}{Count}{0}%
\providecommand\StoreBenchExecResult[7]{\expandafter\newcommand\csname#1#2#3#4#5#6\endcsname{#7}}%
\StoreBenchExecResult{KratosII}{SvcompReachSafetyXCSP}{Status}{All}{}{Score}{0}%
\StoreBenchExecResult{KratosII}{SvcompReachSafetyXCSP}{Status}{All}{}{Count}{119}%
\StoreBenchExecResult{KratosII}{SvcompReachSafetyXCSP}{Status}{Correct}{}{Count}{102}%
\StoreBenchExecResult{KratosII}{SvcompReachSafetyXCSP}{Status}{Correct}{True}{Count}{50}%
\StoreBenchExecResult{KratosII}{SvcompReachSafetyXCSP}{Status}{Correct}{False}{Count}{52}%
\StoreBenchExecResult{KratosII}{SvcompReachSafetyXCSP}{Status}{Wrong}{}{Count}{0}%
\StoreBenchExecResult{KratosII}{SvcompReachSafetyXCSP}{Status}{Wrong}{True}{Count}{0}%
\StoreBenchExecResult{KratosII}{SvcompReachSafetyXCSP}{Status}{Wrong}{False}{Count}{0}%
\providecommand\StoreBenchExecResult[7]{\expandafter\newcommand\csname#1#2#3#4#5#6\endcsname{#7}}%
\StoreBenchExecResult{Symbiotic}{SvcompReachSafetyXCSP}{Status}{All}{}{Score}{0}%
\StoreBenchExecResult{Symbiotic}{SvcompReachSafetyXCSP}{Status}{All}{}{Count}{119}%
\StoreBenchExecResult{Symbiotic}{SvcompReachSafetyXCSP}{Status}{Correct}{}{Count}{91}%
\StoreBenchExecResult{Symbiotic}{SvcompReachSafetyXCSP}{Status}{Correct}{True}{Count}{50}%
\StoreBenchExecResult{Symbiotic}{SvcompReachSafetyXCSP}{Status}{Correct}{False}{Count}{41}%
\StoreBenchExecResult{Symbiotic}{SvcompReachSafetyXCSP}{Status}{Wrong}{}{Count}{0}%
\StoreBenchExecResult{Symbiotic}{SvcompReachSafetyXCSP}{Status}{Wrong}{True}{Count}{0}%
\StoreBenchExecResult{Symbiotic}{SvcompReachSafetyXCSP}{Status}{Wrong}{False}{Count}{0}%
\providecommand\StoreBenchExecResult[7]{\expandafter\newcommand\csname#1#2#3#4#5#6\endcsname{#7}}%
\StoreBenchExecResult{Uautomizer}{DefaultReachSafetyXCSP}{Status}{All}{}{Score}{0}%
\StoreBenchExecResult{Uautomizer}{DefaultReachSafetyXCSP}{Status}{All}{}{Count}{119}%
\StoreBenchExecResult{Uautomizer}{DefaultReachSafetyXCSP}{Status}{Correct}{}{Count}{49}%
\StoreBenchExecResult{Uautomizer}{DefaultReachSafetyXCSP}{Status}{Correct}{True}{Count}{3}%
\StoreBenchExecResult{Uautomizer}{DefaultReachSafetyXCSP}{Status}{Correct}{False}{Count}{46}%
\StoreBenchExecResult{Uautomizer}{DefaultReachSafetyXCSP}{Status}{Wrong}{}{Count}{0}%
\StoreBenchExecResult{Uautomizer}{DefaultReachSafetyXCSP}{Status}{Wrong}{True}{Count}{0}%
\StoreBenchExecResult{Uautomizer}{DefaultReachSafetyXCSP}{Status}{Wrong}{False}{Count}{0}%
\newcommand{\CpvSvcompReachSafetyXCSPBestCount}{5}
\newcommand{\CpvSvcompReachSafetyXCSPUniqCount}{2}
\newcommand{\CpacheckerSvcompReachSafetyXCSPBestCount}{0}
\newcommand{\CpacheckerSvcompReachSafetyXCSPUniqCount}{0}
\newcommand{\EsbmcKindReachSafetyXCSPBestCount}{104}
\newcommand{\EsbmcKindReachSafetyXCSPUniqCount}{3}
\newcommand{\KratosIISvcompReachSafetyXCSPBestCount}{0}
\newcommand{\KratosIISvcompReachSafetyXCSPUniqCount}{0}
\newcommand{\SymbioticSvcompReachSafetyXCSPBestCount}{0}
\newcommand{\SymbioticSvcompReachSafetyXCSPUniqCount}{0}
\newcommand{\UautomizerDefaultReachSafetyXCSPBestCount}{0}
\newcommand{\UautomizerDefaultReachSafetyXCSPUniqCount}{0}
\providecommand\StoreBenchExecResult[7]{\expandafter\newcommand\csname#1#2#3#4#5#6\endcsname{#7}}%
\StoreBenchExecResult{Vb}{SvcompReachSafetyXCSP}{Status}{All}{}{Score}{0}%
\StoreBenchExecResult{Vb}{SvcompReachSafetyXCSP}{Status}{All}{}{Count}{119}%
\StoreBenchExecResult{Vb}{SvcompReachSafetyXCSP}{Status}{Correct}{}{Count}{109}%
\StoreBenchExecResult{Vb}{SvcompReachSafetyXCSP}{Status}{Correct}{True}{Count}{53}%
\StoreBenchExecResult{Vb}{SvcompReachSafetyXCSP}{Status}{Correct}{False}{Count}{56}%
\StoreBenchExecResult{Vb}{SvcompReachSafetyXCSP}{Status}{Wrong}{}{Count}{0}%
\StoreBenchExecResult{Vb}{SvcompReachSafetyXCSP}{Status}{Wrong}{True}{Count}{0}%
\StoreBenchExecResult{Vb}{SvcompReachSafetyXCSP}{Status}{Wrong}{False}{Count}{0}%
\edef\CpvSvcompReachSafetyStatusAllCount{\the\numexpr\CpvSvcompReachSafetyArraysStatusAllCount+\CpvSvcompReachSafetyBitVectorsStatusAllCount+\CpvSvcompReachSafetyCombinationsStatusAllCount+\CpvSvcompReachSafetyControlFlowStatusAllCount+\CpvSvcompReachSafetyECAStatusAllCount+\CpvSvcompReachSafetyFloatsStatusAllCount+\CpvSvcompReachSafetyHardnessStatusAllCount+\CpvSvcompReachSafetyHardwareStatusAllCount+\CpvSvcompReachSafetyHeapStatusAllCount+\CpvSvcompReachSafetyLoopsStatusAllCount+\CpvSvcompReachSafetyProductLinesStatusAllCount+\CpvSvcompReachSafetySequentializedStatusAllCount+\CpvSvcompReachSafetyXCSPStatusAllCount}
\edef\CpvSvcompReachSafetyStatusCorrectCount{\the\numexpr\CpvSvcompReachSafetyArraysStatusCorrectCount+\CpvSvcompReachSafetyBitVectorsStatusCorrectCount+\CpvSvcompReachSafetyCombinationsStatusCorrectCount+\CpvSvcompReachSafetyControlFlowStatusCorrectCount+\CpvSvcompReachSafetyECAStatusCorrectCount+\CpvSvcompReachSafetyFloatsStatusCorrectCount+\CpvSvcompReachSafetyHardnessStatusCorrectCount+\CpvSvcompReachSafetyHardwareStatusCorrectCount+\CpvSvcompReachSafetyHeapStatusCorrectCount+\CpvSvcompReachSafetyLoopsStatusCorrectCount+\CpvSvcompReachSafetyProductLinesStatusCorrectCount+\CpvSvcompReachSafetySequentializedStatusCorrectCount+\CpvSvcompReachSafetyXCSPStatusCorrectCount}
\edef\CpvSvcompReachSafetyStatusCorrectTrueCount{\the\numexpr\CpvSvcompReachSafetyArraysStatusCorrectTrueCount+\CpvSvcompReachSafetyBitVectorsStatusCorrectTrueCount+\CpvSvcompReachSafetyCombinationsStatusCorrectTrueCount+\CpvSvcompReachSafetyControlFlowStatusCorrectTrueCount+\CpvSvcompReachSafetyECAStatusCorrectTrueCount+\CpvSvcompReachSafetyFloatsStatusCorrectTrueCount+\CpvSvcompReachSafetyHardnessStatusCorrectTrueCount+\CpvSvcompReachSafetyHardwareStatusCorrectTrueCount+\CpvSvcompReachSafetyHeapStatusCorrectTrueCount+\CpvSvcompReachSafetyLoopsStatusCorrectTrueCount+\CpvSvcompReachSafetyProductLinesStatusCorrectTrueCount+\CpvSvcompReachSafetySequentializedStatusCorrectTrueCount+\CpvSvcompReachSafetyXCSPStatusCorrectTrueCount}
\edef\CpvSvcompReachSafetyStatusCorrectFalseCount{\the\numexpr\CpvSvcompReachSafetyArraysStatusCorrectFalseCount+\CpvSvcompReachSafetyBitVectorsStatusCorrectFalseCount+\CpvSvcompReachSafetyCombinationsStatusCorrectFalseCount+\CpvSvcompReachSafetyControlFlowStatusCorrectFalseCount+\CpvSvcompReachSafetyECAStatusCorrectFalseCount+\CpvSvcompReachSafetyFloatsStatusCorrectFalseCount+\CpvSvcompReachSafetyHardnessStatusCorrectFalseCount+\CpvSvcompReachSafetyHardwareStatusCorrectFalseCount+\CpvSvcompReachSafetyHeapStatusCorrectFalseCount+\CpvSvcompReachSafetyLoopsStatusCorrectFalseCount+\CpvSvcompReachSafetyProductLinesStatusCorrectFalseCount+\CpvSvcompReachSafetySequentializedStatusCorrectFalseCount+\CpvSvcompReachSafetyXCSPStatusCorrectFalseCount}
\edef\CpvSvcompReachSafetyStatusWrongCount{\the\numexpr\CpvSvcompReachSafetyArraysStatusWrongCount+\CpvSvcompReachSafetyBitVectorsStatusWrongCount+\CpvSvcompReachSafetyCombinationsStatusWrongCount+\CpvSvcompReachSafetyControlFlowStatusWrongCount+\CpvSvcompReachSafetyECAStatusWrongCount+\CpvSvcompReachSafetyFloatsStatusWrongCount+\CpvSvcompReachSafetyHardnessStatusWrongCount+\CpvSvcompReachSafetyHardwareStatusWrongCount+\CpvSvcompReachSafetyHeapStatusWrongCount+\CpvSvcompReachSafetyLoopsStatusWrongCount+\CpvSvcompReachSafetyProductLinesStatusWrongCount+\CpvSvcompReachSafetySequentializedStatusWrongCount+\CpvSvcompReachSafetyXCSPStatusWrongCount}
\edef\CpvSvcompReachSafetyBestCount{\the\numexpr\CpvSvcompReachSafetyArraysBestCount+\CpvSvcompReachSafetyBitVectorsBestCount+\CpvSvcompReachSafetyCombinationsBestCount+\CpvSvcompReachSafetyControlFlowBestCount+\CpvSvcompReachSafetyECABestCount+\CpvSvcompReachSafetyFloatsBestCount+\CpvSvcompReachSafetyHardnessBestCount+\CpvSvcompReachSafetyHardwareBestCount+\CpvSvcompReachSafetyHeapBestCount+\CpvSvcompReachSafetyLoopsBestCount+\CpvSvcompReachSafetyProductLinesBestCount+\CpvSvcompReachSafetySequentializedBestCount+\CpvSvcompReachSafetyXCSPBestCount}
\edef\CpvSvcompReachSafetyUniqCount{\the\numexpr\CpvSvcompReachSafetyArraysUniqCount+\CpvSvcompReachSafetyBitVectorsUniqCount+\CpvSvcompReachSafetyCombinationsUniqCount+\CpvSvcompReachSafetyControlFlowUniqCount+\CpvSvcompReachSafetyECAUniqCount+\CpvSvcompReachSafetyFloatsUniqCount+\CpvSvcompReachSafetyHardnessUniqCount+\CpvSvcompReachSafetyHardwareUniqCount+\CpvSvcompReachSafetyHeapUniqCount+\CpvSvcompReachSafetyLoopsUniqCount+\CpvSvcompReachSafetyProductLinesUniqCount+\CpvSvcompReachSafetySequentializedUniqCount+\CpvSvcompReachSafetyXCSPUniqCount}
\edef\CpacheckerSvcompReachSafetyStatusAllCount{\the\numexpr\CpacheckerSvcompReachSafetyArraysStatusAllCount+\CpacheckerSvcompReachSafetyBitVectorsStatusAllCount+\CpacheckerSvcompReachSafetyCombinationsStatusAllCount+\CpacheckerSvcompReachSafetyControlFlowStatusAllCount+\CpacheckerSvcompReachSafetyECAStatusAllCount+\CpacheckerSvcompReachSafetyFloatsStatusAllCount+\CpacheckerSvcompReachSafetyHardnessStatusAllCount+\CpacheckerSvcompReachSafetyHardwareStatusAllCount+\CpacheckerSvcompReachSafetyHeapStatusAllCount+\CpacheckerSvcompReachSafetyLoopsStatusAllCount+\CpacheckerSvcompReachSafetyProductLinesStatusAllCount+\CpacheckerSvcompReachSafetySequentializedStatusAllCount+\CpacheckerSvcompReachSafetyXCSPStatusAllCount}
\edef\CpacheckerSvcompReachSafetyStatusCorrectCount{\the\numexpr\CpacheckerSvcompReachSafetyArraysStatusCorrectCount+\CpacheckerSvcompReachSafetyBitVectorsStatusCorrectCount+\CpacheckerSvcompReachSafetyCombinationsStatusCorrectCount+\CpacheckerSvcompReachSafetyControlFlowStatusCorrectCount+\CpacheckerSvcompReachSafetyECAStatusCorrectCount+\CpacheckerSvcompReachSafetyFloatsStatusCorrectCount+\CpacheckerSvcompReachSafetyHardnessStatusCorrectCount+\CpacheckerSvcompReachSafetyHardwareStatusCorrectCount+\CpacheckerSvcompReachSafetyHeapStatusCorrectCount+\CpacheckerSvcompReachSafetyLoopsStatusCorrectCount+\CpacheckerSvcompReachSafetyProductLinesStatusCorrectCount+\CpacheckerSvcompReachSafetySequentializedStatusCorrectCount+\CpacheckerSvcompReachSafetyXCSPStatusCorrectCount}
\edef\CpacheckerSvcompReachSafetyStatusCorrectTrueCount{\the\numexpr\CpacheckerSvcompReachSafetyArraysStatusCorrectTrueCount+\CpacheckerSvcompReachSafetyBitVectorsStatusCorrectTrueCount+\CpacheckerSvcompReachSafetyCombinationsStatusCorrectTrueCount+\CpacheckerSvcompReachSafetyControlFlowStatusCorrectTrueCount+\CpacheckerSvcompReachSafetyECAStatusCorrectTrueCount+\CpacheckerSvcompReachSafetyFloatsStatusCorrectTrueCount+\CpacheckerSvcompReachSafetyHardnessStatusCorrectTrueCount+\CpacheckerSvcompReachSafetyHardwareStatusCorrectTrueCount+\CpacheckerSvcompReachSafetyHeapStatusCorrectTrueCount+\CpacheckerSvcompReachSafetyLoopsStatusCorrectTrueCount+\CpacheckerSvcompReachSafetyProductLinesStatusCorrectTrueCount+\CpacheckerSvcompReachSafetySequentializedStatusCorrectTrueCount+\CpacheckerSvcompReachSafetyXCSPStatusCorrectTrueCount}
\edef\CpacheckerSvcompReachSafetyStatusCorrectFalseCount{\the\numexpr\CpacheckerSvcompReachSafetyArraysStatusCorrectFalseCount+\CpacheckerSvcompReachSafetyBitVectorsStatusCorrectFalseCount+\CpacheckerSvcompReachSafetyCombinationsStatusCorrectFalseCount+\CpacheckerSvcompReachSafetyControlFlowStatusCorrectFalseCount+\CpacheckerSvcompReachSafetyECAStatusCorrectFalseCount+\CpacheckerSvcompReachSafetyFloatsStatusCorrectFalseCount+\CpacheckerSvcompReachSafetyHardnessStatusCorrectFalseCount+\CpacheckerSvcompReachSafetyHardwareStatusCorrectFalseCount+\CpacheckerSvcompReachSafetyHeapStatusCorrectFalseCount+\CpacheckerSvcompReachSafetyLoopsStatusCorrectFalseCount+\CpacheckerSvcompReachSafetyProductLinesStatusCorrectFalseCount+\CpacheckerSvcompReachSafetySequentializedStatusCorrectFalseCount+\CpacheckerSvcompReachSafetyXCSPStatusCorrectFalseCount}
\edef\CpacheckerSvcompReachSafetyStatusWrongCount{\the\numexpr\CpacheckerSvcompReachSafetyArraysStatusWrongCount+\CpacheckerSvcompReachSafetyBitVectorsStatusWrongCount+\CpacheckerSvcompReachSafetyCombinationsStatusWrongCount+\CpacheckerSvcompReachSafetyControlFlowStatusWrongCount+\CpacheckerSvcompReachSafetyECAStatusWrongCount+\CpacheckerSvcompReachSafetyFloatsStatusWrongCount+\CpacheckerSvcompReachSafetyHardnessStatusWrongCount+\CpacheckerSvcompReachSafetyHardwareStatusWrongCount+\CpacheckerSvcompReachSafetyHeapStatusWrongCount+\CpacheckerSvcompReachSafetyLoopsStatusWrongCount+\CpacheckerSvcompReachSafetyProductLinesStatusWrongCount+\CpacheckerSvcompReachSafetySequentializedStatusWrongCount+\CpacheckerSvcompReachSafetyXCSPStatusWrongCount}
\edef\CpacheckerSvcompReachSafetyBestCount{\the\numexpr\CpacheckerSvcompReachSafetyArraysBestCount+\CpacheckerSvcompReachSafetyBitVectorsBestCount+\CpacheckerSvcompReachSafetyCombinationsBestCount+\CpacheckerSvcompReachSafetyControlFlowBestCount+\CpacheckerSvcompReachSafetyECABestCount+\CpacheckerSvcompReachSafetyFloatsBestCount+\CpacheckerSvcompReachSafetyHardnessBestCount+\CpacheckerSvcompReachSafetyHardwareBestCount+\CpacheckerSvcompReachSafetyHeapBestCount+\CpacheckerSvcompReachSafetyLoopsBestCount+\CpacheckerSvcompReachSafetyProductLinesBestCount+\CpacheckerSvcompReachSafetySequentializedBestCount+\CpacheckerSvcompReachSafetyXCSPBestCount}
\edef\CpacheckerSvcompReachSafetyUniqCount{\the\numexpr\CpacheckerSvcompReachSafetyArraysUniqCount+\CpacheckerSvcompReachSafetyBitVectorsUniqCount+\CpacheckerSvcompReachSafetyCombinationsUniqCount+\CpacheckerSvcompReachSafetyControlFlowUniqCount+\CpacheckerSvcompReachSafetyECAUniqCount+\CpacheckerSvcompReachSafetyFloatsUniqCount+\CpacheckerSvcompReachSafetyHardnessUniqCount+\CpacheckerSvcompReachSafetyHardwareUniqCount+\CpacheckerSvcompReachSafetyHeapUniqCount+\CpacheckerSvcompReachSafetyLoopsUniqCount+\CpacheckerSvcompReachSafetyProductLinesUniqCount+\CpacheckerSvcompReachSafetySequentializedUniqCount+\CpacheckerSvcompReachSafetyXCSPUniqCount}
\edef\EsbmcKindReachSafetyStatusAllCount{\the\numexpr\EsbmcKindReachSafetyArraysStatusAllCount+\EsbmcKindReachSafetyBitVectorsStatusAllCount+\EsbmcKindReachSafetyCombinationsStatusAllCount+\EsbmcKindReachSafetyControlFlowStatusAllCount+\EsbmcKindReachSafetyECAStatusAllCount+\EsbmcKindReachSafetyFloatsStatusAllCount+\EsbmcKindReachSafetyHardnessStatusAllCount+\EsbmcKindReachSafetyHardwareStatusAllCount+\EsbmcKindReachSafetyHeapStatusAllCount+\EsbmcKindReachSafetyLoopsStatusAllCount+\EsbmcKindReachSafetyProductLinesStatusAllCount+\EsbmcKindReachSafetySequentializedStatusAllCount+\EsbmcKindReachSafetyXCSPStatusAllCount}
\edef\EsbmcKindReachSafetyStatusCorrectCount{\the\numexpr\EsbmcKindReachSafetyArraysStatusCorrectCount+\EsbmcKindReachSafetyBitVectorsStatusCorrectCount+\EsbmcKindReachSafetyCombinationsStatusCorrectCount+\EsbmcKindReachSafetyControlFlowStatusCorrectCount+\EsbmcKindReachSafetyECAStatusCorrectCount+\EsbmcKindReachSafetyFloatsStatusCorrectCount+\EsbmcKindReachSafetyHardnessStatusCorrectCount+\EsbmcKindReachSafetyHardwareStatusCorrectCount+\EsbmcKindReachSafetyHeapStatusCorrectCount+\EsbmcKindReachSafetyLoopsStatusCorrectCount+\EsbmcKindReachSafetyProductLinesStatusCorrectCount+\EsbmcKindReachSafetySequentializedStatusCorrectCount+\EsbmcKindReachSafetyXCSPStatusCorrectCount}
\edef\EsbmcKindReachSafetyStatusCorrectTrueCount{\the\numexpr\EsbmcKindReachSafetyArraysStatusCorrectTrueCount+\EsbmcKindReachSafetyBitVectorsStatusCorrectTrueCount+\EsbmcKindReachSafetyCombinationsStatusCorrectTrueCount+\EsbmcKindReachSafetyControlFlowStatusCorrectTrueCount+\EsbmcKindReachSafetyECAStatusCorrectTrueCount+\EsbmcKindReachSafetyFloatsStatusCorrectTrueCount+\EsbmcKindReachSafetyHardnessStatusCorrectTrueCount+\EsbmcKindReachSafetyHardwareStatusCorrectTrueCount+\EsbmcKindReachSafetyHeapStatusCorrectTrueCount+\EsbmcKindReachSafetyLoopsStatusCorrectTrueCount+\EsbmcKindReachSafetyProductLinesStatusCorrectTrueCount+\EsbmcKindReachSafetySequentializedStatusCorrectTrueCount+\EsbmcKindReachSafetyXCSPStatusCorrectTrueCount}
\edef\EsbmcKindReachSafetyStatusCorrectFalseCount{\the\numexpr\EsbmcKindReachSafetyArraysStatusCorrectFalseCount+\EsbmcKindReachSafetyBitVectorsStatusCorrectFalseCount+\EsbmcKindReachSafetyCombinationsStatusCorrectFalseCount+\EsbmcKindReachSafetyControlFlowStatusCorrectFalseCount+\EsbmcKindReachSafetyECAStatusCorrectFalseCount+\EsbmcKindReachSafetyFloatsStatusCorrectFalseCount+\EsbmcKindReachSafetyHardnessStatusCorrectFalseCount+\EsbmcKindReachSafetyHardwareStatusCorrectFalseCount+\EsbmcKindReachSafetyHeapStatusCorrectFalseCount+\EsbmcKindReachSafetyLoopsStatusCorrectFalseCount+\EsbmcKindReachSafetyProductLinesStatusCorrectFalseCount+\EsbmcKindReachSafetySequentializedStatusCorrectFalseCount+\EsbmcKindReachSafetyXCSPStatusCorrectFalseCount}
\edef\EsbmcKindReachSafetyStatusWrongCount{\the\numexpr\EsbmcKindReachSafetyArraysStatusWrongCount+\EsbmcKindReachSafetyBitVectorsStatusWrongCount+\EsbmcKindReachSafetyCombinationsStatusWrongCount+\EsbmcKindReachSafetyControlFlowStatusWrongCount+\EsbmcKindReachSafetyECAStatusWrongCount+\EsbmcKindReachSafetyFloatsStatusWrongCount+\EsbmcKindReachSafetyHardnessStatusWrongCount+\EsbmcKindReachSafetyHardwareStatusWrongCount+\EsbmcKindReachSafetyHeapStatusWrongCount+\EsbmcKindReachSafetyLoopsStatusWrongCount+\EsbmcKindReachSafetyProductLinesStatusWrongCount+\EsbmcKindReachSafetySequentializedStatusWrongCount+\EsbmcKindReachSafetyXCSPStatusWrongCount}
\edef\EsbmcKindReachSafetyBestCount{\the\numexpr\EsbmcKindReachSafetyArraysBestCount+\EsbmcKindReachSafetyBitVectorsBestCount+\EsbmcKindReachSafetyCombinationsBestCount+\EsbmcKindReachSafetyControlFlowBestCount+\EsbmcKindReachSafetyECABestCount+\EsbmcKindReachSafetyFloatsBestCount+\EsbmcKindReachSafetyHardnessBestCount+\EsbmcKindReachSafetyHardwareBestCount+\EsbmcKindReachSafetyHeapBestCount+\EsbmcKindReachSafetyLoopsBestCount+\EsbmcKindReachSafetyProductLinesBestCount+\EsbmcKindReachSafetySequentializedBestCount+\EsbmcKindReachSafetyXCSPBestCount}
\edef\EsbmcKindReachSafetyUniqCount{\the\numexpr\EsbmcKindReachSafetyArraysUniqCount+\EsbmcKindReachSafetyBitVectorsUniqCount+\EsbmcKindReachSafetyCombinationsUniqCount+\EsbmcKindReachSafetyControlFlowUniqCount+\EsbmcKindReachSafetyECAUniqCount+\EsbmcKindReachSafetyFloatsUniqCount+\EsbmcKindReachSafetyHardnessUniqCount+\EsbmcKindReachSafetyHardwareUniqCount+\EsbmcKindReachSafetyHeapUniqCount+\EsbmcKindReachSafetyLoopsUniqCount+\EsbmcKindReachSafetyProductLinesUniqCount+\EsbmcKindReachSafetySequentializedUniqCount+\EsbmcKindReachSafetyXCSPUniqCount}
\edef\KratosIISvcompReachSafetyStatusAllCount{\the\numexpr\KratosIISvcompReachSafetyArraysStatusAllCount+\KratosIISvcompReachSafetyBitVectorsStatusAllCount+\KratosIISvcompReachSafetyCombinationsStatusAllCount+\KratosIISvcompReachSafetyControlFlowStatusAllCount+\KratosIISvcompReachSafetyECAStatusAllCount+\KratosIISvcompReachSafetyFloatsStatusAllCount+\KratosIISvcompReachSafetyHardnessStatusAllCount+\KratosIISvcompReachSafetyHardwareStatusAllCount+\KratosIISvcompReachSafetyHeapStatusAllCount+\KratosIISvcompReachSafetyLoopsStatusAllCount+\KratosIISvcompReachSafetyProductLinesStatusAllCount+\KratosIISvcompReachSafetySequentializedStatusAllCount+\KratosIISvcompReachSafetyXCSPStatusAllCount}
\edef\KratosIISvcompReachSafetyStatusCorrectCount{\the\numexpr\KratosIISvcompReachSafetyArraysStatusCorrectCount+\KratosIISvcompReachSafetyBitVectorsStatusCorrectCount+\KratosIISvcompReachSafetyCombinationsStatusCorrectCount+\KratosIISvcompReachSafetyControlFlowStatusCorrectCount+\KratosIISvcompReachSafetyECAStatusCorrectCount+\KratosIISvcompReachSafetyFloatsStatusCorrectCount+\KratosIISvcompReachSafetyHardnessStatusCorrectCount+\KratosIISvcompReachSafetyHardwareStatusCorrectCount+\KratosIISvcompReachSafetyHeapStatusCorrectCount+\KratosIISvcompReachSafetyLoopsStatusCorrectCount+\KratosIISvcompReachSafetyProductLinesStatusCorrectCount+\KratosIISvcompReachSafetySequentializedStatusCorrectCount+\KratosIISvcompReachSafetyXCSPStatusCorrectCount}
\edef\KratosIISvcompReachSafetyStatusCorrectTrueCount{\the\numexpr\KratosIISvcompReachSafetyArraysStatusCorrectTrueCount+\KratosIISvcompReachSafetyBitVectorsStatusCorrectTrueCount+\KratosIISvcompReachSafetyCombinationsStatusCorrectTrueCount+\KratosIISvcompReachSafetyControlFlowStatusCorrectTrueCount+\KratosIISvcompReachSafetyECAStatusCorrectTrueCount+\KratosIISvcompReachSafetyFloatsStatusCorrectTrueCount+\KratosIISvcompReachSafetyHardnessStatusCorrectTrueCount+\KratosIISvcompReachSafetyHardwareStatusCorrectTrueCount+\KratosIISvcompReachSafetyHeapStatusCorrectTrueCount+\KratosIISvcompReachSafetyLoopsStatusCorrectTrueCount+\KratosIISvcompReachSafetyProductLinesStatusCorrectTrueCount+\KratosIISvcompReachSafetySequentializedStatusCorrectTrueCount+\KratosIISvcompReachSafetyXCSPStatusCorrectTrueCount}
\edef\KratosIISvcompReachSafetyStatusCorrectFalseCount{\the\numexpr\KratosIISvcompReachSafetyArraysStatusCorrectFalseCount+\KratosIISvcompReachSafetyBitVectorsStatusCorrectFalseCount+\KratosIISvcompReachSafetyCombinationsStatusCorrectFalseCount+\KratosIISvcompReachSafetyControlFlowStatusCorrectFalseCount+\KratosIISvcompReachSafetyECAStatusCorrectFalseCount+\KratosIISvcompReachSafetyFloatsStatusCorrectFalseCount+\KratosIISvcompReachSafetyHardnessStatusCorrectFalseCount+\KratosIISvcompReachSafetyHardwareStatusCorrectFalseCount+\KratosIISvcompReachSafetyHeapStatusCorrectFalseCount+\KratosIISvcompReachSafetyLoopsStatusCorrectFalseCount+\KratosIISvcompReachSafetyProductLinesStatusCorrectFalseCount+\KratosIISvcompReachSafetySequentializedStatusCorrectFalseCount+\KratosIISvcompReachSafetyXCSPStatusCorrectFalseCount}
\edef\KratosIISvcompReachSafetyStatusWrongCount{\the\numexpr\KratosIISvcompReachSafetyArraysStatusWrongCount+\KratosIISvcompReachSafetyBitVectorsStatusWrongCount+\KratosIISvcompReachSafetyCombinationsStatusWrongCount+\KratosIISvcompReachSafetyControlFlowStatusWrongCount+\KratosIISvcompReachSafetyECAStatusWrongCount+\KratosIISvcompReachSafetyFloatsStatusWrongCount+\KratosIISvcompReachSafetyHardnessStatusWrongCount+\KratosIISvcompReachSafetyHardwareStatusWrongCount+\KratosIISvcompReachSafetyHeapStatusWrongCount+\KratosIISvcompReachSafetyLoopsStatusWrongCount+\KratosIISvcompReachSafetyProductLinesStatusWrongCount+\KratosIISvcompReachSafetySequentializedStatusWrongCount+\KratosIISvcompReachSafetyXCSPStatusWrongCount}
\edef\KratosIISvcompReachSafetyBestCount{\the\numexpr\KratosIISvcompReachSafetyArraysBestCount+\KratosIISvcompReachSafetyBitVectorsBestCount+\KratosIISvcompReachSafetyCombinationsBestCount+\KratosIISvcompReachSafetyControlFlowBestCount+\KratosIISvcompReachSafetyECABestCount+\KratosIISvcompReachSafetyFloatsBestCount+\KratosIISvcompReachSafetyHardnessBestCount+\KratosIISvcompReachSafetyHardwareBestCount+\KratosIISvcompReachSafetyHeapBestCount+\KratosIISvcompReachSafetyLoopsBestCount+\KratosIISvcompReachSafetyProductLinesBestCount+\KratosIISvcompReachSafetySequentializedBestCount+\KratosIISvcompReachSafetyXCSPBestCount}
\edef\KratosIISvcompReachSafetyUniqCount{\the\numexpr\KratosIISvcompReachSafetyArraysUniqCount+\KratosIISvcompReachSafetyBitVectorsUniqCount+\KratosIISvcompReachSafetyCombinationsUniqCount+\KratosIISvcompReachSafetyControlFlowUniqCount+\KratosIISvcompReachSafetyECAUniqCount+\KratosIISvcompReachSafetyFloatsUniqCount+\KratosIISvcompReachSafetyHardnessUniqCount+\KratosIISvcompReachSafetyHardwareUniqCount+\KratosIISvcompReachSafetyHeapUniqCount+\KratosIISvcompReachSafetyLoopsUniqCount+\KratosIISvcompReachSafetyProductLinesUniqCount+\KratosIISvcompReachSafetySequentializedUniqCount+\KratosIISvcompReachSafetyXCSPUniqCount}
\edef\SymbioticSvcompReachSafetyStatusAllCount{\the\numexpr\SymbioticSvcompReachSafetyArraysStatusAllCount+\SymbioticSvcompReachSafetyBitVectorsStatusAllCount+\SymbioticSvcompReachSafetyCombinationsStatusAllCount+\SymbioticSvcompReachSafetyControlFlowStatusAllCount+\SymbioticSvcompReachSafetyECAStatusAllCount+\SymbioticSvcompReachSafetyFloatsStatusAllCount+\SymbioticSvcompReachSafetyHardnessStatusAllCount+\SymbioticSvcompReachSafetyHardwareStatusAllCount+\SymbioticSvcompReachSafetyHeapStatusAllCount+\SymbioticSvcompReachSafetyLoopsStatusAllCount+\SymbioticSvcompReachSafetyProductLinesStatusAllCount+\SymbioticSvcompReachSafetySequentializedStatusAllCount+\SymbioticSvcompReachSafetyXCSPStatusAllCount}
\edef\SymbioticSvcompReachSafetyStatusCorrectCount{\the\numexpr\SymbioticSvcompReachSafetyArraysStatusCorrectCount+\SymbioticSvcompReachSafetyBitVectorsStatusCorrectCount+\SymbioticSvcompReachSafetyCombinationsStatusCorrectCount+\SymbioticSvcompReachSafetyControlFlowStatusCorrectCount+\SymbioticSvcompReachSafetyECAStatusCorrectCount+\SymbioticSvcompReachSafetyFloatsStatusCorrectCount+\SymbioticSvcompReachSafetyHardnessStatusCorrectCount+\SymbioticSvcompReachSafetyHardwareStatusCorrectCount+\SymbioticSvcompReachSafetyHeapStatusCorrectCount+\SymbioticSvcompReachSafetyLoopsStatusCorrectCount+\SymbioticSvcompReachSafetyProductLinesStatusCorrectCount+\SymbioticSvcompReachSafetySequentializedStatusCorrectCount+\SymbioticSvcompReachSafetyXCSPStatusCorrectCount}
\edef\SymbioticSvcompReachSafetyStatusCorrectTrueCount{\the\numexpr\SymbioticSvcompReachSafetyArraysStatusCorrectTrueCount+\SymbioticSvcompReachSafetyBitVectorsStatusCorrectTrueCount+\SymbioticSvcompReachSafetyCombinationsStatusCorrectTrueCount+\SymbioticSvcompReachSafetyControlFlowStatusCorrectTrueCount+\SymbioticSvcompReachSafetyECAStatusCorrectTrueCount+\SymbioticSvcompReachSafetyFloatsStatusCorrectTrueCount+\SymbioticSvcompReachSafetyHardnessStatusCorrectTrueCount+\SymbioticSvcompReachSafetyHardwareStatusCorrectTrueCount+\SymbioticSvcompReachSafetyHeapStatusCorrectTrueCount+\SymbioticSvcompReachSafetyLoopsStatusCorrectTrueCount+\SymbioticSvcompReachSafetyProductLinesStatusCorrectTrueCount+\SymbioticSvcompReachSafetySequentializedStatusCorrectTrueCount+\SymbioticSvcompReachSafetyXCSPStatusCorrectTrueCount}
\edef\SymbioticSvcompReachSafetyStatusCorrectFalseCount{\the\numexpr\SymbioticSvcompReachSafetyArraysStatusCorrectFalseCount+\SymbioticSvcompReachSafetyBitVectorsStatusCorrectFalseCount+\SymbioticSvcompReachSafetyCombinationsStatusCorrectFalseCount+\SymbioticSvcompReachSafetyControlFlowStatusCorrectFalseCount+\SymbioticSvcompReachSafetyECAStatusCorrectFalseCount+\SymbioticSvcompReachSafetyFloatsStatusCorrectFalseCount+\SymbioticSvcompReachSafetyHardnessStatusCorrectFalseCount+\SymbioticSvcompReachSafetyHardwareStatusCorrectFalseCount+\SymbioticSvcompReachSafetyHeapStatusCorrectFalseCount+\SymbioticSvcompReachSafetyLoopsStatusCorrectFalseCount+\SymbioticSvcompReachSafetyProductLinesStatusCorrectFalseCount+\SymbioticSvcompReachSafetySequentializedStatusCorrectFalseCount+\SymbioticSvcompReachSafetyXCSPStatusCorrectFalseCount}
\edef\SymbioticSvcompReachSafetyStatusWrongCount{\the\numexpr\SymbioticSvcompReachSafetyArraysStatusWrongCount+\SymbioticSvcompReachSafetyBitVectorsStatusWrongCount+\SymbioticSvcompReachSafetyCombinationsStatusWrongCount+\SymbioticSvcompReachSafetyControlFlowStatusWrongCount+\SymbioticSvcompReachSafetyECAStatusWrongCount+\SymbioticSvcompReachSafetyFloatsStatusWrongCount+\SymbioticSvcompReachSafetyHardnessStatusWrongCount+\SymbioticSvcompReachSafetyHardwareStatusWrongCount+\SymbioticSvcompReachSafetyHeapStatusWrongCount+\SymbioticSvcompReachSafetyLoopsStatusWrongCount+\SymbioticSvcompReachSafetyProductLinesStatusWrongCount+\SymbioticSvcompReachSafetySequentializedStatusWrongCount+\SymbioticSvcompReachSafetyXCSPStatusWrongCount}
\edef\SymbioticSvcompReachSafetyBestCount{\the\numexpr\SymbioticSvcompReachSafetyArraysBestCount+\SymbioticSvcompReachSafetyBitVectorsBestCount+\SymbioticSvcompReachSafetyCombinationsBestCount+\SymbioticSvcompReachSafetyControlFlowBestCount+\SymbioticSvcompReachSafetyECABestCount+\SymbioticSvcompReachSafetyFloatsBestCount+\SymbioticSvcompReachSafetyHardnessBestCount+\SymbioticSvcompReachSafetyHardwareBestCount+\SymbioticSvcompReachSafetyHeapBestCount+\SymbioticSvcompReachSafetyLoopsBestCount+\SymbioticSvcompReachSafetyProductLinesBestCount+\SymbioticSvcompReachSafetySequentializedBestCount+\SymbioticSvcompReachSafetyXCSPBestCount}
\edef\SymbioticSvcompReachSafetyUniqCount{\the\numexpr\SymbioticSvcompReachSafetyArraysUniqCount+\SymbioticSvcompReachSafetyBitVectorsUniqCount+\SymbioticSvcompReachSafetyCombinationsUniqCount+\SymbioticSvcompReachSafetyControlFlowUniqCount+\SymbioticSvcompReachSafetyECAUniqCount+\SymbioticSvcompReachSafetyFloatsUniqCount+\SymbioticSvcompReachSafetyHardnessUniqCount+\SymbioticSvcompReachSafetyHardwareUniqCount+\SymbioticSvcompReachSafetyHeapUniqCount+\SymbioticSvcompReachSafetyLoopsUniqCount+\SymbioticSvcompReachSafetyProductLinesUniqCount+\SymbioticSvcompReachSafetySequentializedUniqCount+\SymbioticSvcompReachSafetyXCSPUniqCount}
\edef\UautomizerDefaultReachSafetyStatusAllCount{\the\numexpr\UautomizerDefaultReachSafetyArraysStatusAllCount+\UautomizerDefaultReachSafetyBitVectorsStatusAllCount+\UautomizerDefaultReachSafetyCombinationsStatusAllCount+\UautomizerDefaultReachSafetyControlFlowStatusAllCount+\UautomizerDefaultReachSafetyECAStatusAllCount+\UautomizerDefaultReachSafetyFloatsStatusAllCount+\UautomizerDefaultReachSafetyHardnessStatusAllCount+\UautomizerDefaultReachSafetyHardwareStatusAllCount+\UautomizerDefaultReachSafetyHeapStatusAllCount+\UautomizerDefaultReachSafetyLoopsStatusAllCount+\UautomizerDefaultReachSafetyProductLinesStatusAllCount+\UautomizerDefaultReachSafetySequentializedStatusAllCount+\UautomizerDefaultReachSafetyXCSPStatusAllCount}
\edef\UautomizerDefaultReachSafetyStatusCorrectCount{\the\numexpr\UautomizerDefaultReachSafetyArraysStatusCorrectCount+\UautomizerDefaultReachSafetyBitVectorsStatusCorrectCount+\UautomizerDefaultReachSafetyCombinationsStatusCorrectCount+\UautomizerDefaultReachSafetyControlFlowStatusCorrectCount+\UautomizerDefaultReachSafetyECAStatusCorrectCount+\UautomizerDefaultReachSafetyFloatsStatusCorrectCount+\UautomizerDefaultReachSafetyHardnessStatusCorrectCount+\UautomizerDefaultReachSafetyHardwareStatusCorrectCount+\UautomizerDefaultReachSafetyHeapStatusCorrectCount+\UautomizerDefaultReachSafetyLoopsStatusCorrectCount+\UautomizerDefaultReachSafetyProductLinesStatusCorrectCount+\UautomizerDefaultReachSafetySequentializedStatusCorrectCount+\UautomizerDefaultReachSafetyXCSPStatusCorrectCount}
\edef\UautomizerDefaultReachSafetyStatusCorrectTrueCount{\the\numexpr\UautomizerDefaultReachSafetyArraysStatusCorrectTrueCount+\UautomizerDefaultReachSafetyBitVectorsStatusCorrectTrueCount+\UautomizerDefaultReachSafetyCombinationsStatusCorrectTrueCount+\UautomizerDefaultReachSafetyControlFlowStatusCorrectTrueCount+\UautomizerDefaultReachSafetyECAStatusCorrectTrueCount+\UautomizerDefaultReachSafetyFloatsStatusCorrectTrueCount+\UautomizerDefaultReachSafetyHardnessStatusCorrectTrueCount+\UautomizerDefaultReachSafetyHardwareStatusCorrectTrueCount+\UautomizerDefaultReachSafetyHeapStatusCorrectTrueCount+\UautomizerDefaultReachSafetyLoopsStatusCorrectTrueCount+\UautomizerDefaultReachSafetyProductLinesStatusCorrectTrueCount+\UautomizerDefaultReachSafetySequentializedStatusCorrectTrueCount+\UautomizerDefaultReachSafetyXCSPStatusCorrectTrueCount}
\edef\UautomizerDefaultReachSafetyStatusCorrectFalseCount{\the\numexpr\UautomizerDefaultReachSafetyArraysStatusCorrectFalseCount+\UautomizerDefaultReachSafetyBitVectorsStatusCorrectFalseCount+\UautomizerDefaultReachSafetyCombinationsStatusCorrectFalseCount+\UautomizerDefaultReachSafetyControlFlowStatusCorrectFalseCount+\UautomizerDefaultReachSafetyECAStatusCorrectFalseCount+\UautomizerDefaultReachSafetyFloatsStatusCorrectFalseCount+\UautomizerDefaultReachSafetyHardnessStatusCorrectFalseCount+\UautomizerDefaultReachSafetyHardwareStatusCorrectFalseCount+\UautomizerDefaultReachSafetyHeapStatusCorrectFalseCount+\UautomizerDefaultReachSafetyLoopsStatusCorrectFalseCount+\UautomizerDefaultReachSafetyProductLinesStatusCorrectFalseCount+\UautomizerDefaultReachSafetySequentializedStatusCorrectFalseCount+\UautomizerDefaultReachSafetyXCSPStatusCorrectFalseCount}
\edef\UautomizerDefaultReachSafetyStatusWrongCount{\the\numexpr\UautomizerDefaultReachSafetyArraysStatusWrongCount+\UautomizerDefaultReachSafetyBitVectorsStatusWrongCount+\UautomizerDefaultReachSafetyCombinationsStatusWrongCount+\UautomizerDefaultReachSafetyControlFlowStatusWrongCount+\UautomizerDefaultReachSafetyECAStatusWrongCount+\UautomizerDefaultReachSafetyFloatsStatusWrongCount+\UautomizerDefaultReachSafetyHardnessStatusWrongCount+\UautomizerDefaultReachSafetyHardwareStatusWrongCount+\UautomizerDefaultReachSafetyHeapStatusWrongCount+\UautomizerDefaultReachSafetyLoopsStatusWrongCount+\UautomizerDefaultReachSafetyProductLinesStatusWrongCount+\UautomizerDefaultReachSafetySequentializedStatusWrongCount+\UautomizerDefaultReachSafetyXCSPStatusWrongCount}
\edef\UautomizerDefaultReachSafetyBestCount{\the\numexpr\UautomizerDefaultReachSafetyArraysBestCount+\UautomizerDefaultReachSafetyBitVectorsBestCount+\UautomizerDefaultReachSafetyCombinationsBestCount+\UautomizerDefaultReachSafetyControlFlowBestCount+\UautomizerDefaultReachSafetyECABestCount+\UautomizerDefaultReachSafetyFloatsBestCount+\UautomizerDefaultReachSafetyHardnessBestCount+\UautomizerDefaultReachSafetyHardwareBestCount+\UautomizerDefaultReachSafetyHeapBestCount+\UautomizerDefaultReachSafetyLoopsBestCount+\UautomizerDefaultReachSafetyProductLinesBestCount+\UautomizerDefaultReachSafetySequentializedBestCount+\UautomizerDefaultReachSafetyXCSPBestCount}
\edef\UautomizerDefaultReachSafetyUniqCount{\the\numexpr\UautomizerDefaultReachSafetyArraysUniqCount+\UautomizerDefaultReachSafetyBitVectorsUniqCount+\UautomizerDefaultReachSafetyCombinationsUniqCount+\UautomizerDefaultReachSafetyControlFlowUniqCount+\UautomizerDefaultReachSafetyECAUniqCount+\UautomizerDefaultReachSafetyFloatsUniqCount+\UautomizerDefaultReachSafetyHardnessUniqCount+\UautomizerDefaultReachSafetyHardwareUniqCount+\UautomizerDefaultReachSafetyHeapUniqCount+\UautomizerDefaultReachSafetyLoopsUniqCount+\UautomizerDefaultReachSafetyProductLinesUniqCount+\UautomizerDefaultReachSafetySequentializedUniqCount+\UautomizerDefaultReachSafetyXCSPUniqCount}
\edef\VbSvcompReachSafetyStatusAllCount{\the\numexpr\VbSvcompReachSafetyArraysStatusAllCount+\VbSvcompReachSafetyBitVectorsStatusAllCount+\VbSvcompReachSafetyCombinationsStatusAllCount+\VbSvcompReachSafetyControlFlowStatusAllCount+\VbSvcompReachSafetyECAStatusAllCount+\VbSvcompReachSafetyFloatsStatusAllCount+\VbSvcompReachSafetyHardnessStatusAllCount+\VbSvcompReachSafetyHardwareStatusAllCount+\VbSvcompReachSafetyHeapStatusAllCount+\VbSvcompReachSafetyLoopsStatusAllCount+\VbSvcompReachSafetyProductLinesStatusAllCount+\VbSvcompReachSafetySequentializedStatusAllCount+\VbSvcompReachSafetyXCSPStatusAllCount}
\edef\VbSvcompReachSafetyStatusCorrectCount{\the\numexpr\VbSvcompReachSafetyArraysStatusCorrectCount+\VbSvcompReachSafetyBitVectorsStatusCorrectCount+\VbSvcompReachSafetyCombinationsStatusCorrectCount+\VbSvcompReachSafetyControlFlowStatusCorrectCount+\VbSvcompReachSafetyECAStatusCorrectCount+\VbSvcompReachSafetyFloatsStatusCorrectCount+\VbSvcompReachSafetyHardnessStatusCorrectCount+\VbSvcompReachSafetyHardwareStatusCorrectCount+\VbSvcompReachSafetyHeapStatusCorrectCount+\VbSvcompReachSafetyLoopsStatusCorrectCount+\VbSvcompReachSafetyProductLinesStatusCorrectCount+\VbSvcompReachSafetySequentializedStatusCorrectCount+\VbSvcompReachSafetyXCSPStatusCorrectCount}
\edef\VbSvcompReachSafetyStatusCorrectTrueCount{\the\numexpr\VbSvcompReachSafetyArraysStatusCorrectTrueCount+\VbSvcompReachSafetyBitVectorsStatusCorrectTrueCount+\VbSvcompReachSafetyCombinationsStatusCorrectTrueCount+\VbSvcompReachSafetyControlFlowStatusCorrectTrueCount+\VbSvcompReachSafetyECAStatusCorrectTrueCount+\VbSvcompReachSafetyFloatsStatusCorrectTrueCount+\VbSvcompReachSafetyHardnessStatusCorrectTrueCount+\VbSvcompReachSafetyHardwareStatusCorrectTrueCount+\VbSvcompReachSafetyHeapStatusCorrectTrueCount+\VbSvcompReachSafetyLoopsStatusCorrectTrueCount+\VbSvcompReachSafetyProductLinesStatusCorrectTrueCount+\VbSvcompReachSafetySequentializedStatusCorrectTrueCount+\VbSvcompReachSafetyXCSPStatusCorrectTrueCount}
\edef\VbSvcompReachSafetyStatusCorrectFalseCount{\the\numexpr\VbSvcompReachSafetyArraysStatusCorrectFalseCount+\VbSvcompReachSafetyBitVectorsStatusCorrectFalseCount+\VbSvcompReachSafetyCombinationsStatusCorrectFalseCount+\VbSvcompReachSafetyControlFlowStatusCorrectFalseCount+\VbSvcompReachSafetyECAStatusCorrectFalseCount+\VbSvcompReachSafetyFloatsStatusCorrectFalseCount+\VbSvcompReachSafetyHardnessStatusCorrectFalseCount+\VbSvcompReachSafetyHardwareStatusCorrectFalseCount+\VbSvcompReachSafetyHeapStatusCorrectFalseCount+\VbSvcompReachSafetyLoopsStatusCorrectFalseCount+\VbSvcompReachSafetyProductLinesStatusCorrectFalseCount+\VbSvcompReachSafetySequentializedStatusCorrectFalseCount+\VbSvcompReachSafetyXCSPStatusCorrectFalseCount}
\edef\VbSvcompReachSafetyStatusWrongCount{\the\numexpr\VbSvcompReachSafetyArraysStatusWrongCount+\VbSvcompReachSafetyBitVectorsStatusWrongCount+\VbSvcompReachSafetyCombinationsStatusWrongCount+\VbSvcompReachSafetyControlFlowStatusWrongCount+\VbSvcompReachSafetyECAStatusWrongCount+\VbSvcompReachSafetyFloatsStatusWrongCount+\VbSvcompReachSafetyHardnessStatusWrongCount+\VbSvcompReachSafetyHardwareStatusWrongCount+\VbSvcompReachSafetyHeapStatusWrongCount+\VbSvcompReachSafetyLoopsStatusWrongCount+\VbSvcompReachSafetyProductLinesStatusWrongCount+\VbSvcompReachSafetySequentializedStatusWrongCount+\VbSvcompReachSafetyXCSPStatusWrongCount}

%% file: eval-results/tex/data-commands.svcomp-term.tex
\providecommand\StoreBenchExecResult[7]{\expandafter\newcommand\csname#1#2#3#4#5#6\endcsname{#7}}%
\StoreBenchExecResult{Cpv}{SvcompTerminationBitVectors}{Status}{All}{}{Score}{0}%
\StoreBenchExecResult{Cpv}{SvcompTerminationBitVectors}{Status}{All}{}{Count}{34}%
\StoreBenchExecResult{Cpv}{SvcompTerminationBitVectors}{Status}{Correct}{}{Count}{24}%
\StoreBenchExecResult{Cpv}{SvcompTerminationBitVectors}{Status}{Correct}{True}{Count}{13}%
\StoreBenchExecResult{Cpv}{SvcompTerminationBitVectors}{Status}{Correct}{False}{Count}{11}%
\StoreBenchExecResult{Cpv}{SvcompTerminationBitVectors}{Status}{Wrong}{}{Count}{0}%
\StoreBenchExecResult{Cpv}{SvcompTerminationBitVectors}{Status}{Wrong}{True}{Count}{0}%
\StoreBenchExecResult{Cpv}{SvcompTerminationBitVectors}{Status}{Wrong}{False}{Count}{0}%
\providecommand\StoreBenchExecResult[7]{\expandafter\newcommand\csname#1#2#3#4#5#6\endcsname{#7}}%
\StoreBenchExecResult{Cpachecker}{SvcompTerminationBitVectors}{Status}{All}{}{Score}{0}%
\StoreBenchExecResult{Cpachecker}{SvcompTerminationBitVectors}{Status}{All}{}{Count}{34}%
\StoreBenchExecResult{Cpachecker}{SvcompTerminationBitVectors}{Status}{Correct}{}{Count}{10}%
\StoreBenchExecResult{Cpachecker}{SvcompTerminationBitVectors}{Status}{Correct}{True}{Count}{3}%
\StoreBenchExecResult{Cpachecker}{SvcompTerminationBitVectors}{Status}{Correct}{False}{Count}{7}%
\StoreBenchExecResult{Cpachecker}{SvcompTerminationBitVectors}{Status}{Wrong}{}{Count}{0}%
\StoreBenchExecResult{Cpachecker}{SvcompTerminationBitVectors}{Status}{Wrong}{True}{Count}{0}%
\StoreBenchExecResult{Cpachecker}{SvcompTerminationBitVectors}{Status}{Wrong}{False}{Count}{0}%
\providecommand\StoreBenchExecResult[7]{\expandafter\newcommand\csname#1#2#3#4#5#6\endcsname{#7}}%
\StoreBenchExecResult{Esbmc}{KindTerminationBitVectors}{Status}{All}{}{Score}{0}%
\StoreBenchExecResult{Esbmc}{KindTerminationBitVectors}{Status}{All}{}{Count}{34}%
\StoreBenchExecResult{Esbmc}{KindTerminationBitVectors}{Status}{Correct}{}{Count}{13}%
\StoreBenchExecResult{Esbmc}{KindTerminationBitVectors}{Status}{Correct}{True}{Count}{13}%
\StoreBenchExecResult{Esbmc}{KindTerminationBitVectors}{Status}{Correct}{False}{Count}{0}%
\StoreBenchExecResult{Esbmc}{KindTerminationBitVectors}{Status}{Wrong}{}{Count}{0}%
\StoreBenchExecResult{Esbmc}{KindTerminationBitVectors}{Status}{Wrong}{True}{Count}{0}%
\StoreBenchExecResult{Esbmc}{KindTerminationBitVectors}{Status}{Wrong}{False}{Count}{0}%
\providecommand\StoreBenchExecResult[7]{\expandafter\newcommand\csname#1#2#3#4#5#6\endcsname{#7}}%
\StoreBenchExecResult{Symbiotic}{SvcompTerminationBitVectors}{Status}{All}{}{Score}{0}%
\StoreBenchExecResult{Symbiotic}{SvcompTerminationBitVectors}{Status}{All}{}{Count}{34}%
\StoreBenchExecResult{Symbiotic}{SvcompTerminationBitVectors}{Status}{Correct}{}{Count}{23}%
\StoreBenchExecResult{Symbiotic}{SvcompTerminationBitVectors}{Status}{Correct}{True}{Count}{13}%
\StoreBenchExecResult{Symbiotic}{SvcompTerminationBitVectors}{Status}{Correct}{False}{Count}{10}%
\StoreBenchExecResult{Symbiotic}{SvcompTerminationBitVectors}{Status}{Wrong}{}{Count}{0}%
\StoreBenchExecResult{Symbiotic}{SvcompTerminationBitVectors}{Status}{Wrong}{True}{Count}{0}%
\StoreBenchExecResult{Symbiotic}{SvcompTerminationBitVectors}{Status}{Wrong}{False}{Count}{0}%
\providecommand\StoreBenchExecResult[7]{\expandafter\newcommand\csname#1#2#3#4#5#6\endcsname{#7}}%
\StoreBenchExecResult{Uautomizer}{DefaultTerminationBitVectors}{Status}{All}{}{Score}{0}%
\StoreBenchExecResult{Uautomizer}{DefaultTerminationBitVectors}{Status}{All}{}{Count}{34}%
\StoreBenchExecResult{Uautomizer}{DefaultTerminationBitVectors}{Status}{Correct}{}{Count}{28}%
\StoreBenchExecResult{Uautomizer}{DefaultTerminationBitVectors}{Status}{Correct}{True}{Count}{22}%
\StoreBenchExecResult{Uautomizer}{DefaultTerminationBitVectors}{Status}{Correct}{False}{Count}{6}%
\StoreBenchExecResult{Uautomizer}{DefaultTerminationBitVectors}{Status}{Wrong}{}{Count}{0}%
\StoreBenchExecResult{Uautomizer}{DefaultTerminationBitVectors}{Status}{Wrong}{True}{Count}{0}%
\StoreBenchExecResult{Uautomizer}{DefaultTerminationBitVectors}{Status}{Wrong}{False}{Count}{0}%
\newcommand{\CpvSvcompTerminationBitVectorsBestCount}{1}
\newcommand{\CpvSvcompTerminationBitVectorsUniqCount}{1}
\newcommand{\CpacheckerSvcompTerminationBitVectorsBestCount}{1}
\newcommand{\CpacheckerSvcompTerminationBitVectorsUniqCount}{0}
\newcommand{\EsbmcKindTerminationBitVectorsBestCount}{12}
\newcommand{\EsbmcKindTerminationBitVectorsUniqCount}{0}
\newcommand{\SymbioticSvcompTerminationBitVectorsBestCount}{11}
\newcommand{\SymbioticSvcompTerminationBitVectorsUniqCount}{0}
\newcommand{\UautomizerDefaultTerminationBitVectorsBestCount}{8}
\newcommand{\UautomizerDefaultTerminationBitVectorsUniqCount}{8}
\providecommand\StoreBenchExecResult[7]{\expandafter\newcommand\csname#1#2#3#4#5#6\endcsname{#7}}%
\StoreBenchExecResult{Vb}{SvcompTerminationBitVectors}{Status}{All}{}{Score}{0}%
\StoreBenchExecResult{Vb}{SvcompTerminationBitVectors}{Status}{All}{}{Count}{34}%
\StoreBenchExecResult{Vb}{SvcompTerminationBitVectors}{Status}{Correct}{}{Count}{33}%
\StoreBenchExecResult{Vb}{SvcompTerminationBitVectors}{Status}{Correct}{True}{Count}{22}%
\StoreBenchExecResult{Vb}{SvcompTerminationBitVectors}{Status}{Correct}{False}{Count}{11}%
\StoreBenchExecResult{Vb}{SvcompTerminationBitVectors}{Status}{Wrong}{}{Count}{0}%
\StoreBenchExecResult{Vb}{SvcompTerminationBitVectors}{Status}{Wrong}{True}{Count}{0}%
\StoreBenchExecResult{Vb}{SvcompTerminationBitVectors}{Status}{Wrong}{False}{Count}{0}%
\providecommand\StoreBenchExecResult[7]{\expandafter\newcommand\csname#1#2#3#4#5#6\endcsname{#7}}%
\StoreBenchExecResult{Cpv}{SvcompTerminationMainControlFlow}{Status}{All}{}{Score}{0}%
\StoreBenchExecResult{Cpv}{SvcompTerminationMainControlFlow}{Status}{All}{}{Count}{253}%
\StoreBenchExecResult{Cpv}{SvcompTerminationMainControlFlow}{Status}{Correct}{}{Count}{82}%
\StoreBenchExecResult{Cpv}{SvcompTerminationMainControlFlow}{Status}{Correct}{True}{Count}{30}%
\StoreBenchExecResult{Cpv}{SvcompTerminationMainControlFlow}{Status}{Correct}{False}{Count}{52}%
\StoreBenchExecResult{Cpv}{SvcompTerminationMainControlFlow}{Status}{Wrong}{}{Count}{0}%
\StoreBenchExecResult{Cpv}{SvcompTerminationMainControlFlow}{Status}{Wrong}{True}{Count}{0}%
\StoreBenchExecResult{Cpv}{SvcompTerminationMainControlFlow}{Status}{Wrong}{False}{Count}{0}%
\providecommand\StoreBenchExecResult[7]{\expandafter\newcommand\csname#1#2#3#4#5#6\endcsname{#7}}%
\StoreBenchExecResult{Cpachecker}{SvcompTerminationMainControlFlow}{Status}{All}{}{Score}{0}%
\StoreBenchExecResult{Cpachecker}{SvcompTerminationMainControlFlow}{Status}{All}{}{Count}{253}%
\StoreBenchExecResult{Cpachecker}{SvcompTerminationMainControlFlow}{Status}{Correct}{}{Count}{178}%
\StoreBenchExecResult{Cpachecker}{SvcompTerminationMainControlFlow}{Status}{Correct}{True}{Count}{130}%
\StoreBenchExecResult{Cpachecker}{SvcompTerminationMainControlFlow}{Status}{Correct}{False}{Count}{48}%
\StoreBenchExecResult{Cpachecker}{SvcompTerminationMainControlFlow}{Status}{Wrong}{}{Count}{1}%
\StoreBenchExecResult{Cpachecker}{SvcompTerminationMainControlFlow}{Status}{Wrong}{True}{Count}{0}%
\StoreBenchExecResult{Cpachecker}{SvcompTerminationMainControlFlow}{Status}{Wrong}{False}{Count}{1}%
\providecommand\StoreBenchExecResult[7]{\expandafter\newcommand\csname#1#2#3#4#5#6\endcsname{#7}}%
\StoreBenchExecResult{Esbmc}{KindTerminationMainControlFlow}{Status}{All}{}{Score}{0}%
\StoreBenchExecResult{Esbmc}{KindTerminationMainControlFlow}{Status}{All}{}{Count}{253}%
\StoreBenchExecResult{Esbmc}{KindTerminationMainControlFlow}{Status}{Correct}{}{Count}{31}%
\StoreBenchExecResult{Esbmc}{KindTerminationMainControlFlow}{Status}{Correct}{True}{Count}{31}%
\StoreBenchExecResult{Esbmc}{KindTerminationMainControlFlow}{Status}{Correct}{False}{Count}{0}%
\StoreBenchExecResult{Esbmc}{KindTerminationMainControlFlow}{Status}{Wrong}{}{Count}{0}%
\StoreBenchExecResult{Esbmc}{KindTerminationMainControlFlow}{Status}{Wrong}{True}{Count}{0}%
\StoreBenchExecResult{Esbmc}{KindTerminationMainControlFlow}{Status}{Wrong}{False}{Count}{0}%
\providecommand\StoreBenchExecResult[7]{\expandafter\newcommand\csname#1#2#3#4#5#6\endcsname{#7}}%
\StoreBenchExecResult{Symbiotic}{SvcompTerminationMainControlFlow}{Status}{All}{}{Score}{0}%
\StoreBenchExecResult{Symbiotic}{SvcompTerminationMainControlFlow}{Status}{All}{}{Count}{253}%
\StoreBenchExecResult{Symbiotic}{SvcompTerminationMainControlFlow}{Status}{Correct}{}{Count}{75}%
\StoreBenchExecResult{Symbiotic}{SvcompTerminationMainControlFlow}{Status}{Correct}{True}{Count}{29}%
\StoreBenchExecResult{Symbiotic}{SvcompTerminationMainControlFlow}{Status}{Correct}{False}{Count}{46}%
\StoreBenchExecResult{Symbiotic}{SvcompTerminationMainControlFlow}{Status}{Wrong}{}{Count}{0}%
\StoreBenchExecResult{Symbiotic}{SvcompTerminationMainControlFlow}{Status}{Wrong}{True}{Count}{0}%
\StoreBenchExecResult{Symbiotic}{SvcompTerminationMainControlFlow}{Status}{Wrong}{False}{Count}{0}%
\providecommand\StoreBenchExecResult[7]{\expandafter\newcommand\csname#1#2#3#4#5#6\endcsname{#7}}%
\StoreBenchExecResult{Uautomizer}{DefaultTerminationMainControlFlow}{Status}{All}{}{Score}{0}%
\StoreBenchExecResult{Uautomizer}{DefaultTerminationMainControlFlow}{Status}{All}{}{Count}{253}%
\StoreBenchExecResult{Uautomizer}{DefaultTerminationMainControlFlow}{Status}{Correct}{}{Count}{205}%
\StoreBenchExecResult{Uautomizer}{DefaultTerminationMainControlFlow}{Status}{Correct}{True}{Count}{156}%
\StoreBenchExecResult{Uautomizer}{DefaultTerminationMainControlFlow}{Status}{Correct}{False}{Count}{49}%
\StoreBenchExecResult{Uautomizer}{DefaultTerminationMainControlFlow}{Status}{Wrong}{}{Count}{0}%
\StoreBenchExecResult{Uautomizer}{DefaultTerminationMainControlFlow}{Status}{Wrong}{True}{Count}{0}%
\StoreBenchExecResult{Uautomizer}{DefaultTerminationMainControlFlow}{Status}{Wrong}{False}{Count}{0}%
\newcommand{\CpvSvcompTerminationMainControlFlowBestCount}{10}
\newcommand{\CpvSvcompTerminationMainControlFlowUniqCount}{0}
\newcommand{\CpacheckerSvcompTerminationMainControlFlowBestCount}{108}
\newcommand{\CpacheckerSvcompTerminationMainControlFlowUniqCount}{3}
\newcommand{\EsbmcKindTerminationMainControlFlowBestCount}{28}
\newcommand{\EsbmcKindTerminationMainControlFlowUniqCount}{1}
\newcommand{\SymbioticSvcompTerminationMainControlFlowBestCount}{44}
\newcommand{\SymbioticSvcompTerminationMainControlFlowUniqCount}{0}
\newcommand{\UautomizerDefaultTerminationMainControlFlowBestCount}{25}
\newcommand{\UautomizerDefaultTerminationMainControlFlowUniqCount}{22}
\providecommand\StoreBenchExecResult[7]{\expandafter\newcommand\csname#1#2#3#4#5#6\endcsname{#7}}%
\StoreBenchExecResult{Vb}{SvcompTerminationMainControlFlow}{Status}{All}{}{Score}{0}%
\StoreBenchExecResult{Vb}{SvcompTerminationMainControlFlow}{Status}{All}{}{Count}{253}%
\StoreBenchExecResult{Vb}{SvcompTerminationMainControlFlow}{Status}{Correct}{}{Count}{215}%
\StoreBenchExecResult{Vb}{SvcompTerminationMainControlFlow}{Status}{Correct}{True}{Count}{161}%
\StoreBenchExecResult{Vb}{SvcompTerminationMainControlFlow}{Status}{Correct}{False}{Count}{54}%
\StoreBenchExecResult{Vb}{SvcompTerminationMainControlFlow}{Status}{Wrong}{}{Count}{0}%
\StoreBenchExecResult{Vb}{SvcompTerminationMainControlFlow}{Status}{Wrong}{True}{Count}{0}%
\StoreBenchExecResult{Vb}{SvcompTerminationMainControlFlow}{Status}{Wrong}{False}{Count}{0}%
\providecommand\StoreBenchExecResult[7]{\expandafter\newcommand\csname#1#2#3#4#5#6\endcsname{#7}}%
\StoreBenchExecResult{Cpv}{SvcompTerminationMainHeap}{Status}{All}{}{Score}{0}%
\StoreBenchExecResult{Cpv}{SvcompTerminationMainHeap}{Status}{All}{}{Count}{180}%
\StoreBenchExecResult{Cpv}{SvcompTerminationMainHeap}{Status}{Correct}{}{Count}{0}%
\StoreBenchExecResult{Cpv}{SvcompTerminationMainHeap}{Status}{Correct}{True}{Count}{0}%
\StoreBenchExecResult{Cpv}{SvcompTerminationMainHeap}{Status}{Correct}{False}{Count}{0}%
\StoreBenchExecResult{Cpv}{SvcompTerminationMainHeap}{Status}{Wrong}{}{Count}{0}%
\StoreBenchExecResult{Cpv}{SvcompTerminationMainHeap}{Status}{Wrong}{True}{Count}{0}%
\StoreBenchExecResult{Cpv}{SvcompTerminationMainHeap}{Status}{Wrong}{False}{Count}{0}%
\providecommand\StoreBenchExecResult[7]{\expandafter\newcommand\csname#1#2#3#4#5#6\endcsname{#7}}%
\StoreBenchExecResult{Cpachecker}{SvcompTerminationMainHeap}{Status}{All}{}{Score}{0}%
\StoreBenchExecResult{Cpachecker}{SvcompTerminationMainHeap}{Status}{All}{}{Count}{180}%
\StoreBenchExecResult{Cpachecker}{SvcompTerminationMainHeap}{Status}{Correct}{}{Count}{5}%
\StoreBenchExecResult{Cpachecker}{SvcompTerminationMainHeap}{Status}{Correct}{True}{Count}{0}%
\StoreBenchExecResult{Cpachecker}{SvcompTerminationMainHeap}{Status}{Correct}{False}{Count}{5}%
\StoreBenchExecResult{Cpachecker}{SvcompTerminationMainHeap}{Status}{Wrong}{}{Count}{0}%
\StoreBenchExecResult{Cpachecker}{SvcompTerminationMainHeap}{Status}{Wrong}{True}{Count}{0}%
\StoreBenchExecResult{Cpachecker}{SvcompTerminationMainHeap}{Status}{Wrong}{False}{Count}{0}%
\providecommand\StoreBenchExecResult[7]{\expandafter\newcommand\csname#1#2#3#4#5#6\endcsname{#7}}%
\StoreBenchExecResult{Esbmc}{KindTerminationMainHeap}{Status}{All}{}{Score}{0}%
\StoreBenchExecResult{Esbmc}{KindTerminationMainHeap}{Status}{All}{}{Count}{180}%
\StoreBenchExecResult{Esbmc}{KindTerminationMainHeap}{Status}{Correct}{}{Count}{11}%
\StoreBenchExecResult{Esbmc}{KindTerminationMainHeap}{Status}{Correct}{True}{Count}{11}%
\StoreBenchExecResult{Esbmc}{KindTerminationMainHeap}{Status}{Correct}{False}{Count}{0}%
\StoreBenchExecResult{Esbmc}{KindTerminationMainHeap}{Status}{Wrong}{}{Count}{0}%
\StoreBenchExecResult{Esbmc}{KindTerminationMainHeap}{Status}{Wrong}{True}{Count}{0}%
\StoreBenchExecResult{Esbmc}{KindTerminationMainHeap}{Status}{Wrong}{False}{Count}{0}%
\providecommand\StoreBenchExecResult[7]{\expandafter\newcommand\csname#1#2#3#4#5#6\endcsname{#7}}%
\StoreBenchExecResult{Symbiotic}{SvcompTerminationMainHeap}{Status}{All}{}{Score}{0}%
\StoreBenchExecResult{Symbiotic}{SvcompTerminationMainHeap}{Status}{All}{}{Count}{180}%
\StoreBenchExecResult{Symbiotic}{SvcompTerminationMainHeap}{Status}{Correct}{}{Count}{12}%
\StoreBenchExecResult{Symbiotic}{SvcompTerminationMainHeap}{Status}{Correct}{True}{Count}{12}%
\StoreBenchExecResult{Symbiotic}{SvcompTerminationMainHeap}{Status}{Correct}{False}{Count}{0}%
\StoreBenchExecResult{Symbiotic}{SvcompTerminationMainHeap}{Status}{Wrong}{}{Count}{0}%
\StoreBenchExecResult{Symbiotic}{SvcompTerminationMainHeap}{Status}{Wrong}{True}{Count}{0}%
\StoreBenchExecResult{Symbiotic}{SvcompTerminationMainHeap}{Status}{Wrong}{False}{Count}{0}%
\providecommand\StoreBenchExecResult[7]{\expandafter\newcommand\csname#1#2#3#4#5#6\endcsname{#7}}%
\StoreBenchExecResult{Uautomizer}{DefaultTerminationMainHeap}{Status}{All}{}{Score}{0}%
\StoreBenchExecResult{Uautomizer}{DefaultTerminationMainHeap}{Status}{All}{}{Count}{180}%
\StoreBenchExecResult{Uautomizer}{DefaultTerminationMainHeap}{Status}{Correct}{}{Count}{76}%
\StoreBenchExecResult{Uautomizer}{DefaultTerminationMainHeap}{Status}{Correct}{True}{Count}{67}%
\StoreBenchExecResult{Uautomizer}{DefaultTerminationMainHeap}{Status}{Correct}{False}{Count}{9}%
\StoreBenchExecResult{Uautomizer}{DefaultTerminationMainHeap}{Status}{Wrong}{}{Count}{1}%
\StoreBenchExecResult{Uautomizer}{DefaultTerminationMainHeap}{Status}{Wrong}{True}{Count}{1}%
\StoreBenchExecResult{Uautomizer}{DefaultTerminationMainHeap}{Status}{Wrong}{False}{Count}{0}%
\newcommand{\CpvSvcompTerminationMainHeapBestCount}{0}
\newcommand{\CpvSvcompTerminationMainHeapUniqCount}{0}
\newcommand{\CpacheckerSvcompTerminationMainHeapBestCount}{5}
\newcommand{\CpacheckerSvcompTerminationMainHeapUniqCount}{0}
\newcommand{\EsbmcKindTerminationMainHeapBestCount}{10}
\newcommand{\EsbmcKindTerminationMainHeapUniqCount}{0}
\newcommand{\SymbioticSvcompTerminationMainHeapBestCount}{3}
\newcommand{\SymbioticSvcompTerminationMainHeapUniqCount}{1}
\newcommand{\UautomizerDefaultTerminationMainHeapBestCount}{62}
\newcommand{\UautomizerDefaultTerminationMainHeapUniqCount}{62}
\providecommand\StoreBenchExecResult[7]{\expandafter\newcommand\csname#1#2#3#4#5#6\endcsname{#7}}%
\StoreBenchExecResult{Vb}{SvcompTerminationMainHeap}{Status}{All}{}{Score}{0}%
\StoreBenchExecResult{Vb}{SvcompTerminationMainHeap}{Status}{All}{}{Count}{180}%
\StoreBenchExecResult{Vb}{SvcompTerminationMainHeap}{Status}{Correct}{}{Count}{80}%
\StoreBenchExecResult{Vb}{SvcompTerminationMainHeap}{Status}{Correct}{True}{Count}{71}%
\StoreBenchExecResult{Vb}{SvcompTerminationMainHeap}{Status}{Correct}{False}{Count}{9}%
\StoreBenchExecResult{Vb}{SvcompTerminationMainHeap}{Status}{Wrong}{}{Count}{0}%
\StoreBenchExecResult{Vb}{SvcompTerminationMainHeap}{Status}{Wrong}{True}{Count}{0}%
\StoreBenchExecResult{Vb}{SvcompTerminationMainHeap}{Status}{Wrong}{False}{Count}{0}%
\providecommand\StoreBenchExecResult[7]{\expandafter\newcommand\csname#1#2#3#4#5#6\endcsname{#7}}%
\StoreBenchExecResult{Cpv}{SvcompTerminationOther}{Status}{All}{}{Score}{0}%
\StoreBenchExecResult{Cpv}{SvcompTerminationOther}{Status}{All}{}{Count}{1526}%
\StoreBenchExecResult{Cpv}{SvcompTerminationOther}{Status}{Correct}{}{Count}{1210}%
\StoreBenchExecResult{Cpv}{SvcompTerminationOther}{Status}{Correct}{True}{Count}{549}%
\StoreBenchExecResult{Cpv}{SvcompTerminationOther}{Status}{Correct}{False}{Count}{661}%
\StoreBenchExecResult{Cpv}{SvcompTerminationOther}{Status}{Wrong}{}{Count}{0}%
\StoreBenchExecResult{Cpv}{SvcompTerminationOther}{Status}{Wrong}{True}{Count}{0}%
\StoreBenchExecResult{Cpv}{SvcompTerminationOther}{Status}{Wrong}{False}{Count}{0}%
\providecommand\StoreBenchExecResult[7]{\expandafter\newcommand\csname#1#2#3#4#5#6\endcsname{#7}}%
\StoreBenchExecResult{Cpachecker}{SvcompTerminationOther}{Status}{All}{}{Score}{0}%
\StoreBenchExecResult{Cpachecker}{SvcompTerminationOther}{Status}{All}{}{Count}{1526}%
\StoreBenchExecResult{Cpachecker}{SvcompTerminationOther}{Status}{Correct}{}{Count}{1006}%
\StoreBenchExecResult{Cpachecker}{SvcompTerminationOther}{Status}{Correct}{True}{Count}{429}%
\StoreBenchExecResult{Cpachecker}{SvcompTerminationOther}{Status}{Correct}{False}{Count}{577}%
\StoreBenchExecResult{Cpachecker}{SvcompTerminationOther}{Status}{Wrong}{}{Count}{0}%
\StoreBenchExecResult{Cpachecker}{SvcompTerminationOther}{Status}{Wrong}{True}{Count}{0}%
\StoreBenchExecResult{Cpachecker}{SvcompTerminationOther}{Status}{Wrong}{False}{Count}{0}%
\providecommand\StoreBenchExecResult[7]{\expandafter\newcommand\csname#1#2#3#4#5#6\endcsname{#7}}%
\StoreBenchExecResult{Esbmc}{KindTerminationOther}{Status}{All}{}{Score}{0}%
\StoreBenchExecResult{Esbmc}{KindTerminationOther}{Status}{All}{}{Count}{1526}%
\StoreBenchExecResult{Esbmc}{KindTerminationOther}{Status}{Correct}{}{Count}{673}%
\StoreBenchExecResult{Esbmc}{KindTerminationOther}{Status}{Correct}{True}{Count}{673}%
\StoreBenchExecResult{Esbmc}{KindTerminationOther}{Status}{Correct}{False}{Count}{0}%
\StoreBenchExecResult{Esbmc}{KindTerminationOther}{Status}{Wrong}{}{Count}{0}%
\StoreBenchExecResult{Esbmc}{KindTerminationOther}{Status}{Wrong}{True}{Count}{0}%
\StoreBenchExecResult{Esbmc}{KindTerminationOther}{Status}{Wrong}{False}{Count}{0}%
\providecommand\StoreBenchExecResult[7]{\expandafter\newcommand\csname#1#2#3#4#5#6\endcsname{#7}}%
\StoreBenchExecResult{Symbiotic}{SvcompTerminationOther}{Status}{All}{}{Score}{0}%
\StoreBenchExecResult{Symbiotic}{SvcompTerminationOther}{Status}{All}{}{Count}{1526}%
\StoreBenchExecResult{Symbiotic}{SvcompTerminationOther}{Status}{Correct}{}{Count}{1138}%
\StoreBenchExecResult{Symbiotic}{SvcompTerminationOther}{Status}{Correct}{True}{Count}{559}%
\StoreBenchExecResult{Symbiotic}{SvcompTerminationOther}{Status}{Correct}{False}{Count}{579}%
\StoreBenchExecResult{Symbiotic}{SvcompTerminationOther}{Status}{Wrong}{}{Count}{0}%
\StoreBenchExecResult{Symbiotic}{SvcompTerminationOther}{Status}{Wrong}{True}{Count}{0}%
\StoreBenchExecResult{Symbiotic}{SvcompTerminationOther}{Status}{Wrong}{False}{Count}{0}%
\providecommand\StoreBenchExecResult[7]{\expandafter\newcommand\csname#1#2#3#4#5#6\endcsname{#7}}%
\StoreBenchExecResult{Uautomizer}{DefaultTerminationOther}{Status}{All}{}{Score}{0}%
\StoreBenchExecResult{Uautomizer}{DefaultTerminationOther}{Status}{All}{}{Count}{1526}%
\StoreBenchExecResult{Uautomizer}{DefaultTerminationOther}{Status}{Correct}{}{Count}{1158}%
\StoreBenchExecResult{Uautomizer}{DefaultTerminationOther}{Status}{Correct}{True}{Count}{630}%
\StoreBenchExecResult{Uautomizer}{DefaultTerminationOther}{Status}{Correct}{False}{Count}{528}%
\StoreBenchExecResult{Uautomizer}{DefaultTerminationOther}{Status}{Wrong}{}{Count}{0}%
\StoreBenchExecResult{Uautomizer}{DefaultTerminationOther}{Status}{Wrong}{True}{Count}{0}%
\StoreBenchExecResult{Uautomizer}{DefaultTerminationOther}{Status}{Wrong}{False}{Count}{0}%
\newcommand{\CpvSvcompTerminationOtherBestCount}{95}
\newcommand{\CpvSvcompTerminationOtherUniqCount}{17}
\newcommand{\CpacheckerSvcompTerminationOtherBestCount}{86}
\newcommand{\CpacheckerSvcompTerminationOtherUniqCount}{5}
\newcommand{\EsbmcKindTerminationOtherBestCount}{641}
\newcommand{\EsbmcKindTerminationOtherUniqCount}{4}
\newcommand{\SymbioticSvcompTerminationOtherBestCount}{576}
\newcommand{\SymbioticSvcompTerminationOtherUniqCount}{8}
\newcommand{\UautomizerDefaultTerminationOtherBestCount}{46}
\newcommand{\UautomizerDefaultTerminationOtherUniqCount}{36}
\providecommand\StoreBenchExecResult[7]{\expandafter\newcommand\csname#1#2#3#4#5#6\endcsname{#7}}%
\StoreBenchExecResult{Vb}{SvcompTerminationOther}{Status}{All}{}{Score}{0}%
\StoreBenchExecResult{Vb}{SvcompTerminationOther}{Status}{All}{}{Count}{1526}%
\StoreBenchExecResult{Vb}{SvcompTerminationOther}{Status}{Correct}{}{Count}{1444}%
\StoreBenchExecResult{Vb}{SvcompTerminationOther}{Status}{Correct}{True}{Count}{755}%
\StoreBenchExecResult{Vb}{SvcompTerminationOther}{Status}{Correct}{False}{Count}{689}%
\StoreBenchExecResult{Vb}{SvcompTerminationOther}{Status}{Wrong}{}{Count}{0}%
\StoreBenchExecResult{Vb}{SvcompTerminationOther}{Status}{Wrong}{True}{Count}{0}%
\StoreBenchExecResult{Vb}{SvcompTerminationOther}{Status}{Wrong}{False}{Count}{0}%
\edef\CpvSvcompTerminationStatusAllCount{\the\numexpr\CpvSvcompTerminationBitVectorsStatusAllCount+\CpvSvcompTerminationMainControlFlowStatusAllCount+\CpvSvcompTerminationMainHeapStatusAllCount+\CpvSvcompTerminationOtherStatusAllCount}
\edef\CpvSvcompTerminationStatusCorrectCount{\the\numexpr\CpvSvcompTerminationBitVectorsStatusCorrectCount+\CpvSvcompTerminationMainControlFlowStatusCorrectCount+\CpvSvcompTerminationMainHeapStatusCorrectCount+\CpvSvcompTerminationOtherStatusCorrectCount}
\edef\CpvSvcompTerminationStatusCorrectTrueCount{\the\numexpr\CpvSvcompTerminationBitVectorsStatusCorrectTrueCount+\CpvSvcompTerminationMainControlFlowStatusCorrectTrueCount+\CpvSvcompTerminationMainHeapStatusCorrectTrueCount+\CpvSvcompTerminationOtherStatusCorrectTrueCount}
\edef\CpvSvcompTerminationStatusCorrectFalseCount{\the\numexpr\CpvSvcompTerminationBitVectorsStatusCorrectFalseCount+\CpvSvcompTerminationMainControlFlowStatusCorrectFalseCount+\CpvSvcompTerminationMainHeapStatusCorrectFalseCount+\CpvSvcompTerminationOtherStatusCorrectFalseCount}
\edef\CpvSvcompTerminationStatusWrongCount{\the\numexpr\CpvSvcompTerminationBitVectorsStatusWrongCount+\CpvSvcompTerminationMainControlFlowStatusWrongCount+\CpvSvcompTerminationMainHeapStatusWrongCount+\CpvSvcompTerminationOtherStatusWrongCount}
\edef\CpvSvcompTerminationBestCount{\the\numexpr\CpvSvcompTerminationBitVectorsBestCount+\CpvSvcompTerminationMainControlFlowBestCount+\CpvSvcompTerminationMainHeapBestCount+\CpvSvcompTerminationOtherBestCount}
\edef\CpvSvcompTerminationUniqCount{\the\numexpr\CpvSvcompTerminationBitVectorsUniqCount+\CpvSvcompTerminationMainControlFlowUniqCount+\CpvSvcompTerminationMainHeapUniqCount+\CpvSvcompTerminationOtherUniqCount}
\edef\CpacheckerSvcompTerminationStatusAllCount{\the\numexpr\CpacheckerSvcompTerminationBitVectorsStatusAllCount+\CpacheckerSvcompTerminationMainControlFlowStatusAllCount+\CpacheckerSvcompTerminationMainHeapStatusAllCount+\CpacheckerSvcompTerminationOtherStatusAllCount}
\edef\CpacheckerSvcompTerminationStatusCorrectCount{\the\numexpr\CpacheckerSvcompTerminationBitVectorsStatusCorrectCount+\CpacheckerSvcompTerminationMainControlFlowStatusCorrectCount+\CpacheckerSvcompTerminationMainHeapStatusCorrectCount+\CpacheckerSvcompTerminationOtherStatusCorrectCount}
\edef\CpacheckerSvcompTerminationStatusCorrectTrueCount{\the\numexpr\CpacheckerSvcompTerminationBitVectorsStatusCorrectTrueCount+\CpacheckerSvcompTerminationMainControlFlowStatusCorrectTrueCount+\CpacheckerSvcompTerminationMainHeapStatusCorrectTrueCount+\CpacheckerSvcompTerminationOtherStatusCorrectTrueCount}
\edef\CpacheckerSvcompTerminationStatusCorrectFalseCount{\the\numexpr\CpacheckerSvcompTerminationBitVectorsStatusCorrectFalseCount+\CpacheckerSvcompTerminationMainControlFlowStatusCorrectFalseCount+\CpacheckerSvcompTerminationMainHeapStatusCorrectFalseCount+\CpacheckerSvcompTerminationOtherStatusCorrectFalseCount}
\edef\CpacheckerSvcompTerminationStatusWrongCount{\the\numexpr\CpacheckerSvcompTerminationBitVectorsStatusWrongCount+\CpacheckerSvcompTerminationMainControlFlowStatusWrongCount+\CpacheckerSvcompTerminationMainHeapStatusWrongCount+\CpacheckerSvcompTerminationOtherStatusWrongCount}
\edef\CpacheckerSvcompTerminationBestCount{\the\numexpr\CpacheckerSvcompTerminationBitVectorsBestCount+\CpacheckerSvcompTerminationMainControlFlowBestCount+\CpacheckerSvcompTerminationMainHeapBestCount+\CpacheckerSvcompTerminationOtherBestCount}
\edef\CpacheckerSvcompTerminationUniqCount{\the\numexpr\CpacheckerSvcompTerminationBitVectorsUniqCount+\CpacheckerSvcompTerminationMainControlFlowUniqCount+\CpacheckerSvcompTerminationMainHeapUniqCount+\CpacheckerSvcompTerminationOtherUniqCount}
\edef\EsbmcKindTerminationStatusAllCount{\the\numexpr\EsbmcKindTerminationBitVectorsStatusAllCount+\EsbmcKindTerminationMainControlFlowStatusAllCount+\EsbmcKindTerminationMainHeapStatusAllCount+\EsbmcKindTerminationOtherStatusAllCount}
\edef\EsbmcKindTerminationStatusCorrectCount{\the\numexpr\EsbmcKindTerminationBitVectorsStatusCorrectCount+\EsbmcKindTerminationMainControlFlowStatusCorrectCount+\EsbmcKindTerminationMainHeapStatusCorrectCount+\EsbmcKindTerminationOtherStatusCorrectCount}
\edef\EsbmcKindTerminationStatusCorrectTrueCount{\the\numexpr\EsbmcKindTerminationBitVectorsStatusCorrectTrueCount+\EsbmcKindTerminationMainControlFlowStatusCorrectTrueCount+\EsbmcKindTerminationMainHeapStatusCorrectTrueCount+\EsbmcKindTerminationOtherStatusCorrectTrueCount}
\edef\EsbmcKindTerminationStatusCorrectFalseCount{\the\numexpr\EsbmcKindTerminationBitVectorsStatusCorrectFalseCount+\EsbmcKindTerminationMainControlFlowStatusCorrectFalseCount+\EsbmcKindTerminationMainHeapStatusCorrectFalseCount+\EsbmcKindTerminationOtherStatusCorrectFalseCount}
\edef\EsbmcKindTerminationStatusWrongCount{\the\numexpr\EsbmcKindTerminationBitVectorsStatusWrongCount+\EsbmcKindTerminationMainControlFlowStatusWrongCount+\EsbmcKindTerminationMainHeapStatusWrongCount+\EsbmcKindTerminationOtherStatusWrongCount}
\edef\EsbmcKindTerminationBestCount{\the\numexpr\EsbmcKindTerminationBitVectorsBestCount+\EsbmcKindTerminationMainControlFlowBestCount+\EsbmcKindTerminationMainHeapBestCount+\EsbmcKindTerminationOtherBestCount}
\edef\EsbmcKindTerminationUniqCount{\the\numexpr\EsbmcKindTerminationBitVectorsUniqCount+\EsbmcKindTerminationMainControlFlowUniqCount+\EsbmcKindTerminationMainHeapUniqCount+\EsbmcKindTerminationOtherUniqCount}
\edef\SymbioticSvcompTerminationStatusAllCount{\the\numexpr\SymbioticSvcompTerminationBitVectorsStatusAllCount+\SymbioticSvcompTerminationMainControlFlowStatusAllCount+\SymbioticSvcompTerminationMainHeapStatusAllCount+\SymbioticSvcompTerminationOtherStatusAllCount}
\edef\SymbioticSvcompTerminationStatusCorrectCount{\the\numexpr\SymbioticSvcompTerminationBitVectorsStatusCorrectCount+\SymbioticSvcompTerminationMainControlFlowStatusCorrectCount+\SymbioticSvcompTerminationMainHeapStatusCorrectCount+\SymbioticSvcompTerminationOtherStatusCorrectCount}
\edef\SymbioticSvcompTerminationStatusCorrectTrueCount{\the\numexpr\SymbioticSvcompTerminationBitVectorsStatusCorrectTrueCount+\SymbioticSvcompTerminationMainControlFlowStatusCorrectTrueCount+\SymbioticSvcompTerminationMainHeapStatusCorrectTrueCount+\SymbioticSvcompTerminationOtherStatusCorrectTrueCount}
\edef\SymbioticSvcompTerminationStatusCorrectFalseCount{\the\numexpr\SymbioticSvcompTerminationBitVectorsStatusCorrectFalseCount+\SymbioticSvcompTerminationMainControlFlowStatusCorrectFalseCount+\SymbioticSvcompTerminationMainHeapStatusCorrectFalseCount+\SymbioticSvcompTerminationOtherStatusCorrectFalseCount}
\edef\SymbioticSvcompTerminationStatusWrongCount{\the\numexpr\SymbioticSvcompTerminationBitVectorsStatusWrongCount+\SymbioticSvcompTerminationMainControlFlowStatusWrongCount+\SymbioticSvcompTerminationMainHeapStatusWrongCount+\SymbioticSvcompTerminationOtherStatusWrongCount}
\edef\SymbioticSvcompTerminationBestCount{\the\numexpr\SymbioticSvcompTerminationBitVectorsBestCount+\SymbioticSvcompTerminationMainControlFlowBestCount+\SymbioticSvcompTerminationMainHeapBestCount+\SymbioticSvcompTerminationOtherBestCount}
\edef\SymbioticSvcompTerminationUniqCount{\the\numexpr\SymbioticSvcompTerminationBitVectorsUniqCount+\SymbioticSvcompTerminationMainControlFlowUniqCount+\SymbioticSvcompTerminationMainHeapUniqCount+\SymbioticSvcompTerminationOtherUniqCount}
\edef\UautomizerDefaultTerminationStatusAllCount{\the\numexpr\UautomizerDefaultTerminationBitVectorsStatusAllCount+\UautomizerDefaultTerminationMainControlFlowStatusAllCount+\UautomizerDefaultTerminationMainHeapStatusAllCount+\UautomizerDefaultTerminationOtherStatusAllCount}
\edef\UautomizerDefaultTerminationStatusCorrectCount{\the\numexpr\UautomizerDefaultTerminationBitVectorsStatusCorrectCount+\UautomizerDefaultTerminationMainControlFlowStatusCorrectCount+\UautomizerDefaultTerminationMainHeapStatusCorrectCount+\UautomizerDefaultTerminationOtherStatusCorrectCount}
\edef\UautomizerDefaultTerminationStatusCorrectTrueCount{\the\numexpr\UautomizerDefaultTerminationBitVectorsStatusCorrectTrueCount+\UautomizerDefaultTerminationMainControlFlowStatusCorrectTrueCount+\UautomizerDefaultTerminationMainHeapStatusCorrectTrueCount+\UautomizerDefaultTerminationOtherStatusCorrectTrueCount}
\edef\UautomizerDefaultTerminationStatusCorrectFalseCount{\the\numexpr\UautomizerDefaultTerminationBitVectorsStatusCorrectFalseCount+\UautomizerDefaultTerminationMainControlFlowStatusCorrectFalseCount+\UautomizerDefaultTerminationMainHeapStatusCorrectFalseCount+\UautomizerDefaultTerminationOtherStatusCorrectFalseCount}
\edef\UautomizerDefaultTerminationStatusWrongCount{\the\numexpr\UautomizerDefaultTerminationBitVectorsStatusWrongCount+\UautomizerDefaultTerminationMainControlFlowStatusWrongCount+\UautomizerDefaultTerminationMainHeapStatusWrongCount+\UautomizerDefaultTerminationOtherStatusWrongCount}
\edef\UautomizerDefaultTerminationBestCount{\the\numexpr\UautomizerDefaultTerminationBitVectorsBestCount+\UautomizerDefaultTerminationMainControlFlowBestCount+\UautomizerDefaultTerminationMainHeapBestCount+\UautomizerDefaultTerminationOtherBestCount}
\edef\UautomizerDefaultTerminationUniqCount{\the\numexpr\UautomizerDefaultTerminationBitVectorsUniqCount+\UautomizerDefaultTerminationMainControlFlowUniqCount+\UautomizerDefaultTerminationMainHeapUniqCount+\UautomizerDefaultTerminationOtherUniqCount}
\edef\VbSvcompTerminationStatusAllCount{\the\numexpr\VbSvcompTerminationBitVectorsStatusAllCount+\VbSvcompTerminationMainControlFlowStatusAllCount+\VbSvcompTerminationMainHeapStatusAllCount+\VbSvcompTerminationOtherStatusAllCount}
\edef\VbSvcompTerminationStatusCorrectCount{\the\numexpr\VbSvcompTerminationBitVectorsStatusCorrectCount+\VbSvcompTerminationMainControlFlowStatusCorrectCount+\VbSvcompTerminationMainHeapStatusCorrectCount+\VbSvcompTerminationOtherStatusCorrectCount}
\edef\VbSvcompTerminationStatusCorrectTrueCount{\the\numexpr\VbSvcompTerminationBitVectorsStatusCorrectTrueCount+\VbSvcompTerminationMainControlFlowStatusCorrectTrueCount+\VbSvcompTerminationMainHeapStatusCorrectTrueCount+\VbSvcompTerminationOtherStatusCorrectTrueCount}
\edef\VbSvcompTerminationStatusCorrectFalseCount{\the\numexpr\VbSvcompTerminationBitVectorsStatusCorrectFalseCount+\VbSvcompTerminationMainControlFlowStatusCorrectFalseCount+\VbSvcompTerminationMainHeapStatusCorrectFalseCount+\VbSvcompTerminationOtherStatusCorrectFalseCount}
\edef\VbSvcompTerminationStatusWrongCount{\the\numexpr\VbSvcompTerminationBitVectorsStatusWrongCount+\VbSvcompTerminationMainControlFlowStatusWrongCount+\VbSvcompTerminationMainHeapStatusWrongCount+\VbSvcompTerminationOtherStatusWrongCount}

%% file: abstract.tex
Formal verification of software programs and hardware designs shares the common goal of reasoning about state-transition systems,
yet the two communities have largely developed separate intermediate representations, verification algorithms, and tools.
This paper investigates sequential circuits as an alternative intermediate representation for software verification,
with the goal of enabling direct application of hardware-model-checking techniques.
We present Circuit-Based Program Verification (\cpv), a modular and extensible framework that
translates C programs into sequential circuits
and employs off-the-shelf hardware model checkers as verification backends.
Unlike traditional software verifiers, which typically rely on path-based exploration,
\cpv reasons over sequential circuits,
where a program's control and data flows are folded into a monolithic transition relation that can be analyzed as a whole.
The framework supports reachability-safety and termination analyses.
For reachability safety, programs are encoded as circuits with a bad-state predicate.
For termination, a liveness-to-safety transformation is applied at the circuit level,
reducing non-termination detection to a state-revisit check.
\cpv integrates a portfolio of state-of-the-art hardware model checkers,
together providing access to a diverse set of verification algorithms,
including bounded model checking, \kinduction, interpolation-based model checking, and IC3/PDR.
Moreover, \cpv further offers different circuit-encoding
strategies that capture program behavior relationally or functionally, catering to different backends.
Counterexamples found by hardware model checkers are automatically translated back into software-verification witnesses
for users to interpret verification results.
We conducted a comprehensive experimental evaluation on the \svcomp 2026 benchmark suite,
covering more than \num{16000} verification tasks.
Our results show that \cpv achieved competitive performance against five well-established software verifiers.
In particular, \cpv exhibited complementary strengths by uniquely solving tasks that other verifiers cannot handle
and solved the second largest number of instances in both reachability-safety and termination benchmark categories.
These results demonstrate that sequential circuits provide a viable intermediate representation for software verification
and that hardware model-checking techniques can be applied effectively with model and witness translation.

%% file: introduction.tex
\section{Introduction}
\label{sect:introduction}

Formal verification plays an important role in ensuring the reliability of software systems.
It is routinely applied to detect property violations
and establish correctness guarantees for safety-critical software.
Over the past decades, the software-verification community has developed a rich ecosystem of verification techniques and tools,
including data-flow analysis~\cite{ValueRangesAnalysis,CPA-DF}, predicate abstraction~\cite{GrafSaidi97,AbstractionsFromProofs,HBMC-predicateabstraction}, and symbolic execution~\cite{SymbolicExecution,KLEE}.
In parallel, the community for hardware model checking has made substantial advances,
leading to highly effective algorithms such as bounded model checking~\cite{BMC,BMCJournal,BMChandbook}, \kinduction~\cite{InductionVerification}, interpolation-based model checking~\cite{McMillanCraig,VizelFMCAD09,ForwardBackwardReachability}, and IC3/PDR~\cite{IC3,EfficientPDR}.

Despite sharing the common goal of reasoning about state-transition systems,
software verification and hardware model checking have largely evolved as separate research communities.
The two domains rely on different modeling formalisms, intermediate representations, benchmark suites, and tool ecosystems.
Software verifiers typically operate on program-centric representations
such as control-flow automata (CFAs) or static single-assignment form,
whereas hardware model checkers reason about transition systems represented as sequential circuits with
parallel execution semantics.
As a consequence, advances in one domain might require substantial technical effort to be transferred to the other,
even when they address rather similar verification problems on an abstract level.

Several efforts have sought to bridge this gap
by translating verification tasks between software and hardware domains~\cite{BTOR2C,Btor2MLIR,Noureddine16,Long17,Edwards01,v2c,MukherjeeSKM16} and
by adopting verification algorithms across communities~\cite{K-Induction,SoftwareIC3,IC3-CFA-STTT,IMC-JAR,DAR-transferability}.
Prior verification approaches based on program-to-circuit translation~\cite{Noureddine16,Long17,Edwards01}, however,
target only bit-level circuit representations for reachability-safety verification,
restricting the scope of problems that can be expressed,
and do not support the backtranslation of verification witnesses.
Studies on algorithm transfer require reimplementing hardware-verification algorithms within a software verifier
rather than reusing existing hardware model checkers.
Taken together, these limitations leave the potential of sequential circuits as an intermediate representation for software verification underexplored.
This work seeks to answer the following question:
\emph{Can sequential circuits serve as a practical representation for software verification
and enable effective application of hardware-model-checking techniques and tools?}

\begin{figure}[t]
    \centering
    \subfloat[CFA]{%
        \begin{minipage}[b]{.45\textwidth}
            \centering
            \scalebox{.9}{\input{figures/cfa}}\label{fig:cfa}%
        \end{minipage}
    }
    \subfloat[Sequential circuit]{%
        \begin{minipage}[b]{.45\textwidth}
            \centering
            \scalebox{.9}{\input{figures/seq-circuit}}\label{fig:seq-circuit}%
        \end{minipage}
    }
    \caption{A CFA~(a) and the encoded sequential circuit~(b) for an example C program}
\end{figure}
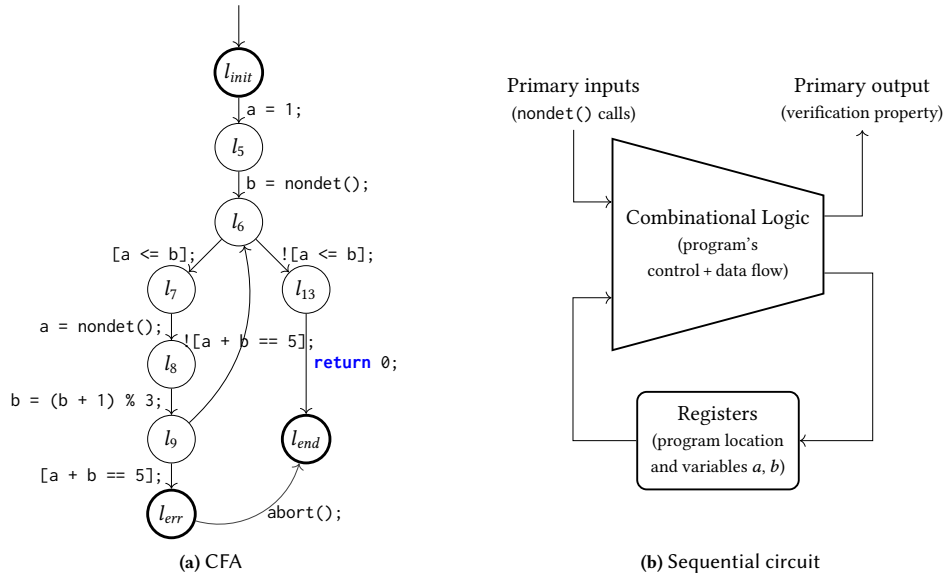

To illustrate, \cref{fig:cfa} shows the CFA representation of a simple C program
(the corresponding C code is given later in \cref{fig:c-prog}),
where each node and edge represent a program location and a statement, respectively.
The function \lstinline[style=cstyle]{nondet()} models an external program input by returning a nondeterministic value on each call.
\Cref{fig:seq-circuit} shows the schematic structure of the sequential circuit obtained after translation.
In the circuit, primary inputs model calls to \lstinline[style=cstyle]{nondet()},
state registers store the program location and the values of program variables,
combinational logic encodes the program's execution semantics,
and the primary output signals the verification property
(e.g., whether the error location $l_{err}$ can ever be reached,
or whether the termination location $l_{end}$ is always eventually reached).
The control flows and data updates in a CFA are thus absorbed into the circuit's next-state functions,
producing a monolithic transition-system representation on which hardware model checkers can operate directly.

We investigate the proposed circuit-based program-verification methodology by developing \cpv,
a verifier for C programs that uses sequential circuits as its intermediate representation.
\cpv translates programs into word-level sequential circuits in the \btortwo format~\cite{Boolector3} and
leverages off-the-shelf hardware model checkers,
namely \abc~\cite{ABC}, \avr~\cite{AVR}, \pono~\cite{Pono2}, and \ricthree~\cite{rIC3},
as verification backends.
Unlike traditional software verifiers, which typically explore programs path-by-path,
\cpv's backends reason over the circuit as a whole, exploiting the monolithic transition relation illustrated in \cref{fig:seq-circuit}.
The framework supports both reachability-safety and termination analyses.
To increase the confidence and interpretability of the verification results,
\cpv additionally translates hardware-level counterexamples~\cite{Boolector3} into standard software-verification witnesses~\cite{WitnessesJournal,VerificationWitnesses-2.0},
making the overall process certifying~\cite{CertifyingAlgorithms}.

%
%

Altogether, \cpv combines several techniques within a unified yet modular verification framework:
property-aware instrumentation,
large-block encoding~\cite{LBE} to extract the program's transition relation,
transition-function synthesis to construct the sequential circuit,
portfolio-based hardware model checking,
and witness reconstruction.

In our comprehensive evaluation on over \num{16000} verification tasks,
we demonstrate the feasibility of the proposed circuit-based verification methodology.
More than \SI{70}{\%} of tasks can be successfully translated into sequential circuits.
The remaining failures were primarily due to implementation limitations in the frontend rather than inherent restrictions of the approach,
suggesting that a more robust program-to-circuit translator is a direction for making circuit-based program verification even more effective.
Subsequent experiments further show that hardware model checkers can effectively verify the translated circuits.
The circuit representation also makes it straightforward to apply transformations
such as liveness-to-safety reduction~\cite{LivenessAsSafetyFinite} and logic optimization~\cite{MonyDATE2009}.
When compared against five state-of-the-art software verifiers,
\cpv exhibited complementary strengths and solved the second-most instances on both the reachability-safety and termination benchmark sets.
The performance of \cpv has also been evaluated in the annual International Competition on Software Verification (\svcomp)~\cite{SVCOMP26}.
\cpv first participated in \svcomp 2024~\cite{CPV-SVCOMP24} and has since achieved strong performance against other participants.
For example, in \svcomp 2025, \cpv ranked third in the ReachSafety category.
These results suggest that sequential circuits constitute a viable intermediate representation for software verification and
that the proposed circuit-based verification framework is competitive with established software verifiers.

\subsection{Contributions}

The \emph{novelty} of the proposed circuit-based program-verification methodology lies in
its unified integration of program instrumentation, circuit translation, property encoding, and witness reconstruction around sequential circuits as an intermediate representation.
Its \emph{significance} stems from complementing existing program-centric intermediate representations by establishing sequential circuits as a practical alternative,
thereby enabling the direct application of state-of-the-art hardware model checkers to software verification.
In summary, this work makes the following contributions:

\begin{itemize}[leftmargin=*]
    \item We systematically investigate sequential circuits as an alternative intermediate representation for software verification
        and study their potential to bridge software verification and hardware model checking.

    \item We present \cpv, a modular and extensible framework for software verification that
        translates C programs into word-level sequential circuits
        and reuses off-the-shelf hardware model checkers as verification backends.

    \item We develop relational and functional circuit-encoding strategies
        for constructing sequential circuits from a program's transition relation,
        catering to different model-checking backends.

    \item We propose program-instrumentation and property-encoding techniques for reachability-safety and termination verification,
        which facilitate a mechanism for reconstructing software-verification witnesses from hardware-model-checking results.

    \item We perform a comprehensive experimental evaluation on the \svcomp benchmark suite,
        assessing translation feasibility, circuit-encoding strategies, and witness validation.

    \item We demonstrate that the proposed circuit-based approach is competitive with state-of-the-art software verifiers,
        including \cpachecker~\cite{CPACHECKER-SVCOMP24}, \esbmc~\cite{ESBMC-SVCOMP25}, \kratostwo~\cite{Kratos2}, \symbiotic~\cite{SYMBIOTIC-SVCOMP26}, and \ultimateautomizer~\cite{UAUTOMIZER-SVCOMP26},
        while exhibiting complementary strengths by solving benchmark tasks that existing approaches struggle to verify.

    \item We release \cpv as open-source software on GitLab\footnote{\url{https://gitlab.com/sosy-lab/software/cpv}}
        and archive stable releases on Zenodo~\cite{CPV-latest} to support reproducibility and long-term availability.
    
    \item As a byproduct of this work, we also contribute a large collection of sequential circuits
        translated from the \svcomp benchmark suite to the Hardware Model Checking Competition (HWMCC)~\cite{HWMCC25},%
        \footnote{%
            The translated sequential circuits are available at \url{https://gitlab.com/sosy-lab/research/data/svcomp25-to-btor2}.
            They were derived from the \svcomp 2025 benchmark set~\cite{SVCOMP25-SVBENCHMARKS-artifact} using \cpv at \commiturl{https://gitlab.com/sosy-lab/software/cpv/-/tree}{2b20529bf4cd49922a14e0514631a148ce69236f}.}
        enabling future cross-domain benchmarking and research.
\end{itemize}

\subsection{Outline}

The remainder of this paper is organized as follows.
\Cref{sect:related-work} reviews related work,
covering approaches that bridge software and hardware verification and intermediate representations used in software verification.
\Cref{sect:background} introduces the necessary background on control-flow automata, sequential circuits, and the specification of safety and liveness properties.
\Cref{sect:pipeline} presents the overall architecture of \cpv.
\Cref{sect:encoding} explains the translation of C programs into sequential circuits.
\Cref{sect:verify-reachsafety,sect:verify-termination} describe property encoding, program instrumentation, and witness translation for verifying reachability-safety and termination properties, respectively.
Throughout the above technical sections,
we use a running example to illustrate the presented concepts and the translation process.
\Cref{sect:implementation} covers implementation details,
including the translation pipeline and the integration of hardware model checkers.
\Cref{sect:evaluation} presents the research questions and discusses the experimental evaluation.
Finally, \cref{sect:conclusion} concludes the paper.

%% file: figures/cfa.tex
\begin{tikzpicture}
    \tikzset{
        cfanode/.style={
            draw,
            circle,
            minimum size=.7cm,
            inner sep=0.03cm,
        }
    }
    \node (s) {};
    \node (linit) [below of = s, node distance=1.1cm, very thick, cfanode]{$l_{\mathit{init}}$};
    \node (l5) [below of = linit, node distance=1.1cm, cfanode]{$l_5$};
    \node (l6) [below of = l5, node distance=1.1cm, cfanode]{$l_6$};
    \node (l7) [below left of = l6, node distance=1.4cm, cfanode]{$l_7$};
    \node (l8) [below of = l7, node distance=1.1cm, cfanode]{$l_8$};
    \node (l9) [below of = l8, node distance=1.1cm, cfanode]{$l_9$};
    \node (lerr) [below of = l9, node distance=1.1cm, very thick, cfanode]{$l_{\mathit{err}}$};
    \node (l13) [below right of = l6, node distance=1.4cm, cfanode]{$l_{13}$};
    \node (lend) [below of = l13, node distance=2.2cm, very thick, cfanode]{$l_{\mathit{end}}$};

    \draw[->] (s) --(linit);
    \draw[->] (linit) to node[right] {\lstinline[style=cstyle]{a = 1;}} (l5);
    \draw[->] (l5) to node[right] {\lstinline[style=cstyle]{b = nondet();}} (l6);
    \draw[->] (l6) to node[left] {\lstinline[style=cstyle]{[a <= b];}} (l7);
    \draw[->] (l7) to node[left] {\lstinline[style=cstyle]{a = nondet();}} (l8);
    \draw[->] (l8) to node[left] {\lstinline[style=cstyle]{b = (b + 1) \% 3;}} (l9);
    \draw[->] (l9) to node[left] {\lstinline[style=cstyle]{[a + b == 5];}} (lerr);
    \draw[->] (lerr) edge[bend right=45, looseness=1, draw=darkgray] node[right] {\lstinline[style=cstyle]{abort();}} (lend);
    \draw[->] (l9) edge[bend right=30, looseness=1] node[right,xshift=-10mm] {\lstinline[style=cstyle]{![a + b == 5];}} (l6);
    \draw[->] (l6) to node[right] {\lstinline[style=cstyle]{![a <= b];}} (l13);
    \draw[->] (l13) to node[right] {\lstinline[style=cstyle]{return 0;}} (lend);
\end{tikzpicture}

%% file: figures/seq-circuit.tex
\begin{tikzpicture}

\node[draw, thick, trapezium, trapezium left angle=75, trapezium right angle=75,
  trapezium stretches, shape border rotate=270, align=center, inner sep=2mm,
] (logic) {Combinational Logic\\ {\smaller (program's}\\{\smaller control\,+\,data flow)}};
\node[draw, thick, rounded corners, inner sep=2mm,
  align=center, below =1cm of logic,
] (regs) {Registers\\{\smaller (program location}\\{\smaller and variables $a$, $b$)}};
\node[align=center, above left of = logic, node distance=3cm] (pi)
  {Primary inputs\\{\smaller (\lstinline[style=cstyle]{nondet()} calls)}};
\node[align=center, above right of = logic, node distance=3cm] (po)
  {Primary output\\{\smaller(verification property)}};
\node[below =5mm of regs] {};

\draw[->] (pi.south) |- ($(logic.north west)!0.3!(logic.south west)$);
\draw[->] (logic.15) -| (po.south);
\draw[->] (regs.west) -- ++(-9.3mm,0) |- ($(logic.north west)!0.75!(logic.south west)$);
\draw[->] (logic.345) -- ++(7.1mm,0) |- (regs.east);

\end{tikzpicture}

%% file: related-work.tex
\section{Related Work}
\label{sect:related-work}

Our work sits at the intersection of two active research communities,
software verification and hardware model checking.
We organize the discussion of related work around four themes:
efforts to bridge the two verification communities through task translation~(\cref{sect:related-work:translate}) and algorithm transfer~(\cref{sect:related-work:transfer}),
and common intermediate representations used in software verification~(\cref{sect:related-work:ir}).

\subsection{Bridging Hardware and Software Verification Through Translation}
\label{sect:related-work:translate}

Numerous approaches have explored translations between software and hardware representations for verification and related applications~\cite{TransformationGame}.
The most closely related work also employs hardware model checkers as backends for software verification
by translating programs into circuit representations.
Long~\cite{Long17}, and Noureddine and Zaraket~\cite{Noureddine16} translated C programs into bit-level sequential circuits in the \aiger format~\cite{AIGER-1.9}.
Both approaches blast arrays into bit-vectors since \aiger does not natively support arrays,
and use \abc~\cite{ABC} as the sole verification backend.
Edwards, Ma, and Damiano~\cite{Edwards01} translated embedded C programs into logic netlists
and verified them using the \tool{Ketchum} model checker~\cite{Ketchum},
which primarily relies on automatic test pattern generation to produce assertion-violating test vectors.
Their circuit representation supports arrays, but the approach has limited support for modeling external program inputs.

Compared with these approaches, our work extends the scope of circuit-based software verification in several respects.
We translate programs into the word-level \btortwo format~\cite{Boolector3},
which natively supports arrays and serves as the standard format in HWMCC~\cite{HWMCC25},
allowing multiple off-the-shelf hardware model checkers to be used as verification backends.
Beyond reachability-safety verification, \cpv also supports termination analysis.
It can further translate hardware-level counterexamples into standard software-verification witnesses~\cite{WitnessesJournal,VerificationWitnesses-2.0},
enabling independently checkable verification results.
In addition, \cpv supports floating-point variables and dynamic memory allocation, features not supported by the previous circuit-based approaches.
Finally, whereas previous work evaluated either handcrafted examples~\cite{Edwards01} or a small subset of the \svcomp benchmark suite~\cite{Noureddine16,Long17},
we conduct a comprehensive experimental evaluation on more than \num{16000} benchmark programs
using different circuit encodings, hardware model checkers, and verification algorithms.

A complementary line of work focuses on establishing equivalence between hardware implementations and the higher-level software specifications.
Clarke and Kröning~\cite{HV-Cref} use C programs as reference specifications for hardware designs
and check their equivalence via bounded model checking through exhaustive loop unrolling.
Commercial tools such as Jasper C2RTL%
\footnote{\url{https://www.cadence.com/en_US/home/tools/system-design-and-verification/formal-and-static-verification/jasper-c-formal-verification.html}}
pursue the same objective in industrial settings.
These approaches treat C as a specification language rather than as the primary verification target,
and therefore address a different problem than our work.

\tool{CirC}~\cite{CirC} is a compiler framework that lowers high-level languages, including C,
to existentially quantified circuits~(EQCs) for applications in cryptographic proof systems.
Unlike sequential circuits, EQCs are stateless and admit nondeterminism through existential quantification.
This representation is well suited to CirC's intended applications
but is less natural for modeling the sequential semantics of imperative programs.

In the opposite direction, several tools translate hardware circuits into software representations.
\tool{v2c}~\cite{v2c,MukherjeeSKM16} translates Verilog circuits into C programs,
while \tool{Btor2C}~\cite{BTOR2C} and \tool{Btor2MLIR}~\cite{Btor2MLIR} translate \btortwo circuits into C programs and LLVM IR, respectively.
These tools enable the application of software verifiers to hardware designs.

\subsection{Transferring Techniques from Hardware Model Checking to Software Verification}
\label{sect:related-work:transfer}

On the algorithmic side, several works have investigated the transfer of verification algorithms from hardware to software.
Bounded model checking~(BMC)~\cite{BMC,BMCJournal,BMChandbook}, \kinduction~\cite{InductionVerification}, interpolation-based model checking~\cite{McMillanCraig,VizelFMCAD09,ForwardBackwardReachability}, and IC3/PDR~\cite{IC3,EfficientPDR} were originally developed for verifying hardware systems,
where SAT solving plays a central role~\cite{HBMC-SAT-MC},
and have been adapted and applied to software programs.
Today, BMC and \kinduction are widely used in software verification,
and many software verifiers employ them as core engines~\cite{CBMC,BMCEmbeddedC,CPACHECKER-3.0-tutorial}.
Interpolation-based methods have likewise been studied extensively~\cite{IMPACT,IMC-JAR,DAR-transferability}.
IC3/PDR has also been adapted to operate on the abstract reachability tree of an unrolled CFA~\cite{SoftwareIC3}
as well as directly on the CFA itself~\cite{IC3-CFA-STTT}.
These works reimplement hardware-model-checking algorithms within software-verification frameworks
by adapting them to operate on program-centric representations.
Our work takes a fundamentally different approach:
rather than reimplementing the algorithms,
we aim to reuse hardware model checkers directly by translating programs into sequential circuits.
This design choice allows \cpv to benefit immediately from advances in hardware model checking without reimplementing individual algorithms,
and enables a head-to-head comparison of hardware- and software-verification techniques on the same benchmark suite.

\subsection{Intermediate Representations for Software Verification}
\label{sect:related-work:ir}

The choice of intermediate representation (IR) is a fundamental design decision in a verification framework,
as it determines both the properties that can be expressed and the algorithms that can be applied.
In software verification, a variety of program-centric IRs have been developed,
each targeting a different stage of the verification workflow.

LLVM IR~\cite{LLVM-CGO04} is primarily used within the LLVM compiler infrastructure,
but several verification tools~\cite{KLEE,LLBMC,ESBMC-SVCOMP25} also build on it.
For instance
\klee~\cite{KLEE} performs symbolic execution on LLVM bitcode,
and \esbmc~\cite{ESBMC-SVCOMP25} first translates programs to LLVM IR as a frontend step
before lowering them further to its goto-program representation.
As a general-purpose compiler IR,
LLVM IR benefits from mature tooling and broad language-frontend support (e.g., via \tool{Clang}),
but verifiers still need to encode its semantics as logic formulas to reason about it.

Several language-agnostic verification IRs have also been proposed to decouple language frontends from verification backends,
including Boogie~\cite{Boogie}, Why3~\cite{Why3}, and Viper~\cite{Viper}.
At a lower level, many software verifiers represent programs as CFAs,
in which nodes correspond to program locations and edges represent operations such as assignments and assumptions.
\cbmc~\cite{CBMC} and \esbmc~\cite{BMCEmbeddedC} use goto programs,
while \cpachecker~\cite{CPACHECKER-3.0-tutorial} and \ultimateautomizer~\cite{UAUTOMIZER2013} operate on their internal CFA representations.
Although many verifiers implement their own CFA variant,
efforts have been made to standardize CFA-based IR that can be consumed by multiple tools.
For example, K2~\cite{Kratos2} and SV-LIB~\cite{SV-LIB-1.0} are two SMT-LIB-like formats
for describing imperative programs as CFAs.
They inherit the expressions and theories of SMT-LIB~\cite{SMTLIB27}
while extending them with constructs for program-specific features such as
loops, function calls, and annotations for loop invariants.
\cpv uses K2 as a frontend IR before translating programs into sequential circuits.

These IRs are typically lowered further to representations suitable for automated reasoning.
SAT/SMT-based verifiers commonly unroll the CFA along (syntactically) feasible control-flow paths,
rename variables into static single-assignment form,
and encode the resulting verification conditions as SAT or SMT queries.
Constrained Horn Clauses (CHCs)~\cite{CHC} serve as another widely used logical IR.
Rather than explicitly exploring execution paths,
CHC-based approaches encode program semantics and properties as a system of Horn clauses,
whose satisfiability determines whether the property holds.
CHCs also provide a standardized SMT-LIB-based representation that can be consumed by many CHC solvers.

Sequential circuits, as employed by \cpv, represent another alternative.
Here, control flow is encoded using a symbolic program counter,
program variables become state registers,
and program execution is captured by next-state functions.
Effectively, the program's transition behavior over control flow and data update is folded into a monolithic state-transition system,
implicitly representing common program syntactic constructs such as loops or function calls.
This representation enables the direct use of off-the-shelf hardware model checkers as verification backends.

%% file: background.tex
\section{Background}
\label{sect:background}

This section introduces the necessary background to understand the proposed verification approach.
We present control-flow automata as a representation for modeling imperative software programs (\cref{sect:cfa})
and sequential circuits as the standard representation for hardware transition systems (\cref{sect:seq-circuit}).
We further describe the formulation of safety and liveness model-checking problems in the context of each representation (\cref{sect:properties})
and the corresponding verification witnesses (\cref{sect:witnesses}).
Throughout this paper, we use standard logical symbols such as $\land$ (conjunction), $\lor$ (disjunction), and $\neg$ (negation),
along with the temporal operators from linear temporal logic~\cite{HBMC-TemporalLogic},
including $\mathbf{G}$ (globally), $\mathbf{F}$ (eventually), and $\mathbf{U}$ (until).

\subsection{State-Transition Systems}
\label{sect:trans-sys}

Software and hardware systems can be modeled as \emph{symbolic state-transition systems}.
Such a system is defined by an initial-state predicate $\mathit{Init}(X)$ and a transition relation $\mathit{TR}(X, I, X')$,
where $X$ is a finite set of state variables,
$I$ is a finite set of primary input variables,
and $X'$ denotes the next-state version of $X$.
The predicate $\mathit{Init}(X)$ specifies the initial states of the system,
while $\mathit{TR}(X, I, X')$ describes the state transitions under the given input values.
Additional invariant constraints may further restrict the reachable state space.
An execution trace (or path) of the system is a sequence of states, starting from an initial state,
in which every two consecutive states are related by $\mathit{TR}$ under some input values.
The two representations introduced in the following differ mainly in how the transition relation is described.

\subsubsection{Software Representation: Control-Flow Automaton}
\label{sect:cfa}

We model an imperative program as a \emph{control-flow automaton} (CFA).
A CFA is a directed graph $A = (\locs, l_{init}, G)$,
where $\locs$ is a finite set of nodes corresponding to program locations,
$l_{init} \in \locs$ is the initial location (i.e., program entry), and
$G \subseteq (\locs \times \Ops \times \locs)$ is a set of edges.
Each edge $(l, \op, l') \in G$ is annotated with an operation $\op \in \Ops$,
which is either a variable assignment or an assumption (i.e., a Boolean guard).
As we work with C programs, we assume that C operators and data types constitute the operations.

For verification purposes, we further distinguish two special locations in the CFA:
an error location $l_{err} \in \locs$, which represents assertion violations or other program locations that should not be reached,
and a termination location $l_{end} \in \locs$, which represents program termination
(e.g., via \lstinline[style=cstyle]{return} from \lstinline[style=cstyle]{main()}, \lstinline[style=cstyle]{exit()}, or \lstinline[style=cstyle]{abort()}).
These locations are used to encode reachability-safety and termination properties.

A CFA can be viewed as a transition system whose state variables comprise
the program variables together with an auxiliary program-counter variable $pc$ tracking the program location.
Its input variables model the nondeterministic values supplied by the environment, its initial-state predicate constrains $pc$ to the program entry location $l_{init}$, and its transition relation captures the effect of executing CFA edges on both $pc$ and the program variables.
Under this interpretation, reaching a program location becomes reaching a state with the corresponding value of $pc$.
\Cref{sect:cfa-to-tr} details the construction used in this work.

\begin{figure}[t]
    \centering
\begin{lstlisting}[style=cstyle]
extern unsigned int __VERIFIER_nondet_uint();
void reach_error() { /* error label */ abort(); }
int main() {
  unsigned int a = 1;(*@\label{fig:c-prog:init-a}@*)
  unsigned int b = __VERIFIER_nondet_uint();(*@\label{fig:c-prog:init-b}@*)
  while (a <= b) {(*@\label{fig:c-prog:while}@*)
    a = __VERIFIER_nondet_uint();(*@\label{fig:c-prog:loop-body}@*)
    b = (b + 1) % 3;
    if (a + b == 5) {
      reach_error();(*@\label{fig:c-prog:err}@*)
    }
  }
  return 0;(*@\label{fig:c-prog:main-return}@*)
}
\end{lstlisting}
    \vspace*{-3mm}
    \caption{An example C program whose CFA is depicted in \cref{fig:cfa}}
    \label{fig:c-prog}
\end{figure}

\begin{example}
Consider the example C program in \cref{fig:c-prog}.
The program declares two variables \lstinline[style=cstyle]{a} and \lstinline[style=cstyle]{b},
both of type \lstinline[style=cstyle]{unsigned int}.
It starts by initializing \lstinline[style=cstyle]{a} to \lstinline[style=cstyle]{1} (\cref{fig:c-prog:init-a}) and
\lstinline[style=cstyle]{b} to a nondeterministic value
(\cref{fig:c-prog:init-b}; modeled by the return of \lstinline[style=cstyle]{__VERIFIER_nondet_uint()}).
Then, the program enters a loop with the condition \lstinline[style=cstyle]{a <= b},
where the values of \lstinline[style=cstyle]{a} and \lstinline[style=cstyle]{b} are updated in each iteration.

The CFA representation of this program is shown in \cref{fig:cfa},
where \lstinline[style=cstyle]{__VERIFIER_nondet_uint()} is abbreviated as \lstinline[style=cstyle]{nondet()}.
Each node corresponds to a program location, and each edge is annotated with the corresponding operation.
For instance, $l_{init}$ corresponds to program entry at \cref{fig:c-prog:init-a},
$l_5$ to \cref{fig:c-prog:init-b},
and $l_{end}$ to the termination location after returning from \lstinline[style=cstyle]{main()} at \cref{fig:c-prog:main-return};
the edge from $l_{init}$ to $l_5$ represents the assignment initializing variable \lstinline[style=cstyle]{a}.
Assumption edges are marked with square brackets.
For example, the edge from $l_6$ to $l_7$ is an assumption edge with guard \lstinline[style=cstyle]{[a <= b]},
indicating that the loop body (\cref{fig:c-prog:loop-body}) is entered under this condition.
\end{example}

\subsubsection{Hardware Representation: Sequential Circuit}
\label{sect:seq-circuit}

A \emph{sequential circuit} is a special form of a transition system
in which the transition relation can be represented explicitly by next-state functions.
For each state variable $x \in X$, there exists a transition function $x' = f_x(X, I)$,
thus allowing the transition relation to be represented functionally instead of relationally.
Structurally, a sequential circuit consists of registers (state variables),
primary inputs, primary outputs encoding verification properties,
and combinational logic that computes the next-state functions and property outputs.
The combinational logic can be represented as a directed acyclic graph,
where nodes represent state variables, input variables, and circuit operations (e.g., bit-vector arithmetic and Boolean connectives),
and edges encode the data dependencies between nodes.
The computed next-state values are fed back into the registers,
thereby connecting different time frames and making the overall graph structure cyclic.
For a node in the graph, its \emph{cone of influence} (CoI), also referred to as its transitive fan-ins,
is the subset of variables and operations that can transitively affect its value through data dependencies,
potentially crossing register boundaries.

In hardware model checking, variables are finite-width and range over bit-vectors and arrays.
The underlying logic is typically quantifier-free bit-vector logic with arrays (QF_ABV),
which extends quantifier-free bit-vector logic (QF_BV) with the array theory~\cite{McCarthy1962} to model memories.
Common interchange formats for sequential circuits
include the word-level \btortwo format~\cite{Boolector3} (supporting both QF_ABV and QF_BV)
and the bit-level \aiger format~\cite{AIGER-1.9} (supporting QF_BV),
which can be translated from higher-level hardware description languages such as SystemVerilog~\cite{SystemVerilog-2024}.
We follow formal semantics of these logics (QF_BV, QF_ABV) as standardized in SMT-LIB~\cite{SMTLIB27}.

\subsection{Safety and Liveness Properties}
\label{sect:properties}

After introducing the systems of interest
we now continue to describe safety and liveness properties
considered in this work for software verification and hardware model checking.

\subsubsection{Software-Verification Properties}
\label{sect:sv-properties}

We focus on two properties from the C verification track of \svcomp~\cite{SVCOMP26}:
reachability safety (ReachSafety) and Termination.
The ReachSafety property%
\footnote{\url{https://gitlab.com/sosy-lab/benchmarking/sv-benchmarks/-/blob/svcomp26/c/properties/unreach-call.prp}}
states that, starting from the program entry \lstinline[style=cstyle]{main()}, the function \lstinline[style=cstyle]{reach_error()} must never be called during execution.
A counterexample is therefore a finite execution trace from the program entry to a call of \lstinline[style=cstyle]{reach_error()}.
The Termination property%
\footnote{\url{https://gitlab.com/sosy-lab/benchmarking/sv-benchmarks/-/blob/svcomp26/c/properties/termination.prp}}
states that every execution (starting from the program entry \lstinline[style=cstyle]{main()}) must eventually reach a termination point,
e.g., upon \lstinline[style=cstyle]{exit()}, \lstinline[style=cstyle]{abort()}, or returning from the call to \lstinline[style=cstyle]{main()}.
A counterexample is an infinite execution trace, which can be represented as a lasso, assuming the state space to be finite,
that never reaches a termination point.

\subsubsection{Hardware-Model-Checking Properties}
\label{sect:hwmc-properties}

We follow the property specification of the \aiger~\cite{AIGER-1.9} and \btortwo~\cite{Boolector3} formats,
as adopted by HWMCC~\cite{HWMCC25}.
A safety property is encoded using a predicate $\mathit{bad}$ over state and input variables,
monitored by a primary output of the circuit.
The predicate $\mathit{bad}$ describes the condition under which the safety property is violated.
A counterexample is an initialized finite execution path satisfying
$c\ \mathbf{U}\ (c \land bad)$,
where $c$ denotes the invariant constraint.
Liveness properties are represented using a finite set of \emph{justice} constraints $J$ (also known as fairness constraints in some literature).
A counterexample to a liveness property is an initialized infinite path satisfying
$(\mathbf{G}\ c) \land \left(\bigwedge_{j \in J} \mathbf{GF}\ j\right)$,
where $c$ denotes the invariant constraint.
That is, on this path, the invariant constraint is always true and each justice constraint $j \in J$ holds infinitely often.
Equivalently, if the liveness property holds (i.e., no counterexample exists),
then along every execution path, either $c$ eventually becomes false or at least one $j \in J$ eventually becomes false forever.
In this work, we consider only a single justice constraint (i.e., $|J| = 1$), which is sufficient for encoding program termination.

\subsection{Verification Witnesses}
\label{sect:witnesses}

A verifier that reports only a verdict asks its users to trust the verifier itself.
Following the paradigm of certifying algorithms~\cite{CertifyingAlgorithms},
a verifier can instead accompany its verdict with a \emph{witness}:
an independently checkable artifact that justifies the reported result.
The process of checking such a witness, rather than repeating the verification, is called \emph{witness validation}.
Two kinds of witnesses are commonly distinguished.
A \emph{correctness witness} documents a proof,
typically by exposing the invariants inferred by the verifier,
whereas a \emph{violation witness} describes an execution path (a counterexample) that violates the property.
Both \svcomp~\cite{SVCOMP26} and HWMCC~\cite{HWMCC25} require participating verifiers to produce witnesses and validate them as part of their competition workflows.
As generating and checking correctness witnesses is an active research area for both software~\cite{WitnessesTransitionInvariants} and hardware~\cite{CertConstraintsHWMC,LivenessProofsHWMC} model checking,
including the discussion of which formats are appropriate,
we focus in this article on violation witnesses and leave exploring translation of correctness witnesses to future work.
For the concrete syntax of the violation witness formats, we refer the reader to the specifications of software-verification witnesses~\cite{WitnessesJournal,VerificationWitnesses-2.0,WitnessesNonTermination} and \btortwo witnesses~\cite{Boolector3}.

\subsubsection{Software-Verification Witnesses}

For reachability safety,
a violation witness for a program with nondeterministic inputs records the values assigned to those inputs along the error path,
each tied to the program location at which it is consumed.
It may additionally constrain the control flow, e.g., by specifying which branch is taken or which value is returned from a function,
and finally identifies the target location at which the property violation occurs.
Since the recorded constraints need not determine every nondeterministic input or branch taken,
a witness can represent a set of execution paths.
The witness is considered valid if this set contains at least one path violating the property.
If the witness describes a single execution trace,
i.e., all nondeterminism is resolved,
it corresponds to a test vector.

A violation witness for termination is an infinite execution trace represented as a \emph{lasso}
(assuming the program's state space is finite).
A lasso consists of a finite \emph{stem} from the program entry to a recurrent location,
followed by a \emph{cycle} from that location back to itself that can be repeated forever.
The witness records the inputs consumed along both the stem and the cycle and identifies the locations that every represented path must visit infinitely often.
It does not necessarily need to provide a recurrent set of program states.
Instead, establishing the recurrence argument is left to the witness validator~\cite{WitnessesNonTermination}.

Two exchange formats are used in \svcomp~\cite{SVCOMP26}.
The GraphML-based format (v1)~\cite{WitnessesJournal} describes violation witnesses as protocol automata.
The YAML-based format (v2)~\cite{VerificationWitnesses-2.0,WitnessesNonTermination} represents violation witnesses as sequences of \emph{waypoints},
each consisting of a source location together with an associated constraint.

\subsubsection{Hardware-Model-Checking Witnesses}

A violation witness for hardware model checking is represented as a sequence of time frames,
each containing the valuations of state variables and the assignments to input variables at the corresponding time step.
In the initial frame, values must be provided for all state variables that are not initialized by the model.
For subsequent frames, state valuations may be omitted,
as they can be inferred from the preceding state and the provided inputs.
Don't-care input values may be omitted as well.
Such a witness is essentially a concrete test vector over time that can be replayed by simulation.
For a safety property, the witness describes a finite counterexample trace from an initial state to a bad state.
For a justice property, the infinite counterexample is likewise exported as a finite lasso-shaped trace
consisting of a stem followed by one iteration of the cycle.
The successor of the state in the last time frame,
computed from its state and input assignments,
is identical to one of the earlier states.
Both the \aiger~\cite{AIGER-1.9} and \btortwo~\cite{Boolector3} formats define standardized representations for violation witnesses.

%% file: pipeline.tex
\section{Verification Pipeline}
\label{sect:pipeline}
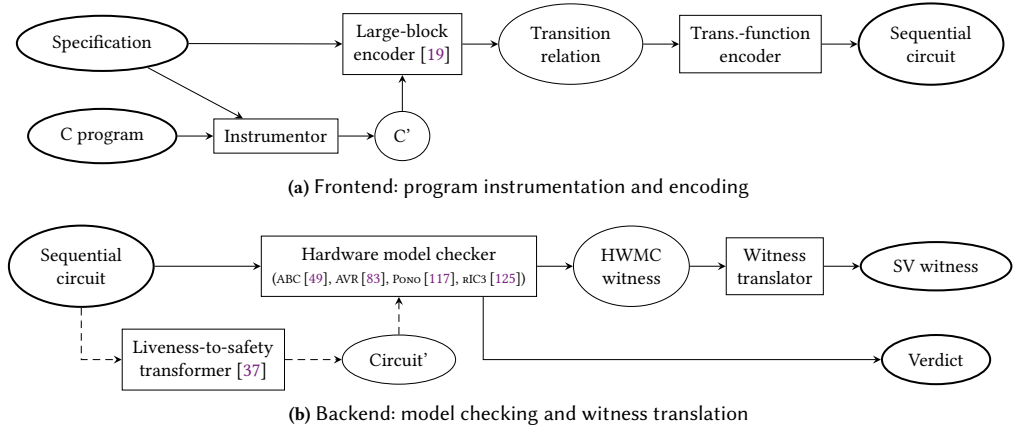
\begin{figure}[t]
    \centering
    \subfloat[Frontend: program instrumentation and encoding]{%
        \scalebox{.95}{\input{figures/frontend}}
        \label{fig:pipeline-frontend}
    }\\[2ex]
    \subfloat[Backend: model checking and witness translation]{%
        \scalebox{.95}{\input{figures/backend}}
        \label{fig:pipeline-backend}
    }
    \caption{Verification pipeline of \cpv (HWMC: hardware-model-checking; SV: software-verification)}
    \label{fig:pipeline}
\end{figure}

The verification pipeline of the proposed circuit-based program-verification framework
is depicted in~\cref{fig:pipeline}.
Its verification workflow is divided into a \emph{frontend} and \emph{backward} stage.

\subsection{Frontend (\cref{fig:pipeline-frontend})}

An input C program with a property to be verified (reachability safety or termination) is first instrumented.
This instrumentation is essential both for encoding the verification problem and for enabling witness translation (\cref{sect:instr-reach,sect:instr-term}).
From the instrumented program, \cpv extracts the transition relation using large-block encoding~\cite{LBE} (\cref{sect:cfa-to-tr}).
The property is encoded as a predicate over the resulting transition system.
A subsequent encoding pass extracts transition functions from the transition relation
and constructs a word-level sequential circuit~\cite{Boolector3} (\cref{sect:tr-to-circuit}),
yielding a transition-function-based representation.

\subsection{Backend (\cref{fig:pipeline-backend})}
Once the frontend produces the sequential circuit,
\cpv invokes hardware model checkers such as \avr~\cite{AVR} and \pono~\cite{Pono2} to verify it (\cref{sect:backend-integration}).
For liveness properties, in addition to using checkers' native liveness checking,
\cpv can apply a liveness-to-safety transformation~\cite{LivenessAsSafetyFinite} prior to verification.
If a property violation is detected, the backend model checker may produce a violation witness~\cite{Boolector3},
that is, a counterexample trace over the translated circuit.
This witness is translated into a software-verification witness~\cite{WitnessesJournal,VerificationWitnesses-2.0} for the original program.

\subsection{Modularity and Extensibility}
\cpv is designed with modularity in mind:
each component is clearly separated and communicates through standardized formats for verification artifacts,
e.g., \btortwo~\cite{Boolector3} for circuits and YAML format~\cite{VerificationWitnesses-2.0} for software-verification witnesses.
This clean interface makes it easy to add new encoders, alternative circuit representations, or additional backend model checkers.
The result is a flexible framework that can be easily extended and adapted for experimenting with diverse circuit-based verification techniques.

%% file: figures/frontend.tex
\begin{tikzpicture}[every node/.style={font=\large}]
    \tikzstyle{data} = [ellipse, text centered, draw=black, align=center, inner sep=1.5mm, font=\smaller]
    \tikzstyle{process} = [rectangle, text centered, draw=black, align=center, inner sep=1.5mm, font=\smaller]
    \tikzstyle{arrow} = [->, >=stealth]

    \node (c) [data, thick] {C program};
    \node (instrumentor) [process, right=5mm of c] {Instrumentor};
    \node (instrc) [data, right=5mm of instrumentor] {C'};
    \node (tr-enc) [process, above=5mm of instrc] {Large-block\\encoder~\cite{LBE}};
    \node (spec) [data, above=5mm of c, thick] {Specification};
    \node (tr) [data, right=5mm of tr-enc] {Transition\\relation};
    \node (tf-enc) [process, right=5mm of tr] {Trans.-function\\encoder};
    \node (tf) [data, right=5mm of tf-enc, thick] {Sequential\\circuit};

    \draw [arrow] (c) -- (instrumentor);
    \draw [arrow] (spec) -- (instrumentor);
    \draw [arrow] (instrumentor) -- (instrc);
    \draw [arrow] (instrc) -- (tr-enc);
    \draw [arrow] (spec) -- (tr-enc);
    \draw [arrow] (tr-enc) -- (tr);
    \draw [arrow] (tr) -- (tf-enc);
    \draw [arrow] (tf-enc) -- (tf);
\end{tikzpicture}

%% file: figures/backend.tex
\begin{tikzpicture}[every node/.style={font=\large}]
    \tikzstyle{data} = [ellipse, text centered, draw=black, align=center, inner sep=1.5mm, font=\smaller]
    \tikzstyle{process} = [rectangle, text centered, draw=black, align=center, inner sep=1.5mm, font=\smaller]
    \tikzstyle{arrow} = [->, >=stealth]

    \node (circuit) [data, thick] {Sequential\\circuit};
    \node (hwmc) [process, right=25mm of circuit.center, minimum width=30mm] {Hardware model checker\\
        {\smaller (\abc{}\,\cite{ABC}, \avr{}\,\cite{AVR}, \pono{}\,\cite{Pono2}, \ricthree{}\,\cite{rIC3})}};
    \node (btorwit) [data, right=5mm of hwmc] {HWMC\\witness};
    \node (wittrans) [process, right=5mm of btorwit] {Witness\\translator};
    \node (svwit) [data, right=5mm of wittrans, thick] {SV witness};
    \node (verdict) [data, below=13mm of svwit.center, anchor=center, thick] {Verdict};
    \node (l2s) [process, below right=13mm and 17mm of circuit.center, anchor=center] {Liveness-to-safety\\transformer~\cite{LivenessAsSafetyFinite}};
    \node (circuit') [data, below=13mm of hwmc.center, anchor=center] {Circuit'};

    \draw [arrow] (circuit) -- (hwmc);
    \draw [arrow] (hwmc) -- (btorwit);
    \draw [arrow] (btorwit) -- (wittrans);
    \draw [arrow] (wittrans) -- (svwit);
    \draw [arrow] (hwmc.340) |- (verdict.west);
    \draw [arrow, densely dashed] (circuit) |- (l2s.west);
    \draw [arrow, densely dashed] (l2s) -- (circuit');
    \draw [arrow, densely dashed] (circuit'.north) -- (hwmc);
\end{tikzpicture}

%% file: encoding.tex
\section{Encoding Programs as Sequential Circuits}
\label{sect:encoding}

We then describe how to encode C programs represented as CFAs into sequential circuits.
A CFA is first translated into a symbolic transition system using large-block encoding (\cref{sect:cfa-to-tr}),
and subsequently encoded as a sequential circuit (\cref{sect:tr-to-circuit}) suitable for hardware model checking.

\subsection{From CFA to Transition Relation via Large-Block Encoding}
\label{sect:cfa-to-tr}

Large-block encoding (LBE)~\cite{LBE} is a technique for extracting a transition relation from a CFA by summarizing multiple CFA edges into a single transition.
LBE first identifies maximum loop-free CFA segments (referred to as \emph{blocks}) and summarizes the edges within each block into a single transition.
That is, in the resulting transition system, all edges within a block are executed atomically in one step.

To construct the transition relation, we enumerate all syntactically feasible block-level transitions within the CFA,
and introduce an auxiliary program-counter variable $pc$ to encode the program locations corresponding to block entries and exits.
A transition corresponding to a block is then encoded as
$(pc = l_{\mathit{in}} \land pc' = l_{out}) \land \varphi(Y, I, Y')$,
where $l_{\mathit{in}}$ and $l_{out}$ are the entry and exit locations of the block, respectively,
and $\varphi(Y, I, Y')$ is a quantifier-free first-order formula encoding the semantics of executing the block.
Here, $Y$ denotes the set of program variables,
while $I$ denotes the set of input variables for modeling nondeterministic inputs from the environment (e.g., \lstinline[style=cstyle]{__VERIFIER_nondet_<type>()} in \svcomp benchmark tasks).
Accordingly, the set of state variables in the transition system is $X \triangleq Y \cup \{pc\}$.
The transition relation is the disjunction over all such formulas,
and the initial-state predicate is $pc = l_{init}$, where $l_{init}$ is the program-entry location.

Instead of encoding every CFA edge as an individual transition step (referred to as \emph{single-block encoding} in the literature~\cite{LBE}),
LBE reduces the number of transition steps required to represent a program execution.
Although each transition becomes more complex,
modern SMT solvers can typically handle the larger formulas efficiently.
This advantage is particularly pronounced for unrolling-based model-checking techniques,
as larger blocks allow more program behavior to be encoded and explored within a single transition step.

\begin{example}\label{ex:lbe}
We demonstrate the application of LBE to the CFA in \cref{fig:cfa}.
The identified blocks have entry and exit\newline
\begin{minipage}{.5\textwidth}
    \setlength{\parindent}{\acmparindent}%
    \setlength{\parskip}{\acmparskip}%
    \noindent locations at $l_{init}$, $l_6$, $l_{\mathit{err}}$, and $l_{\mathit{end}}$.
    The CFA summarized by LBE is shown in \cref{fig:cfa-lbe},
    where each edge corresponds to a block-level transition in the original CFA and may contain multiple program operations.

    The resulting transition system is encoded by state variables $X \triangleq \{pc, a, b\}$
    (where $a$ and $b$ correspond to the program variables \lstinline[style=cstyle]{a} and \lstinline[style=cstyle]{b}, respectively),
    as well as input variables $I \triangleq \{\mathit{nd}_5, \mathit{nd}_7\}$ (modeling return values of \lstinline[style=cstyle]{nondet()} at \cref{fig:c-prog:init-b,fig:c-prog:loop-body}).

    The initial-state predicate $\mathit{Init}$ is $pc = l_{init}$ and
    the transition relation $\mathit{TR}(X, I, X')$ is the disjunction of the following formulas:
\end{minipage}%
\begin{minipage}{.5\textwidth}
    \centering
    \scalebox{.9}{\input{figures/cfa-lbe}}
    
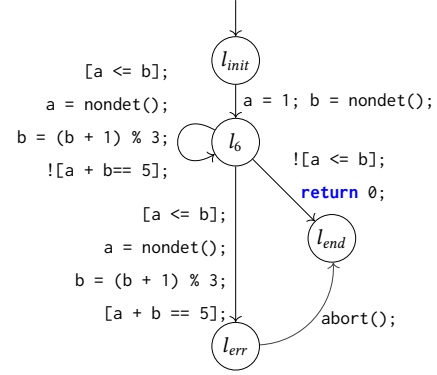
\captionof{figure}{CFA summarized by LBE}
    \label{fig:cfa-lbe}
\end{minipage}

\begin{itemize}
    \item $(pc = l_{init} \land pc' = l_6) \land (a' = 1 \land b' = \mathit{nd}_5)$
    \item $(pc = l_6 \land pc' = l_{\mathit{err}}) \land (a \leq b \land a' = \mathit{nd}_7 \land b' = (b + 1) \% 3 \land \mathit{nd}_7 + ((b + 1) \% 3) = 5)$
    \item $(pc = l_{6} \land pc' = l_6) \land (a \leq b \land a' = \mathit{nd}_7 \land b' = (b + 1) \% 3 \land \mathit{nd}_7 + ((b + 1) \% 3) \neq 5)$
    \item $(pc = l_{6} \land pc' = l_{\mathit{end}}) \land (a > b \land a' = a \land b' = b)$
    \item $(pc = l_{\mathit{err}} \land pc' = l_{\mathit{end}})$
\end{itemize}
\end{example}

\subsection{Constructing Sequential Circuits from Transition Relations}
\label{sect:tr-to-circuit}

Given a transition relation,
we consider two alternative strategies for constructing a sequential circuit.
The resulting circuits differ in how the transition relation is represented:
relational encoding treats the transition relation as a constraint and enforces it through an auxiliary oracle (\cref{sect:rel-enc}),
whereas functional encoding derives explicit next-state functions for each state variable (\cref{sect:func-enc}).
We describe both encoding strategies in the following.

\subsubsection{Relational Encoding}
\label{sect:rel-enc}

Relational encoding utilizes auxiliary variables to represent the next-state valuation.
To this end, we introduce:
(1)~a fresh Boolean state variable $\mathit{valid}$, initialized to $\top$,
which acts as an oracle tracking whether the current execution follows the transition relation $\mathit{TR}$; and
(2)~for each state variable $x \in X$, a fresh input variable $x_{\mathit{in}}$ simulating the next-state value of $x$.

The resulting circuit contains state variables $X \cup \{\mathit{valid}\}$ and input variables $I \cup X_{\mathit{in}}$, where $X_{\mathit{in}} = \{x_{\mathit{in}} \mid x \in X\}$.
The initial-state predicate of the circuit is $\mathit{Init}(X) \land \mathit{valid}$, where $\mathit{Init}(X)$ is the initial-state predicate of the original transition system.
The conjunction ensures that the circuit starts from a valid initial state.
The transition functions are constructed as:
\begin{align*}
\mathit{valid}' &\gets \mathit{valid} \land \mathit{TR}(X, I, X_{\mathit{in}}) \text{, and} \\
x' &\gets x_{\mathit{in}} \text{, for each } x \in X.
\end{align*}

Intuitively, the circuit nondeterministically selects next-state values through the input variables $X_{\mathit{in}}$,
while $\mathit{valid}$ monitors whether the chosen valuation satisfies the transition relation.
To ensure that only valid executions are considered during model checking,
the invariant constraint $\mathit{valid}$ is added to the resulting circuit.

\begin{example}\label{ex:rel-enc}
We apply relational encoding to the transition relation $\mathit{TR}$ obtained in \cref{ex:lbe}.
The resulting circuit has state variables $\{pc, a, b, \mathit{valid}\}$,
and input variables $\{\mathit{nd}_5, \mathit{nd}_7, pc_{\mathit{in}}, a_{\mathit{in}}, b_{\mathit{in}}\}$.
The initial-state predicate is $pc = l_{init} \land \mathit{valid}$, and
the transition functions are:
\begin{align*}
    \mathit{valid}' &\gets \mathit{valid} \land \mathit{TR}(\{pc, a, b\}, \{\mathit{nd}_5, \mathit{nd}_7\}, \{pc_{\mathit{in}}, a_{\mathit{in}}, b_{\mathit{in}}\}), \\
    pc' &\gets pc_{\mathit{in}}, \\
    a' &\gets a_{\mathit{in}} \text{, and} \\
    b' &\gets b_{\mathit{in}}.
\end{align*}
\end{example}

\subsubsection{Functional Encoding}
\label{sect:func-enc}

Functional encoding derives next-state functions for each state variable directly from the transition relation $\mathit{TR}$.
The key idea is to decompose each disjunct in $\mathit{TR}$ into \emph{conditions} and \emph{state updates}
(recall that $\mathit{TR}$ is a disjunction of conjunctions of predicates, where each conjunction corresponds to a block-level transition identified by LBE).
A condition is a predicate that only involves current-state and input variables,
e.g., $pc = l_{init}$ or $a \leq b$.
It captures the current program location and the guard for entering the next block, such as a loop condition.
A state update is a predicate of the form $x' = f(X, I)$ for some state variable $x \in X$,
e.g., $pc' = l_6$ or $b' = (b + 1) \% 3$.

Intuitively, a state update summarizes the net effect of a block on a single variable.
Because the execution of a block is deterministic,
i.e., fixing the current-state and input values uniquely determines the successor value of every state variable,
there exists, for every block and every state variable,
a function mapping the current state and inputs to the corresponding next-state value.
In practice, the LBE process might need to skolemize the formula by introducing static single-assignment indices for the state variables to hold the intermediate state values during a block's execution~\cite{AlgorithmComparison-JAR}.
The indexed variables can then be resolved recursively by substituting their defining equations,
eventually yielding state-update functions that depend only on the current-state and input variables.

For each state variable $x \in X$, the next-state function is obtained by collecting all state updates of $x$ and guarding each update with the conjunction of its corresponding conditions.
The resulting function is encoded as a nested if-then-else expression.
Conceptually, it has the form
\begin{align*}
    x' &\gets
    \begin{cases}
        f_1(X, I), & \text{if } c_1, \\
        f_2(X, I), & \text{if } c_2, \\
        &\vdots \\
        \ast, & \text{otherwise},
    \end{cases}
\end{align*}
where $\ast$ denotes any value or expression, each $f_i(X, I)$ originates from a state-update predicate and each guard $c_i$ is the conjunction of the conditions associated with that update.
The guards are mutually exclusive because they correspond to distinct transitions in the CFA.

To account for the fact that the transition relation extracted via LBE is not right-total
(e.g., the termination location $pc = l_{\mathit{end}}$ has no successor),
we introduce a dedicated sink state $l_{\mathit{sink}}$ for the program counter.
The identified conditions form the guards of the ``if'' and ``else-if'' branches, while the ``else'' branch corresponds to cases not covered by $\mathit{TR}$ and thus represents invalid transitions.
We handle the ``else'' branch as follows:
for $pc$, the next-state value is set to $l_{\mathit{sink}}$;
for all other state variables, the next-state values can be assigned arbitrarily (e.g., to their current values, as done in our implementation),
since execution beyond the current state is considered invalid, making the exact next-state value irrelevant.
Lastly, the invariant constraint $pc \neq l_{\mathit{sink}}$ is added to ensure that only valid executions are considered during model checking.

\begin{example}\label{ex:func-enc}
Applying functional encoding to the transition relation from \cref{ex:lbe} yields the following transition functions:
\begin{align*}
    a' &\gets
    \begin{cases}
        1, & \text{if } pc = l_{init}, \\
        \mathit{nd}_7, & \text{if } pc = l_6 \land a \leq b, \\
        a, & \text{if } pc = l_6 \land a > b, \\
        \ast, & \text{otherwise};
    \end{cases} \\[1ex]
    b' &\gets
    \begin{cases} \mathit{nd}_5, & \text{if } pc = l_{init}, \\
        (b + 1) \% 3, & \text{if } pc = l_6 \land a \leq b, \\
        b, & \text{if } pc = l_6 \land a > b, \\
        \ast, & \text{otherwise};
    \end{cases} \\[1ex]
    pc' &\gets
    \begin{cases}
        l_6, & \text{if } pc = l_{init} \text{ or } pc = l_6 \land a \leq b \land \mathit{nd}_7 + ((b + 1) \% 3) \neq 5, \\
        l_{\mathit{err}}, & \text{if } pc = l_6 \land a \leq b \land \mathit{nd}_7 + ((b + 1) \% 3) = 5, \\
        l_{\mathit{end}}, & \text{if } pc = l_6 \land a > b \text{ or } pc = l_{\mathit{err}}, \\
        l_{\mathit{sink}}, & \text{otherwise}.
    \end{cases}
\end{align*}
We merge the cases with the same next-state value for $pc$ to simplify the expression above.
The initial-state predicate remains unchanged as $pc = l_{init}$.
\end{example}

\subsubsection{Comparison}

Relational and functional encoding represent two different ways of constructing sequential circuits from a transition relation.
Relational encoding represents the transition relation as a constraint over current-state, next-state (simulated by auxiliary inputs $X_{\mathit{in}}$), and input variables
and monitors its validity through an oracle,
whereas functional encoding expresses the next state explicitly as a function of the current state and inputs.

Relational encoding is more general and can be applied to arbitrary transition relations without requiring a particular formula structure.
For example, the same technique has been used to construct sequential circuits from a nondeterministic finite automaton for string analysis~\cite{WangTLYJ16}.
This flexibility comes at the cost of introducing auxiliary variables to represent next-state valuations and validity checks.
Consequently, the generated circuits may be more difficult to verify, as the additional variables enlarge the search space.

Functional encoding, in contrast, relies on a specific characteristic of the transition relation by requiring that the execution of each block is deterministic.
This assumption ensures that state updates admit a functional expression over the current-state and input variables,
and is guaranteed by the transition relation induced by the CFA model.
Because next-state functions are derived directly, no auxiliary next-state variables are required, resulting in more compact circuits.
Moreover, exposing the transition structure through explicit next-state functions
allows more room for logic simplification and enables \emph{functional substitution} during circuit unrolling,
which propagates state information across time frames and simplifies the resulting SAT/SMT formulas.
Such simplifications can be performed during preprocessing by SMT solvers
such as \tool{Boolector}~\cite{Boolector3} and \tool{Bitwuzla}~\cite{Bitwuzla-CAV23} through variable and term substitution.
Hence, functional encoding is often considered more amenable to SAT/SMT-based reasoning than relational encoding,
as shown by previous studies in the context of bounded model checking~\cite{BMC-QBF,BMCLockLessProgram}.
In practice, however, the effectiveness of either encoding also depends on the verification algorithm employed by the backend model checker,
as demonstrated in our evaluation (\cref{sect:evaluation}).

%% file: figures/cfa-lbe.tex
\begin{tikzpicture}
    \tikzset{
        cfanode/.style={
            draw,
            circle,
            minimum size=.7cm,
            inner sep=0.03cm,
        }
    }
    \node (s) {};
    \node (linit) [below of = s, node distance=1cm, cfanode]{$l_{\mathit{init}}$};
    \node (l6) [below of = linit, node distance=1.2cm, cfanode]{$l_6$};
    \node (lerr) [below of = l6, node distance=3cm, cfanode]{$l_{\mathit{err}}$};
    \node (lend) [below right of = l6, node distance=2cm, cfanode]{$l_{\mathit{end}}$};

    \draw[->] (s) --(linit);
    \draw[->] (linit) to node[right] {\lstinline[style=cstyle]{a = 1; b = nondet();}} (l6);
    \draw[->] (l6) edge[out=150, in=210, looseness=6] node[left, align=right, yshift=3mm]
        {\lstinline[style=cstyle]{[a <= b];}\\\lstinline[style=cstyle]{a = nondet();}\\\lstinline[style=cstyle]{b = (b + 1) \% 3;}\\\lstinline[style=cstyle]{![a + b== 5];}} (l6);
    \draw[->] (l6) to node[right, yshift=2mm, align=right] {\lstinline[style=cstyle]{![a <= b];}\\\lstinline[style=cstyle]{return 0;}} (lend);
    \draw[->] (lerr) edge[bend right=45, looseness=1, draw=darkgray] node[right] {\lstinline[style=cstyle]{abort();}} (lend);
    \draw[->] (l6) to node[left, align=right, yshift=-3mm]
        {\lstinline[style=cstyle]{[a <= b];}\\\lstinline[style=cstyle]{a = nondet();}\\\lstinline[style=cstyle]{b = (b + 1) \% 3;}\\\lstinline[style=cstyle]{[a + b == 5];}} (lerr);
\end{tikzpicture}

%% file: reachsafety.tex
\section{Verifying Reachability Safety}
\label{sect:verify-reachsafety}

In this section, we describe how reachability-safety properties are encoded in the sequential circuits (\cref{sect:enc-reach}),
and how we instrument the program to facilitate the extraction of software-verification witnesses from hardware model-checking results (\cref{sect:instr-reach}).

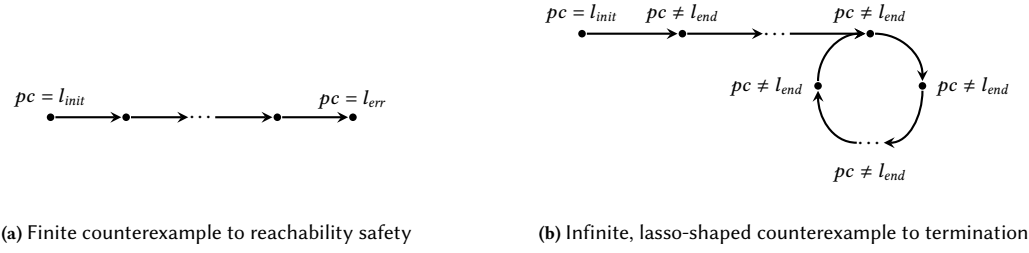
\begin{figure}[t]
    \centering
    \subfloat[Finite counterexample to reachability safety]{%
        \begin{minipage}[b]{.45\textwidth}
            \centering
            \scalebox{1}{\input{figures/cex-bad}}\label{fig:cex-bad}%
            \vspace{1cm}%
        \end{minipage}
    }%
    \subfloat[Infinite, lasso-shaped counterexample to termination]{%
        \begin{minipage}[b]{.55\textwidth}
            \centering
            \scalebox{1}{\input{figures/cex-justice}}\label{fig:cex-justice}%
        \end{minipage}
    }
    \caption{Counterexample structures for reachability-safety~(a) and termination~(b) properties}
    \label{fig:cex}
\end{figure}

\subsection{Encoding Reachability-Safety Property}
\label{sect:enc-reach}

In \svcomp~\cite{SVCOMP26}, error locations in C programs are annotated with calls to the function \lstinline[style=cstyle]{reach_error()},
and later assigned to a designated program-counter value $l_{\mathit{err}}$ during translation.
The reachability-safety property states that error locations are unreachable from the program entry
and can therefore be encoded by asserting that the program counter never equals $l_{\mathit{err}}$,
captured by the bad-state predicate $\mathit{bad} \triangleq (pc = l_{\mathit{err}})$.

Given a translated circuit and a safety property,
a hardware model checker automatically determines whether the property holds.
If the property is violated, the model checker produces a violation witness (counterexample),
namely a trace in the translated circuit that reaches a bad state from an initial state
while satisfying all invariant constraints, as illustrated in \cref{fig:cex-bad}.
Such a trace corresponds to a program execution path that starts from the initial location and reaches an error location.
\Cref{sect:instr-reach} will describe how an error trace can be reconstructed from a hardware-model-checking witness.
Conversely, if the safety property holds, no valid execution path of either the original program or the translated circuit reaches an error location or bad state, respectively.
Some hardware model checkers can additionally produce a correctness witness,
often in the form of an inductive invariant overapproximating the set of reachable states, as a proof of safety.
However, current support for exporting correctness witnesses is less mature and less uniform than that for violation witnesses in hardware model checkers.
Recent editions of HWMCC have taken steps toward standardizing them~\cite{CertificateHWMCC-CAV25}.
In principle, an inductive invariant for the circuit could also be translated back into loop invariants in the original program,
but \cpv currently does not support this translation.

\subsection{Program Instrumentation and Error-Trace Reconstruction}
\label{sect:instr-reach}

Large-block encoding folds program statements at different locations into a single transition-system step.
As a result, a violation witness produced by a hardware model checker may contain values for input variables that are not actually relevant to the property violation.
For instance, when transitioning from $l_{\mathit{init}}$ to $l_6$ in \cref{fig:cfa},
the return value of \lstinline[style=cstyle]{__VERIFIER_nondet_uint()} at \cref{fig:c-prog:loop-body} (encoded as input variable $nd_7$ in the circuit) does not affect the transition.
Nevertheless, a hardware model checker is not able to distinguish the relevant and irrelevant input values
and may assign an arbitrary value to $nd_7$ in the witness.
Such values cannot be mapped back to the original program when reconstructing an error trace.

To easily identify the nondeterministic inputs that are actually consumed along a property-violating execution path,
we instrument each \lstinline[style=cstyle]{__VERIFIER_nondet_<type>()} call with a dedicated counter whose value changes after every invocation.
These counters are translated into state variables in the resulting sequential circuit and are therefore observable in the hardware-model-checking witness.
A change in a counter value between two consecutive time steps indicates that the corresponding nondeterministic input has been consumed during that transition.
Furthermore, the source location of each \lstinline[style=cstyle]{__VERIFIER_nondet_<type>()} call can be encoded in the corresponding counter's variable name
(e.g., using its line and column numbers), as illustrated in \cref{ex:instr-reach}.
Given a hardware violation witness, we reconstruct the corresponding software error trace by tracking changes in the counters and extracting the associated input values.
The recovered inputs are then mapped back to the program locations containing the corresponding \lstinline[style=cstyle]{__VERIFIER_nondet_<type>()} calls.
This allows us to discard irrelevant input assignments introduced by the hardware model checker and produce a witness that reflects the execution of the original program.

\begin{figure}[t]
    \centering
\begin{lstlisting}[style=cstyle]
extern unsigned int __VERIFIER_nondet_uint();
unsigned int __cpv_nondet_uint_5() { // added by CPV
  static unsigned char __cpv_nondet_count_5 = 0;
  __cpv_nondet_count_5 = __cpv_nondet_count_5 + 1;
  return __VERIFIER_nondet_uint();
}
unsigned int __cpv_nondet_uint_7() { // added by CPV
  static unsigned char __cpv_nondet_count_7 = 0;
  __cpv_nondet_count_7 = __cpv_nondet_count_7 + 1;
  return __VERIFIER_nondet_uint();
}
void reach_error() { /* error label */ abort(); }
int main() {
  unsigned int a = 1;
  unsigned int b = __cpv_nondet_uint_5(); // originally at line 5
  while (a <= b) {
    a = __cpv_nondet_uint_7(); // originally at line 7
    b = (b + 1) % 3;
    if (a + b == 5) {
      reach_error();
    }
  }
  return 0;
}
\end{lstlisting}
    \vspace*{-3mm}
    \caption{Instrumented C program for reachability analysis}
    \label{fig:c-instr-reach}
\end{figure}

The instrumentation further improves the modularity of the verification pipeline.
Without it, witness reconstruction would require additional metadata (communicated through a separate artifact) from the task translator,
such as mappings between circuit input variables and the CFA transitions that consume them,
as well as the corresponding source-code locations.
By exposing the consumption of nondeterministic inputs through counters encoded directly in the circuit,
all information required for witness reconstruction becomes part of the verification task itself.
As a result, the witness translator can recover both the relevant input values and their source-code locations directly from the hardware-model-checking witness,
allowing it to operate independently of the task translator and keeping the overall architecture modular.

\begin{example}\label{ex:instr-reach}
We demonstrate the instrumentation technique using the example program in \cref{fig:c-prog}.
The two calls to \lstinline[style=cstyle]{__VERIFIER_nondet_uint()} are instrumented with counters \lstinline[style=cstyle]{__cpv_nondet_count_5} and \lstinline[style=cstyle]{__cpv_nondet_count_7}, as shown in \cref{fig:c-instr-reach}.
The suffixes \lstinline[style=cstyle]{_5} and \lstinline[style=cstyle]{_7} denote the line numbers of the calls in the original program before instrumentation.
For brevity, column numbers are omitted from the counter names shown here.
Both counters are of type \lstinline[style=cstyle]{unsigned char} and are initialized to 0.
As we only need to observe whether a counter value changes, the potential overflow is not a concern.
Using an unsigned type also avoids the undefined behavior associated with signed-integer overflow in C.

After translation, we obtain a sequential circuit similar to those in the previous examples,
with two additional state variables, $\mathit{ndcnt}_5$ and $\mathit{ndcnt}_7$, corresponding to the inserted counters.
The bad-state predicate is $\mathit{bad} \triangleq (pc = l_{\mathit{err}})$,
where $l_{\mathit{err}}$ is the program-counter value assigned to the error location.

A violation witness produced by the hardware model checker is summarized in the following table (``-'' denotes a don't-care value):
\begin{table}[H]
\centering
\begin{tabular}{crrrrrrr}
\toprule
\multirow{2}{*}{Time step}
    & \multicolumn{5}{c}{State}
    & \multicolumn{2}{c}{Input}
    \\
\cmidrule(lr){2-6}\cmidrule(lr){7-8}
    & \multicolumn{1}{c}{$pc$}
    & \multicolumn{1}{c}{$a$}
    & \multicolumn{1}{c}{$b$}
    & \multicolumn{1}{c}{$\mathit{ndcnt}_5$}
    & \multicolumn{1}{c}{$\mathit{ndcnt}_7$}
    & \multicolumn{1}{c}{$nd_5$}
    & \multicolumn{1}{c}{$nd_7$}
    \\
\midrule
0
    & $l_{\mathit{init}}$
    & -
    & -
    & 0
    & 0
    & \texttt{\bfseries UINT_MAX}
    & -
\\
1
    & $l_{6}$
    & 1
    & \texttt{UINT_MAX}
    & 1
    & 0
    & -
    & \textbf{5}
\\
2
    & $l_{\mathit{err}}$
    & 5
    & 0
    & 1
    & 1
    & -
    & -
\\
\bottomrule
\end{tabular}
\end{table}
\noindent The input values to be extracted are highlighted in bold.
The value $nd_5 =$ \lstinline[style=cstyle]{UINT_MAX} at time step~0 is relevant because $\mathit{ndcnt}_5$ changes from 0 to 1 between time steps~0 and~1.
Similarly, we take the value $nd_7 =$ \lstinline[style=cstyle]{5} at time step~1 because $\mathit{ndcnt}_7$ changes from 0 to 1 between time steps~1 and~2.

Mapping these values back to the corresponding nondeterministic calls in the original program yields the following error trace:
\begin{enumerate}
    \item enter \lstinline[style=cstyle]{main()};
    \item \lstinline[style=cstyle]{__VERIFIER_nondet_uint()} returns \lstinline[style=cstyle]{UINT_MAX} at \cref{fig:c-prog:init-b};
    \item enter the \lstinline[style=cstyle]{while} loop at \cref{fig:c-prog:while};
    \item \lstinline[style=cstyle]{__VERIFIER_nondet_uint()} returns \lstinline[style=cstyle]{5} at \cref{fig:c-prog:loop-body};
    \item \lstinline[style=cstyle]{reach_error()} is called at \cref{fig:c-prog:err}.
\end{enumerate}
\end{example}

%% file: figures/cex-bad.tex
\begin{tikzpicture}[>=stealth, thick, every node/.style={draw=none, circle, fill, minimum size=1mm, inner sep = 0pt}]

    \node (s0) {};
    \node (s1) [right of = s0] {};
    \node (s2) [right of = s1, fill=none] {\ldots};
    \node (s3) [right of = s2] {};
    \node (s4) [right of = s3] {};
    \node [above=-3mm of s0, fill=none] {\smaller $pc = l_{\mathit{init}}$};
    \node [above=-3mm of s4, fill=none] {\smaller $pc = l_{\mathit{err}}$};

    \draw[->] (s0) -- (s1);
    \draw[->] (s1) -- (s2);
    \draw[->] (s2) -- (s3);
    \draw[->] (s3) -- (s4);

\end{tikzpicture}

%% file: figures/cex-justice.tex
\begin{tikzpicture}[>=stealth, thick, every node/.style={draw=none, circle, fill, minimum size=1mm, inner sep = 0pt}]

    \node (s1) {};
    \node (s2) [right=1.2cm of s1] {};
    \node (d1) [right= of s2, fill=none] {\ldots};
    \node (s3) [right= of d1] {};
    \node (s4) [below right=.6cm and .6cm of s3] {};
    \node (s5) [below=1.2cm of s3, fill=none] {\ldots};
    \node (s6) [below left=.6cm and .6cm of s3] {};
    \node [above=-3mm of s1, fill=none] {\smaller $pc = l_{\mathit{init}}$};
    \node [above=-3mm of s2, fill=none] {\smaller $pc \neq l_{\mathit{end}}$};
    \node [above=-3mm of s3, fill=none] {\smaller $pc \neq l_{\mathit{end}}$};
    \node [right=1mm of s4, fill=none] {\smaller $pc \neq l_{\mathit{end}}$};
    \node [below=-3mm of s5, fill=none] {\smaller $pc \neq l_{\mathit{end}}$};
    \node [left=1mm of s6, fill=none] {\smaller $pc \neq l_{\mathit{end}}$};

    \draw[->] (s1) -- (s2);
    \draw[->] (s2) -- (d1);
    \draw[->] (d1) -- (s3);
    \draw[->] (s3) to[out=0, in=90] (s4);
    \draw[->] (s4) to[out=-90, in=0] (s5);
    \draw[->] (s5) to[out=180, in=-90] (s6);
    \draw[->] (s6) to[out=90, in=180] (s3);

\end{tikzpicture}

%% file: termination.tex
\section{Verifying Program Termination}
\label{sect:verify-termination}

Here we explain how termination properties are encoded in the translated circuits (\cref{sect:enc-term}),
how the resulting liveness properties can be reduced to safety properties (\cref{sect:termination-l2s}),
and how non-termination witnesses can be extracted from hardware model-checking results (\cref{sect:instr-term}).

\subsection{Encoding Termination Property}
\label{sect:enc-term}

During translation, program-termination locations are assigned a designated program-counter value $l_{\mathit{end}}$.
The termination property requires that every execution starting from the program entry eventually reaches a termination location.
Accordingly, we encode the property using a justice constraint defined as $j \triangleq (pc \neq l_{\mathit{end}})$.
Note, however, that requiring $pc \neq l_{\mathit{end}}$ to hold only infinitely often in a counterexample is not strong enough.
We must additionally enforce that $pc \neq l_{\mathit{end}}$ is true at \emph{every} step.
Therefore, an invariant constraint $pc \neq l_{\mathit{end}}$ is added
in conjunction with the invariant constraint introduced during circuit encoding in \cref{sect:tr-to-circuit}.%
\footnote{Since the justice constraint is implied by the invariant constraint,
it can technically be omitted (i.e., set to $\top$). We retain it here for clarity.}

Recall the semantics of justice constraints from \cref{sect:hwmc-properties}.
A counterexample is an infinite initialized execution in the shape of a lasso,
consisting of a finite prefix (the stem) leading to a cycle that can be traversed indefinitely,
as illustrated in \cref{fig:cex-justice}.
Along this execution, $pc \neq l_{\mathit{end}}$ holds always (and thus infinitely often).
The cycle portion of the lasso represents a recurring program behavior that prevents the execution from reaching a termination location.
Thus, such a counterexample corresponds directly to a non-terminating execution of the original program.
If no counterexample exists, i.e., the termination property holds,
then all executions of the original program eventually reach a termination location.

\subsection{Liveness-to-Safety Transformation}
\label{sect:termination-l2s}

Since many hardware model checkers do not support liveness properties directly,
we apply a standard liveness-to-safety (L2S) transformation~\cite{LivenessAsSafetyFinite}.
The general L2S transformation supports multiple justice constraints.
In our setting, however, only a single justice constraint is present, and it is already implied by the invariant constraint introduced above.
Consequently, the transformation can be simplified to checking whether a previously encountered state can be revisited in the transition system.

The L2S transformation introduces a Boolean input variable $\mathit{save}$ and a Boolean state variable $\mathit{saved}$.
In addition, for each state variable $x$, a corresponding recording state variable $x_r$ of the same sort is introduced.
The input variable $\mathit{save}$ acts as an oracle and may become $\top$ at any time step.
The state variable $\mathit{saved}$ is initialized to $\bot$ and records whether $\mathit{save}$ has been triggered previously.
Its transition function is $\mathit{saved}' \gets \mathit{save} \lor \mathit{saved}$.
When $\mathit{save}$ is triggered for the first time, i.e., when $save \land \neg saved$ holds,
the recording state variables capture the current state.
Thus, the transition function of each recording state variable is defined as:
$x_r' \gets \mathit{ite}(\mathit{save} \land \neg \mathit{saved}, x, x_r)$.
All remaining state variables are updated according to their original transition functions,
unchanged by the L2S transformation.
A state revisit is then detected by the predicate $\mathit{saved} \land \bigwedge_{x \in X} (x = x_r)$,
which checks whether the current state matches the previously saved state.
The transformed safety property is obtained by treating this predicate as the bad-state condition.

A model checker that supports safety properties can then be used to verify the transformed circuit and property.
Besides the verification verdict, model checkers may also produce witnesses.
Similar to \cref{sect:verify-reachsafety},
a violation witness and a correctness witness for the transformed circuit can be translated back
to a non-terminating lasso and a transition invariant~\cite{TransitionInvariants} for the original program, respectively.
The witness translator in \cpv currently supports translating violation witnesses (further described in \cref{sect:instr-term}),
whereas translation of correctness witnesses~\cite{LivenessProofsHWMC,WitnessesTransitionInvariants} as discussed already in \cref{sect:witnesses} is left for future work.

We can further optimize the L2S transformation by excluding from the state-revisit check the auxiliary counter variables introduced during program instrumentation for witness reconstruction (see \cref{sect:instr-term}).
These variables serve solely for bookkeeping purposes and do not influence program execution or termination, and can therefore be safely omitted.
More generally, under functional encoding, any state variable that is not contained in the cone of influence of the justice and invariant constraints
(including the instrumented counters) does not need to be included in the state comparison.
This CoI-based optimization, however, only works for functional encoding,
since in relational encoding, all state variables appear in the transition function of the state variable $\mathit{valid}$.
Hence, CoI analysis cannot identify irrelevant variables and only the instrumented counter variables can be reliably omitted.

The optimization is particularly beneficial for the instrumented counters.
With counters included in the state-revisit check,
the execution would need to revisit not only the same program state but also the same counter values.
The counters are typically set to increase by a certain value upon each invocation.
Re-obtaining the same values may require many additional transitions until wrap-around (due to overflow) occurs.
As a result, the lasso-shaped witness produced by the model checker may become substantially longer, making the verification problem unnecessarily harder.
The effectiveness of this optimization is confirmed by the evaluation results presented later in \cref{sect:evaluation} (\hyperref[sect:rq3.1]{RQ3.1}).

\subsection{Program Instrumentation and Non-Termination Witness Extraction}
\label{sect:instr-term}

As in \cref{sect:instr-reach}, we instrument each \lstinline[style=cstyle]{__VERIFIER_nondet_<type>()} call
with a counter to identify the input values relevant to a non-terminating execution.

Moreover, in order to determine which loop is responsible for non-termination,
one could inspect the program-counter value $pc$ at which a state saving or revisit occurs in the witness
produced by the model checker and map it back to the corresponding program location.
Nevertheless, this requires understanding how program locations are encoded in the circuit,
which may vary across encoders (e.g., one-hot or binary encoding).
One could for example store an explicit mapping between circuit-level $pc$ encodings and program locations
as an auxiliary artifact produced by the encoder, and use it during witness translation.
To reduce coupling between the circuit encoder and the witness translator,
we instead introduce a dedicated counter for each loop in the program,
whose value changes at the beginning of the loop body.
These counters are encoded as state variables in the resulting sequential circuit.
As with the nondeterministic-input counters,
the source location of each loop can be encoded in the corresponding counter variable's name.

Given a violation witness produced by the hardware model checker,
relevant input values can be recovered by tracking changes in the nondeterministic-input counters.
Similarly, the loop responsible for non-termination can be identified via the loop counter that changes at the time the state saving occurs, or equivalently, via the $pc$ value at that point.

\begin{figure}[t]
    \centering
\begin{lstlisting}[style=cstyle]
extern unsigned int __VERIFIER_nondet_uint();
unsigned int __cpv_nondet_uint_5() { ... }   // added by CPV
unsigned int __cpv_nondet_uint_7() { ... }   // added by CPV
unsigned char __cpv_loop_count_6 = 0;        // added by CPV
void reach_error() { /* error label */ abort(); }
int main() {
  unsigned int a = 1;
  unsigned int b = __cpv_nondet_uint_5();
  while (a <= b) {                               // originally at line 6
    __cpv_loop_count_6 = __cpv_loop_count_6 + 1; // added by CPV
    a = __cpv_nondet_uint_7();
    b = (b + 1) % 3;
    if (a + b == 5) {
      reach_error();
    }
  }
  return 0;
}
\end{lstlisting}
    \vspace*{-3mm}
    \caption{Instrumented C program for termination analysis}
    \label{fig:c-instr-term}
\end{figure}

\refstepcounter{footnote}\label{ex:instr-term:footnote}
\footnotetext{In \cpv's implementation,
``\lstinline[style=cstyle]{\&\& ((__cpv_loop_count_<id> = __cpv_loop_count_<id> + 1) || 1)}'' is appended to the original loop condition,
which is equivalent to incrementing the counter at the beginning of the loop body.}

\begin{example}\label{ex:instr-term}
In \cref{fig:c-instr-term}, we illustrate the instrumentation on the example program from \cref{fig:c-prog}.
As in \cref{ex:instr-reach}, the two \lstinline[style=cstyle]{__VERIFIER_nondet_uint()} calls are instrumented with counters to track the consumption of their return values.
In addition, a counter \lstinline[style=cstyle]{__cpv_loop_count_6} is introduced for the \lstinline[style=cstyle]{while} loop at \cref{fig:c-prog:while} to monitor loop entries.
The counter is declared as a global variable, initialized to \lstinline[style=cstyle]{0},
and incremented at the beginning of the loop body.\hyperref[ex:instr-term:footnote]{\footnotemark[\value{footnote}]}
Its name encodes the source location of the \lstinline[style=cstyle]{while} loop in the original program.

After going through the translation pipeline,
we obtain a sequential circuit similar to that in \cref{ex:instr-reach},
with one additional state variable, $\mathit{loopcnt}_6$, corresponding to the inserted loop counter.
The L2S transformation is then applied to the circuit.
For illustration, we assume the functional encoding from \cref{ex:func-enc},
though the transformation itself is independent of the encoding choice and applies also to circuits under relational encoding.
L2S introduces state variables $\mathit{saved}$, $a_r$, $b_r$, and $pc_r$
(the latter three for recording a previously visited program state), together with an input variable $\mathit{save}$.
The liveness property is thereby reduced to checking whether
$\mathit{saved} \land (pc = pc_r) \land (a = a_r) \land (b = b_r)$
can ever become $\top$ while satisfying the invariant constraint ($pc \neq l_{\mathit{end}} \land pc \neq l_{sink}$).

An example violation witness for the transformed problem is shown below (``-'' denotes a don't-care value).
For simplicity, we omit the recording state variables ($a_r$, $b_r$, and $pc_r$) from the table.

\begin{table}[H]
\centering
\begin{tabular}{crrrrrrrrrr}
\toprule
\multirow{2}{*}{Time step}
    & \multicolumn{7}{c}{State}
    & \multicolumn{3}{c}{Input}
    \\
\cmidrule(lr){2-8}\cmidrule(lr){9-11}
    & \multicolumn{1}{c}{$pc$}
    & \multicolumn{1}{c}{$a$}
    & \multicolumn{1}{c}{$b$}
    & \multicolumn{1}{c}{$\mathit{ndcnt}_5$}
    & \multicolumn{1}{c}{$\mathit{ndcnt}_7$}
    & \multicolumn{1}{c}{$\mathit{loopcnt}_6$}
    & \multicolumn{1}{c}{$\mathit{saved}$}
    & \multicolumn{1}{c}{$\mathit{nd}_5$}
    & \multicolumn{1}{c}{$\mathit{nd}_7$}
    & \multicolumn{1}{c}{$\mathit{save}$}
    \\
\midrule
0
    & $l_{\mathit{init}}$
    & -
    & -
    & 0
    & 0
    & 0
    & 0
    & \textbf{1}
    & -
    & 0
    \\
1
    & $\bm{l_6}$
    & \textbf{1}
    & \textbf{1}
    & 1
    & 0
    & 0
    & 0
    & -
    & \textbf{1}
    & 1
    \\
2
    & $l_6$
    & 1
    & 2
    & 1
    & 1
    & 1
    & 1
    & -
    & \textbf{0}
    & -
    \\
3 
    & $l_6$
    & 0
    & 0
    & 1
    & 2
    & 2
    & 1
    & -
    & \textbf{1}
    & -
    \\
4
    & $\bm{l_6}$
    & \textbf{1}
    & \textbf{1}
    & 1
    & 3
    & 3
    & 1
    & -
    & -
    & -
    \\
\bottomrule
\end{tabular}
\end{table}

From the above witness, observe that the values of $pc$, $a$, and $b$ at time step~1 are identical to those at time step~4,
indicating a state revisit and thus the cycle portion of the lasso-shaped counterexample.
These values are captured by the corresponding recording variables from time step~2 onward,
and are subsequently used to detect the revisit condition at time step~4.
The \lstinline[style=cstyle]{while} loop at \cref{fig:c-prog:while} is entered repeatedly between time steps~1 and~4,
as evidenced by the $pc$'s value and the increasing value of the loop counter $\mathit{loopcnt}_6$.
The state-saving event occurs at time step~1, when $\mathit{save}$ is triggered for the first time.
At the same time, $\mathit{loopcnt}_6$ increases from time step~1 to~2,
marking the transition from the stem to the cycle in the lasso-shaped counterexample and
indicating that this \lstinline[style=cstyle]{while} loop is responsible for the non-terminating behavior.
The same procedure described in \cref{sect:instr-reach} can be used to recover the return values of the \lstinline[style=cstyle]{__VERIFIER_nondet_uint()} calls.

After extracting the relevant input values and mapping them back to the original program,
we reconstruct the non-termination witness as follows:

\begin{enumerate}
    \item enter \lstinline[style=cstyle]{main()};
    \item \lstinline[style=cstyle]{__VERIFIER_nondet_uint()} returns \lstinline[style=cstyle]{1} at \cref{fig:c-prog:init-b};
    \item enter the \lstinline[style=cstyle]{while} loop at \cref{fig:c-prog:while}
        with program state \lstinline[style=cstyle]{a = 1} and \lstinline[style=cstyle]{b = 1};
    \item \lstinline[style=cstyle]{__VERIFIER_nondet_uint()} returns \lstinline[style=cstyle]{1} at \cref{fig:c-prog:loop-body};
    \item enter the \lstinline[style=cstyle]{while} loop at \cref{fig:c-prog:while} again
        with \lstinline[style=cstyle]{a = 1} and \lstinline[style=cstyle]{b = 2};
    \item \lstinline[style=cstyle]{__VERIFIER_nondet_uint()} returns \lstinline[style=cstyle]{0} at \cref{fig:c-prog:loop-body};
    \item enter the \lstinline[style=cstyle]{while} loop at \cref{fig:c-prog:while} again
        with \lstinline[style=cstyle]{a = 0} and \lstinline[style=cstyle]{b = 0};
    \item \lstinline[style=cstyle]{__VERIFIER_nondet_uint()} returns \lstinline[style=cstyle]{1} at \cref{fig:c-prog:loop-body};
    \item enter the \lstinline[style=cstyle]{while} loop at \cref{fig:c-prog:while} again
        with the same state as in step~(3), namely \lstinline[style=cstyle]{a = 1} and \lstinline[style=cstyle]{b = 1} (state revisit occurs).
\end{enumerate}

\end{example}

%% file: implementation.tex
\section{Implementation}
\label{sect:implementation}

The previous sections presented the concepts and workflow of circuit-based program verification.
In this section, we describe how they are implemented in \cpv.
%

\cpv orchestrates the execution of the various components involved in the verification workflow.
This orchestration layer is implemented in Python,
with individual components either invoked as subprocesses or integrated as libraries.
The implementation builds upon several external components, including
\kratostwo~\cite{Kratos2} for frontend parsing and parts of the program-to-circuit translation (\cref{sect:impl-c-to-btor2}),
\coveriteam~\cite{COVERITEAM} for invoking backend hardware model checkers (\cref{sect:backend-integration}),
and the hardware model checkers themselves for verifying the translated circuits.
Details on the witness translator (\cref{sect:witness-translation}) and the portfolios used in \svcomp and in our evaluation (\cref{sect:svcomp-portfolios}) are also provided.

\subsection{C-to-\btortwo Encoding}
\label{sect:impl-c-to-btor2}

Depending on the verification task, \cpv first instruments the input program to facilitate witness translation and property encoding.
We then use \kratostwo~\cite{Kratos2} for large-block encoding.
\kratostwo parses the C program and translates it into K2,
an intermediate verification language that essentially represents the program's control-flow automaton in textual form.
Given a program entry and a target function,
it applies large-block encoding and introduces a symbolic program counter to obtain a state-transition system,
from which the corresponding $pc$ values for the program entry and target location can be obtained.
For reachability-safety verification, this workflow can be applied directly,
as the program entry is typically \lstinline[style=cstyle]{main()},
and, in the \svcomp setting, the target error function is \lstinline[style=cstyle]{reach_error()}.
For termination analysis, however, a program may terminate through multiple exit points,
which cannot necessarily be represented by a single target function.
Since \kratostwo supports only one target function,
\cpv instruments the program with a fresh function label \lstinline[style=cstyle]{__cpv_reach_exit()} and uses it as the target.
Calls to \lstinline[style=cstyle]{__cpv_reach_exit()} are inserted immediately \emph{before} every program exit point except returning from \lstinline[style=cstyle]{main()},
including calls to \lstinline[style=cstyle]{exit()} and \lstinline[style=cstyle]{abort()}.
To handle termination through \lstinline[style=cstyle]{main()},
a wrapper function \lstinline[style=cstyle]{__cpv_main()} is introduced,
which invokes the original \lstinline[style=cstyle]{main()} first and then calls \lstinline[style=cstyle]{__cpv_reach_exit()}.
As a result, all program exits are redirected to a single target location represented by \lstinline[style=cstyle]{__cpv_reach_exit()},
and the wrapper function \lstinline[style=cstyle]{__cpv_main()} serves as the new program entry for \kratostwo.

We use \btortwo~\cite{Boolector3} to represent the translated circuit because it is the prevailing format for hardware model checking
and is supported by many tools participating in HWMCC~\cite{HWMCC25}.
Relational encoding is implemented by \kratostwo, which can translate K2 programs directly into \btortwo circuits.
For functional encoding, K2 is first encoded into a VMT-LIB transition system~\cite{VMT-LIB,VMT} by \kratostwo and then to \btortwo using our functional encoder.
The functional encoder is implemented in Python and uses \tool{PyVMT}~\cite{PyVMT} and \tool{PySMT}~\cite{PySMT} for manipulating transition systems and SMT formulas.
Two encoding options for the program-counter variable $pc$ are provided by the functional encoder:
one-hot encoding, which uses $n$ bits for $n$ program locations,
and binary encoding, which uses $\lceil \log_2(n) \rceil$ bits
(relational encoding in \kratostwo currently supports only binary encoding of $pc$).
Although one-hot encoding requires more state registers, it often yields simpler next-state logic and,
consequently, more compact circuits in terms of overall gate count~\cite{SeqLogicDesignBookChapter,TimeFrameFolding}.
One-hot encoding is also argued to provide more fine-grained splitting opportunities for SAT/SMT solvers resulting in more specific learned clauses and thus faster solving.
In our case, since the number of program locations is typically small compared with the program state space,
which is dominated by the program's data variables,
the choice has little effect on the resulting circuits.
For reachability-safety verification, the generated circuit contains a \texttt{bad} signal asserting that $pc$ is equal to the program-counter value assigned to the target error function.
For termination, the generated circuit contains a \texttt{justice} constraint corresponding to the termination encoding described in \cref{sect:verify-termination}.
The L2S transformer in \cpv takes a \btortwo circuit containing a \texttt{justice} constraint as input and produces a \btortwo circuit containing a \texttt{bad} property as output.

During the translation to \btortwo,
primitive data types, including integers, floating-point numbers, and pointers,
in the input program are encoded using fixed-size bit-vectors.
Floating-point values can be encoded by blasting them into bit-vectors,
which is supported by the relational encoding implemented in \kratostwo.
Our functional encoder currently does not support this word-blasting procedure.
Static arrays are naturally represented as \btortwo arrays.
Heap memory is modeled in K2 (and hence in \btortwo) as a large global array,
and pointer dereferences are translated into reads from and writes to this array,
which are subsequently encoded as \btortwo array operations.
Since \btortwo represents transition systems in QF\_ABV, all input and state variables must range over finite domains.
This requirement is naturally satisfied by fixed-width machine types and the finite address space used to model heap memory.
Several language features remain only partially supported.
For instance, support for library functions is limited and programs relying on functions from libraries such as \texttt{math.h} may fail during the C-to-K2 translation stage.
The capabilities and limitations of the frontend translation are evaluated and discussed in \cref{sect:evaluation} (\hyperref[sect:rq1]{RQ1}).
Function calls are inlined during large-block encoding.
Recursive functions are only partially supported,
as \cpv currently relies on \kratostwo to perform bounded unrolling with a user-specified recursion depth,%
\footnote{A possible alternative would be to encode the call stack explicitly as an array in the translated circuit,
although this approach has not yet been explored.}
and support for concurrent programs is currently beyond the scope of \cpv.

While developing \cpv, we uncovered several issues in \kratostwo related to the translation pipeline.
We contributed fixes and reported the issues to the development team of \kratostwo.
All identified problems have since been resolved, improving the robustness of the underlying infrastructure used by \cpv.

\subsection{Integration of Hardware Model Checkers}
\label{sect:backend-integration}

\cpv currently integrates four hardware model checkers:
\abc~\cite{ABC}, \avr~\cite{AVR}, \ricthree~\cite{rIC3}, and \pono~\cite{Pono2}.
Together, these tools provide access to a diverse set of verification algorithms,
including bounded model checking (BMC)~\cite{BMCJournal},
$k$-induction~\cite{InductionVerification},
IC3/PDR~\cite{IC3,EfficientPDR} and its variants such as IC3IA~\cite{PredAbsPDR} and IC3SA~\cite{IC3SA},
as well as interpolation-based model checking~\cite{McMillanCraig,VizelFMCAD09,ForwardBackwardReachability}.
Depending on the selected model checker and engine,
reasoning is performed using SAT solvers such as
\tool{MiniSat}~\cite{MiniSat},
\tool{CaDiCaL}~\cite{CaDiCaL-CAV24},
\tool{GipSAT}~\cite{GipSAT-CAV25},
and \tool{Kissat}~\cite{Kissat-SAT20},
or SMT solvers such as
\mathsat~\cite{MATHSAT5},
\tool{Boolector}~\cite{Boolector3},
\tool{Bitwuzla}~\cite{Bitwuzla-CAV23}, and
\tool{Yices2}~\cite{Yices2}.

The integrated model checkers differ in their input-language support and witness-export capabilities.
While \avr, \ricthree, and \pono natively support \btortwo,
\abc operates on the bit-level \aiger format~\cite{AIGER-1.9}.
When using \abc, \cpv invokes \btortoaiger~\cite{btor2tools-website} to bit-blast the \btortwo circuit into \aiger prior to verification,
and the export of violation witnesses in \btortwo format is thus not supported.
Furthermore, \abc and \ricthree only support QF\_BV and therefore cannot handle sequential circuits involving arrays (QF\_ABV).

The execution of hardware model checkers is orchestrated through \coveriteam~\cite{COVERITEAM},
a library for cooperative verification.
\coveriteam provides a uniform interface for invoking different verification backends
and reuses the mature execution infrastructure of the benchmarking framework \benchexec~\cite{Benchmarking-STTT}.
In particular,
it assembles command lines to run the backend model checkers,
configures isolated containerized environments and enforces resource limits via \tool{runexec},
collects witnesses and verification results,
and parses tool outputs using the so-called \emph{tool-info} modules\footnote{\url{https://github.com/sosy-lab/benchexec/blob/main/doc/tool-integration.md}} from \benchexec.
This design makes it straightforward to integrate additional hardware model checkers into \cpv in the future,
as developers only need to provide the executable of the new verification backend and implement a tool-info module.

\subsection{Witness Translation}
\label{sect:witness-translation}

When a hardware model checker reports a property violation,
\cpv translates the generated hardware-model-checking witness into a software-verification witness.
As a first step, \cpv invokes \btorsim~\cite{btor2tools-website},
a simulator for \btortwo circuits and the official validator for violation witnesses in HWMCC~\cite{HWMCC25}.
This serves two purposes.
First, it independently validates the violation witness produced by the model checker.
Second, it generates a complete simulation trace containing the values of all state and input variables at every time step.
This information is necessary because hardware model checkers typically export only input assignments in a violation witness.
The witness translator then analyzes the simulation trace and identifies value changes in the counters introduced during program instrumentation.
From these changes, it determines the input values relevant to the violation and reconstructs the corresponding software-level execution trace, as described in \cref{sect:instr-reach,sect:instr-term}.
The reconstructed witness can be exported in the GraphML-based witness format (v1)~\cite{WitnessesJournal} or the YAML-based witness format (v2)~\cite{VerificationWitnesses-2.0,WitnessesNonTermination}.

Translation of correctness witnesses is not yet implemented.
At present, among the integrated hardware model checkers,
only \ricthree~\cite{rIC3} is able to export correctness witnesses or inductive invariants in a standardized \btortwo-based format.
Therefore, when a property is proven, \cpv currently emits an empty trivial witness artifact.

\subsection{Sequential Portfolios in \svcomp}
\label{sect:svcomp-portfolios}

We devise sequential portfolios of hardware model checkers for \cpv to participate in \svcomp,
which are also used in the evaluation in \cref{sect:evaluation}.
In the competition, the resource limits are \SI{15}{min} of CPU time and \SI{15}{GB} of memory~\cite{SVCOMP26}.
Due to the tight memory bound,
we adopt sequential portfolios that distribute the available time budget across multiple model-checking engines instead of running them in parallel.

\begin{figure}[t]
    \centering
    \scalebox{.95}{\input{figures/portfolio}}
    \caption{
        Sequential portfolios used in \cpv's \svcomp configuration,
        each consisting of a sequence of hardware model-checking engines
        (denoted as <tool>$\cdot$<engine> with an associated time budget),
        selected based on the verification property and whether the translated circuit contains arrays
    }
    \label{fig:svcomp-portfolios}
\end{figure}
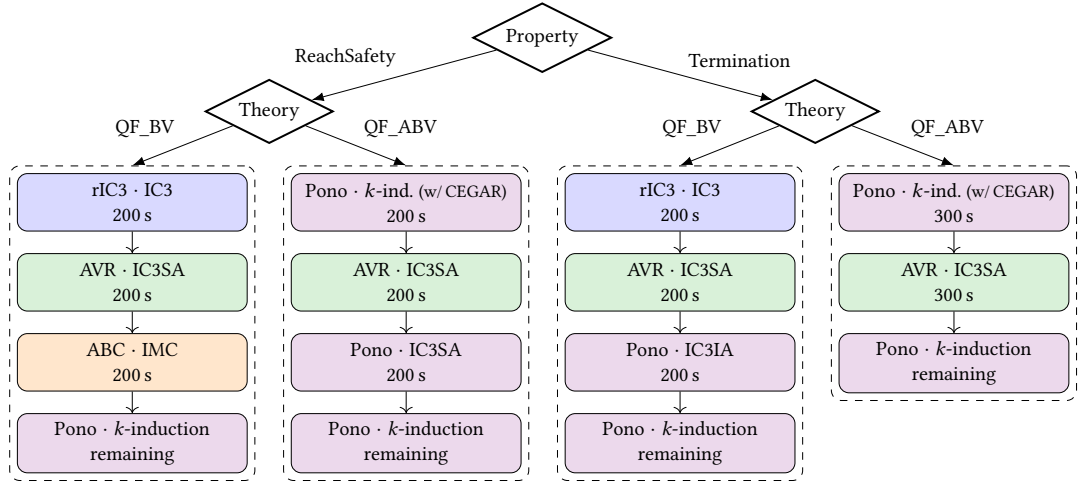

We organize the portfolios along two orthogonal dimensions:
(1) the verification property (ReachSafety vs. Termination) and
(2) the underlying theory that the translated circuit depends on (QF_BV vs.\ QF_ABV).
This yields four configurations, each with its own schedule.
To construct each portfolio, we benchmark individual engines on a sampled subset of verification tasks from \svcomp~2025, and iteratively build a greedy selection that maximizes the number of solved instances.
The resulting schedules are then manually refined to account for complementary strengths of different verification techniques
and to mitigate overfitting to the benchmark set.
The final portfolios are shown in \cref{fig:svcomp-portfolios}.
Since \abc and \ricthree do not support QF\_ABV, they are not included in the portfolios for tasks involving arrays.
Furthermore, given that \abc cannot export violation witnesses in \btortwo format,
we select only one \abc engine for the portfolios, where it is primarily used for proving safety properties.

In addition to backend engines, the portfolio also includes alternative translation strategies,
such as functional and relational encodings,
as well as variants with and without program slicing in the CFA-to-TR translation.
These strategies form additional layers above the backend engine portfolios
and are applied sequentially in the order in which they are mentioned.

%% file: figures/portfolio.tex
\begin{tikzpicture}[
    font=\small,
    decision/.style={
        draw,
        diamond,
        aspect=2,
        thick,
        align=center,
        inner sep=2pt
    },
    theory/.style={
        draw,
        rounded corners,
        thick,
        align=center,
        minimum height=8mm
    },
    portfolio/.style={
        draw,
        dashed,
        rounded corners,
        inner sep=1mm
    },
    tool/.style={
        draw,
        rounded corners,
        minimum width=3.2cm,
        minimum height=8mm,
        align=center
    },
    ric3/.style={
        tool,
        fill=blue!15,
    },
    avr/.style={
        tool,
        fill=green!15,
    },
    abc/.style={
        tool,    
        fill=orange!20,
    },
    pono/.style={
        tool,
        fill=violet!15,
    },
]

\node[decision] (property) {Property};
\node[decision, below left=3mm and 33.5mm of property.south]
      (reachTheory) {Theory};
\node[decision, below right=3mm and 33.5mm of property.south]
      (termTheory) {Theory};

\draw[-{Latex}] (property) -- node[midway, above left] {ReachSafety} (reachTheory);
\draw[-{Latex}] (property) -- node[midway, above right] {Termination} (termTheory);

\node[ric3, below left=4mm and 3mm of reachTheory.south] (r11)
    {rIC3 $\cdot$ IC3\\\SI{200}{s}};
\node[avr, below=3mm of r11] (r12)
    {AVR $\cdot$ IC3SA\\\SI{200}{s}};
\node[abc, below=3mm of r12] (r13)
    {ABC $\cdot$ IMC\\\SI{200}{s}};
\node[pono, below=3mm of r13] (r14)
    {Pono $\cdot$ \kinduction\\remaining};
\node[portfolio, fit=(r11) (r12) (r13) (r14)] (rbv) {};

\draw[->] (r11) -- (r12);
\draw[->] (r12) -- (r13);
\draw[->] (r13) -- (r14);

\node[pono, right=6mm of r11] (r21)
    {Pono $\cdot$ $k$-ind. {\smaller (w/\,CEGAR)}\\\SI{200}{s}};
\node[avr, below=3mm of r21] (r22)
    {AVR $\cdot$ IC3SA\\\SI{200}{s}};
\node[pono, below=3mm of r22] (r23)
    {Pono $\cdot$ IC3SA\\\SI{200}{s}};
\node[pono, below=3mm of r23] (r24)
    {Pono $\cdot$ \kinduction\\remaining};
\node[portfolio, fit=(r21) (r22) (r23) (r24)] (rabv) {};

\draw[->] (r21) -- (r22);
\draw[->] (r22) -- (r23);
\draw[->] (r23) -- (r24);

\node[ric3, right=6mm of r21] (r31)
    {rIC3 $\cdot$ IC3\\\SI{200}{s}};
\node[avr, below=3mm of r31] (r32)
    {AVR $\cdot$ IC3SA\\\SI{200}{s}};
\node[pono, below=3mm of r32] (r33)
    {Pono $\cdot$ IC3IA\\\SI{200}{s}};
\node[pono, below=3mm of r33] (r34)
    {Pono $\cdot$ \kinduction\\remaining};
\node[portfolio, fit=(r31) (r32) (r33) (r34)] (tbv) {};

\draw[->] (r31) -- (r32);
\draw[->] (r32) -- (r33);
\draw[->] (r33) -- (r34);

\node[pono, right=6mm of r31] (r41)
    {Pono $\cdot$ $k$-ind. {\smaller (w/\,CEGAR)}\\\SI{300}{s}};
\node[avr, below=3mm of r41] (r42)
    {AVR $\cdot$ IC3SA\\\SI{300}{s}};
\node[pono, below=3mm of r42] (r43)
    {Pono $\cdot$ \kinduction\\remaining};
\node[portfolio, fit=(r41) (r42) (r43)] (tabv) {};

\draw[->] (r41) -- (r42);
\draw[->] (r42) -- (r43);

\draw[-{Latex}] (reachTheory) -- node[midway,above left] {QF\_BV} (rbv.north);
\draw[-{Latex}] (reachTheory) -- node[midway,above right] {QF\_ABV} (rabv.north);
\draw[-{Latex}] (termTheory) -- node[midway,above left] {QF\_BV} (tbv.north);
\draw[-{Latex}] (termTheory) -- node[midway,above right] {QF\_ABV} (tabv.north);

\end{tikzpicture}

%% file: evaluation.tex
\section{Evaluation}
\label{sect:evaluation}
\input{plots/plot-defs}

\definecolor{rqblue}{RGB}{0,102,204}
\newtcolorbox{rqresult}{
    boxsep=0pt, boxrule=.5pt,
    colframe=rqblue,
    colback=rqblue!8!white
}

We conducted a large-scale and comprehensive evaluation of \cpv on the \svcomp benchmark set~\cite{SVCOMP26}
to demonstrate the feasibility and effectiveness of the proposed circuit-based program-verification approach.
This section first presents the research questions (\cref{sect:rqs}) and the experimental setup (\cref{sect:exp-setup}),
then answers each research question with detailed results and analyses (\cref{sect:answers}),
and concludes with a discussion of potential threats to validity (\cref{sect:threats}).

\subsection{Research Questions}
\label{sect:rqs}

Our evaluation follows the structure of \cpv's verification pipeline (\cref{sect:pipeline}).
We first study the feasibility of translating C programs into sequential circuits~(RQ1),
and then examine how different circuit-encoding choices affect the resulting circuits
and the performance of downstream model checking~(RQ2).
We further investigate \cpv's liveness-to-safety transformation,
evaluating both its internal design choices and how it compares to an alternative source-level transformer~(RQ3).
Moving to the backend, we assess the quality of the violation witnesses exported by \cpv~(RQ4).
Finally, we evaluate \cpv end-to-end against state-of-the-art software verifiers on the \svcomp benchmark suite~(RQ5).

Concretely, we aim to answer the following research questions:

\begin{itemize}
    \item[\textbf{RQ1:}] Can a significant portion of \svcomp benchmark tasks be encoded as sequential circuits? (\hyperref[sect:rq1]{answer})
    \item[\textbf{RQ2:}] What are the effects of different circuit-encoding choices (relational vs. functional)?
    \begin{itemize}
        \item[\textbf{2.1:}] How does the encoding affect the effectiveness of hardware model checkers? (\hyperref[sect:rq2.1]{answer})
        \item[\textbf{2.2:}] How does encoding affect the resulting circuit size and verification run time? (\hyperref[sect:rq2.2]{answer})
        \item[\textbf{2.3:}] What is the impact of sequential logic optimization on circuits under different encodings? (\hyperref[sect:rq2.3]{answer})
    \end{itemize}
    \item[\textbf{RQ3:}] What are the effects of optimizations and implementation choices in \cpv's L2S transformer?
    \begin{itemize}
        \item[\textbf{3.1:}] Does eliminating instrumented variables during L2S significantly improve \cpv's effectiveness? (\hyperref[sect:rq3.1]{answer})
        \item[\textbf{3.2:}] How does \cpv's L2S transformation compare to another source-level L2S transformer, \transver? (\hyperref[sect:rq3.2]{answer})
    \end{itemize}
    \item[\textbf{RQ4:}] How many violation witnesses produced by \cpv can be validated? (\hyperref[sect:rq4]{answer})
    \item[\textbf{RQ5:}] How does \cpv compare with state-of-the-art software verifiers on the \svcomp benchmark set?
    \begin{itemize}
        \item[\textbf{5.1:}] How does \cpv perform on ReachSafety benchmark tasks? (\hyperref[sect:rq5.1]{answer})
        \item[\textbf{5.2:}] How does \cpv perform on Termination benchmark tasks? (\hyperref[sect:rq5.2]{answer})
    \end{itemize}
\end{itemize}

\subsection{Experimental Setup}
\label{sect:exp-setup}

We describe the experimental setup as follows.

\subsubsection{Benchmark set}

We used the benchmark set from \svcomp 2026~\cite{SVCOMP26,SVCOMP26-SVBENCHMARKS-artifact}.
Each verification task consists of a C program, a property (specification),
and an expected verdict indicating whether the property holds or is violated.

The evaluation focuses on the categories \textit{ReachSafety} and \textit{Termination},
which correspond to the classes of properties currently supported by \cpv.
All programs in these categories are single-threaded.
Programs containing recursive functions are excluded, as recursion is not supported by \cpv.
After filtering, the ReachSafety category contains \num{\CpvTranslateRelReachSafetyStatusAllCount} tasks,
and the Termination category contains \num{\CpvTranslateRelLIIsTerminationStatusAllCount} tasks.
\Cref{tab:translation} (column ``\#Tasks'') summarizes the number of tasks in each subcategory of the benchmark set.

\subsubsection{Evaluated Tools}
\label{sect:eval-tools}

\cpv version \href{https://gitlab.com/sosy-lab/software/cpv/-/tree/1.1}{1.1}~\cite{CPV-1.1} was used in the evaluation.
It is bundled with the following hardware model checkers as backends:
\abc~\cite{ABC} at commit \commiturl{https://github.com/berkeley-abc/abc/tree}{474e7fbec23cded25456c85842c5e6404baa8c0e},
\avr~\cite{AVR} at commit \commiturl{https://github.com/nianzelee/avr/tree}{82af52efdf52f2d3f0363682b09680e35a3540fe}
(from a forked repository; with some modifications for easier result parsing),
\pono~\cite{Pono2} at commit \commiturl{https://github.com/stanford-centaur/pono/tree}{f7e3a6cf916bcf6656cc2ced659a5f4802677239}, and
\ricthree~\cite{rIC3} version \href{https://github.com/gipsyh/rIC3/tree/v1.5.1}{1.5.1}.
\cpv utilizes \btortoaiger to bit-blast \btortwo circuits into \aiger format for \abc,
and \btorsim to simulate witnesses in order to reconstruct full counterexample traces.
Both tools are compiled from \tool{Btor2Tools} at commit \commiturl{https://github.com/hwmcc/btor2tools/tree}{d33c73ff1d173f1bfac8ba6b1c6d68ba62c55f8e}.

For RQ3.2, \transver~\cite{TRANSVER} at commit \commiturl{https://gitlab.com/sosy-lab/software/transver/-/tree}{a8dfc6b9ea59f9326ed606ef003f81d7562c536d} was used for comparison.

For RQ4, violation witnesses produced by \cpv were validated using
\cpachecker~\cite{CPACHECKER-VALIDATOR-SVCOMP25} version \href{https://gitlab.com/sosy-lab/software/cpachecker/-/tree/cpachecker-4.2.2}{4.2.2}~\cite{CPAchecker-4.2.2} and
\witch~\cite{WITCH-VALIDATOR-SVCOMP24} at commit \commiturl{https://github.com/ayazip/witch/tree}{ffc643de17ef994a237eee0eca966f305b8ed9e2}~\cite{Witch-SVCOMP26-archive}.
Both tools support the v2 witness format
and were among the top two participants in the C Validation track for the categories ReachSafety and Termination of \svcomp 2026~\cite{SVCOMP26}.
For validation of v1 witnesses, the same version of \cpachecker and
\symwitch~\cite{SYMBIOTICWITCH-VALIDATOR-SVCOMP23},
the predecessor of \witch, 
at commit \commiturl{https://github.com/ayazip/witch/tree}{43ee4b7f5355cc6267054d0d4c8235f9fa5e3fdb}~\cite{SWitch-SVCOMP25-archive} were used.%
\footnote{\symwitch and \witch support only v1 and v2 witness formats, respectively, whereas \cpachecker supports both.}

For RQ5, we compared \cpv with several software verifiers.
These include the leading tools from \svcomp 2026~\cite{SVCOMP26},
evaluated in the versions and configurations used in the competition~\cite{SVCOMP26-FMTOOLS-artifact}:
\cpachecker~\cite{CPACHECKER-3.0-tutorial,CPACHECKER-SVCOMP24} version \href{https://gitlab.com/sosy-lab/software/cpachecker/-/tree/cpachecker-4.2.2}{4.2.2}~\cite{CPAchecker-4.2.2},
\esbmc~\cite{BMCEmbeddedC,ESBMC-SVCOMP25} version \href{https://github.com/esbmc/esbmc/releases/tag/v7.11}{7.11}~\cite{ESBMC-SVCOMP26-archive},
\symbiotic~\cite{SymbioticApproach,SYMBIOTIC-SVCOMP26} at commit \commiturl{https://github.com/staticafi/symbiotic/tree}{98667541ff6eb439f9b40da20120b6834b1e58f7}~\cite{Symbiotic-SVCOMP26-archive}, and
\ultimateautomizer~\cite{UAUTOMIZER2013,UAUTOMIZER-SVCOMP26} at commit \commiturl{https://github.com/ultimate-pa/ultimate/tree}{00d43373f5bb77c515d3b5a3769daa519304183e}~\cite{UAutomizer-SVCOMP26-archive}.
We selected these verifiers based on their strong performance in \svcomp 2026.
In particular, \uautomizer ranked first overall in the C verification track and second in the Termination category,
while \cpachecker, \esbmc, and \symbiotic achieved the top three positions in the ReachSafety category.
\cpachecker also ranked second overall in the C verification track.
Moreover, since \cpv relies on \kratostwo in its frontend,
we evaluated \kratostwo version \href{https://kratos.fbk.eu/releases/kratos-2.3-linux64.tar.gz}{2.3}
in combination with the wrapper script provided in the reproduction artifact for its CAV 2023 paper~\cite{Kratos2,Kratos2-CAV23-artifact}.

The evaluated verifiers cover a broad range of verification techniques, including
configurable program analysis~\cite{CPA},
counterexample-guided abstraction refinement~\cite{ClarkeCEGAR},
bounded model checking~\cite{BMCJournal},
\kinduction~\cite{K-Induction,kInduction},
interpolation-based model checking~\cite{IMC-JAR,DAR-transferability},
symbolic execution~\cite{SymbolicExecution},
automata-based methods~\cite{UAUTOMIZER2013},
and predicate abstraction~\cite{AbstractionsFromProofs},
reflecting both the diversity and the state of the art in software verification.

\subsubsection{Benchmarking Environment}

All experiments were performed on a cluster of machines running Ubuntu 24.04~(64 bit),
each equipped with a \SI{3.4}{GHz} Intel Xeon E3-1230 v5 CPU (8 processing units) and \SI{33}{GB} of RAM.
The experimental environment used Python 3.12 and \tool{pycparser} \href{https://launchpad.net/ubuntu/+source/pycparser/2.21-1}{2.21}.
For each benchmark task, \SI{900}{s} of CPU time, \SI{15}{GB} of memory, and 2 CPU cores were allocated.
We employed \benchexec~\cite{Benchmarking-STTT} and \tool{BenchCloud}~\cite{BENCHCLOUD} to ensure reliable and scalable benchmarking.

\subsection{Answers to Research Questions}
\label{sect:answers}

In this section, we present the experiments conducted for each research question,
along with the corresponding results, analyses, and discussions.
Throughout, \emph{proofs} and \emph{alarms} produced by a verifier
refer to correct results confirming that the specified property holds and that the property is violated, respectively.
A brief summary is highlighted in a blue box at the end of each RQ.

\subsubsection*{RQ1: Translating C Verification Tasks to Sequential Circuits}
\label{sect:rq1}

\begin{table}[t]
    \centering
    \caption{Summary of \svcomp benchmark tasks translated into K2 CFAs and \btortwo circuits}
    \label{tab:translation}
    \input{tables/translation}
\end{table}

To assess the feasibility of encoding C verification tasks as sequential circuits,
we translated the ReachSafety and Termination benchmark tasks into \btortwo circuits
using \cpv's translation pipeline under both relational and functional encodings (\cref{sect:tr-to-circuit}).
Since \cpv's frontend largely relies on \kratostwo,
which first represents C programs as CFAs in K2 IR,
the number of tasks successfully translated into K2 CFAs was also recorded.
\Cref{tab:translation} summarizes the translation results,
showing that a substantial portion of the benchmark tasks can be successfully translated into both K2 CFAs and \btortwo circuits under both encodings.

Translation from C to K2 was largely possible,
with over \SI{80}{\%} of all benchmark tasks successfully encoded as K2 CFAs.
Some subcategories exhibited lower success rates.
In ReachSafety-ControlFlow, most failures involved the use of \lstinline[style=cstyle]{longjmp()},
which is currently unsupported.
In ReachSafety-Floats, a portion of failures arose from unsupported floating-point library functions
such as \lstinline[style=cstyle]{sin()}, \lstinline[style=cstyle]{ceil()}, \lstinline[style=cstyle]{fmod()}, and \lstinline[style=cstyle]{floor()}.
In subcategories such as ReachSafety-Heap, ReachSafety-Sequentialized, and Termination-MainHeap,
as well as parts of ReachSafety-Floats,
failures were caused by limitations of \tool{pycparser} or unsupported non-constant initializers in array declarations in \kratostwo.
These limitations will be addressed in future work by contributing patches to \kratostwo and \tool{pycparser},
as well as extending \cpv's frontend, as already demonstrated in \cref{sect:impl-c-to-btor2}.

In subcategories such as Arrays, BitVectors, Hardware, and Loops in ReachSafety,
and BitVectors and MainControlFlow in Termination,
over \SI{90}{\%} of tasks were successfully translated into \btortwo circuits under both encodings.
For relational encoding, a drop in success rate was observed in some subcategories.
In ReachSafety-ECA, approximately \num{200} tasks failed due to out-of-memory errors.
In ReachSafety-Floats and ReachSafety-Heap,
\kratostwo encountered internal errors when encoding program semantics as SMT formulas using \mathsat.
In ReachSafety-ProductLines, failures arose from unsupported encodings of pointers to constant string literals.
The subcategory Termination-Other shares programs with ReachSafety-Floats and ReachSafety-ProductLines and thus exhibits similar issues.

For functional encoding, failures mirrored those of relational encoding,
as both share the same large-block encoding component.%
\footnote{The relational encoder is more developed.
All tasks translated by the functional encoder can also be handled by the relational one.}
Additionally, the current functional encoding does not support floating-point data types,
contributing to lower success rates in subcategories such as
ReachSafety-Combinations, ReachSafety-Floats, and ReachSafety-Hardness.
In ReachSafety-ECA and other subcategories, some failures were caused by reaching Python's recursion limit when traversing SMT formula trees.
We note that this limitation stems from the current implementation rather than the encoding technique itself.
Future versions of \cpv will address this issue, for example, by adopting an iterative traversal of SMT formulas.

The median translation time for ReachSafety tasks was
\SI[round-mode=figures, round-precision=3]{\CpvTranslateRelTranslationDoneReachSafetyRelCputimeAllMedian}{s} for relational encoding and
\SI[round-mode=figures, round-precision=3]{\CpvTranslateFuncTranslationDoneReachSafetyFuncCputimeAllMedian}{s} for functional encoding.
For Termination tasks, the median translation time was
\SI[round-mode=figures, round-precision=3]{\CpvTranslateRelLIIsTranslationDoneTerminationRelCputimeAllMedian}{s} and
\SI[round-mode=figures, round-precision=3]{\CpvTranslateFuncLIIsTranslationDoneTerminationFuncCputimeAllMedian}{s},
respectively (computed over successfully translated tasks).
Profiling revealed that importing Python libraries
(including \tool{CoVeriTeam}, \tool{NetworkX}, \tool{PySMT}, and \tool{PyVMT})
took approximately \SI{2.2}{s} per run on our machines.
This is because the benchmarking environment did not include all required Python dependencies.
To simplify deployment across the cluster,
these packages were bundled with \cpv and imported via Python wheel files,
which introduced a small startup overhead.
When installed locally, the import time can be significantly reduced.
Nevertheless, given the per-task time limit of \SI{900}{s},
the translation overhead remained modest and was not a bottleneck in the overall verification pipeline,
leaving sufficient time for model checking.

Overall, translating C programs into sequential circuits is feasible.
Although the frontend of \cpv (via \kratostwo) is not as mature as those of long-established software verifiers,
\cpv nevertheless achieves remarkable effectiveness in end-to-end verification,
as further demonstrated in \hyperref[sect:rq5.1]{RQ5.1} and \hyperref[sect:rq5.2]{RQ5.2}.

\begin{rqresult}
    A large portion (over \SI{70}{\%}) of \num{\the\numexpr\CpvTranslateRelReachSafetyStatusAllCount+\CpvTranslateRelLIIsTerminationStatusAllCount} evaluated benchmark tasks can be successfully translated into sequential circuits.
    The translation overhead is low relative to the overall time budget, leaving sufficient resources for model checking.
    The remaining limitations are primarily due to implementation constraints in the frontend (e.g., unsupported floating-point library functions)
    rather than inherent restrictions of the proposed approach.
\end{rqresult}

\subsubsection*{RQ2.1: Encoding's effect on Model-Checking Effectiveness}
\label{sect:rq2.1}

To study the impact of different encoding choices on model checkers' effectiveness,
the four primary backend tools (\abc, \avr, \pono, and \ricthree) in \cpv
were evaluated on \btortwo circuits generated under relational and functional encodings.
We focused on the subset of benchmark tasks that can be successfully translated into \btortwo circuits under both encodings.
To reduce computation time while maintaining a representative sample across subcategories,
up to \num{150} tasks were randomly sampled from each ReachSafety subcategory,
with a balance between tasks with and without property violations,
resulting in a total of \num{\AvrKindReachSafetySampledFuncStatusAllCount} tasks for the category ReachSafety.
For Termination, all \num{\AvrKindTerminationSampledFuncStatusAllCount} tasks that were successfully translated under both encodings were considered,
as the total number is manageable.
Since some model-checking engines only support bit-vector sorts without arrays,
we further identified \num{\AbcImcReachSafetySampledBvFuncStatusAllCount} ReachSafety tasks and \num{\AbcImcTerminationSampledBvFuncStatusAllCount} Termination tasks
that do not contain arrays.

\begin{table}[t]
    \centering
    \caption{Results for hardware model checkers on translated \btortwo tasks under relational and functional encodings}
    \label{tab:encoding}
    \subfloat[On bit-vector-only tasks]{\input{tables/encoding.bv}\label{tab:encoding-bv}}\\[3mm]
    \subfloat[On tasks with both array and bit-vector sorts]{\input{tables/encoding.arr-bv}\label{tab:encoding-arr}}
\end{table}

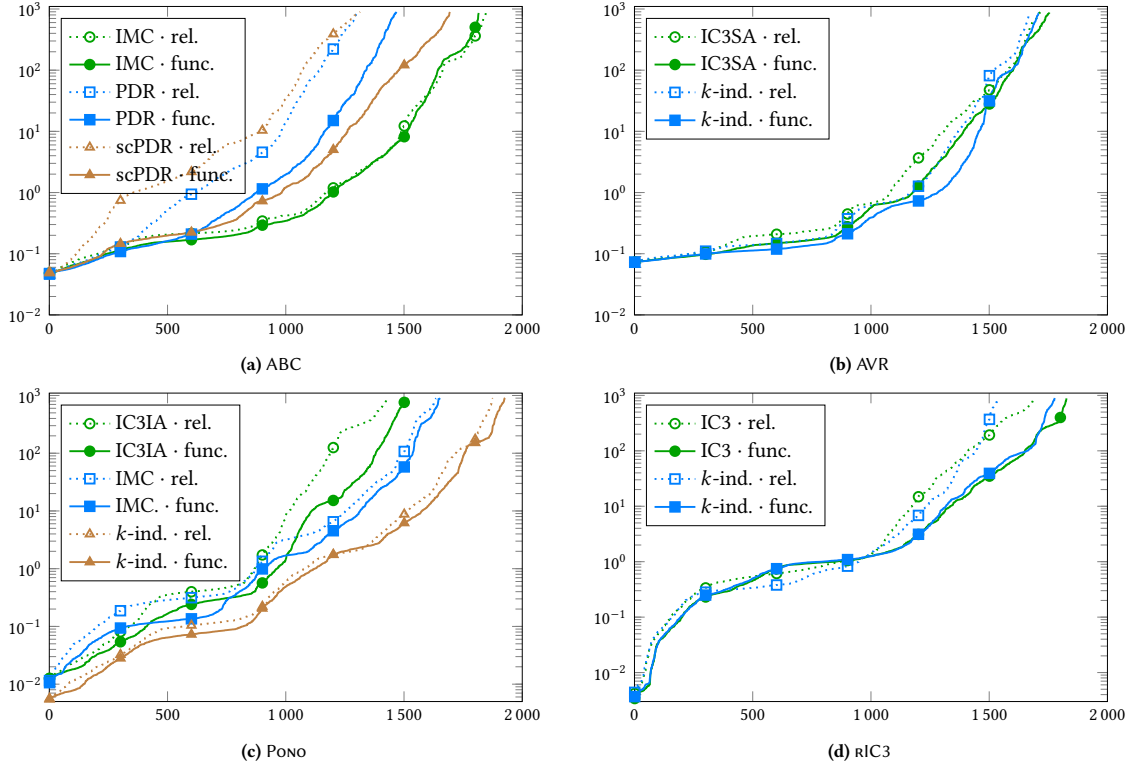
\begin{figure}[t]
    \centering
    \tikzexternalenable
    \subfloat[\abc\label{fig:encoding-abc}]{\scalebox{.9}{\input{plots/abc.svcomp-sampled.cputime.quantile}}}%
    \hfill%
    \subfloat[\avr]{\scalebox{.9}{\input{plots/avr.svcomp-sampled.cputime.quantile.tex}}}\\
    \subfloat[\pono]{\scalebox{.9}{\input{plots/pono.svcomp-sampled.cputime.quantile}}}%
    \hfill%
    \subfloat[\ricthree]{\scalebox{.9}{\input{plots/ric3.svcomp-sampled.cputime.quantile.tex}}}\\
    \tikzexternaldisable
    \caption{Cactus plots comparing model-checking engines on
        \btortwo tasks translated from the \svcomp benchmark set under relational and functional encodings
        ($x$: n-th fastest correct result; $y$: CPU time (s))}
    \label{fig:encoding}
\end{figure}

\Cref{tab:encoding} summarizes the results of each model-checking engine on the sampled subsets,
including bit-vector-only~(\cref{tab:encoding-bv}) and array-manipulating tasks~(\cref{tab:encoding-arr}),
under both relational and functional encodings.
\Cref{fig:encoding} further visualizes the results using cactus plots.%
\footnote{We use a slightly modified cactus plot in which time is plotted on the $y$-axis, following common practice in the \svcomp community.}
A data point $(x,y)$ in the plots indicates that the respective engine correctly solved $x$ tasks under the encoding shown in the legend,
each within $y$ seconds of CPU time.
In the cactus plots, results for ReachSafety and Termination are combined.

The results show that most model-checking engines favor the functional encoding over the relational one in terms of the number of solved tasks,
with improvements in both the number of proofs and alarms.
The cactus plots further indicate that many engines require less CPU time to solve the same number of tasks under functional encoding.
There are a few exceptions, though.
\abc's IMC engine performed slightly better on tasks using relational encoding, and \pono's IMC engine showed little to no difference between the two encodings.
In contrast, \abc's PDR, \ricthree's IC3, and \pono's IC3IA demonstrated clear improvements with functional encoding,
whereas \avr's IC3SA did not exhibit such a trend.

These differences may be attributed to factors beyond the encoding itself,
such as the implementation of the engines,
their internal representations of transition systems,
and the underlying SAT/SMT solvers.
We have submitted the translated \btortwo circuits to HWMCC~\cite{HWMCC24,HWMCC25} and
encourage the developers of the model checkers to further investigate the performance differences.
In summary, the results suggest that offering different encoding choices in \cpv is valuable,
as this caters to different model-checking engines.

\begin{rqresult}
    The evaluation shows that encoding choices have a noticeable impact on model-checking performance,
    with functional encoding generally yielding better results in terms of the number of tasks solved and run time.
    However, the effect is not uniform across all engines, and some tools perform similarly or even slightly better with relational encoding.
    These results highlight the value of offering multiple encoding options in \cpv to better accommodate different model-checking backends.
\end{rqresult}

\subsubsection*{RQ2.2: Encoding's effect on Circuit Size and Run Time}
\label{sect:rq2.2}

In light of the performance differences observed between relational and functional encodings in \hyperref[sect:rq2.1]{RQ2.1},
we further analyzed the resulting circuit sizes under both encodings.
\abc was chosen as the subject of study because it provides circuit statistics such as the number of gates (AIG nodes).

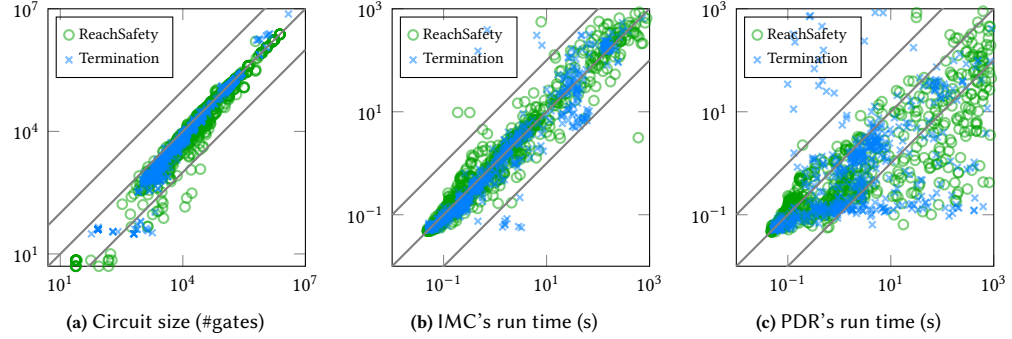
\begin{figure}[t]
    \centering
    \tikzexternalenable
    \subfloat[Circuit size (\#gates)\label{fig:scatter-encoding-gates}]{\scalebox{1}{\input{plots/encoding.gates.scatter}}}%
    \subfloat[IMC's run time (s)\label{fig:scatter-encoding-pdr-cputime}]{\scalebox{1}{\input{plots/encoding.imc.cputime.scatter}}}%
    \subfloat[PDR's run time (s)\label{fig:scatter-encoding-imc-cputime}]{\scalebox{1}{\input{plots/encoding.pdr.cputime.scatter}}}%
    \tikzexternaldisable
    \caption{Scatter plots comparing circuit size and verification run time of \abc
        under relational~($x$) and functional~($y$) encodings
        (gray lines indicate equal performance and one order-of-magnitude differences)}
    \label{fig:scatter-encoding}
\end{figure}

\begin{table}[t]
    \centering
    \caption{Summary of median circuit size and \abc's run time under relational and functional encodings
        on ReachSafety~(a) and Termination~(b) tasks,
        as well as the number of tasks where each encoding yielded better results}
    \label{tab:encoding-size-and-time}
    \subfloat[On ReachSafety tasks]{\input{tables/encoding.gates.reach}\label{tab:encoding-size-and-time-reach}}\\[3mm]
    \subfloat[On Termination tasks]{\input{tables/encoding.gates.term}\label{tab:encoding-size-and-time-term}}
\end{table}

\Cref{fig:scatter-encoding-gates} compares the number of gates in circuits translated from the same C verification task under relational and functional encodings.
A data point $(x,y)$ in the scatter plots indicates that the circuit generated under relational encoding has $x$ gates while the one under functional encoding has $y$ gates.
The color of a data point denotes the task category.
A point below the diagonal corresponds to an instance where functional encoding yields a smaller circuit, and vice versa.
We observe that functional encoding generally resulted in smaller circuits than relational encoding,
with some cases demonstrating an order-of-magnitude reduction in the number of gates.
In addition, \cref{tab:encoding-size-and-time} reports the median circuit sizes under both encodings,
as well as the number of tasks for which each encoding yielded a more compact circuit when applied to the same benchmark task.
Functional encoding achieved smaller median sizes for both benchmark categories,
and produced smaller circuits in
\SI[round-mode=figures, round-precision=3]{\fpeval{\AbcPdrGateReachSafetySampledBvSignTestPlusCount / \AbcPdrGateReachSafetySampledBvSignTestAllCount * 100}}{\%} and
\SI[round-mode=figures, round-precision=3]{\fpeval{\AbcPdrGateTerminationSampledBvSignTestPlusCount / \AbcPdrGateTerminationSampledBvSignTestAllCount * 100}}{\%}
of the ReachSafety and Termination tasks, respectively.

We next compare the run time of \abc's IMC and PDR engines on tasks solved under both encodings in \cref{fig:scatter-encoding-imc-cputime,fig:scatter-encoding-pdr-cputime}, and \cref{tab:encoding-size-and-time}.
For IMC, functional encoding led to an increase in the median run time on the \num{\AbcImcCputimeReachSafetySampledBvSignTestAllCount} commonly solved ReachSafety tasks,
despite solving a larger number of tasks in shorter CPU time (see \cref{tab:encoding-size-and-time-reach}).
On the \num{\AbcImcCputimeTerminationSampledBvSignTestAllCount} commonly solved Termination tasks (see \cref{tab:encoding-size-and-time-term}),
IMC exhibited slightly better run-time efficiency under functional encoding.
However, recall that IMC solved more tasks under relational encoding when considering all evaluated benchmark tasks across both categories (see \cref{tab:encoding-bv}).
On the other hand, \abc's PDR demonstrated a clearer efficiency advantage for circuits generated by functional encoding.
As illustrated in \cref{fig:scatter-encoding-pdr-cputime},
many data points lie below the diagonal, with some showing order-of-magnitude improvements.
The median verification run time was also lower under functional encoding for both categories.
On the commonly solved tasks, PDR was faster under functional encoding in
\SI[round-mode=figures, round-precision=3]{\fpeval{\AbcPdrCputimeReachSafetySampledBvSignTestPlusCount / \AbcPdrCputimeReachSafetySampledBvSignTestAllCount * 100}}{\%} and
\SI[round-mode=figures, round-precision=3]{\fpeval{\AbcPdrCputimeTerminationSampledBvSignTestPlusCount / \AbcPdrCputimeTerminationSampledBvSignTestAllCount * 100}}{\%}
of the ReachSafety and Termination tasks, respectively.
These observations are consistent with \hyperref[sect:rq2.1]{RQ2.1}:
\abc's PDR shows a clear preference for functional encoding, whereas IMC exhibits more mixed behavior.

From the collected experimental results, we observe a positive correlation between circuit size and the run time of both \abc's engines.
However, circuit size (and run time) is influenced not only by the encoding but also by the intrinsic complexity of the original C verification task,
for which no standard quantitative metric exists.
Moreover, structural differences introduced by the encoding may affect solver performance independently of circuit size.
In summary, while functional encoding can often produce more compact circuits,
the performance improvements in subsequent verification runs are not uniform across different backend engines.

\begin{rqresult}
    Functional encoding generally produces smaller circuits than relational encoding.
    From the analysis of \abc's results, we observe that its PDR engine performs more efficiently on circuits generated by functional encoding,
    whereas its IMC engine shows less difference in run time between the two encodings.
\end{rqresult}

\subsubsection*{RQ2.3: Combined Effect of Encoding and Circuit Optimization}
\label{sect:rq2.3}

In \hyperref[sect:rq2.1]{RQ2.1} and \hyperref[sect:rq2.2]{RQ2.2},
we observe that \abc's PDR engine was more effective and efficient on circuits generated by functional encoding.
This raises an interesting research question of whether circuit optimization can further improve PDR's performance and
whether its effects differ between the two encodings.
We continue to use \abc as the primary subject of study,
as it implements several powerful sequential logic optimization techniques such as speculative reduction~\cite{MonyDATE2009}
(referred to as \texttt{scorr} in \abc).
We applied \texttt{scorr} before running PDR and compared the results to those of PDR alone.
The time limit and measurements included both optimization and verification phases.

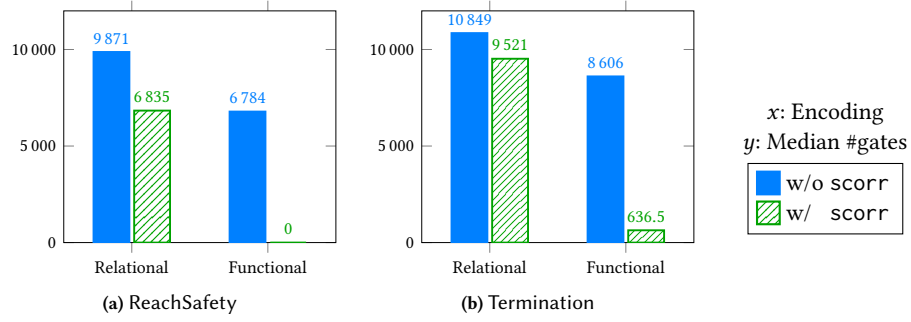
\begin{figure}[t]
    \centering
    \subfloat[ReachSafety]{\scalebox{.9}{\input{plots/scorr-gate.reach.bar}}}%
    \hspace{2mm}%
    \tikzexternalenable
    \subfloat[Termination]{\scalebox{.9}{\input{plots/scorr-gate.term.bar}}}%
    \tikzexternaldisable
    \subfloat{\makebox[.22\linewidth][c]{\raisebox{.04\linewidth}{\shortstack{%
        $x$: Encoding\\$y$: Median \#gates\\
        \pgfplotslegendfromname{legend:scorr-gate-reach}}}}}
    \caption{Bar charts comparing the median number of gates of the encoded circuits with and without \texttt{scorr}
        for commonly solved ReachSafety~(a) and Termination~(b) tasks,
        under relational and functional encodings}
    \label{fig:scorr-gate}
\end{figure}

\begin{figure}[t]
    \centering
    \sisetup{
        round-mode = figures,
        round-precision = 3,
    }
    \subfloat[ReachSafety]{\scalebox{.9}{\input{plots/scorr-time.reach.bar}}}%
    \hspace{2mm}%
    \tikzexternalenable
    \subfloat[Termination]{\scalebox{.9}{\input{plots/scorr-time.term.bar}}}%
    \tikzexternaldisable
    \subfloat{\makebox[.22\linewidth][c]{\raisebox{.04\linewidth}{\shortstack{%
        $x$: Encoding\\$y$: Median time (s)\\
        \pgfplotslegendfromname{legend:scorr-time-reach}}}}}
    \caption{Bar charts comparing the median run time of different PDR configurations of \abc
        for commonly solved ReachSafety~(a) and Termination~(b) tasks,
        under relational and functional encodings
        (``\textcolor{gray}{(\texttt{scorr}+)}PDR'' denotes the run time of PDR subsequent to \texttt{scorr})}
    \label{fig:scorr-time}
\end{figure}
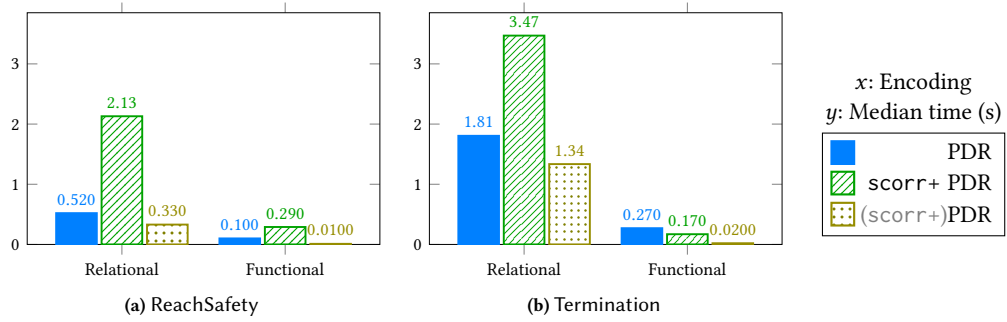

\begin{table}[t]
    \centering
    \caption{Summary of run time and gate counts for \abc's PDR engine on commonly solved tasks,
        across different encodings and with/without \texttt{scorr}
        (reduction rates are computed from the corresponding mean values;
        ``\textcolor{gray}{(\texttt{sc}.+)}PDR'' denotes the run time of PDR subsequent to \texttt{scorr})}
    \label{tab:scorr-reduct}
    \input{tables/scorr.reduct}
\end{table}

\begin{table}[t]
    \centering
    \caption{Results for \abc's PDR engine with and without sequential logic optimization (\texttt{scorr})}
    \label{tab:scorr-solved}
    \input{tables/scorr.solved}
\end{table}

\Cref{fig:scorr-gate,fig:scorr-time} compare the median number of gates and the run time of \abc's PDR engine on commonly solved tasks across all evaluated configurations (encoding and circuit optimization).
In \cref{tab:scorr-reduct}, the mean values and the corresponding reduction rates are reported.
The number of solved tasks is further listed in \cref{tab:scorr-solved}.
The run time reported here is measured internally by \abc, which provides separate timings for \texttt{scorr} and PDR,
rather than the overall CPU time measured by \benchexec in the other RQs.
For reference, the cactus plot in \cref{fig:encoding-abc} also includes the combined configuration running both \texttt{scorr} and PDR (denoted as ``scPDR''),
showing its performance across all sampled tasks.

From \cref{fig:scorr-gate,tab:scorr-reduct}, we see that \texttt{scorr} reduced the number of gates for circuits under both encodings,
with a more pronounced reduction for functional encoding.
Notably, \num{\ScPdrReachSafetyFuncReducedToZeroCount} ReachSafety tasks under functional encoding were reduced to zero gates after applying \texttt{scorr}
(more than half of the \num{\AbcPdrScPdrReachSafetyRelCommonStatusAllCount} considered tasks),
which explains the median circuit size of zero.
Note that, however, this does not imply that the verification problem itself is trivial,
as \texttt{scorr} is a high-effort optimization procedure that involves constraint solving.

When comparing the entire verification run (PDR vs. \texttt{scorr}\,+\,PDR) in \cref{fig:scorr-time,tab:scorr-reduct},
we notice that the efficiency did not improve with the application of \texttt{scorr},
despite the simplification in the circuit.
Nevertheless, effectiveness, in terms of the number of solved tasks, was improved (see \cref{tab:scorr-solved}).
In addition, we observe that the run time of PDR following \texttt{scorr} (denoted by \textcolor{gray}{(\texttt{sc}.+)}PDR) was generally lower than that of PDR alone,
indicating that PDR indeed benefited from the simplified circuit.
This suggests that while circuit optimization can reduce the size of the circuit and help improve the downstream model-checking performance,
it does not always translate to faster end-to-end verification times due to the overhead of the optimization process itself.

\begin{rqresult}
    Sequential logic optimization in \abc effectively reduces circuit size for both encodings,
    with stronger reductions under functional encoding.
    While this simplification can improve the effectiveness of PDR,
    it does not necessarily lead to faster end-to-end verification due to the non-negligible run-time overhead of the optimization itself.
\end{rqresult}

\subsubsection*{RQ3.1: Effect of Eliminating Instrumented Variables During L2S}
\label{sect:rq3.1}

In this RQ, we evaluate the optimization described in \cref{sect:termination-l2s}.
Instead of introducing recording state variables for all state variables in the transformed circuit,
the optimization omits recording variables for auxiliary counter variables introduced during program instrumentation.
In \cpv, this behavior can be toggled via a command-line flag.
The full verification pipeline of \cpv was executed using \ricthree's IC3 and \pono's \kinduction engines as backends,
considering both relational and functional encodings on \svcomp Termination tasks.

\begin{table}[t]
    \centering
    \caption{Effect of L2S variable elimination on \svcomp Termination tasks}
    \label{tab:l2s-simp}
    \input{tables/l2s-simp}
\end{table}

\begin{figure}[t]
    \centering
    \tikzexternalenable
    \subfloat[Relational encoding]{\scalebox{.9}{\input{plots/cpv.rel.l2s-simp.cputime.quantile}}}%
    \hfill%
    \subfloat[Functional encoding]{\scalebox{.9}{\input{plots/cpv.func.l2s-simp.cputime.quantile}}}
    \tikzexternaldisable
    \caption{Cactus plots for \cpv using relational~(a) and functional~(b) encodings,
        each comparing different backends with and without L2S variable elimination on \svcomp Termination tasks}
    \label{fig:l2s-simp}
\end{figure}
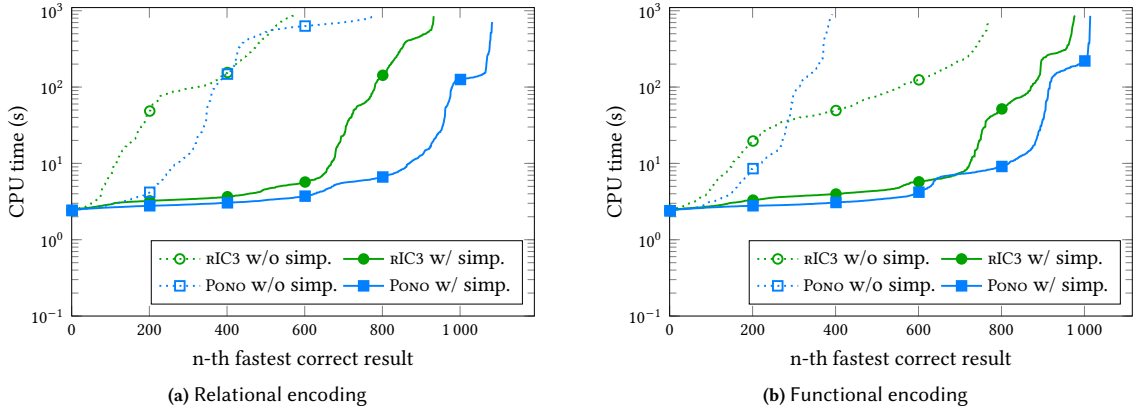

\Cref{tab:l2s-simp} summarizes the number of tasks solved by \cpv with and without L2S variable elimination,
and \cref{fig:l2s-simp} further visualizes the results using cactus plots.%
\footnote{\Cref{tab:l2s-simp} shows more Termination tasks solved by \ricthree's IC3 than \cref{tab:encoding-bv}.
There are two reasons:
First, only tasks that can be successfully translated under both encodings are considered in \cref{tab:encoding-bv}.
Second, when filtering for bit-vector-only tasks, we checked for the presence of array sorts in the \btortwo file.
However, some \btortwo circuits containing arrays can still be handled by \ricthree if the arrays are not in the cone of influence of the property.}
The optimization yields substantial improvements for both relational and functional encodings, as well as for both backend engines.
The increase in the number of detected alarms (non-terminating lassos) is particularly notable.
We compared the lasso lengths of the commonly solved instances by \pono's \kinduction engine with and without variable elimination.%
\footnote{We did not compare lasso lengths for \ricthree's IC3 engine, as IC3 does not guarantee shortest counterexamples.}
With relational encoding, the optimization reduces the lasso length by an average of
\SI[round-mode=figures, round-precision=3]{\CpvLIIsPonoKindRelReductRateAvg}{\%};
with functional encoding, the average reduction is
\SI[round-mode=figures, round-precision=3]{\CpvLIIsPonoKindFuncReductRateAvg}{\%}.
These results confirm the intuition presented in \cref{sect:termination-l2s} that excluding instrumented counter variables from the state-revisit check substantially shortens the lasso witnesses,
making non-termination easier for hardware model checkers to detect.

\begin{rqresult}
    Elimination of instrumented variables during L2S helps increase the number of solved Termination tasks across different encodings and backends.
    The increase is largely due to a higher number of detected alarms, as the optimization makes the state-revisit condition easier to satisfy.
\end{rqresult}

\subsubsection*{RQ3.2: Comparison with \transver}
\label{sect:rq3.2}

We further evaluate the capability of \cpv's L2S transformer by comparing it with \transver~\cite{TRANSVER},
a source-level L2S transformation tool for C programs.
\transver first transforms termination tasks into reachability tasks, after which \cpv verifies the resulting programs.
This combined pipeline is referred to as \transvercpv.
In our experiment, \cpv was executed using functional encoding with \pono's \kinduction as the backend model-checking engine.

\begin{table}[t]
    \centering
    \caption{Comparison of \transver and \cpv on \svcomp Termination tasks
    (\greencmark: \transver's L2S transformation succeeded)}
    \label{tab:transver}
    \input{tables/transver}
\end{table}

The evaluation results are summarized in \cref{tab:transver}.
Column (a) reports the number of tasks for which \transver successfully performed the L2S transformation,
while column (b) shows the number of those tasks subsequently solved by \cpv.
Column (c) reports the number of tasks solved by \cpv using its own L2S transformation.
Column (a)\,$\cap$\,(c) further provides the number of tasks
that are both successfully transformed by \transver and solved by \cpv using its own L2S transformation.

\cpv's L2S transformer outperformed \transver in two aspects:
(1) \cpv was more robust and was capable of handling a larger number of the benchmark tasks.
\transver successfully transformed only about half of the termination tasks (see column (a) in \cref{tab:transver}),
as it does not support programs with arrays or dynamic memory allocation.
By contrast, \cpv's L2S transformation can handle all instances that have been successfully translated into \btortwo circuits (see \cref{tab:translation}).
Moreover, \transvercpv inherits the same frontend limitation (see \hyperref[sect:rq1]{RQ1}) from \cpv.
Therefore, the set of tasks \transvercpv can successfully translate is ultimately a subset of those supported by \cpv alone, which limits its overall effectiveness (see columns (b) and (c) in \cref{tab:transver}).
(2) Circuits produced by \cpv's L2S transformation were more amenable to \pono's \kinduction.
Among the tasks that were successfully transformed by both \transvercpv and \cpv,
the latter solved more instances (see columns (b) and (a)\,$\cap$\,(c) in \cref{tab:transver}),
suggesting that \cpv's L2S transformation yields circuits that are better suited for model checking in our experiments.

It is worth noting, however, that \transver and \cpv's L2S transformer serve different purposes.
\transver is designed as a source-level transformation tool that can be combined with off-the-shelf software verifiers, whereas \cpv's L2S transformer is an integral part of an end-to-end circuit-based verification pipeline.
This design difference leads to distinct trade-offs in supported language features, robustness, and the structure of the generated representations.
It also reflects differences in implementation complexity:
\transver relies on \cpachecker for CFA construction and implements custom instrumentation automata and sequentialization operators,
whereas \cpv's L2S transformer operates directly at the circuit level, which is simpler to manipulate,
and is implemented in approximately 200 lines of Python code.

\begin{rqresult}
    \cpv's L2S transformer outperforms \transver in both robustness and effectiveness.
    It supports a broader range of tasks and produces circuits that are more amenable to model checking,
    resulting in a higher number of solved instances.
    While the two approaches serve different purposes,
    the results demonstrate the advantage of performing L2S at the circuit level in \cpv's pipeline.
\end{rqresult}

\subsubsection*{RQ4: Witness Generation and Validation}
\label{sect:rq4}

In this RQ, we aim to assess the quality and the validity of the violation witnesses generated by \cpv.
We first executed \cpv using functional encoding and \pono's BMC as the backend on the \svcomp benchmark set
to produce the witnesses in both v1 (GraphML) and v2 (YAML) formats.
Subsequently, \cpachecker, \symwitch, and \witch were used to validate the generated witnesses in the formats they support.

\begin{table}[t]
    \centering
    \caption{Validation results of \cpv's violation witnesses}
    \label{tab:wit-val}
    \input{tables/wit-val}
\end{table}

\Cref{tab:wit-val} summarizes the validation results.
Column ``\cpv'' reports the number of detected alarms
(and hence the number of generated violation witnesses).
The subsequent columns report the number of witnesses confirmed by each validator,
followed by the combined validation results and the overall confirmation rate.

For the ReachSafety category, validation results were similar for v1 and v2 witnesses.
A slight drop in confirmed cases by \cpachecker is observed when comparing v1 to v2,
even though witnesses in both formats contain essentially the same information.
This issue has been reported to the \cpachecker development team and is currently under investigation.%
\footnote{\url{https://gitlab.com/sosy-lab/software/cpachecker/-/work_items/1494}}

For the Termination category, validation results were notably better for v2 witnesses.
In particular, \symwitch was unable to confirm any v1 witnesses.
This is because, in the v1 format,
certain structural requirements for non-termination witnesses are not well defined,
and their semantics with respect to the original programs are ambiguous.
This limitation lies in the witness format rather than in \cpv's witness translator,
and has been addressed in the newer v2 format~\cite{WitnessesNonTermination}.

Overall, \cpv produces high-quality violation witnesses that can be confirmed by multiple validators,
achieving a combined confirmation rate of approximately \SI{95}{\%} across both categories.
The unconfirmed cases are mainly due to validators reaching resource limits (CPU time or memory).

\begin{rqresult}
    Violation witnesses exported by \cpv can be reliably validated by multiple witness validators,
    achieving a high overall confirmation rate (about \SI{95}{\%}).
    Validation results are consistent across formats for ReachSafety,
    while the newer v2 format significantly improves validation for Termination tasks.
\end{rqresult}

\subsubsection*{RQ5.1: Comparison with SOTA Software Verifiers on ReachSafety Tasks}
\label{sect:rq5.1}

Finally, we evaluated the end-to-end capability of \cpv on the \svcomp benchmark set and
compared it with state-of-the-art software verifiers.
\cpv was executed in its \svcomp configuration, running the sequential portfolios described in \cref{sect:svcomp-portfolios};
the compared tool versions and configurations are specified in \cref{sect:eval-tools}.

\begin{table}[t]
    \centering
    \caption{\svcomp ReachSafety and Termination results}
    \input{tables/svcomp-overall}
    \label{tab:svcomp-summary}
\end{table}

\begin{table}[t]
    \centering
    \caption{\svcomp ReachSafety results for each subcategory (Part 1)}
    \input{tables/svcomp-reach.1}
    \label{tab:svcomp-reachsafety-1}
\end{table}

\begin{table}[t]
    \centering
    \caption{\svcomp ReachSafety results for each subcategory (Part 2)}
    \input{tables/svcomp-reach.2}
    \label{tab:svcomp-reachsafety-2}
\end{table}

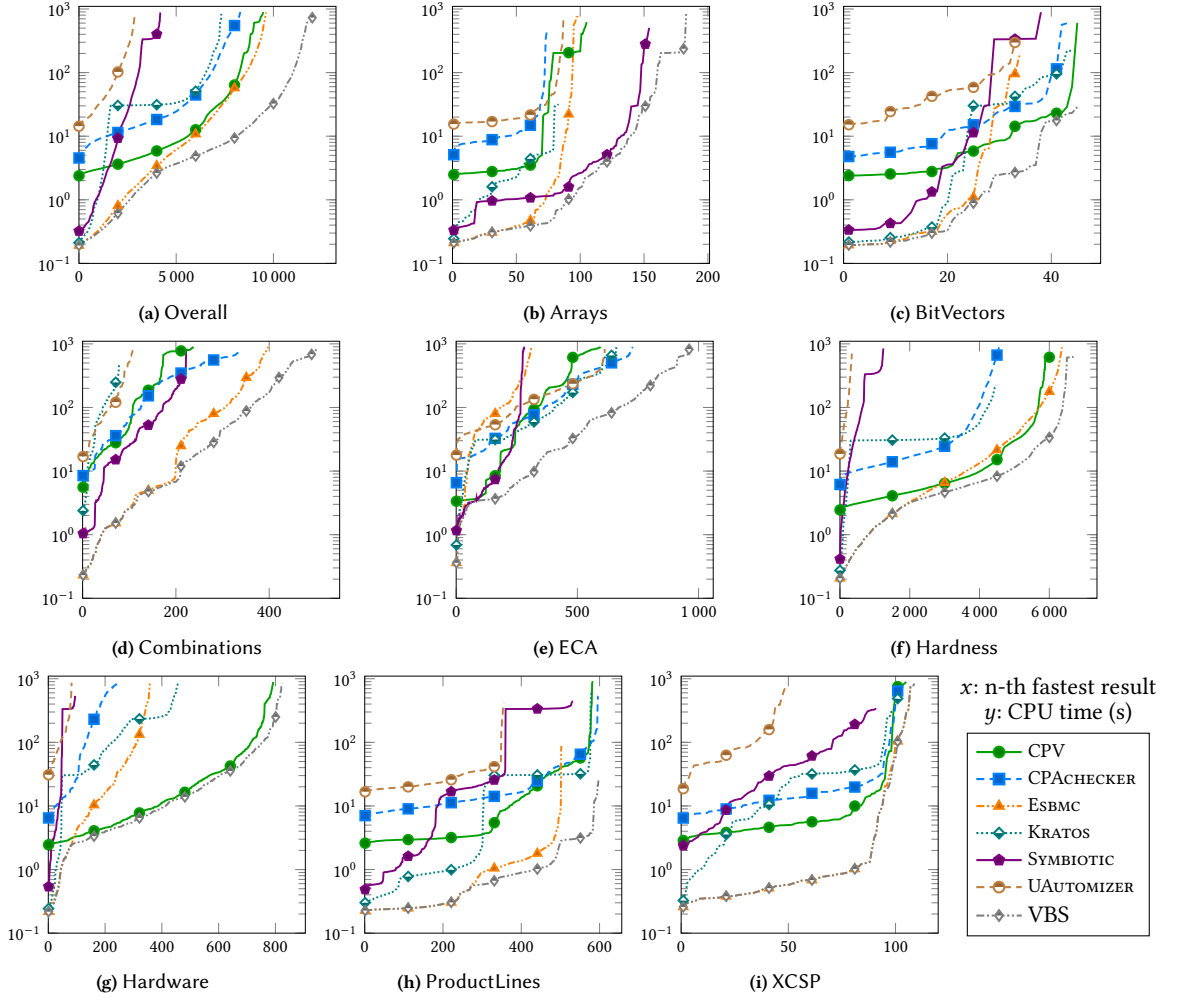
\begin{figure}[t]
    \centering
    \tikzexternalenable
    \subfloat[Overall]{\scalebox{.9}{\input{plots/svcomp.ReachSafety.cputime.quantile}}}\hspace{.05\linewidth}%
    \subfloat[Arrays]{\scalebox{.9}{\input{plots/svcomp.ReachSafety-Arrays.cputime.quantile}}}\hspace{.05\linewidth}%
    \tikzexternaldisable 
    \subfloat[BitVectors]{\scalebox{.9}{\input{plots/svcomp.ReachSafety-BitVectors.cputime.quantile}}}\\
    \tikzexternalenable
    \subfloat[Combinations]{\scalebox{.9}{\input{plots/svcomp.ReachSafety-Combinations.cputime.quantile}}}\hspace{.05\linewidth}%
    \subfloat[ECA]{\scalebox{.9}{\input{plots/svcomp.ReachSafety-ECA.cputime.quantile}}}\hspace{.05\linewidth}%
    \subfloat[Hardness]{\scalebox{.9}{\input{plots/svcomp.ReachSafety-Hardness.cputime.quantile}}}\\
    \subfloat[Hardware]{\scalebox{.9}{\input{plots/svcomp.ReachSafety-Hardware.cputime.quantile}}}%
    \subfloat[ProductLines]{\scalebox{.9}{\input{plots/svcomp.ReachSafety-ProductLines.cputime.quantile}}}%
    \subfloat[XCSP]{\scalebox{.9}{\input{plots/svcomp.ReachSafety-XCSP.cputime.quantile}}}%
    \tikzexternalenable
    \subfloat{\makebox[.2\linewidth][c]{\raisebox{.02\linewidth}{\shortstack{%
        $x$: n-th fastest result\\$y$: CPU time (s)\\
        \pgfplotslegendfromname{legend:svcomp-reach}}}}}
    \caption{Cactus plots for \svcomp ReachSafety category and its subcategories,
        where \cpv is highlighted at least once in \cref{tab:svcomp-reachsafety-1,tab:svcomp-reachsafety-2}}
    \label{fig:svcomp-reachsafety}
\end{figure}

The upper part of \cref{tab:svcomp-summary} summarizes the results for ReachSafety tasks and
reports, for each verifier, the number of ``best'' results (in terms of CPU time) and uniquely solved tasks.
The column ``VBS'' denotes the virtual best solver,
a hypothetical solver that always achieves the best-performing result (in terms of computational resources) among all compared tools for each task.
\Cref{tab:svcomp-reachsafety-1,tab:svcomp-reachsafety-2} further break down the results by subcategory.
In the tables, yellow-highlighted cells indicate the best results, while light-gray cells indicate the second-best results; both are shown in boldface.
Cells colored in red mark incorrect results.
The cactus plots in \cref{fig:svcomp-reachsafety} visualize the CPU time of the compared tools on the benchmark tasks,
focusing on subcategories where \cpv is highlighted at least once in the tables.

On ReachSafety, \cpv achieved competitive results,
producing the highest number of proofs and solving the second-largest number of tasks overall.
It did not produce any incorrect results on the benchmark set, indicating reliable verification behavior.
Moreover, \cpv solved the most unique instances and achieved the best performance on the second-largest number of tasks.
Since subcategory sizes vary significantly and may bias aggregated results,
we additionally examined performance at the subcategory level.
As expected, \cpv performed particularly well on hardware-like programs,
including the Hardware subcategory translated from hardware model checking tasks using \tool{Btor2C}~\cite{BTOR2C-STTT}
and the ECA subcategory derived from Event-Condition-Action reactive systems~\cite{RERS12}.
Strong performance was also observed on subcategories ProductLines (derived from product-line systems) and XCSP (originating from constraint programming),
as well as on benchmark tasks involving common C constructs,
such as subcategories Arrays, BitVectors, and Hardness~\cite{HardnessBenchmarks,Berger24}.
These results indicate that the proposed circuit-based approach is effective across a diverse range of benchmarks
and can outperform many established software-verification techniques.

The subcategory where \cpv showed comparatively weaker performance is Heap,
primarily due to frontend limitations (see \hyperref[sect:rq1]{RQ1}).%
\footnote{More tasks are reported as correctly solved by \cpv's \svcomp portfolio in \cref{tab:svcomp-reachsafety-1}
than those successfully translated in \cref{tab:translation},
because the portfolio incorporates multiple translation configurations (e.g., with program slicing),
whereas in \hyperref[sect:rq1]{RQ1}, only \cpv's default translation mode was evaluated.}
\kratostwo, on which \cpv's frontend relies, also exhibited weaker performance on this subcategory.
That said, even in subcategories where \cpv is not highlighted in \cref{tab:svcomp-reachsafety-1,tab:svcomp-reachsafety-2},
it still delivered some best or uniquely solved results,
demonstrating its complementary strengths to the other evaluated verifiers.

\cpv adopts a fundamentally different verification approach from the other tools in the comparison:
it ``folds'' programs into circuits with a monolithic transition relation,
whereas most other verifiers rely on path-based exploration.
This distinction enables \cpv to complement existing techniques.
Moreover, advances in hardware model checking allow \cpv to leverage powerful engines to effectively analyze the resulting circuits,
despite their potential complexity due to the monolithic encoding.
Overall, \cpv demonstrates competitive performance on ReachSafety tasks in \svcomp.
As mentioned previously, we plan to enhance its frontend capability in order to further improve its effectiveness.

\begin{rqresult}
    \cpv demonstrates competitive performance on the \svcomp ReachSafety benchmark set,
    solving the second-highest number of tasks and achieving strong results across diverse subcategories.
    More importantly, it complements existing software verifiers by uniquely solving tasks that others cannot handle.
    Its current weaknesses are partly due to frontend limitations, which present opportunities for further improvement.
\end{rqresult}

\subsubsection*{RQ5.2: Comparison with SOTA Software Verifiers on Termination Tasks}
\label{sect:rq5.2}

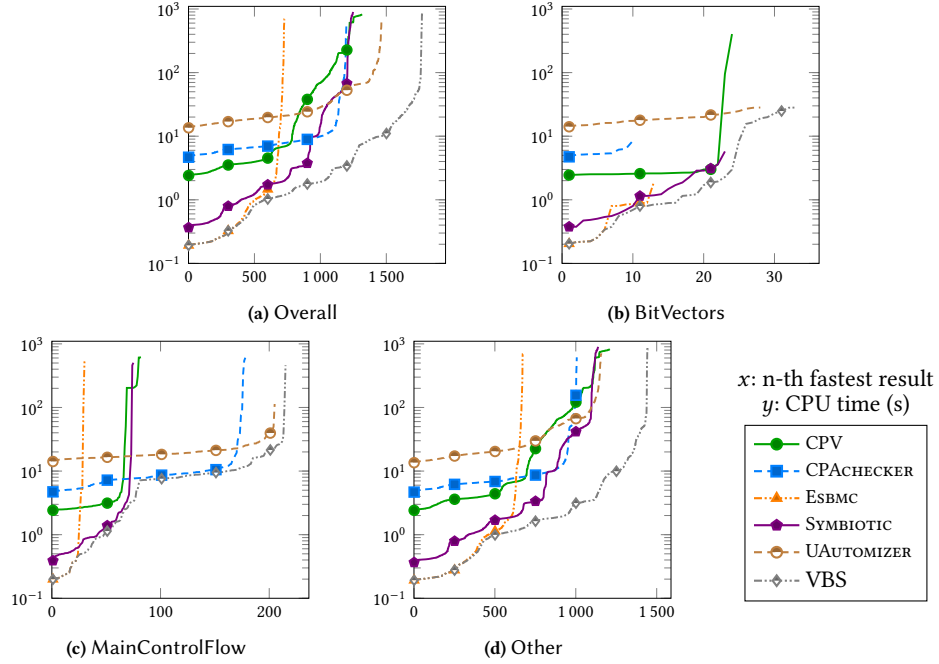
\begin{figure}[t]
    \centering
    \tikzexternalenable
    \subfloat[Overall]{\scalebox{.9}{\input{plots/svcomp.Termination.cputime.quantile}}}\hspace{.05\linewidth}%
    \tikzexternaldisable 
    \subfloat[BitVectors]{\scalebox{.9}{\input{plots/svcomp.Termination-BitVectors.cputime.quantile}}}\hspace{.05\linewidth}\\
    \tikzexternalenable
    \subfloat[MainControlFlow]{\scalebox{.9}{\input{plots/svcomp.Termination-MainControlFlow.cputime.quantile}}}\hspace{.04\linewidth}%
    \subfloat[Other]{\scalebox{.9}{\input{plots/svcomp.Termination-Other.cputime.quantile}}}\hspace{.02\linewidth}%
    \tikzexternaldisable
    \subfloat{\makebox[.22\linewidth][c]{\raisebox{.02\linewidth}{\shortstack{%
        $x$: n-th fastest result\\$y$: CPU time (s)\\
        \pgfplotslegendfromname{legend:svcomp-term}}}}}
    \caption{Cactus plots for \svcomp Termination category and its subcategories,
        where \cpv is highlighted at least once in \cref{tab:svcomp-termination}}
    \label{fig:svcomp-termination}
\end{figure}

Continuing from \hyperref[sect:rq5.1]{RQ5.1},
we follow a similar evaluation procedure to compare \cpv with state-of-the-art software verifiers on the \svcomp Termination benchmark set.
\kratostwo does not support termination analysis for C programs and is therefore excluded from this comparison.
The results are summarized in the lower part of \cref{tab:svcomp-summary}, with \cref{fig:svcomp-termination,tab:svcomp-termination} providing a breakdown by subcategory.

\begin{table}[t]
    \centering
    \caption{\svcomp Termination results for each subcategory}
    \input{tables/svcomp-term}
    \label{tab:svcomp-termination}
\end{table}

\cpv again performed competitively on Termination,
solving the second-largest number of tasks overall and producing no incorrect results.
It also detects the highest number of non-terminating lassos across nearly all subcategories, except for MainHeap,
where the frontend limitations (see \hyperref[sect:rq1]{RQ1}) prevent it from translating the verification tasks into circuits.
\cpachecker and \symbiotic, which employ techniques similar to L2S by checking for state revisits,
also perform well in this regard, detecting the second- and third-largest numbers of non-terminating executions, respectively.
These results suggest that approaches that explicitly check for state revisits are effective for lasso detection in termination analysis.

For proving termination, however, \cpv is less effective than the top-performing tools.
\uautomizer, the strongest verifier in this category,
relies on ranking-function synthesis~\cite{RankingFunctions,ULTIMATE-TERMINATION,RankingTemplates} and is particularly effective at proving termination on the benchmark set,
especially in the subcategory MainControlFlow.
\esbmc, on the other hand, attempts to prove termination by exhaustively exploring all program execution paths,
for example by checking whether loops can be fully unrolled to a finite bound.
L2S-based approaches introduce auxiliary variables (to record state history) and additional constraints (to encode state-revisit conditions),
which may increase the complexity of the verification problem.
Nevertheless, \cpv still achieves reasonable performance.

Most word-level hardware model checkers in recent editions of HWMCC do not directly support termination analysis.
Consequently, \cpv mainly relies on its L2S transformation to reduce termination to reachability problems.
Our results reveal a limitation of this approach:
while simple and general, it is less effective than specialized techniques for proving termination.
This observation opens up interesting research directions,
such as incorporating more liveness-checking techniques for transition systems,
including FAIR~\cite{FAIR-FMCAD11}, $k$-liveness~\cite{kliveness}, and rlive~\cite{rlive,rlive-inf}.
Currently, to the best of our knowledge,
\pono~\cite{Pono2} is the only hardware model checker that directly supports checking justice constraints in \btortwo circuits.
Several existing model checkers implement the aforementioned liveness-checking algorithms,
but they are not readily applicable in our setting because they target different input formats and do not export witnesses in \btortwo format.%
\footnote{%
    \tool{nuXmv}~\cite{NuXMV}, which implements FAIR, $k$-liveness, and rlive,
    accepts SMV~\cite{SMV:McMillan1993}, VMT-LIB~\cite{VMT-LIB}, and \aiger~\cite{AIGER-1.9} as input,
    \tool{IC3ia}~\cite{PredAbsPDR} accepts VMT-LIB,
    and \abc~\cite{ABC} accepts \aiger.
    Although \btortoaiger~\cite{btor2tools-website}, \tool{vmt-tools}~\cite{VMT-LIB},
    and utility scripts shipped with \tool{nuXmv} releases provide translations from \btortwo to these formats,
    none of them currently supports translating justice constraints.
}
By contributing translated \btortwo circuits to HWMCC,
we aim to encourage the development of more advanced liveness-checking capabilities,
which could subsequently be integrated into \cpv to further improve its performance on termination analysis.

\begin{rqresult}
    \cpv achieves competitive results on the \svcomp Termination benchmark set,
    solving the second-largest number of tasks without producing incorrect results.
    It is particularly strong in finding non-terminating executions via state-revisit detection,
    while being less effective at proving termination compared to tools based on ranking function synthesis.
\end{rqresult}

\subsubsection*{Summary}

Across RQ1--RQ4, we examined the feasibility of the proposed methodology,
including encoding C programs as sequential circuits,
reusing off-the-shelf hardware model checkers for verification,
and exporting software-level violation witnesses.
In RQ5, competitive performance against state-of-the-art software verifiers further strengthens the practicality and significance of the proposed framework.
Taken together, these findings show that:

\begin{rqresult}
    Sequential circuits are a practical intermediate representation for software verification,
    and hardware model checking can be applied effectively to software-verification problems.
\end{rqresult}

\subsection{Threats to Validity}
\label{sect:threats}

We discuss the potential threats to the validity of our study and the measures taken to mitigate them.

\subsubsection*{External Validity}

The conclusions of our study are drawn from the evaluation results.
We used the \svcomp benchmark suite,
which is the largest and most diverse open-source collection of verification tasks for the programming language~C.
Programs containing recursive functions were excluded because recursion is not currently supported by \cpv.

For RQ2, evaluating the complete benchmark suite would have been prohibitively expensive.
Therefore, we used a sampled benchmark set and ensured that all benchmark subcategories were sufficiently represented in the sample.
The observations in RQ2 are also limited to benchmark tasks that can be successfully translated into \btortwo circuits by \cpv.
Consequently, the reported effects of different encoding strategies may not generalize to programs that are currently unsupported by the frontend.
Nevertheless, the successfully translated instances span diverse benchmark categories and program characteristics, providing a solid basis for the study.

For RQ5, we compared \cpv against the top-performing tools in \svcomp~2026.
These tools employ SMT-based reasoning but differ substantially
in their architectures, verification algorithms, and areas of strength.
We believe that the selected competitors reflect the current state of the art
and allow a meaningful assessment of \cpv's relative performance.

\cpv's backend consists of several state-of-the-art hardware model checkers,
covering both bit-level (SAT-based) and word-level (SMT-based) reasoning,
as well as a broad range of configurations and verification engines.
This provides representative coverage of mainstream approaches in hardware model checking.
In several experiments, \cpv ran fixed sequential portfolios selected from preliminary experiments.
Different portfolio configurations may lead to different performance results.
To mitigate this threat, the portfolios were selected using a subset of \svcomp~2025 benchmark tasks
and combine engines with complementary strengths rather than being optimized for individual benchmark families.

\subsubsection*{Internal Validity}

The experimental results depend on the correctness of \cpv and the integrated verification workflow.
During the development of \cpv, we discovered several issues in \kratostwo,
one of the key components used for program translation.
We developed patches and reported the issues to the developers of \kratostwo,
and all identified issues were subsequently resolved.
To detect potential unsoundness in backend hardware model checkers,
every violation witness produced is validated using \btorsim before being translated into a software-verification witness.
This validation step ensures that the reported counterexample is consistent with regard to the translated sequential circuit.
The correctness of the overall translation pipeline is another potential source of error.
To increase confidence in the implementation, \cpv emits software-verification witnesses for the detected violations,
which have been validated by multiple independent witness validators.
While we do not claim that \cpv is free of bugs, we did not observe any incorrect verification results in our experiments.

The generated sequential circuits also depend on the underlying frontend implementation.
In particular, the relational encoding is provided by \kratostwo, which is available only as a binary executable.
As a result, certain implementation details, such as the encoding of the symbolic program counter, cannot be controlled or modified by \cpv.
Different encodings may produce structurally different circuits.
However, the program-counter state variable and its update logic make up only a small fraction of the overall transition system
and are largely independent of the encoding of the program's data variables and their corresponding transition logic.
We expect the choice of program-counter encoding to have only a limited impact on circuit size and verification performance.
This expectation is consistent with our preliminary experiments comparing binary and one-hot program-counter encodings in the functional encoder.
We therefore do not expect this implementation detail to significantly affect the conclusions drawn from our comparison of relational and functional encodings.

All experiments were executed via the widely used benchmarking framework \benchexec~\cite{Benchmarking-STTT}.
\benchexec utilizes features in the modern Linux kernel such as cgroups for resource measurement and limitation,
namespaces for process isolation,
and overlay file systems to prevent benchmark runs from modifying the host environment.
To reduce interference from shared hardware resources (e.g., through hyper-threading),
each benchmark run was restricted to two virtual CPU cores pinned to the same physical core,
and that physical core was dedicated exclusively to the run.

%% file: plots/plot-defs.tex
\pgfplotsset{quantile plot/.style={
    width=.25\linewidth,
    height=.25\linewidth,
    scale only axis,
    /pgf/number format/1000 sep={\,},
    /pgfplots/table/x index=0,
    /pgfplots/table/y index=4,
    /pgfplots/table/header=false,
    ticklabel style={font={\smaller}},
    xmin=0,
    ymin=0.1,
    ymax=1100,
    legend cell align={left},
    legend style={at={(0,1)}, anchor=north west, outer xsep=5pt, outer ysep=5pt, fill=none},
    legend columns=1,
    cycle multiindex list={
      green, blue!50!cyan, orange, teal, violet, brown, gray, olive, black\nextlist
      mark list*\nextlist
      solid, densely dashed, densely dashdotdotted, densely dotted, solid, densely dashed, densely dashdotdotted, densely dotted},
  },
  every axis plot/.append style={thick}
}

\onlyifstandalone{
  \pgfplotsset{quantile plot/.append style={
      ylabel=CPU time (\second),
    },
  }
}

\newcommand\addgraph[2]{{
  \newcommand\csvfile{\detokenize{#2}}
  \IfFileExists\csvfile{
    \addplot+ table {\csvfile}; \addlegendentry{#1}
  }{
    \addplot coordinates {};
  }
}}

%% file: tables/translation.tex
\begin{tabular}{llS[table-format=5]S[table-format=5]S[round-mode=figures, round-precision=3]S[table-format=5]S[round-mode=figures, round-precision=3]S[table-format=5]S[round-mode=figures, round-precision=3]}
\toprule
\multicolumn{2}{c}{\multirow{2}{*}{Category}}
    & \multicolumn{1}{c}{\multirow{2}{*}{\#Tasks}}
    & \multicolumn{2}{c}{CFA (K2)}
    & \multicolumn{2}{c}{Relational (\btortwo)}
    & \multicolumn{2}{c}{Functional (\btortwo)}
    \\
\cmidrule(lr){4-5}\cmidrule(lr){6-7}\cmidrule(lr){8-9}
    &
    &
    & \multicolumn{1}{c}{\#}
    & \multicolumn{1}{c}{\%}
    & \multicolumn{1}{c}{\#}
    & \multicolumn{1}{c}{\%}
    & \multicolumn{1}{c}{\#}
    & \multicolumn{1}{c}{\%}
    \\
\midrule
\multirow{14}{*}{ReachSafety}
    & Arrays
    & \CpvTranslateRelReachSafetyArraysStatusAllCount
    & \CpvTranslateRelReachSafetyArraysKIIAllSum
    & \fpeval{\CpvTranslateRelReachSafetyArraysKIIAllSum / \CpvTranslateRelReachSafetyArraysStatusAllCount * 100}
    & \CpvTranslateRelReachSafetyArraysBtorIIAllSum
    & \fpeval{\CpvTranslateRelReachSafetyArraysBtorIIAllSum / \CpvTranslateRelReachSafetyArraysStatusAllCount *100}
    & \CpvTranslateFuncReachSafetyArraysBtorIIAllSum
    & \fpeval{\CpvTranslateFuncReachSafetyArraysBtorIIAllSum / \CpvTranslateRelReachSafetyArraysStatusAllCount *100}
    \\
    & BitVectors
    & \CpvTranslateRelReachSafetyBitVectorsStatusAllCount
    & \CpvTranslateRelReachSafetyBitVectorsKIIAllSum
    & \fpeval{\CpvTranslateRelReachSafetyBitVectorsKIIAllSum / \CpvTranslateRelReachSafetyBitVectorsStatusAllCount * 100}
    & \CpvTranslateRelReachSafetyBitVectorsBtorIIAllSum
    & \fpeval{\CpvTranslateRelReachSafetyBitVectorsBtorIIAllSum / \CpvTranslateRelReachSafetyBitVectorsStatusAllCount *100}
    & \CpvTranslateFuncReachSafetyBitVectorsBtorIIAllSum
    & \fpeval{\CpvTranslateFuncReachSafetyBitVectorsBtorIIAllSum / \CpvTranslateRelReachSafetyBitVectorsStatusAllCount *100}
    \\
    & Combinations
    & \CpvTranslateRelReachSafetyCombinationsStatusAllCount
    & \CpvTranslateRelReachSafetyCombinationsKIIAllSum
    & \fpeval{\CpvTranslateRelReachSafetyCombinationsKIIAllSum / \CpvTranslateRelReachSafetyCombinationsStatusAllCount * 100}
    & \CpvTranslateRelReachSafetyCombinationsBtorIIAllSum
    & \fpeval{\CpvTranslateRelReachSafetyCombinationsBtorIIAllSum / \CpvTranslateRelReachSafetyCombinationsStatusAllCount *100}
    & \CpvTranslateFuncReachSafetyCombinationsBtorIIAllSum
    & \fpeval{\CpvTranslateFuncReachSafetyCombinationsBtorIIAllSum / \CpvTranslateRelReachSafetyCombinationsStatusAllCount *100}
    \\
    & ControlFlow
    & \CpvTranslateRelReachSafetyControlFlowStatusAllCount
    & \CpvTranslateRelReachSafetyControlFlowKIIAllSum
    & \fpeval{\CpvTranslateRelReachSafetyControlFlowKIIAllSum / \CpvTranslateRelReachSafetyControlFlowStatusAllCount * 100}
    & \CpvTranslateRelReachSafetyControlFlowBtorIIAllSum
    & \fpeval{\CpvTranslateRelReachSafetyControlFlowBtorIIAllSum / \CpvTranslateRelReachSafetyControlFlowStatusAllCount *100}
    & \CpvTranslateFuncReachSafetyControlFlowBtorIIAllSum
    & \fpeval{\CpvTranslateFuncReachSafetyControlFlowBtorIIAllSum / \CpvTranslateRelReachSafetyControlFlowStatusAllCount *100}
    \\
    & ECA
    & \CpvTranslateRelReachSafetyECAStatusAllCount
    & \CpvTranslateRelReachSafetyECAKIIAllSum
    & \fpeval{\CpvTranslateRelReachSafetyECAKIIAllSum / \CpvTranslateRelReachSafetyECAStatusAllCount * 100}
    & \CpvTranslateRelReachSafetyECABtorIIAllSum
    & \fpeval{\CpvTranslateRelReachSafetyECABtorIIAllSum / \CpvTranslateRelReachSafetyECAStatusAllCount *100}
    & \CpvTranslateFuncReachSafetyECABtorIIAllSum
    & \fpeval{\CpvTranslateFuncReachSafetyECABtorIIAllSum / \CpvTranslateRelReachSafetyECAStatusAllCount *100}
    \\
    & Floats
    & \CpvTranslateRelReachSafetyFloatsStatusAllCount
    & \CpvTranslateRelReachSafetyFloatsKIIAllSum
    & \fpeval{\CpvTranslateRelReachSafetyFloatsKIIAllSum / \CpvTranslateRelReachSafetyFloatsStatusAllCount * 100}
    & \CpvTranslateRelReachSafetyFloatsBtorIIAllSum
    & \fpeval{\CpvTranslateRelReachSafetyFloatsBtorIIAllSum / \CpvTranslateRelReachSafetyFloatsStatusAllCount *100}
    & \CpvTranslateFuncReachSafetyFloatsBtorIIAllSum
    & \fpeval{\CpvTranslateFuncReachSafetyFloatsBtorIIAllSum / \CpvTranslateRelReachSafetyFloatsStatusAllCount *100}
    \\
    & Hardness
    & \CpvTranslateRelReachSafetyHardnessStatusAllCount
    & \CpvTranslateRelReachSafetyHardnessKIIAllSum
    & \fpeval{\CpvTranslateRelReachSafetyHardnessKIIAllSum / \CpvTranslateRelReachSafetyHardnessStatusAllCount * 100}
    & \CpvTranslateRelReachSafetyHardnessBtorIIAllSum
    & \fpeval{\CpvTranslateRelReachSafetyHardnessBtorIIAllSum / \CpvTranslateRelReachSafetyHardnessStatusAllCount *100}
    & \CpvTranslateFuncReachSafetyHardnessBtorIIAllSum
    & \fpeval{\CpvTranslateFuncReachSafetyHardnessBtorIIAllSum / \CpvTranslateRelReachSafetyHardnessStatusAllCount *100}
    \\
    & Hardware
    & \CpvTranslateRelReachSafetyHardwareStatusAllCount
    & \CpvTranslateRelReachSafetyHardwareKIIAllSum
    & \fpeval{\CpvTranslateRelReachSafetyHardwareKIIAllSum / \CpvTranslateRelReachSafetyHardwareStatusAllCount * 100}
    & \CpvTranslateRelReachSafetyHardwareBtorIIAllSum
    & \fpeval{\CpvTranslateRelReachSafetyHardwareBtorIIAllSum / \CpvTranslateRelReachSafetyHardwareStatusAllCount *100}
    & \CpvTranslateFuncReachSafetyHardwareBtorIIAllSum
    & \fpeval{\CpvTranslateFuncReachSafetyHardwareBtorIIAllSum / \CpvTranslateRelReachSafetyHardwareStatusAllCount *100}
    \\
    & Heap
    & \CpvTranslateRelReachSafetyHeapStatusAllCount
    & \CpvTranslateRelReachSafetyHeapKIIAllSum
    & \fpeval{\CpvTranslateRelReachSafetyHeapKIIAllSum / \CpvTranslateRelReachSafetyHeapStatusAllCount * 100}
    & \CpvTranslateRelReachSafetyHeapBtorIIAllSum
    & \fpeval{\CpvTranslateRelReachSafetyHeapBtorIIAllSum / \CpvTranslateRelReachSafetyHeapStatusAllCount *100}
    & \CpvTranslateFuncReachSafetyHeapBtorIIAllSum
    & \fpeval{\CpvTranslateFuncReachSafetyHeapBtorIIAllSum / \CpvTranslateRelReachSafetyHeapStatusAllCount *100}
    \\
    & Loops
    & \CpvTranslateRelReachSafetyLoopsStatusAllCount
    & \CpvTranslateRelReachSafetyLoopsKIIAllSum
    & \fpeval{\CpvTranslateRelReachSafetyLoopsKIIAllSum / \CpvTranslateRelReachSafetyLoopsStatusAllCount * 100}
    & \CpvTranslateRelReachSafetyLoopsBtorIIAllSum
    & \fpeval{\CpvTranslateRelReachSafetyLoopsBtorIIAllSum / \CpvTranslateRelReachSafetyLoopsStatusAllCount *100}
    & \CpvTranslateFuncReachSafetyLoopsBtorIIAllSum
    & \fpeval{\CpvTranslateFuncReachSafetyLoopsBtorIIAllSum / \CpvTranslateRelReachSafetyLoopsStatusAllCount *100}
    \\
    & ProductLines
    & \CpvTranslateRelReachSafetyProductLinesStatusAllCount
    & \CpvTranslateRelReachSafetyProductLinesKIIAllSum
    & \fpeval{\CpvTranslateRelReachSafetyProductLinesKIIAllSum / \CpvTranslateRelReachSafetyProductLinesStatusAllCount * 100}
    & \CpvTranslateRelReachSafetyProductLinesBtorIIAllSum
    & \fpeval{\CpvTranslateRelReachSafetyProductLinesBtorIIAllSum / \CpvTranslateRelReachSafetyProductLinesStatusAllCount *100}
    & \CpvTranslateFuncReachSafetyProductLinesBtorIIAllSum
    & \fpeval{\CpvTranslateFuncReachSafetyProductLinesBtorIIAllSum / \CpvTranslateRelReachSafetyProductLinesStatusAllCount *100}
    \\
    & Sequentialized
    & \CpvTranslateRelReachSafetySequentializedStatusAllCount
    & \CpvTranslateRelReachSafetySequentializedKIIAllSum
    & \fpeval{\CpvTranslateRelReachSafetySequentializedKIIAllSum / \CpvTranslateRelReachSafetySequentializedStatusAllCount * 100}
    & \CpvTranslateRelReachSafetySequentializedBtorIIAllSum
    & \fpeval{\CpvTranslateRelReachSafetySequentializedBtorIIAllSum / \CpvTranslateRelReachSafetySequentializedStatusAllCount *100}
    & \CpvTranslateFuncReachSafetySequentializedBtorIIAllSum
    & \fpeval{\CpvTranslateFuncReachSafetySequentializedBtorIIAllSum / \CpvTranslateRelReachSafetySequentializedStatusAllCount *100}
    \\
    & XCSP
    & \CpvTranslateRelReachSafetyXCSPStatusAllCount
    & \CpvTranslateRelReachSafetyXCSPKIIAllSum
    & \fpeval{\CpvTranslateRelReachSafetyXCSPKIIAllSum / \CpvTranslateRelReachSafetyXCSPStatusAllCount * 100}
    & \CpvTranslateRelReachSafetyXCSPBtorIIAllSum
    & \fpeval{\CpvTranslateRelReachSafetyXCSPBtorIIAllSum / \CpvTranslateRelReachSafetyXCSPStatusAllCount *100}
    & \CpvTranslateFuncReachSafetyXCSPBtorIIAllSum
    & \fpeval{\CpvTranslateFuncReachSafetyXCSPBtorIIAllSum / \CpvTranslateRelReachSafetyXCSPStatusAllCount *100}
    \\
\cmidrule{2-9}
    & Overall
    & \CpvTranslateRelReachSafetyStatusAllCount
    & \CpvTranslateRelReachSafetyKIIAllSum
    & \fpeval{\CpvTranslateRelReachSafetyKIIAllSum / \CpvTranslateRelReachSafetyStatusAllCount * 100}
    & \CpvTranslateRelReachSafetyBtorIIAllSum
    & \fpeval{\CpvTranslateRelReachSafetyBtorIIAllSum / \CpvTranslateRelReachSafetyStatusAllCount *100}
    & \CpvTranslateFuncReachSafetyBtorIIAllSum
    & \fpeval{\CpvTranslateFuncReachSafetyBtorIIAllSum / \CpvTranslateRelReachSafetyStatusAllCount *100}
    \\
\midrule
\multirow{5}{*}{Termination}
    & BitVectors
    & \CpvTranslateRelLIIsTerminationBitVectorsStatusAllCount
    & \CpvTranslateRelLIIsTerminationBitVectorsKIIAllSum
    & \fpeval{\CpvTranslateRelLIIsTerminationBitVectorsKIIAllSum / \CpvTranslateRelLIIsTerminationBitVectorsStatusAllCount * 100}
    & \CpvTranslateRelLIIsTerminationBitVectorsBtorIIAllSum
    & \fpeval{\CpvTranslateRelLIIsTerminationBitVectorsBtorIIAllSum / \CpvTranslateRelLIIsTerminationBitVectorsStatusAllCount * 100}
    & \CpvTranslateFuncLIIsTerminationBitVectorsBtorIIAllSum
    & \fpeval{\CpvTranslateFuncLIIsTerminationBitVectorsBtorIIAllSum / \CpvTranslateRelLIIsTerminationBitVectorsStatusAllCount * 100}
    \\
    & MainControlFlow
    & \CpvTranslateRelLIIsTerminationMainControlFlowStatusAllCount
    & \CpvTranslateRelLIIsTerminationMainControlFlowKIIAllSum
    & \fpeval{\CpvTranslateRelLIIsTerminationMainControlFlowKIIAllSum / \CpvTranslateRelLIIsTerminationMainControlFlowStatusAllCount * 100}
    & \CpvTranslateRelLIIsTerminationMainControlFlowBtorIIAllSum
    & \fpeval{\CpvTranslateRelLIIsTerminationMainControlFlowBtorIIAllSum / \CpvTranslateRelLIIsTerminationMainControlFlowStatusAllCount * 100}
    & \CpvTranslateFuncLIIsTerminationMainControlFlowBtorIIAllSum
    & \fpeval{\CpvTranslateFuncLIIsTerminationMainControlFlowBtorIIAllSum / \CpvTranslateRelLIIsTerminationMainControlFlowStatusAllCount * 100}
    \\
    & MainHeap
    & \CpvTranslateRelLIIsTerminationMainHeapStatusAllCount
    & \CpvTranslateRelLIIsTerminationMainHeapKIIAllSum
    & \fpeval{\CpvTranslateRelLIIsTerminationMainHeapKIIAllSum / \CpvTranslateRelLIIsTerminationMainHeapStatusAllCount * 100}
    & \CpvTranslateRelLIIsTerminationMainHeapBtorIIAllSum
    & \fpeval{\CpvTranslateRelLIIsTerminationMainHeapBtorIIAllSum / \CpvTranslateRelLIIsTerminationMainHeapStatusAllCount * 100}
    & \CpvTranslateFuncLIIsTerminationMainHeapBtorIIAllSum
    & \fpeval{\CpvTranslateFuncLIIsTerminationMainHeapBtorIIAllSum / \CpvTranslateRelLIIsTerminationMainHeapStatusAllCount * 100}
    \\
    & Other
    & \CpvTranslateRelLIIsTerminationOtherStatusAllCount
    & \CpvTranslateRelLIIsTerminationOtherKIIAllSum
    & \fpeval{\CpvTranslateRelLIIsTerminationOtherKIIAllSum / \CpvTranslateRelLIIsTerminationOtherStatusAllCount * 100}
    & \CpvTranslateRelLIIsTerminationOtherBtorIIAllSum
    & \fpeval{\CpvTranslateRelLIIsTerminationOtherBtorIIAllSum / \CpvTranslateRelLIIsTerminationOtherStatusAllCount * 100}
    & \CpvTranslateFuncLIIsTerminationOtherBtorIIAllSum
    & \fpeval{\CpvTranslateFuncLIIsTerminationOtherBtorIIAllSum / \CpvTranslateRelLIIsTerminationOtherStatusAllCount * 100}
    \\
\cmidrule{2-9}
    & Overall
    & \CpvTranslateRelLIIsTerminationStatusAllCount
    & \CpvTranslateRelLIIsTerminationKIIAllSum
    & \fpeval{\CpvTranslateRelLIIsTerminationKIIAllSum / \CpvTranslateRelLIIsTerminationStatusAllCount * 100}
    & \CpvTranslateRelLIIsTerminationBtorIIAllSum
    & \fpeval{\CpvTranslateRelLIIsTerminationBtorIIAllSum / \CpvTranslateRelLIIsTerminationStatusAllCount * 100}
    & \CpvTranslateFuncLIIsTerminationBtorIIAllSum
    & \fpeval{\CpvTranslateFuncLIIsTerminationBtorIIAllSum / \CpvTranslateRelLIIsTerminationStatusAllCount * 100}
    \\
\bottomrule
\end{tabular}

%% file: tables/encoding.bv.tex
\begin{tabular}{cclS[table-format=5]S[table-format=5]S[round-mode=figures, round-precision=2]S[table-format=5]S[table-format=5]S[round-mode=figures, round-precision=2]}
\toprule
\multicolumn{2}{c}{HWMC}
    & \multirow{2}{*}{Results}
    & \multicolumn{3}{c}{ReachSafety (\num{\AbcImcReachSafetySampledBvFuncStatusAllCount})}
    & \multicolumn{3}{c}{Termination (\num{\AbcImcTerminationSampledBvFuncStatusAllCount})}
    \\
\cmidrule(lr){1-2}\cmidrule(lr){4-6}\cmidrule(lr){7-9}
Tool
    & Config.
    &
    & \multicolumn{1}{c}{Relational}
    & \multicolumn{1}{c}{Functional}
    & \multicolumn{1}{c}{$\Delta$\,(\%)}
    & \multicolumn{1}{c}{Relational}
    & \multicolumn{1}{c}{Functional}
    & \multicolumn{1}{c}{$\Delta$\,(\%)}
    \\
\midrule
\multirow{6}{*}{\abc}
    & \multirow{3}{*}{IMC}
    & \#Solved
    & \boldnum{\AbcImcReachSafetySampledBvRelStatusCorrectCount}
    & \AbcImcReachSafetySampledBvFuncStatusCorrectCount
    & \fpeval{(\AbcImcReachSafetySampledBvFuncStatusCorrectCount - \AbcImcReachSafetySampledBvRelStatusCorrectCount) / \AbcImcReachSafetySampledBvRelStatusCorrectCount * 100}
    & \boldnum{\AbcImcTerminationSampledBvRelStatusCorrectCount}
    & \AbcImcTerminationSampledBvFuncStatusCorrectCount
    & \fpeval{(\AbcImcTerminationSampledBvFuncStatusCorrectCount - \AbcImcTerminationSampledBvRelStatusCorrectCount) / \AbcImcTerminationSampledBvRelStatusCorrectCount * 100}
    \\
    & 
    & \quad proofs
    & \boldnum{\AbcImcReachSafetySampledBvRelStatusCorrectTrueCount}
    & \boldnum{\AbcImcReachSafetySampledBvFuncStatusCorrectTrueCount}
    & \fpeval{(\AbcImcReachSafetySampledBvFuncStatusCorrectTrueCount - \AbcImcReachSafetySampledBvRelStatusCorrectTrueCount) / \AbcImcReachSafetySampledBvRelStatusCorrectTrueCount * 100}
    & \AbcImcTerminationSampledBvRelStatusCorrectTrueCount
    & \boldnum{\AbcImcTerminationSampledBvFuncStatusCorrectTrueCount}
    & \fpeval{(\AbcImcTerminationSampledBvFuncStatusCorrectTrueCount - \AbcImcTerminationSampledBvRelStatusCorrectTrueCount) / \AbcImcTerminationSampledBvRelStatusCorrectTrueCount * 100}
    \\
    & 
    & \quad alarms
    & \boldnum{\AbcImcReachSafetySampledBvRelStatusCorrectFalseCount}
    & \AbcImcReachSafetySampledBvFuncStatusCorrectFalseCount
    & \fpeval{(\AbcImcReachSafetySampledBvFuncStatusCorrectFalseCount - \AbcImcReachSafetySampledBvRelStatusCorrectFalseCount) / \AbcImcReachSafetySampledBvRelStatusCorrectFalseCount * 100}
    & \boldnum{\AbcImcTerminationSampledBvRelStatusCorrectFalseCount}
    & \AbcImcTerminationSampledBvFuncStatusCorrectFalseCount
    & \fpeval{(\AbcImcTerminationSampledBvFuncStatusCorrectFalseCount - \AbcImcTerminationSampledBvRelStatusCorrectFalseCount) / \AbcImcTerminationSampledBvRelStatusCorrectFalseCount * 100}
    \\
\cmidrule(lr){2-9}
    & \multirow{3}{*}{PDR}
    & \#Solved
    & \AbcPdrReachSafetySampledBvRelStatusCorrectCount
    & \boldnum{\AbcPdrReachSafetySampledBvFuncStatusCorrectCount}
    & \fpeval{(\AbcPdrReachSafetySampledBvFuncStatusCorrectCount - \AbcPdrReachSafetySampledBvRelStatusCorrectCount) / \AbcPdrReachSafetySampledBvRelStatusCorrectCount * 100}
    & \AbcPdrTerminationSampledBvRelStatusCorrectCount
    & \boldnum{\AbcPdrTerminationSampledBvFuncStatusCorrectCount}
    & \fpeval{(\AbcPdrTerminationSampledBvFuncStatusCorrectCount - \AbcPdrTerminationSampledBvRelStatusCorrectCount) / \AbcPdrTerminationSampledBvRelStatusCorrectCount * 100}
    \\
    & 
    & \quad proofs
    & \AbcPdrReachSafetySampledBvRelStatusCorrectTrueCount
    & \boldnum{\AbcPdrReachSafetySampledBvFuncStatusCorrectTrueCount}
    & \fpeval{(\AbcPdrReachSafetySampledBvFuncStatusCorrectTrueCount - \AbcPdrReachSafetySampledBvRelStatusCorrectTrueCount) / \AbcPdrReachSafetySampledBvRelStatusCorrectTrueCount * 100}
    & \AbcPdrTerminationSampledBvRelStatusCorrectTrueCount
    & \boldnum{\AbcPdrTerminationSampledBvFuncStatusCorrectTrueCount}
    & \fpeval{(\AbcPdrTerminationSampledBvFuncStatusCorrectTrueCount - \AbcPdrTerminationSampledBvRelStatusCorrectTrueCount) / \AbcPdrTerminationSampledBvRelStatusCorrectTrueCount * 100}
    \\
    & 
    & \quad alarms
    & \AbcPdrReachSafetySampledBvRelStatusCorrectFalseCount
    & \boldnum{\AbcPdrReachSafetySampledBvFuncStatusCorrectFalseCount}
    & \fpeval{(\AbcPdrReachSafetySampledBvFuncStatusCorrectFalseCount - \AbcPdrReachSafetySampledBvRelStatusCorrectFalseCount) / \AbcPdrReachSafetySampledBvRelStatusCorrectFalseCount * 100}
    & \AbcPdrTerminationSampledBvRelStatusCorrectFalseCount
    & \boldnum{\AbcPdrTerminationSampledBvFuncStatusCorrectFalseCount}
    & \fpeval{(\AbcPdrTerminationSampledBvFuncStatusCorrectFalseCount - \AbcPdrTerminationSampledBvRelStatusCorrectFalseCount) / \AbcPdrTerminationSampledBvRelStatusCorrectFalseCount * 100}
    \\
\cmidrule{1-9}
\multirow{6}{*}{\ricthree}
    & \multirow{3}{*}{IC3}
    & \#Solved
    & \RicIIIIcIIIReachSafetySampledBvRelStatusCorrectCount
    & \boldnum{\RicIIIIcIIIReachSafetySampledBvFuncStatusCorrectCount}
    & \fpeval{(\RicIIIIcIIIReachSafetySampledBvFuncStatusCorrectCount - \RicIIIIcIIIReachSafetySampledBvRelStatusCorrectCount) / \RicIIIIcIIIReachSafetySampledBvRelStatusCorrectCount * 100}
    & \RicIIIIcIIITerminationSampledBvRelStatusCorrectCount
    & \boldnum{\RicIIIIcIIITerminationSampledBvFuncStatusCorrectCount}
    & \fpeval{(\RicIIIIcIIITerminationSampledBvFuncStatusCorrectCount - \RicIIIIcIIITerminationSampledBvRelStatusCorrectCount) / \RicIIIIcIIITerminationSampledBvRelStatusCorrectCount * 100}
    \\
    & 
    & \quad proofs
    & \RicIIIIcIIIReachSafetySampledBvRelStatusCorrectTrueCount
    & \boldnum{\RicIIIIcIIIReachSafetySampledBvFuncStatusCorrectTrueCount}
    & \fpeval{(\RicIIIIcIIIReachSafetySampledBvFuncStatusCorrectTrueCount - \RicIIIIcIIIReachSafetySampledBvRelStatusCorrectTrueCount) / \RicIIIIcIIIReachSafetySampledBvRelStatusCorrectTrueCount * 100}
    & \RicIIIIcIIITerminationSampledBvRelStatusCorrectTrueCount
    & \boldnum{\RicIIIIcIIITerminationSampledBvFuncStatusCorrectTrueCount}
    & \fpeval{(\RicIIIIcIIITerminationSampledBvFuncStatusCorrectTrueCount - \RicIIIIcIIITerminationSampledBvRelStatusCorrectTrueCount) / \RicIIIIcIIITerminationSampledBvRelStatusCorrectTrueCount * 100}
    \\
    & 
    & \quad alarms
    & \RicIIIIcIIIReachSafetySampledBvRelStatusCorrectFalseCount
    & \boldnum{\RicIIIIcIIIReachSafetySampledBvFuncStatusCorrectFalseCount}
    & \fpeval{(\RicIIIIcIIIReachSafetySampledBvFuncStatusCorrectFalseCount - \RicIIIIcIIIReachSafetySampledBvRelStatusCorrectFalseCount) / \RicIIIIcIIIReachSafetySampledBvRelStatusCorrectFalseCount * 100}
    & \RicIIIIcIIITerminationSampledBvRelStatusCorrectFalseCount
    & \boldnum{\RicIIIIcIIITerminationSampledBvFuncStatusCorrectFalseCount}
    & \fpeval{(\RicIIIIcIIITerminationSampledBvFuncStatusCorrectFalseCount - \RicIIIIcIIITerminationSampledBvRelStatusCorrectFalseCount) / \RicIIIIcIIITerminationSampledBvRelStatusCorrectFalseCount * 100}
    \\
\cmidrule(lr){2-9}
    & \multirow{3}{*}{\kinduction}
    & \#Solved
    & \RicIIIKindReachSafetySampledBvRelStatusCorrectCount
    & \boldnum{\RicIIIKindReachSafetySampledBvFuncStatusCorrectCount}
    & \fpeval{(\RicIIIKindReachSafetySampledBvFuncStatusCorrectCount - \RicIIIKindReachSafetySampledBvRelStatusCorrectCount) / \RicIIIKindReachSafetySampledBvRelStatusCorrectCount * 100}
    & \RicIIIKindTerminationSampledBvRelStatusCorrectCount
    & \boldnum{\RicIIIKindTerminationSampledBvFuncStatusCorrectCount}
    & \fpeval{(\RicIIIKindTerminationSampledBvFuncStatusCorrectCount - \RicIIIKindTerminationSampledBvRelStatusCorrectCount) / \RicIIIKindTerminationSampledBvRelStatusCorrectCount * 100}
    \\
    & 
    & \quad proofs
    & \RicIIIKindReachSafetySampledBvRelStatusCorrectTrueCount
    & \boldnum{\RicIIIKindReachSafetySampledBvFuncStatusCorrectTrueCount}
    & \fpeval{(\RicIIIKindReachSafetySampledBvFuncStatusCorrectTrueCount - \RicIIIKindReachSafetySampledBvRelStatusCorrectTrueCount) / \RicIIIKindReachSafetySampledBvRelStatusCorrectTrueCount * 100}
    & \RicIIIKindTerminationSampledBvRelStatusCorrectTrueCount
    & \boldnum{\RicIIIKindTerminationSampledBvFuncStatusCorrectTrueCount}
    & \fpeval{(\RicIIIKindTerminationSampledBvFuncStatusCorrectTrueCount - \RicIIIKindTerminationSampledBvRelStatusCorrectTrueCount) / \RicIIIKindTerminationSampledBvRelStatusCorrectTrueCount * 100}
    \\
    & 
    & \quad alarms
    & \RicIIIKindReachSafetySampledBvRelStatusCorrectFalseCount
    & \boldnum{\RicIIIKindReachSafetySampledBvFuncStatusCorrectFalseCount}
    & \fpeval{(\RicIIIKindReachSafetySampledBvFuncStatusCorrectFalseCount - \RicIIIKindReachSafetySampledBvRelStatusCorrectFalseCount) / \RicIIIKindReachSafetySampledBvRelStatusCorrectFalseCount * 100}
    & \boldnum{\RicIIIKindTerminationSampledBvRelStatusCorrectFalseCount}
    & \RicIIIKindTerminationSampledBvFuncStatusCorrectFalseCount
    & \fpeval{(\RicIIIKindTerminationSampledBvFuncStatusCorrectFalseCount - \RicIIIKindTerminationSampledBvRelStatusCorrectFalseCount) / \RicIIIKindTerminationSampledBvRelStatusCorrectFalseCount * 100}
    \\
\cmidrule{1-9}
\multirow{3}{*}{\pono}
    & \multirow{3}{*}{IMC}
    & \#Solved
    & \PonoImcBzlaReachSafetySampledBvRelStatusCorrectCount
    & \boldnum{\PonoImcBzlaReachSafetySampledBvFuncStatusCorrectCount}
    & \fpeval{(\PonoImcBzlaReachSafetySampledBvFuncStatusCorrectCount - \PonoImcBzlaReachSafetySampledBvRelStatusCorrectCount) / \PonoImcBzlaReachSafetySampledBvRelStatusCorrectCount * 100}
    & \boldnum{\PonoImcBzlaTerminationSampledBvRelStatusCorrectCount}
    & \PonoImcBzlaTerminationSampledBvFuncStatusCorrectCount
    & \fpeval{(\PonoImcBzlaTerminationSampledBvFuncStatusCorrectCount - \PonoImcBzlaTerminationSampledBvRelStatusCorrectCount) / \PonoImcBzlaTerminationSampledBvRelStatusCorrectCount * 100}
    \\
    & 
    & \quad proofs
    & \PonoImcBzlaReachSafetySampledBvRelStatusCorrectTrueCount
    & \boldnum{\PonoImcBzlaReachSafetySampledBvFuncStatusCorrectTrueCount}
    & \fpeval{(\PonoImcBzlaReachSafetySampledBvFuncStatusCorrectTrueCount - \PonoImcBzlaReachSafetySampledBvRelStatusCorrectTrueCount) / \PonoImcBzlaReachSafetySampledBvRelStatusCorrectTrueCount * 100}
    & \PonoImcBzlaTerminationSampledBvRelStatusCorrectTrueCount
    & \boldnum{\PonoImcBzlaTerminationSampledBvFuncStatusCorrectTrueCount}
    & \fpeval{(\PonoImcBzlaTerminationSampledBvFuncStatusCorrectTrueCount - \PonoImcBzlaTerminationSampledBvRelStatusCorrectTrueCount) / \PonoImcBzlaTerminationSampledBvRelStatusCorrectTrueCount * 100}
    \\
    & 
    & \quad alarms
    & \PonoImcBzlaReachSafetySampledBvRelStatusCorrectFalseCount
    & \boldnum{\PonoImcBzlaReachSafetySampledBvFuncStatusCorrectFalseCount}
    & \fpeval{(\PonoImcBzlaReachSafetySampledBvFuncStatusCorrectFalseCount - \PonoImcBzlaReachSafetySampledBvRelStatusCorrectFalseCount) / \PonoImcBzlaReachSafetySampledBvRelStatusCorrectFalseCount * 100}
    & \boldnum{\PonoImcBzlaTerminationSampledBvRelStatusCorrectFalseCount}
    & \PonoImcBzlaTerminationSampledBvFuncStatusCorrectFalseCount
    & \fpeval{(\PonoImcBzlaTerminationSampledBvFuncStatusCorrectFalseCount - \PonoImcBzlaTerminationSampledBvRelStatusCorrectFalseCount) / \PonoImcBzlaTerminationSampledBvRelStatusCorrectFalseCount * 100}
    \\
\bottomrule
\end{tabular}

%% file: tables/encoding.arr-bv.tex
\begin{tabular}{cclS[table-format=5]S[table-format=5]S[round-mode=figures, round-precision=2]S[table-format=5]S[table-format=5]S[round-mode=figures, round-precision=2]}
\toprule
\multicolumn{2}{c}{HWMC}
    & \multirow{2}{*}{Results}
    & \multicolumn{3}{c}{ReachSafety (\num{\AvrKindReachSafetySampledFuncStatusAllCount})}
    & \multicolumn{3}{c}{Termination (\num{\AvrKindTerminationSampledFuncStatusAllCount})}
    \\
\cmidrule(lr){1-2}\cmidrule(lr){4-6}\cmidrule(lr){7-9}
Tool
    & Config.
    &
    & \multicolumn{1}{c}{Relational}
    & \multicolumn{1}{c}{Functional}
    & \multicolumn{1}{c}{$\Delta$\,(\%)}
    & \multicolumn{1}{c}{Relational}
    & \multicolumn{1}{c}{Functional}
    & \multicolumn{1}{c}{$\Delta$\,(\%)}
    \\
\midrule
\multirow{6}{*}{\avr}
    & \multirow{3}{*}{IC3SA}
    & \#Solved
    & \boldnum{\AvrIcIIIsaReachSafetySampledRelStatusCorrectCount}
    & \AvrIcIIIsaReachSafetySampledFuncStatusCorrectCount
    & \fpeval{(\AvrIcIIIsaReachSafetySampledFuncStatusCorrectCount - \AvrIcIIIsaReachSafetySampledRelStatusCorrectCount) / \AvrIcIIIsaReachSafetySampledRelStatusCorrectCount * 100}
    & \AvrIcIIIsaTerminationSampledRelStatusCorrectCount
    & \boldnum{\AvrIcIIIsaTerminationSampledFuncStatusCorrectCount}
    & \fpeval{(\AvrIcIIIsaTerminationSampledFuncStatusCorrectCount - \AvrIcIIIsaTerminationSampledRelStatusCorrectCount) / \AvrIcIIIsaTerminationSampledRelStatusCorrectCount * 100}
    \\
    & 
    & \quad proofs
    & \AvrIcIIIsaReachSafetySampledRelStatusCorrectTrueCount
    & \boldnum{\AvrIcIIIsaReachSafetySampledFuncStatusCorrectTrueCount}
    & \fpeval{(\AvrIcIIIsaReachSafetySampledFuncStatusCorrectTrueCount - \AvrIcIIIsaReachSafetySampledRelStatusCorrectTrueCount) / \AvrIcIIIsaReachSafetySampledRelStatusCorrectTrueCount * 100}
    & \AvrIcIIIsaTerminationSampledRelStatusCorrectTrueCount
    & \boldnum{\AvrIcIIIsaTerminationSampledFuncStatusCorrectTrueCount}
    & \fpeval{(\AvrIcIIIsaTerminationSampledFuncStatusCorrectTrueCount - \AvrIcIIIsaTerminationSampledRelStatusCorrectTrueCount) / \AvrIcIIIsaTerminationSampledRelStatusCorrectTrueCount * 100}
    \\
    & 
    & \quad alarms
    & \boldnum{\AvrIcIIIsaReachSafetySampledRelStatusCorrectFalseCount}
    & \AvrIcIIIsaReachSafetySampledFuncStatusCorrectFalseCount
    & \fpeval{(\AvrIcIIIsaReachSafetySampledFuncStatusCorrectFalseCount - \AvrIcIIIsaReachSafetySampledRelStatusCorrectFalseCount) / \AvrIcIIIsaReachSafetySampledRelStatusCorrectFalseCount * 100}
    & \AvrIcIIIsaTerminationSampledRelStatusCorrectFalseCount
    & \boldnum{\AvrIcIIIsaTerminationSampledFuncStatusCorrectFalseCount}
    & \fpeval{(\AvrIcIIIsaTerminationSampledFuncStatusCorrectFalseCount - \AvrIcIIIsaTerminationSampledRelStatusCorrectFalseCount) / \AvrIcIIIsaTerminationSampledRelStatusCorrectFalseCount * 100}
    \\
\cmidrule(lr){2-9}
    & \multirow{3}{*}{\kinduction}
    & \#Solved
    & \AvrKindReachSafetySampledRelStatusCorrectCount
    & \boldnum{\AvrKindReachSafetySampledFuncStatusCorrectCount}
    & \fpeval{(\AvrKindReachSafetySampledFuncStatusCorrectCount - \AvrKindReachSafetySampledRelStatusCorrectCount) / \AvrKindReachSafetySampledRelStatusCorrectCount * 100}
    & \AvrKindTerminationSampledRelStatusCorrectCount
    & \boldnum{\AvrKindTerminationSampledFuncStatusCorrectCount}
    & \fpeval{(\AvrKindTerminationSampledFuncStatusCorrectCount - \AvrKindTerminationSampledRelStatusCorrectCount) / \AvrKindTerminationSampledRelStatusCorrectCount * 100}
    \\
    & 
    & \quad proofs
    & \AvrKindReachSafetySampledRelStatusCorrectTrueCount
    & \boldnum{\AvrKindReachSafetySampledFuncStatusCorrectTrueCount}
    & \fpeval{(\AvrKindReachSafetySampledFuncStatusCorrectTrueCount - \AvrKindReachSafetySampledRelStatusCorrectTrueCount) / \AvrKindReachSafetySampledRelStatusCorrectTrueCount * 100}
    & \AvrKindTerminationSampledRelStatusCorrectTrueCount
    & \boldnum{\AvrKindTerminationSampledFuncStatusCorrectTrueCount}
    & \fpeval{(\AvrKindTerminationSampledFuncStatusCorrectTrueCount - \AvrKindTerminationSampledRelStatusCorrectTrueCount) / \AvrKindTerminationSampledRelStatusCorrectTrueCount * 100}
    \\
    & 
    & \quad alarms
    & \AvrKindReachSafetySampledRelStatusCorrectFalseCount
    & \boldnum{\AvrKindReachSafetySampledFuncStatusCorrectFalseCount}
    & \fpeval{(\AvrKindReachSafetySampledFuncStatusCorrectFalseCount - \AvrKindReachSafetySampledRelStatusCorrectFalseCount) / \AvrKindReachSafetySampledRelStatusCorrectFalseCount * 100}
    & \AvrKindTerminationSampledRelStatusCorrectFalseCount
    & \boldnum{\AvrKindTerminationSampledFuncStatusCorrectFalseCount}
    & \fpeval{(\AvrKindTerminationSampledFuncStatusCorrectFalseCount - \AvrKindTerminationSampledRelStatusCorrectFalseCount) / \AvrKindTerminationSampledRelStatusCorrectFalseCount * 100}
    \\
\cmidrule{1-9}
\multirow{6}{*}{\pono}
    & \multirow{3}{*}{IC3IA}
    & \#Solved
    & \PonoIcIIIiaMsatReachSafetySampledRelStatusCorrectCount
    & \boldnum{\PonoIcIIIiaMsatReachSafetySampledFuncStatusCorrectCount}
    & \fpeval{(\PonoIcIIIiaMsatReachSafetySampledFuncStatusCorrectCount - \PonoIcIIIiaMsatReachSafetySampledRelStatusCorrectCount) / \PonoIcIIIiaMsatReachSafetySampledRelStatusCorrectCount * 100}
    & \PonoIcIIIiaMsatTerminationSampledRelStatusCorrectCount
    & \boldnum{\PonoIcIIIiaMsatTerminationSampledFuncStatusCorrectCount}
    & \fpeval{(\PonoIcIIIiaMsatTerminationSampledFuncStatusCorrectCount - \PonoIcIIIiaMsatTerminationSampledRelStatusCorrectCount) / \PonoIcIIIiaMsatTerminationSampledRelStatusCorrectCount * 100}
    \\
    & 
    & \quad proofs
    & \PonoIcIIIiaMsatReachSafetySampledRelStatusCorrectTrueCount
    & \boldnum{\PonoIcIIIiaMsatReachSafetySampledFuncStatusCorrectTrueCount}
    & \fpeval{(\PonoIcIIIiaMsatReachSafetySampledFuncStatusCorrectTrueCount - \PonoIcIIIiaMsatReachSafetySampledRelStatusCorrectTrueCount) / \PonoIcIIIiaMsatReachSafetySampledRelStatusCorrectTrueCount * 100}
    & \boldnum{\PonoIcIIIiaMsatTerminationSampledRelStatusCorrectTrueCount}
    & \PonoIcIIIiaMsatTerminationSampledFuncStatusCorrectTrueCount
    & \fpeval{(\PonoIcIIIiaMsatTerminationSampledFuncStatusCorrectTrueCount - \PonoIcIIIiaMsatTerminationSampledRelStatusCorrectTrueCount) / \PonoIcIIIiaMsatTerminationSampledRelStatusCorrectTrueCount * 100}
    \\
    & 
    & \quad alarms
    & \PonoIcIIIiaMsatReachSafetySampledRelStatusCorrectFalseCount
    & \boldnum{\PonoIcIIIiaMsatReachSafetySampledFuncStatusCorrectFalseCount}
    & \fpeval{(\PonoIcIIIiaMsatReachSafetySampledFuncStatusCorrectFalseCount - \PonoIcIIIiaMsatReachSafetySampledRelStatusCorrectFalseCount) / \PonoIcIIIiaMsatReachSafetySampledRelStatusCorrectFalseCount * 100}
    & \PonoIcIIIiaMsatTerminationSampledRelStatusCorrectFalseCount
    & \boldnum{\PonoIcIIIiaMsatTerminationSampledFuncStatusCorrectFalseCount}
    & \fpeval{(\PonoIcIIIiaMsatTerminationSampledFuncStatusCorrectFalseCount - \PonoIcIIIiaMsatTerminationSampledRelStatusCorrectFalseCount) / \PonoIcIIIiaMsatTerminationSampledRelStatusCorrectFalseCount * 100}
    \\
\cmidrule(lr){2-9}
    & \multirow{3}{*}{\kinduction}
    & \#Solved
    & \PonoKindReachSafetySampledRelStatusCorrectCount
    & \boldnum{\PonoKindReachSafetySampledFuncStatusCorrectCount}
    & \fpeval{(\PonoKindReachSafetySampledFuncStatusCorrectCount - \PonoKindReachSafetySampledRelStatusCorrectCount) / \PonoKindReachSafetySampledRelStatusCorrectCount * 100}
    & \boldnum{\PonoKindTerminationSampledRelStatusCorrectCount}
    & \PonoKindTerminationSampledFuncStatusCorrectCount
    & \fpeval{(\PonoKindTerminationSampledFuncStatusCorrectCount - \PonoKindTerminationSampledRelStatusCorrectCount) / \PonoKindTerminationSampledRelStatusCorrectCount * 100}
    \\
    & 
    & \quad proofs
    & \PonoKindReachSafetySampledRelStatusCorrectTrueCount
    & \boldnum{\PonoKindReachSafetySampledFuncStatusCorrectTrueCount}
    & \fpeval{(\PonoKindReachSafetySampledFuncStatusCorrectTrueCount - \PonoKindReachSafetySampledRelStatusCorrectTrueCount) / \PonoKindReachSafetySampledRelStatusCorrectTrueCount * 100}
    & \PonoKindTerminationSampledRelStatusCorrectTrueCount
    & \boldnum{\PonoKindTerminationSampledFuncStatusCorrectTrueCount}
    & \fpeval{(\PonoKindTerminationSampledFuncStatusCorrectTrueCount - \PonoKindTerminationSampledRelStatusCorrectTrueCount) / \PonoKindTerminationSampledRelStatusCorrectTrueCount * 100}
    \\
    & 
    & \quad alarms
    & \PonoKindReachSafetySampledRelStatusCorrectFalseCount
    & \boldnum{\PonoKindReachSafetySampledFuncStatusCorrectFalseCount}
    & \fpeval{(\PonoKindReachSafetySampledFuncStatusCorrectFalseCount - \PonoKindReachSafetySampledRelStatusCorrectFalseCount) / \PonoKindReachSafetySampledRelStatusCorrectFalseCount * 100}
    & \boldnum{\PonoKindTerminationSampledRelStatusCorrectFalseCount}
    & \PonoKindTerminationSampledFuncStatusCorrectFalseCount
    & \fpeval{(\PonoKindTerminationSampledFuncStatusCorrectFalseCount - \PonoKindTerminationSampledRelStatusCorrectFalseCount) / \PonoKindTerminationSampledRelStatusCorrectFalseCount * 100}
    \\
\bottomrule
\end{tabular}

%% file: plots/abc.svcomp-sampled.cputime.quantile.tex
\tikzsetnextfilename{abc.svcomp-sampled.cputime.quantile}
\begin{tikzpicture}
\tikzpicturedependsonfile{plots/plot-defs.tex}
\begin{semilogyaxis}[
    quantile plot,
    /pgfplots/table/y index=5,
    ymin=0.01,
    xmax=2000,
    mark repeat=300,
    width=.46\linewidth,
    height=.3\linewidth,
    ]
    \addplot [mark=o, mark options={solid}, dotted, color=green] table {\detokenize{eval-results/csv/abc.imc.rel.svcomp.sampled.cputime.quantile.csv}};
    \addlegendentry{IMC $\cdot$ rel.}
    \addplot [mark=*, mark options={solid}, solid, color=green] table {\detokenize{eval-results/csv/abc.imc.func.svcomp.sampled.cputime.quantile.csv}};
    \addlegendentry{IMC $\cdot$ func.}
    \addplot [mark=square, mark options={solid}, dotted, color=blue!50!cyan] table {\detokenize{eval-results/csv/abc.pdr.rel.svcomp.sampled.cputime.quantile.csv}};
    \addlegendentry{PDR $\cdot$ rel.}
    \addplot [mark=square*, mark options={solid}, solid, color=blue!50!cyan] table {\detokenize{eval-results/csv/abc.pdr.func.svcomp.sampled.cputime.quantile.csv}};
    \addlegendentry{PDR $\cdot$ func.}
    \addplot [mark=triangle, mark options={solid}, dotted, color=brown] table {\detokenize{eval-results/csv/abc.sc-pdr.rel.svcomp.sampled.cputime.quantile.csv}};
    \addlegendentry{scPDR $\cdot$ rel.}
    \addplot [mark=triangle*, mark options={solid}, solid, color=brown] table {\detokenize{eval-results/csv/abc.sc-pdr.func.svcomp.sampled.cputime.quantile.csv}};
    \addlegendentry{scPDR $\cdot$ func.}
\end{semilogyaxis}
\end{tikzpicture}

%% file: plots/avr.svcomp-sampled.cputime.quantile.tex
\tikzsetnextfilename{avr.svcomp-sampled.cputime.quantile}
\begin{tikzpicture}
\tikzpicturedependsonfile{plots/plot-defs.tex}
\begin{semilogyaxis}[
    quantile plot,
    /pgfplots/table/y index=5,
    ymin=0.01,
    xmax=2000,
    mark repeat=300,
    width=.46\linewidth,
    height=.3\linewidth,
    ]
    \addplot [mark=o, mark options={solid}, dotted, color=green] table {\detokenize{eval-results/csv/avr.ic3sa.rel.svcomp.sampled.cputime.quantile.csv}};
    \addlegendentry{IC3SA $\cdot$ rel.}
    \addplot [mark=*, mark options={solid}, solid, color=green] table {\detokenize{eval-results/csv/avr.ic3sa.func.svcomp.sampled.cputime.quantile.csv}};
    \addlegendentry{IC3SA $\cdot$ func.}
    \addplot [mark=square, mark options={solid}, dotted, color=blue!50!cyan] table {\detokenize{eval-results/csv/avr.kind.rel.svcomp.sampled.cputime.quantile.csv}};
    \addlegendentry{$k$-ind. $\cdot$ rel.}
    \addplot [mark=square*, mark options={solid}, solid, color=blue!50!cyan] table {\detokenize{eval-results/csv/avr.kind.func.svcomp.sampled.cputime.quantile.csv}};
    \addlegendentry{$k$-ind. $\cdot$ func.}
\end{semilogyaxis}
\end{tikzpicture}

%% file: plots/pono.svcomp-sampled.cputime.quantile.tex
\tikzsetnextfilename{pono.svcomp-sampled.cputime.quantile}
\begin{tikzpicture}
\tikzpicturedependsonfile{plots/plot-defs.tex}
\begin{semilogyaxis}[
    quantile plot,
    /pgfplots/table/y index=5,
    ymin=0.005,
    xmax=2000,
    mark repeat=300,
    width=.46\linewidth,
    height=.3\linewidth,
    ]
    \addplot [mark=o, mark options={solid}, dotted, color=green] table {\detokenize{eval-results/csv/pono.ic3ia-msat.rel.svcomp.sampled.cputime.quantile.csv}};
    \addlegendentry{IC3IA $\cdot$ rel.}
    \addplot [mark=*, mark options={solid}, solid, color=green] table {\detokenize{eval-results/csv/pono.ic3ia-msat.func.svcomp.sampled.cputime.quantile.csv}};
    \addlegendentry{IC3IA $\cdot$ func.}
    \addplot [mark=square, mark options={solid}, dotted, color=blue!50!cyan] table {\detokenize{eval-results/csv/pono.imc-bzla.rel.svcomp.sampled.cputime.quantile.csv}};
    \addlegendentry{IMC $\cdot$ rel.}
    \addplot [mark=square*, mark options={solid}, solid, color=blue!50!cyan] table {\detokenize{eval-results/csv/pono.imc-bzla.func.svcomp.sampled.cputime.quantile.csv}};
    \addlegendentry{IMC. $\cdot$ func.}
    \addplot [mark=triangle, mark options={solid}, dotted, color=brown] table {\detokenize{eval-results/csv/pono.kind.rel.svcomp.sampled.cputime.quantile.csv}};
    \addlegendentry{$k$-ind. $\cdot$ rel.}
    \addplot [mark=triangle*, mark options={solid}, solid, color=brown] table {\detokenize{eval-results/csv/pono.kind.func.svcomp.sampled.cputime.quantile.csv}};
    \addlegendentry{$k$-ind. $\cdot$ func.}
\end{semilogyaxis}
\end{tikzpicture}

%% file: plots/ric3.svcomp-sampled.cputime.quantile.tex
\tikzsetnextfilename{ric3.svcomp-sampled.cputime.quantile}
\begin{tikzpicture}
\tikzpicturedependsonfile{plots/plot-defs.tex}
\begin{semilogyaxis}[
    quantile plot,
    /pgfplots/table/y index=5,
    ymin=0.003,
    xmax=2000,
    mark repeat=300,
    width=.46\linewidth,
    height=.3\linewidth,
    ]
    \addplot [mark=o, mark options={solid}, dotted, color=green] table {\detokenize{eval-results/csv/ric3.ic3.rel.svcomp.sampled.cputime.quantile.csv}};
    \addlegendentry{IC3 $\cdot$ rel.}
    \addplot [mark=*, mark options={solid}, solid, color=green] table {\detokenize{eval-results/csv/ric3.ic3.func.svcomp.sampled.cputime.quantile.csv}};
    \addlegendentry{IC3 $\cdot$ func.}
    \addplot [mark=square, mark options={solid}, dotted, color=blue!50!cyan] table {\detokenize{eval-results/csv/ric3.kind.rel.svcomp.sampled.cputime.quantile.csv}};
    \addlegendentry{$k$-ind. $\cdot$ rel.}
    \addplot [mark=square*, mark options={solid}, solid, color=blue!50!cyan] table {\detokenize{eval-results/csv/ric3.kind.func.svcomp.sampled.cputime.quantile.csv}};
    \addlegendentry{$k$-ind. $\cdot$ func.}
\end{semilogyaxis}
\end{tikzpicture}

%% file: plots/encoding.gates.scatter.tex
\tikzsetnextfilename{encoding.gates.scatter}
\begin{tikzpicture}
\begin{loglogaxis}[
    width=.33\textwidth,
    height=.33\textwidth,
    xmin=5,
    xmax=10000000,
    ymin=5,
    ymax=10000000,
    domain=5:10000000,
    tick label style={font=\smaller},
    clip mode=individual,
    axis equal image,
    legend pos=north west,
    legend style={font=\scriptsize, fill=none,},
    legend cell align={left},
    mark size=1.8pt,
    ]
    \addplot+[green, mark=o, only marks, opacity=0.5]
         table[
             header=false,
             skip first n=3, 
             x index=2, 
             y index=3, 
             ] {\detokenize{eval-results/csv/abc.pdr.gates.ReachSafety.rel-vs-func.table.csv}};
    \addlegendentry{ReachSafety}
    \addplot+[blue!50!cyan, mark=x, only marks, opacity=0.5]
         table[
             header=false,
             skip first n=3, 
             x index=2, 
             y index=3, 
             ] {\detokenize{eval-results/csv/abc.pdr.gates.Termination.rel-vs-func.table.csv}};
    \addlegendentry{Termination}
    \addplot[gray] {x};
    \addplot[gray] {10*x};
    \addplot[gray] {x/10};
\end{loglogaxis}
\end{tikzpicture}

%% file: plots/encoding.imc.cputime.scatter.tex
\tikzsetnextfilename{encoding.imc.cputime.scatter}
\begin{tikzpicture}
\begin{loglogaxis}[
    width=.33\textwidth,
    height=.33\textwidth,
    xmin=0.01,
    xmax=1000,
    ymin=0.01,
    ymax=1000,
    domain=0.01:1000,
    tick label style={font=\smaller},
    clip mode=individual,
    axis equal image,
    legend pos=north west,
    legend style={font=\scriptsize, fill=none,},
    legend cell align={left},
    mark size=1.8pt,
    ]
    \addplot+[green, mark=o, only marks, opacity=0.5]
         table[
             header=false,
             skip first n=3, 
             x index=2, 
             y index=3, 
             ] {\detokenize{eval-results/csv/abc.imc.cputime.ReachSafety.rel-vs-func.table.csv}};
    \addlegendentry{ReachSafety}
    \addplot+[blue!50!cyan, mark=x, only marks, opacity=0.5]
         table[
             header=false,
             skip first n=3, 
             x index=2, 
             y index=3, 
             ] {\detokenize{eval-results/csv/abc.imc.cputime.Termination.rel-vs-func.table.csv}};
    \addlegendentry{Termination}
    \addplot[gray] {x};
    \addplot[gray] {10*x};
    \addplot[gray] {x/10};
\end{loglogaxis}
\end{tikzpicture}

%% file: plots/encoding.pdr.cputime.scatter.tex
\tikzsetnextfilename{encoding.pdr.cputime.scatter}
\begin{tikzpicture}
\begin{loglogaxis}[
    width=.33\textwidth,
    height=.33\textwidth,
    xmin=0.01,
    xmax=1000,
    ymin=0.01,
    ymax=1000,
    domain=0.01:1000,
    tick label style={font=\smaller},
    clip mode=individual,
    axis equal image,
    legend pos=north west,
    legend style={font=\scriptsize, fill=none,},
    legend cell align={left},
    mark size=1.8pt,
    ]
    \addplot+[green, mark=o, only marks, opacity=0.5]
         table[
             header=false,
             skip first n=3, 
             x index=2, 
             y index=3, 
             ] {\detokenize{eval-results/csv/abc.pdr.cputime.ReachSafety.rel-vs-func.table.csv}};
    \addlegendentry{ReachSafety}
    \addplot+[blue!50!cyan, mark=x, only marks, opacity=0.5]
         table[
             header=false,
             skip first n=3, 
             x index=2, 
             y index=3, 
             ] {\detokenize{eval-results/csv/abc.pdr.cputime.Termination.rel-vs-func.table.csv}};
    \addlegendentry{Termination}
    \addplot[gray] {x};
    \addplot[gray] {10*x};
    \addplot[gray] {x/10};
\end{loglogaxis}
\end{tikzpicture}

%% file: tables/encoding.gates.reach.tex
\begin{tabular}{lS[table-format=5]S[table-format=5]S[round-mode=figures, round-precision=3]S[table-format=5]S[round-mode=figures, round-precision=3]S[table-format=5]}
\toprule
\multirow{2}{*}{Encoding}
    & \multicolumn{2}{c}{Circuit size (\num{\AbcPdrGateReachSafetySampledBvSignTestAllCount} tasks)}
    & \multicolumn{2}{c}{IMC (\num{\AbcImcCputimeReachSafetySampledBvSignTestAllCount} comm. solved)}
    & \multicolumn{2}{c}{PDR (\num{\AbcPdrCputimeReachSafetySampledBvSignTestAllCount} comm. solved)}
    \\
\cmidrule(lr){2-3}\cmidrule(lr){4-5}\cmidrule(lr){6-7}
    & \multicolumn{1}{c}{Med. \#gate}
    & \multicolumn{1}{c}{\#smaller}
    & \multicolumn{1}{c}{Med. time}
    & \multicolumn{1}{c}{\#faster}
    & \multicolumn{1}{c}{Med. time}
    & \multicolumn{1}{c}{\#faster}
    \\
\midrule
Relational
    & \AbcPdrReachSafetySampledBvRelGateAllMedian
    & \AbcPdrGateReachSafetySampledBvSignTestMinusCount
    & \textbf{\tablenum[round-mode=figures, round-precision=3]{\AbcImcCputimeReachSafetySampledBvSignTestGroupOneMedian}}
    & \AbcImcCputimeReachSafetySampledBvSignTestMinusCount
    & \AbcPdrCputimeReachSafetySampledBvSignTestGroupOneMedian
    & \AbcPdrCputimeReachSafetySampledBvSignTestMinusCount
    \\
Functional
    & \boldnum{\AbcPdrReachSafetySampledBvFuncGateAllMedian}
    & \boldnum{\AbcPdrGateReachSafetySampledBvSignTestPlusCount}
    & \AbcImcCputimeReachSafetySampledBvSignTestGroupTwoMedian
    & \boldnum{\AbcImcCputimeReachSafetySampledBvSignTestPlusCount}
    & \textbf{\tablenum[round-mode=figures, round-precision=3]{\AbcPdrCputimeReachSafetySampledBvSignTestGroupTwoMedian}}
    & \boldnum{\AbcPdrCputimeReachSafetySampledBvSignTestPlusCount}
    \\
\bottomrule
\end{tabular}

%% file: tables/encoding.gates.term.tex
\begin{tabular}{lS[table-format=5]S[table-format=5]S[round-mode=figures, round-precision=3]S[table-format=5]S[round-mode=figures, round-precision=3]S[table-format=5]}
\toprule
\multirow{2}{*}{Encoding}
    & \multicolumn{2}{c}{Circuit size (\num{\AbcPdrGateTerminationSampledBvSignTestAllCount} tasks)}
    & \multicolumn{2}{c}{IMC (\num{\AbcImcCputimeTerminationSampledBvSignTestAllCount} comm. solved)}
    & \multicolumn{2}{c}{PDR (\num{\AbcPdrCputimeTerminationSampledBvSignTestAllCount} comm. solved)}
    \\
\cmidrule(lr){2-3}\cmidrule(lr){4-5}\cmidrule(lr){6-7}
    & \multicolumn{1}{c}{Med. \#gate}
    & \multicolumn{1}{c}{\#smaller}
    & \multicolumn{1}{c}{Med. time}
    & \multicolumn{1}{c}{\#faster}
    & \multicolumn{1}{c}{Med. time}
    & \multicolumn{1}{c}{\#faster}
    \\
\midrule
Relational
    & \AbcPdrTerminationSampledBvRelGateAllMedian
    & \AbcPdrGateTerminationSampledBvSignTestMinusCount
    & \AbcImcCputimeTerminationSampledBvSignTestGroupOneMedian
    & \AbcImcCputimeTerminationSampledBvSignTestMinusCount
    & \AbcPdrCputimeTerminationSampledBvSignTestGroupOneMedian
    & \AbcPdrCputimeTerminationSampledBvSignTestMinusCount
    \\
Functional
    & \boldnum{\AbcPdrTerminationSampledBvFuncGateAllMedian}
    & \boldnum{\AbcPdrGateTerminationSampledBvSignTestPlusCount}
    & \textbf{\tablenum[round-mode=figures, round-precision=3]{\AbcImcCputimeTerminationSampledBvSignTestGroupTwoMedian}}
    & \boldnum{\AbcImcCputimeTerminationSampledBvSignTestPlusCount}
    & \textbf{\tablenum[round-mode=figures, round-precision=3]{\AbcPdrCputimeTerminationSampledBvSignTestGroupTwoMedian}}
    & \boldnum{\AbcPdrCputimeTerminationSampledBvSignTestPlusCount}
    \\
\bottomrule
\end{tabular}

%% file: plots/scorr-gate.reach.bar.tex
\tikzsetnextfilename{scorr-gate.reach.bar}
\begin{tikzpicture}
\begin{axis}[
    ybar,
    height=.33\linewidth,
    bar width=15pt,
    symbolic x coords={Relational, Functional},
    xtick=data,
    x=2cm, 
    enlarge x limits=0.5, 
    ymin=0,
    ymax=12000,
    ticklabel style={font={\smaller}},
    legend style={
        at={(1, 1)},
        anchor=north east,
        legend columns=1,
        align=right,
        cells={anchor=east},
    },
    legend to name=legend:scorr-gate-reach,
    legend image code/.code={
        \draw[#1] (0cm,-0.14cm) rectangle (0.3cm,0.16cm);
    },
    nodes near coords,
    nodes near coords style={font=\smaller},
    point meta=explicit symbolic,
]

\addplot+[
    color=blue!50!cyan,
] coordinates {
    (Relational, \AbcPdrScPdrReachSafetyRelCommonGateAllMedian) [\num{\AbcPdrScPdrReachSafetyRelCommonGateAllMedian}]
    (Functional, \AbcPdrScPdrReachSafetyFuncCommonGateAllMedian) [\num{\AbcPdrScPdrReachSafetyFuncCommonGateAllMedian}]
};

\addplot+[
    fill=green,
    draw=green,
    pattern=north east lines,
    pattern color=green,
    text=green,
] coordinates {
    (Relational, \AbcScPdrScPdrReachSafetyRelCommonGateAllMedian) [\num{\AbcScPdrScPdrReachSafetyRelCommonGateAllMedian}]
    (Functional, \AbcScPdrScPdrReachSafetyFuncCommonGateAllMedian) [\num{\AbcScPdrScPdrReachSafetyFuncCommonGateAllMedian}]
};

\legend{
    w/o \texttt{scorr},
    w/\phantom{o} \texttt{scorr},
}

\end{axis}
\end{tikzpicture}

%% file: plots/scorr-gate.term.bar.tex
\tikzsetnextfilename{scorr-gate.term.bar}
\begin{tikzpicture}
\begin{axis}[
    ybar,
    height=.33\linewidth,
    bar width=15pt,
    symbolic x coords={Relational, Functional},
    xtick=data,
    x=2cm, 
    enlarge x limits=0.5, 
    ymin=0,
    ymax=12000,
    ticklabel style={font={\smaller}},
    legend style={
        at={(1, 1)},
        anchor=north east,
        legend columns=1,
        align=right,
        cells={anchor=east},
    },
    legend image code/.code={
        \draw[#1] (0cm,-0.14cm) rectangle (0.3cm,0.16cm);
    },
    nodes near coords,
    nodes near coords style={font=\smaller},
    point meta=explicit symbolic,
]

\addplot+[
    color=blue!50!cyan,
] coordinates {
    (Relational, \AbcPdrScPdrTerminationRelCommonGateAllMedian) [\num{\AbcPdrScPdrTerminationRelCommonGateAllMedian}]
    (Functional, \AbcPdrScPdrTerminationFuncCommonGateAllMedian) [\num{\AbcPdrScPdrTerminationFuncCommonGateAllMedian}]
};

\addplot+[
    fill=green,
    draw=green,
    pattern=north east lines,
    pattern color=green,
    text=green,
] coordinates {
    (Relational, \AbcScPdrScPdrTerminationRelCommonGateAllMedian) [\num{\AbcScPdrScPdrTerminationRelCommonGateAllMedian}]
    (Functional, \AbcScPdrScPdrTerminationFuncCommonGateAllMedian) [\num{\AbcScPdrScPdrTerminationFuncCommonGateAllMedian}]
};


\end{axis}
\end{tikzpicture}

%% file: plots/scorr-time.reach.bar.tex
\tikzsetnextfilename{scorr-time.reach.bar}
\begin{tikzpicture}
\begin{axis}[
    ybar,
    height=.33\linewidth,
    bar width=17pt,
    symbolic x coords={Relational, Functional},
    xtick=data,
    x=2.4cm, 
    enlarge x limits=0.58, 
    ymin=0,
    ymax=3.85,
    ticklabel style={font={\smaller}},
    legend style={
        at={(1, 1)},
        anchor=north east,
        legend columns=1,
        align=right,
        cells={anchor=east},
    },
    legend to name=legend:scorr-time-reach,
    legend image code/.code={
        \draw[#1] (0cm,-0.14cm) rectangle (0.3cm,0.16cm);
    },
    nodes near coords,
    nodes near coords style={font=\smaller},
    point meta=explicit symbolic,
]

\addplot+[
    color=blue!50!cyan,
] coordinates {
    (Relational, \AbcPdrScPdrReachSafetyRelCommonVerifTimeAllMedian) [\num{\AbcPdrScPdrReachSafetyRelCommonVerifTimeAllMedian}]
    (Functional, \AbcPdrScPdrReachSafetyFuncCommonVerifTimeAllMedian) [\num{\AbcPdrScPdrReachSafetyFuncCommonVerifTimeAllMedian}]
};

\addplot+[
    fill=green,
    draw=green,
    pattern=north east lines,
    pattern color=green,
    text=green,
] coordinates {
    (Relational, \AbcScPdrScPdrReachSafetyRelCommonSolveTimeAllMedian) [\num{\AbcScPdrScPdrReachSafetyRelCommonSolveTimeAllMedian}]
    (Functional, \AbcScPdrScPdrReachSafetyFuncCommonSolveTimeAllMedian) [\num{\AbcScPdrScPdrReachSafetyFuncCommonSolveTimeAllMedian}]
};

\addplot+[
    fill=olive,
    draw=olive,
    pattern=dots,
    pattern color=olive,
    text=olive,
] coordinates {
    (Relational, \AbcScPdrScPdrReachSafetyRelCommonVerifTimeAllMedian) [\num{\AbcScPdrScPdrReachSafetyRelCommonVerifTimeAllMedian}]
    (Functional, \AbcScPdrScPdrReachSafetyFuncCommonVerifTimeAllMedian) [\num{\AbcScPdrScPdrReachSafetyFuncCommonVerifTimeAllMedian}]
};

\legend{
    PDR,
    \texttt{scorr}+\phantom{)}PDR,
    \textcolor{gray}{(\texttt{scorr}+)}PDR
}

\end{axis}
\end{tikzpicture}

%% file: plots/scorr-time.term.bar.tex
\tikzsetnextfilename{scorr-time.term.bar}
\begin{tikzpicture}
\begin{axis}[
    ybar,
    height=.33\linewidth,
    bar width=17pt,
    symbolic x coords={Relational, Functional},
    xtick=data,
    x=2.4cm, 
    enlarge x limits=0.58, 
    ymin=0,
    ymax=3.85,
    ticklabel style={font={\smaller}},
    legend style={
        at={(1, 1)},
        anchor=north east,
        legend columns=1,
        align=right,
        cells={anchor=east},
    },
    legend image code/.code={
        \draw[#1] (0cm,-0.14cm) rectangle (0.3cm,0.16cm);
    },
    nodes near coords,
    nodes near coords style={font=\smaller},
    point meta=explicit symbolic,
]

\addplot+[
    color=blue!50!cyan,
] coordinates {
    (Relational, \AbcPdrScPdrTerminationRelCommonVerifTimeAllMedian) [\num{\AbcPdrScPdrTerminationRelCommonVerifTimeAllMedian}]
    (Functional, \AbcPdrScPdrTerminationFuncCommonVerifTimeAllMedian) [\num{\AbcPdrScPdrTerminationFuncCommonVerifTimeAllMedian}]
};

\addplot+[
    fill=green,
    draw=green,
    pattern=north east lines,
    pattern color=green,
    text=green,
] coordinates {
    (Relational, \AbcScPdrScPdrTerminationRelCommonSolveTimeAllMedian) [\num{\AbcScPdrScPdrTerminationRelCommonSolveTimeAllMedian}]
    (Functional, \AbcScPdrScPdrTerminationFuncCommonSolveTimeAllMedian) [\num{\AbcScPdrScPdrTerminationFuncCommonSolveTimeAllMedian}]
};

\addplot+[
    fill=olive,
    draw=olive,
    pattern=dots,
    pattern color=olive,
    text=olive,
] coordinates {
    (Relational, \AbcScPdrScPdrTerminationRelCommonVerifTimeAllMedian) [\num{\AbcScPdrScPdrTerminationRelCommonVerifTimeAllMedian}]
    (Functional, \AbcScPdrScPdrTerminationFuncCommonVerifTimeAllMedian) [\num{\AbcScPdrScPdrTerminationFuncCommonVerifTimeAllMedian}]
};


\end{axis}
\end{tikzpicture}

%% file: tables/scorr.reduct.tex
\begin{tabular}{clrS[round-mode=figures, round-precision=3]S[round-mode=figures, round-precision=3]S[round-mode=figures, round-precision=3]S[table-format=5.1]S[table-format=5.1, round-mode=figures, round-precision=5]S[round-mode=figures, round-precision=3]}
\toprule
\multirow{2}{*}{\shortstack{\\Category\\(\#tasks)}}
    & \multirow{2}{*}{Enc.}
    & \multirow{2}{*}{Config.}
    & \multicolumn{3}{c}{Run time (s)}
    & \multicolumn{3}{c}{\#Gates}
    \\
\cmidrule(lr){4-6}\cmidrule(lr){7-9}
    &
    &
    & \multicolumn{1}{c}{Median}
    & \multicolumn{1}{c}{Geo.\,mean}
    & \multicolumn{1}{c}{Reduct.\,(\%)}
    & \multicolumn{1}{c}{Median}
    & \multicolumn{1}{c}{Arith.\,mean}
    & \multicolumn{1}{c}{Reduct.\,(\%)}
    \\
\midrule
\multirow{6}{*}{\shortstack{\\Reach.\\(\num{\AbcPdrScPdrReachSafetyRelCommonStatusAllCount})}}
    & \multirow{3}{*}{Rel.}
    & PDR 
    & \AbcPdrScPdrReachSafetyRelCommonVerifTimeAllMedian
    & \AbcPdrScPdrReachSafetyRelCommonVerifTimeAllGeoMean
    & \multicolumn{1}{c}{\smaller \textcolor{gray}{(baseline)}}
    & \AbcPdrScPdrReachSafetyRelCommonGateAllMedian
    & \AbcPdrScPdrReachSafetyRelCommonGateAllAvg
    & \multicolumn{1}{c}{\smaller \textcolor{gray}{(baseline)}}
    \\
    &
    & \texttt{sc}.+\phantom{)}PDR
    & \AbcScPdrScPdrReachSafetyRelCommonSolveTimeAllMedian
    & \AbcScPdrScPdrReachSafetyRelCommonSolveTimeAllGeoMean
    & \fpeval{(1 - (\AbcScPdrScPdrReachSafetyRelCommonSolveTimeAllGeoMean / \AbcPdrScPdrReachSafetyRelCommonVerifTimeAllGeoMean)) * 100}
    & \raisebox{-1.4ex}[0ex][0ex]{\tablenum[table-format=5.1]{\AbcScPdrScPdrReachSafetyRelCommonGateAllMedian}}
    & \raisebox{-1.4ex}[0ex][0ex]{\tablenum[table-format=5.1, round-mode=figures, round-precision=5]{\AbcScPdrScPdrReachSafetyRelCommonGateAllAvg}}
    & \raisebox{-1.4ex}[0ex][0ex]{\tablenum[round-mode=figures, round-precision=3]{\fpeval{(1 - (\AbcScPdrScPdrReachSafetyRelCommonGateAllAvg / \AbcPdrScPdrReachSafetyRelCommonGateAllAvg)) * 100}}}
    \\
    &
    & \textcolor{gray}{(\texttt{sc}.+)}PDR
    & \AbcScPdrScPdrReachSafetyRelCommonVerifTimeAllMedian
    & \AbcScPdrScPdrReachSafetyRelCommonVerifTimeAllGeoMean
    & \fpeval{(1 - (\AbcScPdrScPdrReachSafetyRelCommonVerifTimeAllGeoMean / \AbcPdrScPdrReachSafetyRelCommonVerifTimeAllGeoMean)) * 100}
    &
    &
    &
    \\
\cmidrule{2-9}
    & \multirow{3}{*}{Func.}
    & PDR
    & \AbcPdrScPdrReachSafetyFuncCommonVerifTimeAllMedian
    & \AbcPdrScPdrReachSafetyFuncCommonVerifTimeAllGeoMean
    & \fpeval{(1 - (\AbcPdrScPdrReachSafetyFuncCommonVerifTimeAllGeoMean / \AbcPdrScPdrReachSafetyRelCommonVerifTimeAllGeoMean)) * 100}
    & \AbcPdrScPdrReachSafetyFuncCommonGateAllMedian
    & \AbcPdrScPdrReachSafetyFuncCommonGateAllAvg
    & \fpeval{(1 - (\AbcPdrScPdrReachSafetyFuncCommonGateAllAvg / \AbcPdrScPdrReachSafetyRelCommonGateAllAvg)) * 100}
    \\
    &
    & \texttt{sc}.+\phantom{)}PDR
    & \AbcScPdrScPdrReachSafetyFuncCommonSolveTimeAllMedian
    & \AbcScPdrScPdrReachSafetyFuncCommonSolveTimeAllGeoMean
    & \fpeval{(1 - (\AbcScPdrScPdrReachSafetyFuncCommonSolveTimeAllGeoMean / \AbcPdrScPdrReachSafetyRelCommonVerifTimeAllGeoMean)) * 100}
    & \raisebox{-1.4ex}[0ex][0ex]{\tablenum[table-format=5.1]{\AbcScPdrScPdrReachSafetyFuncCommonGateAllMedian}}
    & \raisebox{-1.4ex}[0ex][0ex]{\tablenum[table-format=5.1, round-mode=figures, round-precision=5]{\AbcScPdrScPdrReachSafetyFuncCommonGateAllAvg}}
    & \raisebox{-1.4ex}[0ex][0ex]{\tablenum[round-mode=figures, round-precision=3]{\fpeval{(1 - (\AbcScPdrScPdrReachSafetyFuncCommonGateAllAvg / \AbcPdrScPdrReachSafetyRelCommonGateAllAvg)) * 100}}}
    \\
    &
    & \textcolor{gray}{(\texttt{sc}.+)}PDR
    & \AbcScPdrScPdrReachSafetyFuncCommonVerifTimeAllMedian
    & \AbcScPdrScPdrReachSafetyFuncCommonVerifTimeAllGeoMean
    & \fpeval{(1 - (\AbcScPdrScPdrReachSafetyFuncCommonVerifTimeAllGeoMean / \AbcPdrScPdrReachSafetyRelCommonVerifTimeAllGeoMean)) * 100}
    &
    &
    &
    \\
\cmidrule{1-9}
\multirow{6}{*}{\shortstack{\\Term.\\(\num{\AbcPdrScPdrTerminationRelCommonStatusAllCount})}}
    & \multirow{3}{*}{Rel.}
    & PDR
    & \AbcPdrScPdrTerminationRelCommonVerifTimeAllMedian
    & \AbcPdrScPdrTerminationRelCommonVerifTimeAllGeoMean
    & \multicolumn{1}{c}{\smaller \textcolor{gray}{(baseline)}}
    & \AbcPdrScPdrTerminationRelCommonGateAllMedian
    & \AbcPdrScPdrTerminationRelCommonGateAllAvg
    & \multicolumn{1}{c}{\smaller \textcolor{gray}{(baseline)}}
    \\
    &
    & \texttt{sc}.+\phantom{)}PDR
    & \AbcScPdrScPdrTerminationRelCommonSolveTimeAllMedian
    & \AbcScPdrScPdrTerminationRelCommonSolveTimeAllGeoMean
    & \fpeval{(1 - (\AbcScPdrScPdrTerminationRelCommonSolveTimeAllGeoMean / \AbcPdrScPdrTerminationRelCommonVerifTimeAllGeoMean)) * 100}
    & \raisebox{-1.4ex}[0ex][0ex]{\tablenum[table-format=5.1]{\AbcScPdrScPdrTerminationRelCommonGateAllMedian}}
    & \raisebox{-1.4ex}[0ex][0ex]{\tablenum[table-format=5.1, round-mode=figures, round-precision=5]{\AbcScPdrScPdrTerminationRelCommonGateAllAvg}}
    & \raisebox{-1.4ex}[0ex][0ex]{\tablenum[round-mode=figures, round-precision=3]{\fpeval{(1 - (\AbcScPdrScPdrTerminationRelCommonGateAllAvg / \AbcPdrScPdrTerminationRelCommonGateAllAvg)) * 100}}}
    \\
    &
    & \textcolor{gray}{(\texttt{sc}.+)}PDR
    & \AbcScPdrScPdrTerminationRelCommonVerifTimeAllMedian
    & \AbcScPdrScPdrTerminationRelCommonVerifTimeAllGeoMean
    & \fpeval{(1 - (\AbcScPdrScPdrTerminationRelCommonVerifTimeAllGeoMean / \AbcPdrScPdrTerminationRelCommonVerifTimeAllGeoMean)) * 100}
    &
    &
    &
    \\
\cmidrule{2-9}
    & \multirow{3}{*}{Func.}
    & PDR
    & \AbcPdrScPdrTerminationFuncCommonVerifTimeAllMedian
    & \AbcPdrScPdrTerminationFuncCommonVerifTimeAllGeoMean
    & \fpeval{(1 - (\AbcPdrScPdrTerminationFuncCommonVerifTimeAllGeoMean / \AbcPdrScPdrTerminationRelCommonVerifTimeAllGeoMean)) * 100}
    & \AbcPdrScPdrTerminationFuncCommonGateAllMedian
    & \AbcPdrScPdrTerminationFuncCommonGateAllAvg
    & \fpeval{(1 - (\AbcPdrScPdrTerminationFuncCommonGateAllAvg / \AbcPdrScPdrTerminationRelCommonGateAllAvg)) * 100}
    \\
    &
    & \texttt{sc}.+\phantom{)}PDR
    & \AbcScPdrScPdrTerminationFuncCommonSolveTimeAllMedian
    & \AbcScPdrScPdrTerminationFuncCommonSolveTimeAllGeoMean
    & \fpeval{(1 - (\AbcScPdrScPdrTerminationFuncCommonSolveTimeAllGeoMean / \AbcPdrScPdrTerminationRelCommonVerifTimeAllGeoMean)) * 100}
    & \raisebox{-1.4ex}[0ex][0ex]{\tablenum[table-format=5.1]{\AbcScPdrScPdrTerminationFuncCommonGateAllMedian}}
    & \raisebox{-1.4ex}[0ex][0ex]{\tablenum[table-format=5.1, round-mode=figures, round-precision=5]{\AbcScPdrScPdrTerminationFuncCommonGateAllAvg}}
    & \raisebox{-1.4ex}[0ex][0ex]{\tablenum[round-mode=figures, round-precision=3]{\fpeval{(1 - (\AbcScPdrScPdrTerminationFuncCommonGateAllAvg / \AbcPdrScPdrTerminationRelCommonGateAllAvg)) * 100}}}
    \\
    &
    & \textcolor{gray}{(\texttt{sc}.+)}PDR
    & \AbcScPdrScPdrTerminationFuncCommonVerifTimeAllMedian
    & \AbcScPdrScPdrTerminationFuncCommonVerifTimeAllGeoMean
    & \fpeval{(1 - (\AbcScPdrScPdrTerminationFuncCommonVerifTimeAllGeoMean / \AbcPdrScPdrTerminationRelCommonVerifTimeAllGeoMean)) * 100}
    &
    &
    &
    \\
\bottomrule
\end{tabular}

%% file: tables/scorr.solved.tex
\begin{tabular}{clS[table-format=5]S[table-format=5]S[round-mode=figures, round-precision=2]S[table-format=5]S[table-format=5]S[round-mode=figures, round-precision=2]}
\toprule
\multirow{2}{*}{\shortstack{Category\\(\#tasks)}}
    & \multirow{2}{*}{Results}
    & \multicolumn{3}{c}{Relational}
    & \multicolumn{3}{c}{Functional}
    \\
\cmidrule(lr){3-5}\cmidrule(lr){6-8}
    &
    & \multicolumn{1}{c}{w/o \texttt{scorr}}
    & \multicolumn{1}{c}{w/ \texttt{scorr}}
    & \multicolumn{1}{c}{$\Delta$\,(\%)}
    & \multicolumn{1}{c}{w/o \texttt{scorr}}
    & \multicolumn{1}{c}{w/ \texttt{scorr}}
    & \multicolumn{1}{c}{$\Delta$\,(\%)}
    \\
\midrule
\multirow{3}{*}{\shortstack{ReachSafety\\(\AbcPdrReachSafetySampledBvRelStatusAllCount)}}
    & \#Solved
    & \boldnum{\AbcPdrReachSafetySampledBvRelStatusCorrectCount}
    & \AbcScPdrReachSafetySampledBvRelStatusCorrectCount
    & \fpeval{(\AbcScPdrReachSafetySampledBvRelStatusCorrectCount - \AbcPdrReachSafetySampledBvRelStatusCorrectCount) / \AbcPdrReachSafetySampledBvRelStatusCorrectCount * 100}
    & \AbcPdrReachSafetySampledBvFuncStatusCorrectCount
    & \boldnum{\AbcScPdrReachSafetySampledBvFuncStatusCorrectCount}
    & \fpeval{(\AbcScPdrReachSafetySampledBvFuncStatusCorrectCount - \AbcPdrReachSafetySampledBvFuncStatusCorrectCount) / \AbcPdrReachSafetySampledBvFuncStatusCorrectCount * 100}
    \\
    & \quad proofs
    & \boldnum{\AbcPdrReachSafetySampledBvRelStatusCorrectTrueCount}
    & \AbcScPdrReachSafetySampledBvRelStatusCorrectTrueCount
    & \fpeval{(\AbcScPdrReachSafetySampledBvRelStatusCorrectTrueCount - \AbcPdrReachSafetySampledBvRelStatusCorrectTrueCount) / \AbcPdrReachSafetySampledBvRelStatusCorrectTrueCount * 100}
    & \AbcPdrReachSafetySampledBvFuncStatusCorrectTrueCount
    & \boldnum{\AbcScPdrReachSafetySampledBvFuncStatusCorrectTrueCount}
    & \fpeval{(\AbcScPdrReachSafetySampledBvFuncStatusCorrectTrueCount - \AbcPdrReachSafetySampledBvFuncStatusCorrectTrueCount) / \AbcPdrReachSafetySampledBvFuncStatusCorrectTrueCount * 100}
    \\
    & \quad alarms
    & \boldnum{\AbcPdrReachSafetySampledBvRelStatusCorrectFalseCount}
    & \AbcScPdrReachSafetySampledBvRelStatusCorrectFalseCount
    & \fpeval{(\AbcScPdrReachSafetySampledBvRelStatusCorrectFalseCount - \AbcPdrReachSafetySampledBvRelStatusCorrectFalseCount) / \AbcPdrReachSafetySampledBvRelStatusCorrectFalseCount * 100}
    & \AbcPdrReachSafetySampledBvFuncStatusCorrectFalseCount
    & \boldnum{\AbcScPdrReachSafetySampledBvFuncStatusCorrectFalseCount}
    & \fpeval{(\AbcScPdrReachSafetySampledBvFuncStatusCorrectFalseCount - \AbcPdrReachSafetySampledBvFuncStatusCorrectFalseCount) / \AbcPdrReachSafetySampledBvFuncStatusCorrectFalseCount * 100}
    \\
\cmidrule{1-8}
\multirow{3}{*}{\shortstack{Termination\\(\AbcPdrTerminationSampledBvRelStatusAllCount)}}
    & \#Solved
    & \AbcPdrTerminationSampledBvRelStatusCorrectCount
    & \boldnum{\AbcScPdrTerminationSampledBvRelStatusCorrectCount}
    & \fpeval{(\AbcScPdrTerminationSampledBvRelStatusCorrectCount - \AbcPdrTerminationSampledBvRelStatusCorrectCount) / \AbcPdrTerminationSampledBvRelStatusCorrectCount * 100}
    & \AbcPdrTerminationSampledBvFuncStatusCorrectCount
    & \boldnum{\AbcScPdrTerminationSampledBvFuncStatusCorrectCount}
    & \fpeval{(\AbcScPdrTerminationSampledBvFuncStatusCorrectCount - \AbcPdrTerminationSampledBvFuncStatusCorrectCount) / \AbcPdrTerminationSampledBvFuncStatusCorrectCount * 100}
    \\
    & \quad proofs
    & \AbcPdrTerminationSampledBvRelStatusCorrectTrueCount
    & \boldnum{\AbcScPdrTerminationSampledBvRelStatusCorrectTrueCount}
    & \fpeval{(\AbcScPdrTerminationSampledBvRelStatusCorrectTrueCount - \AbcPdrTerminationSampledBvRelStatusCorrectTrueCount) / \AbcPdrTerminationSampledBvRelStatusCorrectTrueCount * 100}
    & \AbcPdrTerminationSampledBvFuncStatusCorrectTrueCount
    & \boldnum{\AbcScPdrTerminationSampledBvFuncStatusCorrectTrueCount}
    & \fpeval{(\AbcScPdrTerminationSampledBvFuncStatusCorrectTrueCount - \AbcPdrTerminationSampledBvFuncStatusCorrectTrueCount) / \AbcPdrTerminationSampledBvFuncStatusCorrectTrueCount * 100}
    \\
    & \quad alarms
    & \AbcPdrTerminationSampledBvRelStatusCorrectFalseCount
    & \boldnum{\AbcScPdrTerminationSampledBvRelStatusCorrectFalseCount}
    & \fpeval{(\AbcScPdrTerminationSampledBvRelStatusCorrectFalseCount - \AbcPdrTerminationSampledBvRelStatusCorrectFalseCount) / \AbcPdrTerminationSampledBvRelStatusCorrectFalseCount * 100}
    & \AbcPdrTerminationSampledBvFuncStatusCorrectFalseCount
    & \boldnum{\AbcScPdrTerminationSampledBvFuncStatusCorrectFalseCount}
    & \fpeval{(\AbcScPdrTerminationSampledBvFuncStatusCorrectFalseCount - \AbcPdrTerminationSampledBvFuncStatusCorrectFalseCount) / \AbcPdrTerminationSampledBvFuncStatusCorrectFalseCount * 100}
    \\
\bottomrule
\end{tabular}

%% file: tables/l2s-simp.tex
\begin{tabular}{llS[table-format=5]S[table-format=5]S[round-mode=figures, round-precision=2]S[table-format=5]S[table-format=5]S[round-mode=figures, round-precision=2]}
\toprule
\multirow{2}{*}{Encoding}
    & \multirow{2}{*}{Results}
    & \multicolumn{3}{c}{\ricthree $\cdot$ IC3}
    & \multicolumn{3}{c}{\pono $\cdot$ \kinduction}
    \\
\cmidrule(lr){3-5}\cmidrule(lr){6-8}
    &
    & \multicolumn{1}{c}{w/o simp.}
    & \multicolumn{1}{c}{w/ simp.}
    & \multicolumn{1}{c}{$\Delta$\,(\%)}
    & \multicolumn{1}{c}{w/o simp.}
    & \multicolumn{1}{c}{w/ simp.}
    & \multicolumn{1}{c}{$\Delta$\,(\%)}
    \\
\midrule
\multirow{3}{*}{Relational}
    & \#Solved
    & \CpvLIIsRicIIIIcIIIRelNoSimpStatusCorrectCount
    & \boldnum{\CpvLIIsRicIIIIcIIIRelStatusCorrectCount}
    & \fpeval{(\CpvLIIsRicIIIIcIIIRelStatusCorrectCount - \CpvLIIsRicIIIIcIIIRelNoSimpStatusCorrectCount) / \CpvLIIsRicIIIIcIIIRelNoSimpStatusCorrectCount * 100}
    & \CpvLIIsPonoKindRelNoSimpStatusCorrectCount
    & \boldnum{\CpvLIIsPonoKindRelStatusCorrectCount}
    & \fpeval{(\CpvLIIsPonoKindRelStatusCorrectCount - \CpvLIIsPonoKindRelNoSimpStatusCorrectCount) / \CpvLIIsPonoKindRelNoSimpStatusCorrectCount * 100}
    \\
    & \quad proofs
    & \boldnum{\CpvLIIsRicIIIIcIIIRelNoSimpStatusCorrectTrueCount}
    & \CpvLIIsRicIIIIcIIIRelStatusCorrectTrueCount
    & \fpeval{(\CpvLIIsRicIIIIcIIIRelStatusCorrectTrueCount - \CpvLIIsRicIIIIcIIIRelNoSimpStatusCorrectTrueCount) / \CpvLIIsRicIIIIcIIIRelNoSimpStatusCorrectTrueCount * 100}
    & \boldnum{\CpvLIIsPonoKindRelNoSimpStatusCorrectTrueCount}
    & \boldnum{\CpvLIIsPonoKindRelStatusCorrectTrueCount}
    & \fpeval{(\CpvLIIsPonoKindRelStatusCorrectTrueCount - \CpvLIIsPonoKindRelNoSimpStatusCorrectTrueCount) / \CpvLIIsPonoKindRelNoSimpStatusCorrectTrueCount * 100}
    \\
    & \quad alarms
    & \CpvLIIsRicIIIIcIIIRelNoSimpStatusCorrectFalseCount
    & \boldnum{\CpvLIIsRicIIIIcIIIRelStatusCorrectFalseCount}
    & \fpeval{(\CpvLIIsRicIIIIcIIIRelStatusCorrectFalseCount - \CpvLIIsRicIIIIcIIIRelNoSimpStatusCorrectFalseCount) / \CpvLIIsRicIIIIcIIIRelNoSimpStatusCorrectFalseCount * 100}
    & \CpvLIIsPonoKindRelNoSimpStatusCorrectFalseCount
    & \boldnum{\CpvLIIsPonoKindRelStatusCorrectFalseCount}
    & \fpeval{(\CpvLIIsPonoKindRelStatusCorrectFalseCount - \CpvLIIsPonoKindRelNoSimpStatusCorrectFalseCount) / \CpvLIIsPonoKindRelNoSimpStatusCorrectFalseCount * 100}
    \\
\cmidrule{1-8}
\multirow{3}{*}{Functional}
    & \#Solved
    & \CpvLIIsRicIIIIcIIIFuncNoSimpStatusCorrectCount
    & \boldnum{\CpvLIIsRicIIIIcIIIFuncStatusCorrectCount}
    & \fpeval{(\CpvLIIsRicIIIIcIIIFuncStatusCorrectCount - \CpvLIIsRicIIIIcIIIFuncNoSimpStatusCorrectCount) / \CpvLIIsRicIIIIcIIIFuncNoSimpStatusCorrectCount * 100}
    & \CpvLIIsPonoKindFuncNoSimpStatusCorrectCount
    & \boldnum{\CpvLIIsPonoKindFuncStatusCorrectCount}
    & \fpeval{(\CpvLIIsPonoKindFuncStatusCorrectCount - \CpvLIIsPonoKindFuncNoSimpStatusCorrectCount) / \CpvLIIsPonoKindFuncNoSimpStatusCorrectCount * 100}
    \\
    & \quad proofs
    & \CpvLIIsRicIIIIcIIIFuncNoSimpStatusCorrectTrueCount
    & \boldnum{\CpvLIIsRicIIIIcIIIFuncStatusCorrectTrueCount}
    & \fpeval{(\CpvLIIsRicIIIIcIIIFuncStatusCorrectTrueCount - \CpvLIIsRicIIIIcIIIFuncNoSimpStatusCorrectTrueCount) / \CpvLIIsRicIIIIcIIIFuncNoSimpStatusCorrectTrueCount * 100}
    & \CpvLIIsPonoKindFuncNoSimpStatusCorrectTrueCount
    & \boldnum{\CpvLIIsPonoKindFuncStatusCorrectTrueCount}
    & \fpeval{(\CpvLIIsPonoKindFuncStatusCorrectTrueCount - \CpvLIIsPonoKindFuncNoSimpStatusCorrectTrueCount) / \CpvLIIsPonoKindFuncNoSimpStatusCorrectTrueCount * 100}
    \\
    & \quad alarms
    & \CpvLIIsRicIIIIcIIIFuncNoSimpStatusCorrectFalseCount
    & \boldnum{\CpvLIIsRicIIIIcIIIFuncStatusCorrectFalseCount}
    & \fpeval{(\CpvLIIsRicIIIIcIIIFuncStatusCorrectFalseCount - \CpvLIIsRicIIIIcIIIFuncNoSimpStatusCorrectFalseCount) / \CpvLIIsRicIIIIcIIIFuncNoSimpStatusCorrectFalseCount * 100}
    & \CpvLIIsPonoKindFuncNoSimpStatusCorrectFalseCount
    & \boldnum{\CpvLIIsPonoKindFuncStatusCorrectFalseCount}
    & \fpeval{(\CpvLIIsPonoKindFuncStatusCorrectFalseCount - \CpvLIIsPonoKindFuncNoSimpStatusCorrectFalseCount) / \CpvLIIsPonoKindFuncNoSimpStatusCorrectFalseCount * 100}
    \\
\bottomrule
\end{tabular}

%% file: plots/cpv.rel.l2s-simp.cputime.quantile.tex
\tikzsetnextfilename{cpv.rel.l2s-simp.cputime.quantile}
\begin{tikzpicture}
\tikzpicturedependsonfile{plots/plot-defs.tex}
\begin{semilogyaxis}[
    quantile plot,
    xlabel={n-th fastest correct result},
    ylabel={CPU time (s)},
    ylabel style={yshift=-5pt},
    mark repeat=200,
    width=.45\linewidth,
    height=.3\linewidth,
    legend style={at={(1,0)}, anchor=south east},
    legend columns=2,
    ]
    \addplot [mark=o, mark options={solid}, dotted, color=green] table {\detokenize{eval-results/csv/cpv-l2s.ric3-ic3.rel.no-simp.cputime.quantile.csv}};
    \addlegendentry{\ricthree w/o simp.}
    \addplot [mark=*, mark options={solid}, solid, color=green] table {\detokenize{eval-results/csv/cpv-l2s.ric3-ic3.rel.cputime.quantile.csv}};
    \addlegendentry{\ricthree w/ simp.}
    \addplot [mark=square, mark options={solid}, dotted, color=blue!50!cyan] table {\detokenize{eval-results/csv/cpv-l2s.pono-kind.rel.no-simp.cputime.quantile.csv}};
    \addlegendentry{\pono w/o simp.}
    \addplot [mark=square*, mark options={solid}, solid, color=blue!50!cyan] table {\detokenize{eval-results/csv/cpv-l2s.pono-kind.rel.cputime.quantile.csv}};
    \addlegendentry{\pono w/ simp.}
\end{semilogyaxis}
\end{tikzpicture}

%% file: plots/cpv.func.l2s-simp.cputime.quantile.tex
\tikzsetnextfilename{cpv.func.l2s-simp.cputime.quantile}
\begin{tikzpicture}
\tikzpicturedependsonfile{plots/plot-defs.tex}
\begin{semilogyaxis}[
    quantile plot,
    xlabel={n-th fastest correct result},
    ylabel={CPU time (s)},
    ylabel style={yshift=-5pt},
    mark repeat=200,
    width=.45\linewidth,
    height=.3\linewidth,
    legend style={at={(1,0)}, anchor=south east},
    legend columns=2,
    ]
    \addplot [mark=o, mark options={solid}, dotted, color=green] table {\detokenize{eval-results/csv/cpv-l2s.ric3-ic3.func.no-simp.cputime.quantile.csv}};
    \addlegendentry{\ricthree w/o simp.}
    \addplot [mark=*, mark options={solid}, solid, color=green] table {\detokenize{eval-results/csv/cpv-l2s.ric3-ic3.func.cputime.quantile.csv}};
    \addlegendentry{\ricthree w/ simp.}
    \addplot [mark=square, mark options={solid}, dotted, color=blue!50!cyan] table {\detokenize{eval-results/csv/cpv-l2s.pono-kind.func.no-simp.cputime.quantile.csv}};
    \addlegendentry{\pono w/o simp.}
    \addplot [mark=square*, mark options={solid}, solid, color=blue!50!cyan] table {\detokenize{eval-results/csv/cpv-l2s.pono-kind.func.cputime.quantile.csv}};
    \addlegendentry{\pono w/ simp.}
\end{semilogyaxis}
\end{tikzpicture}

%% file: tables/transver.tex
\begin{tabular}{lS[table-format=5]S[table-format=5]S[table-format=5]S[table-format=5]S[round-mode=figures, round-precision=2]S[table-format=5]S[round-mode=figures, round-precision=2]}
\toprule
\multirow{2}{*}{Category}
    & \multicolumn{1}{c}{\multirow{2}{*}{\#Tasks}}
    & \multicolumn{2}{c}{\transvercpv}
    & \multicolumn{4}{c}{\cpv}
    \\
\cmidrule(lr){3-4}\cmidrule(lr){5-8}
    &
    & \multicolumn{1}{c}{(a)~\transver \greencmark}
    & \multicolumn{1}{c}{(b)~Solved}
    & \multicolumn{1}{c}{(c)~Solved}
    & \multicolumn{1}{c}{$\Delta_{\scriptscriptstyle \text{(b)}}^{\scriptscriptstyle \text{(c)}}$\,(\%)}
    & \multicolumn{1}{c}{(a)\,$\cap$\,(c)}
    & \multicolumn{1}{c}{$\Delta_{\scriptscriptstyle \text{(b)}}^{\scriptscriptstyle \text{(a)} \cap \text{(c)}}$\,(\%)}
    \\
\midrule
BitVectors
    & \TransverCpvPonoKindFuncTerminationBitVectorsStatusAllCount
    & \CpvPonoKindFuncLIIsTransVerSucceededTerminationBitVectorsStatusAllCount
    & \TransverCpvPonoKindFuncTerminationBitVectorsStatusCorrectCount
    & \CpvPonoKindFuncLIIsTerminationBitVectorsStatusCorrectCount
    & \fpeval{(\CpvPonoKindFuncLIIsTerminationBitVectorsStatusCorrectCount - \TransverCpvPonoKindFuncTerminationBitVectorsStatusCorrectCount) / \TransverCpvPonoKindFuncTerminationBitVectorsStatusCorrectCount * 100}
    & \CpvPonoKindFuncLIIsTransVerSucceededTerminationBitVectorsStatusCorrectCount
    & \fpeval{(\CpvPonoKindFuncLIIsTransVerSucceededTerminationBitVectorsStatusCorrectCount - \TransverCpvPonoKindFuncTerminationBitVectorsStatusCorrectCount) / \TransverCpvPonoKindFuncTerminationBitVectorsStatusCorrectCount * 100}
    \\
MainControlFlow
    & \TransverCpvPonoKindFuncTerminationMainControlFlowStatusAllCount
    & \CpvPonoKindFuncLIIsTransVerSucceededTerminationMainControlFlowStatusAllCount
    & \TransverCpvPonoKindFuncTerminationMainControlFlowStatusCorrectCount
    & \CpvPonoKindFuncLIIsTerminationMainControlFlowStatusCorrectCount
    & \fpeval{(\CpvPonoKindFuncLIIsTerminationMainControlFlowStatusCorrectCount - \TransverCpvPonoKindFuncTerminationMainControlFlowStatusCorrectCount) / \TransverCpvPonoKindFuncTerminationMainControlFlowStatusCorrectCount * 100}
    & \CpvPonoKindFuncLIIsTransVerSucceededTerminationMainControlFlowStatusCorrectCount
    & \fpeval{(\CpvPonoKindFuncLIIsTransVerSucceededTerminationMainControlFlowStatusCorrectCount - \TransverCpvPonoKindFuncTerminationMainControlFlowStatusCorrectCount) / \TransverCpvPonoKindFuncTerminationMainControlFlowStatusCorrectCount * 100}
    \\
Other
    & \TransverCpvPonoKindFuncTerminationOtherStatusAllCount
    & \CpvPonoKindFuncLIIsTransVerSucceededTerminationOtherStatusAllCount
    & \TransverCpvPonoKindFuncTerminationOtherStatusCorrectCount
    & \CpvPonoKindFuncLIIsTerminationOtherStatusCorrectCount
    & \fpeval{(\CpvPonoKindFuncLIIsTerminationOtherStatusCorrectCount - \TransverCpvPonoKindFuncTerminationOtherStatusCorrectCount) / \TransverCpvPonoKindFuncTerminationOtherStatusCorrectCount * 100}
    & \CpvPonoKindFuncLIIsTransVerSucceededTerminationOtherStatusCorrectCount
    & \fpeval{(\CpvPonoKindFuncLIIsTransVerSucceededTerminationOtherStatusCorrectCount - \TransverCpvPonoKindFuncTerminationOtherStatusCorrectCount) / \TransverCpvPonoKindFuncTerminationOtherStatusCorrectCount * 100}
    \\
\cmidrule{1-8}
Overall
    & \TransverCpvPonoKindFuncTerminationStatusAllCount
    & \CpvPonoKindFuncLIIsTransVerSucceededTerminationStatusAllCount
    & \TransverCpvPonoKindFuncTerminationStatusCorrectCount
    & \CpvPonoKindFuncLIIsTerminationStatusCorrectCount
    & \fpeval{(\CpvPonoKindFuncLIIsTerminationStatusCorrectCount - \TransverCpvPonoKindFuncTerminationStatusCorrectCount) / \TransverCpvPonoKindFuncTerminationStatusCorrectCount * 100}
    & \CpvPonoKindFuncLIIsTransVerSucceededTerminationStatusCorrectCount
    & \fpeval{(\CpvPonoKindFuncLIIsTransVerSucceededTerminationStatusCorrectCount - \TransverCpvPonoKindFuncTerminationStatusCorrectCount) / \TransverCpvPonoKindFuncTerminationStatusCorrectCount * 100}
    \\
\bottomrule
\end{tabular}

%% file: tables/wit-val.tex
\begin{tabular}{lS[table-format=5]S[table-format=5]S[table-format=5]S[table-format=5]S[table-format=5]S[table-format=5]S[round-mode=figures, round-precision=3]}
\toprule
\multirow{2}{*}{Category}
    & \multicolumn{1}{c}{\multirow{2}{*}{\cpv}}
    & \multicolumn{2}{c}{Witness v1}
    & \multicolumn{2}{c}{Witness v2}
    & \multicolumn{1}{c}{\multirow{2}{*}{\shortstack{\\Combined\\\#validated}}}
    & \multicolumn{1}{c}{\multirow{2}{*}{\shortstack{\\Confirmation\\rate\,(\%)}}}
    \\
\cmidrule(lr){3-4}\cmidrule(lr){5-6}
    &
    & \multicolumn{1}{c}{\cpachecker}
    & \multicolumn{1}{c}{\symwitch}
    & \multicolumn{1}{c}{\cpachecker}
    & \multicolumn{1}{c}{\witch}
    &
    &
    \\
\midrule
ReachSafety
    & \CpvWitgenPonoBmcFuncStatusCorrectCount
    & \CpaValidateReachValVIStatusCorrectCount
    & \boldnum{\WitchValidateReachValVIStatusCorrectCount}
    & \CpaValidateReachValVIIStatusCorrectCount
    & \WitchValidateReachValVIIStatusCorrectCount
    & \VbWitValReachSafetyStatusCorrectCount
    & \fpeval{\VbWitValReachSafetyStatusCorrectCount / \CpvWitgenPonoBmcFuncStatusCorrectCount * 100}
    \\
Termination
    & \CpvWitgenPonoBmcFuncLIIsStatusCorrectCount
    & \CpaValidateTermValVIStatusCorrectCount
    & \WitchValidateTermValVIStatusCorrectCount
    & \CpaValidateTermValVIIStatusCorrectCount
    & \boldnum{\WitchValidateTermValVIIStatusCorrectCount}
    & \VbWitValTerminationStatusCorrectCount
    & \fpeval{\VbWitValTerminationStatusCorrectCount / \CpvWitgenPonoBmcFuncLIIsStatusCorrectCount * 100}
    \\
\bottomrule
\end{tabular}

%% file: tables/svcomp-overall.tex
\begin{tabular}{llS[table-format=5]S[table-format=5]S[table-format=5]S[table-format=5]S[table-format=5]S[table-format=5]S[table-format=5]}
\toprule
Category
    & \multicolumn{1}{c}{Results}
    & \multicolumn{1}{c}{\cpv}
    & \multicolumn{1}{c}{\cpachecker}
    & \multicolumn{1}{c}{\esbmc}
    & \multicolumn{1}{c}{\kratostwo}
    & \multicolumn{1}{c}{\symbiotic}
    & \multicolumn{1}{c}{\uautomizer}
    & \multicolumn{1}{c}{VBS}
    \\
\midrule
\parbox[t]{0mm}{\multirow{6}{*}{\rotatebox[origin=c]{90}{\shortstack{\\\\ReachSafety\\(\num{\CpvSvcompReachSafetyStatusAllCount})}}}}
    & Correct
    & \second{\CpvSvcompReachSafetyStatusCorrectCount}
    & \CpacheckerSvcompReachSafetyStatusCorrectCount
    & \best{\EsbmcKindReachSafetyStatusCorrectCount}
    & \KratosIISvcompReachSafetyStatusCorrectCount
    & \SymbioticSvcompReachSafetyStatusCorrectCount
    & \UautomizerDefaultReachSafetyStatusCorrectCount
    & \VbSvcompReachSafetyStatusCorrectCount
    \\
    & \quad proofs
    & \best{\CpvSvcompReachSafetyStatusCorrectTrueCount}
    & \CpacheckerSvcompReachSafetyStatusCorrectTrueCount
    & \second{\EsbmcKindReachSafetyStatusCorrectTrueCount}
    & \KratosIISvcompReachSafetyStatusCorrectTrueCount
    & \SymbioticSvcompReachSafetyStatusCorrectTrueCount
    & \UautomizerDefaultReachSafetyStatusCorrectTrueCount
    & \VbSvcompReachSafetyStatusCorrectTrueCount
    \\
    & \quad alarms
    & \CpvSvcompReachSafetyStatusCorrectFalseCount
    & \second{\CpacheckerSvcompReachSafetyStatusCorrectFalseCount}
    & \best{\EsbmcKindReachSafetyStatusCorrectFalseCount}
    & \KratosIISvcompReachSafetyStatusCorrectFalseCount
    & \SymbioticSvcompReachSafetyStatusCorrectFalseCount
    & \UautomizerDefaultReachSafetyStatusCorrectFalseCount
    & \VbSvcompReachSafetyStatusCorrectFalseCount
    \\
    & Incorrect
    & \CpvSvcompReachSafetyStatusWrongCount
    & \wrong{\CpacheckerSvcompReachSafetyStatusWrongCount}
    & \EsbmcKindReachSafetyStatusWrongCount
    & \KratosIISvcompReachSafetyStatusWrongCount
    & \wrong{\SymbioticSvcompReachSafetyStatusWrongCount}
    & \UautomizerDefaultReachSafetyStatusWrongCount
    & \VbSvcompReachSafetyStatusWrongCount
    \\
    & Best
    & \second{\CpvSvcompReachSafetyBestCount}
    & \CpacheckerSvcompReachSafetyBestCount
    & \best{\EsbmcKindReachSafetyBestCount}
    & \KratosIISvcompReachSafetyBestCount
    & \SymbioticSvcompReachSafetyBestCount
    & \UautomizerDefaultReachSafetyBestCount
    & \multicolumn{1}{r}{-}
    \\
    & Unique
    & \best{\CpvSvcompReachSafetyUniqCount}
    & \CpacheckerSvcompReachSafetyUniqCount
    & \second{\EsbmcKindReachSafetyUniqCount}
    & \KratosIISvcompReachSafetyUniqCount
    & \SymbioticSvcompReachSafetyUniqCount
    & \UautomizerDefaultReachSafetyUniqCount
    & \multicolumn{1}{r}{-}
    \\
\cmidrule{1-9}
\parbox[t]{0mm}{\multirow{6}{*}{\rotatebox[origin=c]{90}{\shortstack{\\\\Termination\\(\num{\CpvSvcompTerminationStatusAllCount})}}}}
    & Correct
    & \second{\CpvSvcompTerminationStatusCorrectCount}
    & \CpacheckerSvcompTerminationStatusCorrectCount
    & \EsbmcKindTerminationStatusCorrectCount
    & \multicolumn{1}{r}{-}
    & \SymbioticSvcompTerminationStatusCorrectCount
    & \best{\UautomizerDefaultTerminationStatusCorrectCount}
    & \VbSvcompTerminationStatusCorrectCount
    \\
    & \quad proofs
    & \CpvSvcompTerminationStatusCorrectTrueCount
    & \CpacheckerSvcompTerminationStatusCorrectTrueCount
    & \second{\EsbmcKindTerminationStatusCorrectTrueCount}
    & \multicolumn{1}{r}{-}
    & \SymbioticSvcompTerminationStatusCorrectTrueCount
    & \best{\UautomizerDefaultTerminationStatusCorrectTrueCount}
    & \VbSvcompTerminationStatusCorrectTrueCount
    \\
    & \quad alarms
    & \best{\CpvSvcompTerminationStatusCorrectFalseCount}
    & \second{\CpacheckerSvcompTerminationStatusCorrectFalseCount}
    & \EsbmcKindTerminationStatusCorrectFalseCount
    & \multicolumn{1}{r}{-}
    & \SymbioticSvcompTerminationStatusCorrectFalseCount
    & \UautomizerDefaultTerminationStatusCorrectFalseCount
    & \VbSvcompTerminationStatusCorrectFalseCount
    \\
    & Incorrect
    & \CpvSvcompTerminationStatusWrongCount
    & \wrong{\CpacheckerSvcompTerminationStatusWrongCount}
    & \EsbmcKindTerminationStatusWrongCount
    & \multicolumn{1}{r}{-}
    & \SymbioticSvcompTerminationStatusWrongCount
    & \wrong{\UautomizerDefaultTerminationStatusWrongCount}
    & \VbSvcompTerminationStatusWrongCount
    \\
    & Best
    & \CpvSvcompTerminationBestCount
    & \CpacheckerSvcompTerminationBestCount
    & \best{\EsbmcKindTerminationBestCount}
    & \multicolumn{1}{r}{-}
    & \second{\SymbioticSvcompTerminationBestCount}
    & \UautomizerDefaultTerminationBestCount
    & \multicolumn{1}{r}{-}
    \\
    & Unique
    & \second{\CpvSvcompTerminationUniqCount}
    & \CpacheckerSvcompTerminationUniqCount
    & \EsbmcKindTerminationUniqCount
    & \multicolumn{1}{r}{-}
    & \SymbioticSvcompTerminationUniqCount
    & \best{\UautomizerDefaultTerminationUniqCount}
    & \multicolumn{1}{r}{-}
    \\
\bottomrule
\end{tabular}

%% file: tables/svcomp-reach.1.tex
\begin{tabular}{llS[table-format=5]S[table-format=5]S[table-format=5]S[table-format=5]S[table-format=5]S[table-format=5]S[table-format=5]}
\toprule
Category
    & \multicolumn{1}{c}{Results}
    & \multicolumn{1}{c}{\cpv}
    & \multicolumn{1}{c}{\cpachecker}
    & \multicolumn{1}{c}{\esbmc}
    & \multicolumn{1}{c}{\kratostwo}
    & \multicolumn{1}{c}{\symbiotic}
    & \multicolumn{1}{c}{\uautomizer}
    & \multicolumn{1}{c}{VBS}
    \\
\midrule
\parbox[t]{0mm}{\multirow{6}{*}{\rotatebox[origin=c]{90}{\shortstack{\\\\Arrays\\(\num{\CpvSvcompReachSafetyArraysStatusAllCount})}}}}
    & Correct
    & \second{\CpvSvcompReachSafetyArraysStatusCorrectCount}
    & \CpacheckerSvcompReachSafetyArraysStatusCorrectCount
    & \EsbmcKindReachSafetyArraysStatusCorrectCount
    & \KratosIISvcompReachSafetyArraysStatusCorrectCount
    & \best{\SymbioticSvcompReachSafetyArraysStatusCorrectCount}
    & \UautomizerDefaultReachSafetyArraysStatusCorrectCount
    & \VbSvcompReachSafetyArraysStatusCorrectCount
    \\
    & \quad proofs
    & \second{\CpvSvcompReachSafetyArraysStatusCorrectTrueCount}
    & \CpacheckerSvcompReachSafetyArraysStatusCorrectTrueCount
    & \EsbmcKindReachSafetyArraysStatusCorrectTrueCount
    & \KratosIISvcompReachSafetyArraysStatusCorrectTrueCount
    & \best{\SymbioticSvcompReachSafetyArraysStatusCorrectTrueCount}
    & \UautomizerDefaultReachSafetyArraysStatusCorrectTrueCount
    & \VbSvcompReachSafetyArraysStatusCorrectTrueCount
    \\
    & \quad alarms
    & \CpvSvcompReachSafetyArraysStatusCorrectFalseCount
    & \CpacheckerSvcompReachSafetyArraysStatusCorrectFalseCount
    & \second{\EsbmcKindReachSafetyArraysStatusCorrectFalseCount}
    & \second{\KratosIISvcompReachSafetyArraysStatusCorrectFalseCount}
    & \best{\SymbioticSvcompReachSafetyArraysStatusCorrectFalseCount}
    & \second{\UautomizerDefaultReachSafetyArraysStatusCorrectFalseCount}
    & \VbSvcompReachSafetyArraysStatusCorrectFalseCount
    \\
    & Incorrect
    & \CpvSvcompReachSafetyArraysStatusWrongCount
    & \CpacheckerSvcompReachSafetyArraysStatusWrongCount
    & \EsbmcKindReachSafetyArraysStatusWrongCount
    & \KratosIISvcompReachSafetyArraysStatusWrongCount
    & \SymbioticSvcompReachSafetyArraysStatusWrongCount
    & \UautomizerDefaultReachSafetyArraysStatusWrongCount
    & \VbSvcompReachSafetyArraysStatusWrongCount
    \\
    & Best
    & \CpvSvcompReachSafetyArraysBestCount
    & \CpacheckerSvcompReachSafetyArraysBestCount
    & \second{\EsbmcKindReachSafetyArraysBestCount}
    & \KratosIISvcompReachSafetyArraysBestCount
    & \best{\SymbioticSvcompReachSafetyArraysBestCount}
    & \UautomizerDefaultReachSafetyArraysBestCount
    & \multicolumn{1}{r}{-}
    \\
    & Unique
    & \second{\CpvSvcompReachSafetyArraysUniqCount}
    & \CpacheckerSvcompReachSafetyArraysUniqCount
    & \EsbmcKindReachSafetyArraysUniqCount
    & \KratosIISvcompReachSafetyArraysUniqCount
    & \best{\SymbioticSvcompReachSafetyArraysUniqCount}
    & \UautomizerDefaultReachSafetyArraysUniqCount
    & \multicolumn{1}{r}{-}
    \\
\cmidrule{1-9}
\parbox[t]{0mm}{\multirow{6}{*}{\rotatebox[origin=c]{90}{\shortstack{\\\\BitVectors\\(\num{\CpvSvcompReachSafetyBitVectorsStatusAllCount})}}}}
    & Correct
    & \best{\CpvSvcompReachSafetyBitVectorsStatusCorrectCount}
    & \CpacheckerSvcompReachSafetyBitVectorsStatusCorrectCount
    & \EsbmcKindReachSafetyBitVectorsStatusCorrectCount
    & \second{\KratosIISvcompReachSafetyBitVectorsStatusCorrectCount}
    & \SymbioticSvcompReachSafetyBitVectorsStatusCorrectCount
    & \UautomizerDefaultReachSafetyBitVectorsStatusCorrectCount
    & \VbSvcompReachSafetyBitVectorsStatusCorrectCount
    \\
    & \quad proofs
    & \best{\CpvSvcompReachSafetyBitVectorsStatusCorrectTrueCount}
    & \CpacheckerSvcompReachSafetyBitVectorsStatusCorrectTrueCount
    & \EsbmcKindReachSafetyBitVectorsStatusCorrectTrueCount
    & \second{\KratosIISvcompReachSafetyBitVectorsStatusCorrectTrueCount}
    & \SymbioticSvcompReachSafetyBitVectorsStatusCorrectTrueCount
    & \UautomizerDefaultReachSafetyBitVectorsStatusCorrectTrueCount
    & \VbSvcompReachSafetyBitVectorsStatusCorrectTrueCount
    \\
    & \quad alarms
    & \best{\CpvSvcompReachSafetyBitVectorsStatusCorrectFalseCount}
    & \best{\CpacheckerSvcompReachSafetyBitVectorsStatusCorrectFalseCount}
    & \best{\EsbmcKindReachSafetyBitVectorsStatusCorrectFalseCount}
    & \best{\KratosIISvcompReachSafetyBitVectorsStatusCorrectFalseCount}
    & \best{\SymbioticSvcompReachSafetyBitVectorsStatusCorrectFalseCount}
    & \UautomizerDefaultReachSafetyBitVectorsStatusCorrectFalseCount
    & \VbSvcompReachSafetyBitVectorsStatusCorrectFalseCount
    \\
    & Incorrect
    & \CpvSvcompReachSafetyBitVectorsStatusWrongCount
    & \CpacheckerSvcompReachSafetyBitVectorsStatusWrongCount
    & \EsbmcKindReachSafetyBitVectorsStatusWrongCount
    & \KratosIISvcompReachSafetyBitVectorsStatusWrongCount
    & \SymbioticSvcompReachSafetyBitVectorsStatusWrongCount
    & \UautomizerDefaultReachSafetyBitVectorsStatusWrongCount
    & \VbSvcompReachSafetyBitVectorsStatusWrongCount
    \\
    & Best
    & \second{\CpvSvcompReachSafetyBitVectorsBestCount}
    & \CpacheckerSvcompReachSafetyBitVectorsBestCount
    & \best{\EsbmcKindReachSafetyBitVectorsBestCount}
    & \KratosIISvcompReachSafetyBitVectorsBestCount
    & \SymbioticSvcompReachSafetyBitVectorsBestCount
    & \UautomizerDefaultReachSafetyBitVectorsBestCount
    & \multicolumn{1}{r}{-}
    \\
    & Unique
    & \best{\CpvSvcompReachSafetyBitVectorsUniqCount}
    & \CpacheckerSvcompReachSafetyBitVectorsUniqCount
    & \EsbmcKindReachSafetyBitVectorsUniqCount
    & \KratosIISvcompReachSafetyBitVectorsUniqCount
    & \SymbioticSvcompReachSafetyBitVectorsUniqCount
    & \UautomizerDefaultReachSafetyBitVectorsUniqCount
    & \multicolumn{1}{r}{-}
    \\
\cmidrule{1-9}
\parbox[t]{0mm}{\multirow{6}{*}{\rotatebox[origin=c]{90}{\shortstack{\\\\Combinations\\(\num{\CpvSvcompReachSafetyCombinationsStatusAllCount})}}}}
    & Correct
    & \CpvSvcompReachSafetyCombinationsStatusCorrectCount
    & \second{\CpacheckerSvcompReachSafetyCombinationsStatusCorrectCount}
    & \best{\EsbmcKindReachSafetyCombinationsStatusCorrectCount}
    & \KratosIISvcompReachSafetyCombinationsStatusCorrectCount
    & \SymbioticSvcompReachSafetyCombinationsStatusCorrectCount
    & \UautomizerDefaultReachSafetyCombinationsStatusCorrectCount
    & \VbSvcompReachSafetyCombinationsStatusCorrectCount
    \\
    & \quad proofs
    & \second{\CpvSvcompReachSafetyCombinationsStatusCorrectTrueCount}
    & \CpacheckerSvcompReachSafetyCombinationsStatusCorrectTrueCount
    & \best{\EsbmcKindReachSafetyCombinationsStatusCorrectTrueCount}
    & \KratosIISvcompReachSafetyCombinationsStatusCorrectTrueCount
    & \SymbioticSvcompReachSafetyCombinationsStatusCorrectTrueCount
    & \UautomizerDefaultReachSafetyCombinationsStatusCorrectTrueCount
    & \VbSvcompReachSafetyCombinationsStatusCorrectTrueCount
    \\
    & \quad alarms
    & \CpvSvcompReachSafetyCombinationsStatusCorrectFalseCount
    & \second{\CpacheckerSvcompReachSafetyCombinationsStatusCorrectFalseCount}
    & \best{\EsbmcKindReachSafetyCombinationsStatusCorrectFalseCount}
    & \KratosIISvcompReachSafetyCombinationsStatusCorrectFalseCount
    & \SymbioticSvcompReachSafetyCombinationsStatusCorrectFalseCount
    & \UautomizerDefaultReachSafetyCombinationsStatusCorrectFalseCount
    & \VbSvcompReachSafetyCombinationsStatusCorrectFalseCount
    \\
    & Incorrect
    & \CpvSvcompReachSafetyCombinationsStatusWrongCount
    & \CpacheckerSvcompReachSafetyCombinationsStatusWrongCount
    & \EsbmcKindReachSafetyCombinationsStatusWrongCount
    & \KratosIISvcompReachSafetyCombinationsStatusWrongCount
    & \SymbioticSvcompReachSafetyCombinationsStatusWrongCount
    & \UautomizerDefaultReachSafetyCombinationsStatusWrongCount
    & \VbSvcompReachSafetyCombinationsStatusWrongCount
    \\
    & Best
    & \CpvSvcompReachSafetyCombinationsBestCount
    & \CpacheckerSvcompReachSafetyCombinationsBestCount
    & \best{\EsbmcKindReachSafetyCombinationsBestCount}
    & \KratosIISvcompReachSafetyCombinationsBestCount
    & \second{\SymbioticSvcompReachSafetyCombinationsBestCount}
    & \UautomizerDefaultReachSafetyCombinationsBestCount
    & \multicolumn{1}{r}{-}
    \\
    & Unique
    & \CpvSvcompReachSafetyCombinationsUniqCount
    & \best{\CpacheckerSvcompReachSafetyCombinationsUniqCount}
    & \EsbmcKindReachSafetyCombinationsUniqCount
    & \KratosIISvcompReachSafetyCombinationsUniqCount
    & \second{\SymbioticSvcompReachSafetyCombinationsUniqCount}
    & \UautomizerDefaultReachSafetyCombinationsUniqCount
    & \multicolumn{1}{r}{-}
    \\
\cmidrule{1-9}
\parbox[t]{0mm}{\multirow{6}{*}{\rotatebox[origin=c]{90}{\shortstack{\\\\ControlFlow\\(\num{\CpvSvcompReachSafetyControlFlowStatusAllCount})}}}}
    & Correct
    & \CpvSvcompReachSafetyControlFlowStatusCorrectCount
    & \best{\CpacheckerSvcompReachSafetyControlFlowStatusCorrectCount}
    & \EsbmcKindReachSafetyControlFlowStatusCorrectCount
    & \KratosIISvcompReachSafetyControlFlowStatusCorrectCount
    & \SymbioticSvcompReachSafetyControlFlowStatusCorrectCount
    & \best{\UautomizerDefaultReachSafetyControlFlowStatusCorrectCount}
    & \VbSvcompReachSafetyControlFlowStatusCorrectCount
    \\
    & \quad proofs
    & \CpvSvcompReachSafetyControlFlowStatusCorrectTrueCount
    & \best{\CpacheckerSvcompReachSafetyControlFlowStatusCorrectTrueCount}
    & \EsbmcKindReachSafetyControlFlowStatusCorrectTrueCount
    & \KratosIISvcompReachSafetyControlFlowStatusCorrectTrueCount
    & \SymbioticSvcompReachSafetyControlFlowStatusCorrectTrueCount
    & \best{\UautomizerDefaultReachSafetyControlFlowStatusCorrectTrueCount}
    & \VbSvcompReachSafetyControlFlowStatusCorrectTrueCount
    \\
    & \quad alarms
    & \CpvSvcompReachSafetyControlFlowStatusCorrectFalseCount
    & \second{\CpacheckerSvcompReachSafetyControlFlowStatusCorrectFalseCount}
    & \EsbmcKindReachSafetyControlFlowStatusCorrectFalseCount
    & \KratosIISvcompReachSafetyControlFlowStatusCorrectFalseCount
    & \best{\SymbioticSvcompReachSafetyControlFlowStatusCorrectFalseCount}
    & \second{\UautomizerDefaultReachSafetyControlFlowStatusCorrectFalseCount}
    & \VbSvcompReachSafetyControlFlowStatusCorrectFalseCount
    \\
    & Incorrect
    & \CpvSvcompReachSafetyControlFlowStatusWrongCount
    & \CpacheckerSvcompReachSafetyControlFlowStatusWrongCount
    & \EsbmcKindReachSafetyControlFlowStatusWrongCount
    & \KratosIISvcompReachSafetyControlFlowStatusWrongCount
    & \SymbioticSvcompReachSafetyControlFlowStatusWrongCount
    & \UautomizerDefaultReachSafetyControlFlowStatusWrongCount
    & \VbSvcompReachSafetyControlFlowStatusWrongCount
    \\
    & Best
    & \CpvSvcompReachSafetyControlFlowBestCount
    & \CpacheckerSvcompReachSafetyControlFlowBestCount
    & \best{\EsbmcKindReachSafetyControlFlowBestCount}
    & \KratosIISvcompReachSafetyControlFlowBestCount
    & \second{\SymbioticSvcompReachSafetyControlFlowBestCount}
    & \UautomizerDefaultReachSafetyControlFlowBestCount
    & \multicolumn{1}{r}{-}
    \\
    & Unique
    & \CpvSvcompReachSafetyControlFlowUniqCount
    & \second{\CpacheckerSvcompReachSafetyControlFlowUniqCount}
    & \best{\EsbmcKindReachSafetyControlFlowUniqCount}
    & \KratosIISvcompReachSafetyControlFlowUniqCount
    & \SymbioticSvcompReachSafetyControlFlowUniqCount
    & \second{\UautomizerDefaultReachSafetyControlFlowUniqCount}
    & \multicolumn{1}{r}{-}
    \\
\cmidrule{1-9}
\parbox[t]{0mm}{\multirow{6}{*}{\rotatebox[origin=c]{90}{\shortstack{\\\\ECA\\(\num{\CpvSvcompReachSafetyECAStatusAllCount})}}}}
    & Correct
    & \CpvSvcompReachSafetyECAStatusCorrectCount
    & \best{\CpacheckerSvcompReachSafetyECAStatusCorrectCount}
    & \EsbmcKindReachSafetyECAStatusCorrectCount
    & \second{\KratosIISvcompReachSafetyECAStatusCorrectCount}
    & \SymbioticSvcompReachSafetyECAStatusCorrectCount
    & \UautomizerDefaultReachSafetyECAStatusCorrectCount
    & \VbSvcompReachSafetyECAStatusCorrectCount
    \\
    & \quad proofs
    & \second{\CpvSvcompReachSafetyECAStatusCorrectTrueCount}
    & \best{\CpacheckerSvcompReachSafetyECAStatusCorrectTrueCount}
    & \EsbmcKindReachSafetyECAStatusCorrectTrueCount
    & \KratosIISvcompReachSafetyECAStatusCorrectTrueCount
    & \SymbioticSvcompReachSafetyECAStatusCorrectTrueCount
    & \UautomizerDefaultReachSafetyECAStatusCorrectTrueCount
    & \VbSvcompReachSafetyECAStatusCorrectTrueCount
    \\
    & \quad alarms
    & \CpvSvcompReachSafetyECAStatusCorrectFalseCount
    & \CpacheckerSvcompReachSafetyECAStatusCorrectFalseCount
    & \best{\EsbmcKindReachSafetyECAStatusCorrectFalseCount}
    & \second{\KratosIISvcompReachSafetyECAStatusCorrectFalseCount}
    & \SymbioticSvcompReachSafetyECAStatusCorrectFalseCount
    & \UautomizerDefaultReachSafetyECAStatusCorrectFalseCount
    & \VbSvcompReachSafetyECAStatusCorrectFalseCount
    \\
    & Incorrect
    & \CpvSvcompReachSafetyECAStatusWrongCount
    & \CpacheckerSvcompReachSafetyECAStatusWrongCount
    & \EsbmcKindReachSafetyECAStatusWrongCount
    & \KratosIISvcompReachSafetyECAStatusWrongCount
    & \SymbioticSvcompReachSafetyECAStatusWrongCount
    & \UautomizerDefaultReachSafetyECAStatusWrongCount
    & \VbSvcompReachSafetyECAStatusWrongCount
    \\
    & Best
    & \best{\CpvSvcompReachSafetyECABestCount}
    & \CpacheckerSvcompReachSafetyECABestCount
    & \EsbmcKindReachSafetyECABestCount
    & \KratosIISvcompReachSafetyECABestCount
    & \second{\SymbioticSvcompReachSafetyECABestCount}
    & \UautomizerDefaultReachSafetyECABestCount
    & \multicolumn{1}{r}{-}
    \\
    & Unique
    & \CpvSvcompReachSafetyECAUniqCount
    & \second{\CpacheckerSvcompReachSafetyECAUniqCount}
    & \EsbmcKindReachSafetyECAUniqCount
    & \KratosIISvcompReachSafetyECAUniqCount
    & \SymbioticSvcompReachSafetyECAUniqCount
    & \best{\UautomizerDefaultReachSafetyECAUniqCount}
    & \multicolumn{1}{r}{-}
    \\
\cmidrule{1-9}
\parbox[t]{0mm}{\multirow{6}{*}{\rotatebox[origin=c]{90}{\shortstack{\\\\Floats\\(\num{\CpvSvcompReachSafetyFloatsStatusAllCount})}}}}
    & Correct
    & \CpvSvcompReachSafetyFloatsStatusCorrectCount
    & \second{\CpacheckerSvcompReachSafetyFloatsStatusCorrectCount}
    & \best{\EsbmcKindReachSafetyFloatsStatusCorrectCount}
    & \KratosIISvcompReachSafetyFloatsStatusCorrectCount
    & \SymbioticSvcompReachSafetyFloatsStatusCorrectCount
    & \UautomizerDefaultReachSafetyFloatsStatusCorrectCount
    & \VbSvcompReachSafetyFloatsStatusCorrectCount
    \\
    & \quad proofs
    & \CpvSvcompReachSafetyFloatsStatusCorrectTrueCount
    & \second{\CpacheckerSvcompReachSafetyFloatsStatusCorrectTrueCount}
    & \best{\EsbmcKindReachSafetyFloatsStatusCorrectTrueCount}
    & \KratosIISvcompReachSafetyFloatsStatusCorrectTrueCount
    & \SymbioticSvcompReachSafetyFloatsStatusCorrectTrueCount
    & \UautomizerDefaultReachSafetyFloatsStatusCorrectTrueCount
    & \VbSvcompReachSafetyFloatsStatusCorrectTrueCount
    \\
    & \quad alarms
    & \CpvSvcompReachSafetyFloatsStatusCorrectFalseCount
    & \CpacheckerSvcompReachSafetyFloatsStatusCorrectFalseCount
    & \best{\EsbmcKindReachSafetyFloatsStatusCorrectFalseCount}
    & \KratosIISvcompReachSafetyFloatsStatusCorrectFalseCount
    & \SymbioticSvcompReachSafetyFloatsStatusCorrectFalseCount
    & \second{\UautomizerDefaultReachSafetyFloatsStatusCorrectFalseCount}
    & \VbSvcompReachSafetyFloatsStatusCorrectFalseCount
    \\
    & Incorrect
    & \CpvSvcompReachSafetyFloatsStatusWrongCount
    & \CpacheckerSvcompReachSafetyFloatsStatusWrongCount
    & \EsbmcKindReachSafetyFloatsStatusWrongCount
    & \KratosIISvcompReachSafetyFloatsStatusWrongCount
    & \wrong{\SymbioticSvcompReachSafetyFloatsStatusWrongCount}
    & \UautomizerDefaultReachSafetyFloatsStatusWrongCount
    & \VbSvcompReachSafetyFloatsStatusWrongCount
    \\
    & Best
    & \CpvSvcompReachSafetyFloatsBestCount
    & \CpacheckerSvcompReachSafetyFloatsBestCount
    & \best{\EsbmcKindReachSafetyFloatsBestCount}
    & \KratosIISvcompReachSafetyFloatsBestCount
    & \second{\SymbioticSvcompReachSafetyFloatsBestCount}
    & \UautomizerDefaultReachSafetyFloatsBestCount
    & \multicolumn{1}{r}{-}
    \\
    & Unique
    & \CpvSvcompReachSafetyFloatsUniqCount
    & \CpacheckerSvcompReachSafetyFloatsUniqCount
    & \best{\EsbmcKindReachSafetyFloatsUniqCount}
    & \KratosIISvcompReachSafetyFloatsUniqCount
    & \second{\SymbioticSvcompReachSafetyFloatsUniqCount}
    & \UautomizerDefaultReachSafetyFloatsUniqCount
    & \multicolumn{1}{r}{-}
    \\
\cmidrule{1-9}
\parbox[t]{0mm}{\multirow{6}{*}{\rotatebox[origin=c]{90}{\shortstack{\\\\Hardness\\(\num{\CpvSvcompReachSafetyHardnessStatusAllCount})}}}}
    & Correct
    & \second{\CpvSvcompReachSafetyHardnessStatusCorrectCount}
    & \CpacheckerSvcompReachSafetyHardnessStatusCorrectCount
    & \best{\EsbmcKindReachSafetyHardnessStatusCorrectCount}
    & \KratosIISvcompReachSafetyHardnessStatusCorrectCount
    & \SymbioticSvcompReachSafetyHardnessStatusCorrectCount
    & \UautomizerDefaultReachSafetyHardnessStatusCorrectCount
    & \VbSvcompReachSafetyHardnessStatusCorrectCount
    \\
    & \quad proofs
    & \second{\CpvSvcompReachSafetyHardnessStatusCorrectTrueCount}
    & \CpacheckerSvcompReachSafetyHardnessStatusCorrectTrueCount
    & \best{\EsbmcKindReachSafetyHardnessStatusCorrectTrueCount}
    & \KratosIISvcompReachSafetyHardnessStatusCorrectTrueCount
    & \SymbioticSvcompReachSafetyHardnessStatusCorrectTrueCount
    & \UautomizerDefaultReachSafetyHardnessStatusCorrectTrueCount
    & \VbSvcompReachSafetyHardnessStatusCorrectTrueCount
    \\
    & \quad alarms
    & \CpvSvcompReachSafetyHardnessStatusCorrectFalseCount
    & \CpacheckerSvcompReachSafetyHardnessStatusCorrectFalseCount
    & \EsbmcKindReachSafetyHardnessStatusCorrectFalseCount
    & \KratosIISvcompReachSafetyHardnessStatusCorrectFalseCount
    & \SymbioticSvcompReachSafetyHardnessStatusCorrectFalseCount
    & \UautomizerDefaultReachSafetyHardnessStatusCorrectFalseCount
    & \VbSvcompReachSafetyHardnessStatusCorrectFalseCount
    \\
    & Incorrect
    & \CpvSvcompReachSafetyHardnessStatusWrongCount
    & \CpacheckerSvcompReachSafetyHardnessStatusWrongCount
    & \EsbmcKindReachSafetyHardnessStatusWrongCount
    & \KratosIISvcompReachSafetyHardnessStatusWrongCount
    & \SymbioticSvcompReachSafetyHardnessStatusWrongCount
    & \UautomizerDefaultReachSafetyHardnessStatusWrongCount
    & \VbSvcompReachSafetyHardnessStatusWrongCount
    \\
    & Best
    & \best{\CpvSvcompReachSafetyHardnessBestCount}
    & \CpacheckerSvcompReachSafetyHardnessBestCount
    & \second{\EsbmcKindReachSafetyHardnessBestCount}
    & \KratosIISvcompReachSafetyHardnessBestCount
    & \SymbioticSvcompReachSafetyHardnessBestCount
    & \UautomizerDefaultReachSafetyHardnessBestCount
    & \multicolumn{1}{r}{-}
    \\
    & Unique
    & \best{\CpvSvcompReachSafetyHardnessUniqCount}
    & \CpacheckerSvcompReachSafetyHardnessUniqCount
    & \second{\EsbmcKindReachSafetyHardnessUniqCount}
    & \KratosIISvcompReachSafetyHardnessUniqCount
    & \SymbioticSvcompReachSafetyHardnessUniqCount
    & \UautomizerDefaultReachSafetyHardnessUniqCount
    & \multicolumn{1}{r}{-}
    \\
\bottomrule
\end{tabular}

%% file: tables/svcomp-reach.2.tex
\begin{tabular}{llS[table-format=5]S[table-format=5]S[table-format=5]S[table-format=5]S[table-format=5]S[table-format=5]S[table-format=5]}
\toprule
Category
    & \multicolumn{1}{c}{Results}
    & \multicolumn{1}{c}{\cpv}
    & \multicolumn{1}{c}{\cpachecker}
    & \multicolumn{1}{c}{\esbmc}
    & \multicolumn{1}{c}{\kratostwo}
    & \multicolumn{1}{c}{\symbiotic}
    & \multicolumn{1}{c}{\uautomizer}
    & \multicolumn{1}{c}{VBS}
    \\\midrule
\parbox[t]{0mm}{\multirow{6}{*}{\rotatebox[origin=c]{90}{\shortstack{\\\\Hardware\\(\num{\CpvSvcompReachSafetyHardwareStatusAllCount})}}}}
    & Correct
    & \best{\CpvSvcompReachSafetyHardwareStatusCorrectCount}
    & \CpacheckerSvcompReachSafetyHardwareStatusCorrectCount
    & \EsbmcKindReachSafetyHardwareStatusCorrectCount
    & \second{\KratosIISvcompReachSafetyHardwareStatusCorrectCount}
    & \SymbioticSvcompReachSafetyHardwareStatusCorrectCount
    & \UautomizerDefaultReachSafetyHardwareStatusCorrectCount
    & \VbSvcompReachSafetyHardwareStatusCorrectCount
    \\
    & \quad proofs
    & \best{\CpvSvcompReachSafetyHardwareStatusCorrectTrueCount}
    & \CpacheckerSvcompReachSafetyHardwareStatusCorrectTrueCount
    & \EsbmcKindReachSafetyHardwareStatusCorrectTrueCount
    & \second{\KratosIISvcompReachSafetyHardwareStatusCorrectTrueCount}
    & \SymbioticSvcompReachSafetyHardwareStatusCorrectTrueCount
    & \UautomizerDefaultReachSafetyHardwareStatusCorrectTrueCount
    & \VbSvcompReachSafetyHardwareStatusCorrectTrueCount
    \\
    & \quad alarms
    & \second{\CpvSvcompReachSafetyHardwareStatusCorrectFalseCount}
    & \CpacheckerSvcompReachSafetyHardwareStatusCorrectFalseCount
    & \best{\EsbmcKindReachSafetyHardwareStatusCorrectFalseCount}
    & \KratosIISvcompReachSafetyHardwareStatusCorrectFalseCount
    & \SymbioticSvcompReachSafetyHardwareStatusCorrectFalseCount
    & \UautomizerDefaultReachSafetyHardwareStatusCorrectFalseCount
    & \VbSvcompReachSafetyHardwareStatusCorrectFalseCount
    \\
    & Incorrect
    & \CpvSvcompReachSafetyHardwareStatusWrongCount
    & \CpacheckerSvcompReachSafetyHardwareStatusWrongCount
    & \EsbmcKindReachSafetyHardwareStatusWrongCount
    & \KratosIISvcompReachSafetyHardwareStatusWrongCount
    & \SymbioticSvcompReachSafetyHardwareStatusWrongCount
    & \UautomizerDefaultReachSafetyHardwareStatusWrongCount
    & \VbSvcompReachSafetyHardwareStatusWrongCount
    \\
    & Best
    & \best{\CpvSvcompReachSafetyHardwareBestCount}
    & \CpacheckerSvcompReachSafetyHardwareBestCount
    & \second{\EsbmcKindReachSafetyHardwareBestCount}
    & \KratosIISvcompReachSafetyHardwareBestCount
    & \SymbioticSvcompReachSafetyHardwareBestCount
    & \UautomizerDefaultReachSafetyHardwareBestCount
    & \multicolumn{1}{r}{-}
    \\
    & Unique
    & \best{\CpvSvcompReachSafetyHardwareUniqCount}
    & \CpacheckerSvcompReachSafetyHardwareUniqCount
    & \second{\EsbmcKindReachSafetyHardwareUniqCount}
    & \KratosIISvcompReachSafetyHardwareUniqCount
    & \SymbioticSvcompReachSafetyHardwareUniqCount
    & \UautomizerDefaultReachSafetyHardwareUniqCount
    & \multicolumn{1}{r}{-}
    \\
\cmidrule{1-9}
\parbox[t]{0mm}{\multirow{6}{*}{\rotatebox[origin=c]{90}{\shortstack{\\\\Heap\\(\num{\CpvSvcompReachSafetyHeapStatusAllCount})}}}}
    & Correct
    & \CpvSvcompReachSafetyHeapStatusCorrectCount
    & \second{\CpacheckerSvcompReachSafetyHeapStatusCorrectCount}
    & \best{\EsbmcKindReachSafetyHeapStatusCorrectCount}
    & \KratosIISvcompReachSafetyHeapStatusCorrectCount
    & \second{\SymbioticSvcompReachSafetyHeapStatusCorrectCount}
    & \UautomizerDefaultReachSafetyHeapStatusCorrectCount
    & \VbSvcompReachSafetyHeapStatusCorrectCount
    \\
    & \quad proofs
    & \CpvSvcompReachSafetyHeapStatusCorrectTrueCount
    & \CpacheckerSvcompReachSafetyHeapStatusCorrectTrueCount
    & \best{\EsbmcKindReachSafetyHeapStatusCorrectTrueCount}
    & \KratosIISvcompReachSafetyHeapStatusCorrectTrueCount
    & \second{\SymbioticSvcompReachSafetyHeapStatusCorrectTrueCount}
    & \UautomizerDefaultReachSafetyHeapStatusCorrectTrueCount
    & \VbSvcompReachSafetyHeapStatusCorrectTrueCount
    \\
    & \quad alarms
    & \CpvSvcompReachSafetyHeapStatusCorrectFalseCount
    & \best{\CpacheckerSvcompReachSafetyHeapStatusCorrectFalseCount}
    & \best{\EsbmcKindReachSafetyHeapStatusCorrectFalseCount}
    & \KratosIISvcompReachSafetyHeapStatusCorrectFalseCount
    & \SymbioticSvcompReachSafetyHeapStatusCorrectFalseCount
    & \UautomizerDefaultReachSafetyHeapStatusCorrectFalseCount
    & \VbSvcompReachSafetyHeapStatusCorrectFalseCount
    \\
    & Incorrect
    & \CpvSvcompReachSafetyHeapStatusWrongCount
    & \CpacheckerSvcompReachSafetyHeapStatusWrongCount
    & \EsbmcKindReachSafetyHeapStatusWrongCount
    & \KratosIISvcompReachSafetyHeapStatusWrongCount
    & \SymbioticSvcompReachSafetyHeapStatusWrongCount
    & \UautomizerDefaultReachSafetyHeapStatusWrongCount
    & \VbSvcompReachSafetyHeapStatusWrongCount
    \\
    & Best
    & \CpvSvcompReachSafetyHeapBestCount
    & \CpacheckerSvcompReachSafetyHeapBestCount
    & \best{\EsbmcKindReachSafetyHeapBestCount}
    & \KratosIISvcompReachSafetyHeapBestCount
    & \second{\SymbioticSvcompReachSafetyHeapBestCount}
    & \UautomizerDefaultReachSafetyHeapBestCount
    & \multicolumn{1}{r}{-}
    \\
    & Unique
    & \CpvSvcompReachSafetyHeapUniqCount
    & \CpacheckerSvcompReachSafetyHeapUniqCount
    & \best{\EsbmcKindReachSafetyHeapUniqCount}
    & \KratosIISvcompReachSafetyHeapUniqCount
    & \second{\SymbioticSvcompReachSafetyHeapUniqCount}
    & \second{\UautomizerDefaultReachSafetyHeapUniqCount}
    & \multicolumn{1}{r}{-}
    \\
\cmidrule{1-9}
\parbox[t]{0mm}{\multirow{6}{*}{\rotatebox[origin=c]{90}{\shortstack{\\\\Loops\\(\num{\CpvSvcompReachSafetyLoopsStatusAllCount})}}}}
    & Correct
    & \CpvSvcompReachSafetyLoopsStatusCorrectCount
    & \CpacheckerSvcompReachSafetyLoopsStatusCorrectCount
    & \EsbmcKindReachSafetyLoopsStatusCorrectCount
    & \KratosIISvcompReachSafetyLoopsStatusCorrectCount
    & \best{\SymbioticSvcompReachSafetyLoopsStatusCorrectCount}
    & \second{\UautomizerDefaultReachSafetyLoopsStatusCorrectCount}
    & \VbSvcompReachSafetyLoopsStatusCorrectCount
    \\
    & \quad proofs
    & \CpvSvcompReachSafetyLoopsStatusCorrectTrueCount
    & \CpacheckerSvcompReachSafetyLoopsStatusCorrectTrueCount
    & \EsbmcKindReachSafetyLoopsStatusCorrectTrueCount
    & \KratosIISvcompReachSafetyLoopsStatusCorrectTrueCount
    & \best{\SymbioticSvcompReachSafetyLoopsStatusCorrectTrueCount}
    & \second{\UautomizerDefaultReachSafetyLoopsStatusCorrectTrueCount}
    & \VbSvcompReachSafetyLoopsStatusCorrectTrueCount
    \\
    & \quad alarms
    & \CpvSvcompReachSafetyLoopsStatusCorrectFalseCount
    & \CpacheckerSvcompReachSafetyLoopsStatusCorrectFalseCount
    & \second{\EsbmcKindReachSafetyLoopsStatusCorrectFalseCount}
    & \KratosIISvcompReachSafetyLoopsStatusCorrectFalseCount
    & \best{\SymbioticSvcompReachSafetyLoopsStatusCorrectFalseCount}
    & \UautomizerDefaultReachSafetyLoopsStatusCorrectFalseCount
    & \VbSvcompReachSafetyLoopsStatusCorrectFalseCount
    \\
    & Incorrect
    & \CpvSvcompReachSafetyLoopsStatusWrongCount
    & \wrong{\CpacheckerSvcompReachSafetyLoopsStatusWrongCount}
    & \EsbmcKindReachSafetyLoopsStatusWrongCount
    & \KratosIISvcompReachSafetyLoopsStatusWrongCount
    & \SymbioticSvcompReachSafetyLoopsStatusWrongCount
    & \UautomizerDefaultReachSafetyLoopsStatusWrongCount
    & \VbSvcompReachSafetyLoopsStatusWrongCount
    \\
    & Best
    & \CpvSvcompReachSafetyLoopsBestCount
    & \CpacheckerSvcompReachSafetyLoopsBestCount
    & \second{\EsbmcKindReachSafetyLoopsBestCount}
    & \best{\KratosIISvcompReachSafetyLoopsBestCount}
    & \SymbioticSvcompReachSafetyLoopsBestCount
    & \UautomizerDefaultReachSafetyLoopsBestCount
    & \multicolumn{1}{r}{-}
    \\
    & Unique
    & \CpvSvcompReachSafetyLoopsUniqCount
    & \CpacheckerSvcompReachSafetyLoopsUniqCount
    & \EsbmcKindReachSafetyLoopsUniqCount
    & \KratosIISvcompReachSafetyLoopsUniqCount
    & \best{\SymbioticSvcompReachSafetyLoopsUniqCount}
    & \second{\UautomizerDefaultReachSafetyLoopsUniqCount}
    & \multicolumn{1}{r}{-}
    \\
\cmidrule{1-9}
\parbox[t]{0mm}{\multirow{6}{*}{\rotatebox[origin=c]{90}{\shortstack{\\\\ProductLines\\(\num{\CpvSvcompReachSafetyProductLinesStatusAllCount})}}}}
    & Correct
    & \second{\CpvSvcompReachSafetyProductLinesStatusCorrectCount}
    & \best{\CpacheckerSvcompReachSafetyProductLinesStatusCorrectCount}
    & \EsbmcKindReachSafetyProductLinesStatusCorrectCount
    & \KratosIISvcompReachSafetyProductLinesStatusCorrectCount
    & \SymbioticSvcompReachSafetyProductLinesStatusCorrectCount
    & \UautomizerDefaultReachSafetyProductLinesStatusCorrectCount
    & \VbSvcompReachSafetyProductLinesStatusCorrectCount
    \\
    & \quad proofs
    & \second{\CpvSvcompReachSafetyProductLinesStatusCorrectTrueCount}
    & \best{\CpacheckerSvcompReachSafetyProductLinesStatusCorrectTrueCount}
    & \EsbmcKindReachSafetyProductLinesStatusCorrectTrueCount
    & \KratosIISvcompReachSafetyProductLinesStatusCorrectTrueCount
    & \SymbioticSvcompReachSafetyProductLinesStatusCorrectTrueCount
    & \UautomizerDefaultReachSafetyProductLinesStatusCorrectTrueCount
    & \VbSvcompReachSafetyProductLinesStatusCorrectTrueCount
    \\
    & \quad alarms
    & \CpvSvcompReachSafetyProductLinesStatusCorrectFalseCount
    & \best{\CpacheckerSvcompReachSafetyProductLinesStatusCorrectFalseCount}
    & \best{\EsbmcKindReachSafetyProductLinesStatusCorrectFalseCount}
    & \best{\KratosIISvcompReachSafetyProductLinesStatusCorrectFalseCount}
    & \best{\SymbioticSvcompReachSafetyProductLinesStatusCorrectFalseCount}
    & \UautomizerDefaultReachSafetyProductLinesStatusCorrectFalseCount
    & \VbSvcompReachSafetyProductLinesStatusCorrectFalseCount
    \\
    & Incorrect
    & \CpvSvcompReachSafetyProductLinesStatusWrongCount
    & \CpacheckerSvcompReachSafetyProductLinesStatusWrongCount
    & \EsbmcKindReachSafetyProductLinesStatusWrongCount
    & \KratosIISvcompReachSafetyProductLinesStatusWrongCount
    & \SymbioticSvcompReachSafetyProductLinesStatusWrongCount
    & \UautomizerDefaultReachSafetyProductLinesStatusWrongCount
    & \VbSvcompReachSafetyProductLinesStatusWrongCount
    \\
    & Best
    & \CpvSvcompReachSafetyProductLinesBestCount
    & \CpacheckerSvcompReachSafetyProductLinesBestCount
    & \best{\EsbmcKindReachSafetyProductLinesBestCount}
    & \second{\KratosIISvcompReachSafetyProductLinesBestCount}
    & \SymbioticSvcompReachSafetyProductLinesBestCount
    & \UautomizerDefaultReachSafetyProductLinesBestCount
    & \multicolumn{1}{r}{-}
    \\
    & Unique
    & \CpvSvcompReachSafetyProductLinesUniqCount
    & \best{\CpacheckerSvcompReachSafetyProductLinesUniqCount}
    & \EsbmcKindReachSafetyProductLinesUniqCount
    & \KratosIISvcompReachSafetyProductLinesUniqCount
    & \SymbioticSvcompReachSafetyProductLinesUniqCount
    & \UautomizerDefaultReachSafetyProductLinesUniqCount
    & \multicolumn{1}{r}{-}
    \\
\cmidrule{1-9}
\parbox[t]{0mm}{\multirow{6}{*}{\rotatebox[origin=c]{90}{\shortstack{\\\\Sequentialized\\(\num{\CpvSvcompReachSafetySequentializedStatusAllCount})}}}}
    & Correct
    & \CpvSvcompReachSafetySequentializedStatusCorrectCount
    & \best{\CpacheckerSvcompReachSafetySequentializedStatusCorrectCount}
    & \EsbmcKindReachSafetySequentializedStatusCorrectCount
    & \KratosIISvcompReachSafetySequentializedStatusCorrectCount
    & \second{\SymbioticSvcompReachSafetySequentializedStatusCorrectCount}
    & \UautomizerDefaultReachSafetySequentializedStatusCorrectCount
    & \VbSvcompReachSafetySequentializedStatusCorrectCount
    \\
    & \quad proofs
    & \CpvSvcompReachSafetySequentializedStatusCorrectTrueCount
    & \best{\CpacheckerSvcompReachSafetySequentializedStatusCorrectTrueCount}
    & \EsbmcKindReachSafetySequentializedStatusCorrectTrueCount
    & \KratosIISvcompReachSafetySequentializedStatusCorrectTrueCount
    & \second{\SymbioticSvcompReachSafetySequentializedStatusCorrectTrueCount}
    & \UautomizerDefaultReachSafetySequentializedStatusCorrectTrueCount
    & \VbSvcompReachSafetySequentializedStatusCorrectTrueCount
    \\
    & \quad alarms
    & \CpvSvcompReachSafetySequentializedStatusCorrectFalseCount
    & \best{\CpacheckerSvcompReachSafetySequentializedStatusCorrectFalseCount}
    & \EsbmcKindReachSafetySequentializedStatusCorrectFalseCount
    & \KratosIISvcompReachSafetySequentializedStatusCorrectFalseCount
    & \second{\SymbioticSvcompReachSafetySequentializedStatusCorrectFalseCount}
    & \UautomizerDefaultReachSafetySequentializedStatusCorrectFalseCount
    & \VbSvcompReachSafetySequentializedStatusCorrectFalseCount
    \\
    & Incorrect
    & \CpvSvcompReachSafetySequentializedStatusWrongCount
    & \CpacheckerSvcompReachSafetySequentializedStatusWrongCount
    & \EsbmcKindReachSafetySequentializedStatusWrongCount
    & \KratosIISvcompReachSafetySequentializedStatusWrongCount
    & \SymbioticSvcompReachSafetySequentializedStatusWrongCount
    & \UautomizerDefaultReachSafetySequentializedStatusWrongCount
    & \VbSvcompReachSafetySequentializedStatusWrongCount
    \\
    & Best
    & \CpvSvcompReachSafetySequentializedBestCount
    & \second{\CpacheckerSvcompReachSafetySequentializedBestCount}
    & \best{\EsbmcKindReachSafetySequentializedBestCount}
    & \KratosIISvcompReachSafetySequentializedBestCount
    & \SymbioticSvcompReachSafetySequentializedBestCount
    & \UautomizerDefaultReachSafetySequentializedBestCount
    & \multicolumn{1}{r}{-}
    \\
    & Unique
    & \CpvSvcompReachSafetySequentializedUniqCount
    & \best{\CpacheckerSvcompReachSafetySequentializedUniqCount}
    & \second{\EsbmcKindReachSafetySequentializedUniqCount}
    & \KratosIISvcompReachSafetySequentializedUniqCount
    & \SymbioticSvcompReachSafetySequentializedUniqCount
    & \UautomizerDefaultReachSafetySequentializedUniqCount
    & \multicolumn{1}{r}{-}
    \\
\cmidrule{1-9}
\parbox[t]{0mm}{\multirow{6}{*}{\rotatebox[origin=c]{90}{\shortstack{\\\\XCSP\\(\num{\CpvSvcompReachSafetyXCSPStatusAllCount})}}}}
    & Correct
    & \second{\CpvSvcompReachSafetyXCSPStatusCorrectCount}
    & \CpacheckerSvcompReachSafetyXCSPStatusCorrectCount
    & \best{\EsbmcKindReachSafetyXCSPStatusCorrectCount}
    & \KratosIISvcompReachSafetyXCSPStatusCorrectCount
    & \SymbioticSvcompReachSafetyXCSPStatusCorrectCount
    & \UautomizerDefaultReachSafetyXCSPStatusCorrectCount
    & \VbSvcompReachSafetyXCSPStatusCorrectCount
    \\
    & \quad proofs
    & \second{\CpvSvcompReachSafetyXCSPStatusCorrectTrueCount}
    & \second{\CpacheckerSvcompReachSafetyXCSPStatusCorrectTrueCount}
    & \best{\EsbmcKindReachSafetyXCSPStatusCorrectTrueCount}
    & \KratosIISvcompReachSafetyXCSPStatusCorrectTrueCount
    & \SymbioticSvcompReachSafetyXCSPStatusCorrectTrueCount
    & \UautomizerDefaultReachSafetyXCSPStatusCorrectTrueCount
    & \VbSvcompReachSafetyXCSPStatusCorrectTrueCount
    \\
    & \quad alarms
    & \second{\CpvSvcompReachSafetyXCSPStatusCorrectFalseCount}
    & \CpacheckerSvcompReachSafetyXCSPStatusCorrectFalseCount
    & \best{\EsbmcKindReachSafetyXCSPStatusCorrectFalseCount}
    & \KratosIISvcompReachSafetyXCSPStatusCorrectFalseCount
    & \SymbioticSvcompReachSafetyXCSPStatusCorrectFalseCount
    & \UautomizerDefaultReachSafetyXCSPStatusCorrectFalseCount
    & \VbSvcompReachSafetyXCSPStatusCorrectFalseCount
    \\
    & Incorrect
    & \CpvSvcompReachSafetyXCSPStatusWrongCount
    & \CpacheckerSvcompReachSafetyXCSPStatusWrongCount
    & \EsbmcKindReachSafetyXCSPStatusWrongCount
    & \KratosIISvcompReachSafetyXCSPStatusWrongCount
    & \SymbioticSvcompReachSafetyXCSPStatusWrongCount
    & \UautomizerDefaultReachSafetyXCSPStatusWrongCount
    & \VbSvcompReachSafetyXCSPStatusWrongCount
    \\
    & Best
    & \second{\CpvSvcompReachSafetyXCSPBestCount}
    & \CpacheckerSvcompReachSafetyXCSPBestCount
    & \best{\EsbmcKindReachSafetyXCSPBestCount}
    & \KratosIISvcompReachSafetyXCSPBestCount
    & \SymbioticSvcompReachSafetyXCSPBestCount
    & \UautomizerDefaultReachSafetyXCSPBestCount
    & \multicolumn{1}{r}{-}
    \\
    & Unique
    & \second{\CpvSvcompReachSafetyXCSPUniqCount}
    & \CpacheckerSvcompReachSafetyXCSPUniqCount
    & \best{\EsbmcKindReachSafetyXCSPUniqCount}
    & \KratosIISvcompReachSafetyXCSPUniqCount
    & \SymbioticSvcompReachSafetyXCSPUniqCount
    & \UautomizerDefaultReachSafetyXCSPUniqCount
    & \multicolumn{1}{r}{-}
    \\
\bottomrule
\end{tabular}

%% file: plots/svcomp.ReachSafety.cputime.quantile.tex
\tikzsetnextfilename{svcomp.ReachSafety.cputime.quantile}
\begin{tikzpicture}
\tikzpicturedependsonfile{plots/plot-defs.tex}
\begin{semilogyaxis}[
    quantile plot,
    mark repeat=250,
    ]
    \addgraph{\cpv}{eval-results/csv/cpv.svcomp.ReachSafety.cputime.quantile.csv}
    \addgraph{\cpachecker}{eval-results/csv/cpachecker.svcomp.ReachSafety.cputime.quantile.csv}
    \addgraph{\esbmc}{eval-results/csv/esbmc.kind.ReachSafety.cputime.quantile.csv}
    \addgraph{\kratos}{eval-results/csv/kratos2.svcomp.ReachSafety.cputime.quantile.csv}
    \addgraph{\symbiotic}{eval-results/csv/symbiotic.svcomp.ReachSafety.cputime.quantile.csv}
    \addgraph{\uautomizer}{eval-results/csv/uautomizer.default.ReachSafety.cputime.quantile.csv}
    \addplot+ table[y index=3] {eval-results/csv/vb.svcomp.ReachSafety.cputime.quantile.csv}; \addlegendentry{VBS}
    \legend{} 
\end{semilogyaxis}
\end{tikzpicture}

%% file: plots/svcomp.ReachSafety-Arrays.cputime.quantile.tex
\tikzsetnextfilename{svcomp.ReachSafety-Arrays.cputime.quantile}
\begin{tikzpicture}
\tikzpicturedependsonfile{plots/plot-defs.tex}
\begin{semilogyaxis}[
    quantile plot,
    mark repeat=30,
    ]
    \addgraph{\cpv}{eval-results/csv/cpv.svcomp.ReachSafety-Arrays.cputime.quantile.csv}
    \addgraph{\cpachecker}{eval-results/csv/cpachecker.svcomp.ReachSafety-Arrays.cputime.quantile.csv}
    \addgraph{\esbmc}{eval-results/csv/esbmc.kind.ReachSafety-Arrays.cputime.quantile.csv}
    \addgraph{\kratos}{eval-results/csv/kratos2.svcomp.ReachSafety-Arrays.cputime.quantile.csv}
    \addgraph{\symbiotic}{eval-results/csv/symbiotic.svcomp.ReachSafety-Arrays.cputime.quantile.csv}
    \addgraph{\uautomizer}{eval-results/csv/uautomizer.default.ReachSafety-Arrays.cputime.quantile.csv}
    \addplot+ table[y index=3] {eval-results/csv/vb.svcomp.ReachSafety-Arrays.cputime.quantile.csv}; \addlegendentry{VBS}
    \legend{} 
\end{semilogyaxis}
\end{tikzpicture}

%% file: plots/svcomp.ReachSafety-BitVectors.cputime.quantile.tex
\tikzsetnextfilename{svcomp.ReachSafety-BitVectors.cputime.quantile}
\begin{tikzpicture}
\tikzpicturedependsonfile{plots/plot-defs.tex}
\begin{semilogyaxis}[
    quantile plot,
    mark repeat=8,
    ]
    \pgfplotsset{legend to name=legend:svcomp-reach}
    \addgraph{\cpv}{eval-results/csv/cpv.svcomp.ReachSafety-BitVectors.cputime.quantile.csv}
    \addgraph{\cpachecker}{eval-results/csv/cpachecker.svcomp.ReachSafety-BitVectors.cputime.quantile.csv}
    \addgraph{\esbmc}{eval-results/csv/esbmc.kind.ReachSafety-BitVectors.cputime.quantile.csv}
    \addgraph{\kratos}{eval-results/csv/kratos2.svcomp.ReachSafety-BitVectors.cputime.quantile.csv}
    \addgraph{\symbiotic}{eval-results/csv/symbiotic.svcomp.ReachSafety-BitVectors.cputime.quantile.csv}
    \addgraph{\uautomizer}{eval-results/csv/uautomizer.default.ReachSafety-BitVectors.cputime.quantile.csv}
    \addplot+ table[y index=3] {eval-results/csv/vb.svcomp.ReachSafety-BitVectors.cputime.quantile.csv}; \addlegendentry{VBS}
\end{semilogyaxis}
\end{tikzpicture}

%% file: plots/svcomp.ReachSafety-Combinations.cputime.quantile.tex
\tikzsetnextfilename{svcomp.ReachSafety-Combinations.cputime.quantile}
\begin{tikzpicture}
\tikzpicturedependsonfile{plots/plot-defs.tex}
\begin{semilogyaxis}[
    quantile plot,
    mark repeat=70,
    ]
    \addgraph{\cpv}{eval-results/csv/cpv.svcomp.ReachSafety-Combinations.cputime.quantile.csv}
    \addgraph{\cpachecker}{eval-results/csv/cpachecker.svcomp.ReachSafety-Combinations.cputime.quantile.csv}
    \addgraph{\esbmc}{eval-results/csv/esbmc.kind.ReachSafety-Combinations.cputime.quantile.csv}
    \addgraph{\kratos}{eval-results/csv/kratos2.svcomp.ReachSafety-Combinations.cputime.quantile.csv}
    \addgraph{\symbiotic}{eval-results/csv/symbiotic.svcomp.ReachSafety-Combinations.cputime.quantile.csv}
    \addgraph{\uautomizer}{eval-results/csv/uautomizer.default.ReachSafety-Combinations.cputime.quantile.csv}
    \addplot+ table[y index=3] {eval-results/csv/vb.svcomp.ReachSafety-Combinations.cputime.quantile.csv}; \addlegendentry{VBS}
    \legend{} 
\end{semilogyaxis}
\end{tikzpicture}

%% file: plots/svcomp.ReachSafety-ECA.cputime.quantile.tex
\tikzsetnextfilename{svcomp.ReachSafety-ECA.cputime.quantile}
\begin{tikzpicture}
\tikzpicturedependsonfile{plots/plot-defs.tex}
\begin{semilogyaxis}[
    quantile plot,
    mark repeat=160,
    ]
    \addgraph{\cpv}{eval-results/csv/cpv.svcomp.ReachSafety-ECA.cputime.quantile.csv}
    \addgraph{\cpachecker}{eval-results/csv/cpachecker.svcomp.ReachSafety-ECA.cputime.quantile.csv}
    \addgraph{\esbmc}{eval-results/csv/esbmc.kind.ReachSafety-ECA.cputime.quantile.csv}
    \addgraph{\kratos}{eval-results/csv/kratos2.svcomp.ReachSafety-ECA.cputime.quantile.csv}
    \addgraph{\symbiotic}{eval-results/csv/symbiotic.svcomp.ReachSafety-ECA.cputime.quantile.csv}
    \addgraph{\uautomizer}{eval-results/csv/uautomizer.default.ReachSafety-ECA.cputime.quantile.csv}
    \addplot+ table[y index=3] {eval-results/csv/vb.svcomp.ReachSafety-ECA.cputime.quantile.csv}; \addlegendentry{VBS}
    \legend{} 
\end{semilogyaxis}
\end{tikzpicture}

%% file: plots/svcomp.ReachSafety-Hardness.cputime.quantile.tex
\tikzsetnextfilename{svcomp.ReachSafety-Hardness.cputime.quantile}
\begin{tikzpicture}
\tikzpicturedependsonfile{plots/plot-defs.tex}
\begin{semilogyaxis}[
    quantile plot,
    mark repeat=250,
    /pgfplots/table/y index=3,
    ]
    \addgraph{\cpv}{eval-results/csv/cpv.svcomp.ReachSafety-Hardness.cputime.quantile.csv}
    \addgraph{\cpachecker}{eval-results/csv/cpachecker.svcomp.ReachSafety-Hardness.cputime.quantile.csv}
    \addgraph{\esbmc}{eval-results/csv/esbmc.kind.ReachSafety-Hardness.cputime.quantile.csv}
    \addgraph{\kratos}{eval-results/csv/kratos2.svcomp.ReachSafety-Hardness.cputime.quantile.csv}
    \addgraph{\symbiotic}{eval-results/csv/symbiotic.svcomp.ReachSafety-Hardness.cputime.quantile.csv}
    \addgraph{\uautomizer}{eval-results/csv/uautomizer.default.ReachSafety-Hardness.cputime.quantile.csv}
    \addplot+ table[y index=3] {eval-results/csv/vb.svcomp.ReachSafety-Hardness.cputime.quantile.csv}; \addlegendentry{VBS}
    \legend{} 
\end{semilogyaxis}
\end{tikzpicture}

%% file: plots/svcomp.ReachSafety-Hardware.cputime.quantile.tex
\tikzsetnextfilename{svcomp.ReachSafety-Hardware.cputime.quantile}
\begin{tikzpicture}
\tikzpicturedependsonfile{plots/plot-defs.tex}
\begin{semilogyaxis}[
    quantile plot,
    mark repeat=160,
    ]
    \addgraph{\cpv}{eval-results/csv/cpv.svcomp.ReachSafety-Hardware.cputime.quantile.csv}
    \addgraph{\cpachecker}{eval-results/csv/cpachecker.svcomp.ReachSafety-Hardware.cputime.quantile.csv}
    \addgraph{\esbmc}{eval-results/csv/esbmc.kind.ReachSafety-Hardware.cputime.quantile.csv}
    \addgraph{\kratos}{eval-results/csv/kratos2.svcomp.ReachSafety-Hardware.cputime.quantile.csv}
    \addgraph{\symbiotic}{eval-results/csv/symbiotic.svcomp.ReachSafety-Hardware.cputime.quantile.csv}
    \addgraph{\uautomizer}{eval-results/csv/uautomizer.default.ReachSafety-Hardware.cputime.quantile.csv}
    \addplot+ table[y index=3] {eval-results/csv/vb.svcomp.ReachSafety-Hardware.cputime.quantile.csv}; \addlegendentry{VBS}
    \legend{} 
\end{semilogyaxis}
\end{tikzpicture}

%% file: plots/svcomp.ReachSafety-ProductLines.cputime.quantile.tex
\tikzsetnextfilename{svcomp.ReachSafety-ProductLines.cputime.quantile}
\begin{tikzpicture}
\tikzpicturedependsonfile{plots/plot-defs.tex}
\begin{semilogyaxis}[
    quantile plot,
    mark repeat=110,
    ]
    \addgraph{\cpv}{eval-results/csv/cpv.svcomp.ReachSafety-ProductLines.cputime.quantile.csv}
    \addgraph{\cpachecker}{eval-results/csv/cpachecker.svcomp.ReachSafety-ProductLines.cputime.quantile.csv}
    \addgraph{\esbmc}{eval-results/csv/esbmc.kind.ReachSafety-ProductLines.cputime.quantile.csv}
    \addgraph{\kratos}{eval-results/csv/kratos2.svcomp.ReachSafety-ProductLines.cputime.quantile.csv}
    \addgraph{\symbiotic}{eval-results/csv/symbiotic.svcomp.ReachSafety-ProductLines.cputime.quantile.csv}
    \addgraph{\uautomizer}{eval-results/csv/uautomizer.default.ReachSafety-ProductLines.cputime.quantile.csv}
    \addplot+ table[y index=3] {eval-results/csv/vb.svcomp.ReachSafety-ProductLines.cputime.quantile.csv}; \addlegendentry{VBS}
    \legend{} 
\end{semilogyaxis}
\end{tikzpicture}

%% file: plots/svcomp.ReachSafety-XCSP.cputime.quantile.tex
\tikzsetnextfilename{svcomp.ReachSafety-XCSP.cputime.quantile}
\begin{tikzpicture}
\tikzpicturedependsonfile{plots/plot-defs.tex}
\begin{semilogyaxis}[
    quantile plot,
    mark repeat=20,
    ]
    \addgraph{\cpv}{eval-results/csv/cpv.svcomp.ReachSafety-XCSP.cputime.quantile.csv}
    \addgraph{\cpachecker}{eval-results/csv/cpachecker.svcomp.ReachSafety-XCSP.cputime.quantile.csv}
    \addgraph{\esbmc}{eval-results/csv/esbmc.kind.ReachSafety-XCSP.cputime.quantile.csv}
    \addgraph{\kratos}{eval-results/csv/kratos2.svcomp.ReachSafety-XCSP.cputime.quantile.csv}
    \addgraph{\symbiotic}{eval-results/csv/symbiotic.svcomp.ReachSafety-XCSP.cputime.quantile.csv}
    \addgraph{\uautomizer}{eval-results/csv/uautomizer.default.ReachSafety-XCSP.cputime.quantile.csv}
    \addplot+ table[y index=3] {eval-results/csv/vb.svcomp.ReachSafety-XCSP.cputime.quantile.csv}; \addlegendentry{VBS}
    \legend{} 
\end{semilogyaxis}
\end{tikzpicture}

%% file: tables/svcomp-term.tex
\begin{tabular}{llS[table-format=5]S[table-format=5]S[table-format=5]S[table-format=5]S[table-format=5]S[table-format=5]}
\toprule
Category
    & \multicolumn{1}{c}{Results}
    & \multicolumn{1}{c}{\cpv}
    & \multicolumn{1}{c}{\cpachecker}
    & \multicolumn{1}{c}{\esbmc}
    & \multicolumn{1}{c}{\symbiotic}
    & \multicolumn{1}{c}{\uautomizer}
    & \multicolumn{1}{c}{VBS}
    \\\midrule
\parbox[t]{0mm}{\multirow{6}{*}{\rotatebox[origin=c]{90}{\shortstack{\\\\BitVectors\\(\num{\CpvSvcompTerminationBitVectorsStatusAllCount})}}}}
    & Correct
    & \second{\CpvSvcompTerminationBitVectorsStatusCorrectCount}
    & \CpacheckerSvcompTerminationBitVectorsStatusCorrectCount
    & \EsbmcKindTerminationBitVectorsStatusCorrectCount
    & \SymbioticSvcompTerminationBitVectorsStatusCorrectCount
    & \best{\UautomizerDefaultTerminationBitVectorsStatusCorrectCount}
    & \VbSvcompTerminationBitVectorsStatusCorrectCount
    \\
    & \quad proofs
    & \second{\CpvSvcompTerminationBitVectorsStatusCorrectTrueCount}
    & \CpacheckerSvcompTerminationBitVectorsStatusCorrectTrueCount
    & \second{\EsbmcKindTerminationBitVectorsStatusCorrectTrueCount}
    & \second{\SymbioticSvcompTerminationBitVectorsStatusCorrectTrueCount}
    & \best{\UautomizerDefaultTerminationBitVectorsStatusCorrectTrueCount}
    & \VbSvcompTerminationBitVectorsStatusCorrectTrueCount
    \\
    & \quad alarms
    & \best{\CpvSvcompTerminationBitVectorsStatusCorrectFalseCount}
    & \CpacheckerSvcompTerminationBitVectorsStatusCorrectFalseCount
    & \EsbmcKindTerminationBitVectorsStatusCorrectFalseCount
    & \second{\SymbioticSvcompTerminationBitVectorsStatusCorrectFalseCount}
    & \UautomizerDefaultTerminationBitVectorsStatusCorrectFalseCount
    & \VbSvcompTerminationBitVectorsStatusCorrectFalseCount
    \\
    & Incorrect
    & \CpvSvcompTerminationBitVectorsStatusWrongCount
    & \CpacheckerSvcompTerminationBitVectorsStatusWrongCount
    & \EsbmcKindTerminationBitVectorsStatusWrongCount
    & \SymbioticSvcompTerminationBitVectorsStatusWrongCount
    & \UautomizerDefaultTerminationBitVectorsStatusWrongCount
    & \VbSvcompTerminationBitVectorsStatusWrongCount
    \\
    & Best
    & \CpvSvcompTerminationBitVectorsBestCount
    & \CpacheckerSvcompTerminationBitVectorsBestCount
    & \best{\EsbmcKindTerminationBitVectorsBestCount}
    & \second{\SymbioticSvcompTerminationBitVectorsBestCount}
    & \UautomizerDefaultTerminationBitVectorsBestCount
    & \multicolumn{1}{r}{-}
    \\
    & Unique
    & \second{\CpvSvcompTerminationBitVectorsUniqCount}
    & \CpacheckerSvcompTerminationBitVectorsUniqCount
    & \EsbmcKindTerminationBitVectorsUniqCount
    & \SymbioticSvcompTerminationBitVectorsUniqCount
    & \best{\UautomizerDefaultTerminationBitVectorsUniqCount}
    & \multicolumn{1}{r}{-}
    \\
\cmidrule{1-8}
\parbox[t]{0mm}{\multirow{6}{*}{\rotatebox[origin=c]{90}{\shortstack{\\\\MainControlFlow\\(\num{\CpvSvcompTerminationMainControlFlowStatusAllCount})}}}}
    & Correct
    & \CpvSvcompTerminationMainControlFlowStatusCorrectCount
    & \second{\CpacheckerSvcompTerminationMainControlFlowStatusCorrectCount}
    & \EsbmcKindTerminationMainControlFlowStatusCorrectCount
    & \SymbioticSvcompTerminationMainControlFlowStatusCorrectCount
    & \best{\UautomizerDefaultTerminationMainControlFlowStatusCorrectCount}
    & \VbSvcompTerminationMainControlFlowStatusCorrectCount
    \\
    & \quad proofs
    & \CpvSvcompTerminationMainControlFlowStatusCorrectTrueCount
    & \second{\CpacheckerSvcompTerminationMainControlFlowStatusCorrectTrueCount}
    & \EsbmcKindTerminationMainControlFlowStatusCorrectTrueCount
    & \SymbioticSvcompTerminationMainControlFlowStatusCorrectTrueCount
    & \best{\UautomizerDefaultTerminationMainControlFlowStatusCorrectTrueCount}
    & \VbSvcompTerminationMainControlFlowStatusCorrectTrueCount
    \\
    & \quad alarms
    & \best{\CpvSvcompTerminationMainControlFlowStatusCorrectFalseCount}
    & \CpacheckerSvcompTerminationMainControlFlowStatusCorrectFalseCount
    & \EsbmcKindTerminationMainControlFlowStatusCorrectFalseCount
    & \SymbioticSvcompTerminationMainControlFlowStatusCorrectFalseCount
    & \second{\UautomizerDefaultTerminationMainControlFlowStatusCorrectFalseCount}
    & \VbSvcompTerminationMainControlFlowStatusCorrectFalseCount
    \\
    & Incorrect
    & \CpvSvcompTerminationMainControlFlowStatusWrongCount
    & \wrong{\CpacheckerSvcompTerminationMainControlFlowStatusWrongCount}
    & \EsbmcKindTerminationMainControlFlowStatusWrongCount
    & \SymbioticSvcompTerminationMainControlFlowStatusWrongCount
    & \UautomizerDefaultTerminationMainControlFlowStatusWrongCount
    & \VbSvcompTerminationMainControlFlowStatusWrongCount
    \\
    & Best
    & \CpvSvcompTerminationMainControlFlowBestCount
    & \best{\CpacheckerSvcompTerminationMainControlFlowBestCount}
    & \EsbmcKindTerminationMainControlFlowBestCount
    & \second{\SymbioticSvcompTerminationMainControlFlowBestCount}
    & \UautomizerDefaultTerminationMainControlFlowBestCount
    & \multicolumn{1}{r}{-}
    \\
    & Unique
    & \CpvSvcompTerminationMainControlFlowUniqCount
    & \second{\CpacheckerSvcompTerminationMainControlFlowUniqCount}
    & \EsbmcKindTerminationMainControlFlowUniqCount
    & \SymbioticSvcompTerminationMainControlFlowUniqCount
    & \best{\UautomizerDefaultTerminationMainControlFlowUniqCount}
    & \multicolumn{1}{r}{-}
    \\
\cmidrule{1-8}
\parbox[t]{0mm}{\multirow{6}{*}{\rotatebox[origin=c]{90}{\shortstack{\\\\MainHeap\\(\num{\CpvSvcompTerminationMainHeapStatusAllCount})}}}}
    & Correct
    & \CpvSvcompTerminationMainHeapStatusCorrectCount
    & \CpacheckerSvcompTerminationMainHeapStatusCorrectCount
    & \EsbmcKindTerminationMainHeapStatusCorrectCount
    & \second{\SymbioticSvcompTerminationMainHeapStatusCorrectCount}
    & \best{\UautomizerDefaultTerminationMainHeapStatusCorrectCount}
    & \VbSvcompTerminationMainHeapStatusCorrectCount
    \\
    & \quad proofs
    & \CpvSvcompTerminationMainHeapStatusCorrectTrueCount
    & \CpacheckerSvcompTerminationMainHeapStatusCorrectTrueCount
    & \EsbmcKindTerminationMainHeapStatusCorrectTrueCount
    & \second{\SymbioticSvcompTerminationMainHeapStatusCorrectTrueCount}
    & \best{\UautomizerDefaultTerminationMainHeapStatusCorrectTrueCount}
    & \VbSvcompTerminationMainHeapStatusCorrectTrueCount
    \\
    & \quad alarms
    & \CpvSvcompTerminationMainHeapStatusCorrectFalseCount
    & \second{\CpacheckerSvcompTerminationMainHeapStatusCorrectFalseCount}
    & \EsbmcKindTerminationMainHeapStatusCorrectFalseCount
    & \SymbioticSvcompTerminationMainHeapStatusCorrectFalseCount
    & \best{\UautomizerDefaultTerminationMainHeapStatusCorrectFalseCount}
    & \VbSvcompTerminationMainHeapStatusCorrectFalseCount
    \\
    & Incorrect
    & \CpvSvcompTerminationMainHeapStatusWrongCount
    & \CpacheckerSvcompTerminationMainHeapStatusWrongCount
    & \EsbmcKindTerminationMainHeapStatusWrongCount
    & \SymbioticSvcompTerminationMainHeapStatusWrongCount
    & \wrong{\UautomizerDefaultTerminationMainHeapStatusWrongCount}
    & \VbSvcompTerminationMainHeapStatusWrongCount
    \\
    & Best
    & \CpvSvcompTerminationMainHeapBestCount
    & \CpacheckerSvcompTerminationMainHeapBestCount
    & \second{\EsbmcKindTerminationMainHeapBestCount}
    & \SymbioticSvcompTerminationMainHeapBestCount
    & \best{\UautomizerDefaultTerminationMainHeapBestCount}
    & \multicolumn{1}{r}{-}
    \\
    & Unique
    & \CpvSvcompTerminationMainHeapUniqCount
    & \CpacheckerSvcompTerminationMainHeapUniqCount
    & \EsbmcKindTerminationMainHeapUniqCount
    & \second{\SymbioticSvcompTerminationMainHeapUniqCount}
    & \best{\UautomizerDefaultTerminationMainHeapUniqCount}
    & \multicolumn{1}{r}{-}
    \\
\cmidrule{1-8}
\parbox[t]{0mm}{\multirow{6}{*}{\rotatebox[origin=c]{90}{\shortstack{\\\\Other\\(\num{\CpvSvcompTerminationOtherStatusAllCount})}}}}
    & Correct
    & \best{\CpvSvcompTerminationOtherStatusCorrectCount}
    & \CpacheckerSvcompTerminationOtherStatusCorrectCount
    & \EsbmcKindTerminationOtherStatusCorrectCount
    & \SymbioticSvcompTerminationOtherStatusCorrectCount
    & \second{\UautomizerDefaultTerminationOtherStatusCorrectCount}
    & \VbSvcompTerminationOtherStatusCorrectCount
    \\
    & \quad proofs
    & \CpvSvcompTerminationOtherStatusCorrectTrueCount
    & \CpacheckerSvcompTerminationOtherStatusCorrectTrueCount
    & \best{\EsbmcKindTerminationOtherStatusCorrectTrueCount}
    & \SymbioticSvcompTerminationOtherStatusCorrectTrueCount
    & \second{\UautomizerDefaultTerminationOtherStatusCorrectTrueCount}
    & \VbSvcompTerminationOtherStatusCorrectTrueCount
    \\
    & \quad alarms
    & \best{\CpvSvcompTerminationOtherStatusCorrectFalseCount}
    & \CpacheckerSvcompTerminationOtherStatusCorrectFalseCount
    & \EsbmcKindTerminationOtherStatusCorrectFalseCount
    & \second{\SymbioticSvcompTerminationOtherStatusCorrectFalseCount}
    & \UautomizerDefaultTerminationOtherStatusCorrectFalseCount
    & \VbSvcompTerminationOtherStatusCorrectFalseCount
    \\
    & Incorrect
    & \CpvSvcompTerminationOtherStatusWrongCount
    & \CpacheckerSvcompTerminationOtherStatusWrongCount
    & \EsbmcKindTerminationOtherStatusWrongCount
    & \SymbioticSvcompTerminationOtherStatusWrongCount
    & \UautomizerDefaultTerminationOtherStatusWrongCount
    & \VbSvcompTerminationOtherStatusWrongCount
    \\
    & Best
    & \CpvSvcompTerminationOtherBestCount
    & \CpacheckerSvcompTerminationOtherBestCount
    & \best{\EsbmcKindTerminationOtherBestCount}
    & \second{\SymbioticSvcompTerminationOtherBestCount}
    & \UautomizerDefaultTerminationOtherBestCount
    & \multicolumn{1}{r}{-}
    \\
    & Unique
    & \second{\CpvSvcompTerminationOtherUniqCount}
    & \CpacheckerSvcompTerminationOtherUniqCount
    & \EsbmcKindTerminationOtherUniqCount
    & \SymbioticSvcompTerminationOtherUniqCount
    & \best{\UautomizerDefaultTerminationOtherUniqCount}
    & \multicolumn{1}{r}{-}
    \\
\bottomrule
\end{tabular}

%% file: plots/svcomp.Termination.cputime.quantile.tex
\tikzsetnextfilename{svcomp.Termination.cputime.quantile}
\begin{tikzpicture}
\tikzpicturedependsonfile{plots/plot-defs.tex}
\begin{semilogyaxis}[
    quantile plot,
    mark repeat=300,
    ]
    \addgraph{\cpv}{eval-results/csv/cpv.svcomp.Termination.cputime.quantile.csv}
    \addgraph{\cpachecker}{eval-results/csv/cpachecker.svcomp.Termination.cputime.quantile.csv}
    \addplot+ table[y index=3] {eval-results/csv/esbmc.kind.Termination.cputime.quantile.csv}; \addlegendentry{\esbmc}
    \addgraph{dummy}{dummy}
    \addgraph{\symbiotic}{eval-results/csv/symbiotic.svcomp.Termination.cputime.quantile.csv}
    \addgraph{\uautomizer}{eval-results/csv/uautomizer.default.Termination.cputime.quantile.csv}
    \addplot+ table[y index=3] {eval-results/csv/vb.svcomp.Termination.cputime.quantile.csv}; \addlegendentry{VBS}
    \legend{} 
\end{semilogyaxis}
\end{tikzpicture}

%% file: plots/svcomp.Termination-BitVectors.cputime.quantile.tex
\tikzsetnextfilename{svcomp.Termination-BitVectors.cputime.quantile}
\begin{tikzpicture}
\tikzpicturedependsonfile{plots/plot-defs.tex}
\begin{semilogyaxis}[
    quantile plot,
    mark repeat=10,
    ]
    \pgfplotsset{legend to name=legend:svcomp-term}
    \addgraph{\cpv}{eval-results/csv/cpv.svcomp.Termination-BitVectors.cputime.quantile.csv}
    \addgraph{\cpachecker}{eval-results/csv/cpachecker.svcomp.Termination-BitVectors.cputime.quantile.csv}
    \addplot+ table[y index=3] {eval-results/csv/esbmc.kind.Termination-BitVectors.cputime.quantile.csv}; \addlegendentry{\esbmc}
    \addgraph{dummy}{dummy} 
    \addgraph{\symbiotic}{eval-results/csv/symbiotic.svcomp.Termination-BitVectors.cputime.quantile.csv}
    \addgraph{\uautomizer}{eval-results/csv/uautomizer.default.Termination-BitVectors.cputime.quantile.csv}
    \addplot+ table[y index=3] {eval-results/csv/vb.svcomp.Termination-BitVectors.cputime.quantile.csv}; \addlegendentry{VBS}
\end{semilogyaxis}
\end{tikzpicture}

%% file: plots/svcomp.Termination-MainControlFlow.cputime.quantile.tex
\tikzsetnextfilename{svcomp.Termination-MainControlFlow.cputime.quantile}
\begin{tikzpicture}
\tikzpicturedependsonfile{plots/plot-defs.tex}
\begin{semilogyaxis}[
    quantile plot,
    mark repeat=50,
    ]
    \addgraph{\cpv}{eval-results/csv/cpv.svcomp.Termination-MainControlFlow.cputime.quantile.csv}
    \addgraph{\cpachecker}{eval-results/csv/cpachecker.svcomp.Termination-MainControlFlow.cputime.quantile.csv}
    \addplot+ table[y index=3] {eval-results/csv/esbmc.kind.Termination-MainControlFlow.cputime.quantile.csv}; \addlegendentry{\esbmc}
    \addgraph{dummy}{dummy} 
    \addgraph{\symbiotic}{eval-results/csv/symbiotic.svcomp.Termination-MainControlFlow.cputime.quantile.csv}
    \addgraph{\uautomizer}{eval-results/csv/uautomizer.default.Termination-MainControlFlow.cputime.quantile.csv}
    \addplot+ table[y index=3] {eval-results/csv/vb.svcomp.Termination-MainControlFlow.cputime.quantile.csv}; \addlegendentry{VBS}
    \legend{} 
\end{semilogyaxis}
\end{tikzpicture}

%% file: plots/svcomp.Termination-Other.cputime.quantile.tex
\tikzsetnextfilename{svcomp.Termination-Other.cputime.quantile}
\begin{tikzpicture}
\tikzpicturedependsonfile{plots/plot-defs.tex}
\begin{semilogyaxis}[
    quantile plot,
    mark repeat=250,
    ]
    \addgraph{\cpv}{eval-results/csv/cpv.svcomp.Termination-Other.cputime.quantile.csv}
    \addgraph{\cpachecker}{eval-results/csv/cpachecker.svcomp.Termination-Other.cputime.quantile.csv}
    \addplot+ table[y index=3] {eval-results/csv/esbmc.kind.Termination-Other.cputime.quantile.csv}; \addlegendentry{\esbmc}
    \addgraph{dummy}{dummy} 
    \addgraph{\symbiotic}{eval-results/csv/symbiotic.svcomp.Termination-Other.cputime.quantile.csv}
    \addgraph{\uautomizer}{eval-results/csv/uautomizer.default.Termination-Other.cputime.quantile.csv}
    \addplot+ table[y index=3] {eval-results/csv/vb.svcomp.Termination-Other.cputime.quantile.csv}; \addlegendentry{VBS}
    \legend{} 
\end{semilogyaxis}
\end{tikzpicture}

%% file: conclusion.tex
\section{Conclusion and Future Work}
\label{sect:conclusion}

We present \cpv, a circuit-based program-verification framework that
translates C programs into word-level sequential circuits and leverages off-the-shelf hardware model checkers as verification backends.
The central premise of this work is that sequential circuits constitute a viable intermediate representation for software verification,
allowing the direct reuse of hardware model checkers without algorithmic reimplementation.
Our comprehensive evaluation on the \svcomp~2026 benchmark set supports this premise.
More than \SI{70}{\%} of the benchmark tasks can be successfully translated into sequential circuits.
Among all evaluated tools, \cpv solved the second-largest number of tasks in both the ReachSafety and Termination categories while complementing existing software verifiers by uniquely solving many benchmark instances.
Approximately \SI{95}{\%} of the generated violation witnesses were confirmed by independent witness validators.
The results further show that functional encoding produces more compact circuits
and generally improves model-checking performance.
They also demonstrate the effectiveness of the circuit-level liveness-to-safety transformation
and that eliminating auxiliary instrumentation variables can further improve the lasso-finding capability.

The current limitations of \cpv are primarily in its frontend.
The functional encoder does not yet support floating-point data types,
and the underlying C-to-K2 translation does not support several C language features, such as non-constant initializers in array declarations.
Replacing K2 with SV-LIB~\cite{SV-LIB-1.0}, a recently proposed SMT-LIB-based exchange format for software-verification tasks,
is a promising direction once a robust C-to-SV-LIB translator becomes available.
On the proof side, \cpv currently emits only trivial correctness witnesses when a property is proven.
Extracting the inductive invariants produced by hardware model checkers and translating them into program-level loop invariants
would enable the generation of independently checkable correctness witnesses.
Another direction for future work is to improve the selection of verification backends.
Currently, \cpv employs a fixed sequential portfolio of model-checker engines determined by the verification property and theories involved.
A more adaptive strategy could adopt machine-learning-based algorithm selection to predict suitable backend configurations for individual verification tasks~\cite{Btor2-Select},
potentially improving overall performance and reducing unnecessary computational effort.

More broadly, this work contributes a large collection of \btortwo circuits translated from the \svcomp benchmark suite,
which have been made publicly available and used in recent editions of HWMCC.
The utility of these circuits is already evidenced by the development of the model checker \pono,
which has shown significant performance improvements on these translated \btortwo tasks~\cite{Pono2} and now serves as the primary backend for \cpv.
We hope these translated benchmark tasks will encourage broader support for liveness properties in hardware model checkers,
which could directly benefit \cpv's termination analysis,
and further strengthen the connection between the hardware- and software-verification communities.

%% file: main.bbl

\providecommand{\serysort}{}\providecommand{\svejdasort}{}
\begin{thebibliography}{133}


\ifx \showCODEN    \undefined \def \showCODEN     #1{\unskip}     \fi
\ifx \showISBNx    \undefined \def \showISBNx     #1{\unskip}     \fi
\ifx \showISBNxiii \undefined \def \showISBNxiii  #1{\unskip}     \fi
\ifx \showISSN     \undefined \def \showISSN      #1{\unskip}     \fi
\ifx \showLCCN     \undefined \def \showLCCN      #1{\unskip}     \fi
\ifx \shownote     \undefined \def \shownote      #1{#1}          \fi
\ifx \showarticletitle \undefined \def \showarticletitle #1{#1}   \fi
\ifx \showURL      \undefined \def \showURL       {\relax}        \fi
\providecommand\bibfield[2]{#2}
\providecommand\bibinfo[2]{#2}
\providecommand\natexlab[1]{#1}
\providecommand\showeprint[2][]{arXiv:#2}

\bibitem[Ates et~al\mbox{.}(2026)]%
        {BTOR2C-STTT}
\bibfield{author}{\bibinfo{person}{S. Ates}, \bibinfo{person}{D. Beyer},
  \bibinfo{person}{P.-C. Chien}, {and} \bibinfo{person}{N.-Z. Lee}.}
  \bibinfo{year}{2026}\natexlab{}.
\newblock \showarticletitle{Bridging Hardware and Software Analysis with
  \textsc{Btor2C}: {A} Word-Level-Circuit-to-{C} Translator (Extended
  Version)}.
\newblock \bibinfo{journal}{\emph{Int. J. Softw. Tools Technol. Transf.}}
  (\bibinfo{year}{2026}).
\newblock
\href{https://doi.org/10.1007/s10009-026-00847-z}{doi:\nolinkurl{10.1007/s10009-026-00847-z}}


\bibitem[Ayaziová et~al\mbox{.}(2024)]%
        {VerificationWitnesses-2.0}
\bibfield{author}{\bibinfo{person}{P. Ayaziová}, \bibinfo{person}{D. Beyer},
  \bibinfo{person}{M. Lingsch-Rosenfeld}, \bibinfo{person}{M. Spiessl}, {and}
  \bibinfo{person}{J. Strejček}.} \bibinfo{year}{2024}\natexlab{}.
\newblock \showarticletitle{Software Verification Witnesses~2.0}. In
  \bibinfo{booktitle}{\emph{Proc.\ SPIN}}
  \emph{(\bibinfo{series}{LNCS~14624})}. \bibinfo{publisher}{Springer},
  \bibinfo{pages}{184--203}.
\newblock
\href{https://doi.org/10.1007/978-3-031-66149-5_11}{doi:\nolinkurl{10.1007/978-3-031-66149-5_11}}


\bibitem[Ayaziová et~al\mbox{.}(2025)]%
        {Symbiotic-SVCOMP26-archive}
\bibfield{author}{\bibinfo{person}{P. Ayaziová}, \bibinfo{person}{M. Jonáš},
  \bibinfo{person}{V. Mihalkovič}, \bibinfo{person}{J. Sedláček}, {and}
  \bibinfo{person}{J. Strejček}.} \bibinfo{year}{2025}\natexlab{}.
\newblock \bibinfo{title}{{Symbiotic}: Submission to {SV-COMP} 2026}.
\newblock \bibinfo{howpublished}{Zenodo}.
\newblock
\href{https://doi.org/10.5281/zenodo.17698724}{doi:\nolinkurl{10.5281/zenodo.17698724}}


\bibitem[Ayaziová et~al\mbox{.}(2026)]%
        {SYMBIOTIC-SVCOMP26}
\bibfield{author}{\bibinfo{person}{P. Ayaziová}, \bibinfo{person}{M. Jonáš},
  \bibinfo{person}{V. Mihalkovič}, \bibinfo{person}{J. Sedláček}, {and}
  \bibinfo{person}{J. Strejček}.} \bibinfo{year}{2026}\natexlab{}.
\newblock \showarticletitle{\textsc{Symbiotic 11}: {Predicate} Abstraction
  Joins the Party (Competition Contribution)}. In
  \bibinfo{booktitle}{\emph{Proc.\ TACAS~(2)}}
  \emph{(\bibinfo{series}{LNCS~16506})}. \bibinfo{publisher}{Springer},
  \bibinfo{pages}{583--588}.
\newblock
\href{https://doi.org/10.1007/978-3-032-22749-2_36}{doi:\nolinkurl{10.1007/978-3-032-22749-2_36}}


\bibitem[Ayaziová and Strejček(2023)]%
        {SYMBIOTICWITCH-VALIDATOR-SVCOMP23}
\bibfield{author}{\bibinfo{person}{P. Ayaziová} {and} \bibinfo{person}{J.
  Strejček}.} \bibinfo{year}{2023}\natexlab{}.
\newblock \showarticletitle{\textsc{Symbiotic-Witch 2}: {More} Efficient
  Algorithm and Witness Refutation (Competition Contribution)}. In
  \bibinfo{booktitle}{\emph{Proc.\ TACAS~(2)}}
  \emph{(\bibinfo{series}{LNCS~13994})}. \bibinfo{publisher}{Springer},
  \bibinfo{pages}{523--528}.
\newblock
\href{https://doi.org/10.1007/978-3-031-30820-8_30}{doi:\nolinkurl{10.1007/978-3-031-30820-8_30}}


\bibitem[Ayaziová and Strejček(2024a)]%
        {SWitch-SVCOMP25-archive}
\bibfield{author}{\bibinfo{person}{P. Ayaziová} {and} \bibinfo{person}{J.
  Strejček}.} \bibinfo{year}{2024}\natexlab{a}.
\newblock \bibinfo{title}{{Symbiotic-Witch} 2: {SV-COMP} 2025}.
\newblock \bibinfo{howpublished}{Zenodo}.
\newblock
\href{https://doi.org/10.5281/zenodo.18462288}{doi:\nolinkurl{10.5281/zenodo.18462288}}


\bibitem[Ayaziová and Strejček(2024b)]%
        {WITCH-VALIDATOR-SVCOMP24}
\bibfield{author}{\bibinfo{person}{P. Ayaziová} {and} \bibinfo{person}{J.
  Strejček}.} \bibinfo{year}{2024}\natexlab{b}.
\newblock \showarticletitle{\textsc{Witch 3}: {Validation} of Violation
  Witnesses in the Witness Format~2.0 (Competition Contribution)}. In
  \bibinfo{booktitle}{\emph{Proc.\ TACAS~(3)}}
  \emph{(\bibinfo{series}{LNCS~14572})}. \bibinfo{publisher}{Springer},
  \bibinfo{pages}{341--346}.
\newblock
\href{https://doi.org/10.1007/978-3-031-57256-2_18}{doi:\nolinkurl{10.1007/978-3-031-57256-2_18}}


\bibitem[Ayaziová and Strejček(2025)]%
        {Witch-SVCOMP26-archive}
\bibfield{author}{\bibinfo{person}{P. Ayaziová} {and} \bibinfo{person}{J.
  Strejček}.} \bibinfo{year}{2025}\natexlab{}.
\newblock \bibinfo{title}{{Witch} 4: {SV-COMP} 2026}.
\newblock \bibinfo{howpublished}{Zenodo}.
\newblock
\href{https://doi.org/10.5281/zenodo.17697224}{doi:\nolinkurl{10.5281/zenodo.17697224}}


\bibitem[Baier et~al\mbox{.}(2024a)]%
        {CPACHECKER-3.0-tutorial}
\bibfield{author}{\bibinfo{person}{D. Baier}, \bibinfo{person}{D. Beyer},
  \bibinfo{person}{P.-C. Chien}, \bibinfo{person}{M.-C. Jakobs},
  \bibinfo{person}{M. Jankola}, \bibinfo{person}{M. Kettl},
  \bibinfo{person}{N.-Z. Lee}, \bibinfo{person}{T. Lemberger},
  \bibinfo{person}{M. Lingsch-Rosenfeld}, \bibinfo{person}{H. Wachowitz}, {and}
  \bibinfo{person}{P. Wendler}.} \bibinfo{year}{2024}\natexlab{a}.
\newblock \showarticletitle{Software Verification with \textsc{CPAchecker} 3.0:
  {Tutorial} and User Guide}. In \bibinfo{booktitle}{\emph{Proc.\ FM}}
  \emph{(\bibinfo{series}{LNCS~14934})}. \bibinfo{publisher}{Springer},
  \bibinfo{pages}{543--570}.
\newblock
\href{https://doi.org/10.1007/978-3-031-71177-0_30}{doi:\nolinkurl{10.1007/978-3-031-71177-0_30}}


\bibitem[Baier et~al\mbox{.}(2024b)]%
        {CPACHECKER-SVCOMP24}
\bibfield{author}{\bibinfo{person}{D. Baier}, \bibinfo{person}{D. Beyer},
  \bibinfo{person}{P.-C. Chien}, \bibinfo{person}{M. Jankola},
  \bibinfo{person}{M. Kettl}, \bibinfo{person}{N.-Z. Lee}, \bibinfo{person}{T.
  Lemberger}, \bibinfo{person}{M. Lingsch-Rosenfeld}, \bibinfo{person}{M.
  Spiessl}, \bibinfo{person}{H. Wachowitz}, {and} \bibinfo{person}{P.
  Wendler}.} \bibinfo{year}{2024}\natexlab{b}.
\newblock \showarticletitle{\textsc{CPAchecker} 2.3 with Strategy Selection
  (Competition Contribution)}. In \bibinfo{booktitle}{\emph{Proc.\ TACAS~(3)}}
  \emph{(\bibinfo{series}{LNCS~14572})}. \bibinfo{publisher}{Springer},
  \bibinfo{pages}{359--364}.
\newblock
\href{https://doi.org/10.1007/978-3-031-57256-2_21}{doi:\nolinkurl{10.1007/978-3-031-57256-2_21}}


\bibitem[Barrett et~al\mbox{.}(2025)]%
        {SMTLIB27}
\bibfield{author}{\bibinfo{person}{C. Barrett}, \bibinfo{person}{P. Fontaine},
  {and} \bibinfo{person}{C. Tinelli}.} \bibinfo{year}{2025}\natexlab{}.
\newblock \bibinfo{booktitle}{\emph{{The SMT-LIB Standard: Version 2.7}}}.
\newblock \bibinfo{type}{{T}echnical {R}eport}.
  \bibinfo{institution}{University of Iowa}.
\newblock
\urldef\tempurl%
\url{https://smt-lib.org/papers/smt-lib-reference-v2.7-r2026-03-27.pdf}
\showURL{%
\tempurl}


\bibitem[Bentele et~al\mbox{.}(2026)]%
        {UAUTOMIZER-SVCOMP26}
\bibfield{author}{\bibinfo{person}{M. Bentele}, \bibinfo{person}{M. Barth},
  \bibinfo{person}{M. Ebbinghaus}, \bibinfo{person}{J. Körner},
  \bibinfo{person}{D. Dietsch}, \bibinfo{person}{M. Heizmann},
  \bibinfo{person}{D. Klumpp}, \bibinfo{person}{F. Schüssele}, {and}
  \bibinfo{person}{A. Podelski}.} \bibinfo{year}{2026}\natexlab{}.
\newblock \showarticletitle{\textsc{Ultimate Automizer} with a One-Dimensional
  Memory Model (Competition Contribution)}. In \bibinfo{booktitle}{\emph{Proc.\
  TACAS~(2)}} \emph{(\bibinfo{series}{LNCS~16506})}.
  \bibinfo{publisher}{Springer}, \bibinfo{pages}{589--594}.
\newblock
\href{https://doi.org/10.1007/978-3-032-22749-2_37}{doi:\nolinkurl{10.1007/978-3-032-22749-2_37}}


\bibitem[Berger(2024)]%
        {Berger24}
\bibfield{author}{\bibinfo{person}{P. Berger}.}
  \bibinfo{year}{2024}\natexlab{}.
\newblock \emph{\bibinfo{title}{Applying Software Model Checking: Experiences
  and Advancements}}.
\newblock \bibinfo{thesistype}{Ph.\,D. Dissertation}. \bibinfo{school}{RWTH
  Aachen University}.
\newblock
\href{https://doi.org/10.18154/RWTH-2024-10081}{doi:\nolinkurl{10.18154/RWTH-2024-10081}}


\bibitem[Beyer(2026)]%
        {SVCOMP26-FMTOOLS-artifact}
\bibfield{author}{\bibinfo{person}{D. Beyer}.} \bibinfo{year}{2026}\natexlab{}.
\newblock \bibinfo{title}{{FM-Tools Release~2.3}: {Data} Set of Metadata about
  Tools for Formal Methods ({SV-COMP~2026}, {Test-Comp~2026})}.
\newblock \bibinfo{howpublished}{Zenodo}.
\newblock
\href{https://doi.org/10.5281/zenodo.18650756}{doi:\nolinkurl{10.5281/zenodo.18650756}}


\bibitem[Beyer et~al\mbox{.}(2024a)]%
        {BENCHCLOUD}
\bibfield{author}{\bibinfo{person}{D. Beyer}, \bibinfo{person}{P.-C. Chien},
  {and} \bibinfo{person}{M. Jankola}.} \bibinfo{year}{2024}\natexlab{a}.
\newblock \showarticletitle{\textsc{BenchCloud}: {A} Platform for Scalable
  Performance Benchmarking}. In \bibinfo{booktitle}{\emph{Proc.\ ASE}}.
  \bibinfo{publisher}{ACM}, \bibinfo{pages}{2386--2389}.
\newblock
\href{https://doi.org/10.1145/3691620.3695358}{doi:\nolinkurl{10.1145/3691620.3695358}}


\bibitem[Beyer et~al\mbox{.}(2024b)]%
        {DAR-transferability}
\bibfield{author}{\bibinfo{person}{D. Beyer}, \bibinfo{person}{P.-C. Chien},
  \bibinfo{person}{M. Jankola}, {and} \bibinfo{person}{N.-Z. Lee}.}
  \bibinfo{year}{2024}\natexlab{b}.
\newblock \showarticletitle{A Transferability Study of Interpolation-Based
  Hardware Model Checking for Software Verification}.
\newblock \bibinfo{journal}{\emph{Proc. ACM Softw. Eng.}} \bibinfo{volume}{1},
  \bibinfo{number}{FSE}, Article \bibinfo{articleno}{90}
  (\bibinfo{year}{2024}), \bibinfo{numpages}{23}~pages.
\newblock
\href{https://doi.org/10.1145/3660797}{doi:\nolinkurl{10.1145/3660797}}


\bibitem[Beyer et~al\mbox{.}(2023a)]%
        {BTOR2C}
\bibfield{author}{\bibinfo{person}{D. Beyer}, \bibinfo{person}{P.-C. Chien},
  {and} \bibinfo{person}{N.-Z. Lee}.} \bibinfo{year}{2023}\natexlab{a}.
\newblock \showarticletitle{Bridging Hardware and Software Analysis with
  \textsc{Btor2C}: {A} Word-Level-Circuit-to-{C} Translator}. In
  \bibinfo{booktitle}{\emph{Proc.\ TACAS~(2)}}
  \emph{(\bibinfo{series}{LNCS~13994})}. \bibinfo{publisher}{Springer},
  \bibinfo{pages}{152--172}.
\newblock
\href{https://doi.org/10.1007/978-3-031-30820-8_12}{doi:\nolinkurl{10.1007/978-3-031-30820-8_12}}


\bibitem[Beyer et~al\mbox{.}(2023b)]%
        {CPA-DF}
\bibfield{author}{\bibinfo{person}{D. Beyer}, \bibinfo{person}{P.-C. Chien},
  {and} \bibinfo{person}{N.-Z. Lee}.} \bibinfo{year}{2023}\natexlab{b}.
\newblock \showarticletitle{{CPA-DF}: {A} Tool for Configurable Interval
  Analysis to Boost Program Verification}. In \bibinfo{booktitle}{\emph{Proc.\
  ASE}}. \bibinfo{publisher}{IEEE}, \bibinfo{pages}{2050--2053}.
\newblock
\href{https://doi.org/10.1109/ASE56229.2023.00213}{doi:\nolinkurl{10.1109/ASE56229.2023.00213}}


\bibitem[Beyer et~al\mbox{.}(2009)]%
        {LBE}
\bibfield{author}{\bibinfo{person}{D. Beyer}, \bibinfo{person}{A. Cimatti},
  \bibinfo{person}{A. Griggio}, \bibinfo{person}{M.~E. Keremoglu}, {and}
  \bibinfo{person}{R. Sebastiani}.} \bibinfo{year}{2009}\natexlab{}.
\newblock \showarticletitle{Software Model Checking via Large-Block Encoding}.
  In \bibinfo{booktitle}{\emph{Proc.\ FMCAD}}. \bibinfo{publisher}{IEEE},
  \bibinfo{pages}{25--32}.
\newblock
\href{https://doi.org/10.1109/FMCAD.2009.5351147}{doi:\nolinkurl{10.1109/FMCAD.2009.5351147}}


\bibitem[Beyer et~al\mbox{.}(2022)]%
        {WitnessesJournal}
\bibfield{author}{\bibinfo{person}{D. Beyer}, \bibinfo{person}{M. Dangl},
  \bibinfo{person}{D. Dietsch}, \bibinfo{person}{M. Heizmann},
  \bibinfo{person}{T. Lemberger}, {and} \bibinfo{person}{M. Tautschnig}.}
  \bibinfo{year}{2022}\natexlab{}.
\newblock \showarticletitle{Verification Witnesses}.
\newblock \bibinfo{journal}{\emph{ACM Trans. Softw. Eng. Methodol.}}
  \bibinfo{volume}{31}, \bibinfo{number}{4} (\bibinfo{year}{2022}),
  \bibinfo{pages}{57:1--57:69}.
\newblock
\href{https://doi.org/10.1145/3477579}{doi:\nolinkurl{10.1145/3477579}}


\bibitem[Beyer et~al\mbox{.}(2015)]%
        {kInduction}
\bibfield{author}{\bibinfo{person}{D. Beyer}, \bibinfo{person}{M. Dangl}, {and}
  \bibinfo{person}{P. Wendler}.} \bibinfo{year}{2015}\natexlab{}.
\newblock \showarticletitle{Boosting k-Induction with Continuously-Refined
  Invariants}. In \bibinfo{booktitle}{\emph{Proc.\ CAV}}
  \emph{(\bibinfo{series}{LNCS~9206})}. \bibinfo{publisher}{Springer},
  \bibinfo{pages}{622--640}.
\newblock
\href{https://doi.org/10.1007/978-3-319-21690-4_42}{doi:\nolinkurl{10.1007/978-3-319-21690-4_42}}


\bibitem[Beyer et~al\mbox{.}(2018)]%
        {AlgorithmComparison-JAR}
\bibfield{author}{\bibinfo{person}{D. Beyer}, \bibinfo{person}{M. Dangl}, {and}
  \bibinfo{person}{P. Wendler}.} \bibinfo{year}{2018}\natexlab{}.
\newblock \showarticletitle{A Unifying View on {SMT}-Based Software
  Verification}.
\newblock \bibinfo{journal}{\emph{J. Autom. Reasoning}} \bibinfo{volume}{60},
  \bibinfo{number}{3} (\bibinfo{year}{2018}), \bibinfo{pages}{299--335}.
\newblock
\showISSN{1573-0670}
\href{https://doi.org/10.1007/s10817-017-9432-6}{doi:\nolinkurl{10.1007/s10817-017-9432-6}}


\bibitem[Beyer et~al\mbox{.}(2025a)]%
        {SV-LIB-1.0}
\bibfield{author}{\bibinfo{person}{D. Beyer}, \bibinfo{person}{G. Ernst},
  \bibinfo{person}{M. Jonáš}, {and} \bibinfo{person}{M. Lingsch-Rosenfeld}.}
  \bibinfo{year}{2025}\natexlab{a}.
\newblock \showarticletitle{{SV-LIB} 1.0: {A} Standard Exchange Format for
  Software-Verification Tasks}.
\newblock \bibinfo{journal}{\emph{arXiv/CoRR}} \bibinfo{volume}{2511},
  \bibinfo{number}{21509} (\bibinfo{date}{December} \bibinfo{year}{2025}).
\newblock
\href{https://doi.org/10.48550/arXiv.2511.21509}{doi:\nolinkurl{10.48550/arXiv.2511.21509}}


\bibitem[Beyer et~al\mbox{.}(2007)]%
        {CPA}
\bibfield{author}{\bibinfo{person}{D. Beyer}, \bibinfo{person}{T.~A.
  Henzinger}, {and} \bibinfo{person}{G. Th{\'e}oduloz}.}
  \bibinfo{year}{2007}\natexlab{}.
\newblock \showarticletitle{Configurable Software Verification: Concretizing
  the Convergence of Model Checking and Program Analysis}. In
  \bibinfo{booktitle}{\emph{Proc.\ CAV}} \emph{(\bibinfo{series}{LNCS~4590})}.
  \bibinfo{publisher}{Springer}, \bibinfo{pages}{504--518}.
\newblock
\href{https://doi.org/10.1007/978-3-540-73368-3_51}{doi:\nolinkurl{10.1007/978-3-540-73368-3_51}}


\bibitem[Beyer et~al\mbox{.}(2026)]%
        {WitnessesTransitionInvariants}
\bibfield{author}{\bibinfo{person}{D. Beyer}, \bibinfo{person}{M. Jankola},
  {and} \bibinfo{person}{M. Lingsch-Rosenfeld}.}
  \bibinfo{year}{2026}\natexlab{}.
\newblock \bibinfo{title}{Transition Invariants Revisited: {Termination}
  Witnesses and Their Validation}.
\newblock \bibinfo{numpages}{28--53}~pages.
\newblock
\href{https://doi.org/10.1007/978-3-032-32537-2_2}{doi:\nolinkurl{10.1007/978-3-032-32537-2_2}}


\bibitem[Beyer et~al\mbox{.}(2025b)]%
        {TRANSVER}
\bibfield{author}{\bibinfo{person}{D. Beyer}, \bibinfo{person}{M. Jankola},
  \bibinfo{person}{M. Lingsch-Rosenfeld}, \bibinfo{person}{T. Xia}, {and}
  \bibinfo{person}{X. Zheng}.} \bibinfo{year}{2025}\natexlab{b}.
\newblock \showarticletitle{\textsc{TransVer}: A Modular Program-Transformation
  Framework for Reduction to Reachability}. In \bibinfo{booktitle}{\emph{Proc.\
  SPIN}} \emph{(\bibinfo{series}{LNCS~15945})}. \bibinfo{publisher}{Springer},
  \bibinfo{pages}{1--24}.
\newblock
\href{https://doi.org/10.1007/978-3-032-06847-7_1}{doi:\nolinkurl{10.1007/978-3-032-06847-7_1}}


\bibitem[Beyer and Kanav(2022)]%
        {COVERITEAM}
\bibfield{author}{\bibinfo{person}{D. Beyer} {and} \bibinfo{person}{S. Kanav}.}
  \bibinfo{year}{2022}\natexlab{}.
\newblock \showarticletitle{\textsc{CoVeriTeam}: {O}n-Demand Composition of
  Cooperative Verification Systems}. In \bibinfo{booktitle}{\emph{Proc.\
  TACAS}} \emph{(\bibinfo{series}{LNCS~13243})}. \bibinfo{publisher}{Springer},
  \bibinfo{pages}{561--579}.
\newblock
\href{https://doi.org/10.1007/978-3-030-99524-9_31}{doi:\nolinkurl{10.1007/978-3-030-99524-9_31}}


\bibitem[Beyer and Lee(2024)]%
        {TransformationGame}
\bibfield{author}{\bibinfo{person}{D. Beyer} {and} \bibinfo{person}{N.-Z.
  Lee}.} \bibinfo{year}{2024}\natexlab{}.
\newblock \showarticletitle{The Transformation Game: Joining Forces for
  Verification}. In \bibinfo{booktitle}{\emph{Principles of Verification:
  Cycling the Probabilistic Landscape}} \emph{(\bibinfo{series}{LNCS~15262})}.
  \bibinfo{publisher}{Springer}, \bibinfo{pages}{175--205}.
\newblock
\href{https://doi.org/10.1007/978-3-031-75778-5_9}{doi:\nolinkurl{10.1007/978-3-031-75778-5_9}}


\bibitem[Beyer et~al\mbox{.}(2025c)]%
        {IMC-JAR}
\bibfield{author}{\bibinfo{person}{D. Beyer}, \bibinfo{person}{N.-Z. Lee},
  {and} \bibinfo{person}{P. Wendler}.} \bibinfo{year}{2025}\natexlab{c}.
\newblock \showarticletitle{Interpolation and {SAT}-Based Model Checking
  Revisited: {Adoption} to Software Verification}.
\newblock \bibinfo{journal}{\emph{J. Autom. Reasoning}} \bibinfo{volume}{69},
  \bibinfo{number}{1} (\bibinfo{year}{2025}), \bibinfo{pages}{5}.
\newblock
\href{https://doi.org/10.1007/s10817-024-09702-9}{doi:\nolinkurl{10.1007/s10817-024-09702-9}}


\bibitem[Beyer and Lingsch-Rosenfeld(2025)]%
        {CPACHECKER-VALIDATOR-SVCOMP25}
\bibfield{author}{\bibinfo{person}{D. Beyer} {and} \bibinfo{person}{M.
  Lingsch-Rosenfeld}.} \bibinfo{year}{2025}\natexlab{}.
\newblock \showarticletitle{\textsc{CPAchecker 4.0} as Witness Validator
  (Competition Contribution)}. In \bibinfo{booktitle}{\emph{Proc.\ TACAS~(3)}}
  \emph{(\bibinfo{series}{LNCS~15698})}. \bibinfo{publisher}{Springer},
  \bibinfo{pages}{192--198}.
\newblock
\href{https://doi.org/10.1007/978-3-031-90660-2_11}{doi:\nolinkurl{10.1007/978-3-031-90660-2_11}}


\bibitem[Beyer et~al\mbox{.}(2019)]%
        {Benchmarking-STTT}
\bibfield{author}{\bibinfo{person}{D. Beyer}, \bibinfo{person}{S. Löwe}, {and}
  \bibinfo{person}{P. Wendler}.} \bibinfo{year}{2019}\natexlab{}.
\newblock \showarticletitle{Reliable Benchmarking: {R}equirements and
  Solutions}.
\newblock \bibinfo{journal}{\emph{Int.\ J.\ Softw.\ Tools Technol.\ Transfer}}
  \bibinfo{volume}{21}, \bibinfo{number}{1} (\bibinfo{year}{2019}),
  \bibinfo{pages}{1--29}.
\newblock
\href{https://doi.org/10.1007/s10009-017-0469-y}{doi:\nolinkurl{10.1007/s10009-017-0469-y}}


\bibitem[Beyer and Strejček(2025)]%
        {SVCOMP25-SVBENCHMARKS-artifact}
\bibfield{author}{\bibinfo{person}{D. Beyer} {and} \bibinfo{person}{J.
  Strejček}.} \bibinfo{year}{2025}\natexlab{}.
\newblock \bibinfo{title}{{SV-Benchmarks}: {Benchmark} Set for Software
  Verification ({SV-COMP~2025})}.
\newblock \bibinfo{howpublished}{Zenodo}.
\newblock
\href{https://doi.org/10.5281/zenodo.15012096}{doi:\nolinkurl{10.5281/zenodo.15012096}}


\bibitem[Beyer and Strejček(2026a)]%
        {SVCOMP26}
\bibfield{author}{\bibinfo{person}{D. Beyer} {and} \bibinfo{person}{J.
  Strejček}.} \bibinfo{year}{2026}\natexlab{a}.
\newblock \showarticletitle{Evaluating Software Verifiers for {C}, {Java}, and
  {SV-LIB} ({Report} on {SV-COMP 2026})}. In \bibinfo{booktitle}{\emph{Proc.\
  TACAS~(2)}} \emph{(\bibinfo{series}{LNCS~16506})}.
  \bibinfo{publisher}{Springer}, \bibinfo{pages}{461--502}.
\newblock
\href{https://doi.org/10.1007/978-3-032-22749-2_23}{doi:\nolinkurl{10.1007/978-3-032-22749-2_23}}


\bibitem[Beyer and Strejček(2026b)]%
        {SVCOMP26-SVBENCHMARKS-artifact}
\bibfield{author}{\bibinfo{person}{D. Beyer} {and} \bibinfo{person}{J.
  Strejček}.} \bibinfo{year}{2026}\natexlab{b}.
\newblock \bibinfo{title}{{SV-Benchmarks}: {Benchmark} Set for Software
  Verification and Testing ({SV-COMP~2026}, {Test-Comp~2026})}.
\newblock \bibinfo{howpublished}{Zenodo}.
\newblock
\href{https://doi.org/10.5281/zenodo.18650775}{doi:\nolinkurl{10.5281/zenodo.18650775}}


\bibitem[Beyer and Wendler(2025)]%
        {CPAchecker-4.2.2}
\bibfield{author}{\bibinfo{person}{D. Beyer} {and} \bibinfo{person}{P.
  Wendler}.} \bibinfo{year}{2025}\natexlab{}.
\newblock \bibinfo{title}{{CPAchecker} Release 4.2.2 (unix)}.
\newblock \bibinfo{howpublished}{Zenodo}.
\newblock
\href{https://doi.org/10.5281/zenodo.17777566}{doi:\nolinkurl{10.5281/zenodo.17777566}}


\bibitem[Biere(2021)]%
        {BMChandbook}
\bibfield{author}{\bibinfo{person}{A. Biere}.} \bibinfo{year}{2021}\natexlab{}.
\newblock \showarticletitle{Bounded Model Checking}.
\newblock In \bibinfo{booktitle}{\emph{Handbook of Satisfiability - Second
  Edition}}. \bibinfo{series}{Frontiers in Artificial Intelligence and
  Applications}, Vol.~\bibinfo{volume}{336}. \bibinfo{publisher}{{IOS} Press},
  \bibinfo{pages}{739--764}.
\newblock
\href{https://doi.org/10.3233/FAIA201002}{doi:\nolinkurl{10.3233/FAIA201002}}


\bibitem[Biere et~al\mbox{.}(2002)]%
        {LivenessAsSafetyFinite}
\bibfield{author}{\bibinfo{person}{A. Biere}, \bibinfo{person}{C. Artho}, {and}
  \bibinfo{person}{V. Schuppan}.} \bibinfo{year}{2002}\natexlab{}.
\newblock \showarticletitle{Liveness Checking as Safety Checking}. In
  \bibinfo{booktitle}{\emph{Proc.\ FMICS}} \emph{(\bibinfo{series}{ENTSC~66},
  \bibinfo{number}{2})}. \bibinfo{publisher}{Elsevier},
  \bibinfo{pages}{160--177}.
\newblock
\href{https://doi.org/10.1016/S1571-0661(04)80410-9}{doi:\nolinkurl{10.1016/S1571-0661(04)80410-9}}


\bibitem[Biere et~al\mbox{.}(2003)]%
        {BMCJournal}
\bibfield{author}{\bibinfo{person}{A. Biere}, \bibinfo{person}{A. Cimatti},
  \bibinfo{person}{E.~M. Clarke}, \bibinfo{person}{O. Strichman}, {and}
  \bibinfo{person}{Y. Zhu}.} \bibinfo{year}{2003}\natexlab{}.
\newblock \showarticletitle{Bounded model checking}.
\newblock \bibinfo{journal}{\emph{Advances in Computers}}  \bibinfo{volume}{58}
  (\bibinfo{year}{2003}), \bibinfo{pages}{117--148}.
\newblock
\href{https://doi.org/10.1016/S0065-2458(03)58003-2}{doi:\nolinkurl{10.1016/S0065-2458(03)58003-2}}


\bibitem[Biere et~al\mbox{.}(1999)]%
        {BMC}
\bibfield{author}{\bibinfo{person}{A. Biere}, \bibinfo{person}{A. Cimatti},
  \bibinfo{person}{E.~M. Clarke}, {and} \bibinfo{person}{Y. Zhu}.}
  \bibinfo{year}{1999}\natexlab{}.
\newblock \showarticletitle{Symbolic Model Checking without {BDD}s}. In
  \bibinfo{booktitle}{\emph{Proc.\ TACAS}}
  \emph{(\bibinfo{series}{LNCS~1579})}. \bibinfo{publisher}{Springer},
  \bibinfo{pages}{193--207}.
\newblock
\href{https://doi.org/10.1007/3-540-49059-0_14}{doi:\nolinkurl{10.1007/3-540-49059-0_14}}


\bibitem[Biere et~al\mbox{.}(2024a)]%
        {CaDiCaL-CAV24}
\bibfield{author}{\bibinfo{person}{A. Biere}, \bibinfo{person}{T. Faller},
  \bibinfo{person}{K. Fazekas}, \bibinfo{person}{M. Fleury},
  \bibinfo{person}{N. Froleyks}, {and} \bibinfo{person}{F. Pollitt}.}
  \bibinfo{year}{2024}\natexlab{a}.
\newblock \showarticletitle{{CaDiCaL} 2.0}. In \bibinfo{booktitle}{\emph{Proc.\
  CAV~(1)}} \emph{(\bibinfo{series}{LNCS~14681})}.
  \bibinfo{publisher}{Springer}, \bibinfo{pages}{133--152}.
\newblock
\href{https://doi.org/10.1007/978-3-031-65627-9\_7}{doi:\nolinkurl{10.1007/978-3-031-65627-9\_7}}


\bibitem[Biere et~al\mbox{.}(2020)]%
        {Kissat-SAT20}
\bibfield{author}{\bibinfo{person}{A. Biere}, \bibinfo{person}{K. Fazekas},
  \bibinfo{person}{M. Fleury}, {and} \bibinfo{person}{M. Heisinger}.}
  \bibinfo{year}{2020}\natexlab{}.
\newblock \showarticletitle{{CaDiCaL}, {Kissat}, {Paracooba}, {Plingeling} and
  {Treengeling} Entering the {SAT Competition 2020}}. In
  \bibinfo{booktitle}{\emph{Proc.~ {SAT Competition} -- Solver and Benchmark
  Descriptions}} \emph{(\bibinfo{series}{Department of Computer Science Report
  Series B}, Vol.~\bibinfo{volume}{B-2020-1})}. \bibinfo{publisher}{University
  of Helsinki}, \bibinfo{pages}{51--53}.
\newblock
\urldef\tempurl%
\url{https://tuhat.helsinki.fi/ws/files/142452772/sc2020_proceedings.pdf.#page=50}
\showURL{%
\tempurl}


\bibitem[Biere et~al\mbox{.}(2024b)]%
        {HWMCC24}
\bibfield{author}{\bibinfo{person}{A. Biere}, \bibinfo{person}{N. Froleyks},
  {and} \bibinfo{person}{M. Preiner}.} \bibinfo{year}{2024}\natexlab{b}.
\newblock \showarticletitle{Hardware Model Checking Competition 2024}. In
  \bibinfo{booktitle}{\emph{Proc.\ FMCAD}}. \bibinfo{publisher}{TU Wien
  Academic Press}, \bibinfo{pages}{7--7}.
\newblock
\href{https://doi.org/10.34727/2024/isbn.978-3-85448-065-5_6}{doi:\nolinkurl{10.34727/2024/isbn.978-3-85448-065-5_6}}


\bibitem[Biere et~al\mbox{.}(2025)]%
        {HWMCC25}
\bibfield{author}{\bibinfo{person}{A. Biere}, \bibinfo{person}{N. Froleyks},
  {and} \bibinfo{person}{M. Preiner}.} \bibinfo{year}{2025}\natexlab{}.
\newblock \showarticletitle{Hardware Model Checking Competition 2025}. In
  \bibinfo{booktitle}{\emph{Proc.\ FMCAD}}. \bibinfo{publisher}{TU Wien
  Academic Press}, \bibinfo{pages}{7}.
\newblock
\href{https://doi.org/10.34727/2025/isbn.978-3-85448-084-6\_6}{doi:\nolinkurl{10.34727/2025/isbn.978-3-85448-084-6\_6}}


\bibitem[Biere et~al\mbox{.}(2011)]%
        {AIGER-1.9}
\bibfield{author}{\bibinfo{person}{A. Biere}, \bibinfo{person}{K. Heljanko},
  {and} \bibinfo{person}{S. Wieringa}.} \bibinfo{year}{2011}\natexlab{}.
\newblock \bibinfo{booktitle}{\emph{{AIGER 1.9} And Beyond}}.
\newblock \bibinfo{type}{{T}echnical {R}eport} 11/2.
  \bibinfo{institution}{Institute for Formal Models and Verification, Johannes
  Kepler University}.
\newblock
\href{https://doi.org/10.35011/fmvtr.2011-2}{doi:\nolinkurl{10.35011/fmvtr.2011-2}}


\bibitem[Biere and Kröning(2018)]%
        {HBMC-SAT-MC}
\bibfield{author}{\bibinfo{person}{A. Biere} {and} \bibinfo{person}{D.
  Kröning}.} \bibinfo{year}{2018}\natexlab{}.
\newblock \showarticletitle{{SAT}-Based Model Checking}.
\newblock In \bibinfo{booktitle}{\emph{Handbook of Model Checking}}.
  \bibinfo{publisher}{Springer}, \bibinfo{pages}{277--303}.
\newblock
\href{https://doi.org/10.1007/978-3-319-10575-8_10}{doi:\nolinkurl{10.1007/978-3-319-10575-8_10}}


\bibitem[Bj{\o}rner et~al\mbox{.}(2015)]%
        {CHC}
\bibfield{author}{\bibinfo{person}{N.~S. Bj{\o}rner}, \bibinfo{person}{A.
  Gurfinkel}, \bibinfo{person}{K.~L. McMillan}, {and} \bibinfo{person}{A.
  Rybalchenko}.} \bibinfo{year}{2015}\natexlab{}.
\newblock \showarticletitle{Horn Clause Solvers for Program Verification}. In
  \bibinfo{booktitle}{\emph{Fields of Logic and Computation {II} - Essays
  Dedicated to Yuri Gurevich on the Occasion of His 75th Birthday}}
  \emph{(\bibinfo{series}{LNCS~9300})}. \bibinfo{publisher}{Springer},
  \bibinfo{pages}{24--51}.
\newblock
\href{https://doi.org/10.1007/978-3-319-23534-9_2}{doi:\nolinkurl{10.1007/978-3-319-23534-9_2}}


\bibitem[Bradley(2011)]%
        {IC3}
\bibfield{author}{\bibinfo{person}{A.~R. Bradley}.}
  \bibinfo{year}{2011}\natexlab{}.
\newblock \showarticletitle{{SAT}-Based model checking without unrolling}. In
  \bibinfo{booktitle}{\emph{Proc.\ VMCAI}}
  \emph{(\bibinfo{series}{LNCS~6538})}. \bibinfo{publisher}{Springer},
  \bibinfo{pages}{70--87}.
\newblock
\href{https://doi.org/10.1007/978-3-642-18275-4_7}{doi:\nolinkurl{10.1007/978-3-642-18275-4_7}}


\bibitem[Bradley et~al\mbox{.}(2011)]%
        {FAIR-FMCAD11}
\bibfield{author}{\bibinfo{person}{A.~R. Bradley}, \bibinfo{person}{F.
  Somenzi}, \bibinfo{person}{Z. Hassan}, {and} \bibinfo{person}{Y. Zhang}.}
  \bibinfo{year}{2011}\natexlab{}.
\newblock \showarticletitle{An Incremental Approach to Model Checking Progress
  Properties}. In \bibinfo{booktitle}{\emph{Proc.\ FMCAD}}.
  \bibinfo{publisher}{{FMCAD} Inc.}, \bibinfo{pages}{144--153}.
\newblock
\urldef\tempurl%
\url{https://dl.acm.org/doi/10.5555/2157654.2157677}
\showURL{%
\tempurl}


\bibitem[Brayton and Mishchenko(2010)]%
        {ABC}
\bibfield{author}{\bibinfo{person}{R. Brayton} {and} \bibinfo{person}{A.
  Mishchenko}.} \bibinfo{year}{2010}\natexlab{}.
\newblock \showarticletitle{{ABC}: An Academic Industrial-Strength Verification
  Tool}. In \bibinfo{booktitle}{\emph{Proc.\ CAV}}
  \emph{(\bibinfo{series}{LNCS~6174})}. \bibinfo{publisher}{Springer},
  \bibinfo{pages}{24--40}.
\newblock
\href{https://doi.org/10.1007/978-3-642-14295-6_5}{doi:\nolinkurl{10.1007/978-3-642-14295-6_5}}


\bibitem[Cadar et~al\mbox{.}(2008)]%
        {KLEE}
\bibfield{author}{\bibinfo{person}{C. Cadar}, \bibinfo{person}{D. Dunbar},
  {and} \bibinfo{person}{D.~R. Engler}.} \bibinfo{year}{2008}\natexlab{}.
\newblock \showarticletitle{\textsc{Klee:} {U}nassisted and Automatic
  Generation of High-Coverage Tests for Complex Systems Programs}. In
  \bibinfo{booktitle}{\emph{Proc.\ {OSDI}}}. \bibinfo{publisher}{{USENIX}
  Association}, \bibinfo{pages}{209--224}.
\newblock
\urldef\tempurl%
\url{https://dl.acm.org/doi/10.5555/1855741.1855756}
\showURL{%
\tempurl}


\bibitem[Cavada et~al\mbox{.}(2014)]%
        {NuXMV}
\bibfield{author}{\bibinfo{person}{R. Cavada}, \bibinfo{person}{A. Cimatti},
  \bibinfo{person}{M. Dorigatti}, \bibinfo{person}{A. Griggio},
  \bibinfo{person}{A. Mariotti}, \bibinfo{person}{A. Micheli},
  \bibinfo{person}{S. Mover}, \bibinfo{person}{M. Roveri}, {and}
  \bibinfo{person}{S. Tonetta}.} \bibinfo{year}{2014}\natexlab{}.
\newblock \showarticletitle{The \textsc{nuxmv} Symbolic Model Checker}. In
  \bibinfo{booktitle}{\emph{Proc.\ CAV}} \emph{(\bibinfo{series}{LNCS~8559})}.
  \bibinfo{publisher}{Springer}, \bibinfo{pages}{334--342}.
\newblock
\href{https://doi.org/10.1007/978-3-319-08867-9_22}{doi:\nolinkurl{10.1007/978-3-319-08867-9_22}}


\bibitem[Chien and Jiang(2019)]%
        {TimeFrameFolding}
\bibfield{author}{\bibinfo{person}{P.-C. Chien} {and} \bibinfo{person}{J.-H.~R.
  Jiang}.} \bibinfo{year}{2019}\natexlab{}.
\newblock \showarticletitle{Time-Frame Folding: {Back} to the Sequentiality}.
  In \bibinfo{booktitle}{\emph{Proc.\ ICCAD}}. \bibinfo{publisher}{{IEEE}},
  \bibinfo{pages}{1--8}.
\newblock
\href{https://doi.org/10.1109/ICCAD45719.2019.8942078}{doi:\nolinkurl{10.1109/ICCAD45719.2019.8942078}}


\bibitem[Chien and Lee({[n.\,d.]})]%
        {CPV-latest}
\bibfield{author}{\bibinfo{person}{P.-C. Chien} {and} \bibinfo{person}{N.-Z.
  Lee}.} \bibinfo{year}{[n.\,d.]}\natexlab{}.
\newblock \bibinfo{title}{CPV Releases}.
\newblock \bibinfo{howpublished}{Zenodo}.
\newblock
\href{https://doi.org/10.5281/zenodo.10063681}{doi:\nolinkurl{10.5281/zenodo.10063681}}


\bibitem[Chien and Lee(2024)]%
        {CPV-SVCOMP24}
\bibfield{author}{\bibinfo{person}{P.-C. Chien} {and} \bibinfo{person}{N.-Z.
  Lee}.} \bibinfo{year}{2024}\natexlab{}.
\newblock \showarticletitle{\textsc{CPV}: {A} Circuit-Based Program Verifier
  (Competition Contribution)}. In \bibinfo{booktitle}{\emph{Proc.\ TACAS~(3)}}
  \emph{(\bibinfo{series}{LNCS~14572})}. \bibinfo{publisher}{Springer},
  \bibinfo{pages}{365--370}.
\newblock
\href{https://doi.org/10.1007/978-3-031-57256-2_22}{doi:\nolinkurl{10.1007/978-3-031-57256-2_22}}


\bibitem[Chien and Lee(2026)]%
        {CPV-1.1}
\bibfield{author}{\bibinfo{person}{P.-C. Chien} {and} \bibinfo{person}{N.-Z.
  Lee}.} \bibinfo{year}{2026}\natexlab{}.
\newblock \bibinfo{title}{CPV Release 1.1}.
\newblock \bibinfo{howpublished}{Zenodo}.
\newblock
\href{https://doi.org/10.5281/zenodo.18773927}{doi:\nolinkurl{10.5281/zenodo.18773927}}


\bibitem[.Cimatti and Griggio(2012)]%
        {SoftwareIC3}
\bibfield{author}{\bibinfo{person}{A .Cimatti} {and} \bibinfo{person}{A.
  Griggio}.} \bibinfo{year}{2012}\natexlab{}.
\newblock \showarticletitle{Software Model Checking via {IC3}}. In
  \bibinfo{booktitle}{\emph{Proc.\ CAV}} \emph{(\bibinfo{series}{LNCS~7358})}.
  \bibinfo{publisher}{Springer}, \bibinfo{pages}{277--293}.
\newblock
\href{https://doi.org/10.1007/978-3-642-31424-7_23}{doi:\nolinkurl{10.1007/978-3-642-31424-7_23}}


\bibitem[Cimatti et~al\mbox{.}(2025)]%
        {rlive-inf}
\bibfield{author}{\bibinfo{person}{A. Cimatti}, \bibinfo{person}{A. Griggio},
  \bibinfo{person}{C. Johannsen}, \bibinfo{person}{K.~Y. Rozier}, {and}
  \bibinfo{person}{S. Tonetta}.} \bibinfo{year}{2025}\natexlab{}.
\newblock \showarticletitle{Infinite-State Liveness Checking with rlive}. In
  \bibinfo{booktitle}{\emph{Proc.\ CAV~(1)}}
  \emph{(\bibinfo{series}{LNCS~15931})}. \bibinfo{publisher}{Springer},
  \bibinfo{pages}{215--236}.
\newblock
\href{https://doi.org/10.1007/978-3-031-98668-0_11}{doi:\nolinkurl{10.1007/978-3-031-98668-0_11}}


\bibitem[Cimatti et~al\mbox{.}(2022b)]%
        {VMT}
\bibfield{author}{\bibinfo{person}{A. Cimatti}, \bibinfo{person}{A. Griggio},
  \bibinfo{person}{S. Mover}, \bibinfo{person}{M. Roveri}, {and}
  \bibinfo{person}{S. Tonetta}.} \bibinfo{year}{2022}\natexlab{b}.
\newblock \showarticletitle{Verification modulo theories}.
\newblock \bibinfo{journal}{\emph{Formal Methods Syst. Des.}}
  \bibinfo{volume}{60}, \bibinfo{number}{3} (\bibinfo{year}{2022}),
  \bibinfo{pages}{452--481}.
\newblock
\href{https://doi.org/10.1007/S10703-023-00434-X}{doi:\nolinkurl{10.1007/S10703-023-00434-X}}


\bibitem[Cimatti et~al\mbox{.}(2014)]%
        {PredAbsPDR}
\bibfield{author}{\bibinfo{person}{A. Cimatti}, \bibinfo{person}{A. Griggio},
  \bibinfo{person}{S. Mover}, {and} \bibinfo{person}{S. Tonetta}.}
  \bibinfo{year}{2014}\natexlab{}.
\newblock \showarticletitle{{IC3} Modulo Theories via Implicit Predicate
  Abstraction}. In \bibinfo{booktitle}{\emph{Proc.\ TACAS}}
  \emph{(\bibinfo{series}{LNCS~8413})}. \bibinfo{publisher}{Springer},
  \bibinfo{pages}{46--61}.
\newblock
\href{https://doi.org/10.1007/978-3-642-54862-8_4}{doi:\nolinkurl{10.1007/978-3-642-54862-8_4}}


\bibitem[Cimatti et~al\mbox{.}(2013)]%
        {MATHSAT5}
\bibfield{author}{\bibinfo{person}{A. Cimatti}, \bibinfo{person}{A. Griggio},
  \bibinfo{person}{B.~J. Schaafsma}, {and} \bibinfo{person}{R. Sebastiani}.}
  \bibinfo{year}{2013}\natexlab{}.
\newblock \showarticletitle{The \textsc{MathSAT5} {SMT} Solver}. In
  \bibinfo{booktitle}{\emph{Proc.\ TACAS}}
  \emph{(\bibinfo{series}{LNCS~7795})}. \bibinfo{publisher}{Springer},
  \bibinfo{pages}{93--107}.
\newblock
\href{https://doi.org/10.1007/978-3-642-36742-7_7}{doi:\nolinkurl{10.1007/978-3-642-36742-7_7}}


\bibitem[Cimatti et~al\mbox{.}(2022a)]%
        {VMT-LIB}
\bibfield{author}{\bibinfo{person}{A. Cimatti}, \bibinfo{person}{A. Griggio},
  {and} \bibinfo{person}{S. Tonetta}.} \bibinfo{year}{2022}\natexlab{a}.
\newblock \showarticletitle{The {VMT-LIB} Language and Tools}. In
  \bibinfo{booktitle}{\emph{Proc.\ SMT}} \emph{(\bibinfo{series}{{CEUR}
  Workshop Proceedings}, Vol.~\bibinfo{volume}{3185})}.
  \bibinfo{publisher}{CEUR-WS.org}, \bibinfo{pages}{80--89}.
\newblock
\urldef\tempurl%
\url{https://ceur-ws.org/Vol-3185/extended9547.pdf}
\showURL{%
\tempurl}


\bibitem[Claessen and Sörensson(2012)]%
        {kliveness}
\bibfield{author}{\bibinfo{person}{K. Claessen} {and} \bibinfo{person}{N.
  Sörensson}.} \bibinfo{year}{2012}\natexlab{}.
\newblock \showarticletitle{A Liveness Checking Algorithm that Counts}. In
  \bibinfo{booktitle}{\emph{Proc.\ FMCAD}}. \bibinfo{publisher}{IEEE},
  \bibinfo{pages}{52--59}.
\newblock
\urldef\tempurl%
\url{https://ieeexplore.ieee.org/document/6462555}
\showURL{%
\tempurl}


\bibitem[Clarke and Kröning(2003)]%
        {HV-Cref}
\bibfield{author}{\bibinfo{person}{E. Clarke} {and} \bibinfo{person}{D.
  Kröning}.} \bibinfo{year}{2003}\natexlab{}.
\newblock \showarticletitle{Hardware verification using {ANSI-C} programs as a
  reference}. In \bibinfo{booktitle}{\emph{Proc.\ ASP-DAC}}.
  \bibinfo{publisher}{ACM}, \bibinfo{pages}{308--311}.
\newblock
\href{https://doi.org/10.1145/1119772.1119831}{doi:\nolinkurl{10.1145/1119772.1119831}}


\bibitem[Clarke et~al\mbox{.}(2003)]%
        {ClarkeCEGAR}
\bibfield{author}{\bibinfo{person}{E.~M. Clarke}, \bibinfo{person}{O.
  Grumberg}, \bibinfo{person}{S. Jha}, \bibinfo{person}{Y. Lu}, {and}
  \bibinfo{person}{H. Veith}.} \bibinfo{year}{2003}\natexlab{}.
\newblock \showarticletitle{Counterexample-guided abstraction refinement for
  symbolic model checking}.
\newblock \bibinfo{journal}{\emph{J. {ACM}}} \bibinfo{volume}{50},
  \bibinfo{number}{5} (\bibinfo{year}{2003}), \bibinfo{pages}{752--794}.
\newblock
\href{https://doi.org/10.1145/876638.876643}{doi:\nolinkurl{10.1145/876638.876643}}


\bibitem[Clarke et~al\mbox{.}(2004)]%
        {CBMC}
\bibfield{author}{\bibinfo{person}{E.~M. Clarke}, \bibinfo{person}{D.
  Kröning}, {and} \bibinfo{person}{F. Lerda}.}
  \bibinfo{year}{2004}\natexlab{}.
\newblock \showarticletitle{A Tool for Checking {ANSI-C} Programs}. In
  \bibinfo{booktitle}{\emph{Proc.\ TACAS}}
  \emph{(\bibinfo{series}{LNCS~2988})}. \bibinfo{publisher}{Springer},
  \bibinfo{pages}{168--176}.
\newblock
\href{https://doi.org/10.1007/978-3-540-24730-2_15}{doi:\nolinkurl{10.1007/978-3-540-24730-2_15}}


\bibitem[Cordeiro et~al\mbox{.}(2012)]%
        {BMCEmbeddedC}
\bibfield{author}{\bibinfo{person}{L.~C. Cordeiro}, \bibinfo{person}{B.
  Fischer}, {and} \bibinfo{person}{J. Marques-Silva}.}
  \bibinfo{year}{2012}\natexlab{}.
\newblock \showarticletitle{{SMT}-Based Bounded Model Checking for Embedded
  {ANSI-C} Software}.
\newblock \bibinfo{journal}{\emph{{IEEE} Trans. Software Eng.}}
  \bibinfo{volume}{38}, \bibinfo{number}{4} (\bibinfo{year}{2012}),
  \bibinfo{pages}{957--974}.
\newblock
\href{https://doi.org/10.1109/TSE.2011.59}{doi:\nolinkurl{10.1109/TSE.2011.59}}


\bibitem[DeLine and Leino(2005)]%
        {Boogie}
\bibfield{author}{\bibinfo{person}{R. DeLine} {and} \bibinfo{person}{R.
  Leino}.} \bibinfo{year}{2005}\natexlab{}.
\newblock \bibinfo{booktitle}{\emph{{BoogiePL}: {A} Typed Procedural Language
  for Checking Object-Oriented Programs}}.
\newblock \bibinfo{type}{{T}echnical {R}eport} MSR-TR-2005-70.
  \bibinfo{institution}{Microsoft Research}.
\newblock
\urldef\tempurl%
\url{https://www.microsoft.com/en-us/research/publication/boogiepl-a-typed-procedural-language-for-checking-object-oriented-programs/}
\showURL{%
\tempurl}


\bibitem[Dietsch et~al\mbox{.}(2025)]%
        {UAutomizer-SVCOMP26-archive}
\bibfield{author}{\bibinfo{person}{D. Dietsch}, \bibinfo{person}{M. Bentele},
  \bibinfo{person}{M. Ebbinghaus}, \bibinfo{person}{M. Heizmann},
  \bibinfo{person}{D. Klumpp}, \bibinfo{person}{A. Podelski}, {and}
  \bibinfo{person}{F. Schüssele}.} \bibinfo{year}{2025}\natexlab{}.
\newblock \bibinfo{title}{{Ultimate Automizer} {SV-COMP} 2026}.
\newblock \bibinfo{howpublished}{Zenodo}.
\newblock
\href{https://doi.org/10.5281/zenodo.17735224}{doi:\nolinkurl{10.5281/zenodo.17735224}}


\bibitem[Donaldson et~al\mbox{.}(2011)]%
        {K-Induction}
\bibfield{author}{\bibinfo{person}{A.~F. Donaldson}, \bibinfo{person}{L.
  Haller}, \bibinfo{person}{D. Kröning}, {and} \bibinfo{person}{P.
  R{\"{u}}mmer}.} \bibinfo{year}{2011}\natexlab{}.
\newblock \showarticletitle{Software Verification Using $k$-Induction}. In
  \bibinfo{booktitle}{\emph{Proc.\ SAS}} \emph{(\bibinfo{series}{LNCS~6887})}.
  \bibinfo{publisher}{Springer}, \bibinfo{pages}{351--368}.
\newblock
\href{https://doi.org/10.1007/978-3-642-23702-7_26}{doi:\nolinkurl{10.1007/978-3-642-23702-7_26}}


\bibitem[Dutertre(2014)]%
        {Yices2}
\bibfield{author}{\bibinfo{person}{B. Dutertre}.}
  \bibinfo{year}{2014}\natexlab{}.
\newblock \showarticletitle{\textsc{Yices}\ 2.2}. In
  \bibinfo{booktitle}{\emph{Proc.\ {CAV}}}
  \emph{(\bibinfo{series}{LNCS~8559})}. \bibinfo{publisher}{Springer},
  \bibinfo{pages}{737--744}.
\newblock
\href{https://doi.org/10.1007/978-3-319-08867-9_49}{doi:\nolinkurl{10.1007/978-3-319-08867-9_49}}


\bibitem[Edwards et~al\mbox{.}(2001)]%
        {Edwards01}
\bibfield{author}{\bibinfo{person}{S.~A. Edwards}, \bibinfo{person}{T. Ma},
  {and} \bibinfo{person}{R. Damiano}.} \bibinfo{year}{2001}\natexlab{}.
\newblock \showarticletitle{Using a hardware model checker to verify software}.
  In \bibinfo{booktitle}{\emph{Proc.\ ASICON}}. \bibinfo{publisher}{IEEE},
  \bibinfo{pages}{85--90}.
\newblock
\href{https://doi.org/10.1109/ICASIC.2001.982504}{doi:\nolinkurl{10.1109/ICASIC.2001.982504}}


\bibitem[E{\'{e}}n et~al\mbox{.}(2011)]%
        {EfficientPDR}
\bibfield{author}{\bibinfo{person}{N. E{\'{e}}n}, \bibinfo{person}{A.
  Mishchenko}, {and} \bibinfo{person}{R.~K. Brayton}.}
  \bibinfo{year}{2011}\natexlab{}.
\newblock \showarticletitle{Efficient Implementation of Property Directed
  Reachability}. In \bibinfo{booktitle}{\emph{Proc.\ FMCAD}}.
  \bibinfo{publisher}{{FMCAD} Inc.}, \bibinfo{pages}{125--134}.
\newblock
\urldef\tempurl%
\url{https://dl.acm.org/doi/10.5555/2157654.2157675}
\showURL{%
\tempurl}


\bibitem[E{\'{e}}n and S{\"{o}}rensson(2003)]%
        {MiniSat}
\bibfield{author}{\bibinfo{person}{N. E{\'{e}}n} {and} \bibinfo{person}{N.
  S{\"{o}}rensson}.} \bibinfo{year}{2003}\natexlab{}.
\newblock \showarticletitle{An Extensible {SAT}-solver}. In
  \bibinfo{booktitle}{\emph{Proc.\ {SAT}}}
  \emph{(\bibinfo{series}{LNCS~2919})}. \bibinfo{publisher}{Springer},
  \bibinfo{pages}{502--518}.
\newblock
\href{https://doi.org/10.1007/978-3-540-24605-3_37}{doi:\nolinkurl{10.1007/978-3-540-24605-3_37}}


\bibitem[Falke et~al\mbox{.}(2013)]%
        {LLBMC}
\bibfield{author}{\bibinfo{person}{S. Falke}, \bibinfo{person}{F. Merz}, {and}
  \bibinfo{person}{C .Sinz}.} \bibinfo{year}{2013}\natexlab{}.
\newblock \showarticletitle{The bounded model checker \textsc{LLBMC}}. In
  \bibinfo{booktitle}{\emph{Proc.\ ASE}}. \bibinfo{publisher}{{IEEE}},
  \bibinfo{pages}{706--709}.
\newblock
\href{https://doi.org/10.1109/ASE.2013.6693138}{doi:\nolinkurl{10.1109/ASE.2013.6693138}}


\bibitem[Filli{\^a}tre and Paskevich(2013)]%
        {Why3}
\bibfield{author}{\bibinfo{person}{J.-C. Filli{\^a}tre} {and}
  \bibinfo{person}{A. Paskevich}.} \bibinfo{year}{2013}\natexlab{}.
\newblock \showarticletitle{Why3: Where Programs Meet Provers}. In
  \bibinfo{booktitle}{\emph{Programming Languages and Systems}}.
  \bibinfo{publisher}{Springer}, \bibinfo{pages}{125--128}.
\newblock
\href{https://doi.org/10.1007/978-3-642-37036-6_8}{doi:\nolinkurl{10.1007/978-3-642-37036-6_8}}


\bibitem[Fink et~al\mbox{.}(2022)]%
        {HardnessBenchmarks}
\bibfield{author}{\bibinfo{person}{X. Fink}, \bibinfo{person}{P. Berger}, {and}
  \bibinfo{person}{J.-P. Katoen}.} \bibinfo{year}{2022}\natexlab{}.
\newblock \showarticletitle{Configurable Benchmarks for {C} Model Checkers}. In
  \bibinfo{booktitle}{\emph{Proc.\ NFM}} \emph{(\bibinfo{series}{LNCS~13260})}.
  \bibinfo{publisher}{Springer}, \bibinfo{pages}{338--354}.
\newblock
\href{https://doi.org/10.1007/978-3-031-06773-0_18}{doi:\nolinkurl{10.1007/978-3-031-06773-0_18}}


\bibitem[Froleyks et~al\mbox{.}(2026a)]%
        {CertConstraintsHWMC}
\bibfield{author}{\bibinfo{person}{N. Froleyks}, \bibinfo{person}{E. Yu},
  \bibinfo{person}{A. Biere}, {and} \bibinfo{person}{K. Heljanko}.}
  \bibinfo{year}{2026}\natexlab{a}.
\newblock \showarticletitle{Certifying Constraints in Hardware Model Checking}.
  In \bibinfo{booktitle}{\emph{Proc.\ FM~(1)}}
  \emph{(\bibinfo{series}{LNCS~16556})}. \bibinfo{publisher}{Springer},
  \bibinfo{pages}{170--188}.
\newblock
\href{https://doi.org/10.1007/978-3-032-26204-2_9}{doi:\nolinkurl{10.1007/978-3-032-26204-2_9}}


\bibitem[Froleyks et~al\mbox{.}(2026b)]%
        {LivenessProofsHWMC}
\bibfield{author}{\bibinfo{person}{N. Froleyks}, \bibinfo{person}{E. Yu},
  \bibinfo{person}{B. Bogaerts}, \bibinfo{person}{A. Biere}, {and}
  \bibinfo{person}{K. Heljanko}.} \bibinfo{year}{2026}\natexlab{b}.
\newblock \showarticletitle{Liveness Proofs for Hardware Model Checking}. In
  \bibinfo{booktitle}{\emph{Proc.\ CAV~(3)}}
  \emph{(\bibinfo{series}{LNCS~16684})}. \bibinfo{publisher}{Springer},
  \bibinfo{pages}{3--27}.
\newblock
\href{https://doi.org/10.1007/978-3-032-32537-2_1}{doi:\nolinkurl{10.1007/978-3-032-32537-2_1}}


\bibitem[Froleyks et~al\mbox{.}(2025)]%
        {CertificateHWMCC-CAV25}
\bibfield{author}{\bibinfo{person}{N. Froleyks}, \bibinfo{person}{E. Yu},
  \bibinfo{person}{M. Preiner}, \bibinfo{person}{A. Biere}, {and}
  \bibinfo{person}{K. Heljanko}.} \bibinfo{year}{2025}\natexlab{}.
\newblock \showarticletitle{Introducing Certificates to the Hardware Model
  Checking Competition}. In \bibinfo{booktitle}{\emph{Proc.\ CAV~(1)}}
  \emph{(\bibinfo{series}{LNCS~15931})}. \bibinfo{publisher}{Springer},
  \bibinfo{pages}{281--295}.
\newblock
\href{https://doi.org/10.1007/978-3-031-98668-0\_14}{doi:\nolinkurl{10.1007/978-3-031-98668-0\_14}}


\bibitem[Gario and Micheli(2015)]%
        {PySMT}
\bibfield{author}{\bibinfo{person}{M. Gario} {and} \bibinfo{person}{A.
  Micheli}.} \bibinfo{year}{2015}\natexlab{}.
\newblock \showarticletitle{\textsc{PySMT}: {A} solver-agnostic library for
  fast prototyping of {SMT}-Based algorithms}. In
  \bibinfo{booktitle}{\emph{Proc.\ SMT}}.
\newblock


\bibitem[Girardi et~al\mbox{.}({[n.\,d.]})]%
        {PyVMT}
\bibfield{author}{\bibinfo{person}{A. Girardi}, \bibinfo{person}{A
  .Bombardelli}, \bibinfo{person}{A. Griggio}, {and} \bibinfo{person}{J.
  Tomassi}.} \bibinfo{year}{[n.\,d.]}\natexlab{}.
\newblock \bibinfo{title}{{PyVMT}: {A} {Python} library to interact with
  transition systems}.
\newblock
\shownote{Accessed: 2025-10-15}.
\newblock
\urldef\tempurl%
\url{https://github.com/pyvmt/pyvmt}
\showURL{%
\tempurl}


\bibitem[Goel and Sakallah(2019)]%
        {IC3SA}
\bibfield{author}{\bibinfo{person}{A. Goel} {and} \bibinfo{person}{K.
  Sakallah}.} \bibinfo{year}{2019}\natexlab{}.
\newblock \showarticletitle{Model Checking of {Verilog} {RTL} Using {IC3} with
  Syntax-Guided Abstraction}. In \bibinfo{booktitle}{\emph{Proc.\ NFM}}.
  \bibinfo{publisher}{Springer}, \bibinfo{pages}{166--185}.
\newblock
\href{https://doi.org/10.1007/978-3-030-20652-9_11}{doi:\nolinkurl{10.1007/978-3-030-20652-9_11}}


\bibitem[Goel and Sakallah(2020)]%
        {AVR}
\bibfield{author}{\bibinfo{person}{A. Goel} {and} \bibinfo{person}{K.
  Sakallah}.} \bibinfo{year}{2020}\natexlab{}.
\newblock \showarticletitle{{AVR}: Abstractly Verifying Reachability}. In
  \bibinfo{booktitle}{\emph{Proc.\ TACAS}}
  \emph{(\bibinfo{series}{LNCS~12078})}. \bibinfo{publisher}{Springer},
  \bibinfo{pages}{413--422}.
\newblock
\href{https://doi.org/10.1007/978-3-030-45190-5_23}{doi:\nolinkurl{10.1007/978-3-030-45190-5_23}}


\bibitem[Graf and Sa{\"\i}di(1997)]%
        {GrafSaidi97}
\bibfield{author}{\bibinfo{person}{S. Graf} {and} \bibinfo{person}{H.
  Sa{\"\i}di}.} \bibinfo{year}{1997}\natexlab{}.
\newblock \showarticletitle{Construction of Abstract State Graphs with
  \textsc{Pvs}}. In \bibinfo{booktitle}{\emph{Proc.\ CAV}}
  \emph{(\bibinfo{series}{LNCS~1254})}. \bibinfo{publisher}{Springer},
  \bibinfo{pages}{72--83}.
\newblock
\href{https://doi.org/10.1007/3-540-63166-6_10}{doi:\nolinkurl{10.1007/3-540-63166-6_10}}


\bibitem[Griggio and Jon\'{a}\v{s}(2023)]%
        {Kratos2}
\bibfield{author}{\bibinfo{person}{A. Griggio} {and} \bibinfo{person}{M.
  Jon\'{a}\v{s}}.} \bibinfo{year}{2023}\natexlab{}.
\newblock \showarticletitle{\textsc{Kratos2}: An {SMT}-Based Model Checker for
  Imperative Programs}. In \bibinfo{booktitle}{\emph{Proc.\ CAV}}.
  \bibinfo{publisher}{Springer}, \bibinfo{pages}{423--436}.
\newblock
\showISBNx{978-3-031-37708-2}
\href{https://doi.org/10.1007/978-3-031-37709-9_20}{doi:\nolinkurl{10.1007/978-3-031-37709-9_20}}


\bibitem[Griggio and Jonáš(2023)]%
        {Kratos2-CAV23-artifact}
\bibfield{author}{\bibinfo{person}{A. Griggio} {and} \bibinfo{person}{M.
  Jonáš}.} \bibinfo{year}{2023}\natexlab{}.
\newblock \bibinfo{title}{Artifact for the {CAV}'23 tool paper ``{Kratos2}: an
  {SMT}-Based Model Checker for Imperative Programs''}.
\newblock \bibinfo{howpublished}{Zenodo}.
\newblock
\href{https://doi.org/10.5281/zenodo.7890411}{doi:\nolinkurl{10.5281/zenodo.7890411}}


\bibitem[Harris and Harris(2007)]%
        {SeqLogicDesignBookChapter}
\bibfield{author}{\bibinfo{person}{D.~M. Harris} {and} \bibinfo{person}{S.~L.
  Harris}.} \bibinfo{year}{2007}\natexlab{}.
\newblock \showarticletitle{Sequential Logic Design}.
\newblock In \bibinfo{booktitle}{\emph{Digital Design and Computer
  Architecture}}. \bibinfo{publisher}{Morgan Kaufmann},
  \bibinfo{pages}{103--165}.
\newblock
\showISBNx{978-0-12-370497-9}
\href{https://doi.org/10.1016/B978-012370497-9/50004-4}{doi:\nolinkurl{10.1016/B978-012370497-9/50004-4}}


\bibitem[Harrison(1977)]%
        {ValueRangesAnalysis}
\bibfield{author}{\bibinfo{person}{W.~H. Harrison}.}
  \bibinfo{year}{1977}\natexlab{}.
\newblock \showarticletitle{Compiler Analysis of the Value Ranges for
  Variables}.
\newblock \bibinfo{journal}{\emph{IEEE Trans.\ Softw.\ Eng.}}
  \bibinfo{volume}{SE-3}, \bibinfo{number}{3} (\bibinfo{year}{1977}),
  \bibinfo{pages}{243--250}.
\newblock
\href{https://doi.org/10.1109/TSE.1977.231133}{doi:\nolinkurl{10.1109/TSE.1977.231133}}


\bibitem[Heizmann et~al\mbox{.}(2013b)]%
        {ULTIMATE-TERMINATION}
\bibfield{author}{\bibinfo{person}{M. Heizmann}, \bibinfo{person}{J. Hoenicke},
  \bibinfo{person}{J. Leike}, {and} \bibinfo{person}{A. Podelski}.}
  \bibinfo{year}{2013}\natexlab{b}.
\newblock \showarticletitle{Linear Ranking for Linear Lasso Programs}. In
  \bibinfo{booktitle}{\emph{Proc.\ ATVA}} \emph{(\bibinfo{series}{LNCS~8172})}.
  \bibinfo{publisher}{Springer}, \bibinfo{pages}{365--380}.
\newblock
\href{https://doi.org/10.1007/978-3-319-02444-8_26}{doi:\nolinkurl{10.1007/978-3-319-02444-8_26}}


\bibitem[Heizmann et~al\mbox{.}(2013a)]%
        {UAUTOMIZER2013}
\bibfield{author}{\bibinfo{person}{M. Heizmann}, \bibinfo{person}{J. Hoenicke},
  {and} \bibinfo{person}{A. Podelski}.} \bibinfo{year}{2013}\natexlab{a}.
\newblock \showarticletitle{Software Model Checking for People Who Love
  Automata}. In \bibinfo{booktitle}{\emph{Proc.\ CAV}}
  \emph{(\bibinfo{series}{LNCS~8044})}. \bibinfo{publisher}{Springer},
  \bibinfo{pages}{36--52}.
\newblock
\href{https://doi.org/10.1007/978-3-642-39799-8_2}{doi:\nolinkurl{10.1007/978-3-642-39799-8_2}}


\bibitem[Henzinger et~al\mbox{.}(2004)]%
        {AbstractionsFromProofs}
\bibfield{author}{\bibinfo{person}{T.~A. Henzinger}, \bibinfo{person}{R.
  Jhala}, \bibinfo{person}{R. Majumdar}, {and} \bibinfo{person}{K.~L.
  McMillan}.} \bibinfo{year}{2004}\natexlab{}.
\newblock \showarticletitle{Abstractions from proofs}. In
  \bibinfo{booktitle}{\emph{Proc.\ POPL}}. \bibinfo{publisher}{{ACM}},
  \bibinfo{pages}{232--244}.
\newblock
\href{https://doi.org/10.1145/964001.964021}{doi:\nolinkurl{10.1145/964001.964021}}


\bibitem[Ho et~al\mbox{.}(2000)]%
        {Ketchum}
\bibfield{author}{\bibinfo{person}{P.-H. Ho}, \bibinfo{person}{T.~R. Shiple},
  \bibinfo{person}{K. Harer}, \bibinfo{person}{J.~H. Kukula},
  \bibinfo{person}{R.~F. Damiano}, \bibinfo{person}{V. Bertacco},
  \bibinfo{person}{J. Taylor}, {and} \bibinfo{person}{J. Long}.}
  \bibinfo{year}{2000}\natexlab{}.
\newblock \showarticletitle{Smart Simulation Using Collaborative Formal and
  Simulation Engines}. In \bibinfo{booktitle}{\emph{Proc. ICCAD}}.
  \bibinfo{publisher}{{IEEE}}, \bibinfo{pages}{120--126}.
\newblock
\href{https://doi.org/10.1109/ICCAD.2000.896461}{doi:\nolinkurl{10.1109/ICCAD.2000.896461}}


\bibitem[Howar et~al\mbox{.}(2012)]%
        {RERS12}
\bibfield{author}{\bibinfo{person}{F. Howar}, \bibinfo{person}{M. Isberner},
  \bibinfo{person}{M. Merten}, \bibinfo{person}{B. Steffen}, {and}
  \bibinfo{person}{D. Beyer}.} \bibinfo{year}{2012}\natexlab{}.
\newblock \showarticletitle{The {RERS} Grey-Box Challenge 2012: Analysis of
  Event-Condition-Action Systems}. In \bibinfo{booktitle}{\emph{Proc.\ ISoLA}}
  \emph{(\bibinfo{series}{LNCS~7609})}. \bibinfo{publisher}{Springer},
  \bibinfo{pages}{608--614}.
\newblock
\href{https://doi.org/10.1007/978-3-642-34026-0_45}{doi:\nolinkurl{10.1007/978-3-642-34026-0_45}}


\bibitem[Jhala et~al\mbox{.}(2018)]%
        {HBMC-predicateabstraction}
\bibfield{author}{\bibinfo{person}{R. Jhala}, \bibinfo{person}{A. Podelski},
  {and} \bibinfo{person}{A. Rybalchenko}.} \bibinfo{year}{2018}\natexlab{}.
\newblock \showarticletitle{Predicate Abstraction for Program Verification}.
\newblock In \bibinfo{booktitle}{\emph{Handbook of Model Checking}}.
  \bibinfo{publisher}{Springer}, \bibinfo{pages}{447--491}.
\newblock
\href{https://doi.org/10.1007/978-3-319-10575-8_15}{doi:\nolinkurl{10.1007/978-3-319-10575-8_15}}


\bibitem[Jussila and Biere(2007)]%
        {BMC-QBF}
\bibfield{author}{\bibinfo{person}{T. Jussila} {and} \bibinfo{person}{A.
  Biere}.} \bibinfo{year}{2007}\natexlab{}.
\newblock \showarticletitle{Compressing {BMC} Encodings with {QBF}}.
\newblock \bibinfo{journal}{\emph{Electronic Notes in Theoretical Computer
  Science}} \bibinfo{volume}{174}, \bibinfo{number}{3} (\bibinfo{year}{2007}),
  \bibinfo{pages}{45--56}.
\newblock
\showISSN{1571-0661}
\href{https://doi.org/10.1016/j.entcs.2006.12.022}{doi:\nolinkurl{10.1016/j.entcs.2006.12.022}}


\bibitem[King(1976)]%
        {SymbolicExecution}
\bibfield{author}{\bibinfo{person}{J.~C. King}.}
  \bibinfo{year}{1976}\natexlab{}.
\newblock \showarticletitle{Symbolic Execution and Program Testing}.
\newblock \bibinfo{journal}{\emph{Commun. ACM}} \bibinfo{volume}{19},
  \bibinfo{number}{7} (\bibinfo{year}{1976}), \bibinfo{pages}{385--394}.
\newblock
\href{https://doi.org/10.1145/360248.360252}{doi:\nolinkurl{10.1145/360248.360252}}


\bibitem[Lange et~al\mbox{.}(2020)]%
        {IC3-CFA-STTT}
\bibfield{author}{\bibinfo{person}{T. Lange}, \bibinfo{person}{M.~R.
  Neuhäußer}, \bibinfo{person}{T. Noll}, {and} \bibinfo{person}{J.-P.
  Katoen}.} \bibinfo{year}{2020}\natexlab{}.
\newblock \showarticletitle{{IC3} software model checking}.
\newblock \bibinfo{journal}{\emph{Int. J. Softw. Tools Technol. Transf.}}
  \bibinfo{volume}{22}, \bibinfo{number}{2} (\bibinfo{year}{2020}),
  \bibinfo{pages}{135--161}.
\newblock
\href{https://doi.org/10.1007/S10009-019-00547-X}{doi:\nolinkurl{10.1007/S10009-019-00547-X}}


\bibitem[Lattner and Adve(2004)]%
        {LLVM-CGO04}
\bibfield{author}{\bibinfo{person}{C. Lattner} {and} \bibinfo{person}{V.~S.
  Adve}.} \bibinfo{year}{2004}\natexlab{}.
\newblock \showarticletitle{\textsc{Llvm}: {A} Compilation Framework for
  Lifelong Program Analysis and Transformation}. In
  \bibinfo{booktitle}{\emph{Proc.\ CGO}}. \bibinfo{publisher}{IEEE},
  \bibinfo{pages}{75--88}.
\newblock
\href{https://doi.org/10.1109/CGO.2004.1281665}{doi:\nolinkurl{10.1109/CGO.2004.1281665}}


\bibitem[Leike and Heizmann(2015)]%
        {RankingTemplates}
\bibfield{author}{\bibinfo{person}{J. Leike} {and} \bibinfo{person}{M.
  Heizmann}.} \bibinfo{year}{2015}\natexlab{}.
\newblock \showarticletitle{Ranking Templates for Linear Loops}.
\newblock \bibinfo{journal}{\emph{Logical Methods in Computer Science}}
  \bibinfo{volume}{11}, \bibinfo{number}{1} (\bibinfo{year}{2015}).
\newblock
\href{https://doi.org/10.2168/LMCS-11(1:16)2015}{doi:\nolinkurl{10.2168/LMCS-11(1:16)2015}}


\bibitem[Li(2025)]%
        {ESBMC-SVCOMP26-archive}
\bibfield{author}{\bibinfo{person}{X. Li}.} \bibinfo{year}{2025}\natexlab{}.
\newblock \bibinfo{title}{ESBMC submission for the pre-run of {SV-COMP}'26}.
\newblock \bibinfo{howpublished}{Zenodo}.
\newblock
\href{https://doi.org/10.5281/zenodo.17741226}{doi:\nolinkurl{10.5281/zenodo.17741226}}


\bibitem[Long(2017)]%
        {Long17}
\bibfield{author}{\bibinfo{person}{J. Long}.} \bibinfo{year}{2017}\natexlab{}.
\newblock \emph{\bibinfo{title}{Reasoning about High-Level Constructs in
  Hardware/Software Formal Verification}}.
\newblock \bibinfo{thesistype}{Ph.\,D. Dissertation}.
  \bibinfo{school}{University of California, Berkeley}.
\newblock
\urldef\tempurl%
\url{http://www2.eecs.berkeley.edu/Pubs/TechRpts/2017/EECS-2017-150.html}
\showURL{%
\tempurl}


\bibitem[Lu et~al\mbox{.}(2025)]%
        {Btor2-Select}
\bibfield{author}{\bibinfo{person}{Z. Lu}, \bibinfo{person}{P.-C. Chien},
  \bibinfo{person}{N.-Z. Lee}, \bibinfo{person}{A. Gurfinkel}, {and}
  \bibinfo{person}{V. Ganesh}.} \bibinfo{year}{2025}\natexlab{}.
\newblock \showarticletitle{\textsc{Btor2-Select}: {Machine} Learning Based
  Algorithm Selection for Hardware Model Checking}. In
  \bibinfo{booktitle}{\emph{Proc.\ CAV}} \emph{(\bibinfo{series}{LNCS~15931})}.
  \bibinfo{publisher}{Springer}, \bibinfo{pages}{296--311}.
\newblock
\href{https://doi.org/10.1007/978-3-031-98668-0_15}{doi:\nolinkurl{10.1007/978-3-031-98668-0_15}}


\bibitem[McCarthy(1962)]%
        {McCarthy1962}
\bibfield{author}{\bibinfo{person}{J. McCarthy}.}
  \bibinfo{year}{1962}\natexlab{}.
\newblock \showarticletitle{Towards a Mathematical Science of Computation}. In
  \bibinfo{booktitle}{\emph{Information Processing, Proc.\ DFIP Congress}}.
  \bibinfo{publisher}{North-Holland}, \bibinfo{pages}{21--28}.
\newblock
\shownote{{Republished at}
  \href{https://doi.org/10.1007/978-94-011-1793-7_2}{doi:\nolinkurl{10.1007/978-94-011-1793-7_2}}}.
\newblock


\bibitem[McConnell et~al\mbox{.}(2011)]%
        {CertifyingAlgorithms}
\bibfield{author}{\bibinfo{person}{R.~M. McConnell}, \bibinfo{person}{K.
  Mehlhorn}, \bibinfo{person}{S. N{\"{a}}her}, {and} \bibinfo{person}{P.
  Schweitzer}.} \bibinfo{year}{2011}\natexlab{}.
\newblock \showarticletitle{Certifying Algorithms}.
\newblock \bibinfo{journal}{\emph{Computer Science Review}}
  \bibinfo{volume}{5}, \bibinfo{number}{2} (\bibinfo{year}{2011}),
  \bibinfo{pages}{119--161}.
\newblock
\href{https://doi.org/10.1016/j.cosrev.2010.09.009}{doi:\nolinkurl{10.1016/j.cosrev.2010.09.009}}


\bibitem[McMillan(1993)]%
        {SMV:McMillan1993}
\bibfield{author}{\bibinfo{person}{K.~L. McMillan}.}
  \bibinfo{year}{1993}\natexlab{}.
\newblock \showarticletitle{The {SMV} System}.
\newblock In \bibinfo{booktitle}{\emph{Symbolic Model Checking}}.
  \bibinfo{publisher}{Springer}, \bibinfo{pages}{61--85}.
\newblock
\href{https://doi.org/10.1007/978-1-4615-3190-6_4}{doi:\nolinkurl{10.1007/978-1-4615-3190-6_4}}


\bibitem[McMillan(2003)]%
        {McMillanCraig}
\bibfield{author}{\bibinfo{person}{K.~L. McMillan}.}
  \bibinfo{year}{2003}\natexlab{}.
\newblock \showarticletitle{Interpolation and {SAT}-Based Model Checking}. In
  \bibinfo{booktitle}{\emph{Proc.\ CAV}} \emph{(\bibinfo{series}{LNCS~2725})}.
  \bibinfo{publisher}{Springer}, \bibinfo{pages}{1--13}.
\newblock
\href{https://doi.org/10.1007/978-3-540-45069-6_1}{doi:\nolinkurl{10.1007/978-3-540-45069-6_1}}


\bibitem[McMillan(2006)]%
        {IMPACT}
\bibfield{author}{\bibinfo{person}{K.~L. McMillan}.}
  \bibinfo{year}{2006}\natexlab{}.
\newblock \showarticletitle{Lazy Abstraction with Interpolants}. In
  \bibinfo{booktitle}{\emph{Proc.\ CAV}} \emph{(\bibinfo{series}{LNCS~4144})}.
  \bibinfo{publisher}{Springer}, \bibinfo{pages}{123--136}.
\newblock
\href{https://doi.org/10.1007/11817963_14}{doi:\nolinkurl{10.1007/11817963_14}}


\bibitem[Mony et~al\mbox{.}(2009)]%
        {MonyDATE2009}
\bibfield{author}{\bibinfo{person}{H. Mony}, \bibinfo{person}{J. Baumgartner},
  \bibinfo{person}{A. Mishchenko}, {and} \bibinfo{person}{R. Brayton}.}
  \bibinfo{year}{2009}\natexlab{}.
\newblock \showarticletitle{Speculative reduction-based scalable redundancy
  identification}. In \bibinfo{booktitle}{\emph{Proc.\ DATE}}.
  \bibinfo{publisher}{IEEE}, \bibinfo{pages}{1674--1679}.
\newblock
\href{https://doi.org/10.1109/DATE.2009.5090932}{doi:\nolinkurl{10.1109/DATE.2009.5090932}}


\bibitem[Mukherjee et~al\mbox{.}(2016a)]%
        {MukherjeeSKM16}
\bibfield{author}{\bibinfo{person}{R. Mukherjee}, \bibinfo{person}{P.
  Schrammel}, \bibinfo{person}{D. Kröning}, {and} \bibinfo{person}{T.
  Melham}.} \bibinfo{year}{2016}\natexlab{a}.
\newblock \showarticletitle{Unbounded safety verification for hardware using
  software analyzers}. In \bibinfo{booktitle}{\emph{Proc.\ DATE}}.
  \bibinfo{publisher}{IEEE}, \bibinfo{pages}{1152--1155}.
\newblock
\shownote{\url{https://ieeexplore.ieee.org/document/7459484}}.
\newblock


\bibitem[Mukherjee et~al\mbox{.}(2016b)]%
        {v2c}
\bibfield{author}{\bibinfo{person}{R. Mukherjee}, \bibinfo{person}{M.
  Tautschnig}, {and} \bibinfo{person}{D. Kröning}.}
  \bibinfo{year}{2016}\natexlab{b}.
\newblock \showarticletitle{v2c: {A} {Verilog} to {C} Translator}. In
  \bibinfo{booktitle}{\emph{Proc.\ TACAS}}
  \emph{(\bibinfo{series}{LNCS~9636})}. \bibinfo{publisher}{Springer},
  \bibinfo{pages}{580--586}.
\newblock
\href{https://doi.org/10.1007/978-3-662-49674-9_38}{doi:\nolinkurl{10.1007/978-3-662-49674-9_38}}


\bibitem[Müller et~al\mbox{.}(2016)]%
        {Viper}
\bibfield{author}{\bibinfo{person}{P. Müller}, \bibinfo{person}{M.
  Schwerhoff}, {and} \bibinfo{person}{A.~J. Summers}.}
  \bibinfo{year}{2016}\natexlab{}.
\newblock \showarticletitle{Viper: {A} Verification Infrastructure for
  Permission-Based Reasoning}. In \bibinfo{booktitle}{\emph{Proc.\ VMCAI}}
  \emph{(\bibinfo{series}{LNCS~9583})}. \bibinfo{publisher}{Springer},
  \bibinfo{pages}{41--62}.
\newblock
\href{https://doi.org/10.1007/978-3-662-49122-5_2}{doi:\nolinkurl{10.1007/978-3-662-49122-5_2}}


\bibitem[Niemetz and Preiner(2023)]%
        {Bitwuzla-CAV23}
\bibfield{author}{\bibinfo{person}{A. Niemetz} {and} \bibinfo{person}{M.
  Preiner}.} \bibinfo{year}{2023}\natexlab{}.
\newblock \showarticletitle{Bitwuzla}. In \bibinfo{booktitle}{\emph{Proc.\
  CAV~(2)}} \emph{(\bibinfo{series}{LNCS~13965})}.
  \bibinfo{publisher}{Springer}, \bibinfo{pages}{3--17}.
\newblock
\href{https://doi.org/10.1007/978-3-031-37703-7\_1}{doi:\nolinkurl{10.1007/978-3-031-37703-7\_1}}


\bibitem[Niemetz et~al\mbox{.}({[n.\,d.]})]%
        {btor2tools-website}
\bibfield{author}{\bibinfo{person}{A. Niemetz}, \bibinfo{person}{M. Preiner},
  \bibinfo{person}{C. Wolf}, {and} \bibinfo{person}{A. Biere}.}
  \bibinfo{year}{[n.\,d.]}\natexlab{}.
\newblock \bibinfo{title}{A generic parser and tool package for the
  \textsc{Btor2} format}.
\newblock \bibinfo{howpublished}{\url{https://github.com/hwmcc/btor2tools}}.
\newblock
\shownote{Accessed: 2026-02-14}.
\newblock


\bibitem[Niemetz et~al\mbox{.}(2018)]%
        {Boolector3}
\bibfield{author}{\bibinfo{person}{A. Niemetz}, \bibinfo{person}{M. Preiner},
  \bibinfo{person}{C. Wolf}, {and} \bibinfo{person}{A. Biere}.}
  \bibinfo{year}{2018}\natexlab{}.
\newblock \showarticletitle{\textsc{Btor2}, \textsc{BtorMC}, and
  \textsc{Boolector}\ 3.0}. In \bibinfo{booktitle}{\emph{Proc.\ {CAV}}}
  \emph{(\bibinfo{series}{LNCS~10981})}. \bibinfo{publisher}{Springer},
  \bibinfo{pages}{587--595}.
\newblock
\href{https://doi.org/10.1007/978-3-319-96145-3_32}{doi:\nolinkurl{10.1007/978-3-319-96145-3_32}}


\bibitem[Noureddine and Zaraket(2016)]%
        {Noureddine16}
\bibfield{author}{\bibinfo{person}{M.~A. Noureddine} {and}
  \bibinfo{person}{F.~A. Zaraket}.} \bibinfo{year}{2016}\natexlab{}.
\newblock \showarticletitle{Model Checking Software with First Order Logic
  Specifications Using {AIG} Solvers}.
\newblock \bibinfo{journal}{\emph{IEEE Trans.\ Softw.\ Eng.}}
  \bibinfo{volume}{42}, \bibinfo{number}{8} (\bibinfo{year}{2016}),
  \bibinfo{pages}{741--763}.
\newblock
\href{https://doi.org/10.1109/TSE.2016.2520468}{doi:\nolinkurl{10.1109/TSE.2016.2520468}}


\bibitem[Ozdemir et~al\mbox{.}(2022)]%
        {CirC}
\bibfield{author}{\bibinfo{person}{A. Ozdemir}, \bibinfo{person}{F. Brown},
  {and} \bibinfo{person}{R.~S. Wahby}.} \bibinfo{year}{2022}\natexlab{}.
\newblock \showarticletitle{{CirC}: {Compiler} infrastructure for proof
  systems, software verification, and more}. In
  \bibinfo{booktitle}{\emph{Proc.\ SP}}. \bibinfo{publisher}{{IEEE}},
  \bibinfo{pages}{2248--2266}.
\newblock
\href{https://doi.org/10.1109/SP46214.2022.9833782}{doi:\nolinkurl{10.1109/SP46214.2022.9833782}}


\bibitem[Perez-Lopez et~al\mbox{.}(2026)]%
        {Pono2}
\bibfield{author}{\bibinfo{person}{Á.~R. Perez-Lopez}, \bibinfo{person}{P.-C.
  Chien}, \bibinfo{person}{F. Lonsing}, \bibinfo{person}{S. Archer},
  \bibinfo{person}{A. Irfan}, {and} \bibinfo{person}{C. Barrett}.}
  \bibinfo{year}{2026}\natexlab{}.
\newblock \showarticletitle{{Pono} 2.0: {A} Versatile {SMT}-Based Model Checker
  for Safety and Liveness (Long Tool Paper)}. In
  \bibinfo{booktitle}{\emph{Proc.\ FM}} \emph{(\bibinfo{series}{LNCS~16557})}.
  \bibinfo{publisher}{Springer}, \bibinfo{pages}{3--26}.
\newblock
\href{https://doi.org/10.1007/978-3-032-26220-2_1}{doi:\nolinkurl{10.1007/978-3-032-26220-2_1}}


\bibitem[Piterman and Pnueli(2018)]%
        {HBMC-TemporalLogic}
\bibfield{author}{\bibinfo{person}{N. Piterman} {and} \bibinfo{person}{A.
  Pnueli}.} \bibinfo{year}{2018}\natexlab{}.
\newblock \showarticletitle{Temporal Logic and Fair Discrete Systems}.
\newblock In \bibinfo{booktitle}{\emph{Handbook of Model Checking}}.
  \bibinfo{publisher}{Springer}, \bibinfo{pages}{27--73}.
\newblock
\href{https://doi.org/10.1007/978-3-319-10575-8_2}{doi:\nolinkurl{10.1007/978-3-319-10575-8_2}}


\bibitem[Podelski and Rybalchenko(2004a)]%
        {RankingFunctions}
\bibfield{author}{\bibinfo{person}{A. Podelski} {and} \bibinfo{person}{A.
  Rybalchenko}.} \bibinfo{year}{2004}\natexlab{a}.
\newblock \showarticletitle{A Complete Method for the Synthesis of Linear
  Ranking Functions}. In \bibinfo{booktitle}{\emph{Proc.\ VMCAI}}
  \emph{(\bibinfo{series}{LNCS~2937})}. \bibinfo{publisher}{Springer},
  \bibinfo{pages}{239--251}.
\newblock
\href{https://doi.org/10.1007/978-3-540-24622-0_20}{doi:\nolinkurl{10.1007/978-3-540-24622-0_20}}


\bibitem[Podelski and Rybalchenko(2004b)]%
        {TransitionInvariants}
\bibfield{author}{\bibinfo{person}{A. Podelski} {and} \bibinfo{person}{A.
  Rybalchenko}.} \bibinfo{year}{2004}\natexlab{b}.
\newblock \showarticletitle{Transition Invariants}. In
  \bibinfo{booktitle}{\emph{Proc.\ LICS}}. \bibinfo{publisher}{IEEE},
  \bibinfo{pages}{32--41}.
\newblock
\href{https://doi.org/10.1109/LICS.2004.1319598}{doi:\nolinkurl{10.1109/LICS.2004.1319598}}


\bibitem[Schrögendorfer(2021)]%
        {BMCLockLessProgram}
\bibfield{author}{\bibinfo{person}{F. Schrögendorfer}.}
  \bibinfo{year}{2021}\natexlab{}.
\newblock \emph{\bibinfo{title}{Bounded Model Checking of Lockless Programs}}.
\newblock \bibinfo{thesistype}{Master's\ thesis}. \bibinfo{school}{Johannes
  Kepler University Linz}.
\newblock
\urldef\tempurl%
\url{https://resolver.obvsg.at/urn:nbn:at:at-ubl:1-45203}
\showURL{%
\tempurl}


\bibitem[Sheeran et~al\mbox{.}(2000)]%
        {InductionVerification}
\bibfield{author}{\bibinfo{person}{M. Sheeran}, \bibinfo{person}{S. Singh},
  {and} \bibinfo{person}{G. Stålmarck}.} \bibinfo{year}{2000}\natexlab{}.
\newblock \showarticletitle{Checking Safety Properties Using Induction and a
  {SAT}-Solver}.
\newblock In \bibinfo{booktitle}{\emph{Proc.\ FMCAD}}.
  \bibinfo{publisher}{Springer}, \bibinfo{pages}{127--144}.
\newblock
\href{https://doi.org/10.1007/3-540-40922-X_8}{doi:\nolinkurl{10.1007/3-540-40922-X_8}}


\bibitem[Slabý et~al\mbox{.}(2012)]%
        {SymbioticApproach}
\bibfield{author}{\bibinfo{person}{J. Slabý}, \bibinfo{person}{J. Strejček},
  {and} \bibinfo{person}{M. Trtík}.} \bibinfo{year}{2012}\natexlab{}.
\newblock \showarticletitle{Checking Properties Described by State Machines:
  {On} Synergy of Instrumentation, Slicing, and Symbolic Execution}. In
  \bibinfo{booktitle}{\emph{Proc.\ FMICS}}
  \emph{(\bibinfo{series}{LNCS~7437})}. \bibinfo{publisher}{Springer},
  \bibinfo{pages}{207--221}.
\newblock
\href{https://doi.org/10.1007/978-3-642-32469-7_14}{doi:\nolinkurl{10.1007/978-3-642-32469-7_14}}


\bibitem[Society(2024)]%
        {SystemVerilog-2024}
\bibfield{author}{\bibinfo{person}{IEEE~Computer Society}.}
  \bibinfo{year}{2024}\natexlab{}.
\newblock \showarticletitle{IEEE Standard for SystemVerilog--Unified Hardware
  Design, Specification, and Verification Language}.
\newblock \bibinfo{journal}{\emph{IEEE Std 1800-2023 (Revision of IEEE Std
  1800-2017)}} (\bibinfo{year}{2024}), \bibinfo{pages}{1--1354}.
\newblock
\href{https://doi.org/10.1109/IEEESTD.2024.10458102}{doi:\nolinkurl{10.1109/IEEESTD.2024.10458102}}


\bibitem[Su et~al\mbox{.}(2025a)]%
        {rIC3}
\bibfield{author}{\bibinfo{person}{Y. Su}, \bibinfo{person}{Q. Yang},
  \bibinfo{person}{Y. Ci}, \bibinfo{person}{T. Bu}, {and} \bibinfo{person}{Z.
  Huang}.} \bibinfo{year}{2025}\natexlab{a}.
\newblock \showarticletitle{The {rIC3} Hardware Model Checker}. In
  \bibinfo{booktitle}{\emph{Proc.\ CAV~(1)}}
  \emph{(\bibinfo{series}{LNCS~15931})}. \bibinfo{publisher}{Springer},
  \bibinfo{pages}{185--199}.
\newblock
\href{https://doi.org/10.1007/978-3-031-98668-0\_9}{doi:\nolinkurl{10.1007/978-3-031-98668-0\_9}}


\bibitem[Su et~al\mbox{.}(2025b)]%
        {GipSAT-CAV25}
\bibfield{author}{\bibinfo{person}{Y. Su}, \bibinfo{person}{Q. Yang},
  \bibinfo{person}{Y. Ci}, \bibinfo{person}{Y. Li}, \bibinfo{person}{T. Bu},
  {and} \bibinfo{person}{Z. Huang}.} \bibinfo{year}{2025}\natexlab{b}.
\newblock \showarticletitle{Deeply Optimizing the {SAT} Solver for the {IC3}
  Algorithm}. In \bibinfo{booktitle}{\emph{Proc.\ CAV~(1)}}
  \emph{(\bibinfo{series}{LNCS~15931})}. \bibinfo{publisher}{Springer},
  \bibinfo{pages}{237--257}.
\newblock
\href{https://doi.org/10.1007/978-3-031-98668-0\_12}{doi:\nolinkurl{10.1007/978-3-031-98668-0\_12}}


\bibitem[Tafese et~al\mbox{.}(2023)]%
        {Btor2MLIR}
\bibfield{author}{\bibinfo{person}{J. Tafese}, \bibinfo{person}{I.
  Garcia-Contreras}, {and} \bibinfo{person}{A. Gurfinkel}.}
  \bibinfo{year}{2023}\natexlab{}.
\newblock \showarticletitle{\textsc{Btor2MLIR}: {A} Format and Toolchain for
  Hardware Verification}. In \bibinfo{booktitle}{\emph{Proc.\ FMCAD}}.
  \bibinfo{publisher}{TU Wien Academic Press}, \bibinfo{pages}{55--63}.
\newblock
\href{https://doi.org/10.34727/2023/ISBN.978-3-85448-060-0_13}{doi:\nolinkurl{10.34727/2023/ISBN.978-3-85448-060-0_13}}


\bibitem[Vizel and Grumberg(2009)]%
        {VizelFMCAD09}
\bibfield{author}{\bibinfo{person}{Y. Vizel} {and} \bibinfo{person}{O.
  Grumberg}.} \bibinfo{year}{2009}\natexlab{}.
\newblock \showarticletitle{Interpolation-Sequence Based Model Checking}. In
  \bibinfo{booktitle}{\emph{Proc.\ FMCAD}}. \bibinfo{publisher}{{IEEE}},
  \bibinfo{pages}{1--8}.
\newblock
\href{https://doi.org/10.1109/FMCAD.2009.5351148}{doi:\nolinkurl{10.1109/FMCAD.2009.5351148}}


\bibitem[Vizel et~al\mbox{.}(2013)]%
        {ForwardBackwardReachability}
\bibfield{author}{\bibinfo{person}{Y. Vizel}, \bibinfo{person}{O. Grumberg},
  {and} \bibinfo{person}{S. Shoham}.} \bibinfo{year}{2013}\natexlab{}.
\newblock \showarticletitle{Intertwined Forward-Backward Reachability Analysis
  Using Interpolants}. In \bibinfo{booktitle}{\emph{Proc.\ TACAS}}
  \emph{(\bibinfo{series}{LNCS~7795})}. \bibinfo{publisher}{Springer},
  \bibinfo{pages}{308--323}.
\newblock
\href{https://doi.org/10.1007/978-3-642-36742-7_22}{doi:\nolinkurl{10.1007/978-3-642-36742-7_22}}


\bibitem[Wang et~al\mbox{.}(2016)]%
        {WangTLYJ16}
\bibfield{author}{\bibinfo{person}{H.-E. Wang}, \bibinfo{person}{T.-L. Tsai},
  \bibinfo{person}{C.-H. Lin}, \bibinfo{person}{F. Yu}, {and}
  \bibinfo{person}{J.-H.~R. Jiang}.} \bibinfo{year}{2016}\natexlab{}.
\newblock \showarticletitle{String Analysis via Automata Manipulation with
  Logic Circuit Representation}. In \bibinfo{booktitle}{\emph{Proc.\ CAV~(1)}}
  \emph{(\bibinfo{series}{LNCS~9779})}. \bibinfo{publisher}{Springer},
  \bibinfo{pages}{241--260}.
\newblock
\href{https://doi.org/10.1007/978-3-319-41528-4\_13}{doi:\nolinkurl{10.1007/978-3-319-41528-4\_13}}


\bibitem[Wu et~al\mbox{.}(2025)]%
        {ESBMC-SVCOMP25}
\bibfield{author}{\bibinfo{person}{T. Wu}, \bibinfo{person}{X. Li},
  \bibinfo{person}{E. Manino}, \bibinfo{person}{R. Menezes},
  \bibinfo{person}{M. Gadelha}, \bibinfo{person}{S. Xiong}, \bibinfo{person}{N.
  Tihanyi}, \bibinfo{person}{P. Petoumenos}, {and} \bibinfo{person}{L.
  Cordeiro}.} \bibinfo{year}{2025}\natexlab{}.
\newblock \showarticletitle{\textsc{ESBMC v7.7}: {Efficient} Concurrent
  Software Verification with Scheduling, Incremental {SMT} and Partial Order
  Reduction (Competition Contribution)}. In \bibinfo{booktitle}{\emph{Proc.\
  TACAS~(3)}} \emph{(\bibinfo{series}{LNCS~15698})}.
  \bibinfo{publisher}{Springer}, \bibinfo{pages}{223--228}.
\newblock
\href{https://doi.org/10.1007/978-3-031-90660-2_16}{doi:\nolinkurl{10.1007/978-3-031-90660-2_16}}


\bibitem[Xia et~al\mbox{.}(2024)]%
        {rlive}
\bibfield{author}{\bibinfo{person}{Y. Xia}, \bibinfo{person}{A. Cimatti},
  \bibinfo{person}{A. Griggio}, {and} \bibinfo{person}{J. Li}.}
  \bibinfo{year}{2024}\natexlab{}.
\newblock \showarticletitle{Avoiding the Shoals - {A} New Approach to Liveness
  Checking}. In \bibinfo{booktitle}{\emph{Proc.\ CAV~(1)}}
  \emph{(\bibinfo{series}{LNCS~14681})}. \bibinfo{publisher}{Springer},
  \bibinfo{pages}{234--254}.
\newblock
\href{https://doi.org/10.1007/978-3-031-65627-9_12}{doi:\nolinkurl{10.1007/978-3-031-65627-9_12}}


\bibitem[{Ádám} et~al\mbox{.}(2025)]%
        {WitnessesNonTermination}
\bibfield{author}{\bibinfo{person}{{Zs.} {Ádám}}, \bibinfo{person}{P.
  Ayaziová}, \bibinfo{person}{L. Bajczi}, \bibinfo{person}{D. Beyer},
  \bibinfo{person}{M. Jankola}, \bibinfo{person}{M. Lingsch-Rosenfeld}, {and}
  \bibinfo{person}{J. Strejček}.} \bibinfo{year}{2025}\natexlab{}.
\newblock \showarticletitle{Non-termination Witnesses and Their Validation}. In
  \bibinfo{booktitle}{\emph{Proc.\ ASE~2025}}. \bibinfo{publisher}{IEEE},
  \bibinfo{pages}{1969--1981}.
\newblock
\href{https://doi.org/10.1109/ASE63991.2025.00164}{doi:\nolinkurl{10.1109/ASE63991.2025.00164}}


\end{thebibliography}
